%% file: pdr_arXiv.tex
\documentclass[11pt,a4paper,twoside]{book}
\usepackage[english]{babel}
\usepackage{mathpazo}
\usepackage{a4wide}
\usepackage[textsize=small]{todonotes}
\usepackage{subfigure}
\usepackage{fancyhdr}
\usepackage{sidecap}
\usepackage{graphicx}\graphicspath{{figs/}}
\usepackage[tableposition=top,font=small]{caption}
\usepackage{upgreek}
\usepackage{xspace}
\usepackage{amsmath,amssymb,bm}
\usepackage{booktabs}
\usepackage{eurosym}
\usepackage{lscape}
\usepackage{cite}
\usepackage{paralist}
\usepackage{multirow}
\usepackage{units}
\usepackage{lipsum}
\usepackage{colortbl}
\usepackage[switch,modulo]{lineno}
\usepackage[ansinew]{inputenc}
\usepackage[today,nofancy]{svninfo}
\usepackage{xcolor}
\usepackage{floatflt}
\definecolor{darkred}{rgb}{0.5,0,0}
\definecolor{darkblue}{rgb}{0,0,0.5}
\definecolor{firebrick}{rgb}{0.75,0.125,0.125}
\definecolor{darkgreen}{rgb}{0,0.5,0}
\usepackage[colorlinks=true,linkcolor=black,citecolor=darkgreen,urlcolor=darkblue]{hyperref}

\usepackage{varwidth}
\usepackage{verbatim}
\usepackage{footnote}
\usepackage{textcomp}

\renewcommand{\chaptermark}[1]%
                 {\ifnum\value{chapter}>0
           \markboth{\chaptername \thechapter:~ #1}{}
             \else
             \markboth{#1}{}%
           \fi
}
\renewcommand{\sectionmark}[1]%
                 {\markright{\thesection\ #1}{}}

\def\xmax{\ensuremath{X_\text{max}}\xspace}
\def\xmumax{\ensuremath{X^\mu_\text{max}}\xspace}
\def\nmu{\ensuremath{N_\mu}\xspace}

\def\gcm{\ensuremath{\unit{g/cm^2}}\xspace}

\def\Offline{\mbox{$\overline{\textrm%
{Off}}$\hspace{.05em}\protect\raisebox{.4ex}%
{$\protect\underline{\textrm{line}}$}}\xspace}

\begin{document}

\pagenumbering{roman}
\include{title}
\cleardoublepage

\pagestyle{plain}
\thispagestyle{plain}
\include{authors}

\include{foreword}

\include{executive_summary}

\tableofcontents
\cleardoublepage

\setcounter{page}{1}\pagenumbering{arabic}
\pagestyle{headings}
\thispagestyle{headings}

\include{introduction}

\include{scientific_achievements_and_goals}

\include{expected_physics_performance}
\include{surface_detector}
\include{underground_muon_detector}

\include{fluorescence_detector}

\include{comms_and_daq}
\include{dpa_offline}
\include{assembly_tests_maintenance}

\include{organization_and_management}

\include{cost_schedule_funding}

\include{outreach_and_education}

\include{acknowledgements}

\appendix

\include{appendix-wbs_schedule}

\include{appendix-current_state}

\bibliographystyle{utcaps}

{\sloppy
\bibliography{references}
}

\end{document}

%% file: title.tex
\thispagestyle{empty}

\begin{titlepage}

$~$
\begin{center}
\vspace{10mm}
\LARGE
{\huge\bf The Pierre Auger Observatory Upgrade}
\\
\vspace{7mm}
{\huge\bf ``AugerPrime''}
\\
\vspace{10mm}
\textbf{Preliminary Design Report}
\\
\vspace{20mm}
\includegraphics[width=0.5\textwidth]{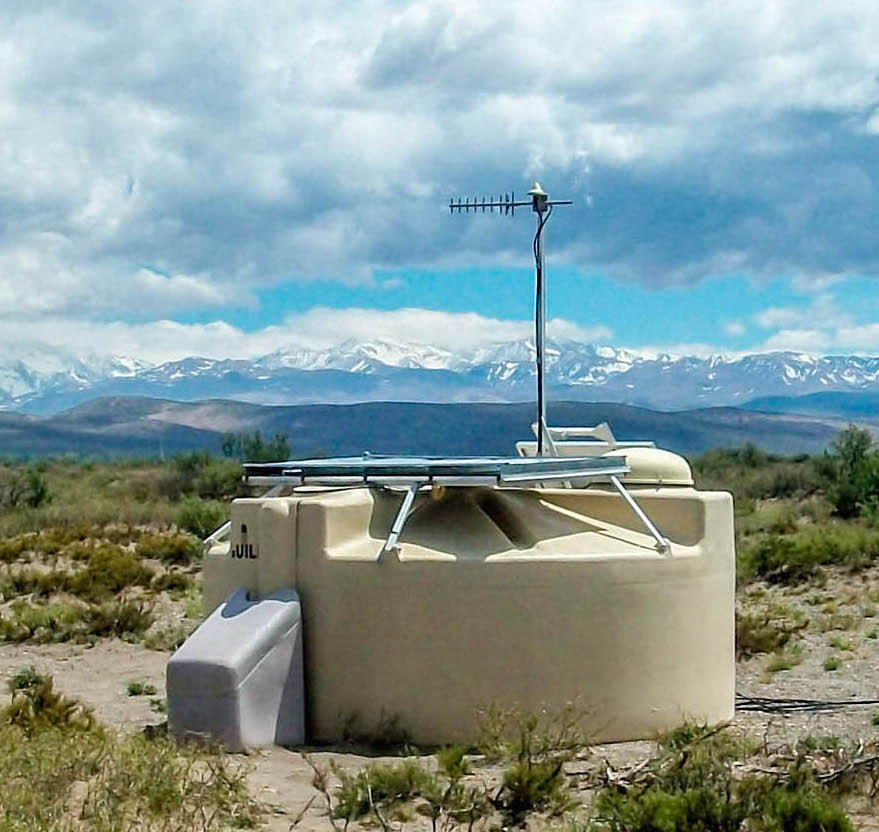}

\large
\vspace{20mm}
The Pierre Auger Collaboration\\
April, 2015 
\\
\vspace{15mm}
\begin{center}
\begin{tabular}{rll}
\multirow{4}{*}{~~~~~\includegraphics[width=13mm]{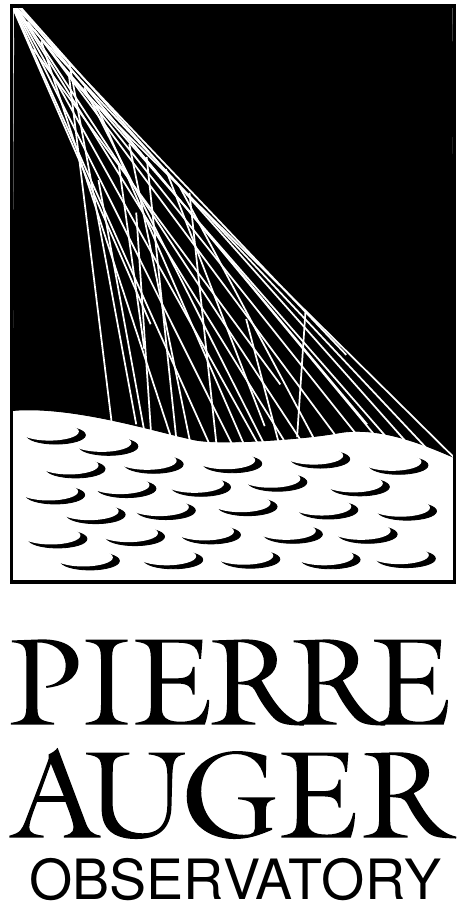}}
\\
              & Observatorio Pierre Auger,
\\
              & Av.\ San Mart\'in Norte 304,
\\
              & 5613 Malarg\"ue, Argentina
\end{tabular}
\end{center}
\end{center}

\end{titlepage}

%% file: authors.tex
\par\noindent
{\bf The Pierre Auger Collaboration, April 2015} \\
A.~Aab$^{41}$, 
P.~Abreu$^{65}$, 
M.~Aglietta$^{52}$, 
E.J.~Ahn$^{82}$, 
I.~Al Samarai$^{28}$, 
I.F.M.~Albuquerque$^{16}$, 
I.~Allekotte$^{1}$, 
P.~Allison$^{87}$, 
A.~Almela$^{11,\: 8}$, 
J.~Alvarez Castillo$^{58}$
% ...

\par\noindent
$^{1}$ Centro At\'{o}mico Bariloche and Instituto Balseiro (CNEA-UNCuyo-CONICET), San 
Carlos de Bariloche, 
Argentina \\
% ...
$^{8}$ Instituto de Tecnolog\'{\i}as en Detecci\'{o}n y Astropart\'{\i}culas (CNEA, CONICET, UNSAM), 
Buenos Aires, 
Argentina \\
% ...
$^{11}$ Universidad Tecnol\'{o}gica Nacional - Facultad Regional Buenos Aires, Buenos Aires,
Argentina \\
% ...
$^{16}$ Universidade de S\~{a}o Paulo, Instituto de F\'{\i}sica, S\~{a}o Paulo, SP, 
Brazil \\
% ...
$^{28}$ Institut de Physique Nucl\'{e}aire d'Orsay (IPNO), Universit\'{e} Paris 11, CNRS-IN2P3, 
Orsay, 
France \\
% ...
$^{41}$ Universit\"{a}t Siegen, Siegen, 
Germany \\
% ...
$^{52}$ Osservatorio Astrofisico di Torino  (INAF), Universit\`{a} di Torino and Sezione INFN, 
Torino, 
Italy \\
% ...
$^{58}$ Universidad Nacional Aut\'{o}noma de M\'{e}xico, M\'{e}xico, D.F., 
M\'{e}xico \\
% ...
$^{65}$ Laborat\'{o}rio de Instrumenta\c{c}\~{a}o e F\'{\i}sica Experimental de Part\'{\i}culas - LIP and  
Instituto Superior T\'{e}cnico - IST, Universidade de Lisboa - UL, 
Portugal \\
% ...
$^{82}$ Fermilab, Batavia, IL, 
USA \\
% ...
$^{87}$ Ohio State University, Columbus, OH, 
USA \\
% ...
\par\noindent
(\ddag) Deceased \\
(a) Also at Vrije Universiteit Brussels, Belgium \\

%% file: foreword.tex
\chapter*{Foreword}

The Pierre Auger Observatory has begun a major Upgrade of its already
impressive capabilities, with an emphasis on improved mass composition
determination using the surface detectors of the Observatory.  Known
as AugerPrime, the upgrade will include new 4\,m$^2$ plastic
scintillator detectors on top of all 1660 water-Cherenkov detectors,
updated and more flexible surface detector electronics, a large array
of buried muon detectors, and an extended duty cycle for operations of
the fluorescence detectors.

This Preliminary Design Report was produced by the Collaboration in
April 2015 as an internal document and information for funding
agencies.  It outlines the scientific and technical case for
AugerPrime\footnote{As a result of continuing R\&D, slight changes
  have been implemented in the baseline design since this Report was
  written.  These changes will be documented in a forthcoming
  Technical Design Report.}.  We now release it to the public via the
arXiv server.  We invite you to review the large number of fundamental
results already achieved by the Observatory and our plans for the
future.  \\

\vspace{1cm}
\noindent
The Pierre Auger Collaboration\\

%% file: executive_summary.tex

\chapter*{Executive Summary}

\section*{Present Results from the Pierre Auger Observatory}

\vspace{-3mm}
Measurements of the Auger Observatory have dramatically advanced our
understanding of ultra-high energy cosmic rays.  The suppression of the flux around
$5{\times}10^{19}$\,eV  is now confirmed without any doubt. Strong limits
have been placed on the photon and neutrino components of the flux indicating that  ``top-down'' source
processes, such as the decay of super-heavy particles, cannot account
for a significant part of the observed particle flux. 
A large-scale dipole anisotropy of ${\sim}7$\% amplitude has been
found for energies above $8{\times}10^{18}$\,eV. In addition there is
also an indication of the presence of a large scale anisotropy below the
ankle.  Particularly exciting is the observed behavior of the depth of
shower maximum with energy, which changes in an unexpected,
non-trivial way. Around $\unit[3{\times}10^{18}]{eV}$ it shows a
distinct change of slope with energy, and the shower-to-shower
variance decreases. Interpreted with the leading LHC-tuned shower
models, this implies a gradual shift to a heavier composition. A
number of fundamentally different astrophysical model scenarios have
been developed to describe this evolution. The high degree of isotropy
observed in numerous tests of the small-scale angular distribution of
UHECR above $\unit[4{\times}10^{19}]{eV}$ is remarkable, challenging original expectations
that assumed only a few cosmic ray sources with a light composition at
the highest energies. Interestingly, the largest departures from
isotropy are observed for cosmic rays with $E>\unit[5.8{\times}10^{19}]{eV}$ in ${\sim}20^\circ$
sky-windows.  Due to a duty cycle of ${\sim}15$\% of the fluorescence
telescopes, the data on the depth of shower maximum extend only up to
the flux suppression region, i.e.\ $\unit[4{\times}10^{19}]{eV}$.
Obtaining more information on the composition of cosmic rays at higher
energies will provide crucial means to discriminate between the model
classes and to understand the origin of the observed flux suppression.
Care must be taken, since precision Auger measurements of shower properties, strongly
constrained by the hybrid data, have revealed inconsistencies within present
shower models, opening the possibility that the unexpected behavior is
due to new hadronic interaction physics at energy scales beyond the reach of
the LHC.


\section*{Motivation for the upgrade}

\vspace{-3mm}
It is planned to operate the Pierre Auger Observatory until the end of 2024.
The motivation of the upgrade is to provide additional measurements to allow us
to address the following important questions:

\begin{itemize}
\item 
Elucidate the mass composition and 
the origin of the flux suppression at the highest energies,
i.e.\ the differentiation between
the energy loss effects due to propagation, and the maximum energy of
particles injected by astrophysical sources. 

\item 
Search for a flux contribution of protons up to the highest
energies. We aim to reach a sensitivity to a contribution as small as 10\% in the flux
suppression region. The measurement of the fraction of protons is the
decisive ingredient for estimating the physics potential of existing
and future cosmic ray, neutrino, and gamma-ray detectors; thus prospects
for proton astronomy with future detectors will be
clarified. Moreover, the flux of secondary gamma-rays and neutrinos
due to proton energy loss processes will be predicted.

\item 
Study extensive air showers and
hadronic multiparticle production. This will include the exploration
of fundamental particle physics at energies beyond those accessible at
man-made accelerators, and the derivation of constraints on new
physics phenomena, such as Lorentz invariance violation or extra
dimensions. 
\end{itemize}

With operation planned until 2024, event statistics will more than
double compared with the existing Auger data set, with the critical
added advantage that every event will now have mass information.  
Obtaining additional composition-sensitive
information will not only help to better reconstruct the properties of
the primary particles at the highest energies, but also improve the  measurements
in the important energy range just above the ankle.  Furthermore, measurements with the new detectors will help to reduce
systematic uncertainties related to modeling hadronic showers and to 
limitations of reconstruction algorithms.
This improved knowledge of air shower physics will likely then also allow a re-analysis of existing data -
for improved energy assignments, for mass composition studies, and for
photon and neutrino searches.  

The Auger upgrade promises high-quality
future data, and real scope for new physics uses of existing events.
Furthermore, the addition of scintillator detectors across the entire Observatory
will also make possible direct comparisons of Auger measurements with
those of the surface detectors of the Telescope Array experiment.
This will strengthen the already productive cooperation between the
two collaborations, which has an aim of understanding the highest
energy cosmic ray flux across the entire sky.


\section*{The configuration of the Auger upgrade}

\vspace{-3mm}
The proposed Auger upgrade consists of the following components:
\begin{itemize}

\item A complementary measurement of the shower particles will be provided by
a plastic scintillator plane above the existing Water-Cherenkov Detectors
(WCD). This allows the sampling of the shower particles 
with two detectors having different responses to muons and
electromagnetic particles.  The design of the Surface Scintillator
Detectors (SSD) is simple, reliable and they can be easily deployed over
the full 3000\,km$^2$ area of the Surface Detector.

\item The surface detector stations will be upgraded with new electronics that will process
both WCD and SSD signals. Use of the new electronics also aims to
increase the data quality (with faster sampling of ADC traces, better
timing accuracy, increased dynamic range), to enhance the local
trigger and processing capabilities (with a more powerful local station
processor and FPGA) and to improve calibration and monitoring
capabilities of the surface detector stations. The surface detector electronics upgrade
(SDEU) can be easily deployed, and will have only minimal impact on the
continuous data taking of the Surface Detector.

\item An Underground Muon Detector (UMD) is required in the existing
  SD infill area of 23.5\,km$^2$. The UMD will provide important
  direct measurements of the shower muon content and its time
  structure, while serving as verification and fine-tuning of the
  methods used to extract muon information with the SSD and WCD
  measurements. The performance and characteristics of the AMIGA
  underground muon detectors, now being deployed, match these
  requirements, and thus the completed AMIGA array will serve as the
  UMD.
 
\item In parallel with the Surface Detector upgrade, the operation mode of the
Fluorescence Detector (FD) will be changed to extend
measurements into periods with higher night sky background.
This will allow an increase of about 50\% in the current duty cycle of the FD.

\end{itemize}
The Auger upgrade will not affect the continuous data taking and
maintenance of the existing detectors. The current communication system and
solar power system for the Surface Detector will remain
unchanged. Only minor software changes are required for the central
data acquisition system (CDAS) and for the monitoring system.


\section*{Organization, cost, and schedule}

\vspace{-3mm}
The Auger Project Management Plan~\cite{PMP} has provided the basis
for the detailed organization and management of both the construction and
the operation of the Auger Observatory. It has been updated to describe
the organizational and management features of the
upgrade. In particular, the MOUs with each institution participating
in the Auger upgrade will include the commitment of institutional
collaborators to the upgrade effort, their deliverables, and delivery
schedule.

The Pierre Auger Project has a Quality Assurance Plan~\cite{QAP} to
ensure the performance and reliability of the Observatory systems.
This plan will be followed to accommodate the specific quality
assurance/quality control requirements of the upgrade. The risk management
structure will include processes for risk management planning,
identification, analysis, monitoring and control.

The cost and schedule estimates for the Auger upgrade are based on the
upgrade work breakdown structure (WBS). The total cost with contingency, including
infrastructure costs, is currently estimated to be US\$\,15.2\,M.
The total cost with contingency, but without infrastructure costs (the so-called
European accounting) is estimated to be US\$\,12.7\,M. AMIGA muon counters will be used for the UMD and the estimated cost for the completion of the AMIGA counters is US\$\,1.6\,M without infrastructure costs. The overall
increase in operating costs as a result of the upgrade is expected to
be less than five percent.

The components of the baseline design have been tested for their
suitability for the upgrade.  Some specific R\&D is still in progress
to optimize the quality/cost ratio, in particular for the scintillator
detectors and the underground muon detectors. The production of the
detectors and electronics boards will be done in parallel at three or
four production sites. The final validation of the SSD and SDEU
designs will be undertaken in an Engineering Array of 10 detector stations
at the end of 2015. The production and deployment of the SSD, the SDEU
and the UMD will be done in parallel and will extend over the period 2016-18. The
production schedule is mainly driven by the funding profiles in
the countries of the collaboration, not by the production or deployment rates.

%% file: introduction.tex

\chapter{Introduction}

The Pierre Auger Observatory, located on a vast, high plain near
Malarg\"ue in western Argentina, is the world's largest cosmic ray
observatory. The objectives of the Observatory are to probe the origin
and characteristics of cosmic rays above $10^{17}$eV and to study the
interactions of them, the most energetic particles observed in nature,
with the Earth's atmosphere.

\begin{figure}[!b]
\centering
\includegraphics[width=0.62\textwidth]{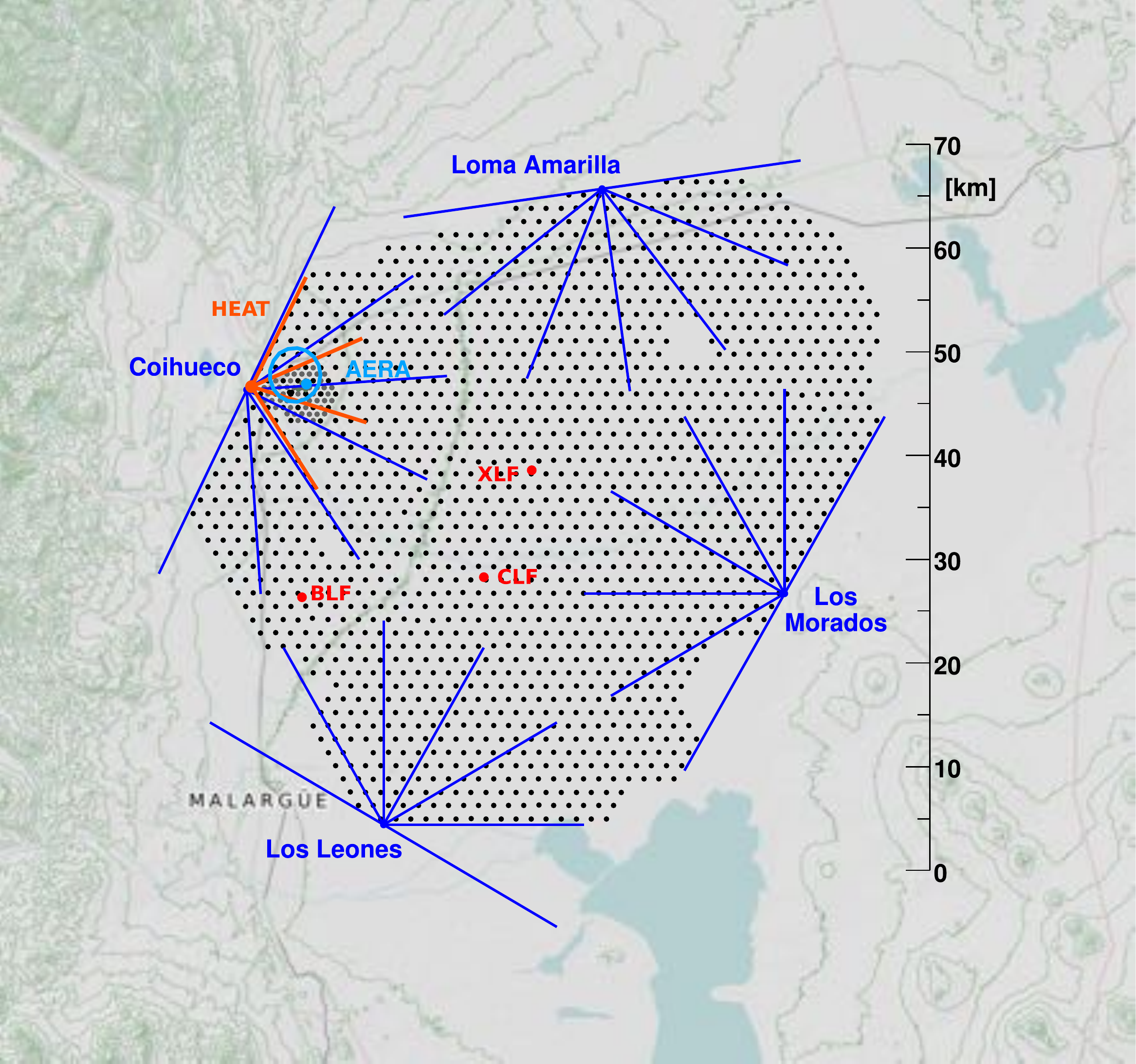}
\caption{The Auger Observatory layout.  Each dot corresponds to one of
  the 1660 surface detector stations.  The four fluorescence detector
  sites are shown, each with the field of view of its six telescopes.
  The Coihueco site hosts three extra high elevation (HEAT)
  telescopes.  Laser (CLF, XLF) and weather balloon launching (BLF)
  facilities are also shown.  The 750\,m array and the AERA
  radio array are located a few kilometers from Coihueco.}
\label{chap1_southern_site}
\end{figure}

Figure~\ref{chap1_southern_site} shows an overview of the
Observatory. It features an array of 1660 water-Cherenkov particle
detector stations spread over 3000\,km$^{2}$ over-looked by 24 air
fluorescence telescopes. In addition, three high elevation
fluorescence telescopes overlook a 23.5\,km$^2$, 61 detector
array with spacing of 750\,m (the Infill).

The genesis of the Observatory was a vision of Jim Cronin and Alan
Watson in 1991 to build an experiment with sufficient size and
precision to answer many of the long-standing questions in this field.
A collaboration was built which now amounts to over 500 scientists
from 16 countries.  Construction of the Observatory began in 2002
after a period of prototyping and a successful engineering array.
Construction was completed in 2008, and the integrated exposure of the
observatory now exceeds 40,000\,km$^2$\,sr\,yr at energies above
$3{\times}10^{18}$\,eV.  A thorough description of the Observatory is
provided in Appendix~\ref{chap:appendix_currentObservatory}.  This
description also includes recent enhancements, including high elevation
fluorescence telescopes (HEAT), the AMIGA underground muon detectors,
and detectors for radio (AERA) and microwave emissions from air showers.

The rich science outcomes from the Auger Observatory up to the present
time are described in Chapter~\ref{chap:science}.  This proposal to upgrade the Auger Observatory follows from a decade
of discovery and a recognition that shower-by-shower measurements of
cosmic ray mass-related properties are essential to further advance
the field.  The proposed solution is to infer the muon component of
the air showers by installing a complementary scintillator detector on
top of every water-Cherenkov detector across the array, coupled with
upgraded station electronics with enhanced capabilities.

The goals of the upgrade are described below in
Chapter~\ref{chap:science}, before the physics performance of the
proposed solution is detailed in Chapter~\ref{chap:performance}.  The
following chapters describe the design of the new scintillator
detectors and electronics, and the underground muon array that will be
used to verify the performance of the scintillator and water-Cherenkov
detector combination.  A description of plans to extend the duty cycle
of the fluorescence detector system follows.  The impact of the
upgrade on key systems of the Observatory, including communications,
data acquisition and data processing is described.  Finally, we
outline plans for deployment and maintenance of the upgrade, and the
structures in place to manage its funding and organization.

%% file: scientific_achievements_and_goals.tex

\chapter{Scientific Achievements and Goals}
\label{chap:science}

\section{Scientific results from the Pierre Auger Observatory}

The data taken with the Pierre Auger Observatory have led to a number
of major breakthroughs in the field of ultra-high energy cosmic rays (UHECRs).
Firstly, a suppression of the cosmic ray flux at energies above
$4{\times}10^{19}$\,eV has been established
unambiguously~\cite{Abraham:2008ru,Abraham:2010mj,ThePierreAuger:2015rha}.
Secondly, due to
the Auger limits on
photon~\cite{Abraham:2006ar,Aglietta:2007yx,Abraham:2009qb,Settimo:2011pf} and
neutrino~\cite{Abraham:2007rj,Abreu:2011zze,Abreu:2012zz,Abreu:2013zbq} fluxes at
ultra-high energy, it is now clear that unusual ``top-down'' source
processes such as the decay of super-heavy particles cannot account
for a significant part of the observed particle flux. 
Thirdly, a large-scale ${\sim}7$\% dipole anisotropy is found at
energies exceeding the ankle,
$E>8{\times}10^{18}$\,eV~\cite{ThePierreAuger:2014nja}.  At lower
energy, however, where the transition from galactic to extragalactic
cosmic rays is expected, the dipolar anisotropy of the particle
arrival directions is quite
small~\cite{Abreu:2011ve,Auger:2012an,Abreu:2012ybu}, which in
combination with the light composition at this energy challenges
models in which cosmic rays are of Galactic origin up to the ankle.
Furthermore, there were indications of an anisotropic distribution of
the arrival directions of particles with energies greater than
$5.5{\times}10^{19}$\,eV~\cite{Abraham:2007bb,Abraham:2007si,Abreu:2010zzj}
but, at the time being, the sensitivity of the Auger Observatory does
not allow source correlations to be
established~\cite{PierreAuger:2014yba}.
The fourth discovery is that of an unexpected evolution of the mass composition of
cosmic rays in the energy range from $10^{18}$ to $10^{19.5}$\,eV.
Interpreting the observed
longitudinal shower profiles~\cite{Aab:2014kda} with LHC-tuned interaction models, we can conclude that 
there is a large fraction of protons present at $10^{18}$\,eV, changing to a heavier composition,
possibly dominated by elements of the CNO group, at $10^{19.5}$\,eV~\cite{Aab:2014aea}.
In addition, it has been demonstrated that the Auger data can be used for
particle physics studies. Examples are the measurement of the 
proton-air and corresponding proton-proton cross sections at
$\unit[57]{TeV}$ c.m.s.\ energy~\cite{Abreu:2012wt} and the finding
that current air shower simulations fail to describe the relationship between
the longitudinal shower profile and the lateral particle densities at
ground level~\cite{FarrarICRC2013,Aab:2014pza}. In the following we will briefly review
these key observations.


\subsection{All-particle flux}
\begin{figure}[t]
\def\fgh{0.37}
\centering
\includegraphics[height=\fgh\columnwidth]{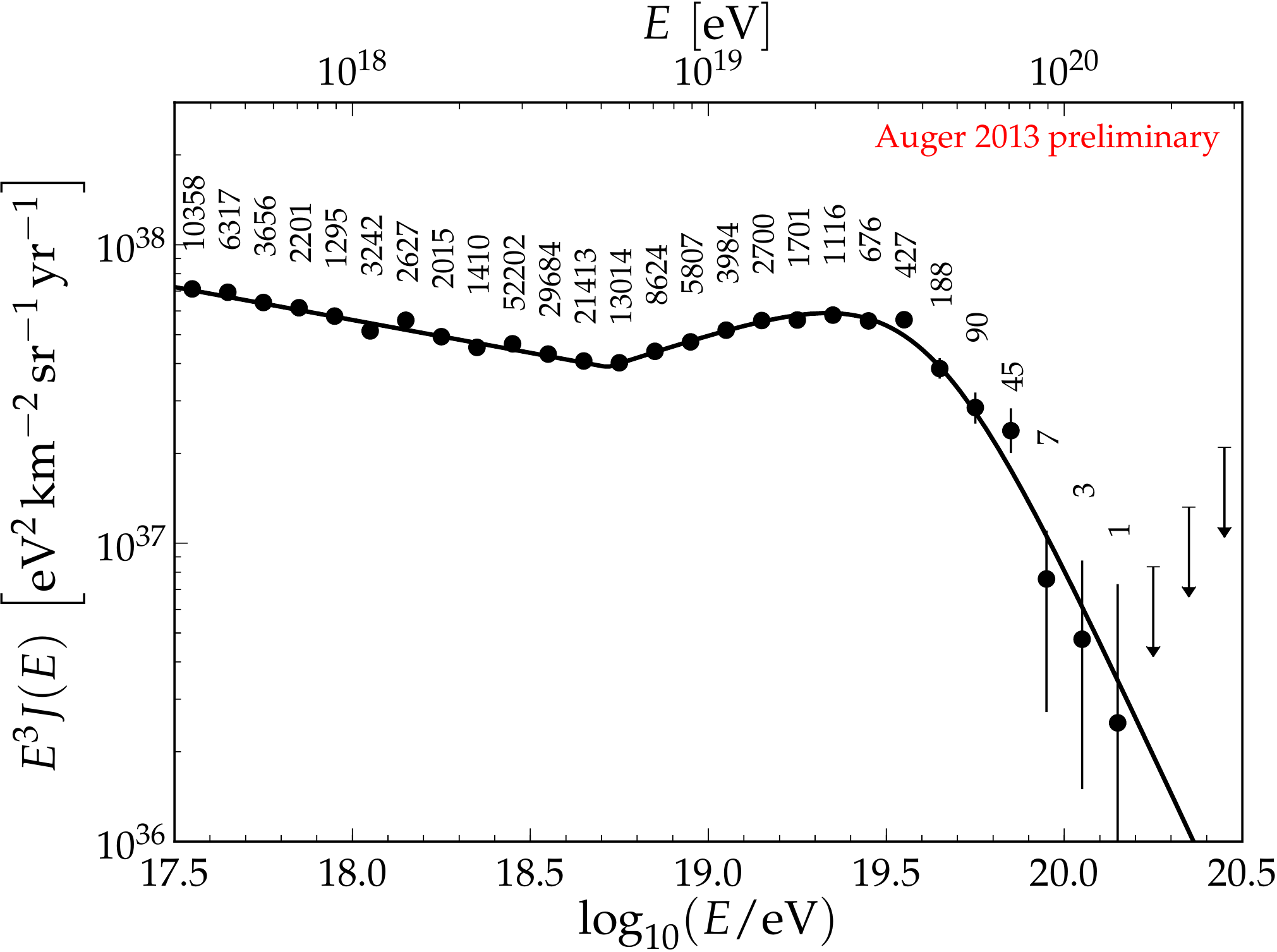}\hfill
\includegraphics[height=\fgh\columnwidth]{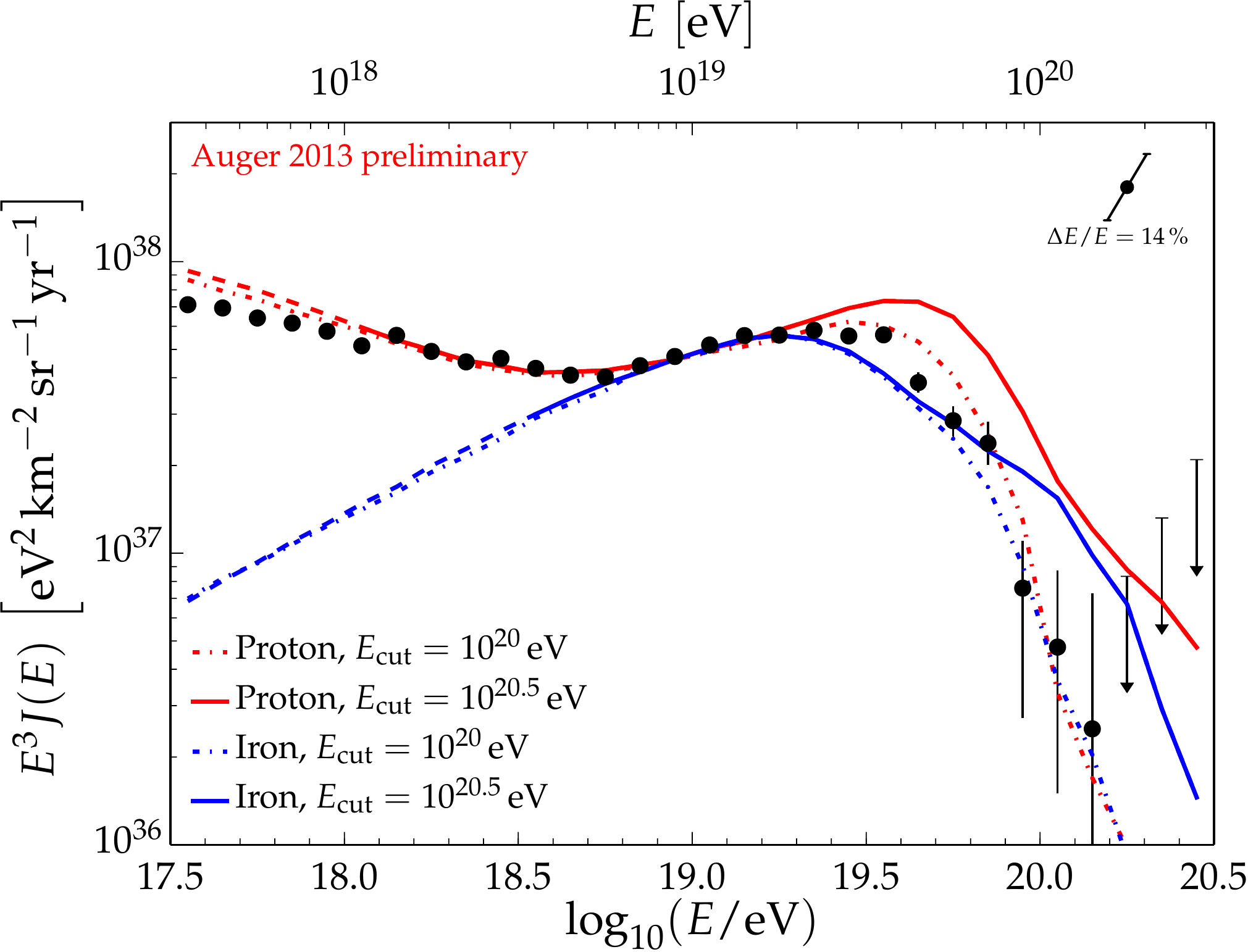}
\caption{
All-particle flux measured with the Auger
Observatory~\cite{SchulzICRC2013}. Left panel: The Auger data are
shown with an empirical fit. In addition the number
of events is given for each energy bin. Right panel: The energy
spectrum is compared to predictions for the idealized scenario of
homogeneously distributed sources injecting either only proton or iron
primaries, see text. The model lines have been calculated using
CRPropa~\cite{Armengaud:2006fx,Kampert:2012fi} and validated with
SimProp~\cite{Aloisio:2012wj}.
}
\label{fig:all-particle-flux}
\end{figure}

\noindent
The all-particle spectrum of the Auger
Observatory~\cite{SchulzICRC2013} is shown in
Fig.~\ref{fig:all-particle-flux}, and compared with two model calculations. The energy
spectrum was obtained by combining the individual energy spectra
derived from the array with \unit[$1500$]{m} spacing, the smaller array
of $\unit[750]{m}$ station separation, and the hybrid data set.
Showers with zenith angles up to $60^\circ$ as well as inclined showers
($\theta > 60^\circ$) have been used from the $\unit[1500]{m}$  array.
The statistics at high energy are dominated by the surface detector
array, having reached, after quality cuts, an exposure of about
$\unit[32,000]{km^2\,sr\,yr}$ by the end of 2012. The suppression of the
flux at high energy is established beyond any doubt.
Compared to a power-law extrapolation, the energy at which the flux
has dropped to 50\% of the value of the extrapolation is
$E_{50\%} \approx \unit[10^{19.6}]{eV} \approx \unit[4{\times}10^{19}]{eV}$.
There are $4$ events above $\unit[10^{20}]{eV}$ in this spectrum.
The low-energy part of the spectrum is
driven by data of the later-built $\unit[750]{m}$ array with an exposure of
$\unit[80]{km^2\,sr\,yr}$~\cite{RavignaniICRC2013}. 
The energy scale of the
spectrum has an overall systematic uncertainty of
14\%~\cite{VerziICRC2013}.

For orientation, the data are compared to two model scenarios, namely
continuously distributed sources that inject either only proton or
iron primaries. The sources are assumed to produce particles with the
energy spectrum ${\rm d}N/{\rm d}E \propto E^{-\beta}$ and the
cosmological evolution of the source luminosity is parameterized as
$(1+z)^m$. The model spectra are assumed to be exponentially
suppressed with the scale parameter $E_{\rm cut}$. The proton (iron)
lines correspond to $m=5$ ($m=0$) and $\beta = 2.35$ ($\beta = 2.3$).
In the case of proton primaries, a significantly better description of
the data is obtained by choosing $E_{\rm cut}=\unit[10^{20}]{eV}$
rather than a higher value as typically done in literature (see, for
example,~\cite{Protheroe:1995ft,Stanev:2000fb,Berezinsky:2002nc,Anchordoqui:2002hs,Hooper:2009fd}).

\begin{figure}[t]
\centering
\includegraphics[width=0.5\columnwidth]{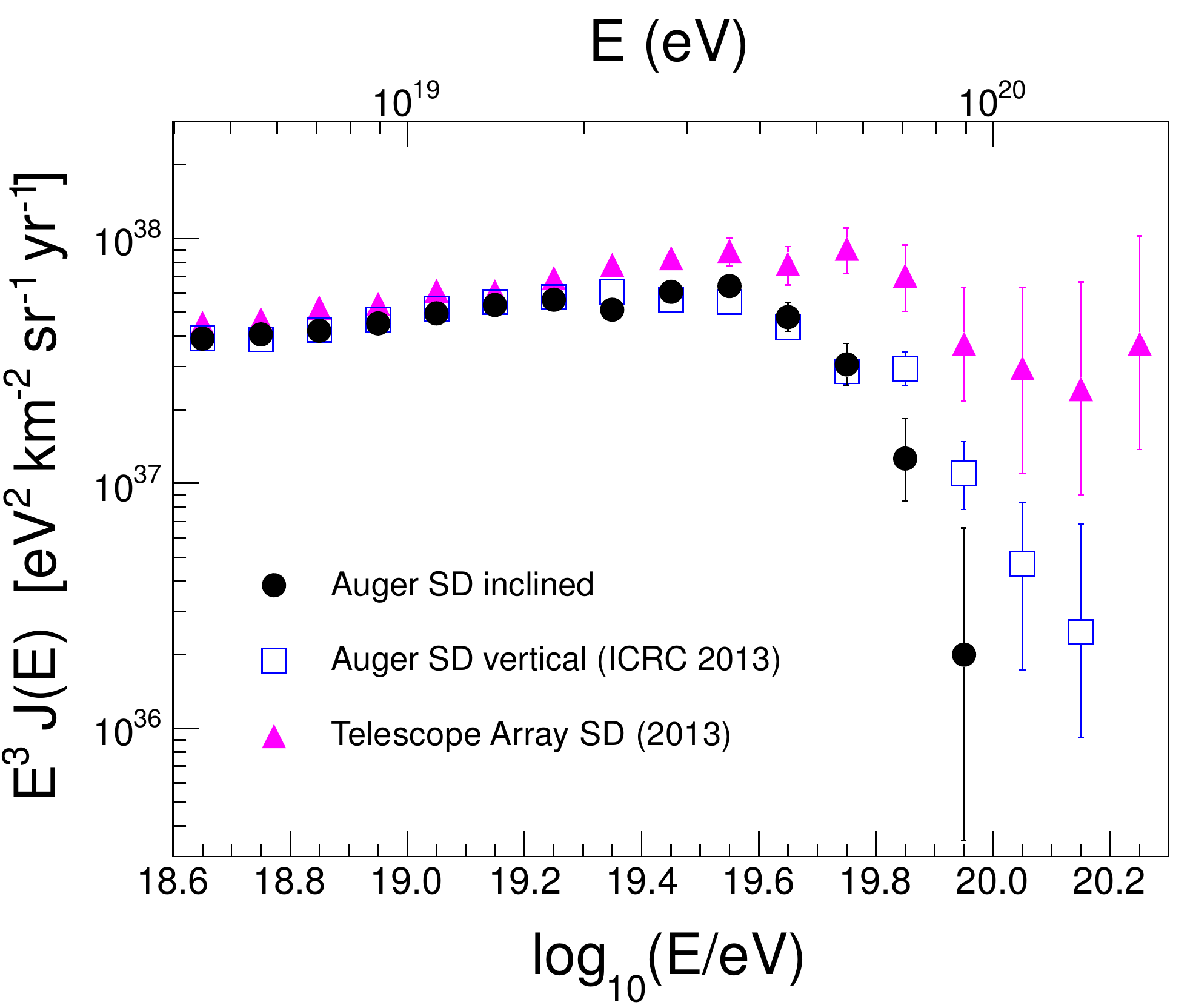}
\caption{Comparison of the two flux measurements of the Auger
Collaboration~\protect\cite{SchulzICRC2013,ThePierreAuger:2015rha}
with that of Telescope Array~\protect\cite{Tinyakov:2014lla}. The differences of
the spectra at the highest energies are intriguing
but still within the systematic uncertainties of the measurements.
}
\label{fig:ComparisonSpectra}
\end{figure} 

The recent, independent measurement of the energy spectrum using inclined
showers ($\theta > 60^\circ$)~\cite{ThePierreAuger:2015rha}
confirms the flux measured with showers up to a zenith angle of $60^\circ$. 
Within the systematic uncertainties, the data of the Auger Observatory are
compatible with the measurements of the Telescope Array
(TA)~\cite{AbuZayyad:2012ru,Tinyakov:2014lla} and
HiRes~\cite{Abbasi:2007sv}. While the energy of the ankle are within 10\% in the TA and Auger data sets, the suppression of the flux is shifted
by ${\sim}70$\% to higher energies in the TA spectrum, see Fig.~\ref{fig:ComparisonSpectra}.
Work is in progress
to determine to what degree, if any, the observed difference of
the spectra measured in the northern and southern hemispheres 
might be a real astrophysical effect.


\subsection{Photon and neutrino limits}

\begin{figure}[t]
\centering
\includegraphics[width=0.489\columnwidth]{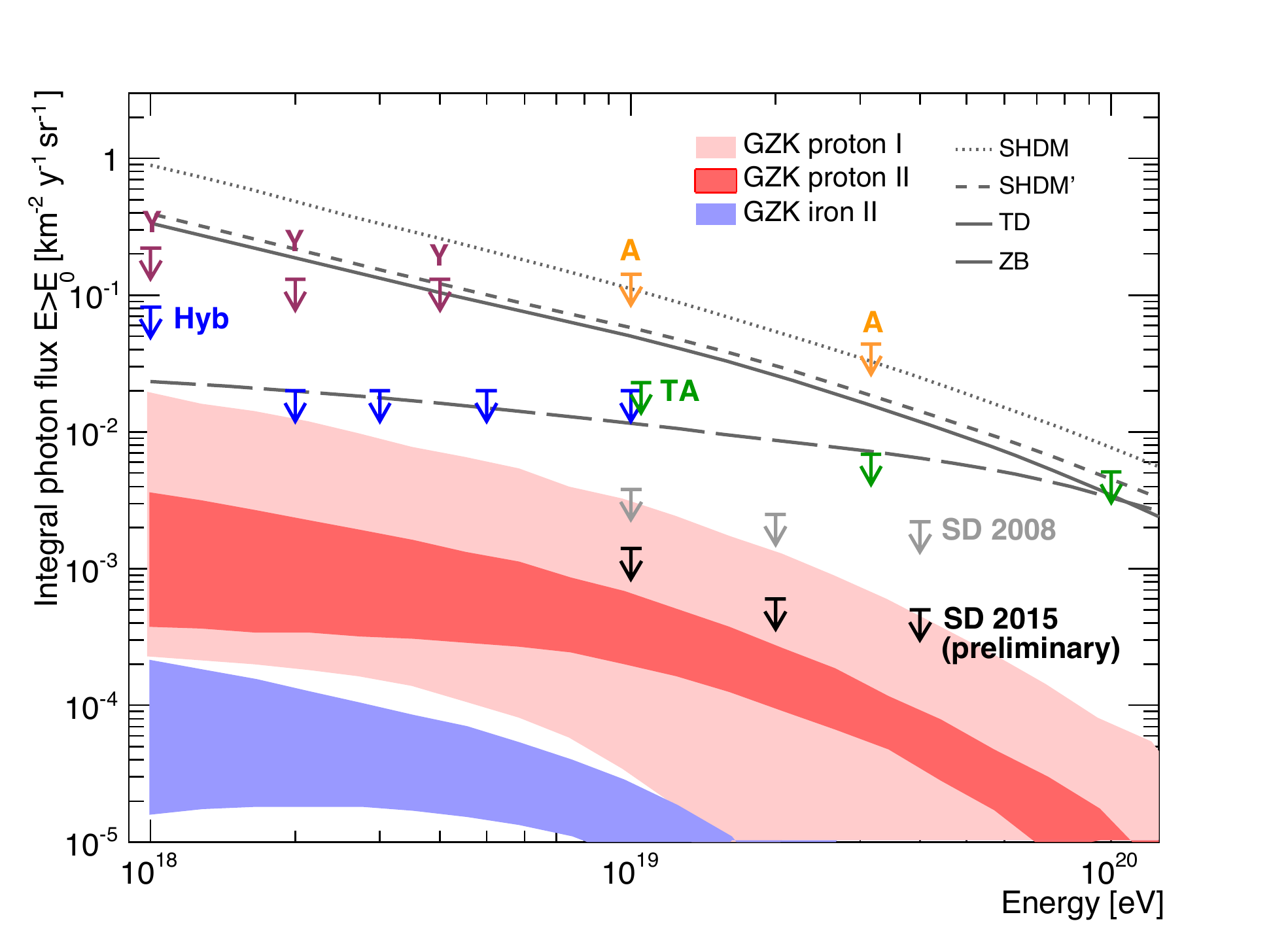}\hfill
\includegraphics[width=0.51\columnwidth]{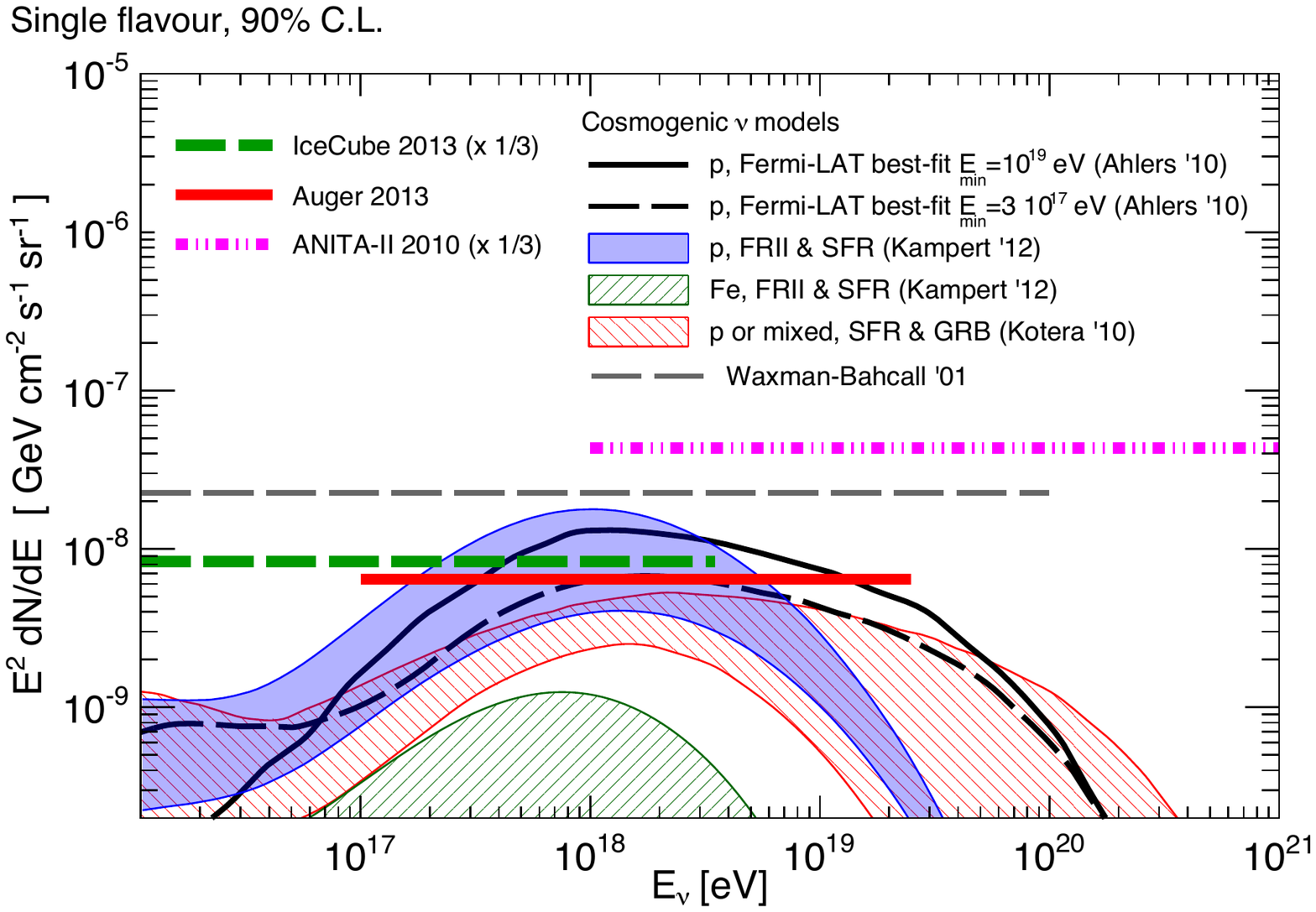}
\caption{
Limits on the flux of
photons (left)~\cite{Abraham:2006ar,Aglietta:2007yx,Abraham:2009qb,Settimo:2011pf} and
neutrinos (right)~\cite{PieroniICRC2013} obtained from the Pierre Auger
Observatory. The data are shown together with the current limits from
other experiments~\cite{Aartsen:2013dsm,Gorham:2008yk,Gorham:2008dv,%
Gorham:2010kv,Kravchenko:2011im,Abbasi:2008hr,Abu-Zayyad:2013dii}
and some examples of predicted fluxes,
see text.
}
\label{fig:photon-neutrino-limits}
\end{figure} 

\noindent
The latest limits on the fluxes of
photons~\cite{Abraham:2006ar,Aglietta:2007yx,Abraham:2009qb,Settimo:2011pf} and
neutrinos~\cite{Abraham:2007rj,Abreu:2011zze,PieroniICRC2013,Abreu:2013zbq} obtained
with the Pierre Auger Observatory are
shown in Fig.~\ref{fig:photon-neutrino-limits}. Model scenarios for
sources of UHECRs, in which the observed particles are produced by the
decay of other particles (top-down models), lead to large secondary
fluxes of photons and neutrinos~\cite{Bhattacharjee:1998qc}. 
In contrast, models in which the production of photons and neutrinos originates
from secondaries generated by the propagation in the cosmic background
(GZK effect) lead to much lower fluxes. Some
representative examples of predicted secondary fluxes of such models
are shown in Fig.~\ref{fig:photon-neutrino-limits} (photons:
GZK~\cite{Ahlers:2010fw,Kotera:2010yn,Kampert:2012mx},
top-down (TD), Z-burst, and super-heavy dark matter
(SHDM)~\cite{Gelmini:2005wu}, SHDM$^\prime$~\cite{Ellis:2005jc};
neutrinos: TD~\cite{Sigl:1998vz}, Z-burst~\cite{Kalashev:2008dh}).
The neutrino flux limit of the Auger Observatory is now lower than the
Waxman-Bahcall flux~\cite{Waxman:1998yy,Bahcall:1999yr}.

The current flux limits rule out, or strongly disfavor, that top-down
models can account for a significant part of the observed UHECR flux.
The bounds are reliable as the photon flux limits in
Fig.~\ref{fig:photon-neutrino-limits} depend only on the simulation of
electromagnetic showers and, hence, are very robust against
assumptions about hadronic interactions at very high
energy~\cite{Risse:2007sd}.

In addition, the flux limits already probe the predicted secondary fluxes
for models in which the suppression of the cosmic ray flux is assumed to originate
entirely from the GZK energy loss process for a proton dominated 
flux~\cite{Gelmini:2005wu,Ahlers:2010fw,Kotera:2010yn,Kampert:2012mx}. 

The photon flux limits have further far-reaching consequences by
providing important constraints on theories of quantum gravity
involving Lorentz invariance violation (LIV), see, for example,
\cite{Galaverni:2008yj,Galaverni:2007tq,Liberati:2011bp,Horvat:2010sr}.
Further, identifying  a single photon shower at ultra-high energy would imply
very strong limits on another set of parameters of LIV
theories~\cite{Klinkhamer:2010pq,Girelli:2012ju,Rubtsov:2013wwa}. Similarly, observing
cosmogenic neutrinos would allow placing constraints on LIV in the 
neutrino sector~\cite{Mattingly:2009jf}.


\subsection{Depth of shower maximum}

\begin{figure}[t]
\centering
\includegraphics[width=\columnwidth]{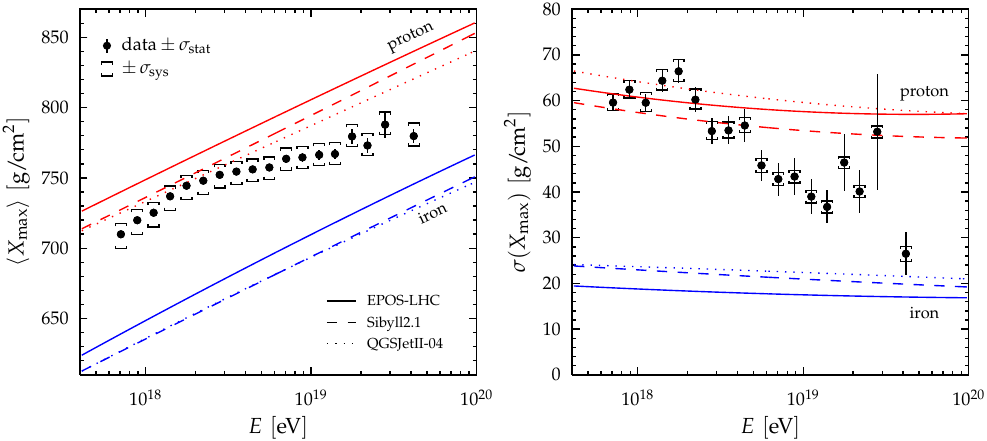}
\caption{
Depth of shower maximum, \xmax, as measured with the Pierre
Auger Observatory~\cite{Aab:2014kda}. The left panel shows the mean
\xmax, and the dispersion is given
in the right panel after correcting for the reconstruction resolution.
The data are compared to model predictions for proton and iron 
primaries~\cite{Werner:2005jf,Pierog:2006qv,Ostapchenko:2010vb,Ahn:2009wx}.
}
\label{fig:Xmax-mean-sigma}
\end{figure}

The Pierre Auger Collaboration has addressed the challenge of
determining the composition of UHECRs by measuring the depth of shower
maximum \xmax~\cite{Aab:2014kda,Aab:2014aea},
the muon production depth~\cite{Aab:2014dua}, and  rise-time
asymmetry of the shower disk at ground level~\cite{Abreu:2011pe}. Out of
these observables, the \xmax measurement using fluorescence
telescopes is currently the one with the smallest systematic uncertainties
and the most direct link to the mass distribution of the primary
particles~\cite{Abraham:2010yv,Engel:2011zzb,Abreu:2013env}. The mean depth of
shower maximum and the fluctuations measured by the shower-to-shower
variation of \xmax, which are a superposition of fluctuations of showers of a given primary and 
differences due to different primary particles, are shown in Fig.~\ref{fig:Xmax-mean-sigma}
together with model predictions for proton and iron primaries. The data of the fluorescence
telescopes cover energies up to the suppression range with good
statistics. The last data point represents all events with
$E>\unit[3{\times}10^{19}]{eV}$.

The comparison of \xmax and $\sigma(X_{\rm max})$ with models, of
which EPOS-LHC~\cite{Pierog:2013ria} and QGSJetII-04~\cite{Ostapchenko:2014mna}
have already been tuned to describe the new LHC data~\cite{dEnterria:2011kw}, shows that
the composition appears to be
predominantly light from $\unit[10^{18}]{eV}$ up to about $\unit[3{\times}10^{18}]{eV}$.
Above this energy the elongation rate changes,
indicating a transition to heavier elements. 
This interpretation as a change of composition is supported by the
change of the variance of \xmax in the same energy region.
The trend in composition is
also confirmed by the other aforementioned composition-sensitive
measurements~\cite{Abreu:2011pe,Aab:2014dua}. If interpreted with current
interaction models, the size of the fluctuations relative to the mean
\xmax implies a very small variance in the masses of the
primary particles contributing at a given energy~\cite{Abreu:2013env}.
Alternatively, these observations could be interpreted as an
unexpected change of the properties of hadronic interactions, most
likely involving new particle physics,
see~\cite{DiasdeDeus:2005sq,Wibig:2009zza,Allen:2013hfa}.

A comparison of the Auger data on \xmax with that from
HiRes~\cite{Abbasi:2004nz} and TA~\cite{Abbasi:2014sfa} is not
straightforward as the latter two data sets are not corrected for
detector acceptance effects. A joint working group of the Auger and TA
collaborations is investigating to what degree the two data sets are
compatible~\cite{Barcikowski:2013nfa}. In a recent study published by
this working group, the mean \xmax of the Auger and TA data has been
found to be in good agreement within current statistical and
systematic errors~\cite{Abbasi:2015xga}.


\subsection{Arrival direction distribution}
 
\noindent
The arrival direction distribution is one of the key observables to
search for the transition from galactic to extragalactic cosmic rays,
and for sources or source regions of UHECRs.

Combining the data set of showers with zenith angles up to $60^\circ$,
which is typically used for anisotropy studies, with that of inclined
showers ($60^\circ < \theta < 80^\circ$), a large-scale dipolar
anisotropy was found for showers with
$\unit[E>8{\times}10^{18}]{eV}$~\cite{ThePierreAuger:2014nja}.  An
amplitude of the first harmonic in right ascension of $r_{1\alpha} =
(4.4\pm 1.0){\times}10^{-2}$ was measured, having a chance probability
of $P(\ge r_{1\alpha}) = 6.4{\times}10^{-5}$.  Under the assumption
that the only significant contribution to the anisotropy is from the
dipolar component, this observation would correspond to a dipole of
amplitude $d = 0.073\pm 0.015$ pointing to $(\alpha,\delta)=
(95^\circ\pm 13^\circ,-39^\circ\pm 13^\circ)$.  The origin of this
anisotropy is subject to ongoing discussions. It could arise, for
example, from an inhomogeneity of the distribution of nearby sources,
see \cite{Harari:2013pea}.

\begin{figure}[t]
\centering
\includegraphics[width=0.49\columnwidth]{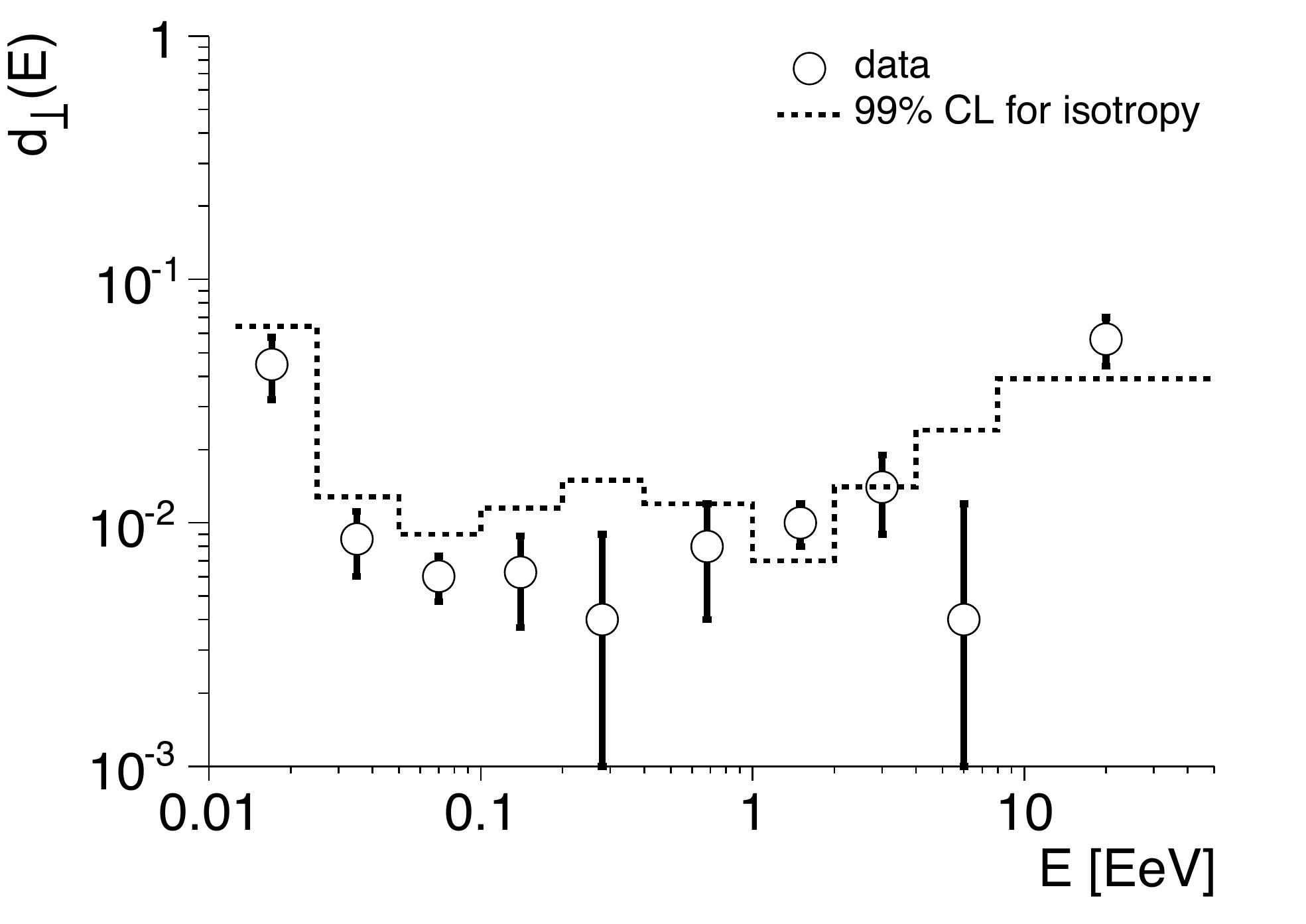}\hfill
\includegraphics[width=0.49\columnwidth]{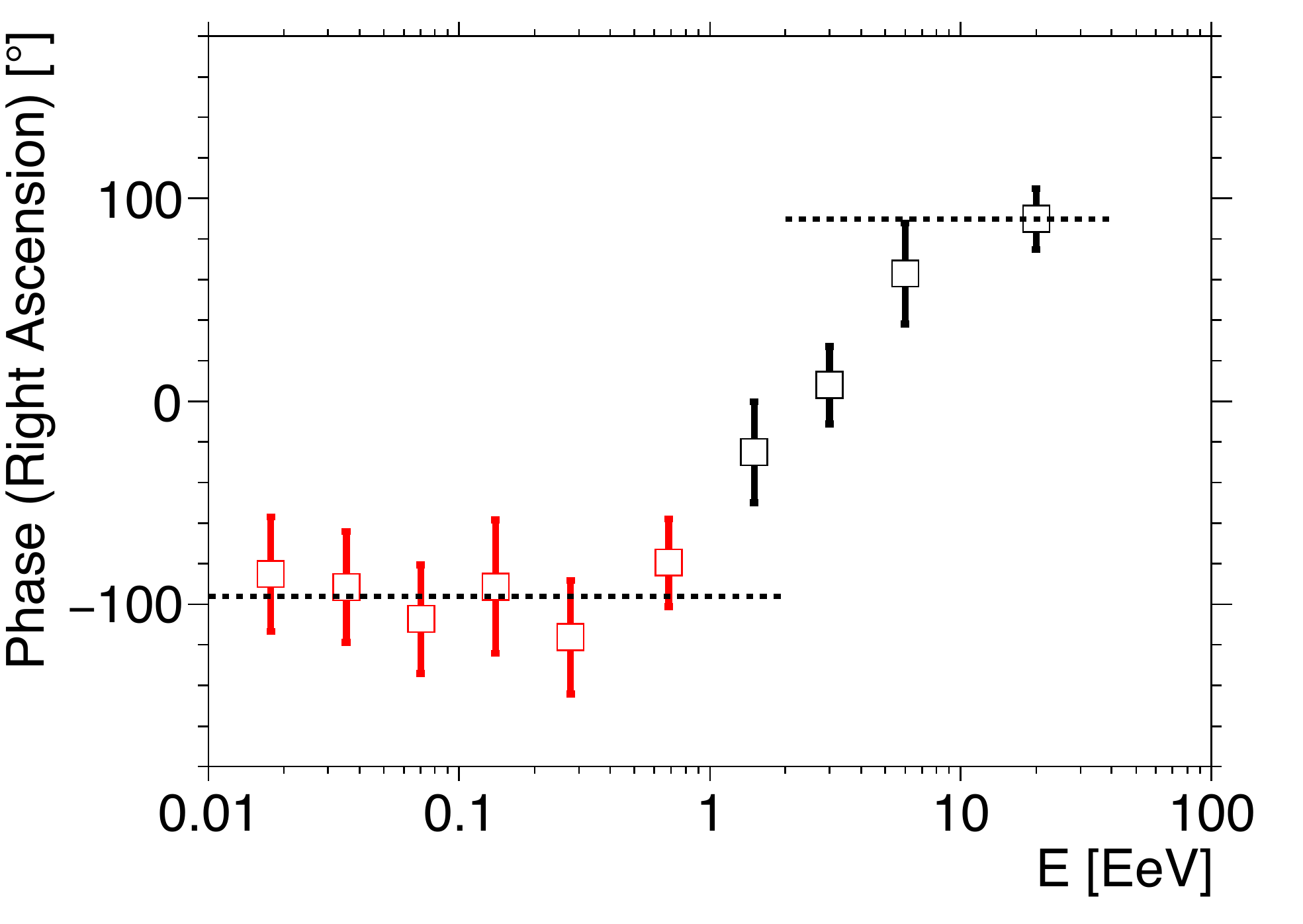}
\caption{ Large scale anisotropy search. Left: 99\% limits on the
  dipole anisotropy in the equatorial plane for the collected
  statistics until end of 2014 (dashed line) and values of the dipole
  amplitude \protect$d_\perp$.  Right: estimated phase angles.
  The red points of the equatorial phase are from
  the analysis of the $\unit[750]{m}$ array.  The data shown is an
  update of the analyses~\cite{Abreu:2011ve,SidelnikICRC2013}, to be
  published at ICRC 2015.
}
\label{fig:large-scale-anisotropy}
\end{figure} 

The full-acceptance threshold of the
$\unit[1500]{m}$-array, having the highest exposure of all detectors
of the Auger Observatory, is in the energy range of the ankle.
Therefore, to search for anisotropies in the
energy range of the transition from Galactic to extragalactic cosmic rays,
it is crucial to combine this data set with that of the
$\unit[750]{m}$ array~\cite{SidelnikICRC2013}. 
Both the Rayleigh~\cite{Linsley:1975fk} and differential 
East-West~\cite{Bonino:2011nx} methods have been applied to the Auger data~\cite{Abreu:2011ve}.
Updated results for the measured amplitude of the dipole
(as well as the corresponding upper limit) and its phase angle
are shown in Fig.~\ref{fig:large-scale-anisotropy} as a function
of energy.
The phase angle exhibits a
smooth change with energy. It points near the Galactic center below
$\unit[10^{18}]{eV}$, suggesting an origin in a galactic component. The phase angle
points in the opposite direction at higher energy, possibly
manifesting a signature of the inhomogeneous distribution of nearby
extragalactic matter. Given that the phase angle is statistically more
sensitive than the dipole amplitude~\cite{Linsley:1975fk}, a prescribed test to
determine the statistical significance of the observed transition in
the phase is being performed, and will run until mid-2015.

\begin{figure}[t]
\centering
\includegraphics[width=0.55\columnwidth]{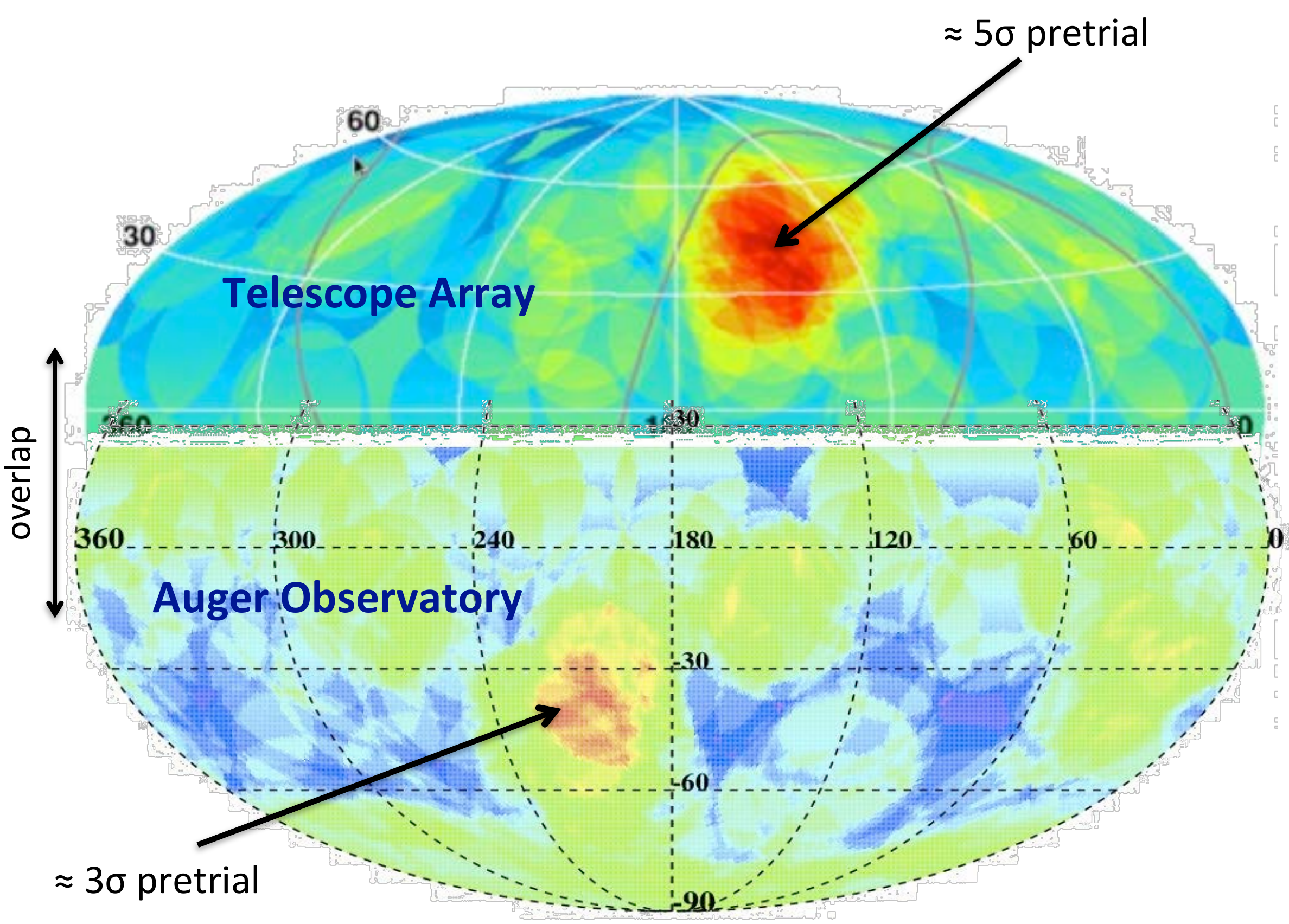}
\caption{
Regions of over-density observed after ${\sim}20^\circ$-smearing of the arrival directions of 
particles with $\unit[E > 5.5{\times}10^{19}]{eV}$. The results from the northern
hemisphere are from the TA Collaboration~\cite{Abbasi:2014lda}.
}
\label{fig:Auger-TA-hot-spots}
\end{figure}

Up to now it has not been possible to establish small-angle
correlations of the arrival direction distribution of Auger data with
possible sources or source regions beyond any
doubt~\cite{PierreAuger:2014yba}, even though there were some
intriguing
indications~\cite{Abraham:2007bb,Abraham:2007si,Abreu:2010zzj}.  Here
we only want to mention the $15^\circ$ region of over-density observed
around the direction of Centaurus
A~\cite{PierreAuger:2014yba}. Although not being a statistically
significant excess beyond $3\sigma$, it is interesting to note that
the TA Collaboration has recently reported a ``hot spot'' of similar
intermediate angular scale~\cite{Abbasi:2014lda}, see
Fig.~\ref{fig:Auger-TA-hot-spots}.


\subsection{Air shower and hadronic interaction physics}

\begin{figure}[t]
\centering
\includegraphics[width=0.6\columnwidth]{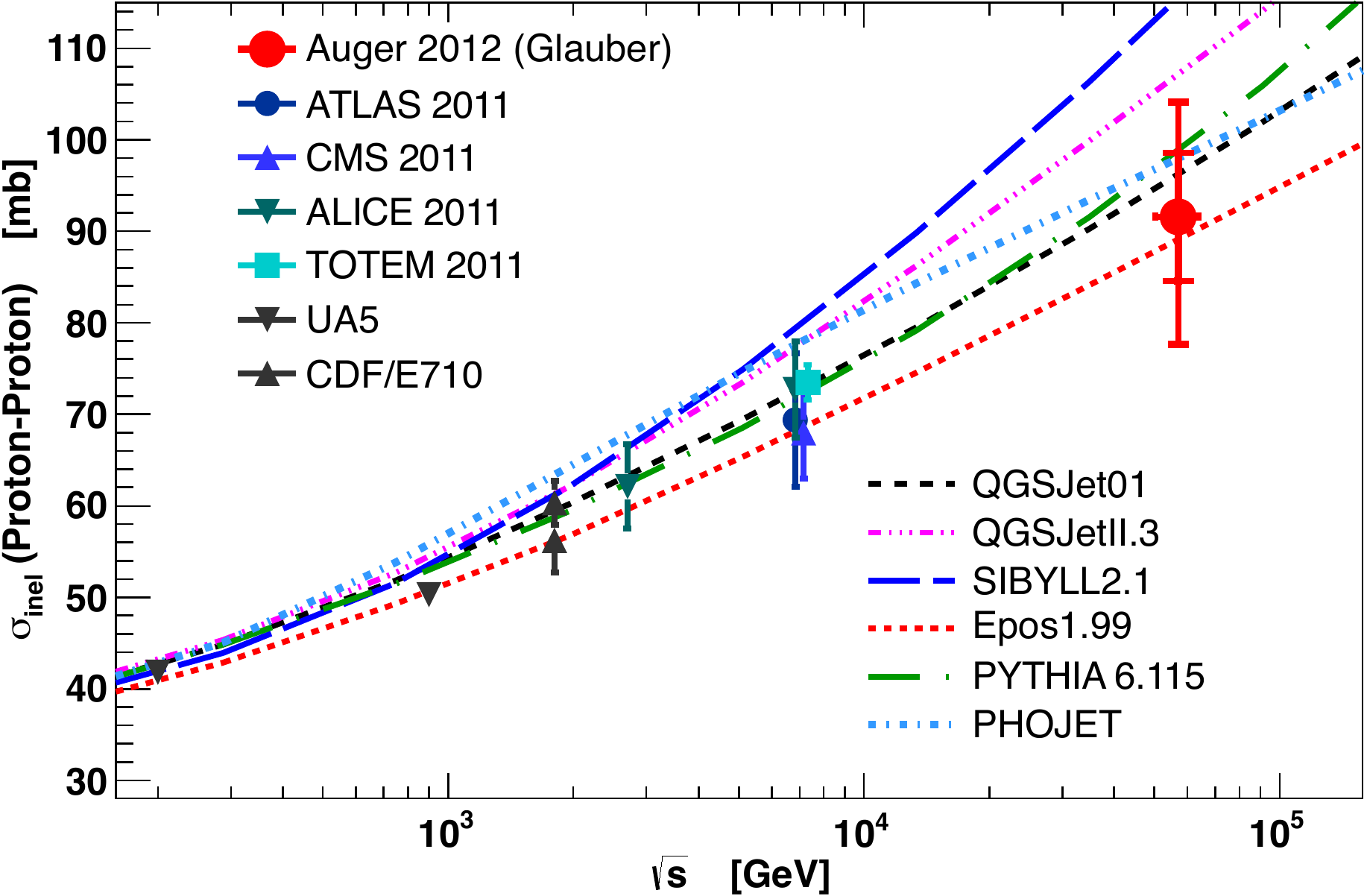}
\caption{
Proton-proton cross section derived from the proton-air cross section
measured with the Pierre Auger Observatory~\cite{Abreu:2012wt}.
The Auger result is shown together with collider measurements
and model extrapolations.
}
\label{fig:sigma-pp}
\end{figure} 

\noindent
The depth of shower maximum is directly related to the depth of the
first interaction of the cosmic ray in the
atmosphere~\cite{Ulrich:2009zq}. Based on this correlation, the 
proton-air cross section has been measured at $\unit[57]{TeV}$ c.m.s.\ energy
using hybrid data of the Auger Observatory~\cite{Abreu:2012wt}.
Applying the Glauber approximation~\cite{Glauber:1970jm} this cross
section can be converted to an equivalent (inelastic) proton-proton
cross section, see Fig.~\ref{fig:sigma-pp}.
The cross section is found to be consistent with model extrapolations
that describe the LHC data, which were becoming available at the same
time as the Auger measurement was published. An unexpected, rapid
increase of the cross section directly above the LHC energy is not evident.

The muonic component of air showers is sensitive to hadronic particle
interactions at all stages in the air shower cascade, and to many
properties of hadronic interactions such as the multiplicity,
elasticity, fraction of neutral secondary pions, and the
baryon-to-pion ratio~\cite{Ulrich:2010rg,Engel:2011zzb}. Currently the
number of muons can only be measured indirectly~\cite{KeglICRC2013}
except at very large
lateral distances~\cite{Cazon:2004zx,Aab:2014dua} and in very
inclined showers~\cite{Ave:2000xs,Aab:2014pza}, where muons
dominate the shower signal at ground level, and for which the electromagnetic
component due to muon decay and interaction is understood~\cite{Valino:2009dv}.

\begin{figure}[t]
\centering
\includegraphics[width=0.45\columnwidth]{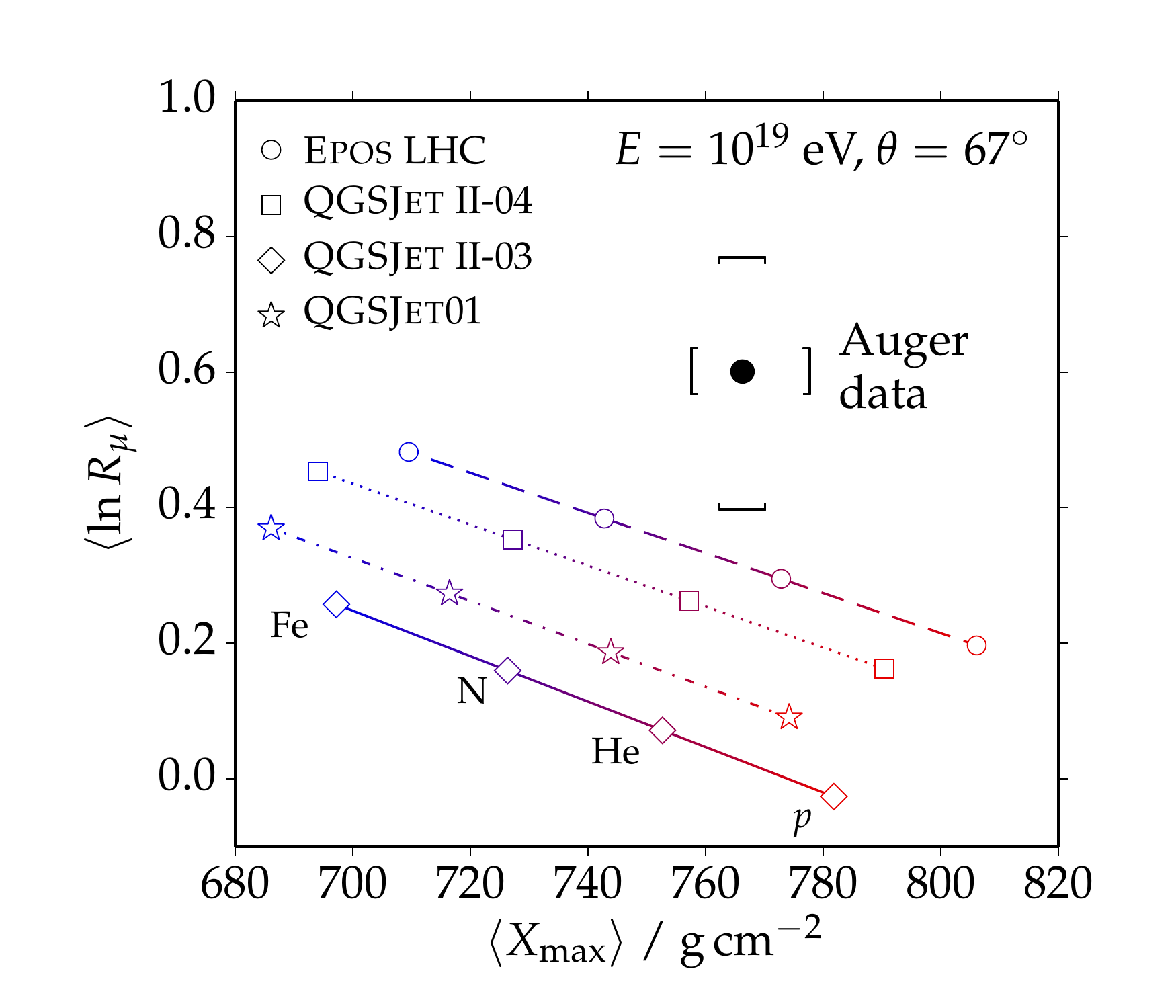}
\includegraphics[width=0.54\columnwidth]{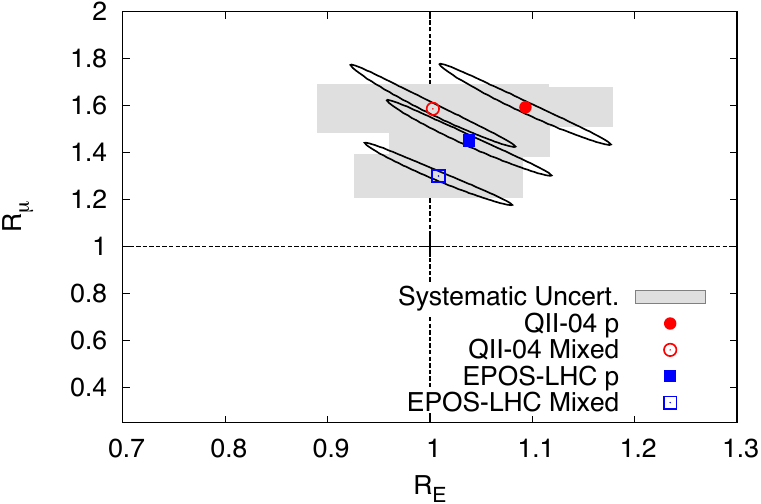}
\caption{
Left: Mean number of muons $R_\mu$ relative to that of proton reference showers, 
and depth of shower maximum at $\unit[10^{19}]{eV}$. The Auger
data point~\cite{Aab:2014pza}, where the
muon number is derived from inclined showers,
is compared with predictions obtained from different interaction models.
Right: Muon discrepancy~\cite{FarrarICRC2013}
observed in showers of $\unit[10^{19}]{eV}$. Shown are
the phenomenological scaling factors $R_E$ and $R_\mu$ for the primary energy 
and the hadronic (primarily muonic) component of the shower that would be needed to bring a
model calculation into agreement with Auger data, see text.
}
\label{fig:muon-discrepancy}
\end{figure}

Despite these limitations, it was possible to
show that current hadronic interaction models do not provide a good description of the
number of muons produced in air showers~\cite{FarrarICRC2013,Aab:2014pza}.
This is illustrated in
Fig.~\ref{fig:muon-discrepancy}~(left) where the observed muon number,
given relative to proton reference showers, is compared with predictions of
commonly used hadronic interaction models. Even though some of these models have
been re-tuned recently to provide an improved description of LHC data
and fixed-target measurements, none of the models can reproduce the muon number.
This conclusion still holds even if different
primary mass compositions, as indicated by \xmax in the plot, are
considered.

Similar results are found in an independent analysis of showers
of $\theta < 60^\circ$ measured
with the surface array in coincidence with fluorescence light
observation, see Fig.~\ref{fig:muon-discrepancy}~(right).
There the scaling
factors needed to obtain a good description of Auger showers of
$\unit[10^{19}]{eV}$ are given for air shower simulations made with
the models QGSJet II-04~\cite{Ostapchenko:2010vb} and
EPOS-LHC~\cite{Werner:2005jf,Pierog:2006qv}, both already tuned to LHC
data. Matching the measured longitudinal shower profile with a
simulated profile of the same energy, the muon signal has been derived
by comparing the surface detector signals of the measured and
simulated showers~\cite{FarrarICRC2013}. 

An observable sensitive to composition and hadronic interactions
is the distribution of the production depths of
muons~\cite{Cazon:2004zx}. Hadronic interaction models can be tested by
comparing the mass estimates derived from the longitudinal shower profiles
with that derived from the muon production profiles~\cite{Aab:2014dua}.

Finally it should be mentioned that, although not directly comparable due
to the different types of surface
detectors, the discrepancy between the fluorescence and surface
detector signals of ${\sim}27$\% reported by the
TA Collaboration~\cite{SagawaICRC2013} is
qualitatively in agreement with the Auger data~\cite{Allen:2013ofa}.


\section{Open questions and goals of upgrading the Observatory}

\noindent
The data of the Pierre Auger Observatory are often considered as
providing strong support for classic models of UHECR sources 
(e.g.\ 
\cite{Berezinsky:2002nc,Berezinsky:2004wx,Wibig:2004ye,Hillas:2005cs,Allard:2005ha}). 
In these models it is typically assumed that particle
acceleration takes place at sites distributed in a similar way to the matter
distribution in the Universe and the effect of energy loss
processes~\cite{Greisen:1966jv,Zatsepin66e}
(either pion-photoproduction or photo-disintegration) causes the
observed flux suppression at energies above $\unit[5{\times}10^{19}]{eV}$,
and some anisotropy of the arrival direction
distribution. It is common to these models that particles
are injected at extra-galactic sources with an energy spectrum
following a power-law $E^{-\beta}$ with $\beta \ge 2$ and a maximum energy exceeding
$\unit[10^{20}]{eV}$. The main differences of the models are related
to the assumptions on the index of the power-law and the mass
composition at the sources. While almost all models attribute the suppression of the
energy spectrum to propagation effects, different interpretations of
the origin of the
ankle~\cite{Berezinsky:2002nc,Berezinsky:2004wx,Wibig:2004ye,%
Hillas:2005cs,Allard:2005ha,Allard:2011aa,Gaisser:2013bla,Fang:2013cba}
are considered.


\subsection{Possible data interpretations and astrophysical scenarios}

\noindent
In the face of these usual assumptions, the measurements of the Pierre Auger Observatory have led to
a number of puzzling observations that indicate a much more complex
astrophysical scenario, with a phenomenology that is far from being understood. In the
following we will discuss some aspects of possible interpretations to
illustrate the wealth of information we have collected, but also to
point out the lack of some key measurements.

\begin{figure}[t]
\centering
\includegraphics[width=0.95\columnwidth]{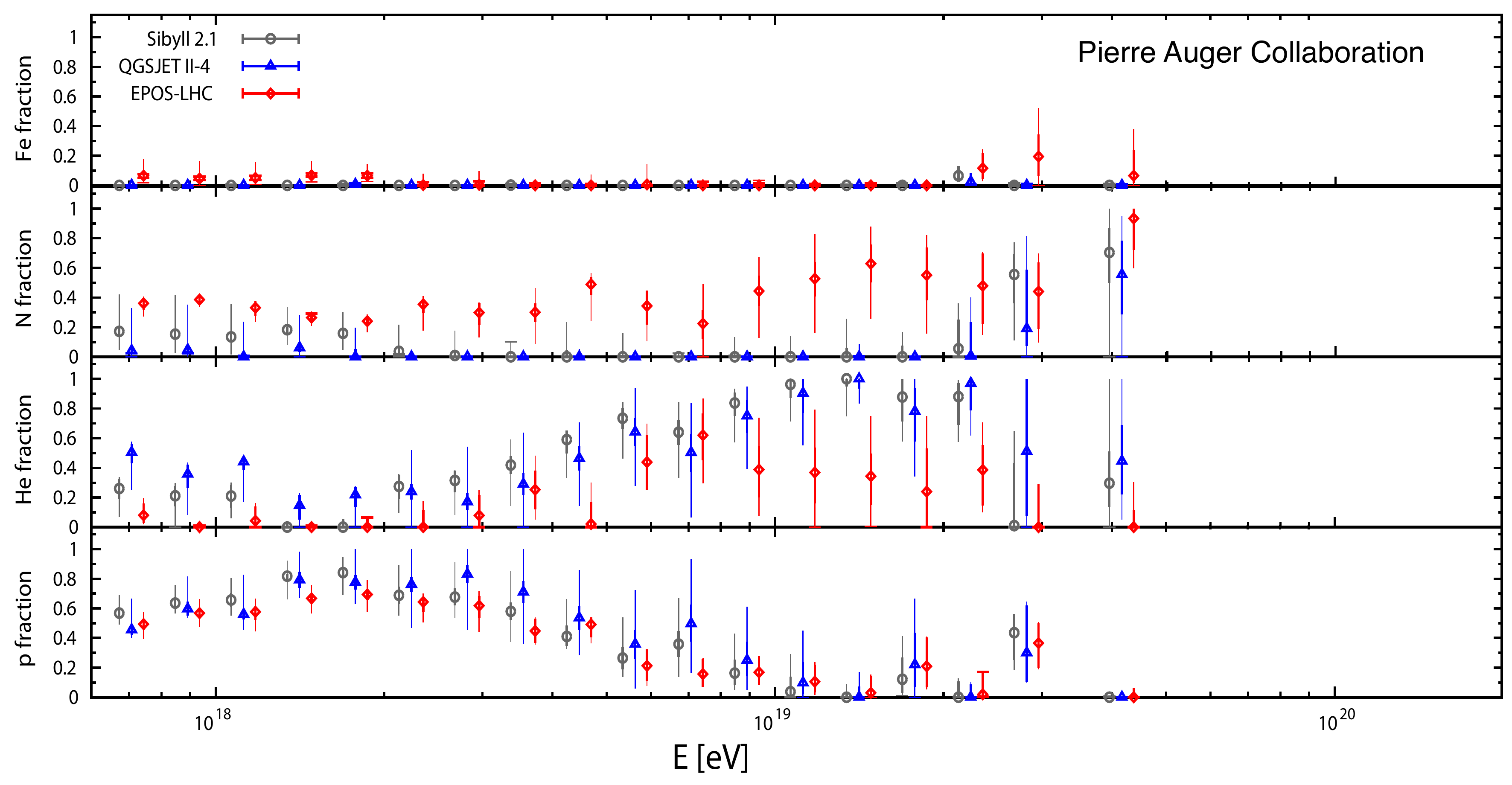}
\caption{Estimate of the composition of ultra-high energy cosmic rays at the
top of the atmosphere~\cite{Aab:2014aea}. The \xmax
distributions measured with the Auger Observatory have been fitted by a
superposition of four mass groups accounting for detector resolution and
acceptance effects. The error bars show the combined
statistical and systematic uncertainties of the mass estimates,
except those related to the choice of the hadronic interaction
models.}
\label{fig:composition}
\end{figure}

The \xmax distributions measured with the fluorescence
telescopes for different energy intervals~\cite{Aab:2014kda}
can be used to estimate the
UHECR composition at Earth. This composition will depend on
the number of mass groups considered and the hadronic interaction
models employed in the simulations.
The result of such analysis, fitting four mass groups to the
measured \xmax distributions~\cite{Aab:2014aea},
is shown in Fig.~\ref{fig:composition}. The
interaction models EPOS-LHC~\cite{Werner:2005jf,Pierog:2006qv}, QGSJet
II.04~\cite{Ostapchenko:2010vb} and Sibyll 2.1~\cite{Ahn:2009wx}
have been used for data
interpretation to get some understanding of the systematic
uncertainties related to the modeling of hadronic interactions.

One striking result is the presence of a large fraction of protons in
the energy range of the ankle. At the same time, according to the
Auger data, the anisotropy of the arrival directions of these protons
cannot be larger than a few percent. This is in contradiction to the
expectations for light particles produced in continuously distributed
Galactic sources, given the current knowledge of propagation in the
Galactic magnetic field~\cite{Candia:2003dk,Giacinti:2011ww}. Thus the
protons at energies as low as $\unit[10^{18}]{eV}$ are most likely of
extragalactic origin, or one has to accept rather extreme assumptions
about the Galactic magnetic field.

Another surprising observation is the disappearance of the proton
component just below $\unit[10^{19}]{eV}$ and, at the same time, the
appearance of a helium component. There are indications that a similar
transition from helium to the nitrogen mass group could take place at
higher energy, but the statistics of the data of the fluorescence
telescopes are not high enough to be conclusive. Furthermore there
is a large correlation between the fractions of the different mass groups.
We will not attempt
here to speculate on the origin of these transitions and only point
out that we do not have enough composition-sensitive data to derive
the composition at energies higher than a few times
$\unit[10^{19}]{eV}$, even if we understood hadronic interactions at
these energies perfectly.

Finally we want to mention that there are indications for a possible
re-appearance of a proton component at high energy that could be
related to the possible anisotropy on the 15 degrees scale. With
respect to the model scenarios we will discuss below, confirming the
existence of a proton population at the highest energies would
indicate another class of sources, possibly distributed over
cosmological distances. These protons are expected to be correlated in
arrival direction with their sources and could open a window to
particle astronomy.

A number of authors have used our data on the all-particle spectrum
and \xmax to develop generic scenarios of UHECR models. We
will discuss some of these scenarios  here to demonstrate 
fundamentally different interpretations of the Auger data
(see also the recent articles~\cite{Taylor:2013gga,Aloisio:2013hya,Kampert:2014eja}).
All the models we consider here are, of course, very much simplified
representations of the complex physics of UHECR sources and propagation. For
example, the inhomogeneous distribution of the sources and the
detailed magnetic structure of the Universe is not accounted for,
and all sources are assumed to inject similar particle spectra
(i.e.\ mass composition and energy distribution). 

\paragraph{Maximum-rigidity scenario}
If one assumes that
the sources accelerate particles to maximum energies proportional to their
charge (i.e.\ to the same maximum rigidity),
one obtains a model in which the proton component around
$\unit[10^{18.5}]{eV}$ is naturally related to similar components of heavier
elements, each shifted in energy by the charge number $Z$. The upper end of the
all-particle spectrum would then be dominated by heavy elements of the iron
group. The observed suppression of the flux at the highest energies would then
be caused by the cutoff of the source spectrum
rather than energy loss processes (photo-disintegration) during propagation.
In Ref.\,\cite{Allard:2008gj,Allard:2011aa}
the mass composition needed to describe both the flux and \xmax data
of the Auger Observatory is assumed similar to that of Galactic cosmic
rays with fluxes having the same spectral index in energy per nucleon, but with a
strong enhancement of heavy elements.
One key feature of this scenario is that the
protons in the energy range of the ankle are injected by the same extragalactic
sources that produce the flux at the highest energies.

\paragraph{Photo-disintegration scenario} If one assumes
that the sources accelerate nuclei to a maximum energy above the energy
threshold for photo-disintegration on CMB photons, the light elements could then be fragments
of heavier nuclei that disintegrated during propagation. In this scenario
(see e.g.\,\cite{Hooper:2009fd,Taylor:2011ta}) the
suppression of the all-particle flux originates mainly from energy loss
processes (photo-disintegration of nuclei). Lighter elements appear at energies
shifted by the ratio of the daughter to parent mass numbers. Again the protons
at the ankle energy are naturally linked to the particles at the highest energy
of the spectrum, and are of extragalactic origin. The heaviest elements need to
be in the mass range between nitrogen and silicon to describe the Auger data,
and it is assumed that almost no light elements are accelerated in the sources.

Neither the maximum-rigidity scenario nor the photo-disintegration scenario provide 
an explanation of the ankle. An extra component, possibly of galactic origin,
is required to generate the ankle in these models.

\paragraph{Proton-dominance model~\cite{Berezinsky:2002nc,Berezinsky:2005cq}} This well-known
model, also often referred to as the dip model, provides a natural explanation of
the ankle as the imprint of $e^+ e^-$ pair production. In this model the
all-particle flux consists mainly of extragalactic protons at all energies higher
than $\unit[10^{18}]{eV}$. The suppression of the
spectrum at the highest energies is attributed solely to pion-photoproduction.
Fig.~\ref{fig:all-particle-flux}~(right) shows the best fit of this model 
to the Auger flux data; it shows that a maximum injection
energy much higher than $\unit[10^{20}]{eV}$ is only marginally compatible with
the Auger data within the systematic uncertainties. A source cutoff energy just
below $\unit[10^{20}]{eV}$ would improve the description of the spectrum data.
Such a low source cutoff energy
would also imply that part of the observed suppression of the all-particle
flux would be related to the details of the upper end of source spectra.
And, of course, new
particle physics would be needed to describe the \xmax data with a
proton-dominated flux.\\[3mm]

\begin{figure}[t]
\centering
\includegraphics[width=0.495\columnwidth]{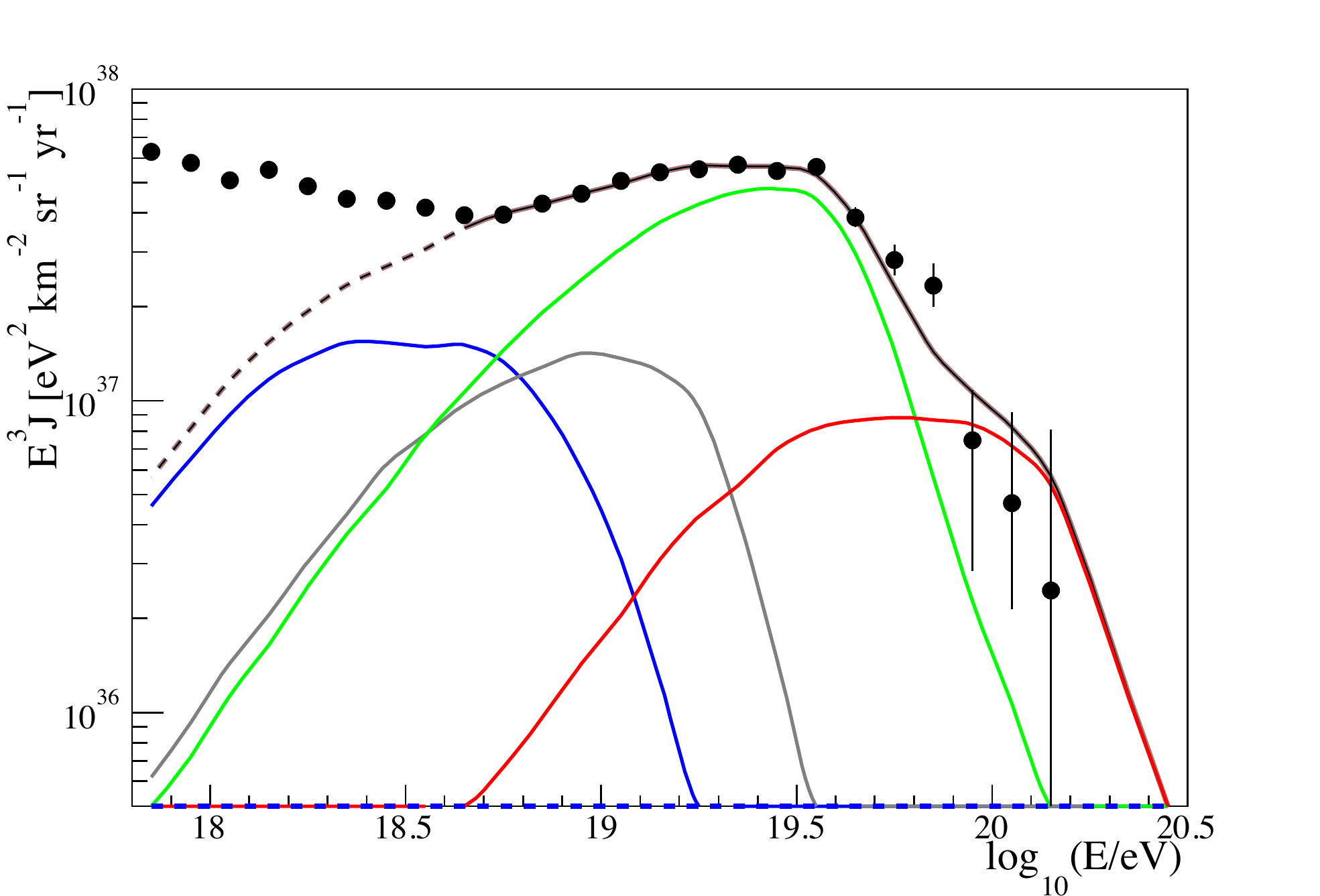}
\includegraphics[width=0.495\columnwidth]{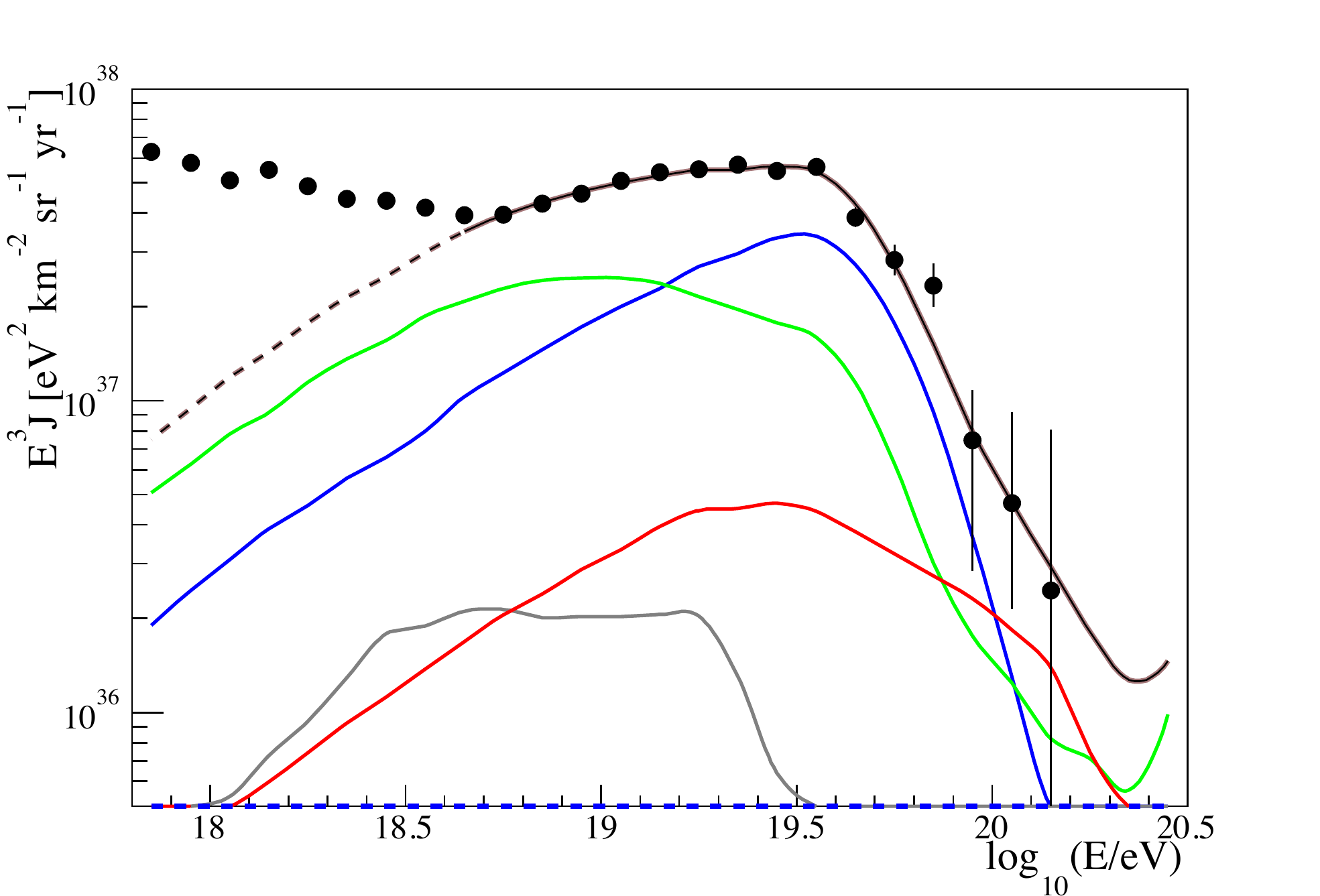}
\caption{
Examples of fluxes of different mass groups for describing the Auger spectrum and
composition data. Shown are the fluxes of different mass groups
that are approximations of one  maximum-rigidity scenario (left panel) and one
photo-disintegration scenario (right panel). The
colors for the different mass groups are protons -- blue, helium --
gray, nitrogen -- green, and iron -- red.  
The model calculations were done with SimProp~\protect\cite{Aloisio:2012wj},
very similar results are obtained with CRPropa~\protect\cite{Kampert:2012fi}.
}
\label{fig:models-flux}
\end{figure}

Representative examples of descriptions of the latest Auger flux data
within the maximum-rigidity and photo-disintegration models are shown in
Fig.~\ref{fig:models-flux}. A numerical fit was made to optimize the
description of the all-particle flux and the \xmax distributions in
the different energy intervals.  For sake of simplicity we have
assumed homogeneously distributed sources injecting identical
power-law spectra of energy-independent mass composition.  The index
of the injection power law, the maximum energy of the particles
injected by the sources, and the source composition were free
parameters.  Even after accounting for the systematic uncertainties,
it is difficult to obtain a satisfactory
description of the flux and composition data of the Auger Observatory
with these approximations.  The best description is obtained for a
hard source spectrum ${\rm d}N/{\rm d}E \sim E^{-1}$ and a low cutoff
energy of $E_{\rm cut}\sim 10^{18.7}$\,eV for protons at the
source. The cutoff energies of the other primaries are taken to scale
in proportion to their charge. This parameter set corresponds to a
good approximation to a ``maximum-rigidity scenario.''  A somewhat
better description of the Auger data, in particular the \xmax
fluctuations at high energy, can be obtained if an additional light
component is assumed to appear in a limited energy range. 

\begin{figure}[htb!]
\centering
\includegraphics[width=0.5\columnwidth]{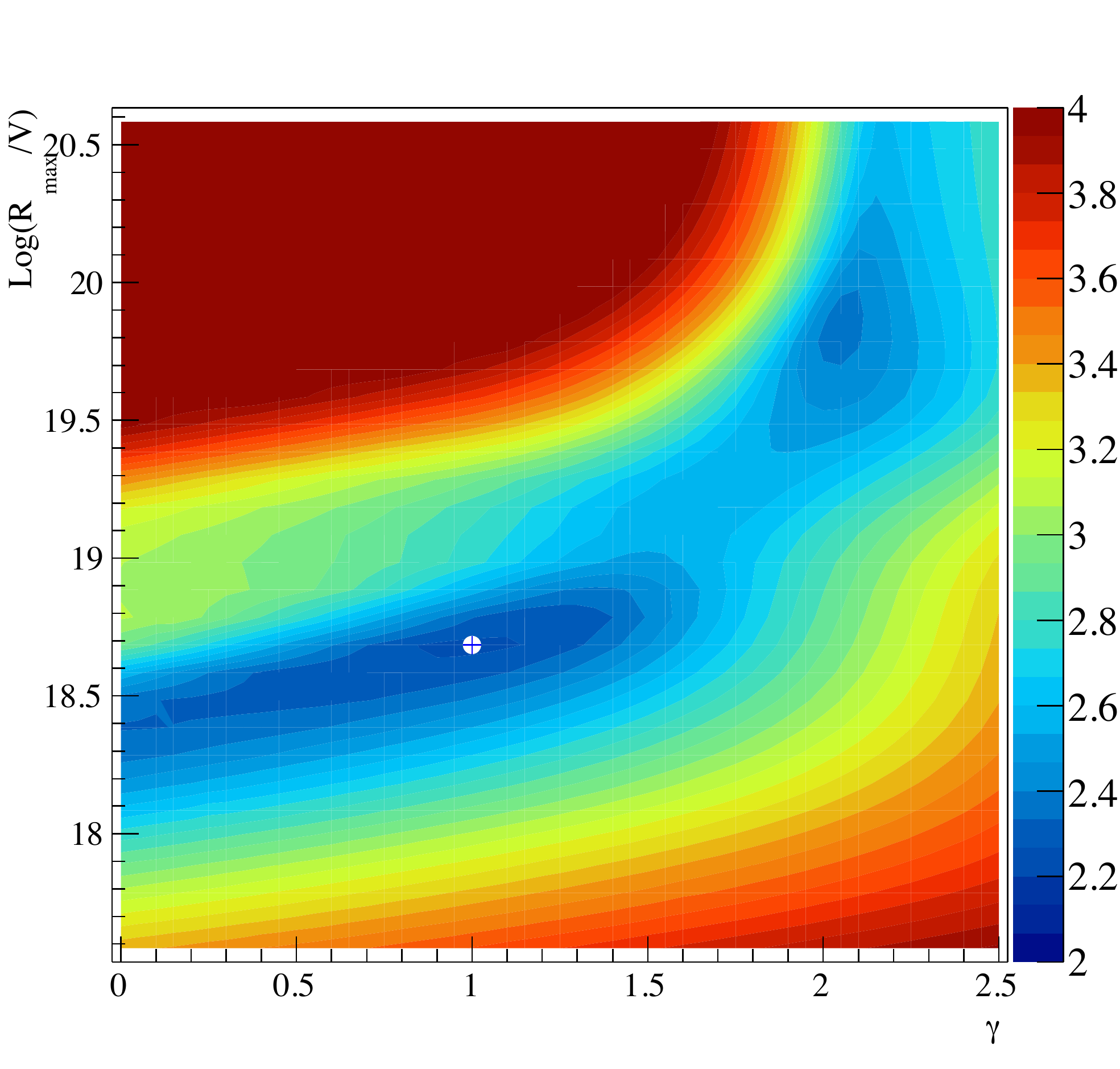}
\caption{ Two-dimensional projection of the parameter space
  illustrating the goodness of data description. The colors denote
  $\log_{10}(D_{\rm min}$), where $D_{\rm min}$ is the minimum
  log-likelihood and is approximately $\chi^2$ distributed with the
  number of degrees of freedom 124.  The $\log_{10}(D_{\rm min})$ for
  a four-component source composition is shown as function of the
  injection index $\gamma$ and the maximum rigidity $R_{\rm max}$
  above which an exponential suppression of the source flux is
  assumed. }
\label{fig:Dmin-gamma-Ecut}
\end{figure}

The quality of data description is shown in
Fig.~\ref{fig:Dmin-gamma-Ecut} as function of the two-di\-men\-sion\-al
parameter space of the injection index and maximum proton energy.
There is a wide range of possible parameter combinations, and given
the simplifications of the source model, one should not over-interpret
the numerical values of the parameters, nor the actual values of
the minimum.  It is interesting to note that there is a second local
minimum, although disfavored in this simple model scenario, which
corresponds to an injection index compatible with ${\rm d}N/{\rm d}E
\sim E^{-2}$, i.e.\ Fermi acceleration.  The second minimum is an
example of the ``photo-disintegration scenario.''

It should be noted that both the maximum-rigidity and the 
photo-dissociation scenarios require a composition of the particles
injected by the sources that is heavier than the composition of
Galactic cosmic rays. While the fraction of heavy elements in the
source flux has to be enhanced by a factor of a few in the case of the
maximum-rigidity model~\cite{Allard:2008gj,Allard:2005ha}, essentially only
nuclei of the nitrogen or silicon groups have to be injected by the
sources to describe the Auger data within a photo-disintegration
scenario~\cite{Hooper:2009fd,Taylor:2011ta}. In other words, the Auger
data require a very unusual metallicity of the sources, or a change of
the properties of hadronic interactions at the highest
energies~\cite{Shaham:2012tx}.

We have presented here very different scenarios for interpreting the
Auger data. Of course, a steady transition between these models, as
well as a superposition of them, is possible.  For example, by
adjusting the maximum injection energy of the sources, the importance
of energy losses during propagation relative to that of the
rigidity-dependent cutoff of the source spectra can be changed.  In
general, it is possible that both the maximum-rigidity effect and the
energy loss processes are important for shaping the flux, composition,
and arrival direction distribution observed at Earth.

There are many other scenarios which we will not discuss here. These
include, for example, models that place transient sources in the
Galaxy~\cite{Calvez:2010uh}, scenarios in which Galactic and
extragalactic neutron stars are the
sources~\cite{Kotera:2011vs,Fang:2013cba}, or
Cen-A~\cite{Biermann:2011wf} (scaling of the maximum energy of
individual elements proportional to $Z$), and models of vacuum
Cherenkov radiation that lead to a flux scaling which is proportional
to the particle velocity and, hence, mass number
$A$~\cite{Klinkhamer:2010pq}.

We conclude from these considerations that the origin of the flux
suppression observed in the all-particle spectrum is not understood.
Furthermore, anisotropy and composition data are compatible with the
hypothesis of an additional proton component appearing at very high energy
($E>\unit[4{\times}10^{19}]{eV}$), but mainly due to the lack of
composition data no conclusion can be drawn.

However, it should not be forgotten that the interpretation of the
Auger data in terms of composition does rely on the accuracy of
modeling  air showers and, in particular, hadronic multiparticle
production. Therefore all these possible astrophysical interpretations
have to be considered in the context of our current understanding of
hadronic interactions. It is not excluded that changes of hadronic
interaction models within the limitations given by accelerator data
can lead to a different  interpretation of our composition-sensitive
measurements~\cite{Ulrich:2010rg,Farrar:2013sfa,Allen:2013hfa}. In addition, it is
possible that the overall features of hadronic interactions are
significantly different at energies, and in phase space regions, not
accessible to current colliders. Such a deviation could be related to
new particle physics or just an unreliable extrapolation of existing
data.


\subsection{Goals of the detector upgrade}

\noindent
It is planned to operate the Pierre Auger Observatory until end of 2024.
This will triple the statistics of the data set presented in this
proposal which contains all data until the end of 2012. However,
increasing the statistics of the measured showers will not be sufficient to answer
the key questions outlined in the previous section. Therefore, we aim
at an upgrade of the Pierre Auger Observatory to ensure that the data
collected after 2017 will provide additional information to allow us
to address the following questions.

\begin{enumerate}

\item
The primary objective of the upgrade of the Auger Observatory is to
elucidate the mass composition and 
the origin of the flux suppression at the highest energies,
i.e.\ the differentiation between
the energy loss effects due to propagation and the maximum energy of
particles injected by astrophysical sources. This is a natural
evolution and major step forward from the original objective of the
Pierre Auger Observatory, which was motivated primarily by the
question of the existence of a GZK-like flux suppression.
Understanding the origin of the flux suppression will provide
fundamental constraints on the astrophysical sources
and will allow much more reliable estimates of neutrino and gamma-ray
fluxes at ultra-high energy for which we will continue to search.

\item
The search for a flux contribution of protons up to the highest
energies will be the second key science objective. We aim to reach a
sensitivity to a contribution as small as 10\% in the flux
suppression region. The measurement of the fraction of protons is the
decisive ingredient for estimating the physics potential of existing
and future cosmic ray, neutrino, and gamma-ray detectors; prospects
for proton astronomy with future detectors will be
clarified. Moreover, the flux of secondary gamma-rays and neutrinos
due to proton energy loss processes will be predicted.

\item
Determining the mass composition of ultra-high energy cosmic rays is
closely related to, and crucially depends on, understanding extensive
air showers and hadronic interactions. When estimating the number of
muons in air showers from Auger data, a discrepancy between the
observed and expected muon numbers is found. Therefore, the third key
science objective will be the study of extensive air showers and
hadronic multiparticle production. This will include the exploration
of fundamental particle physics at energies beyond those accessible at
man-made accelerators, and the derivation of constraints on new
physics phenomena, such as Lorentz invariance violation or extra
dimensions~\cite{PhysRevD.78.085026,PhysRevD.77.016002,Klinkhamer:2010pq}.

\end{enumerate}

To accomplish these science objectives it will be of central
importance to improve the composition sensitivity of the Auger
Observatory and to extend it into the energy region of the flux
suppression. The most promising way to obtain further
composition-sensitive information is the discrimination between the
electromagnetic and muonic components of the shower with ground-array
measurements.

It is clear that obtaining additional composition-sensitive
information will not only help to better reconstruct the properties of
the primary particles at the highest energies. Also, measurements
in the important energy range just above the ankle will greatly profit from
the additional observables. Furthermore, it is expected that the
additional composition-sensitive information will help to reduce
systematic uncertainties related to modeling hadronic showers and to 
limitations of reconstruction algorithms.


\section{Importance of determining the muonic shower component}

\noindent
Optical observations of \xmax can presently be performed only with a
duty cycle of about 15\%. Therefore, the largest boost in performance
towards the aforementioned science goals is expected from an improved
composition sensitivity of the surface detector array with its 100\%
duty cycle. Such an improvement can be reached by measuring, in
addition to the total particle signal, the decomposition between electromagnetic
and muonic components at ground level.

Currently, information on the muonic component of the bulk of air
showers is obtained by employing indirect methods which lack the desired
precision, and moreover require validation through direct observation.
Muons lead to characteristic peaks in the time trace of the water-Cherenkov surface
detectors~\cite{KeglICRC2013}, and their number can be estimated for example by
subtracting the signal of the electromagnetic shower component
obtained from shower universality
considerations~\cite{Schmidt:2007vq,Lafebre:2009en,Ave:2011x1}
(see also below). Only for sufficiently inclined showers, or at large
lateral distance from the shower core, can the muon component be
measured
directly~\cite{Cazon:2004zx,Aab:2014dua,Aab:2014pza}.
A shower-by-shower measurement of the muon component would allow us to
\begin{compactitem} 
\item extend the composition measurement into the
  flux suppression region to eventually distinguish
  different model scenarios for understanding the origin of the flux
  suppression. 
\item estimate the primary mass and charge on a
  shower-by-shower basis. This would enable us to select light
  elements and perform composition-enhanced anisotropy studies.
  Moreover, back-tracking of particles through the Galactic magnetic
  field can also be done for particles of different charges in directions
  of low integrated field strength. 
\item study hadronic interactions at high energy,
  understand the observed muon discrepancy, and discriminate between
  different exotic interaction model scenarios. Furthermore, we could
  study systematic uncertainties by performing measurements with
  different observables and derive consistency checks on models. 
\item
  improve the current photon and neutrino sensitivity not only by
  collecting more statistics, but also by having a much improved
  discrimination power.
\item understand better, and reduce the
  systematic uncertainties of, many different measurements including
  the all-particle flux and the cosmic ray composition measurement.
\end{compactitem}
 The key question is whether we can use additional information
 on the separation between the electromagnetic and muonic shower
 components for improving the estimate of the mass of the primary
 particles.

\begin{figure}[t]
\centering
\def\figh{0.37}
\includegraphics[height=\figh\textwidth]{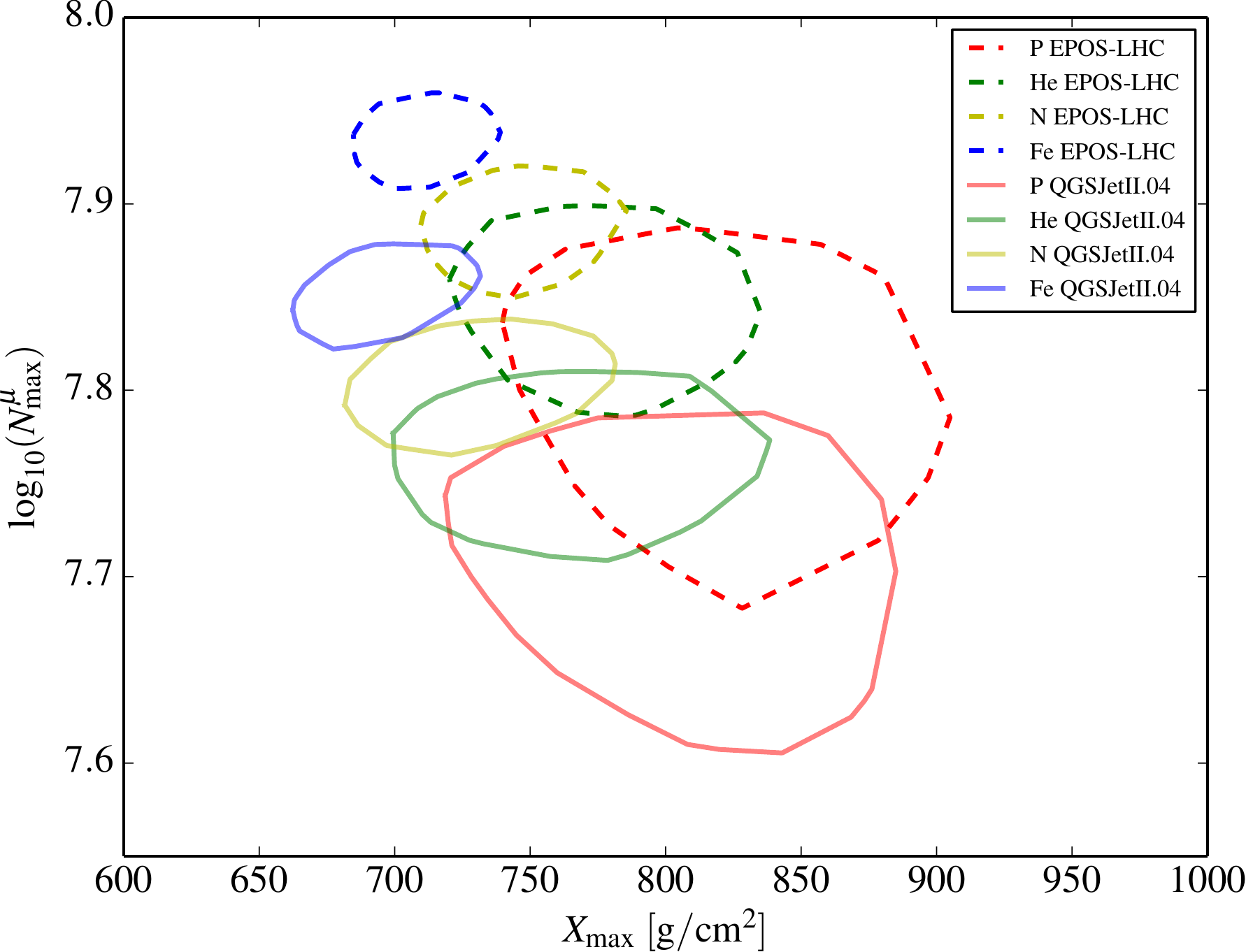}\hfill
\includegraphics[height=\figh\textwidth]{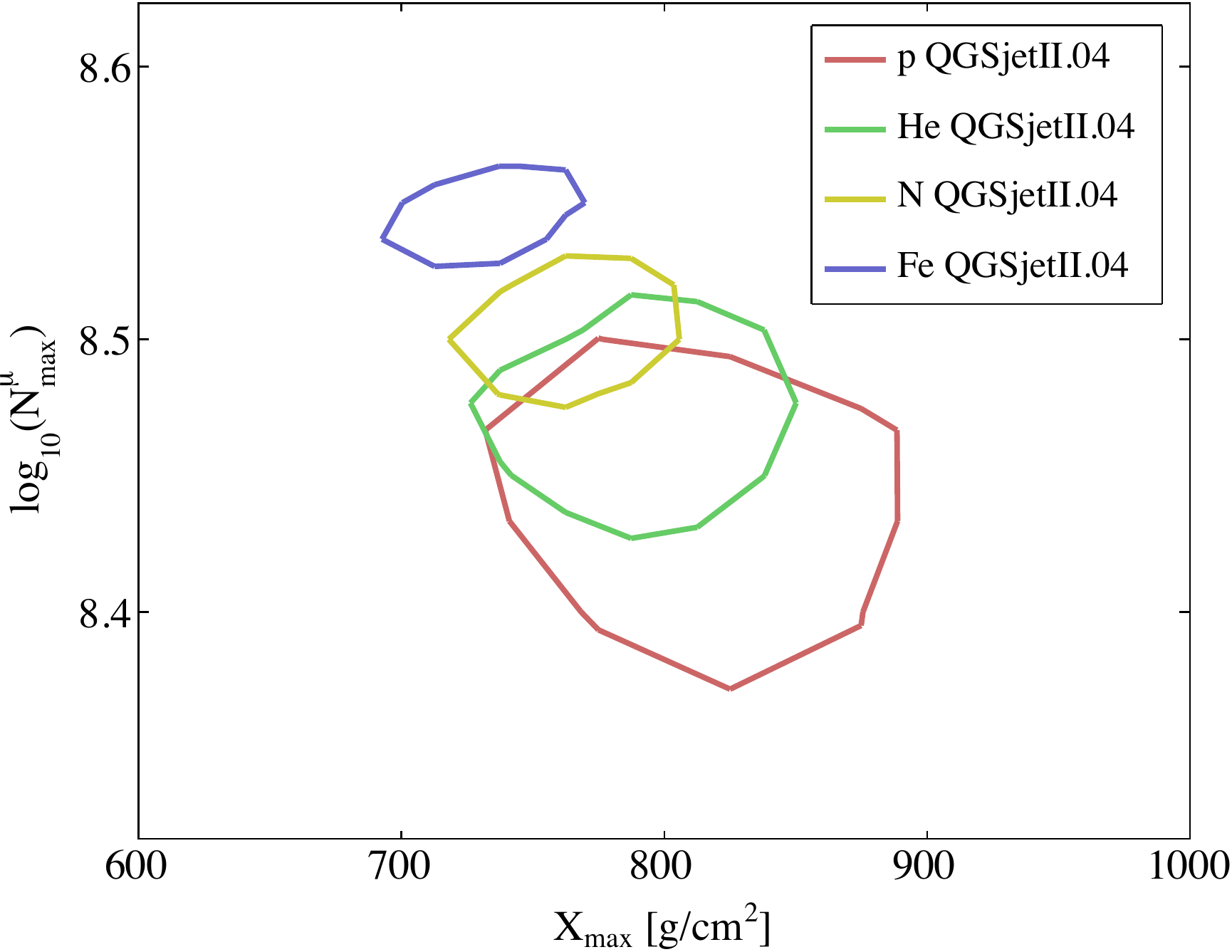}
\caption{The 1$\sigma$
   contour of the number of muons at maximum of the muon shower
   development, $\log_{10}N_\text{max}^\mu$, vs the depth of shower
   maximum, \xmax, for fixed energies, $E=\unit[10^{19}]{eV}$
   (left) and $E=\unit[5{\times}10^{19}]{eV}$ (right), and fixed zenith
   angle, $\theta=38^\circ$.}
\label{fig:NmuVsXmax}
\end{figure}

\begin{figure}[t]
\centering
\def\figh{0.37}
\includegraphics[height=\figh\textwidth]{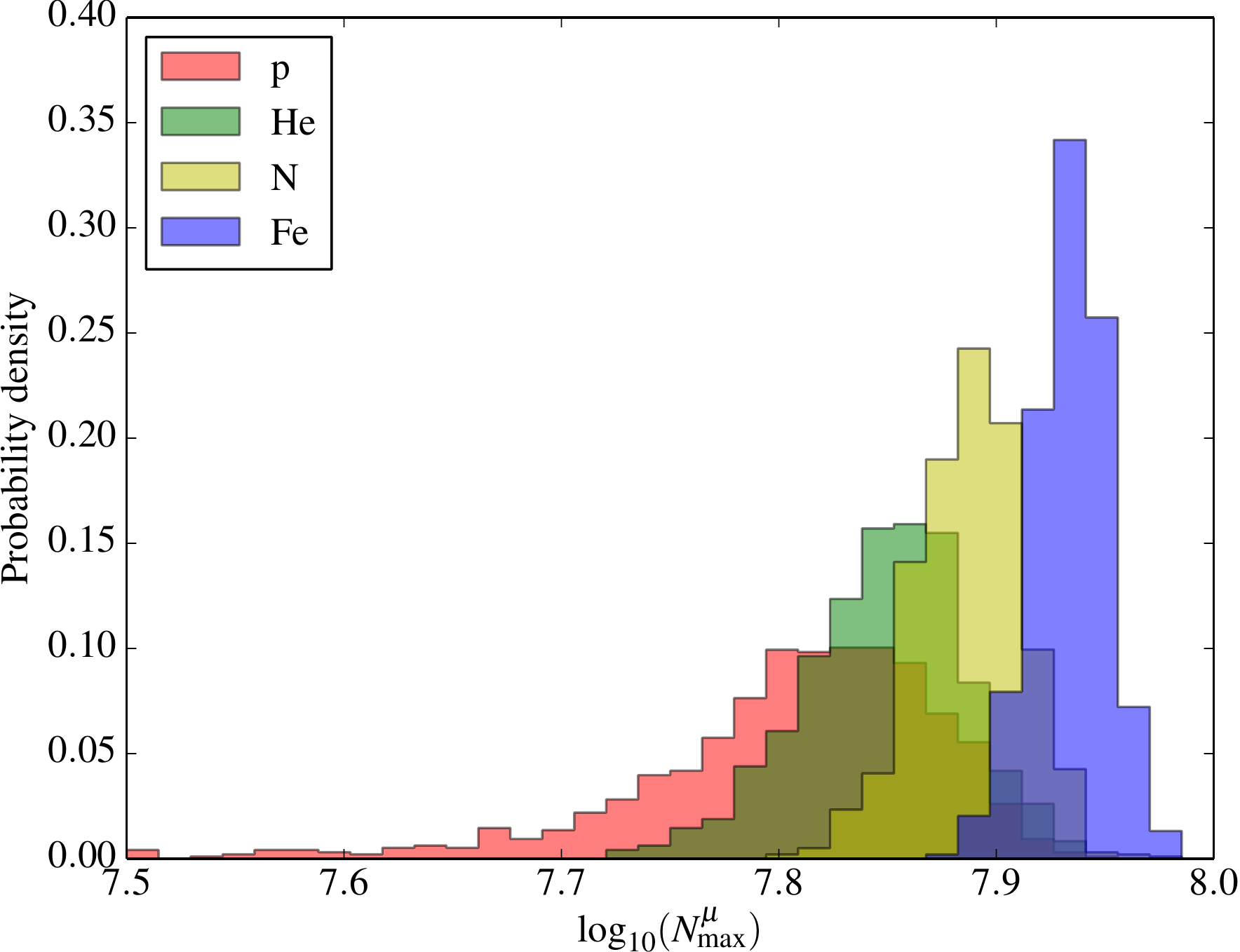}\hfill
\includegraphics[height=\figh\textwidth]{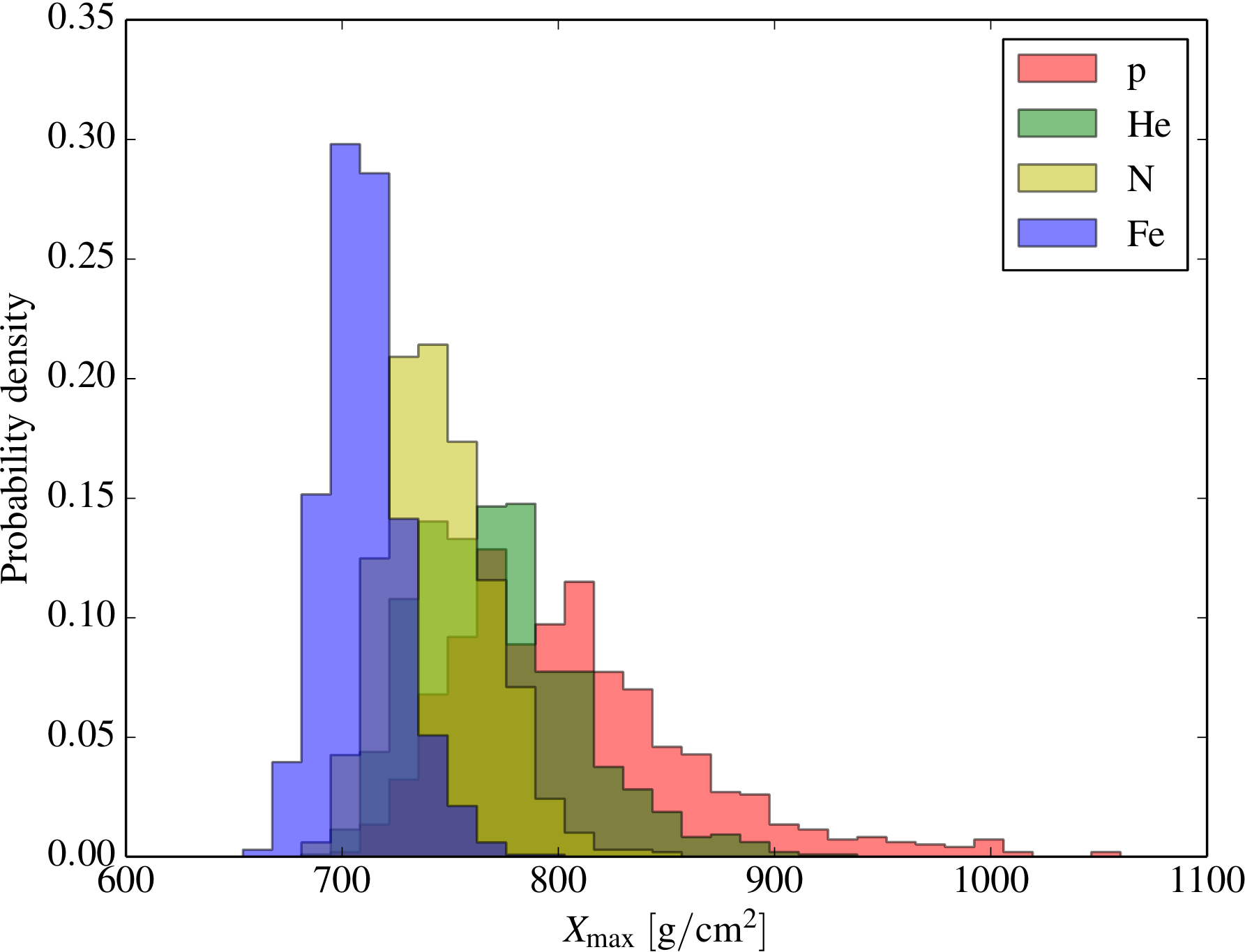}
\caption{Number of muons at maximum of the muon shower development,
   $\log_{10}N_\text{max}^\mu$ (left) and depth of shower maximum,
   \xmax, for fixed energy, $E=\unit[10^{19}]{eV}$, and fixed
   zenith angle $\theta=38^\circ$ (EPOS-LHC is the generator
   for hadronic interactions).}
\label{fig:NmuAndXmax}
\end{figure}

The simulated number of muons at maximum of the muon shower
development,\\ $\log_{10}N_\text{max}^\mu$, versus the shower maximum
\xmax at $\unit[10^{19}]{eV}$ ($\unit[5{\times}10^{19}]{eV}$) and $38^\circ$ of zenith angle, as well as the
marginal distributions are displayed in Figs.~\ref{fig:NmuVsXmax}
and~\ref{fig:NmuAndXmax}. 

The difference in $\log_{10}N_\text{max}^\mu$ and \xmax for
the two most recent incarnations of LHC tuned models (EPOS-LHC and
QGSJetII.04) are of the order of $\Delta
\log_{10}N_\text{max}^\mu\approx 0.1$ and $\Delta
X_\text{max}\approx\unit[15]{g/cm^2}$ independent of the primary.
Within the frame of a single model, a clear separation of light and
heavy primaries seems possible. Even intermediate primaries like
nitrogen can be separated from protons and helium if the recorded
statistics permit. Overall, the expectations from air shower
simulations strongly indicate the feasibility of composition
determination at the highest energies. It can be expected
that, if the detector resolution is smaller or of
the order of the shower fluctuations, the primary mass can be
inferred on an event-by-event basis.

The fact that the average properties of the cascade can, to a large
extent, be described in terms of energy and shower age only is called
\emph{shower universality}, see \cite{Lipari:2008td} and Refs.\ therein. To first
approximation there is no direct dependence on the primary mass nor
zenith angle. This is a very remarkable result. Despite the vast
number of interactions in an air shower, its overall shape as well as
the time profiles of particles reaching ground can be described very
well with very few measurable quantities. In the literature it has been
described for the electromagnetic component of
showers~\cite{Hillas:1982vn,Giller:2005qz,Nerling:2006yt,Lafebre:2009en,Lipari:2008td}.
The concept can also be extended to hadronic showers by introducing
one additional parameter, the muon scale
$N_\mu$~\cite{Schmidt:2007vq,Ave:2011x1,Maurel:ICRC2013}. The result
is a model that describes showers initiated by protons, nuclei up to
iron as well as photon showers using only three parameters: $E$, \xmax
and $N_\mu$. Based on the signal and timing information in individual
SD stations we have encouraging results on event-by-event
determination of the primary mass exploiting shower universality
features to decompose the relative abundances of shower components,
e.g.~the muon content.  Nevertheless, these results are based on Monte
Carlo parameterizations only, which eventuate in large systematic
uncertainties and call for a significant step forward in a direct
measurement of individual components of air-shower events.

In the following we will show some examples to demonstrate the
improved physics power of an enhanced surface detector array.
%
%
\subsection{Mass composition and anisotropy\label{sec:muons-universality}}
We can already use parameterizations based on shower
universality~\cite{Ave:2011x1,Maurel:ICRC2013} to analyze our data at
high energy. A fit of parameterized shower components to the time
traces of the water-Cherenkov detector signals of a high-energy shower
yields estimates of the depth of shower maximum, and either an
unbiased energy estimator or the muon number, without referring to any
data from the fluorescence telescopes (see
Chap.~\ref{chap:performance}).  This approach can extend our
measurements of \xmax to energies higher than the fluorescence
detector. However, due to the lack of direct muon information, the
method currently relies heavily on shower simulations and is
characterized by systematic uncertainties that are difficult to
estimate. In addition, the large correlation between the SD-based
estimates for the shower energy and muon number limits the composition
sensitivity. It is clear, however, that improvements in the systematic
uncertainties and potential model sensitivity -- as addressed by this
upgrade proposal -- are needed to obtain reliable estimates of the
primary mass composition.

Applying cuts on the estimated depth of shower maximum (and muon
number or other composition sensitive parameters if available) one can
pre-select shower candidates of light and heavy primary particles and
perform composition-improved anisotropy studies.

\subsection{Physics of air showers and hadronic interactions}
\begin{figure}[t]
\centering
\def\figh{0.32}
\includegraphics[height=\figh\columnwidth]{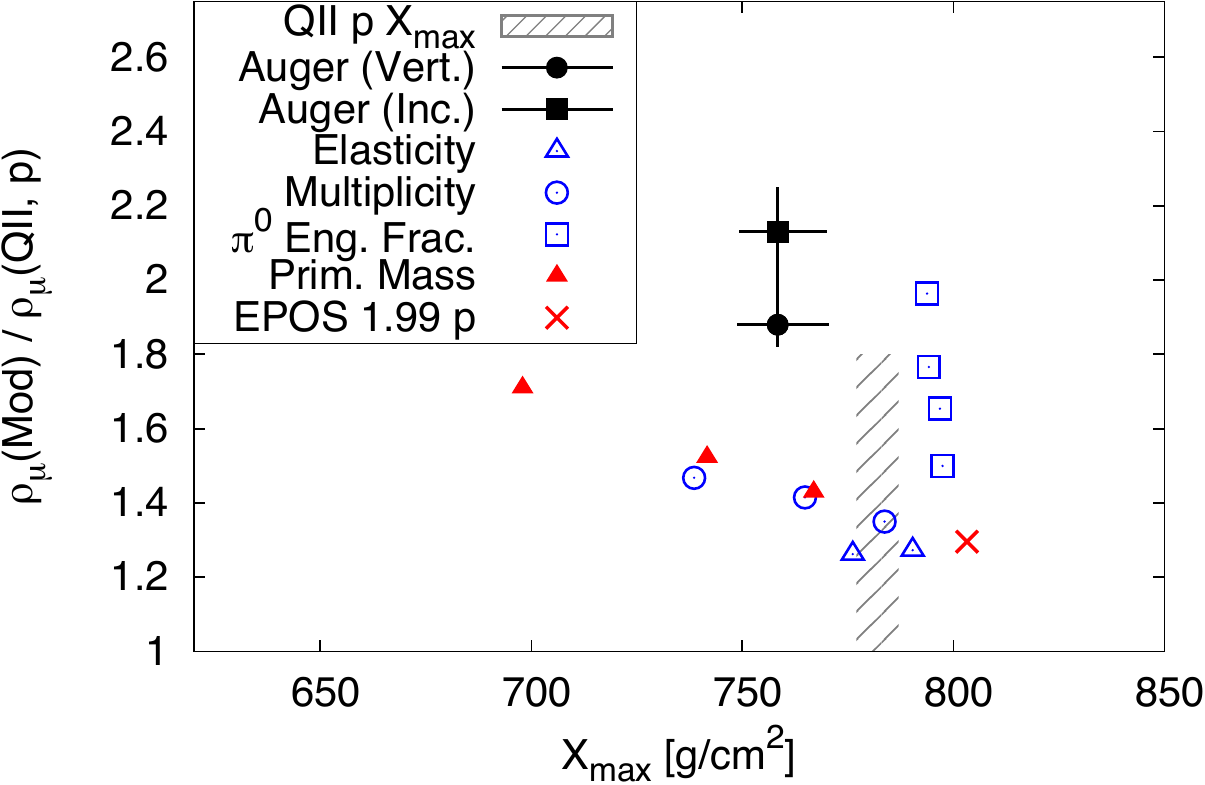}\hfill
\includegraphics[height=\figh\columnwidth]{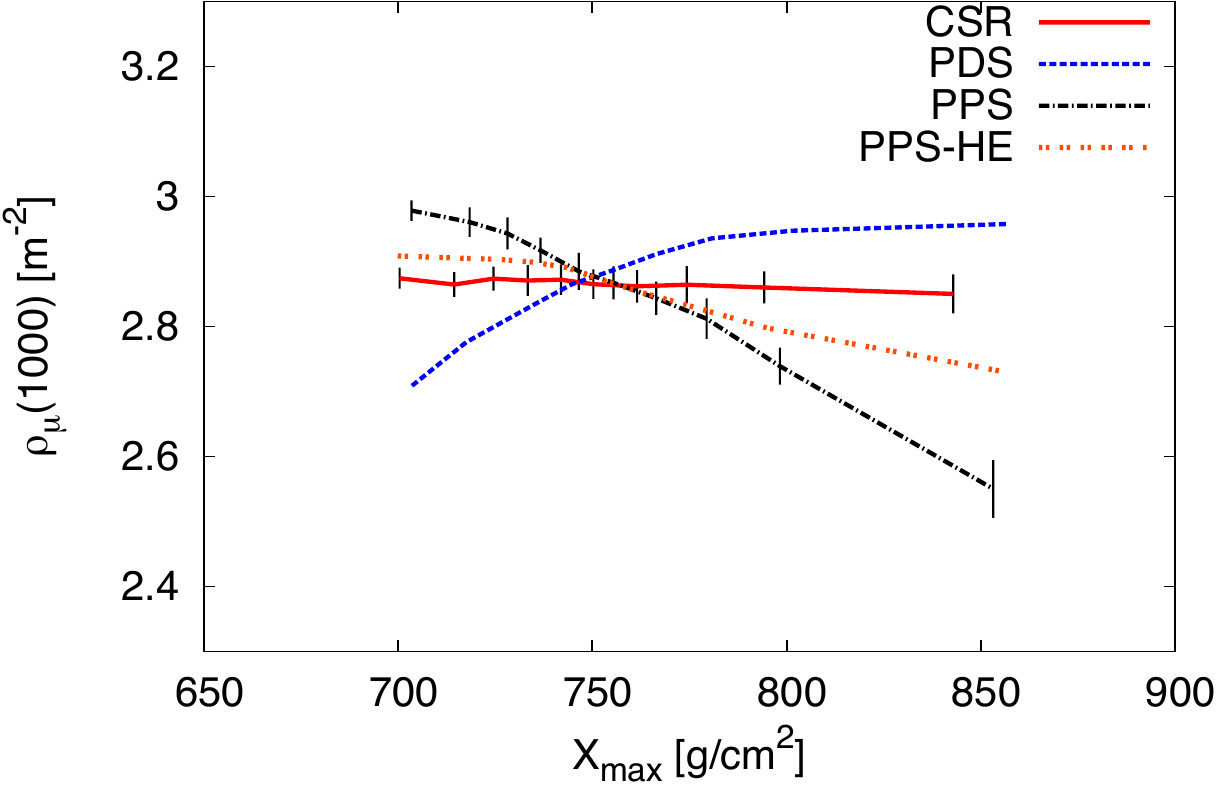}
\caption{Discrimination power of the event-by-event correlation
    between the muonic signal at ground and the depth of shower
    maximum \xmax~\cite{Allen:2013hfa}. Left panel: Relative
    number of muons at $\unit[1000]{m}$ from the shower core and
    \xmax for EPOS~1.99 and QGSJET~II.04 and modified versions of
    it (see text). The Auger data are also shown as derived from
    showers of $\unit[10^{19}]{eV}$ with zenith angles smaller (larger)
    than $60^\circ$. Right panel: Mean shower-by-shower correlation of
    the number of muons and \xmax for different exotic
    interaction model scenarios. The scenarios are CSR -- chiral
    symmetry restoration, PPS -- pion production
    suppression, PDS -- pion decay suppression, and PPS-HE -- pion
    production suppression at high energy~\cite{Farrar:2013sfa}.}
\label{fig:Nmu-Xmax}
\end{figure}

Having direct muon information will also greatly enhance our
capabilities of studying hadronic interactions. In particular the
shower-by-shower correlation of the depth of shower maximum with the
number of muons at ground level has proven to be a very powerful observable
to distinguish different conventional or exotic interaction
scenarios~\cite{Allen:2013hfa}. This can be understood by realizing
that the depth of shower maximum is mainly determined by the secondary
particles of high energy produced in the first few interactions of the
cascade. In contrast, muons are produced only if pions decay, which is
only the case at low energy. A simulation study for different
modifications of hadronic interaction models is shown in
Fig.~\ref{fig:Nmu-Xmax}.

\begin{figure}[th]
\centering
\def\figh{0.35}
\includegraphics[height=0.38\columnwidth]{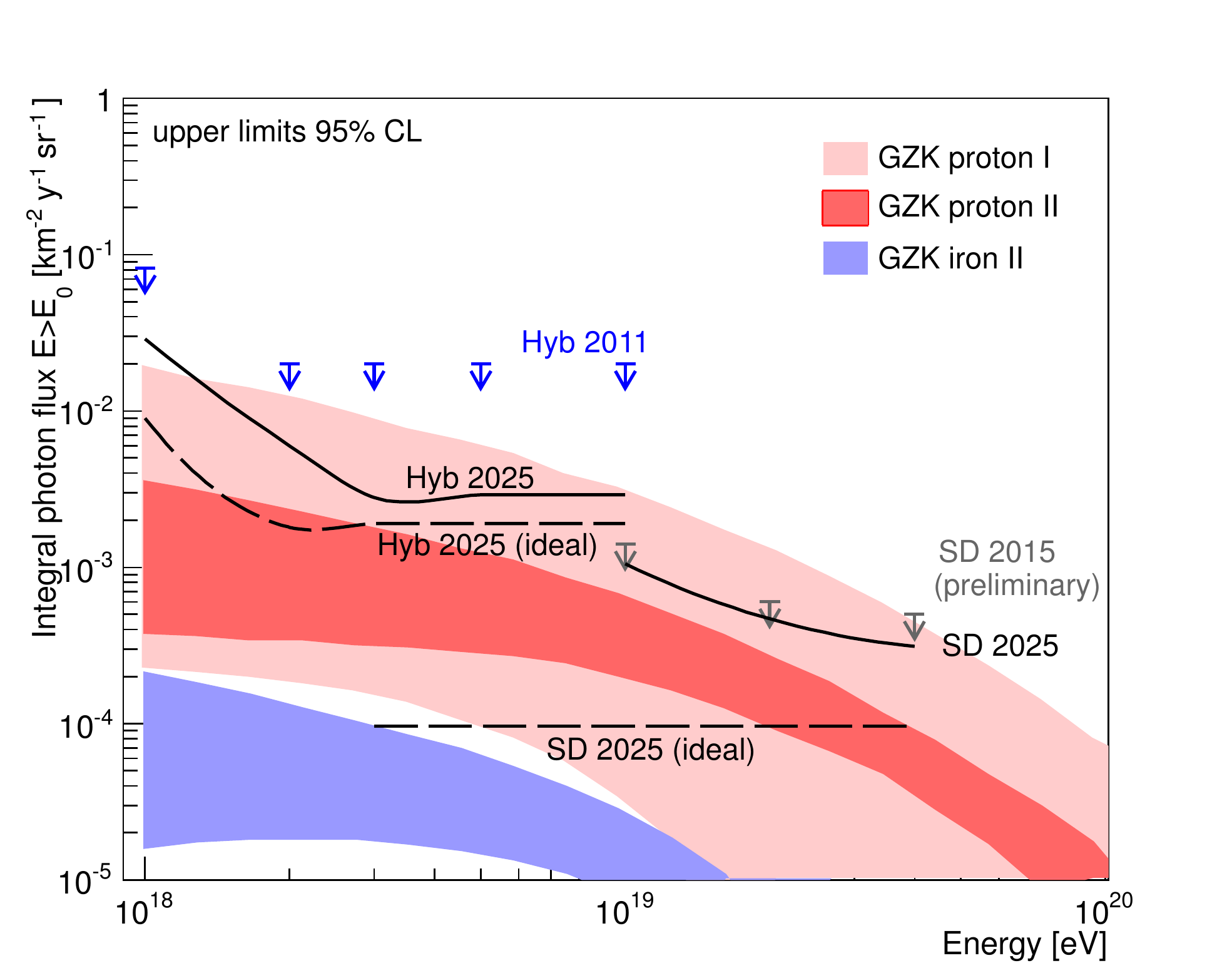}\hfill
\includegraphics[height=\figh\columnwidth]{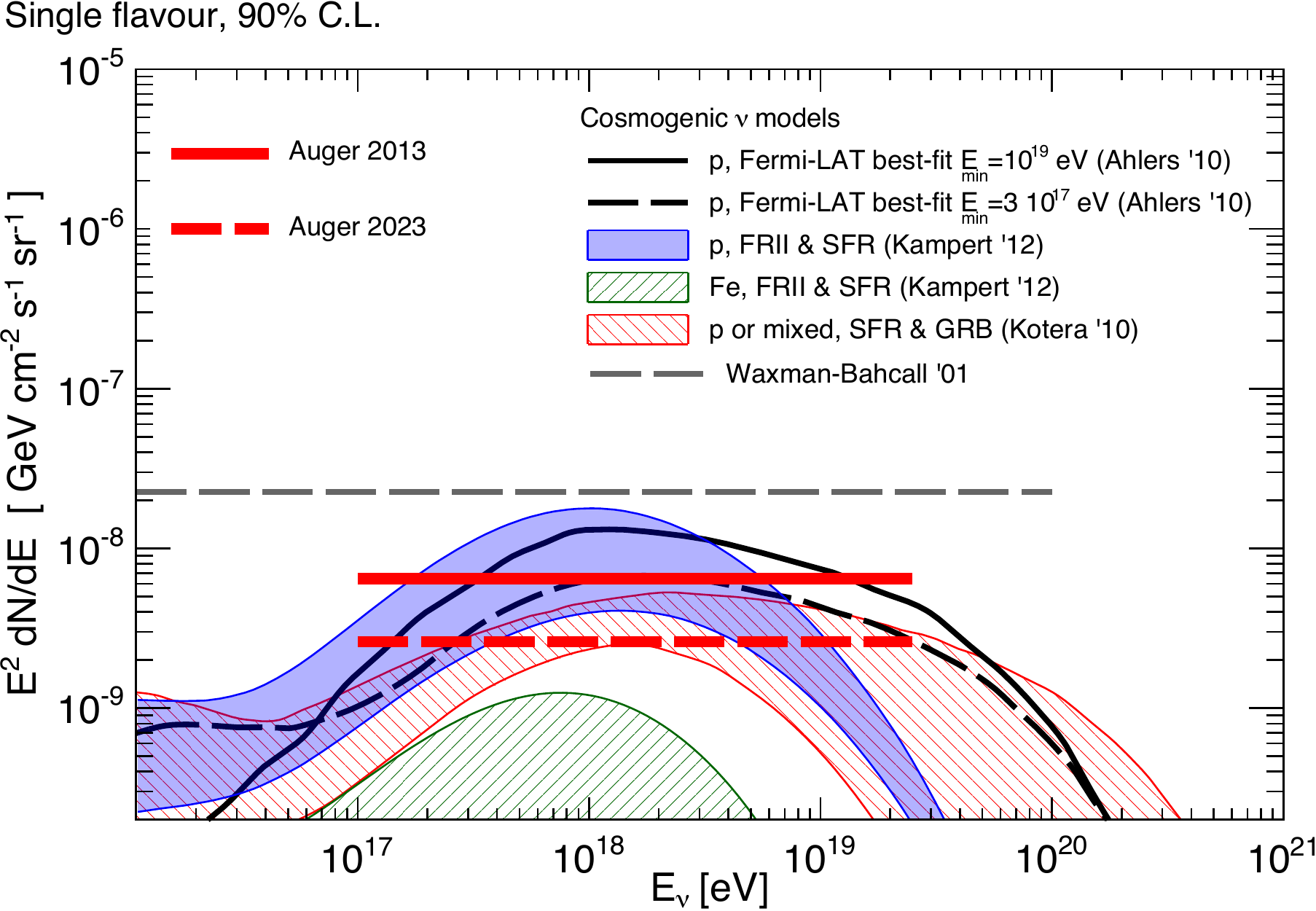}
\caption{ Expected sensitivity on the flux of photons and neutrinos.
  In addition to the conservative estimates based on the increase of
  statistics, also the projected photon sensitivity for the ideal case
  of being able to reject any hadronic background due to the upgraded
  surface detector array is shown.}
\label{fig:photon-neutrino-limits-extrapolated}
\end{figure}

 Already the comparison of the mean depth of
shower maximum with the mean muon number provides strong constraints
on the interaction model. The simulations also demonstrate how
different scenarios of modified hadronic interactions can be
distinguished if the event-by-event correlation of $N_\mu$ and \xmax
can be measured. For details
see~\cite{Farrar:2013sfa,Allen:2013hfa}.

\subsection{Upper limits on photon and neutrino fluxes}

In case no positive photon or neutrino signal is found, the limits will
improve relative to the current values for several reasons:
\begin{itemize}
\item
The statistics of the events available for determining the 
limits will triple relative to the data collected by the end of 2012.
\item
In 2013 two new trigger algorithms (ToTd and MoPS) have been
added to the local station software of the water-Cherenkov stations to lower the
trigger threshold, in particular for signals dominated by
the electromagnetic component. As a result, there will be more stations
contributing to the typical shower footprint, improving
the reconstruction and, for example, photon/hadron separation at low energies in particular.
New station electronics, as foreseen for the upgrade
(see Sec.~\ref{sec:sdeu}), will allow us to improve
the triggering algorithms further.
\item
At the present time, the photon limits are no longer background
free. Improved muon discrimination will help to reduce the background
due to hadronic events in our photon candidate sample, or to identify
photons and neutrinos.
\end{itemize}

The analyses of the impact of the improved triggering algorithms and
composition sensitivity are still underway. Therefore, we show in
Fig.~\ref{fig:photon-neutrino-limits-extrapolated} (dashed lines) the
maximum reach allowed by exposure until 2024. In case of the photon
studies a selection efficiency of 50\% (due to the a-priori cut) is
assumed. Also, the hypothesis that a perfect background rejection
after the detector is upgraded and that the new triggers are fully
exploited is taken. These lines have to be interpreted as a boundary
of what we can do in an optimistic case. The improvement compared to
the simple extension of the current data analysis (solid line) until
2024 is significant. The extension of the energy range for current
limits below 10 EeV is due to the new triggers. The predicted hybrid
limits include the exposure gained with the extended duty cycle.
The limits are compared to theoretical predictions (photons: GZK
fluxes proton I~\cite{Gelmini:2005wu}, proton \& iron
II~\cite{SarkarICRC2011}; neutrinos: AGN~\cite{Becker:2007sv},
Waxman-Bahcall flux~\cite{Waxman:1998yy,Bahcall:1999yr}, cosmogenic
neutrino fluxes~\cite{Ahlers:2010fw,Kotera:2010yn,SarkarICRC2011}).

By 2024 we expect to lower our photon
limits to reach the band of even conservative predictions for GZK photons -- or discover
ultra-high energy photons. It is expected that the limits will improve
further, mainly at the low-energy end, due to optimized trigger
algorithms. If we were able to reject our current photon
candidates due to improved analysis algorithms these limits could be much stronger.


\subsection{Methods for determining the muonic shower component}
\label{sec:muon-methods}

There are different methods of measuring the density of muons as function of
the lateral distance. The most direct method, of using detectors sufficiently
shielded to absorb the electromagnetic shower component by, for example,
burying them under a layer of soil, is not feasible for the 1660 surface detectors of
the Auger Observatory. Therefore, we have chosen to add another particle detector
to each water-Cherenkov detector of the surface array taking advantage of the fact that the two detectors
have a different response to the electromagnetic and muonic shower components.
Comparing the signal traces of the two detectors allows us to derive the
contributions due to electromagnetic and muonic particles and thus to reconstruct
the number of muons.

\begin{figure}[t]
\centering
\includegraphics[width=0.92\columnwidth]{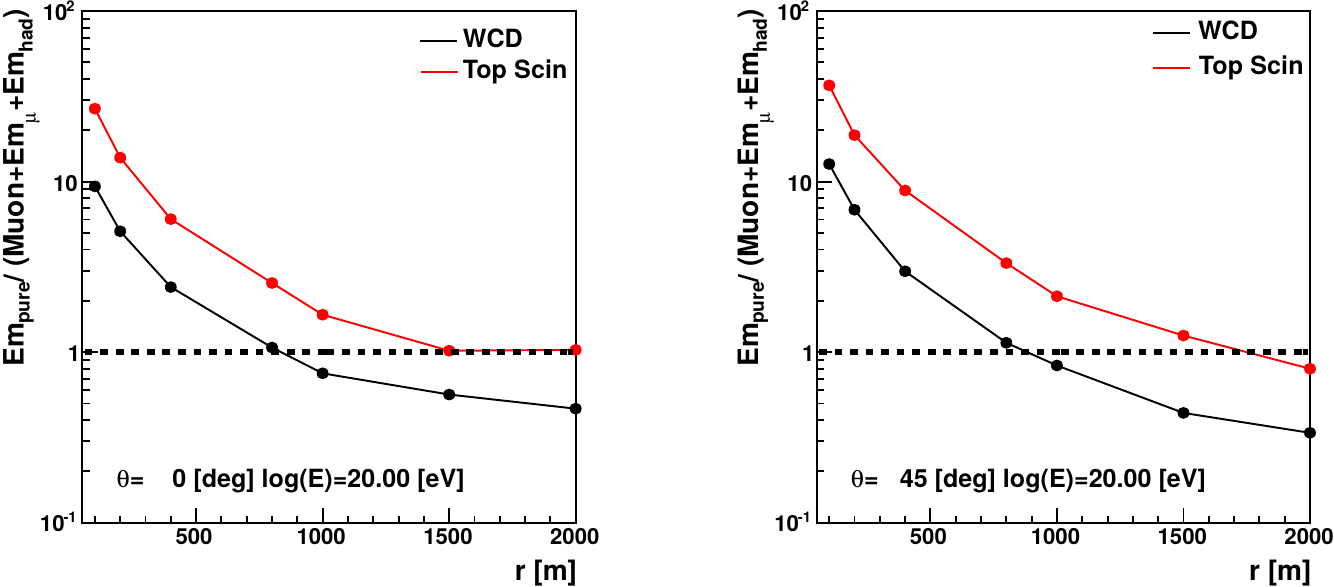}
\caption{ Ratios of different contributions to the integrated signal
  detected for air showers of $\unit[10^{20}]{eV}$ at two zenith
  angles. Shown is the ratio between the electromagnetic component
  (without the part related to muon decay or low-energy hadronic
  interactions) and the muonic part. The electromagnetic particles
  from muon decay are included in the muonic contribution. The
  different terms are explained in the text. The curve labeled ``WCD''
  corresponds to the water-Cherenkov detectors of the Auger array,
  while the red one ``Top Scin'' corresponds to a thin scintillator.  
}
\label{fig:signal-ratios-SSD-WCD}
\end{figure}

A thin scintillation detector, which is mounted above and
triggered by the larger water-Cherenkov detector below it, provides a 
robust and well-understood way of particle detection that is sufficiently 
complementary to obtain a good measurement of the density of muons.
This can be understood by comparing the signal contributions for different shower
components as shown in Fig.~\ref{fig:signal-ratios-SSD-WCD}. Over a wide range in lateral distance,
the ratio between the integrated signal 
of electromagnetic particles (photons and electrons) and that of muons is more than a factor two
higher in a scintillation detector without any shielding than in a water-Cherenkov detector.


\subsubsection{Matrix inversion approach}
\label{sec:matrix-inversion}

The matrix formalism developed in~\cite{Letessier-Selvon:2014sga} for a layered
surface detector can be adapted for the reconstruction of the muonic signal
contribution and, hence, the number of muons detected in a single upgraded
detector station. The motivation for this formalism is to relate intrinsic
shower parameters at ground level, such as energy or particle fluxes, to the
detector signals via a matrix whose coefficients depend only on the shower
geometry but very little on the shower primary mass or on the interaction model
used to describe it.

For the combination of the SSD and the Auger WCD we can relate the signal in
MIP measured with the SSD and the signal in VEM measured by the WCD to the
electromagnetic energy flux $\mathcal{F}_\text{em}$ and the muonic flux $\mathcal{F}_\mu$
at ground, both expressed in VEM/m$^2$, by
\begin{equation}
\begin{pmatrix}
  S_\text{SSD} \\
  S_\text{WCD}
\end{pmatrix} =
\begin{pmatrix}
  \lambda\,\mathcal{A}_\text{SSD} & \mathcal{A}_\text{SSD} \\
  \beta\,\mathcal{A}_\text{WCD} & \mathcal{A}_\text{WCD}
\end{pmatrix} \,
\begin{pmatrix}
  \mathcal{F}_\text{em} \\
  \mathcal{F}_\mu
\end{pmatrix}.
\end{equation}
where $\mathcal{A}_\text{SSD}$ (respectively  $\mathcal{A}_\text{WCD}$) are the horizontal area
of the SSD (respectively WCD), $\beta$ represents the projection factor of the
WCD surface perpendicularly to the electromagnetic flux ($\beta=1$ at vertical
incidence)\footnote{$\beta$ is the ratio of the horizontal WCD surface to its
surface perpendicular to the zenith direction $\theta$. $\beta(\theta)=S(\theta)/S(0)=\cos\theta + 2h/(\pi R)\sin\theta$ where $h$ and $R$ are
the height and radius of the water volume in the WCD. For the Auger WCD
geometry: $\beta(\theta)=\cos\theta + 0.42\,\sin\theta$.}, and
$\lambda$ is the average energy (in units of MIP/VEM/$\cos\theta$) deposited in
the scintillator per VEM of electromagnetic flux. The parameter $\lambda$
depends on the electron to photon ratio and on the energy distribution of the
electromagnetic particles in the shower. Note that because the average energy
deposition of the muons is proportional to their path length which increases
with zenith angle it compensates the decrease of perpendicular surface (this is
also true for electrons above one MIP/$\cos\theta$ in the scintillators). With
those definitions and expressing the fluxes in VEM/m$^2$, $S_\text{SSD}$ will be
expressed in MIP and $S_\text{WCD}$ in VEM.

From these definitions it is straightforward to obtain the electromagnetic and
muon energy fluxes at ground
\begin{align}
\mathcal{F}_\text{em} &= \frac{1}{\lambda -\beta} \left ( \frac{S_\text{SSD}}{\mathcal{A}_\text{SSD}} -  \frac{S_\text{WCD}}{\mathcal{A}_\text{WCD}}\right),
\\
\mathcal{F}_\mu &= \frac{1}{\lambda -\beta} \left ( \lambda\,\frac{S_\text{WCD}}{\mathcal{A}_\text{WCD}} -  \beta\,\frac{S_\text{SSD}}{\mathcal{A}_\text{SSD}}\right).
\end{align}
An approximation of the resolution on those quantities can be obtained assuming
that the variance of the signal in the WCD and the SSD is equal to the signal
itself when using their respective units (VEM and MIP), as expected from a
Poisson law. As an example the relative resolution on the muon energy flux is
\begin{equation}
\sigma_\mu = \frac{\sqrt{\operatorname{Var}[\mathcal{F}_\mu]}}{\mathcal{F}_\mu} = \frac { \sqrt{\lambda^2\frac{S_\text{WCD}}{\mathcal{A}^2_\text{WCD} } + \beta^2\frac{S_\text{SSD}}{\mathcal{A}^2_\text{SSD} }}} {\lambda\,\frac{S_\text{WCD}}{\mathcal{A}_\text{WCD}} - \beta\,\frac{S_\text{SSD}}{\mathcal{A}_\text{SSD}}} \sim \frac{K}{(\lambda-\beta)\sqrt{\bar{s}}}.
\end{equation}

The resolution evolves as expected as $1/\sqrt(\bar{s})$ where $\bar{s}$ is the
average of the signals per unit area in both detectors. From simulations we
find that $\lambda$ is about $2.8\,\beta$ in MIP/VEM  for EAS of energies
above \unit[10]{EeV}. Preliminary MC studies have shown that both $\lambda$ and $\beta$
evolve in similar ways with zenith angle and have little dependence on
interaction models and primary mass. This allows to reconstruct the muon signal
in the WCD in a nearly model independent way, as given by
\begin{equation}
S_{\text{WCD},\mu} = \mathcal{A}_\text{WCD}\,\mathcal{F}_\mu = \frac{\lambda}{\lambda -\beta}\,S_\text{WCD} -  \frac{\beta}{\lambda -\beta}\frac{\mathcal{A}_\text{WCD}}{\mathcal{A}_\text{SSD}}\,S_\text{SSD} = \delta\,S_\text{WCD} - \gamma\,S_\text{SSD}.
\end{equation}

More details on the use of this method to extract $S_{\text{WCD},\mu}$ from $S_\text{WCD}$
and $S_\text{SSD}$ will be discussed in Sec.~\ref{sec:particle-identification}.

\subsubsection{Shower universality approach\label{sec:universality}}
The shower universality method predicts for the entire range of
primary masses the air-shower characteristics on the ground using only
three parameters: $E$, \xmax and $N_\mu$ . Therefore, from the
integrated signal and the temporal structure of the signal measured in
the individual stations, one can estimate these three parameters on an
event-by-event basis.

The main advantage of this method is that it exploits all
experimentally collected information of an air shower for deriving the
physics observables.  In this sense a universality-based
reconstruction can be considered as a very advanced multivariate
analysis of the shower data that employs parameterized physics
relations to combine the different measured quantities.  A weakness 
of the universality approach is that is has systematic uncertainties
that are difficult to control.  The parameterization of the
universality relations between the energy, shower age (i.e.\ \xmax),
and muon number, and the different signal components at ground (for
different lateral distances), can only be derived from libraries of
simulated showers and depends to some degree on the hadronic
interaction models used for the simulations.

To fully exploit the universality features of air showers, four shower
components have to be introduced: (a) the muonic component, (b) the
electromagnetic component stemming from muon interactions and muon
decay, (c) the purely electromagnetic component, and (d) the
electromagnetic component from low-energy hadrons (the collimated beam
component).

\begin{figure}[t]
\centering
\def\figh{0.37}
\includegraphics[height=\figh\columnwidth]{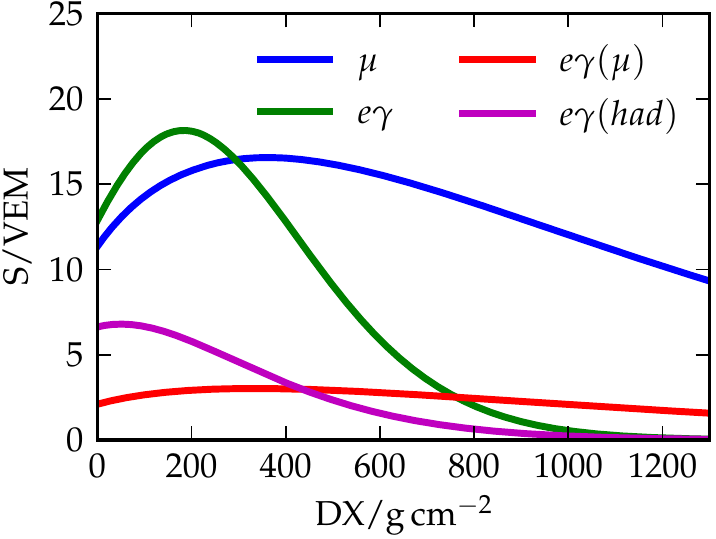}\hfill
\includegraphics[height=\figh\columnwidth]{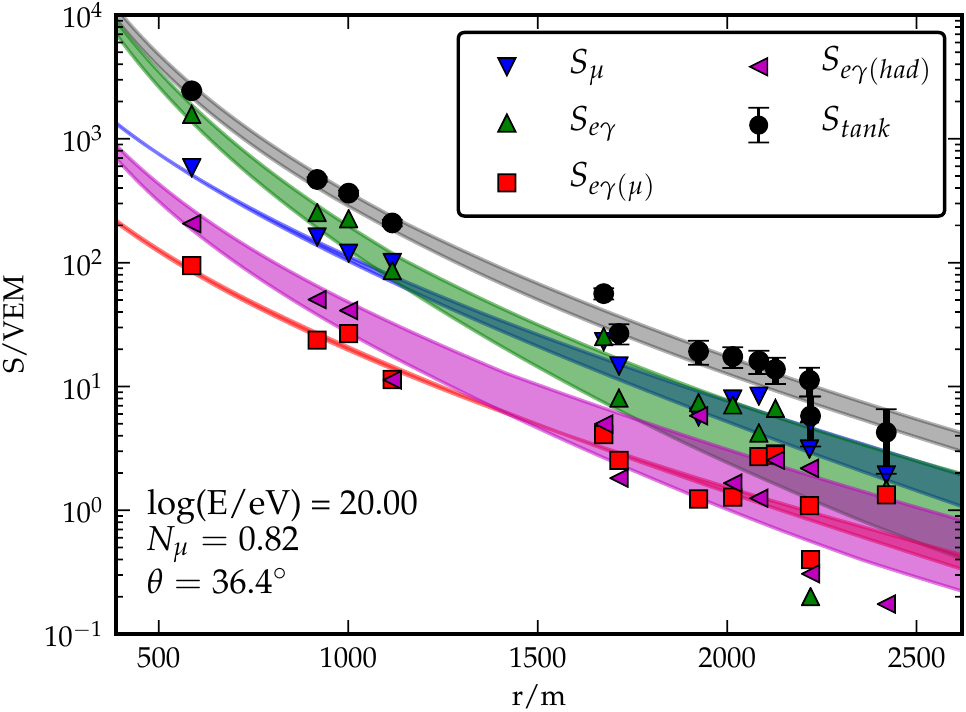}
\caption{
    Left: 
    The simulated average signal of the 4 different components as a
    function of $DX$ (the slant depth between \xmax and ground level) at a distance of \unit[1000]{m} and a zenith
    angle of $36^\circ$ for fixed energy, $E=\unit[10]{EeV}$.  Right:
    The lateral shape of the simulated signal of a $\unit[10^{20}]{eV}$
    shower. The upper band indicates the model prediction based on the
    fitted parameters.  Round markers refer to the total simulated
    signal.  The lower bands show the prediction for the four signal
    components compared to the simulated values. The size of the
    azimuthal signal asymmetry is indicated by the width of the bands.
    }
\label{fig:fitted-ldf-mc}
\end{figure}

Splitting the electromagnetic component into component (c), originating from
the decay products of high energy $\pi^0$ that have been produced in the
first generations of hadronic interactions in a shower, and component (d) which is stemming
from hadronic interactions at low energy taking place close to the individual detectors, allows
us to include the correlation between muons and electromagnetic particles arising
from such low energy interactions.

The contributions of the different universality components to the overall signal are illustrated
in Fig.~\ref{fig:fitted-ldf-mc}. Each component has a different dependence on the observation 
depth $X$ relative to \xmax, $DX = X - \xmax$, and also the slopes of the lateral distributions differ.
In addition, there is a dependence on the azimuthal angle about the shower axis that leads to
an asymmetry of the ground signals. The colored bands reflect
the predicted range of
the asymmetry effect of individual components (black being the sum of
all others). The markers indicate the simulated total signal
(black) and the reconstructed component signals given by the
reconstruction algorithm (see below).

\begin{figure}[t]
\centering
\def\figh{0.36}
\subfigure[Individual component traces]{
\includegraphics[height=\figh\textwidth]{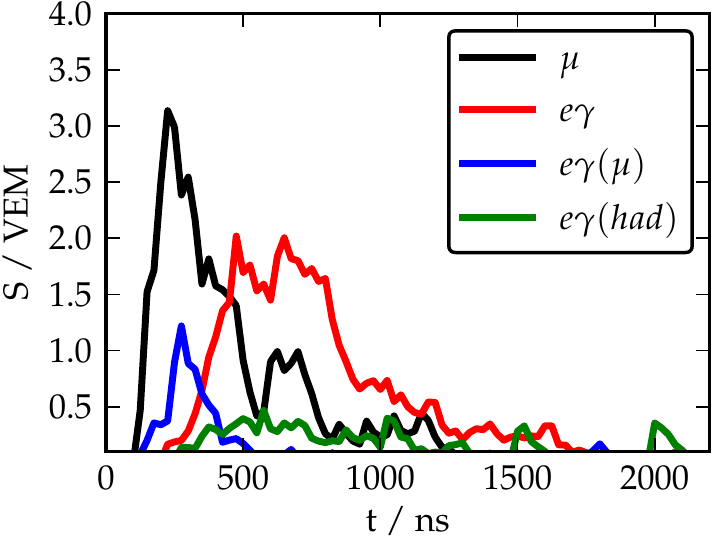}
\label{fig:single-component-trace-example}
}\hfill
\subfigure[Average traces in one $DX$ bin]{
\includegraphics[height=\figh\textwidth]{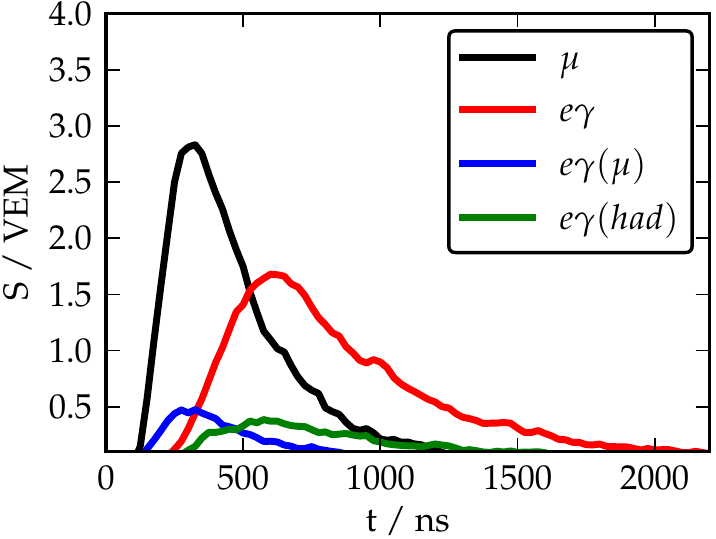}
\label{fig:average-component-trace-example}
}
\caption{Time dependence of the signal of the different universality components in the Auger
    surface detectors for proton showers of $\unit[10^{19.5}]{eV}$ simulated with QGSJet~II-03.
}
\label{fig:component-trace-example}
\end{figure}

For each of the individual universality components the expected arrival time profile of the 
particles can be parameterized. An example of the simulated time response of a
water-Cherenkov detector of the Auger array is shown in~\ref{fig:single-component-trace-example}
for a particular station far from the shower core. The average trace of the components at
the same distance and for a given $DX$ of about $\unit[200]{g/cm^2}$ is shown in
Fig.~\ref{fig:average-component-trace-example}.

The total expected signal in a station at position $r$ and relative depth $DX$ is given by 
\begin{equation}
  S_\text{tot}
  =
  S_\text{em}(r,DX,E)
  +
  N_\mu^\text{rel}
  \left[
    S_\mu^\text{ref}(r,DX,E)
    + S_\text{em}^\mu(r,DX,E)
  \right]
  + (N_\mu^\text{rel})^\alpha 
  S_\text{em}^\text{low-energy}(r,DX,E),
\label{eqn:signal}
\end{equation}
where we have not written the dependence on the azimuthal asymmetry for sake of clarity. The 
muonic contributions (both from muons directly and the decay/interaction products of the muons) are
scaled with the factor $N_\mu^\text{rel}$ relative to the expectation values for proton
primaries. The contribution from low-energy interactions to electromagnetic particles through $\pi^0$
production, $S_\text{em}^\text{low-energy}$, is scaled with $(N_\mu^\text{rel})^\alpha$.
The parameter $\alpha$ is determined in simulations with different
hadronic interaction models, resulting in $\alpha \approx 1$ to a  very good approximation. The integrated
signal as well as the time trace of each individual detector station is fitted by adjusting the
parameters $E$, $DX$, and $N_\mu^\text{rel}$.

Unfortunately, the primary energy $E$ and relative muon number $N_\mu^\text{rel}$
are very strongly correlated and cannot be determined independently from the data of the water-Cherenkov
detectors alone, except for very few high-energy events with an exceptionally large number of stations.
Therefore, two approaches have been developed within the Auger Collaboration. In the first approach,
the energy is taken from the standard reconstruction of $S(1000)$ (the WCD signal 1000\,m from the
shower core, see Sec.\ref{sec:dynamic-resolution-ldf}) and the corresponding conversion
from $S(1000)$ to energy, which is calibrated by events measured independently 
with both the surface detector and the fluorescence
telescopes. Then $DX$ and $N_\mu^\text{rel}$ can be reconstructed.
In the second approach the number of muons is parameterized as a function of energy and \xmax
using hybrid events. Knowing the number of muons, the energy $E$ and the depth of shower maximum \xmax can be reconstructed. The results obtained with the two approaches for \xmax are very encouraging.

\begin{figure}[t]
\centering
\def\figh{0.43}
\includegraphics[height=\figh\columnwidth]{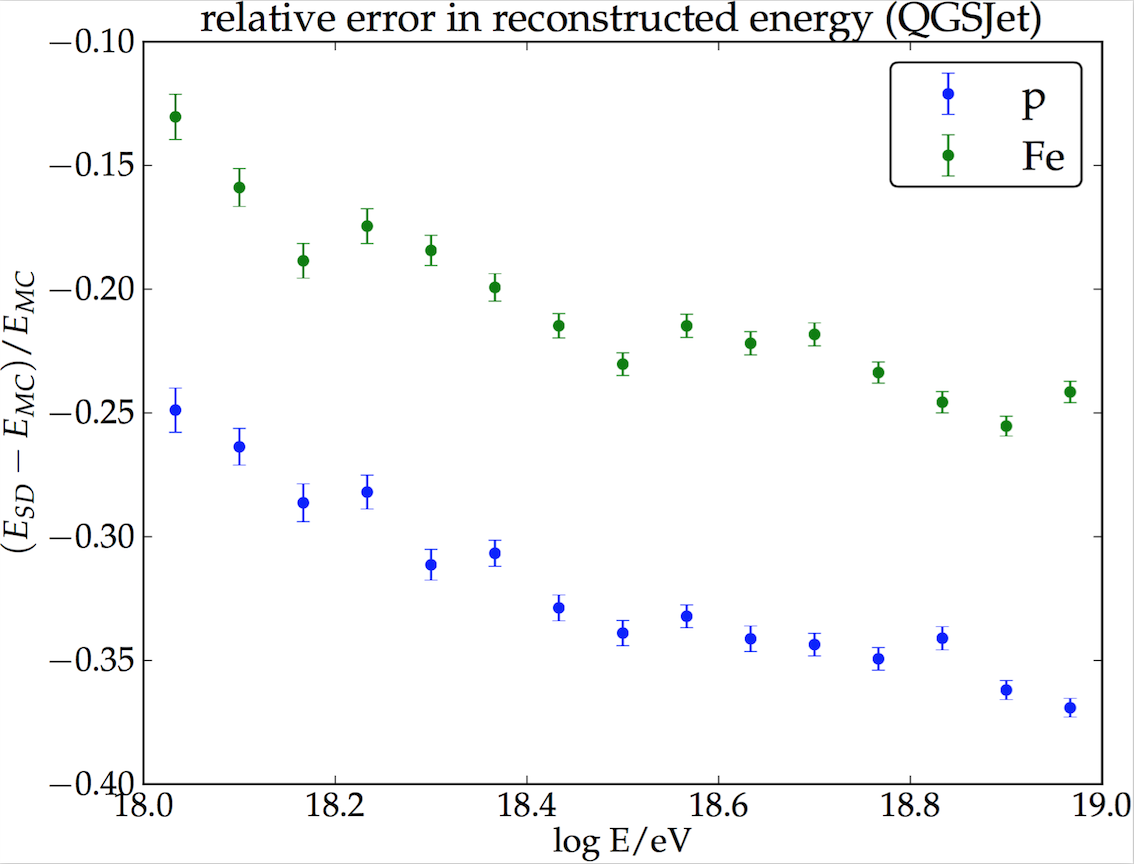}\hfill
\includegraphics[height=\figh\columnwidth]{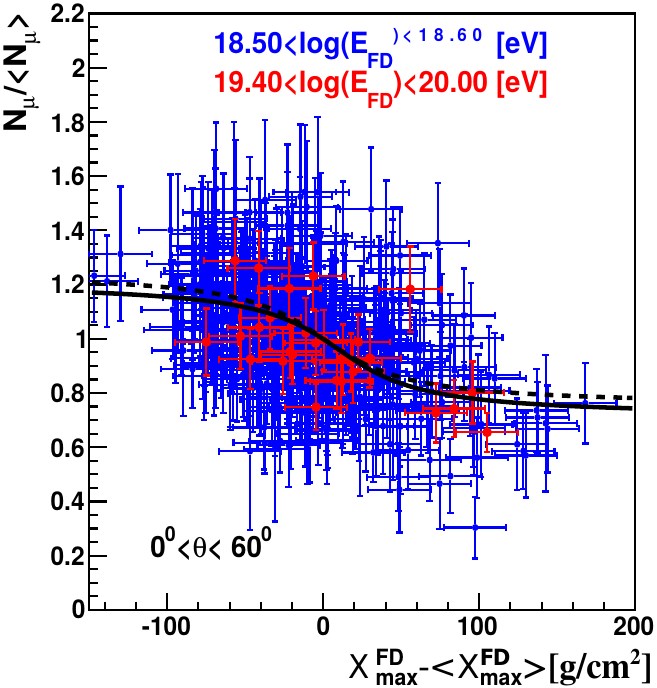}
\caption{ Left panel: Expected bias in the energy reconstruction if
    the signal at $\unit[1000]{m}$ is used as energy estimator. Right
    panel: Correlation between the relative number of muons and
    \protect\xmax derived from the universality interpretation of
    hybrid events.  The line shows a phenomenological parameterization
    for the mean of the distribution.  }
\label{fig:universality-weaknesses}
\end{figure}

An independent determination of the number 
of muons is required for the full potential of these universality methods to be exploited.
Such a measurement will also give a handle on better understanding of the systematic uncertainties.
Considering the first universality approach, 
this can be understood by looking at the bias of the energy reconstruction if $S(1000)$ is
used as energy estimator. The reconstructed energy differs by more than 10\% between proton
and iron primaries, see Fig.~\ref{fig:universality-weaknesses}~(left),
making the corresponding estimate of the muon number
quantitatively unreliable. In the case of the second approach, the key input is the parameterization of
the relative number of muons as function of \xmax, see Fig.~\ref{fig:universality-weaknesses}~(right).
While it is possible to derive such a
parameterization for showers with $\unit[E< 10^{19.5}]{eV}$ from hybrid data with sufficient statistics,
one has to assume that this
relation does not change at higher energies. In particular, any sensitivity to a possible,
unexpected change of the properties of hadronic interactions would be lost without a direct measurement of
the muon densities.
Another general weakness of certain applications of these universality parameterizations
is the discrepancy in the number of muons found in data in comparison to the simulation predictions. It is
not clear whether simply increasing the relative muon number $N_\mu^\text{rel}$ by an overall factor
does indeed properly account for the physics missing in the hadronic interaction
models and possibly air shower simulations.

The extension of the existing universality parameterization for water-Cherenkov detectors to
scintillators is straightforward and has been done in preparation of the Auger Upgrade. Results
obtained by applying this universality-based reconstruction simultaneously to both detectors will be
discussed in Sec.~\ref{sec:particle-identification}.


\subsubsection{Cross-checks using a direct muon detector}

Both the matrix inversion method and the universality reconstruction depend at some level on 
air shower simulations. With one major aim of the Auger upgrade being the reduction of
systematic uncertainties, it is highly desirable to cross-check the performance of both
reconstructions. Such a cross-check will improve our understanding of the
upgraded detectors and will allow us to reduce the systematic uncertainty of
the derived number of muons. For example, discrepancies in the number of muons 
as currently observed at the $2\sigma$ level could be established beyond any doubt.

We will measure muons directly with an Underground Muon Detector.
The AMIGA array of muon counters, covering almost $\unit[24]{km^2}$
with sixty-one  $\unit[30]{m^2}$ detector stations at a spacing of $\unit[750]{m}$, offers an ideal means for 
measuring the number of muons directly and, moreover, in a totally independent way.
However, buried under a layer of $\unit[1.3]{m}$ of soil, as needed for a direct measurement, the 
effective energy threshold for muon detection will be higher than that of the
scintillator-water-Cherenkov detector combination.

%% file: expected_physics_performance.tex

\chapter{Expected Physics Performance}
\label{chap:performance}


\section{Introduction}

In this chapter we give an overview of the physics reach of the
planned Upgraded Detector.  We begin with the Surface Detector, first
describing the three complementary detector elements and their
performance, and then how the observations with these detector
elements will be combined to reconstruct events and estimate
composition, to vet shower modeling and different reconstruction
approaches, and serve as cross-calibrating tools.

In order to assess the physics performance expected for the upgraded
ground array, the expected number of events per year, as well as the
cumulative number for a data taking period from 2018-2024, are shown
in Table~\ref{tab:stat}, for both the \unit[750]{m} array and the
\unit[1500]{m} array. Given a data taking period of the upgraded
detector of \unit[7]{years} we can expect about 700 events above
$\unit[3{\times}10^{19}]{eV}$ and more than 60 above
$\unit[6{\times}10^{19}]{eV}$ for zenith angles less than
$60^\circ$. Horizontal air showers ($60^\circ<\theta<80^\circ$) will
add about 30\,\% to the exposure and thus to the number of expected
events. Accounting for a detector resolution of 15\,\% or better in
determining the number of muons, this would allow e.g.~for a
separation of a fraction as small as 10\,\% of protons from
intermediate and heavy primaries.

Taking data until end of 2024 will double our present SD event statistics and
reduce the total statistical uncertainty significantly at highest
energies. We have recently improved considerably our scheme for
triggering individual stations. Predominantly low electromagnetic
signals of physics events are additionally recorded. Not only do the new triggers
lower the energy threshold of the \unit[750]{m} and \unit[1500]{m}
arrays by half a decade in energy, but we gain additional stations far
from the shower core in individual events, which will give a better
handle on identifying different shower components. The gamma/hadron
and neutrino/hadron separation will clearly benefit from an increased
station multiplicity per event.

\begin{table}[tb!]
\centering
\caption{Expected number of events per year and cumulative number of
  events for a data taking period from 2018 until 2024 for events up
  to zenith angles of $\theta=60^\circ$. Horizontal air showers
  ($60^\circ<\theta<80^\circ$) add about 30\,\% to the exposure.
  Moreover, the new trigger scheme implemented in 2013 will allow us
  to significantly lower the energy threshold of both the
  \unit[750]{m} and \unit[1500]{m} arrays.}
\label{tab:stat}
\begin{tabular}{rrrrr}
\toprule
${\log_{10}(E/\unit{eV})}$ & ${\left.\mathrm{d}N/\mathrm{d}t\right|_\textbf{infill}}$ &
${\left.\mathrm{d}N/\mathrm{d}t\right|_\textbf{SD}}$ & ${\left.N\right|_\textbf{infill}}$ &
${\left.N\right|_\text{\bf SD}}$ \vspace{8pt}\\
 & $[\unit{yr^{-1}}]$ & $[\unit{yr^{-1}}]$ & [2018-2024] & [2018-2024] \\
\midrule
\rowcolor[gray]{.9} 17.5   &    11500       & - &  80700 & - \\
18.0    &    900        & - & 6400 & - \\
\rowcolor[gray]{.9} 18.5    &    80         &   12000  & 530 & 83200 \\
19.0    &    8          &    1500  & 50 & 10200 \\
\rowcolor[gray]{.9} 19.5    &    ${\sim}1$          &    100  & 7 & 700 \\
19.8    &    -    &      9  & - & 60\\
\rowcolor[gray]{.9} 20.0    &    -    & ${\sim}1$ & - & ${\sim}9$ \\
\bottomrule
\end{tabular}
\end{table}

The three elements of the upgraded detector discussed below are
(Sec.\,\ref{sec:dynamic-resolution-ldf}) the WCD which -- apart from the
electronics upgrade -- is Auger's current, well-performing Surface Detector,
essentially untouched.   Data from this detector are now powerful tools for
optimizing the rest of the upgrade.  In Sec.\,\ref{sec:ssd-perf}, we
introduce the main component of the Upgrade, the Surface Scintillator Detector
which consists of a 4\,m$^2$ plastic scintillator detector which will be
mounted on top of every WCD.  Engineering aspects of the SSD are deferred to a
later chapter; here we concentrate on its measurement capabilities and how
those will contribute to the analysis.  In Sec.\,\ref{sec:umdPerformance} we
describe the sub-array of 30\,m$^2$ underground muon detectors, the current
AMIGA project. The UMD, in tandem with the WCD+SSD will both constrain muon
energy spectrum and serve as a cross-check of the muon estimation techniques. 
 
\section{Water Cherenkov Detector global performance parameters}
\label{sec:dynamic-resolution-ldf}

\begin{figure}[t]
\centering
\def\figh{0.33}
\includegraphics[height=\figh\textwidth]{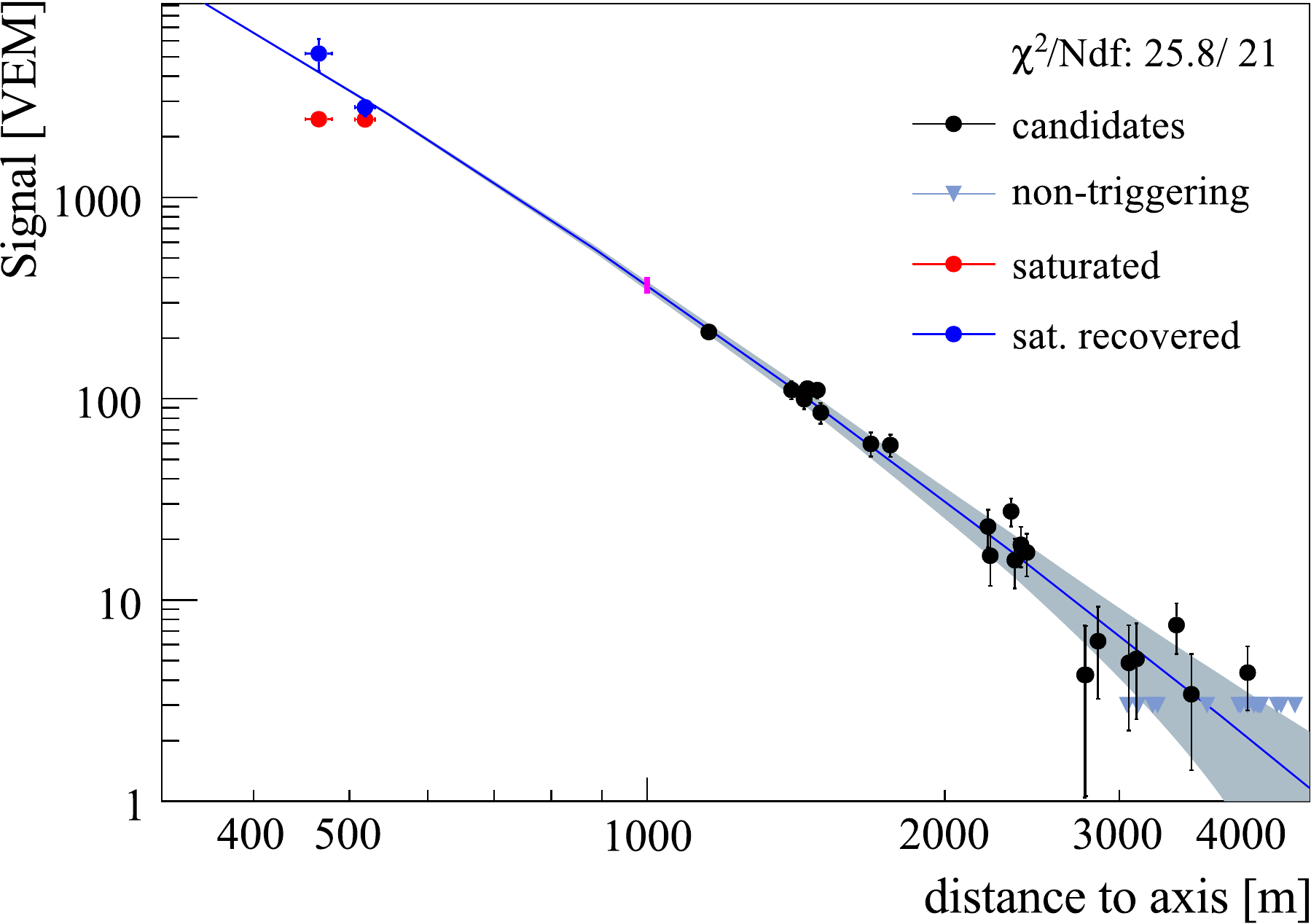}\hfill
\includegraphics[height=\figh\textwidth]{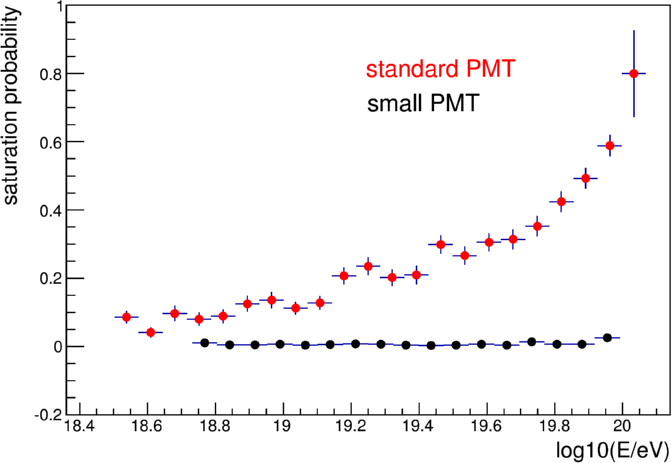}
\caption{Left panel: Lateral distribution of the signal sizes recorded
  in the water-Cherenkov detectors. The two stations closest to the
  shower core are saturated (filled red circles). A procedure to
  recover the signal is applied (filled blue circles). Right panel:
  Probability of having at least one saturated station in an event as 
  function of energy, obtained from simulations, for the standard PMT 
  configuration (red) and for the small PMT option (black).}
\label{fig:SDldfAndSatProb}
\end{figure}

Air-showers produced by cosmic rays with energies above \unit[10]{EeV}
produce a footprint on the ground extending over more than
\unit[20]{km$^2$}. The surface detector samples the arrival times and
the signals produced by the particles reaching the ground. The
reconstruction of the energy of the primary cosmic ray is based on the
determination of the shower size, the signal interpolated at a certain
distance to the air shower axis. In case of the \unit[1500]{m} array
the shower size is defined at \unit[1000]{m}. An example of the
lateral distribution of an event produced by a cosmic ray with an
energy of \unit[100]{EeV} is illustrated in
Fig.~\ref{fig:SDldfAndSatProb}\ (left).

\subsection{Dynamic range and saturation}

A drawback at present is the limited dynamic range of the SD
electronics and PMTs for the very large signals at highest
energies. At energies above $\unit[3{\times}10^{19}]{eV}$ more than
40\% of the events suffer from saturated signals in at least the
station closest to the shower core, see
Fig.~\ref{fig:SDldfAndSatProb}(right). A large fraction of these
signals are recovered based on the PMT and FADC response and the
time dependence of the signal~\cite{Veberic-ICRC:2013}. 

\begin{figure}[t]
\centering
\def\figh{0.325}
\includegraphics[height=\figh\textwidth]{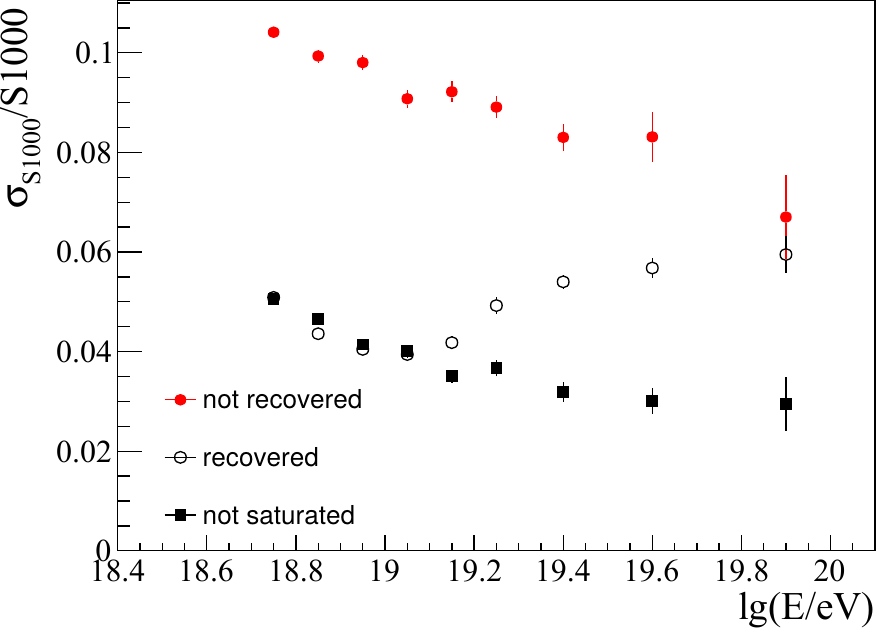}\hfill
\includegraphics[height=\figh\textwidth]{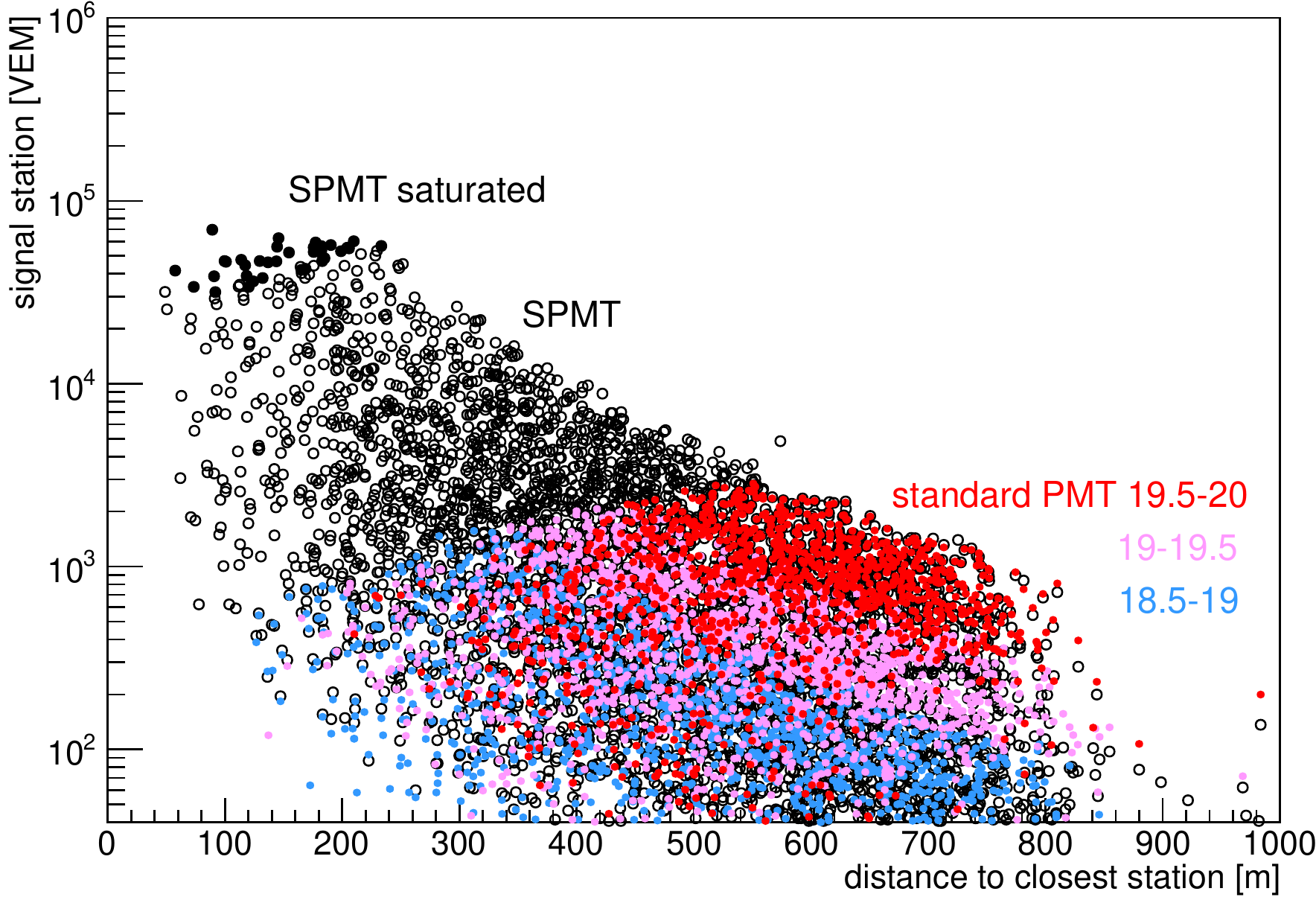}
\caption{ Left panel: Resolution of the reconstructed $S(1000)$ that
    is related to the uncertainty of the lateral distribution
    function. Shown is the estimated resolution for events without any
    saturated station (black squares) and for events with at least one
    saturated station (red circles). The open symbols show the
    improvement obtained if the currently available procedures for
    saturation recovery during reconstruction are applied. Right panel:
    The distribution of simulated signals as a function of the
    distance to the air-shower axis for different energies in the case of
    the standard PMT (opened colored circles) and in the case of the small
    PMT for the same air-showers (black markers).}
\label{fig:DynamicRangeDistance}
\end{figure}

Simulations show that the expectation value of reconstructed
observables such as $S(1000)$, energy, and the arrival direction are
not affected by having a station with a saturated signal trace in the
event. Only the reconstruction resolution is affected to some
extent. This is shown in Fig.~\ref{fig:DynamicRangeDistance}\ (left)
for the resolution of $S(1000)$.  The optimum distance of
\unit[1000]{m} is determined mostly by the spacing between the surface
detectors and it provides a robust estimate of the shower size with
respect to the assumptions about the functional form of the lateral
distribution~\cite{Newton:2007}. In the case of events with a
saturated signal, the optimal distance is \unit[1400]{m} and therefore
the choice of a particular LDF induces a systematic uncertainty in
$S(1000)$.  At $\unit[10^{19.5}]{eV}$, the contribution to the
reconstruction resolution of $S(1000)$ from the lateral distribution
fit alone is less than 4\% for events without saturated stations and
increases to 8\% if a signal trace is saturated.  The software
procedures developed for recovering the saturated signals work very
well for energies below $\unit[10^{19}]{eV}$ and are increasingly less
effective at higher energies.

However, the detailed study of the whole lateral distribution requires
us to reach an accuracy of better than 15\% on the recovered signal,
which can only be obtained with a detailed knowledge of the individual
PMT responses in the non-linear region (a non feasible solution, which
needs a measurement of the deep saturation curve of each of the 5000
PMTs and monitoring of their constancy in time). Due to the current
limited information, the accuracy of the recovered signals larger than
10 kVEM (close to the core) can be worse than 70\%.

The proposed equipping of the water-Cherenkov detector with an extra
small PMT (SPMT) will increase the dynamic range from about
\unit[600]{VEM} to more than \unit[30,000]{VEM} (for details see
Sec.~\ref{sd-dynamics}). With the new configuration we expect less
than 2\% of saturated events at the highest energies.  The
distribution of the expected signals as a function of the distance
between the shower axis and the closest station is shown in
Fig.~\ref{fig:DynamicRangeDistance}\ (right). The predicted measured
signals for the current PMTs (colored filled circles) and for the SPMT
(black circles) were obtained from CORSIKA simulations of air-showers
induced by primary protons with an energy between 3 and
\unit[100]{EeV}. The increased dynamic range will allow measurement of
complete signals at a distance as close as \unit[300]{m} from the
core. The signal variance in the extended dynamic range interval will
be reduced significantly, being dominated by the calibration
uncertainties of 6\%. Event selections based on cuts in energy will be
more accurate and flux corrections of the energy spectrum due to
resolution-dependent migrations will be smaller.

\begin{figure}[t]
\centering
\def\figh{0.38}
\includegraphics[height=\figh\textwidth]{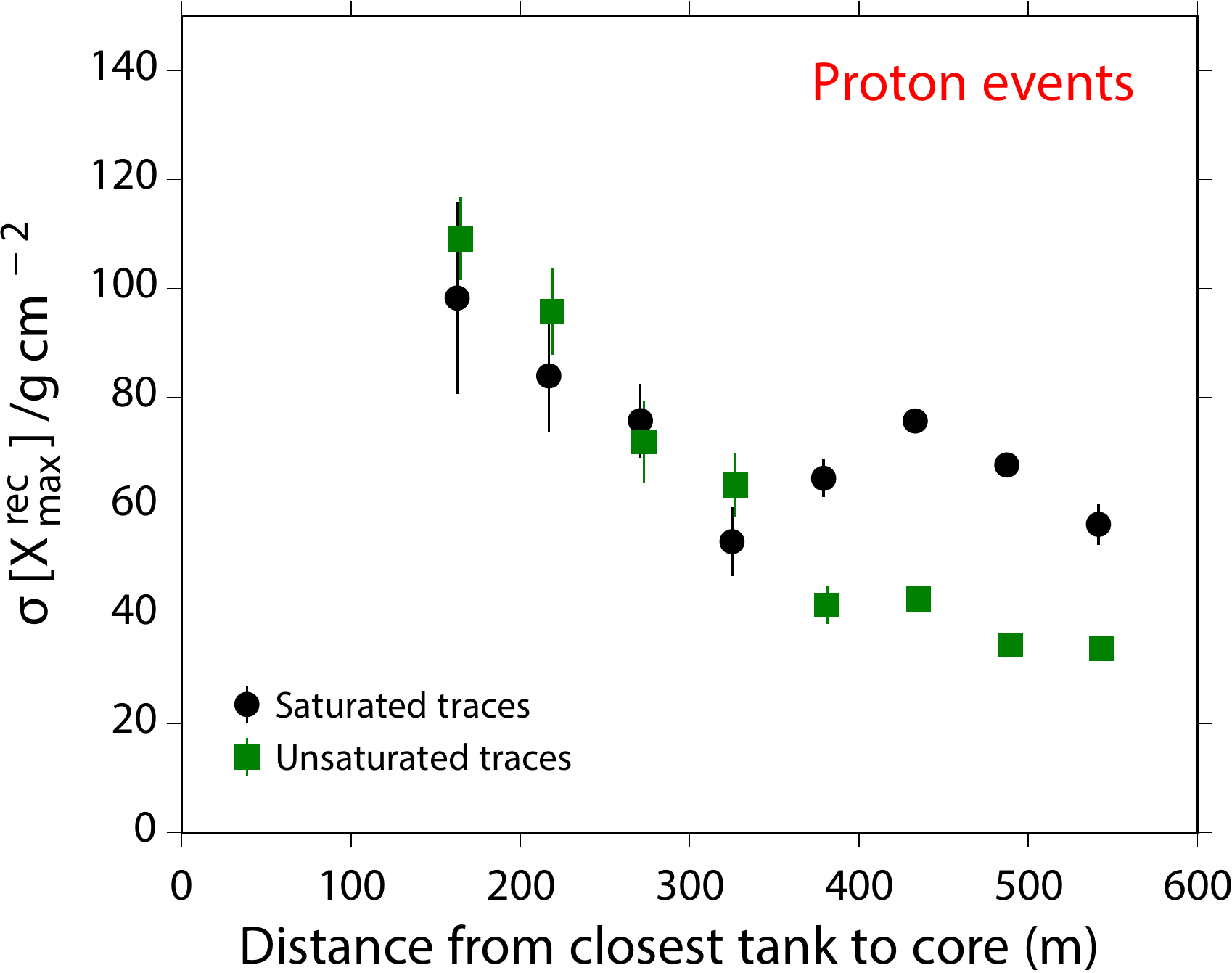}\hfill
\includegraphics[height=\figh\textwidth]{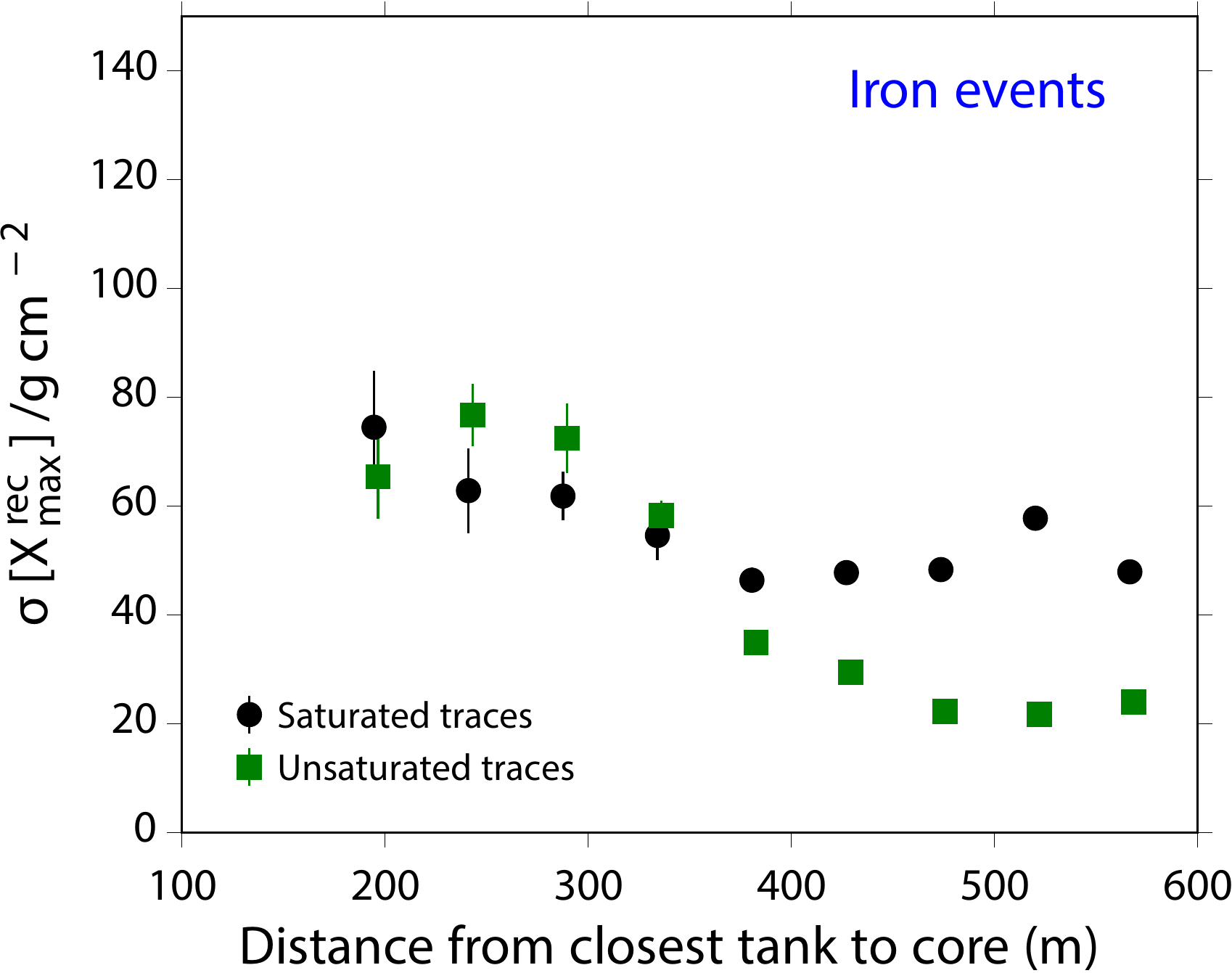}
\caption{Resolution of \xmax reconstructed from data of the 
    water-Cherenkov detectors only. The same simulated events have 
    been reconstructed using the signal trace without saturation effects
    (unsaturated traces) and with saturation (saturated traces). If
    the saturated station is at a distance of more than
    $\unit[300]{m}$ the resolution of the reconstructed \xmax can be
    improved considerably by increasing the dynamic range of the
    detectors.}
\label{fig:xmax-universality-sat}
\end{figure}

Another example of the importance of measuring the signal traces close
to the shower core is shown in
Fig.~\ref{fig:xmax-universality-sat}. In this simulation study the
resolution of the universality reconstructed \xmax is shown
reconstructing the same events twice, once with saturated stations as
one would have with the current surface detector, and once with
increased dynamic range preventing any saturation of the time traces.
The resolution of the reconstructed \xmax is significantly worse for
showers with a saturated station close to the core. It should be noted
that the gain of information by having non-saturated traces is,
however, limited to distances larger than $\unit[300]{m}$.  At smaller
distances the uncertainty of the core position limits the usefulness
of the measured signal.

\subsection{Angular and energy resolution}

\begin{figure}[t]
\centering
\def\figh{0.335}
\includegraphics[height=\figh\textwidth]{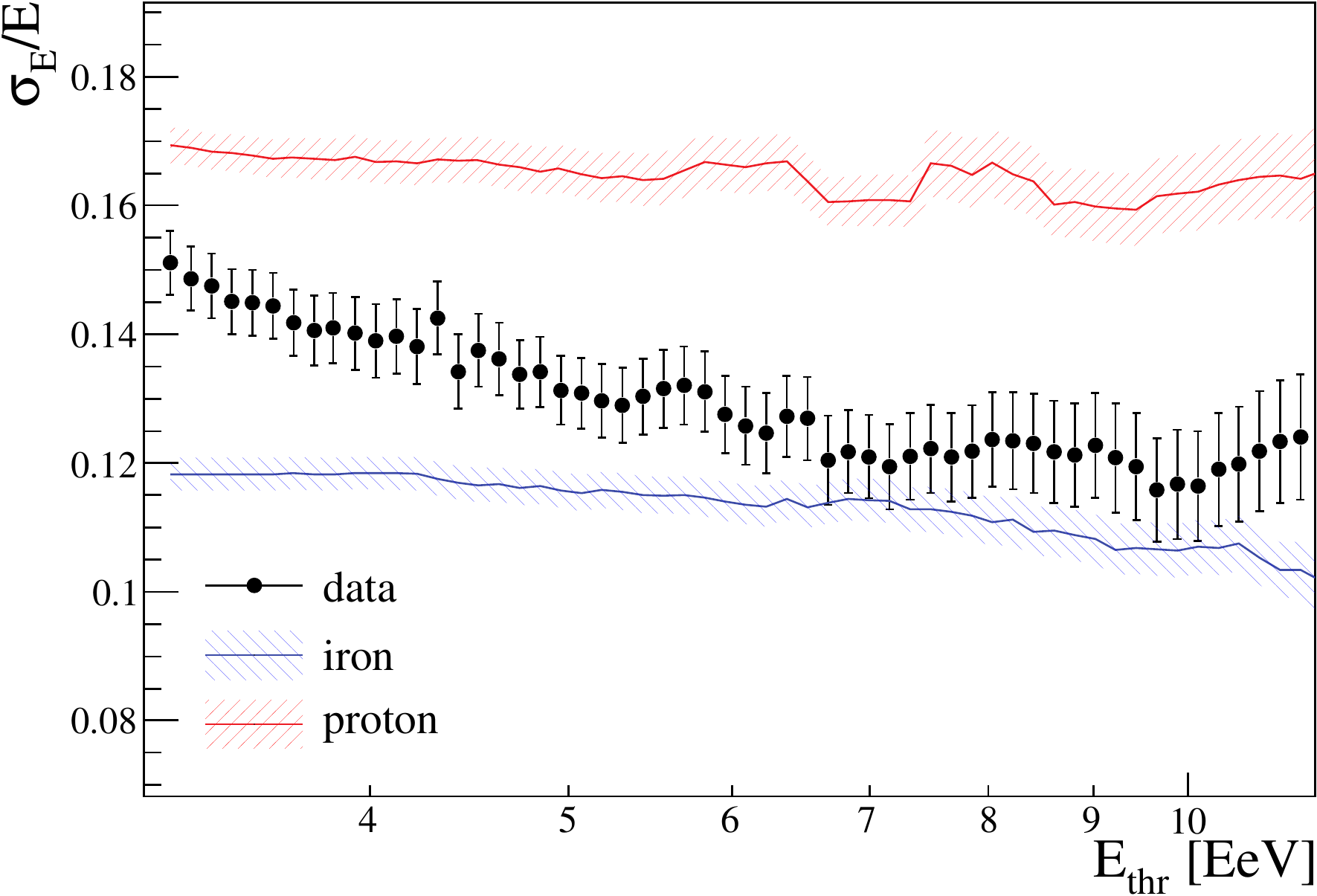}\hfill
\includegraphics[height=\figh\textwidth]{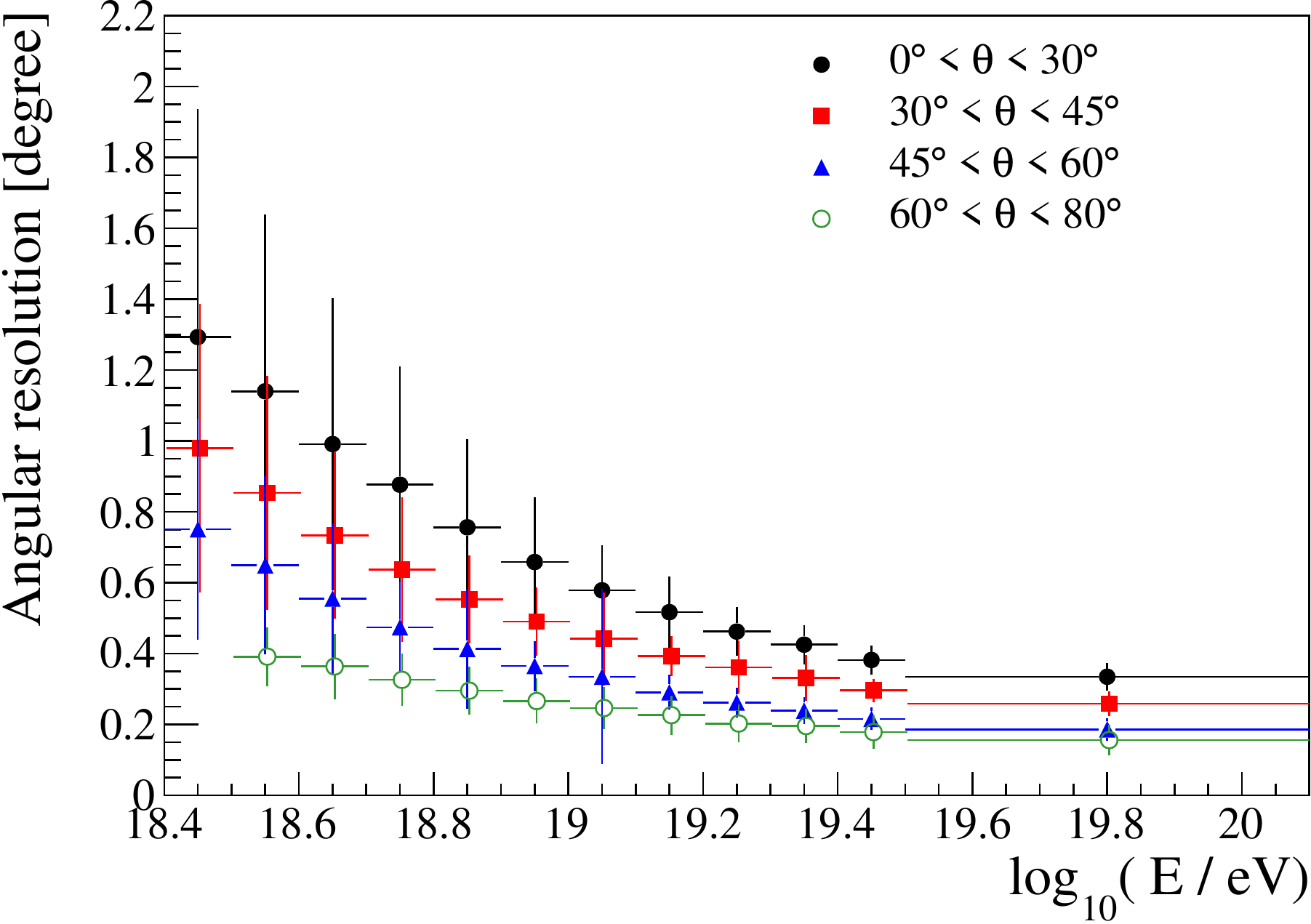}
\caption{Left panel: The cumulative
  energy resolution above a threshold energy $E_\text{thr}$ vs threshold 
  energy for data, compared with CORSIKA (QGSJet) simulations of proton 
  and iron induced air-showers. Right panel: The mean angular resolution of the 
  surface detector as a function of energy for different angular ranges. 
  The error bars represent the RMS of the distributions.}
\label{fig:SDAngularResolution}
\end{figure}

The energy resolution of the surface detector can be retrieved from
the events used for the energy cross-calibration, i.e.\ events with
coincident measurements by the FD and the
SD~\cite{VerziICRC2013,Schulz-ICRC:2013}. It is illustrated in
Fig.~\ref{fig:SDAngularResolution}\ (left) for different energy
thresholds. Above \unit[10]{EeV} the energy resolution is 12\%, with
contributions from the detector resolution, reconstruction algorithms
and from the shower-to-shower fluctuations.  At the highest energies
the resolution of the reconstructed impact point on the ground is
highly correlated with the resolution on the shower size. A larger
dynamic range, used together with the universality principles will
improve the core location and the energy determination (for more
details see Section~\ref{sec:universality}).  In the same figure the
energy resolutions for proton and iron simulations are shown, the
difference between them being the dependence of the intrinsic shower
fluctuations on the primary mass. The proposed upgrade used with the
universality principle will diminish the mass dependency of the energy
assignment.

The incident direction of the cosmic rays is obtained from the
reconstruction of the air-shower front based on the arrival time of
the particles in the detectors. The angular resolution is determined
mainly by the number of the stations participating in the event and by
the variance of the arrival times. In a single station the accuracy
on the determination of the arrival of the shower front is given by
the sum of the intrinsic fluctuations of the air-showers, the absolute
time given by the GPS, and by the FADC sampling. The current angular
resolution as a function of energy for different zenith angle
intervals is depicted in Fig.~\ref{fig:SDAngularResolution}\ (right). For
events above \unit[10]{EeV}, for which the majority have triggered
more than 6 detectors, the current angular resolution is better than 1
degree. Switching to a faster ADC sampling of \unit[120]{MHz} and a GPS
RMS accuracy of about \unit[2]{ns} (see
Section~\ref{sec:timing}) is expected to slightly improve the
reconstruction of the direction of the cosmic rays.


\section{Scintillator Detector performance considerations}
\label{sec:ssd-perf}

\subsection{Dynamic range}

In order to benefit in an optimal way from the surface scintillator
detector (SSD) information together with the water-Cherenkov detector
(WCD) signal, the dynamic range of both detectors should be
similar. The requirements on the SSD dynamic range could be slightly
smaller, since close to the EAS core the signals of both detectors
will be dominated by the electromagnetic component and therefore the
two measurements do not bring as much information as at an
intermediate distance to the core. However, at the highest energies a
significant signal is expected even at intermediate distances from the
core.

For a shower at 10$^{20}$\,eV and a zenith angle of 38 degrees, the
peak signal in a 4\,m$^2$ SSD at 200\,m from the core is expected to
be around 12,000\,MIP (Minimum Ionizing Particles), with the
contribution of the muons to the SSD signal of less than 10\% (and
less than 20\% for the WCD). A maximum signal of 12,000\,MIP seems
therefore a reasonable upper bound for determining the dynamic
range. It has furthermore the advantage of being achievable with a
single PMT, if it is linear enough. See section \ref{sec:ScinPMT} for
more details on the electronics designed to achieve such a dynamic
range.

\subsection{$\boldsymbol{N_\mu}$ and $\boldsymbol{X_\text{max}}$ resolutions using Universality based reconstruction}
\label{sec:ssd-universality}

One of the main issues found when extracting physics quantities from
an event observed by two detectors is the correlation of the signals
in them. For example, if one applies an LDF fit to the signals, the
WCD LDF and the SSD LDF will be strongly correlated, as it is the same
shower. Also, the slope of these LDFs will be correlated with, for
example, the radius of curvature of the shower front, as they both
indicate the age of the EAS.

An appealing framework to solve this issue is shower universality where a
parameterization of the detectors signals is made as a function of EAS
macro-parameters. In the reconstruction process, these macro-parameters are
adjusted and any change in them directly translates to the expected change in
signal in both detectors. This allows therefore to properly take into account
the correlations between the measurements of the SSD and the WCD. Adding an
extra detector adds extra information to the global fit reducing the number of 
parameters to be reconstructed.

A description of the response of the WCD and SSD to EAS has been done,
parameterizing both the integral signal deposited, and the timing
structure of the signals as a function of the geometry and
macro-parameters of the EAS.  More details on universality can be
found in Sec.~\ref{sec:universality}. The parameterization is model
independent but depends on more parameters than just the \xmax and
\nmu. For example, the parameterizations are dependent on \xmumax
instead of \xmax, and while in all models \xmax and \xmumax are
strongly related, the relation between them is dependent on the hadronic interaction model.

\begin{figure}[t]
\centering
\def\figh{0.28}
\includegraphics[height=\figh\textwidth]{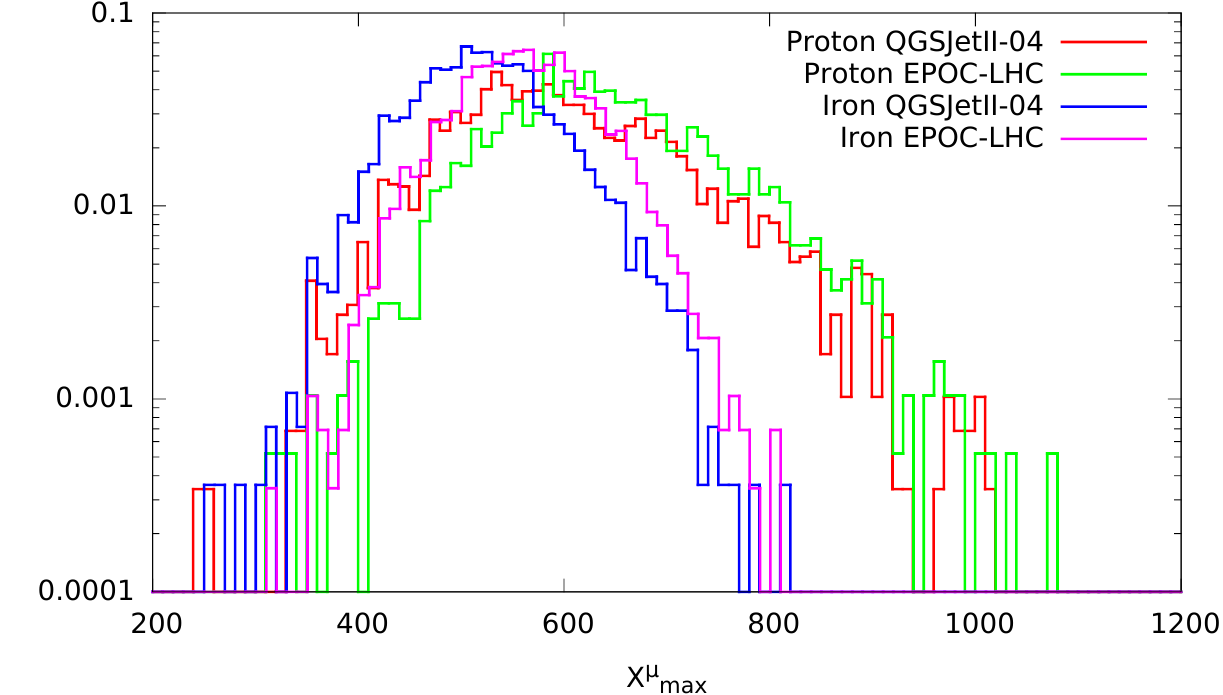}\hfill
\includegraphics[height=\figh\textwidth]{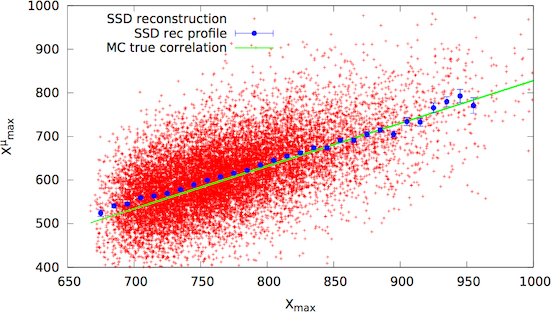}
\caption{\xmumax distribution reconstructed at 10\,EeV for proton and iron showers simulated with
EPOS-LHC and QGSJetII-04 (left), and \xmax-\xmumax correlation obtained using 
SSD reconstructed values of \xmumax (right). Some small systematics can be seen 
for low \xmax (corresponding to lower energy EAS).}
\label{fig:xmax-xmumax-ssd}
\end{figure}

Using a full set of MC showers, based on EPOS-LHC and QGSJetII-04,
including the simulation of the detector response, it was shown that
the combination of WCD and SSD are sensitive enough to reconstruct all
the macro-parameters on a statistical basis. However, to get the best
event-by-event characterization, it is beneficial to fix some of the
very strong correlations between macro-parameters (such as the one
between \xmax and \xmumax) and then let only one of them free in the
reconstruction. The calibration of the correlation can be made by the
WCD+SSD data set, and furthermore cross-checked with hybrid events. In
the $\unit[750]{m}$ array region, FD will provide a direct \xmax measurement, while
AMIGA will deliver a \xmumax measurement applying the muon production
depth (MPD) technique,
allowing a cross-check and a proper calibration of the universality
parameterization. The accuracy in the \xmumax reconstruction using WCD
and SSD data has been found to be 70\,g/cm$^2$ at 10\,EeV and
50\,g/cm$^2$ at 63\,EeV. While these values are somewhat large to be
used for composition on an event by event basis, they are well suited
to make a statistical measurement of the \xmax-\xmumax relation
obtained in real data in order to calibrate properly the universality
reconstruction and perform a reconstruction with only \xmax and \nmu
as free parameters. In Figure~\ref{fig:xmax-xmumax-ssd} the \xmumax
reconstruction based on simulations and the obtained \xmax-\xmumax
correlation is shown.

\begin{figure}[t]
\centering
\def\figh{0.45}
\includegraphics[height=\figh\textwidth]{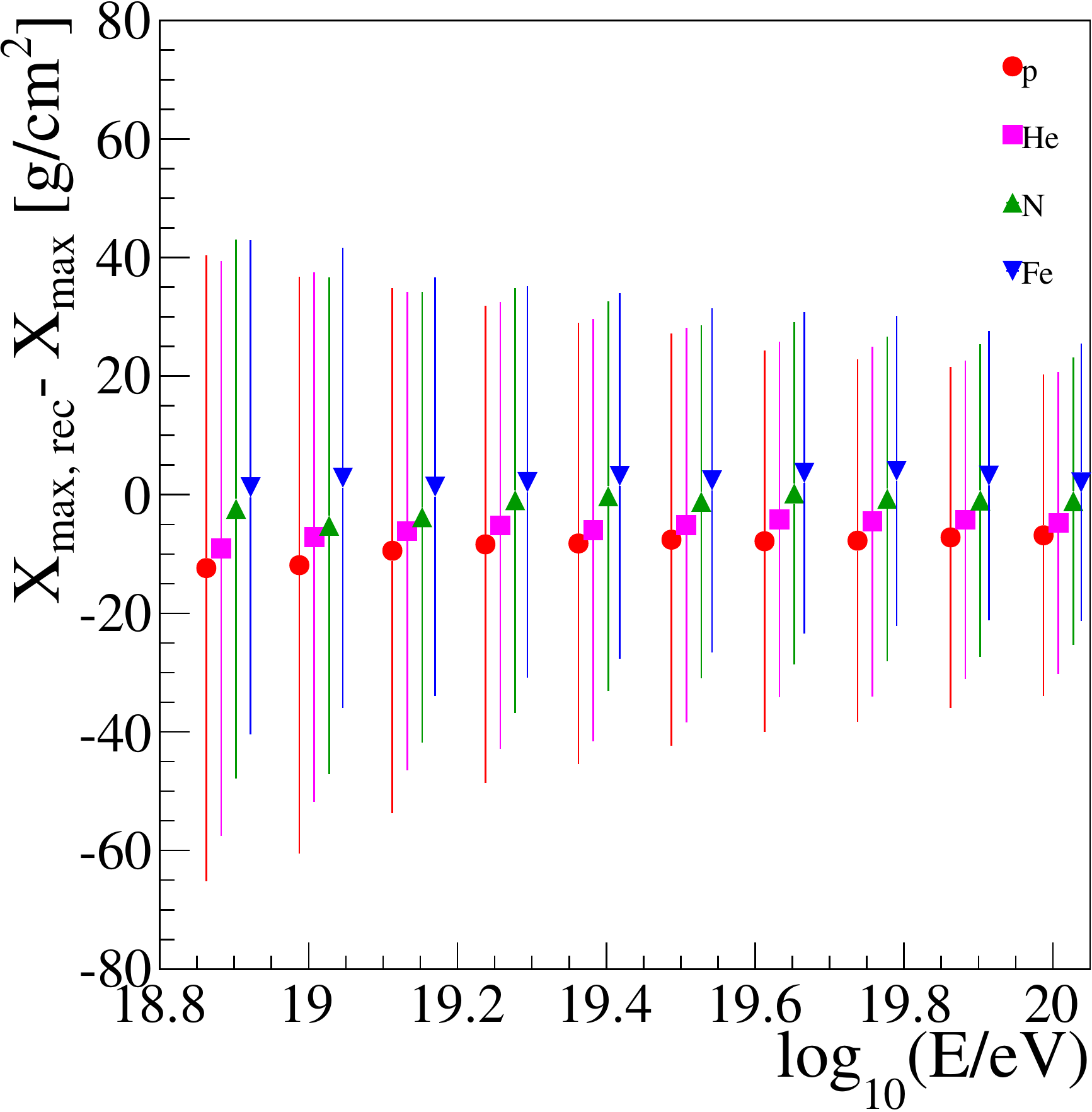}\hfill
\includegraphics[height=\figh\textwidth]{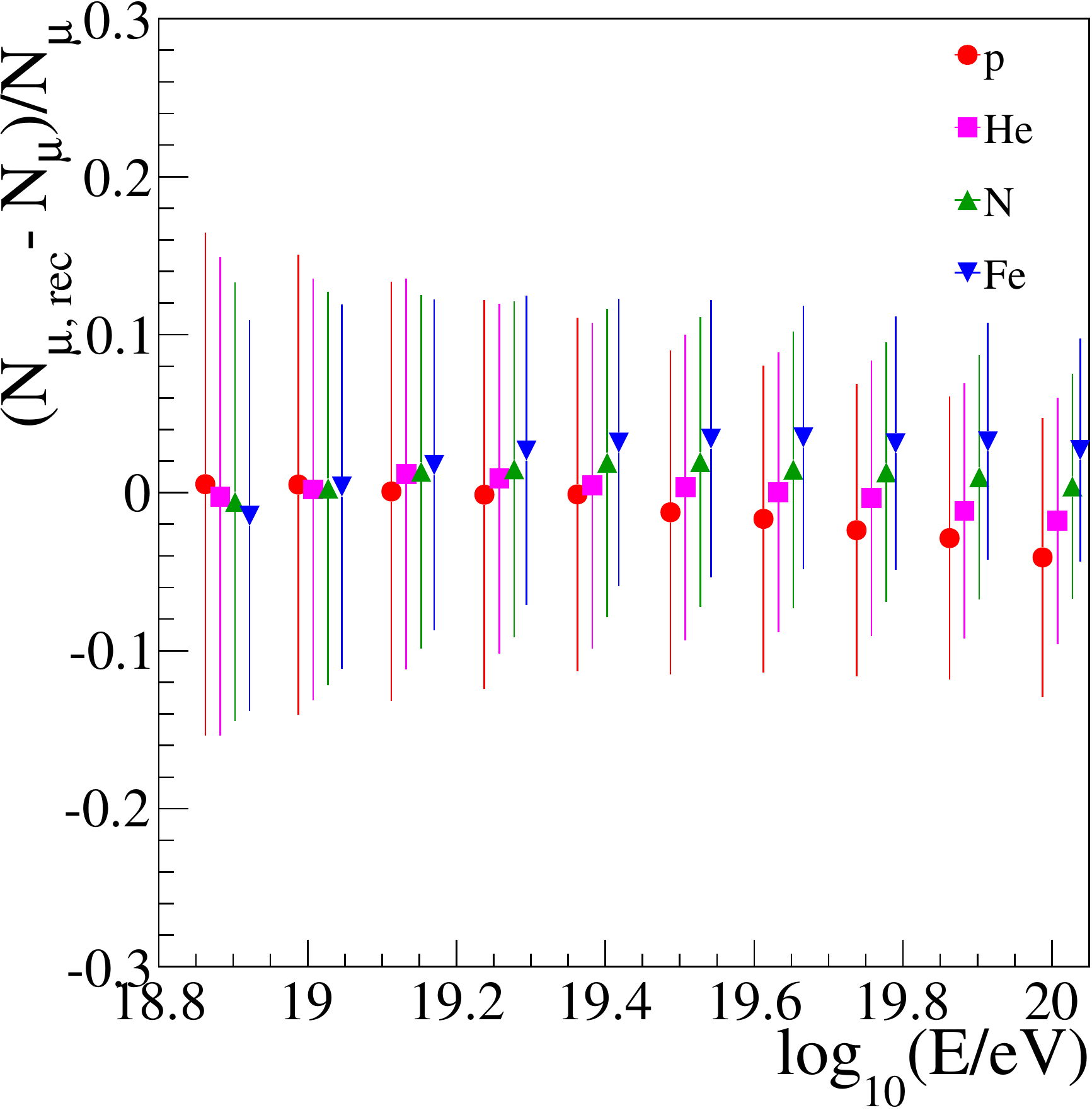}
\caption{Left panel: The reconstructed \xmax compared with the
  true \xmax as a function of energy. Error bars represent the
  RMS of the distributions. Right panel: The reconstructed \nmu
  compared with the true \nmu as a function of energy. Error bars
  represent the RMS of the distributions. The resolutions are obtained
  from parameterizations and interpolations of EPOS-LHC simulations at
  fixed energies and zenith angles and are shown for events up to 60$^\circ$.
}
\label{fig:ASCIIXmaxNmuResolution}
\end{figure}

Once the \xmumax-\xmax relationship is determined from the calibration
described in the previous section, the remaining composition sensitive
parameters to fit are just \xmax and \nmu. In order to properly
determine the resolution of the SSD, the Monte Carlo simulations were
treated as real data, and the \xmumax-\xmax relationship determined
with reconstructed values. The events were then reconstructed again
using this calibration and the resolution on \xmax and \nmu, and
systematic biases, were derived. Figure
\ref{fig:ASCIIXmaxNmuResolution} shows both resolution and bias for
both variables as a function of composition and energy. Biases are
small, below 15\,g/cm$^2$ for \xmax and 5\% for \nmu, and the
resolution is about 40\,g/cm$^2$ at 10\,EeV, down to 25\,g/cm$^2$ at
100\,EeV for \xmax, and 15\% at 10\,EeV down to 8\% at 100\,EeV for
\nmu. Of interest is also the energy resolution for the reconstruction
of around 10\% at 10\,EeV down to 7\% at 100\,EeV.

\subsection{Some results from scintillator (ASCII) prototypes}

SSD units have been operating on top of an Auger WCD since 2010, under
the name of ASCII, standing for Auger Scintillator for Composition
II. At first a 0.25\,m$^2$ scintillator was operated for a year,
mainly to check the concept, by turning off one of the WCD PMTs and
plugging in the SSD PMT instead. 
In 2012 a 2\,m$^2$
detector was installed at the Central Radio Station with an
independent DAQ in order to study the calibration. From its operation
it was checked that the Minimum Ionizing Particle (MIP) peak in the
charge histogram clearly stands out when a trigger is requested in
coincidence with the WCD.  After completion of these tests an array of
7 $\times$ 2\,m$^2$ detectors was deployed in 2014 on the central
$\unit[750]{m}$ array hexagon. All these detectors were built using scintillator bars
similar to those of the MINOS experiment, including the procedures of
gluing the fiber, cutting one of their ends and collecting light only
on one side. They are inferior to the design described in
Sec.~\ref{sec:scintillator}, both in terms of total light collection
and uniformity.  As the upgraded electronics was not available for the
detectors, again it was decided to remove one PMT of the WCD and use
the channels for the SSD PMT. The PMT base had to provide the PMT with
high voltage, and had to extract two signals, a low gain and a high
gain, with a gain ratio of roughly 32, to make it compatible with the
current local station electronics. Furthermore, some signal shaping was
needed due to the frequency of the current FADC, 40\,MHz, slower than
the typical light response of the scintillator. The prototype PMT
(R1463) and base electronics did behave reasonably, but displayed
important non-linearity, limiting the effective dynamic range of the
prototypes to less than 200\,MIP.

The operation of 2\,m$^2$ SSD units for more than one year went
without issue and no intervention was needed. The support and double
roof system was validated in strong wind conditions, and the
temperature of the SSD was found to be similar 
to the temperature on the base of PMTs inside the WCD.  The MIP peak
was found to be clearly determined. More details on the calibration
can be found in Sec.~\ref{sec:ScinCali}, including
Figure~\ref{fig:mip-hist} showing a histogram of a MIP from real data
recorded with the ASCII prototypes.

\begin{figure}[t]
\centering
\def\figh{0.28}
\includegraphics[height=\figh\textwidth]{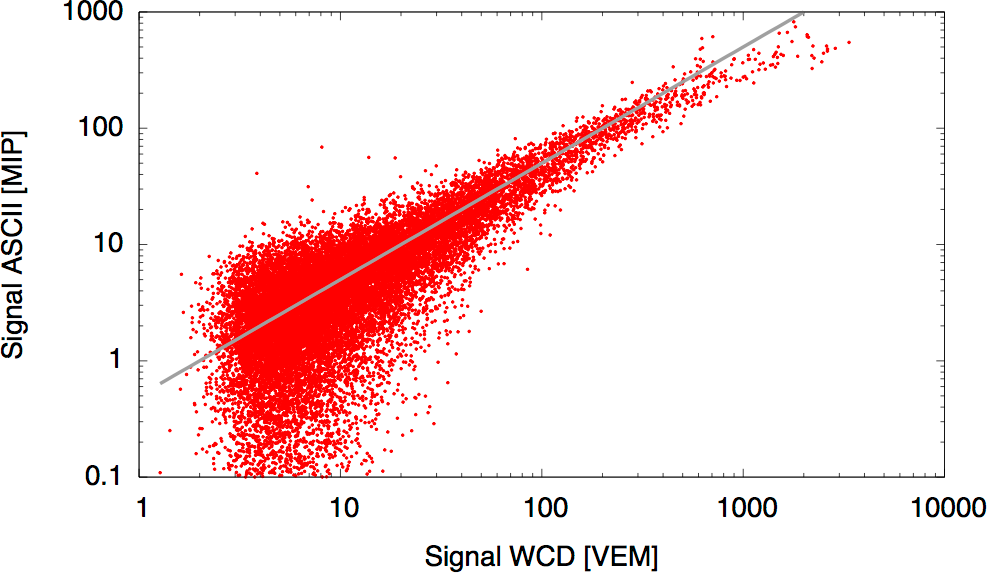}\hfill
\includegraphics[height=\figh\textwidth]{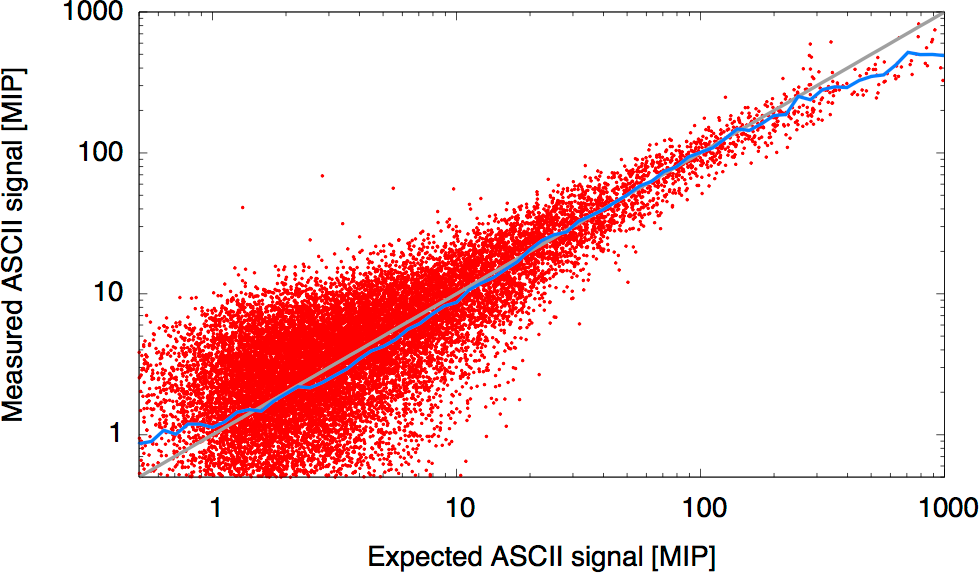}
\\[5mm]
\includegraphics[height=\figh\textwidth]{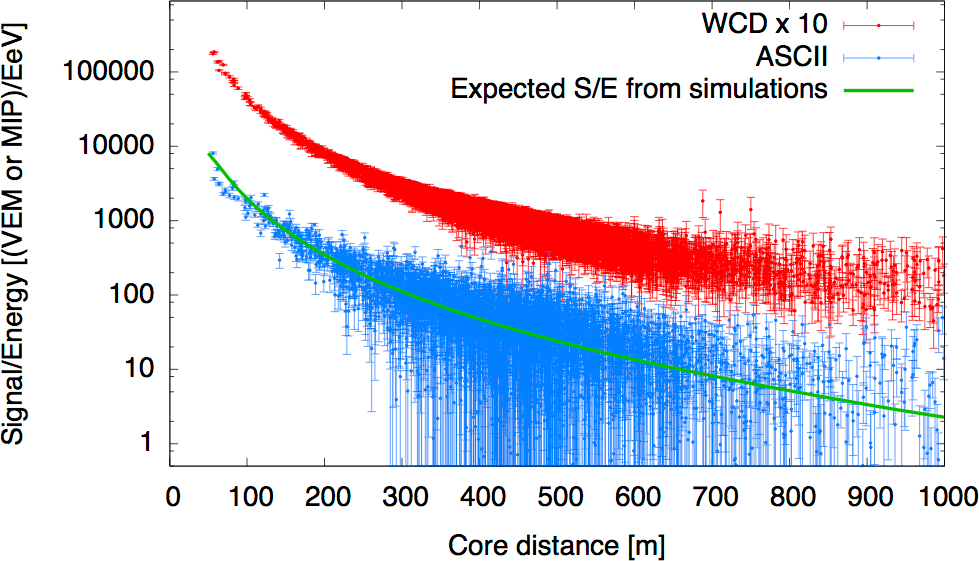}\hfill
\includegraphics[height=\figh\textwidth]{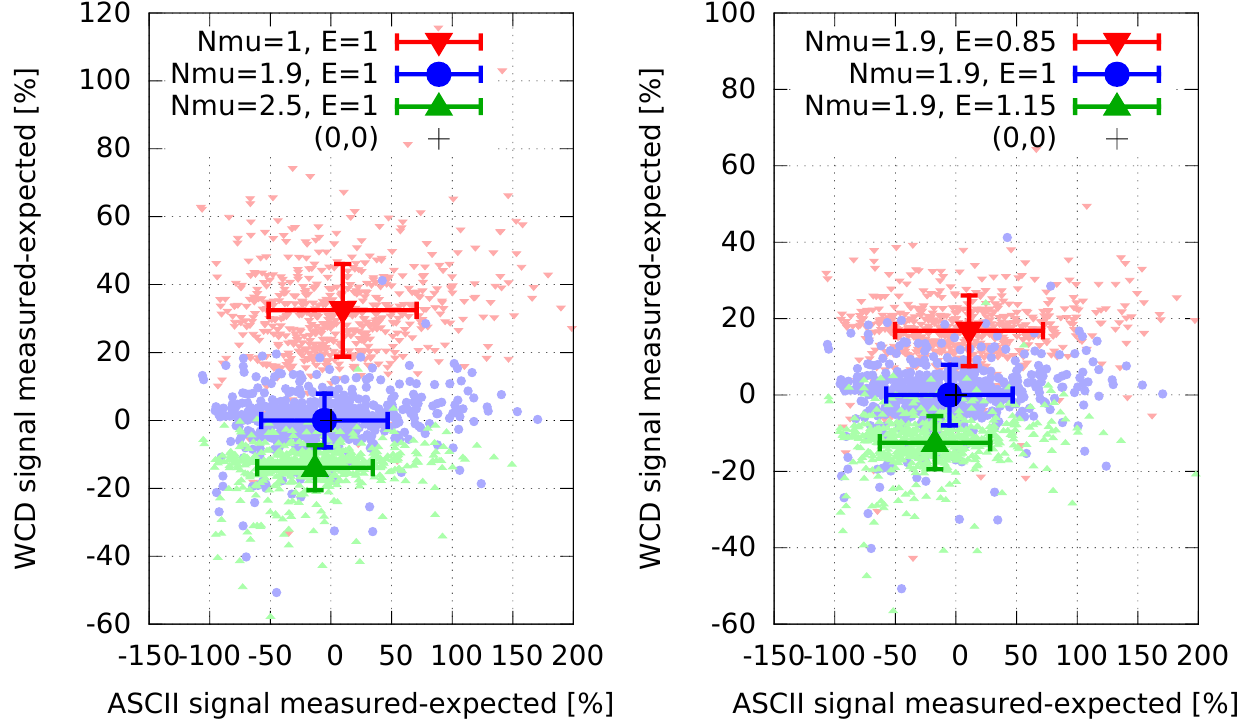}
\caption{Results for the prototype SSD detectors (ASCII). Top left,
  ASCII signal compared to the WCD. The signal of a 2\,m$^2$ SSD is
  roughly half the signal of the WCD. Non-linearity of the prototypes
  detectors can be seen for large WCD signals. Top right: comparison
  of measured signal with expected one from universality and average
  \xmax and \nmu at the reconstructed energy. The agreement is very
  good. The profile of the data points is also indicated. The
  non-linearity at large signals is again visible.  Bottom left: LDF
  from WCD (scaled ${\times}10$) and SSD for low energy events from 25
  to 30 degrees of zenith angle, normalized to reconstructed energy,
  compared to prediction from simulations. Bottom right: measured
  signals in SSD and WCD compared to predicted ones from the
  universality parameterization, changing \nmu or the energy scale. A
  similar study, once systematics are understood, will allow to
  determine the muon numbers of real data and any systematic in the
  FD-based energy scale.}
\label{fig:ascii-res}
\end{figure}

While no real physics output could be obtained in the prototype phase
due to the non-linearity of the detectors, some checks of the data
quality could be performed. The comparison of the signal in ASCII with
the signal in the WCD can be seen in Figure~\ref{fig:ascii-res}. In a
2\,m$^2$ detector, the signal in units of a single MIP is roughly half
the signal of the WCD in VEM. As the signal fluctuations in a WCD are
of the order of $\sqrt{S/\text{VEM}}$, and the fluctuations in an SSD
are of the order of $\sqrt{S/\text{MIP}}$, a size of around 4\,m$^2$
for the SSD would imply a similar accuracy in the measurement for both
detectors, allowing an optimal global reconstruction. Another test of
the data can be done even at low energy, by doing the usual SD
reconstruction of the WCD signals and taking the average \xmax and
\nmu at the energy obtained to estimate the signals expected in the
SSD using the universality parameterization. Doing so allows us to
compare the observed signal to the predicted one, even for low energy
events where the number of stations and the total signals are not
enough to allow a reconstruction based on the universality
parameterization on event-by-event basis. One can also do the same
exercise but using different values for \nmu, or assuming a systematic
error in the energy scale of for example 15\% in one or the other
direction. All these tests are summarized in
Figure~\ref{fig:ascii-res}.

\section{Underground Muon Detector performance considerations}
\label{sec:umdPerformance}

The AMIGA muon detectors are buried scintillator counters optimized
to perform a direct measurement of the air shower muon component
at energies of ${\sim}10^{17.5}$\,eV and higher. 
The detectors will be co-located with the WCDs of the $\unit[750]{m}$ array,
i.e.\ at 750\,m spacing and covering an area of $23.5$\,km$^2$.  A single
station will have an area of $30$\,m$^2$ and consist of 3 modules of
$10$\,m$^2$.  One key element for this direct measurement is the
absorption of the electromagnetic shower component by the
overburden. Results of a detailed simulation study of the
punch-through are given in table \ref{tab:punch-through}.  The optimal
depth is found to be in the range $110-150$\,cm. The detectors will be
deployed at a depth of $\unit[1.3]{m}$. At this depth the effective
energy threshold for muons is $600\,{\rm MeV}/\cos\theta_\mu$ with
$\theta_\mu$ being the zenith angle of the muon.  
As an important cross-check of the absolute efficiency and threshold
energy of detected muons, a small precision muon-counter with several
threshold energies will be installed on the surface at the
Observatory, to monitor reference rates of unaccompanied atmospheric
muons.


\begin{table}[t]
\caption{Relative punch-through (PT) for vertical showers of $10^{19}$\,eV
and number of muons for detectors at different depths in the Pampa soil
(${\sim}$2.4\,g/cm$^3$ average density). }
\label{tab:punch-through}
\centering
\begin{tabular}{r@{\hskip 0.7cm}rr@{\hskip 0.7cm}rr@{\hskip 0.7cm}rr}
\toprule
\multirow{2}{*}{Detector depth} & \multicolumn{2}{c}{70\,cm} & \multicolumn{2}{c}{110\,cm} & \multicolumn{2}{c}{150\,cm}
\\
 & \multicolumn{2}{c}{170\,\gcm} & \multicolumn{2}{c}{265\,\gcm} & \multicolumn{2}{c}{360\,\gcm}
\\
\cmidrule{2-7}
$r_\text{core}$ (m)    & Rel.\ PT & \nmu/10\,m$^2$ & Rel.\ PT & \nmu/10\,m$^2$ & Rel.\ PT  & \nmu/10\,m$^2$   
\\
\midrule
200     & 88\%  & 512 & 14\%   & 493  & 2.3\%    & 461
\\                                                  
600     & 38\%  & 49  &  7\%   & 43   &  1.1\%   & 40 
\\                                                  
1000    & 16\%  & 10  &  3\%   & 9    &  0.5\%   & 8.4   
\\                                                  
1400    & 3\%   & 3   &  0.6\% & 2.6  &  0.1\%   & 2.4   
\\
\bottomrule
\end{tabular}
\end{table}

To validate this design, a Unitary Cell of AMIGA was built, 
consisting of a full 7 station hexagon where
counters of different sizes were installed at depths of 2.3\,m.


Direct counting of the muons through segmentation has the significant
advantage that each channel counts pulses above a given threshold,
without a detailed study of signal structure or peak intensity. This
method is very robust since it does not rely on deconvolving the
number of muons from an integrated signal. It does not depend on the
PMT gain or gain fluctuations nor on the muon hitting position on the
scintillator strip and the corresponding light production and
attenuation along the fiber track. Neither does it require thick
scintillators to control Poisson fluctuations in the number of
photo-electrons per impinging muon.  This {\it one-bit} electronics
design relies on a fine counter segmentation to prevent under-counting
due to simultaneous muon arrivals.  The advantage of counting is
however lost close to the core in the saturation region, where
integration is needed.  AMIGA will therefore include a channel for
signal integration to cope with saturated stations.  

\begin{figure}[t]
\centering
\includegraphics[width=0.6\textwidth]{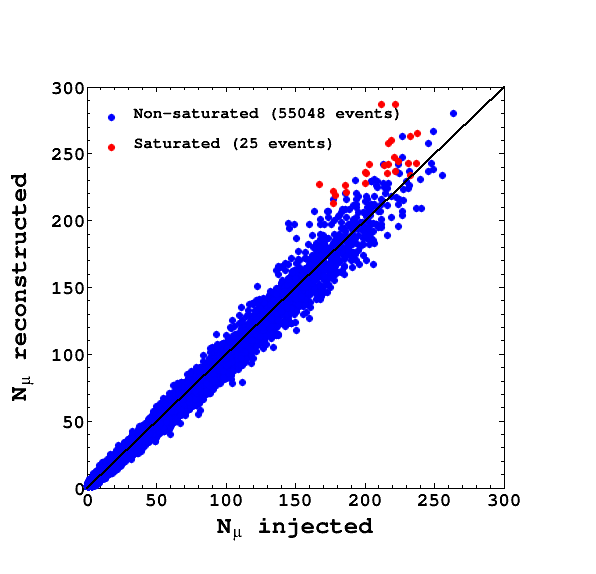}
\caption{Muon counting efficiency of a $\unit[30]{m^2}$ scintillator module for a
proton initiated shower at an energy of $\unit[10^{18.5}]{eV}$ and a zenith
angle of $21^\circ$. }
\label{fig:rec_eff}
\end{figure}

To summarize, the main features of the muon counters are
{\bf (i)} sturdy plastic-scintillator based detectors proven to support up to 2.3\,m of soil load without
any mechanical failure and having long term stability (the first counter was buried in the Pampas in November 2009);
{\bf (ii)} segmented system to count muons on a very reliable and fast basis.  
The counting efficiency 
is shown in Fig.~\ref{fig:rec_eff} for a simulated proton shower.
{\bf (iii)} dedicated channel for a summed analog signal 
to cope with high muon counts close to shower core (resolution $\approx$1.12/$\sqrt{N_\mu}$ for more than 20 muons); 
{\bf (iv)} cross calibration of the summed analog signal to segmented counting in the region 
of overlap (${\sim}$30 to 70 muons).

The AMIGA detectors are described briefly in Chapter~\ref{sec:UMD} and
a detailed description is given in Appendix~\ref{sec:amiga}.

%
%
%
%

\section{Primary Particle Identification with Surface Detector}
\label{sec:particle-identification}

We will exploit several methods of measuring the muon density, which,
combined with other observables, will enable us to estimate the mass
of the primary particle.

The matrix inversion method (see Sec.~\ref{sec:matrix-inversion}) will
allow us to derive the muon density for stations close to the shower
core (i.e.\ below $\unit[1200]{m}$) in a transparent and almost
model-independent way. While being limited to detector stations with
high electromagnetic and muonic signal, this method will enable us to
derive the average number of muons in air showers of different
energies and zenith angles unambiguously. As will be elaborated below,
it is a robust technique for estimating the separate contributions of
the electromagnetic and muonic shower components to the overall
detector signal.

\begin{figure}[t]
\centering
\def\figh{0.39}
\includegraphics[width=0.5\columnwidth]{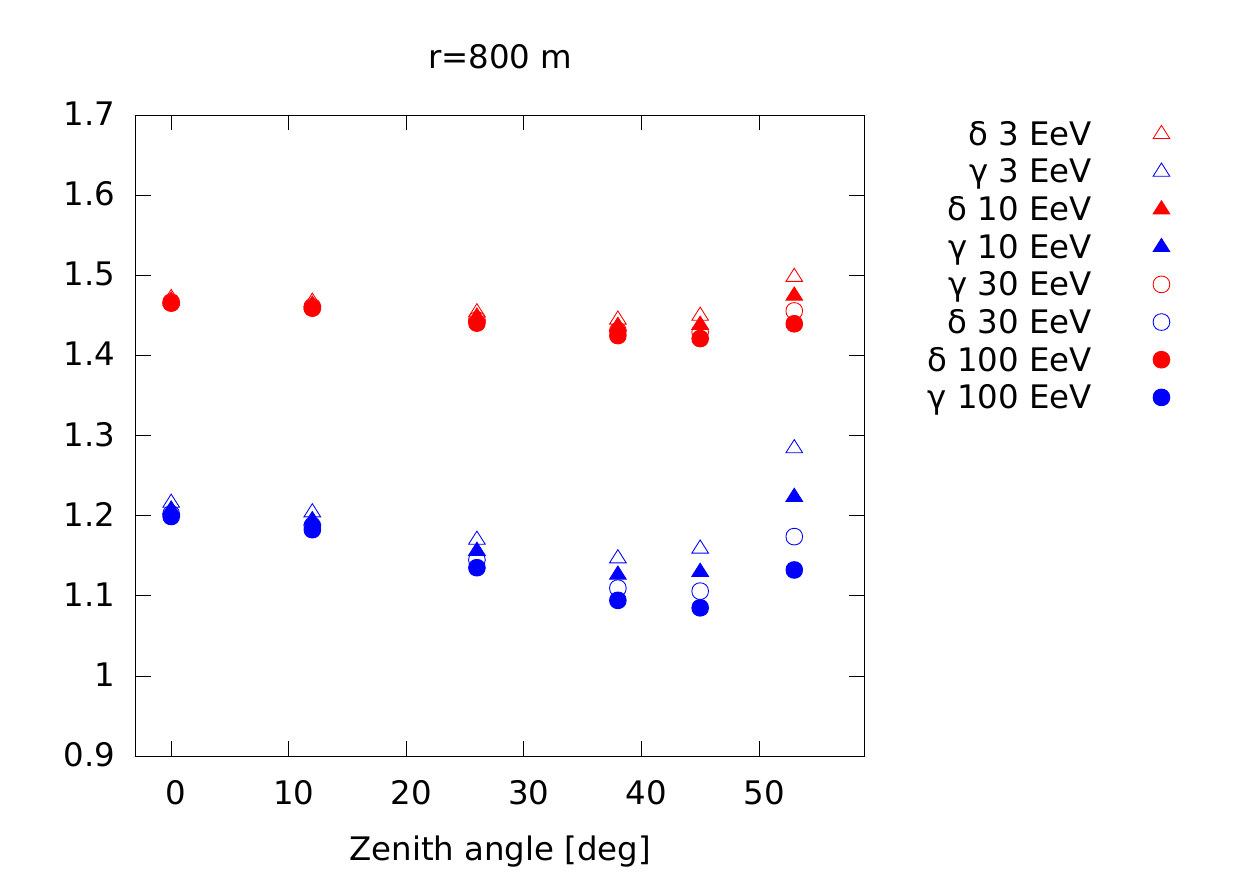}\includegraphics[width=0.5\columnwidth]{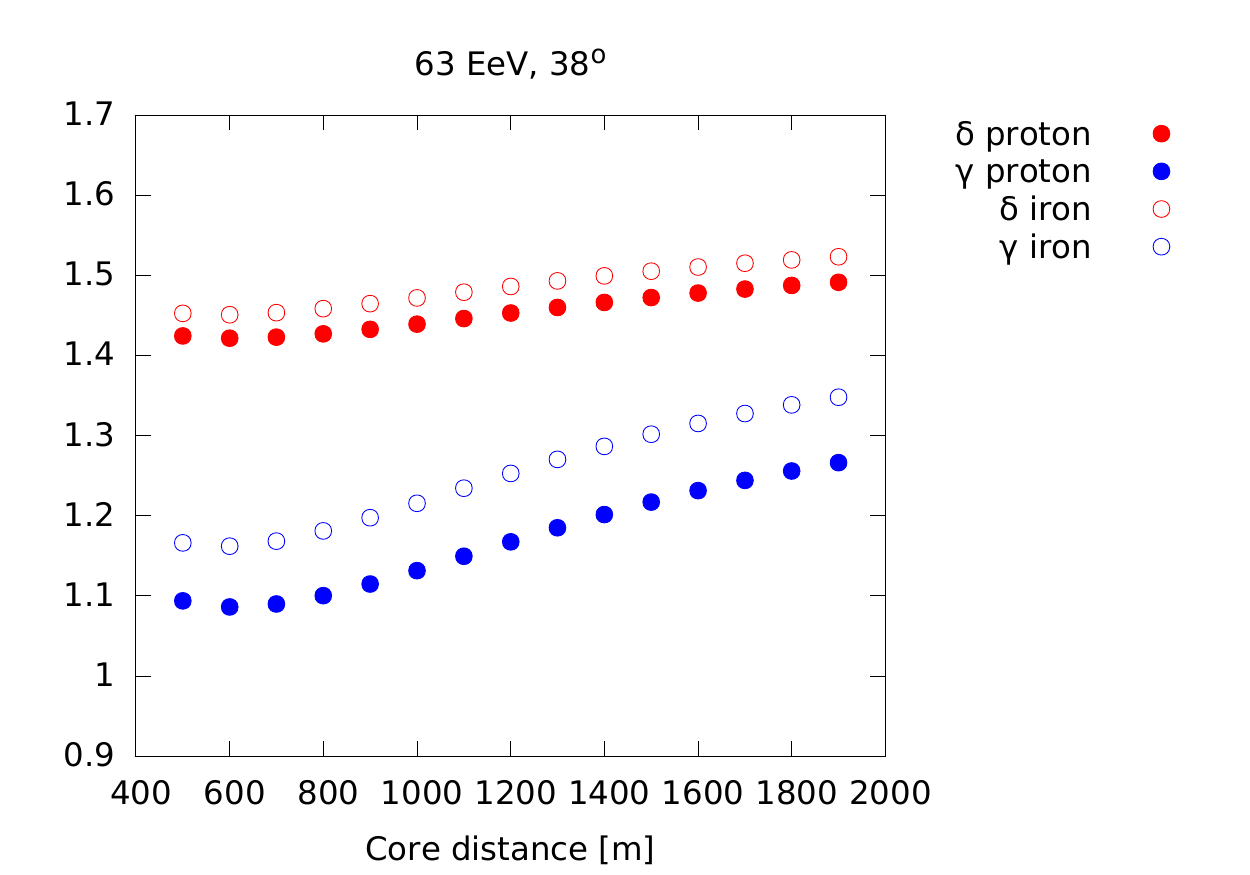}
\caption{Coefficients for the matrix inversion approach. The values of
  $\delta$ and $\gamma$ were calculated from showers simulated with QGSJet II.04
  and EPOS-LHC using the GEANT-based detector simulation of the
  signals in the water-Cherenkov and scintillator detectors. }
\label{fig:matrix-elements}
\end{figure}

\begin{figure}[t]
\centering
\includegraphics[height=0.39\columnwidth]{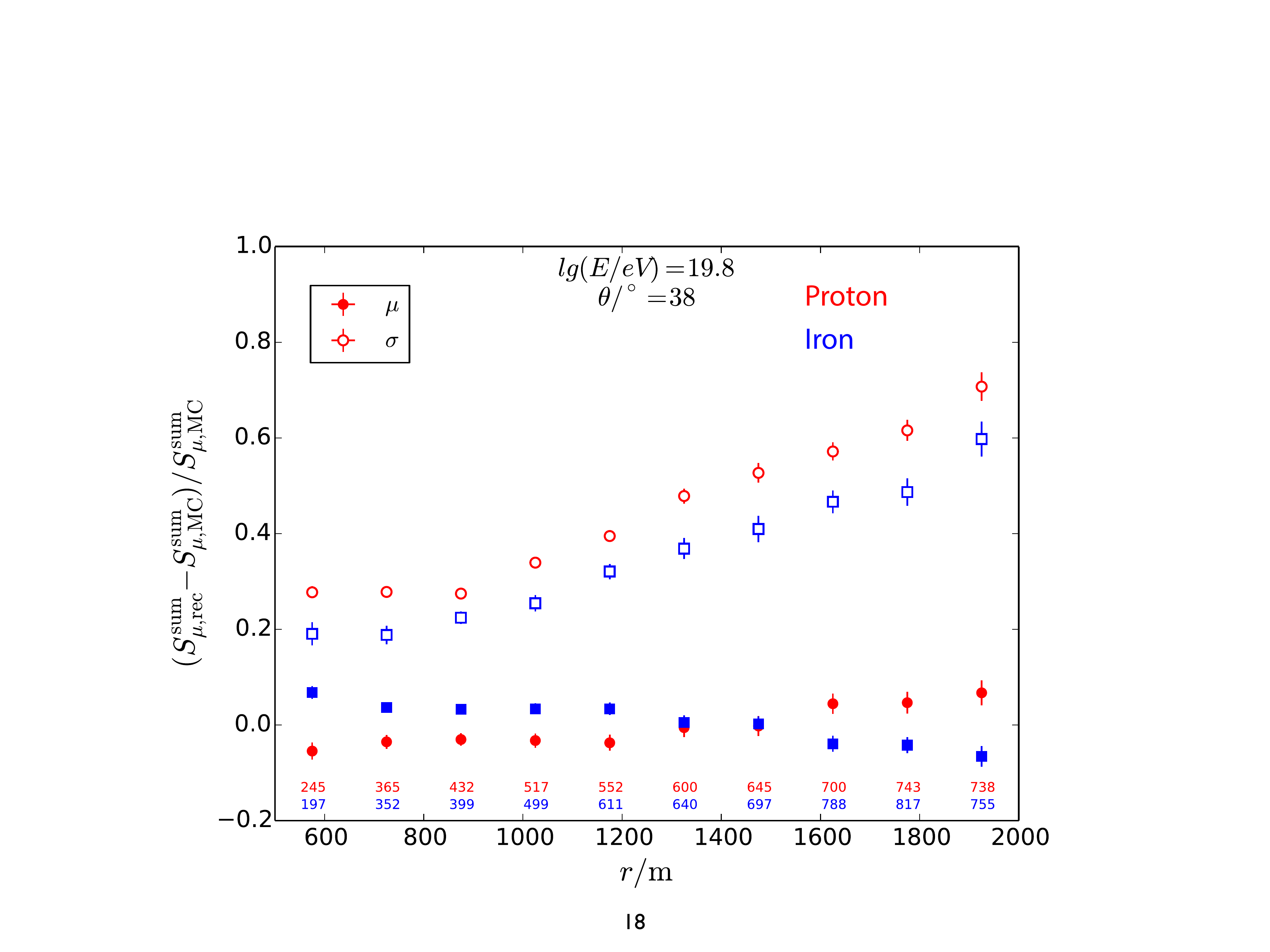}
\caption{Reconstruction bias (solid symbols) and resolution (open symbols) of the muonic signal
contribution for individual detector stations. The results for proton and iron
showers are shown in red and blue, respectively.
The number of detector
stations analyzed for the different lateral distance intervals is also given.
}
\label{fig:matrix-inversion-bias-resolution}
\end{figure}

The coefficients $\delta$ and $\gamma$, introduced in
Sec.~\ref{sec:matrix-inversion}, are shown in
Fig.~\ref{fig:matrix-elements} for different lateral distances, zenith
angles, and primary particles. A small dependence
on the mass of the primary particle and similarly on the hadronic
interaction model can be seen. To check the impact of this small dependence,
the resolution and the mass-dependent bias in the
reconstruction of the muon density for individual detector stations is
given in Fig.~\ref{fig:matrix-inversion-bias-resolution}. The
resolution is about $20-30$\% on the single detector level for showers
in the flux suppression region if stations with a lateral distance up
to $\unit[1000]{m}$ are considered. This value is the result of a
first implementation of the matrix inversion method without any
optimization. It is expected that the resolution can be improved for
large lateral distances by using the constraint that the detector
signal is dominated there by muons.

In the following we will use constant values for $\delta$ and
$\gamma$. Including the dependence of the coefficients on lateral
distance and zenith angle, even if it is small, will help to
improve the results and is foreseen once enough data is available to
characterize the SSD response to the electromagnetic and muonic shower
particles in more detail.


\subsection{Event-based observables and merit factors}

It is convenient to parameterize the discrimination power between the
primary particles $i$ and $j$ in terms of the merit factor, defined as
\begin{equation}
f_{\rm MF} = \frac{|\langle S_i\rangle -\langle S_j\rangle|}{\sqrt{\sigma(S_i)^2 + \sigma(S_j)^2}} .
\label{eq:merit-factor}
\end{equation}
Merit factors of $1.5$ or higher allow a comfortable separation of the respective elements.


\subsubsection{Single station estimate}

The simplest, but at the same time least powerful method for
distinguishing different primaries is the use of the muon density
measured in individual detectors in a given lateral distance range.
Due to the limited resolution of the matrix inversion method the
corresponding merit factors for distinguishing between proton and iron
primaries are only about 0.9 for stations at a lateral distance of
$\unit[800]{m}$ from the core and shower energies $\unit[E>
  10^{19.5}]{eV}$.


\subsubsection{Muon lateral distribution}

\begin{figure}[t]
\centering
\def\figh{0.37}
\includegraphics[height=\figh\columnwidth]{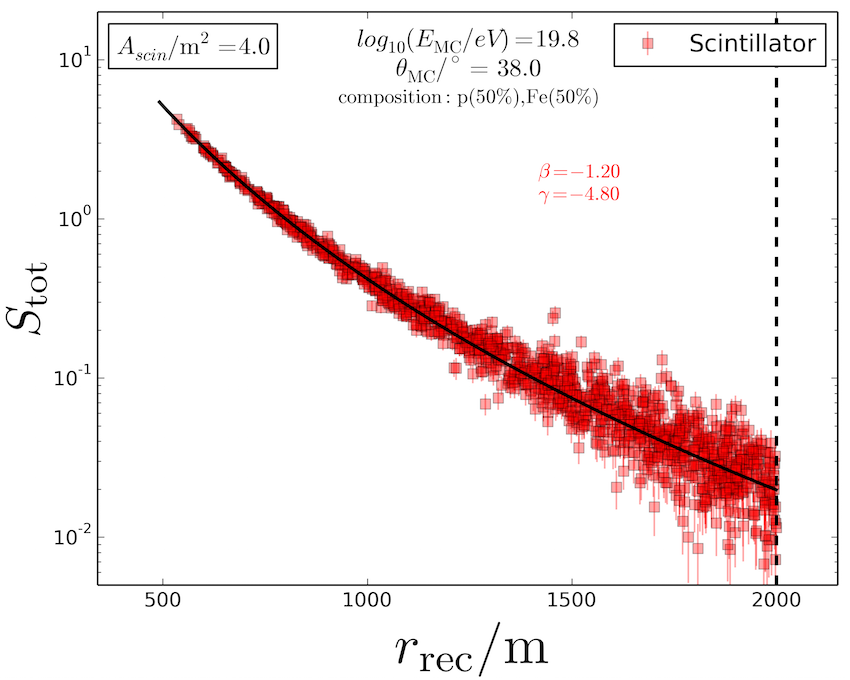}\hfill
\includegraphics[height=\figh\columnwidth]{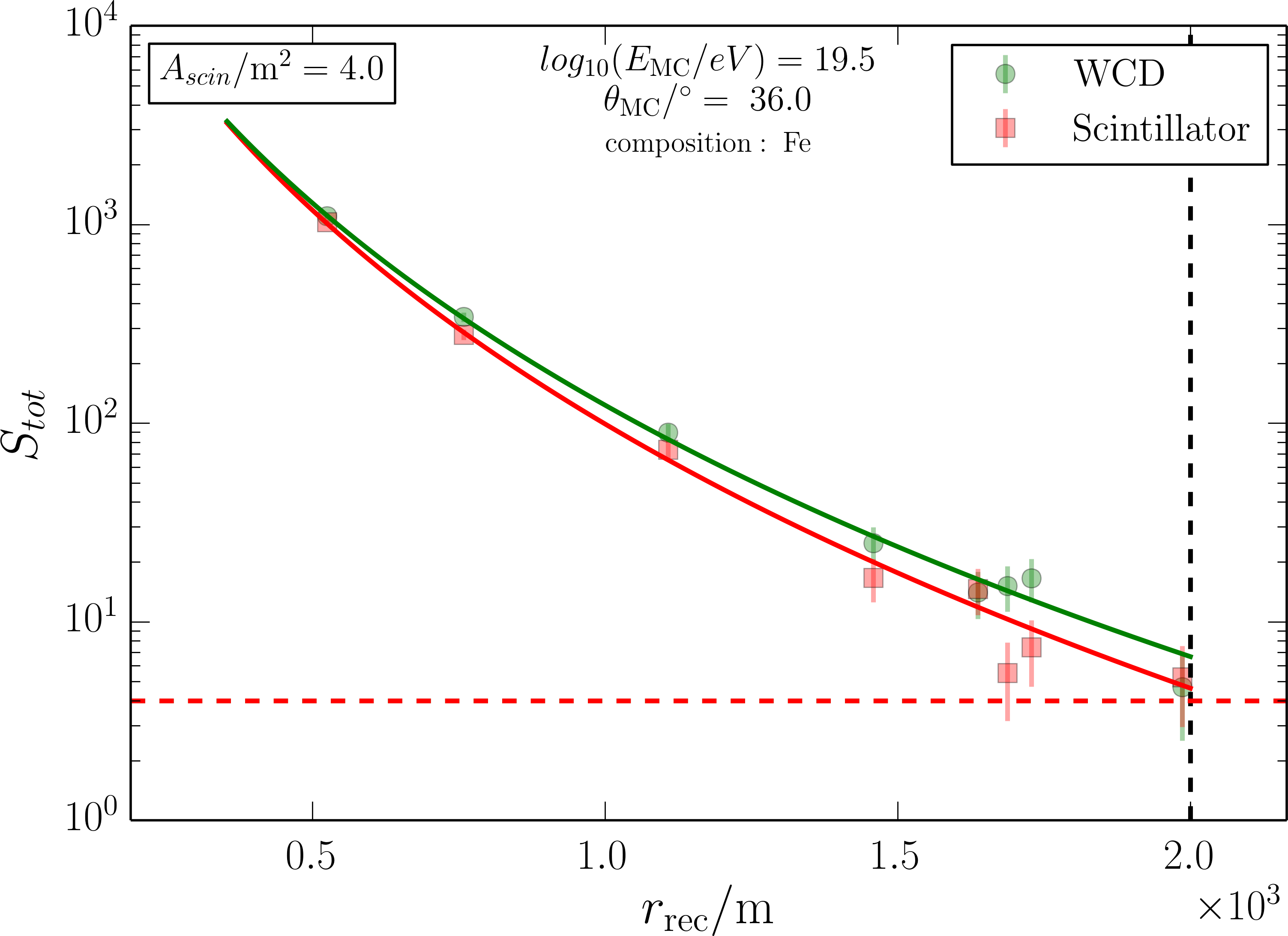}
\caption{Left: Simulated showers for deriving the mean lateral
  distribution of the scintillator detectors. Right: Example of the
  LDFs of one iron shower using the shape parameters derived before.
  Both the results for the WCD and the SSD are shown. The horizontal
  dashed line indicates the single station trigger threshold and the
  vertical line the range up to which stations are used in the LDF
  fit.  }
\label{fig:LDF-examples}
\end{figure}

Fitting first a lateral distribution function (LDF) to the signals of
the scintillator detectors in an event allows the estimate of the muon
density at a given distance with very much reduced statistical
uncertainty. In the following we have used simulated showers ($50$\%
proton and iron primaries) to first derive a parameterization of the
LDF for the SSD.  The slope of this parameterization,
\begin{equation}
\operatorname{LDF}(r) = C\,\left(\frac{r}{800\,{\rm m}}\right)^\beta
\left(\frac{r+700\,{\rm m}}{800\,{\rm m}+700\,{\rm m}}\right)^\gamma ,
\end{equation}
is kept fixed in the subsequent analysis of another, independent set
of Monte Carlo generated showers.  Only the normalization is fitted
on an event-by-event basis. The showers were
selected from a sample simulated with a continuous energy
distribution and the energy derived from $S(1000)$ using
the standard energy conversion of the Auger Observatory.  
The Monte Carlo data for deriving the LDF for SSD and its application
to one example event are shown in Fig.~\ref{fig:LDF-examples}.

The matrix inversion algorithm is then applied to the LDF values for
the WCD and SSD to calculate the muonic signal expected in a
water-Cherenkov detector at $\unit[800]{m}$ core distance,
$S_\mu(800)$.  A reconstruction resolution of the muonic signal of,
for example,
\begin{equation}
\left. \frac{\sigma[S_\mu(800)]}{\langle S_\mu(800)\rangle}\right|_{\rm proton} \approx 22\%
\hspace*{1cm}{\rm and}\hspace*{1cm}
\left. \frac{\sigma[S_\mu(800)]}{\langle S_\mu(800)\rangle}\right|_{\rm iron} \approx 14\%
\end{equation}
is reached at $\unit[E \approx 10^{19.8}]{eV}$ and $\theta = 38^\circ$.
Using $S_\mu(800)$ as composition estimator, the obtained merit factors for
distinguishing between proton and iron
primaries are above $1.5$ at high shower energies
($\unit[E> 10^{19.5}]{eV}$) and small zenith angles.


\subsubsection{Universality and multivariate analyses}

A universality-based analysis, or a sophisticated multivariate
analysis, allows one to correlate the detector signals at different
lateral distances and also takes advantage of the arrival time (shower
front curvature) and temporal structure of the signal measured in the
detectors. At this stage we are only at the beginning of developing a
reconstruction using all these observables. Nevertheless, some results
are given in the following, but it should be kept in mind that the
corresponding merit factors should be considered as lower limits to
what will be reached after having a better understanding of the
detectors.

The universality-based reconstruction provides the depth of shower
maximum with a resolution of about $\unit[35]{g/cm^2}$. If one were to
use only the \xmax derived from universality one would obtain a merit
factor of $1.2$ for the separation of proton and iron primaries over a
wide range in energy and zenith angle. Conversely, if one would use
only the relative muon number from the universality reconstruction the
merit factor would be just above $1.4$. The merit factors derived from
the LDF fit are larger because of the different energy reconstructions
used. In the case of the universality reconstruction the energy is
obtained together with \xmax and $N_\mu^{\rm rel}$. The current
implementation of the universality reconstruction exhibits a
composition-dependent bias in the energy reconstruction that reduces
the merit factors. Work is ongoing to reduce this bias and
correspondingly increase the merit factors.

A summary of merit factors for the separation of different primary
particles at different energies and zenith angles is given in
Tabs.~\ref{tab:mf1} and \ref{tab:mf2}. The merit factors are
consistently higher for showers simulated with EPOS-LHC in comparison
to QGSJET II-04.  This is related to the higher muon multiplicity of
EPOS showers. As even EPOS predictions of the muon number are lower
than those observed in data, the merit factors will be higher when
applying universality to real data. Finally it should be mentioned
that advanced multivariate analysis methods are expected to reach
similar merit factors as shown here.

\begin{table}[t]
\caption{\label{tab:mf1} Fisher discriminant merit factor
  $f_\text{MF}^\text{F}$ for fixed energy simulations. In addition,
  merit factors $f_\text{MF}^\text{Rec}$ are shown after accounting
  for the degradation due to the resolution of the energy
  reconstruction.}
\centering
\begin{tabular}{ccr@{-}lccc}
\toprule
Model & Energy & \multicolumn{2}{c}{Composition} & Zenith angle & $f_\text{MF}^\text{F}$ &
$f_\text{MF}^\text{Rec}$\\
\midrule
QGSJetII-04 & 10\,EeV & Proton & Iron & All & 1.49 & 1.31\\
QGSJetII-04 & 10\,EeV & Proton & Helium & All & 0.47 & 0.34\\
QGSJetII-04 & 10\,EeV & Nitrogen & Iron & All & 0.6 & 0.46\\[2mm]

QGSJetII-04 & 63\,EeV & Proton & Iron & All & 1.86 & 1.59 \\
QGSJetII-04 & 63\,EeV & Proton & Helium & All & 0.46 & 0.38 \\
QGSJetII-04 & 63\,EeV & Nitrogen & Iron & All & 0.94 & 0.66\\[2mm]

EPOS-LHC & 10\,EeV & Proton & Iron & All & 1.59 & 1.35\\
EPOS-LHC & 10\,EeV & Proton & Helium & All & 0.4 & 0.29\\
EPOS-LHC & 10\,EeV & Nitrogen & Iron & All & 0.75 & 0.62\\[2mm]

EPOS-LHC & 63\,EeV & Proton & Iron & All & 1.82 & 1.45 \\
EPOS-LHC & 63\,EeV & Proton & Helium & All & 0.38 & 0.23 \\
EPOS-LHC & 63\,EeV & Nitrogen & Iron & All & 1.07 & 0.79\\[4mm]

QGSJetII-04 & 10\,EeV & Proton & Iron & 21$^\circ$ &  1.52 & 1.15\\
QGSJetII-04 & 10\,EeV & Proton & Iron & 38$^\circ$ &  1.55 & 1.43\\
QGSJetII-04 & 10\,EeV & Proton & Iron & 56$^\circ$ &  1.5 & 1.43\\[2mm]

QGSJetII-04 & 63\,EeV & Proton & Iron & 21$^\circ$ & 2.08 & 1.56 \\
QGSJetII-04 & 63\,EeV & Proton & Iron & 38$^\circ$ &  1.97 & 1.67 \\
QGSJetII-04 & 63\,EeV & Proton & Iron & 56$^\circ$ &  2.14 & 2.1 \\
\bottomrule
\end{tabular}
\end{table}

\begin{table}[t]
\caption{\label{tab:mf2} Fisher discriminant merit factor for 
a set of simulated showers with continuous energy distribution.
The reconstructed showers have a mean energy of $\unit[10^{19.6}]{eV}$.
}
\centering
\begin{tabular}{cr@{-}lc}
\toprule
Model &  \multicolumn{2}{c}{Composition} & Merit Factor (C)\\
\midrule
QGSJetII-04 & Proton & Iron & 1.54\\
QGSJetII-04 & Proton & Helium & 0.41\\
QGSJetII-04 & Nitrogen & Iron & 0.64\\
\bottomrule
\end{tabular}
\end{table}


\subsection{Cross-checks with an underground muon detector}

There are different kinds of cross-checks that can be performed with
an array of additional, independent muon detectors. With the upgraded
surface detectors being fully efficient for muon separation only above
$\unit[10^{19}]{eV}$, however, one would need an array of independent
muon detectors of the size of 200 to $\unit[300]{km^2}$ to collect
enough statistics to do an event-by-event comparison of the muon
content of the showers.

\begin{figure}[t]
\centering
\def\figh{0.39}
\includegraphics[height=\figh\columnwidth]{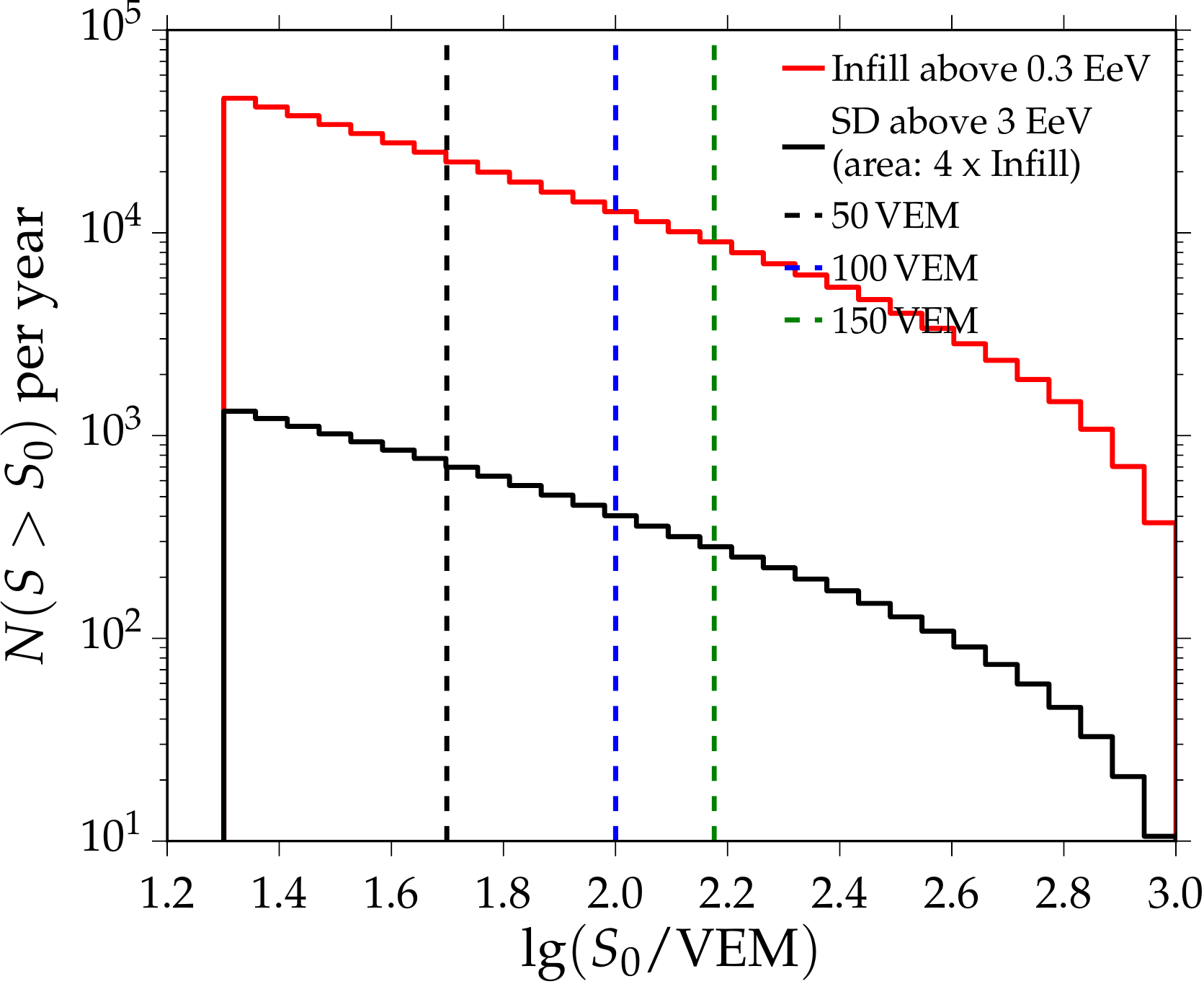}\hfill
\includegraphics[height=\figh\columnwidth]{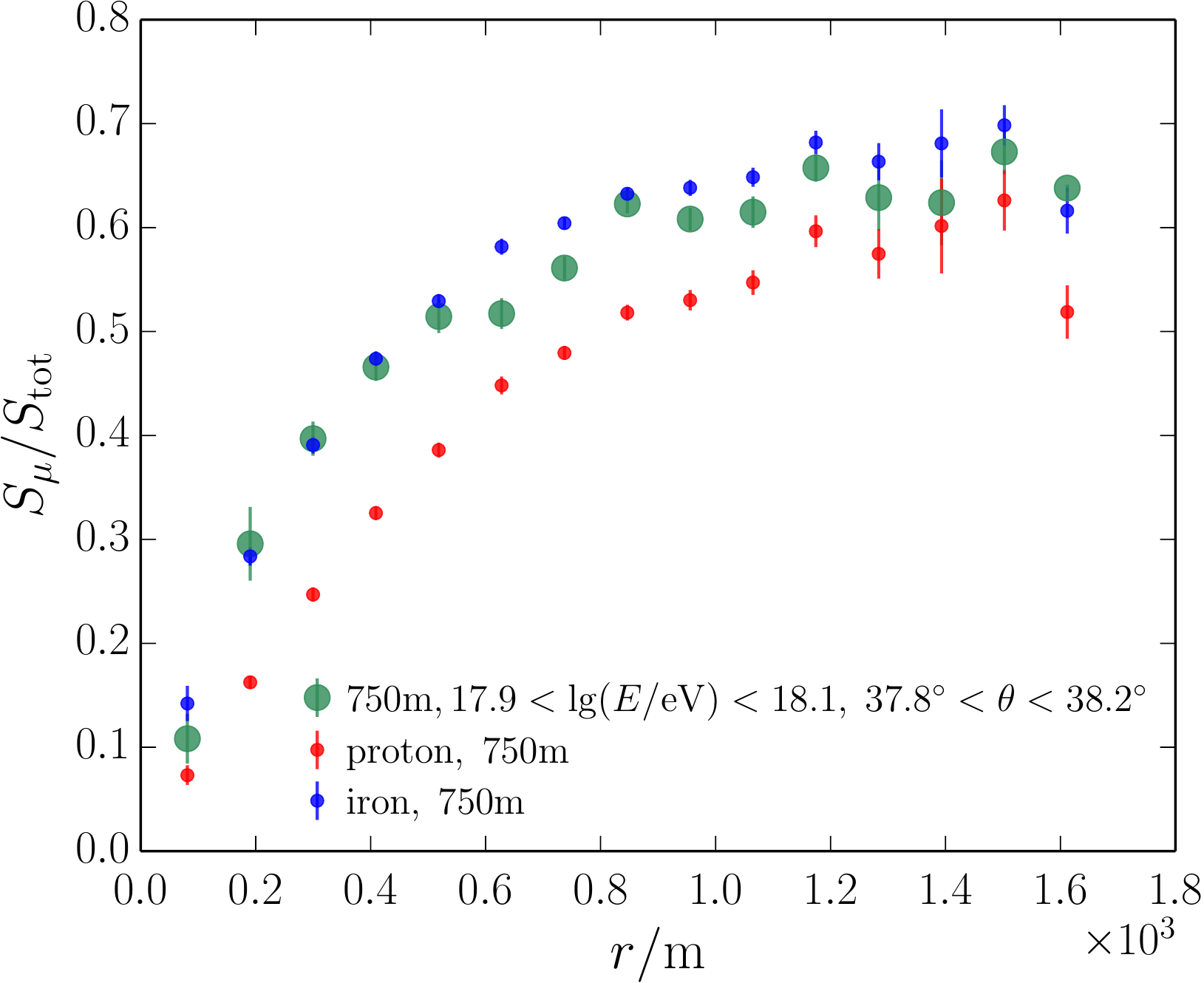}
\caption{Left: Number of stations per year whose signal exceeds
  $S_0$. Shown is the comparison of the rate detected in the $\unit[750]{m}$
  array and the regular array if 61 stations are considered in both
  cases.  Right: Ratio between the muonic and electromagnetic
  contributions to the signal of the WCD as function of the lateral
  distance, and for shower energies typical for the $\unit[750]{m}$
  array. Shown are the expectations for proton and iron showers (small
  red/blue symbols) and the ratio reconstructed for Auger data using
  universality (large green circles).}
\label{fig:HPA-statistics}
\end{figure}

A much more economic option is to limit the cross-checks to individual
detector stations. This can be done with the $\unit[30]{m^2}$ AMIGA
counters foreseen to be deployed in the region of Auger's $\unit[750]{m}$ array. The trigger
threshold for full efficiency of the $\unit[750]{m}$ array is below
$\unit[10^{17.5}]{eV}$, in comparison to $\unit[10^{18.5}]{eV}$ for
the regular array.  A comparison of the statistics of the number of
detectors having a WCD signal larger than $S_0$ in the $\unit[750]{m}$ array region
($61$ detectors with $\unit[750]{m}$ spacing) and in the regular array
($61$ detectors with $\unit[1500]{m}$ spacing) is shown in
Fig.~\ref{fig:HPA-statistics}\ (left). More than 10,000 times per year
a station in a shower will exceed a signal of $\unit[100]{VEM}$ in the
$\unit[750]{m}$ array region. The corresponding rate would be well below 1000 for
the regular array and the same number of additional muon detectors.

Showers detected with the $\unit[750]{m}$ array have typically a much lower
energy, and the stations with the highest signal will be closer to the
core than in the regular array. For example, a station with $S\sim100$
to \unit[200]{VEM} will have a typical lateral distance of
$\unit[1000]{m}$ in the regular array and $\unit[400]{m}$ in the
$\unit[750]{m}$ spacing array.  The different distances and shower energies lead to a
different ratio between the muonic and electromagnetic signal
contributions. This ratio is about 0.5 to 0.6 at $\unit[1000]{m}$ for
showers of $\unit[10^{19.5}]{eV}$.  The corresponding value for the
$\unit[750]{m}$ array is shown in Fig.~\ref{fig:HPA-statistics}\ (right).  The
ratio between the muonic and electromagnetic shower signals will be
about 20 to 30\% smaller than that of ultra-high energy showers. This
difference is not too large and will allow us to use the AMIGA
counters for cross-checking the muon measurement with the upgraded
surface array.


\section{Fluorescence Detector performance parameters}

\subsection{Geometry, energy, $\boldsymbol{X_\text{max}}$ reconstruction}

\begin{figure}[t]
\centering
\includegraphics[width=0.55\textwidth]{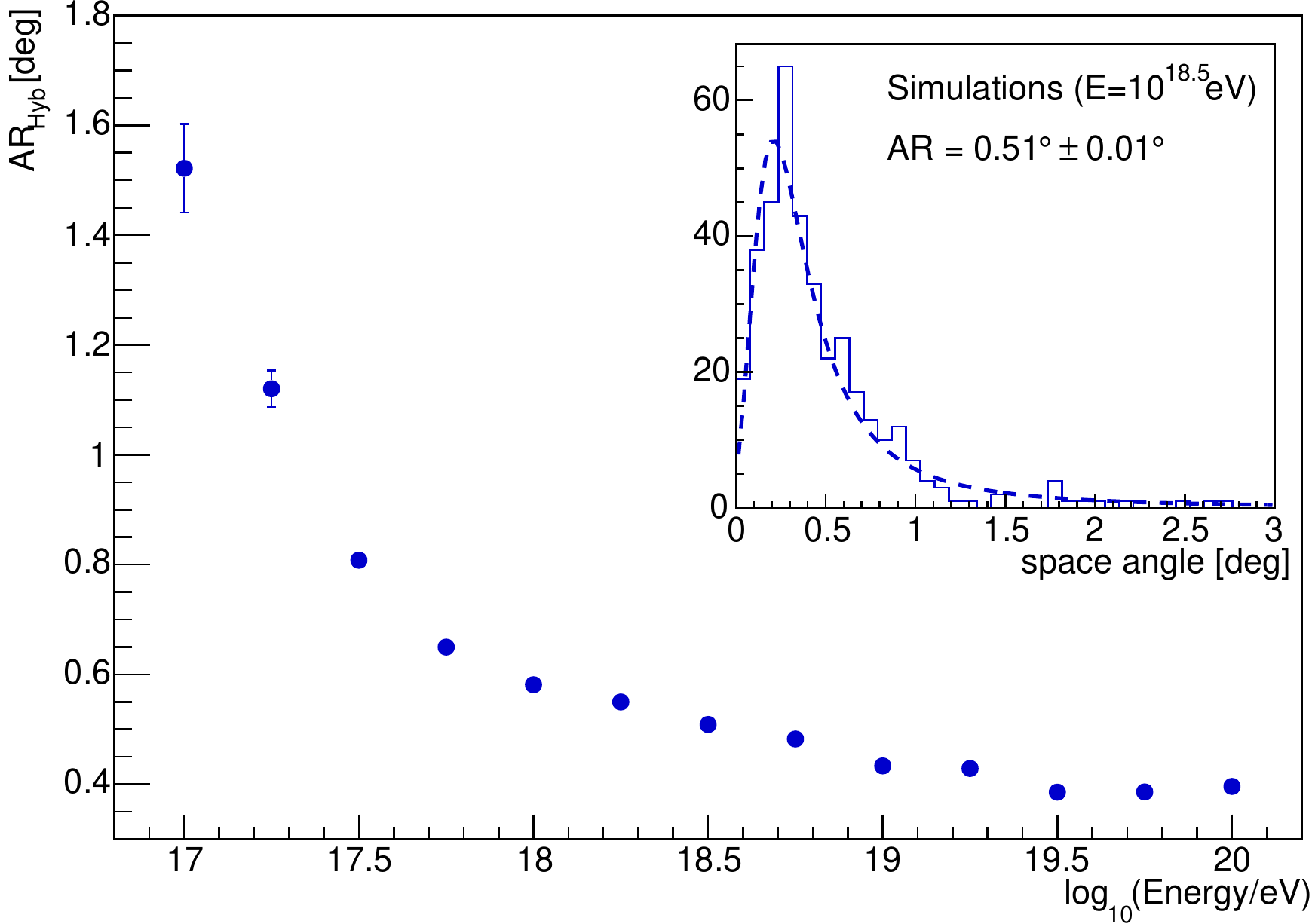}
\caption{Angular resolution for hybrid events.}
\label{fig:FD-AR}
\end{figure}

In the FD, cosmic ray showers are detected as a sequence of triggered
pixels in the camera. The first step in the analysis is the
determination of the shower-detector plane (SDP). The SDP is the plane
that includes the location of the FD station and the line of the shower
axis. Next, the timing information of the pixels is used for
reconstructing the shower axis within the SDP.  The accuracy of the
monocular reconstruction is limited. More accurate reconstruction is
obtained by combining the timing information from the SD stations with
that of the FD telescopes. This is called hybrid reconstruction. A
hybrid detector achieves the best geometrical accuracy by using timing
information from all the detector components, both FD pixels and SD
stations.  As can be seen in Fig.~\ref{fig:FD-AR},
the angular resolution for hybrid events above $10^{18.5}$~eV is
better than $0.5$\textdegree.

The FD provides a nearly calorimetric energy measurement as the
fluorescence light is produced in proportion to the energy dissipation
by a shower in the atmosphere. The reconstruction of the fluorescence
events is a complex process that requires the knowledge of several
parameters, e.g. the absolute calibration of telescopes, the
fluorescence yield, light attenuation and scattering in the
atmosphere, optical properties of telescopes and an estimate of the
invisible energy. The current systematic uncertainty on the energy
scale for hybrid events (i.e. at least one SD station is used in the
reconstruction) is $14\%$.

\begin{figure}[t]
\centering
\includegraphics[width=0.48\textwidth]{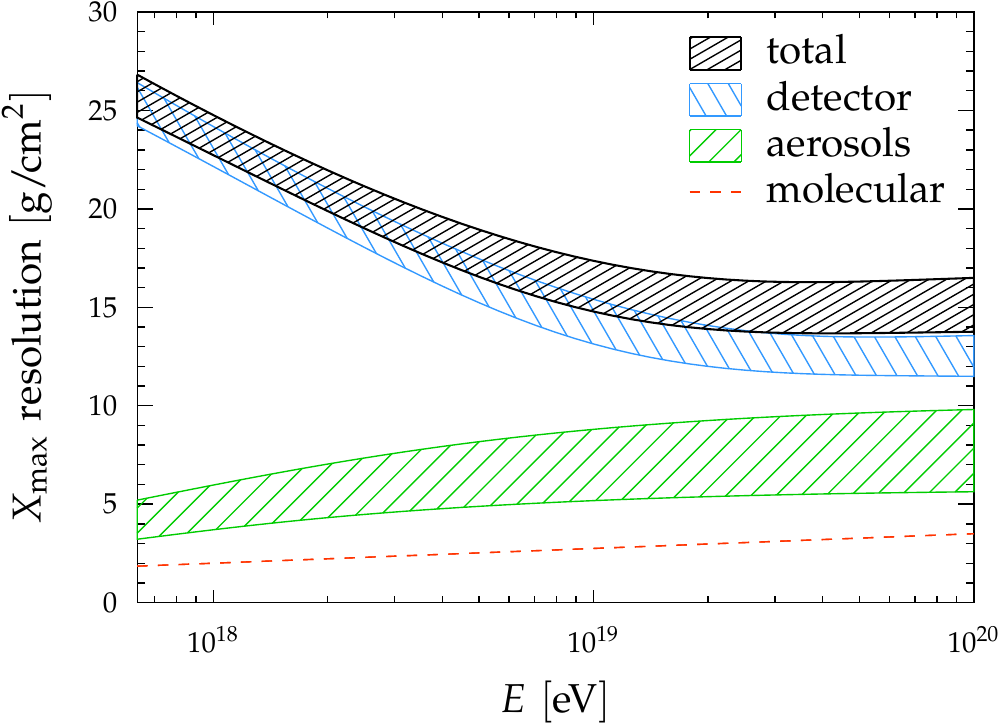}\hfill
\includegraphics[width=0.48\textwidth]{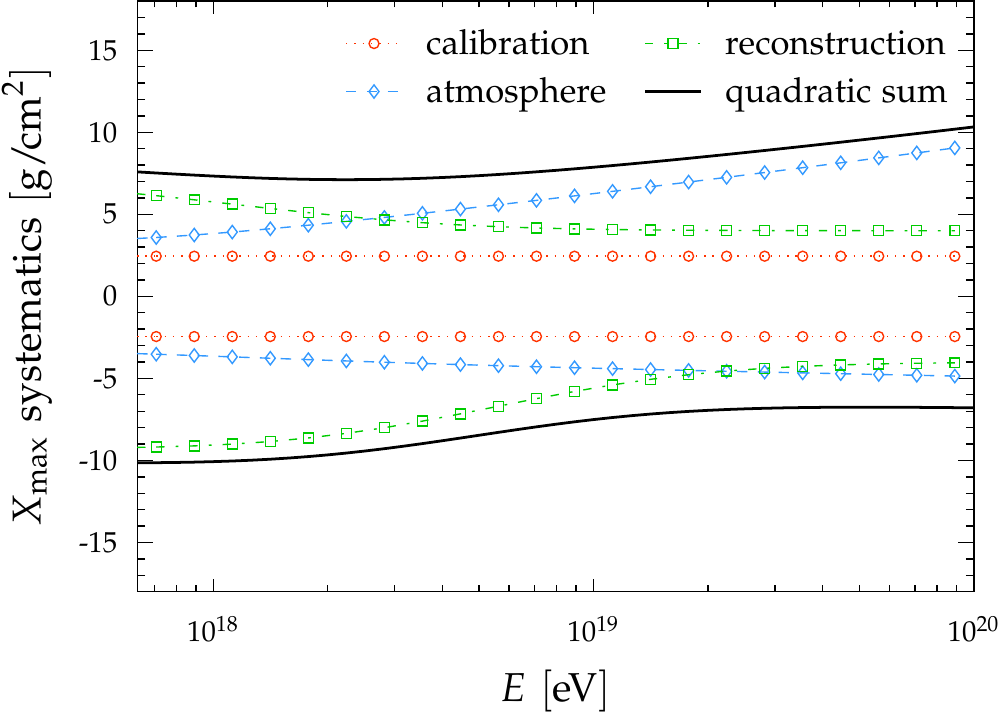}
\caption{ Left panel: \xmax resolution as a function of
  energy. Bands denote the estimated systematic uncertainties.  Right
  panel: Systematic uncertainties in the \xmax scale as a
  function of energy.}
\label{fig:FD-Xmax}
\end{figure}

The energy evolution of the $X_{\text{max}}$ resolution is shown in
the left panel of Fig.~\ref{fig:FD-Xmax}. As can be seen, the total
$X_{\text{max}}$ resolution is better than $26$\,g/cm$^2$ at
$10^{17.8}$\,eV and improves with energy to reach about $15$\,g/cm$^2$
above $10^{19.3}$\,eV. The systematic uncertainty of the
$X_{\text{max}}$ scale, i.e.\ the precision with which the absolute
value of $X_{\text{max}}$ can be measured, is shown in the right panel
of Fig.~\ref{fig:FD-Xmax}. As can be seen, this uncertainty is
$\leq10$\,g/cm$^2$ at all energies. At low energies, the scale
uncertainty is dominated by the uncertainties in the event
reconstruction and at high energies the atmospheric uncertainties
prevail.

\subsection{Estimated event numbers}

\begin{figure}[t]
\centering
\includegraphics[width=0.45\textwidth]{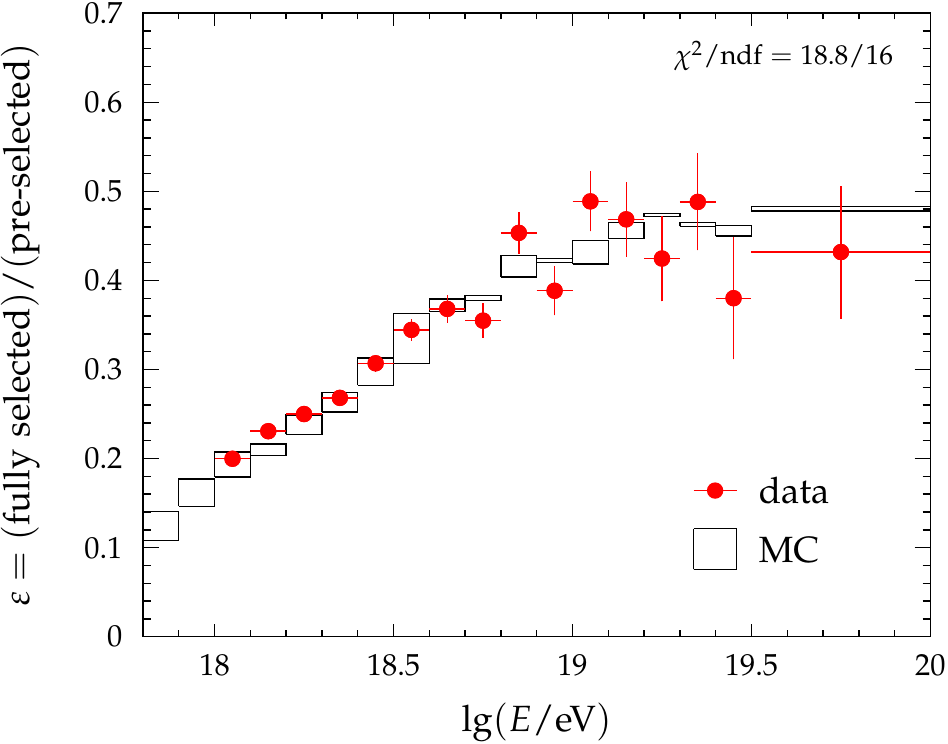}
\caption{
Efficiency of the quality and fiducial selection for data and MC. The $\chi^2$ of the sum of the (data-MC)
residuals is quoted on the top right.
}
\label{fig:FD-Efficiency}
\end{figure}

The selection efficiency after including the quality and fiducial cuts
is between $40$ and $50\%$ above $10^{19}$\,eV, see
Fig.~\ref{fig:FD-Efficiency}. A potential bias from these selection
cuts can be checked by comparing its efficiency as a function of
energy for data and simulated events.  For this purpose, we use the
independent measurement of air showers provided by the SD and measure
the fraction of events surviving the quality and fiducial cuts out of
the total sample of pre-selected events. This estimate of the
selection efficiency is shown in Fig.~\ref{fig:FD-Efficiency} as a
function of SD energy above $10^{18}$\,eV. Below that energy, the SD
trigger efficiency drops below $50\%$. The comparison to the simulated
data shows a good overall agreement and we conclude that the selection
efficiency is fully described by our simulation.

The expected number of events after seven years of the standard FD
measurement is $510$ and $37$ above energy $10^{19}$\,eV and
$10^{19.5}$\,eV, respectively. An additional increase up to about
$40\%$ at the highest energies can be achieved by extending the FD
operation to periods with higher night sky background. For more
details see Chapter~\ref{FDextension}.


\section{Benefits of hybrid (multi-detector) measurements}

There are many benefits of having scintillator detectors in addition
to the existing array of WCD that have not been discussed so far.

First of all, a direct comparison of the data of Telescope Array (TA)
and the Pierre Auger Observatory can be made because it will be
possible to implement the reconstruction procedures of TA also at the
Auger Observatory. For example, the energy calibration of the surface
arrays can be compared directly if mapped to the scintillator signal
at a given lateral distance. Furthermore, the different indications
for muon discrepancies can be compared between both experiments.
Many other comparisons will be possible, including
composition-sensitive quantities such as shower front curvature or
signal rise times.

Another important application of the multi-detector measurements
possible with the upgraded array are detailed investigations of
detector aging effects and possible drifts in the detector
calibration.  In general, the reduction of existing systematic
uncertainties will be one of the main aims of installing the detector
upgrade.

Last but not least, we hope to further improve our understanding of
shower physics and, in particular, of universality based and
multivariate reconstruction techniques in the end by such a degree
that we will be able to re-analyze the data taken with the Auger
Observatory before the upgrade was installed, and derive reliable
composition information from the data taken so far.


\section{Application to physics goals}

In the following we will apply the universality reconstruction to
simulated data of the water-Cherenkov and scintillator detectors 
to demonstrate the ability to derive composition sensitive observables
with the upgraded Auger Observatory.

Without knowing what composition to expect in the GZK suppression
region it is difficult to demonstrate the potential of the upgraded
Auger Observatory. Therefore, we will use the two flux models
introduced in Chap.~\ref{chap:science} and shown in
Fig.~\ref{fig:models-flux} to illustrate the discriminative power of
the additional muon information. Mock data sets were generated for
these scenarios that reproduce the predicted energy distribution and
energy-dependent composition. The event number of each of these
artificial data sets corresponds to that expected for 7 years of data
taking with the upgraded Auger Observatory.

\begin{figure}[t]
\centering
\includegraphics[width=0.49\columnwidth]{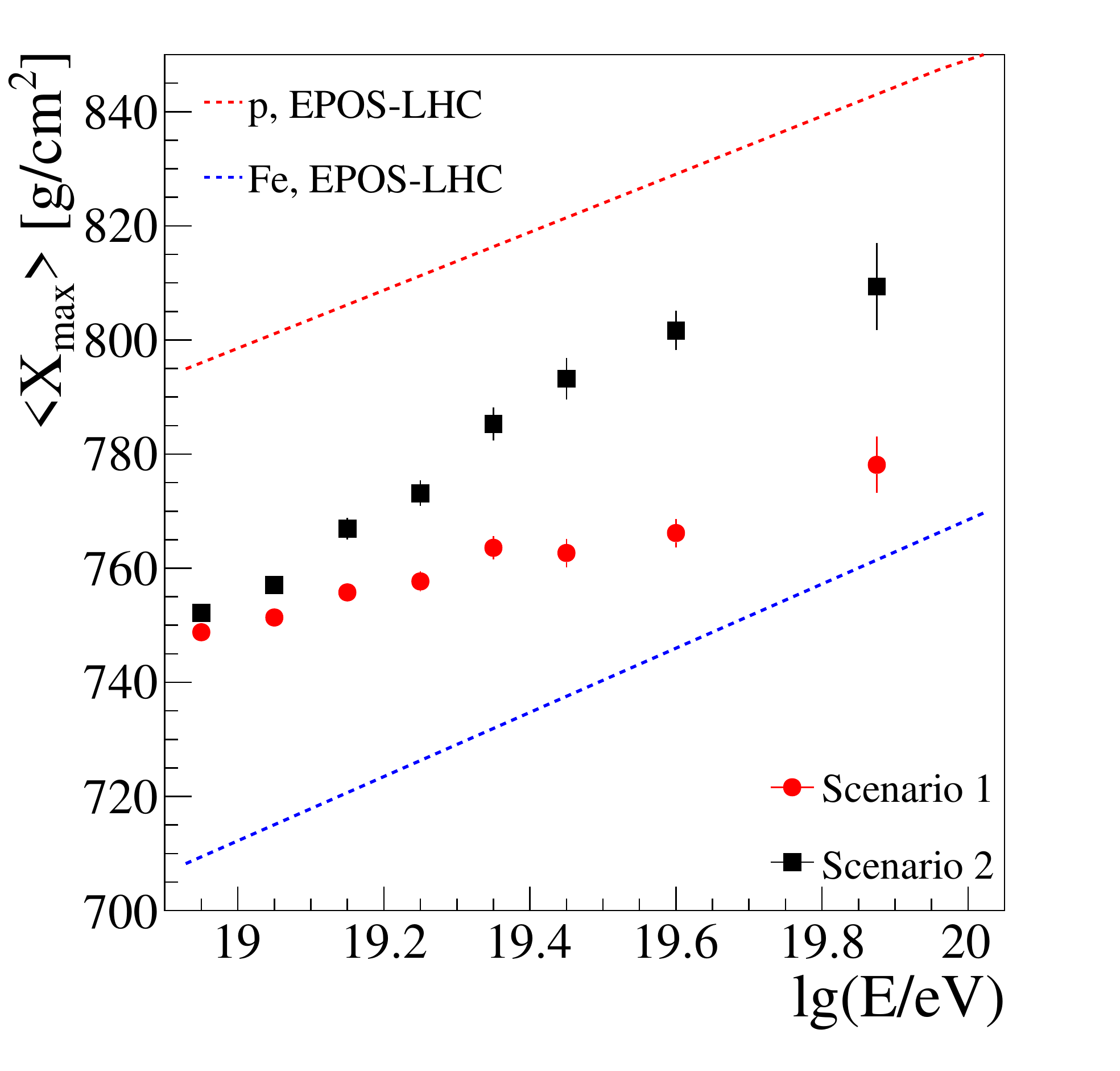}\hfill
\includegraphics[width=0.49\columnwidth]{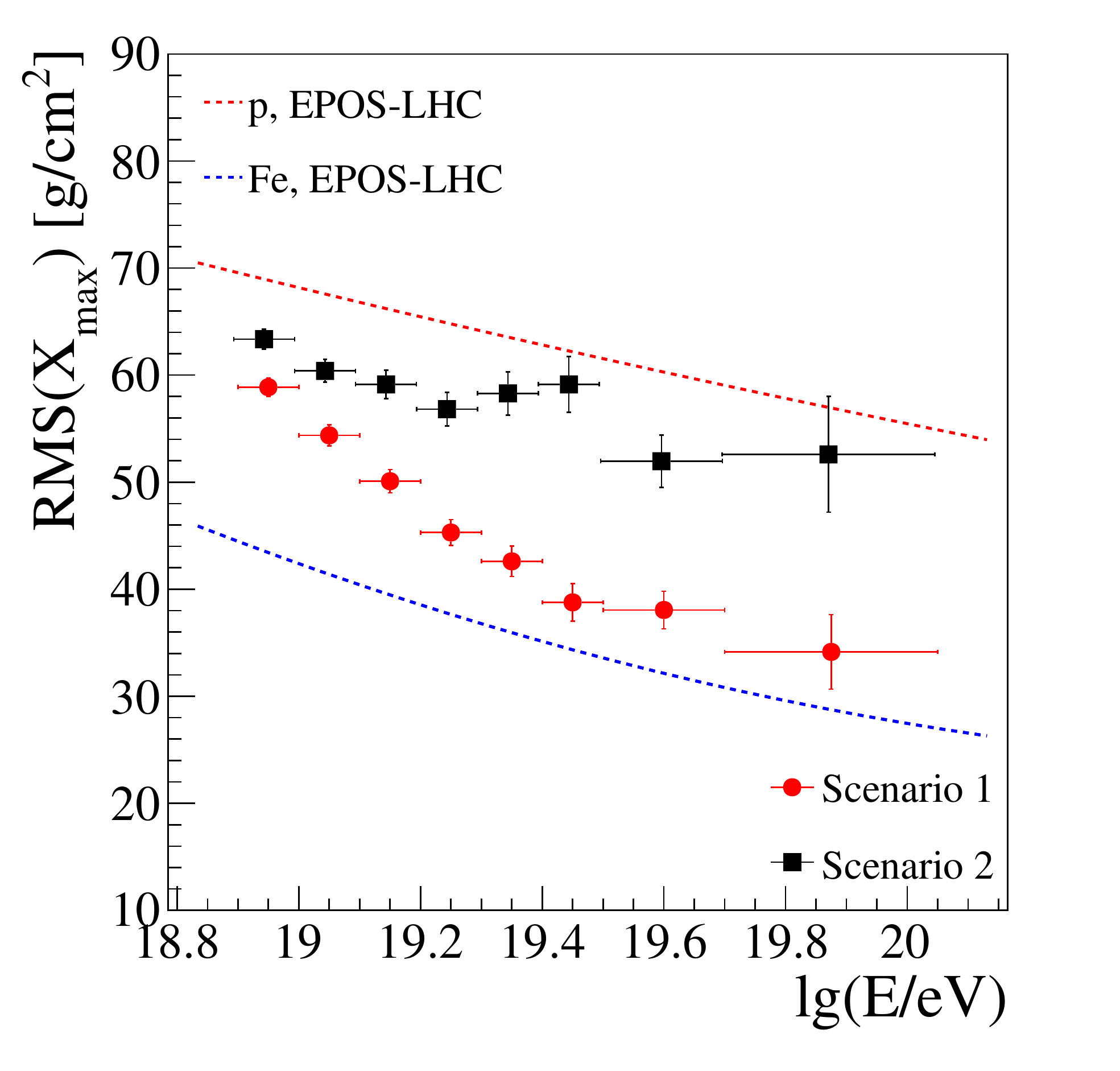}
\caption{Reconstructed mean depth of shower maximum \xmax and its
  fluctuations for the two scenarios: (1) maximum-rigidity model; (2)
  photo-disintegration model. The RMS(\xmax) contains the intrinsic
  air-shower fluctuations and the detector resolution. The same
  quantities as expected for pure p and pure Fe compositions are
  illustrated.  The difference in the evolution of the mass
  compositions of the two models in the energy range of the flux
  suppression can be distinguished with high significance.  }
\label{fig:benchmarks-Xmax}
\end{figure}

\begin{figure}[t]
\centering
\includegraphics[width=0.5\columnwidth]{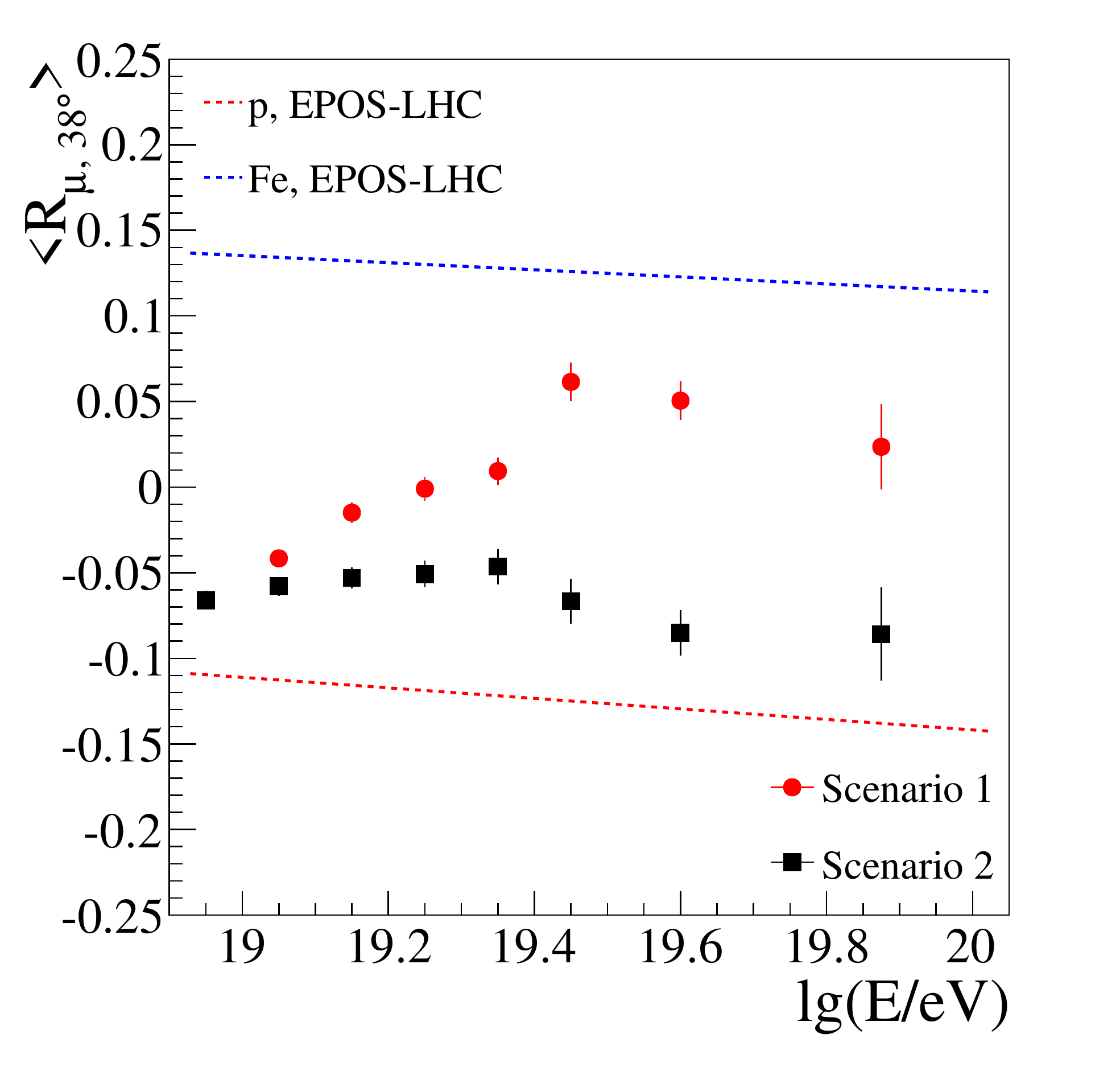}
\caption{Reconstructed mean relative muon number $R_\mu$ for an
  equivalent shower at $38^\circ$ zenith angle. The expectations for 
  pure p and pure Fe compositions are shown together with the two
  scenarios: (1) maximum-rigidity model; (2) photo-disintegration
  model.  In the energy range of the flux suppression the models can
  be distinguished with high significance.}
\label{fig:benchmarks-Nmu}
\end{figure}

\subsection{Composition sensitivity and measurement of the muon number}

One of the key questions of the physics reach of the Auger Upgrade is
that of being able to discriminate different composition and, hence,
physics scenarios in the energy range of the flux suppression. Using
only the surface detector data of the upgraded array we have
reconstructed the number of muons and the depth of shower maximum.
The mean \xmax and the RMS(\xmax) are depicted in
Fig.~\ref{fig:benchmarks-Xmax}. The RMS contains the intrinsic
air-shower fluctuations and the reconstruction resolution.  The number
of muons, $R_\mu$, relative to that expected for an equal mix of
p-He-CNO-Fe as primary particles, is shown in
Fig.~\ref{fig:benchmarks-Nmu}.  The expectations for pure p and pure
Fe compositions for the same variables are also illustrated. While the
mean \xmax, RMS(\xmax) and $R_\mu$ are very similar up to
$\unit[10^{19.2}]{eV}$, the energy range that is well covered by data
of the fluorescence telescopes, the models predict significantly
different extrapolations into the GZK suppression region. This
difference is well reproduced with the reconstructed \xmax, RMS(\xmax)
and $R_\mu$ and the two scenarios can be distinguished with high
significance and statistics.

\subsubsection*{Selection of light primaries}

Another very important feature of the upgraded array will be the
possibility of selecting data sets enriched with light or heavy
primaries. Such a selection will be needed for searching for a proton
component in the primary flux at the highest energies, and for
carrying out anisotropy studies with light/heavy primaries.

\begin{figure}[t]
\centering
\includegraphics[width=0.6\columnwidth]{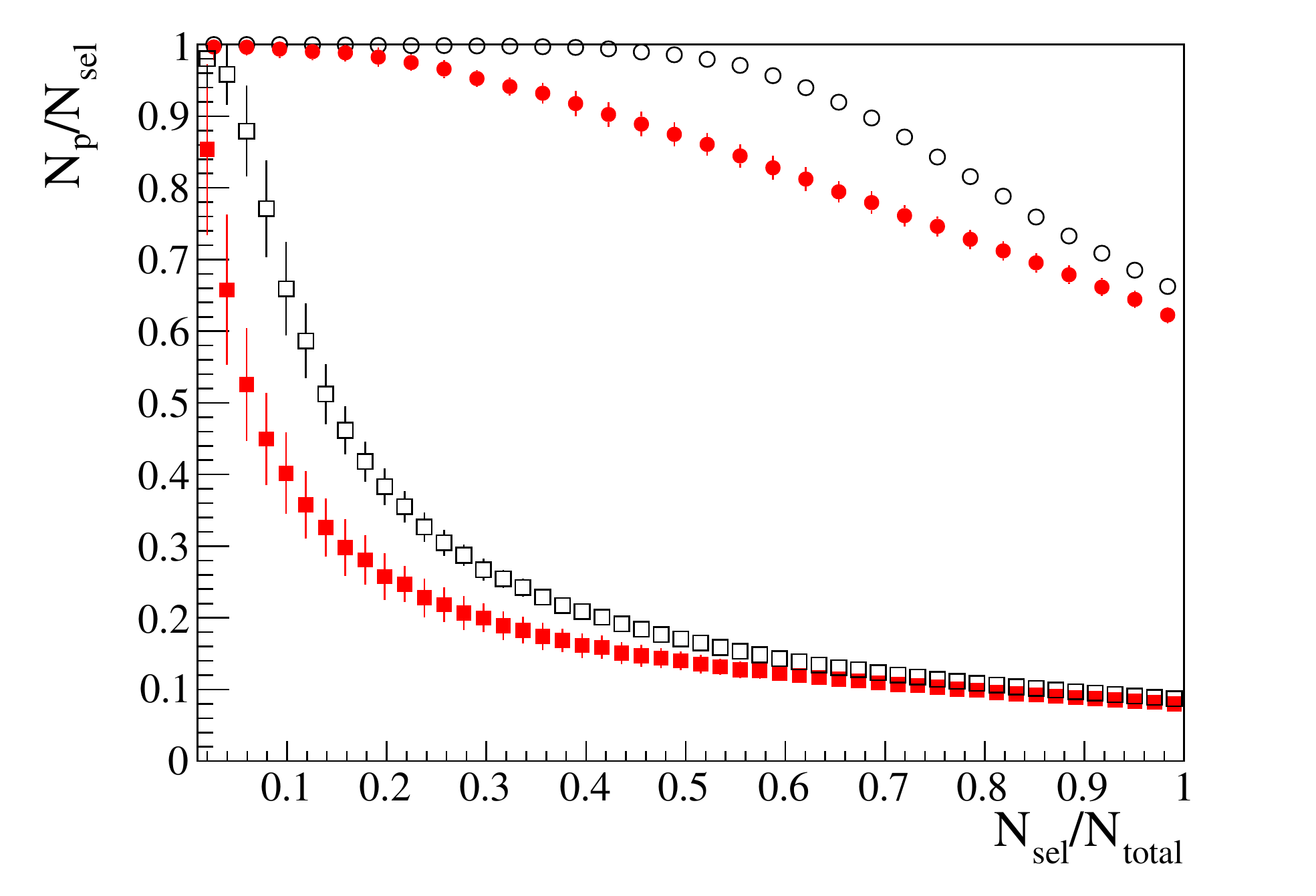}
\caption{Purity of the selected proton-enriched sample ($N_{\rm
    p}/N_{\rm sel}$) as function of the fraction of events selected
  with the cut ($N_{\rm sel}/N_{\rm total}$).  There are 10\% protons
  added to both model scenarios to ensure that there is non-vanishing
  number of protons in the initial data sample. Squares are scenario
  1, and circles are scenario 2.  The solid symbols show the selection
  using the reconstructed observables and the open symbols
  corresponding the theoretical limit, i.e.\ a selection based on the
  generated (true) \xmax and $R_\mu$.  }
\label{fig:Scenarios-purity-efficienty-19p4}
\end{figure}

Using the reconstructed \xmax and $R_\mu$ in a Fisher discriminant
analysis, one can apply event by event a selection cut and study the
purity of the selected data sample as function of the selection
efficiency. The result of such a selection is shown in
Fig.~\ref{fig:Scenarios-purity-efficienty-19p4} for energies greater
than $\unit[10^{19.4}]{eV}$ with the aim of selecting a
proton-enriched data sample. As the reference model 1 (maximum-rigidity
scenario) does not predict any protons at such a high energy we have
added proton events equivalent to 10\% of the flux at all
energies. The open symbols show the result if the selection is applied
to the true \xmax and $R_\mu$ values, i.e. for a perfect detector with
vanishing resolution.  The difference between the solid symbols and
the open symbols shows the effect of the limited detector and
reconstruction resolution. For example, in the model of scenario 1
there are exactly 10\% protons at this energy. Therefore a purity of
0.1 is expected if one selects all events $N_{\rm sel}/N_{\rm total} =
1$. And selecting the 10\% most proton-like events, the purity of the
sample will be 40\%. These numbers are very different for model 2
(photo-disintegration scenario).  Due to the presence of a large
fraction of protons, a purity of 90\% is already reached by selecting
40\% of the most proton-like events.  These studies show that the
upgraded surface array of the Auger Observatory does indeed allow us
to detect a proton component in the primary flux with a relative
contribution of 10\% or more.

\subsection{Composition enhanced anisotropy}

In the following we will consider anisotropy studies applied to a 
data set with the statistics collected by the Auger Observatory
until now~\cite{PierreAuger:2014yba}. To  
illustrate the gain in sensitivity we will have due to adding scintillator detectors to the
surface array we will compare analyses of the same simulated data sets with and
without the additional information provided by the upgraded array.

\begin{table}[t]
\caption{ Examples of parameters for which the correlation of observed
  arrival directions with selected astrophysical objects is the
  strongest (i.e. maximum departure from isotropy).  The energy
  threshold for event selection is $E_\text{th}$ and the maximum
  angular difference is $\Delta\Psi$.  The second section of the table
  shows the parameters for luminosity-selected AGNs with ${\cal
    L}_\text{min}$.  The nominal $p$ value and the chance probability
  accounting for the parameter scans are given as $f_\text{min}$ and
  ${\cal P_{\rm scan}}$.  For details
  see~\protect\cite{PierreAuger:2014yba}.  }
\centering
\begin{tabular}{c c c c c c c }
\toprule
Objects & $E_\text{th}$  & $\Delta\Psi$  &  $D_{\rm max}$ & ${\cal L}_\text{min}$  & $f_\text{min}$ & ${\cal P_{\rm scan}}$ \\
 &  [EeV] &  [$^\circ$] & [Mpc] & [erg/s] & &\\
\midrule
2MRS Galaxies & 52 &  9 &  90 & - &  $1.5{\times} 10^{-3}$ &  24\% \\
Swift AGNs & 58 &  1 &  80 & - &  $6{\times} 10^{-5}$ &  6\% \\
Radio galaxies & 72 &  4.75 &  90 & - &  $2{\times} 10^{-4}$ &  8\% \\[3mm]
Swift AGNs & 58 &  18 &  130 & $10^{44}$ &  $2{\times} 10^{-6}$ &  1.3\% \\
Radio galaxies & 58 &  12 &  90 & $10^{39.33}$ &  $5.6{\times} 10^{-5}$ &  11\% \\[3mm]
Centaurus~A & 58 &  15 & - & - &  $2{\times} 10^{-4}$ &  1.4\% \\
\bottomrule
\end{tabular}
\label{tab:anisotropy-search-parameters}
\end{table}

\begin{figure}[t]
\centering
\def\figh{0.32}
\includegraphics[height=\figh\textwidth]{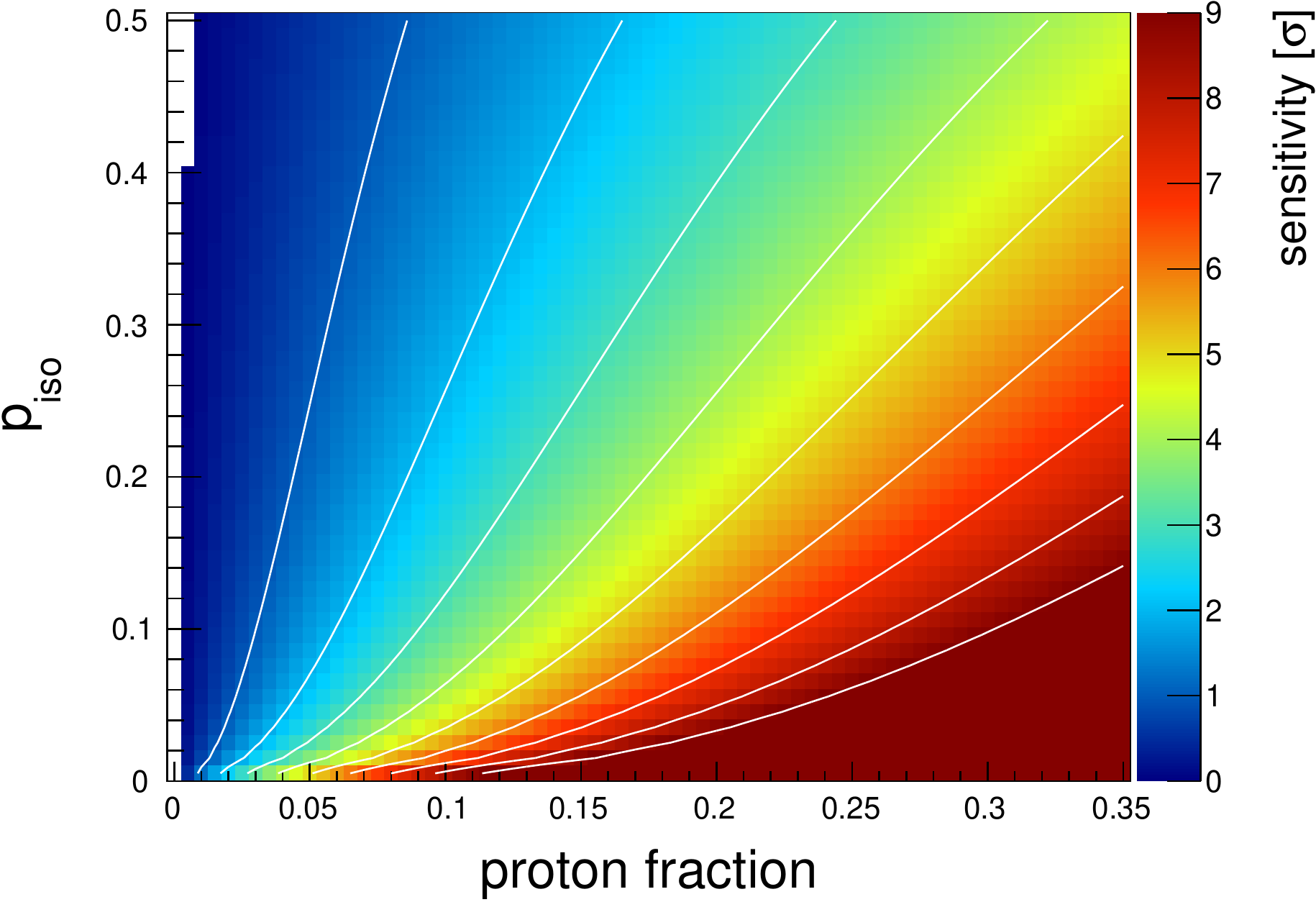}\hfill
\includegraphics[height=\figh\textwidth]{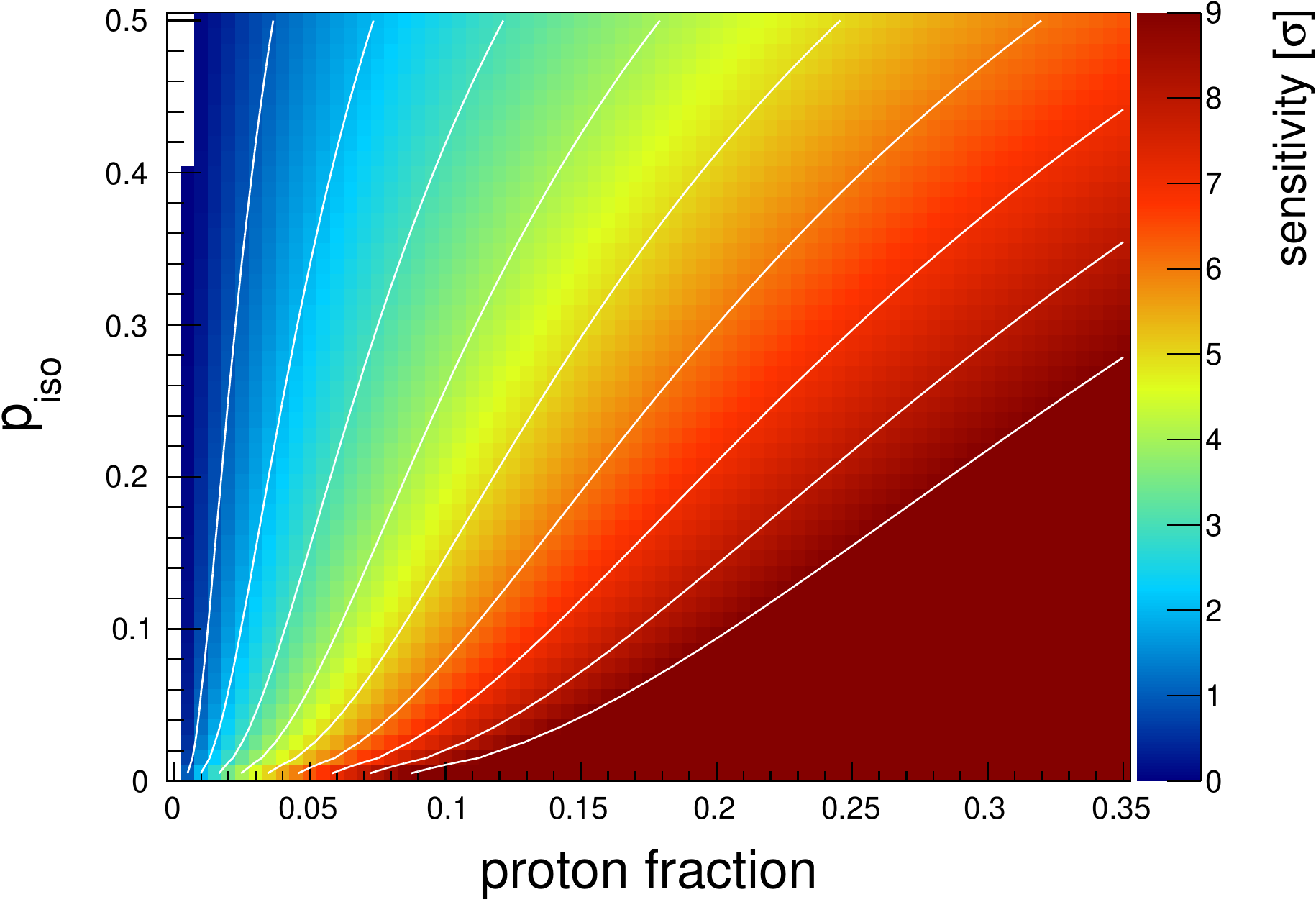}
\\[3mm]
\includegraphics[height=\figh\textwidth]{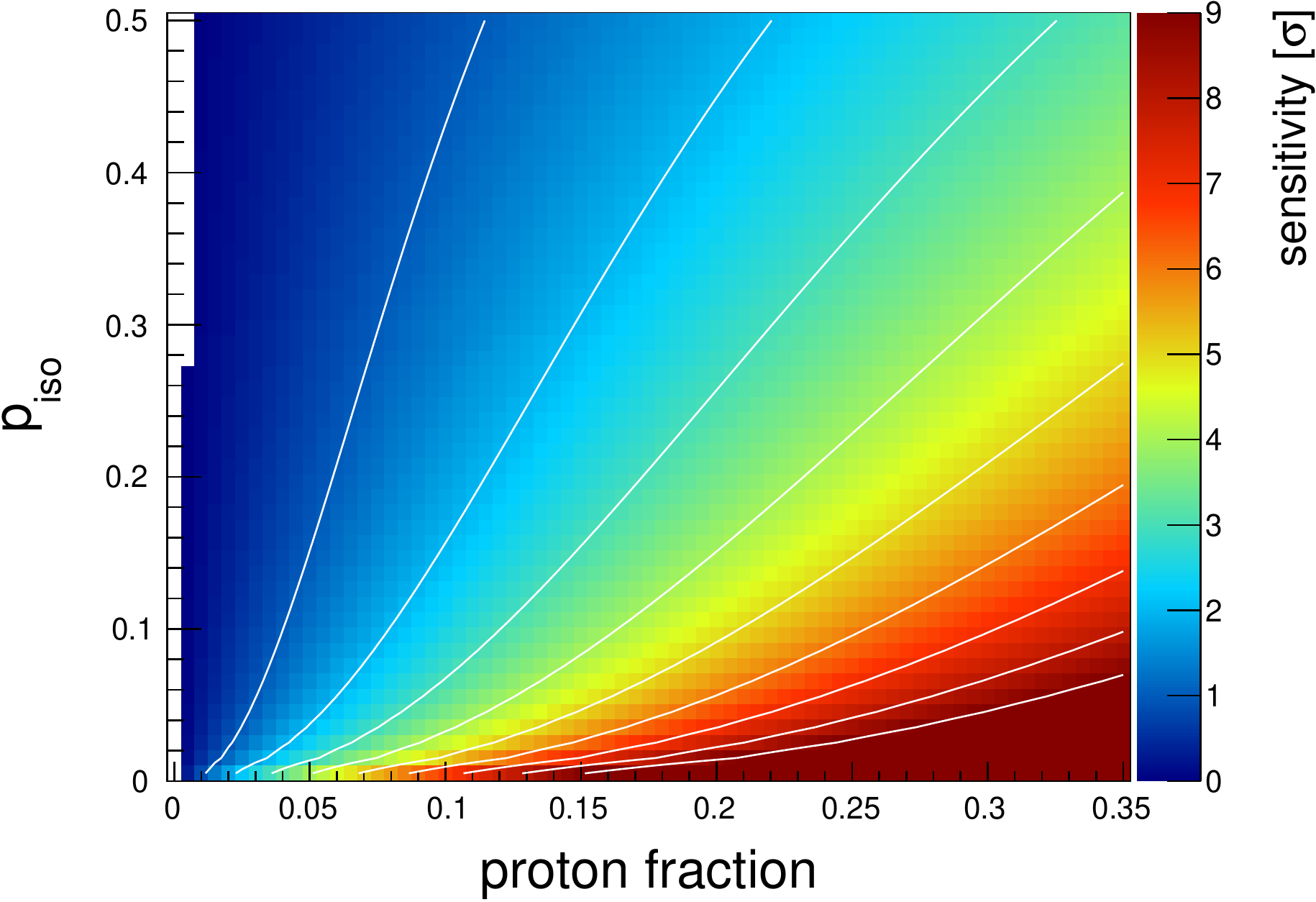}\hfill
\includegraphics[height=\figh\textwidth]{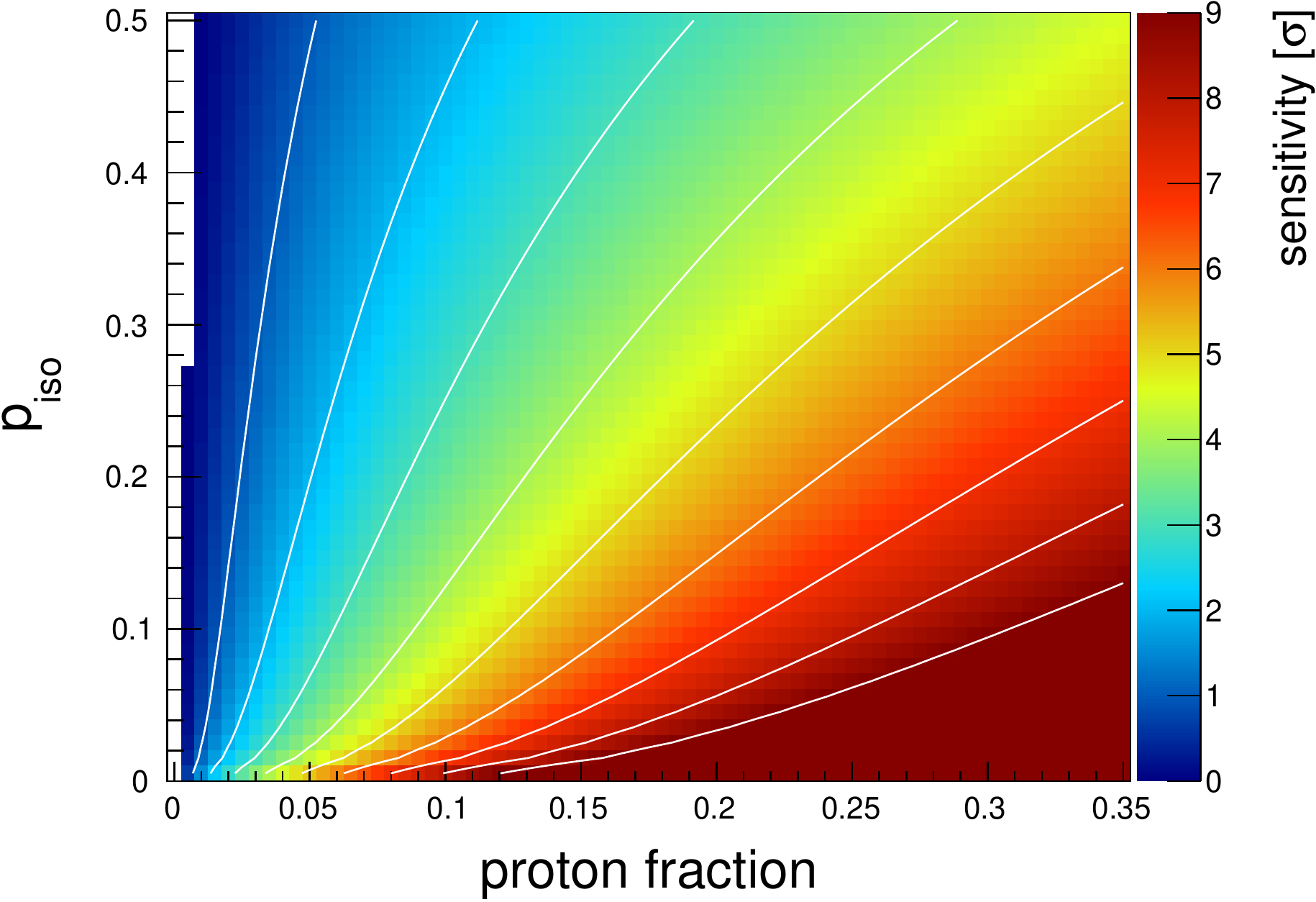}
\\[3mm]
\includegraphics[height=\figh\textwidth]{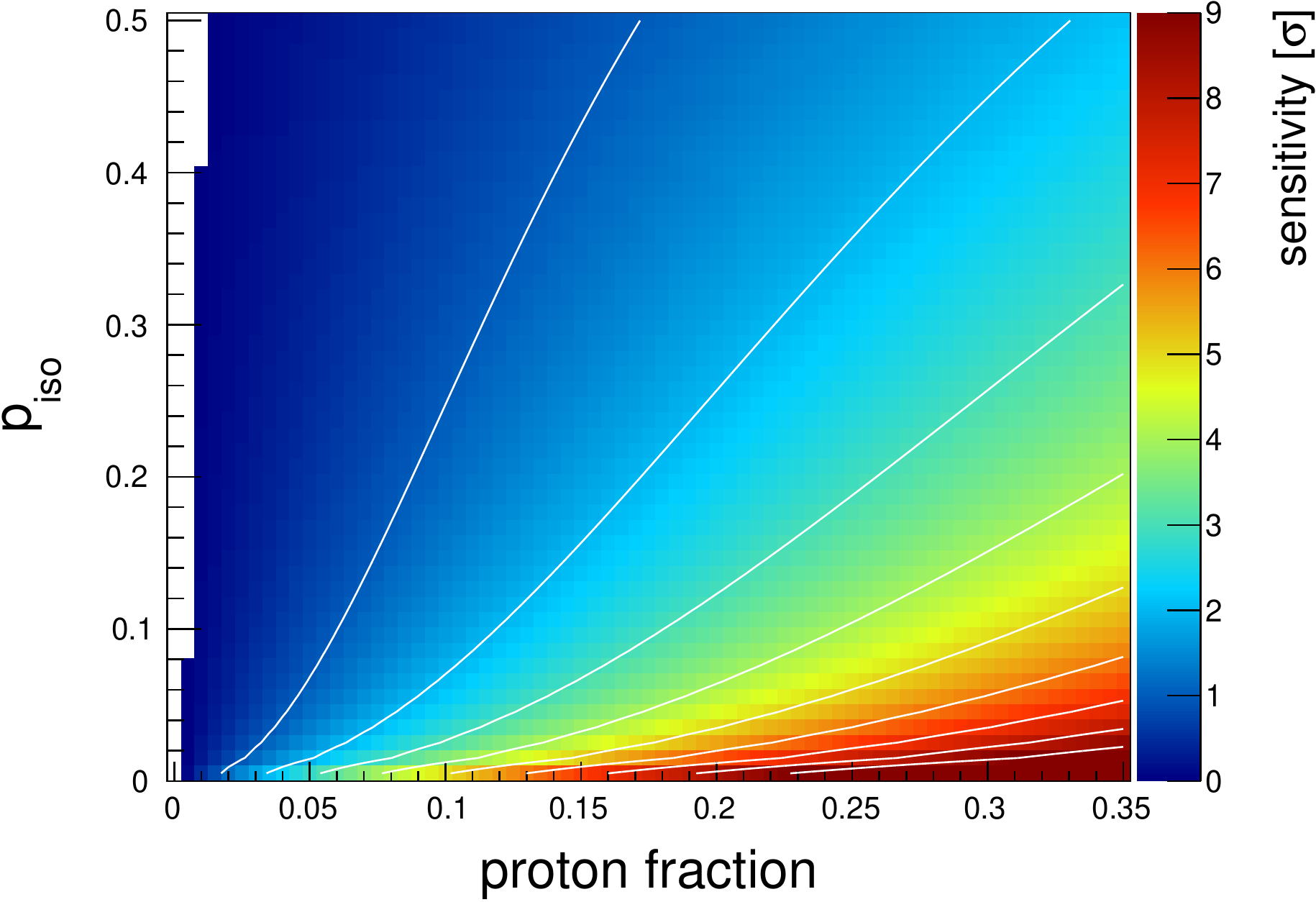}\hfill
\includegraphics[height=\figh\textwidth]{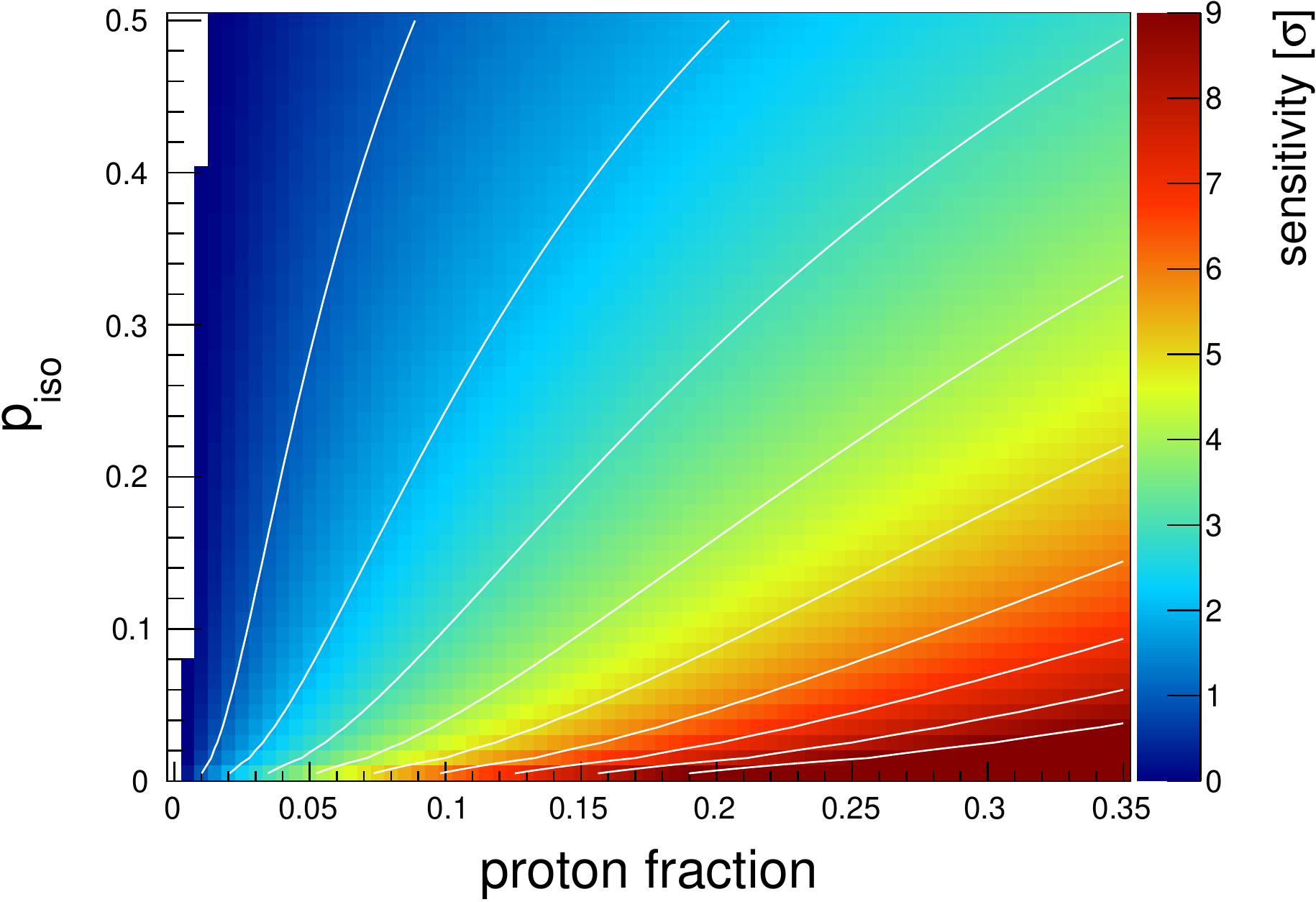}
\caption{Expected correlation of the observed arrival direction
  distribution with a source catalog and selection criteria
  characterized by $p_{\rm iso}$ (see text) to a given proton fraction
  in the data. The upper row shows the scenario in which 100\% of all
  protons are correlated with the sources of the catalog.  The middle
  and lower rows are calculated for 75\% and 50\%, respectively.  The
  plots on the left hand side are showing the correlation expected for
  the current surface array, and the ones on the right hand side
  for the upgraded array, both calculated for the same exposure.
  The white lines show the $1\sigma$ to
  $9\sigma$ thresholds from left to right.}
\label{fig:signi}
\end{figure}

The improvement of the sensitivity of the upgraded Auger Observatory
for searching for source correlations can be demonstrated by studying
the expected correlations with sources, first without making any
specific assumptions.  In this study the arrival direction is
considered to correlate by chance with an object of a reference
catalog of sources with the probability $p_{\rm iso}$, i.e. if the
arrival direction distribution were isotropic. This probability
follows from the number of sources times the solid angle around the
sources that is taken for defining a correlation, and also depends on
the source distribution within the sky exposure of the Auger
Observatory. For example, for AGNs of the V\'eron-Cetty--V\'eron
catalog~\cite{VeronCetty:2003xx} within a distance of up to
$\unit[75]{Mpc}$ ($z < 0.018$) and considering a particle as
correlated if it arrived with an angular distance of less than
$3.1^\circ$ to an AGN, one has $p_{\rm iso} =
0.21$~\cite{Abraham:2007bb}. Other examples taken from the latest
correlation study are given in
Tab.~\ref{tab:anisotropy-search-parameters}, see
also~\cite{PierreAuger:2014yba}.  The results of this generic study
are given in Fig.~\ref{fig:signi} for three scenarios and assuming a
merit factor of $1.5$, see Tabs.~\ref{tab:mf1} and \ref{tab:mf2}.
The assumed statistics of 155 events above an energy threshold of
about $\unit[5.5{\times}10^{19}]{eV}$ corresponds to
the currently collected exposure of the 
Auger Observatory~\cite{PierreAuger:2014yba}. In all cases it is assumed that
heavy elements are distributed isotropically and only protons
contribute to the correlation beyond random coincidences.  The
scenarios differ in the fraction of protons that correlate with the
sources, which are 100\%, 75\% and 50\%. For example, in the 75\%
scenario it is assumed that 25\% of all protons are distributed
isotropically. In all cases the sensitivity to sources is improved by
being able to select light elements with the separation power given by
the merit factor of the proposed upgrade. The improvement is the
largest for the 100\% scenario. For example, having a 10\% fraction of
protons correlating with sources corresponding to $p_{\rm iso} = 0.2$
will result in a deviation from isotropy of about $2.5\sigma$ if the
primary particles masses cannot be discriminated. Having a merit
factor of 1.5 for discriminating light from heavy particles will
increase this signal to more than $4.5\sigma$.

As a next step we want to illustrate the increased sensitivity of the
upgraded Auger Observatory with a more specific example. We use the
arrival directions of the 454 measured events with $\theta < 60^\circ$
and energy higher than $\unit[4{\times}10^{19}]{eV}$
(see~\cite{PierreAuger:2014yba}) and randomly assign each event an
\xmax value according to model 1 (maximum-rigidity scenario). To
implement a 10\% proton contribution we assigned 10\% of the events a
proton-like \xmax. Half of these randomly chosen, proton-like events
were given arrival directions that correlate with AGNs with a distance
of less than $\unit[100]{Mpc}$ of the Swift-BAT
catalog~\cite{Baumgartner:2012qx} within $3^\circ$. The other half
were chosen with larger angular distances. By construction, this
artificial data set reproduces many arrival direction features found
in the Auger data while at the same time having a model-predicted mass
composition.

\begin{figure}[t]
\centering
\def\figh{0.28}
\includegraphics[height=\figh\columnwidth]{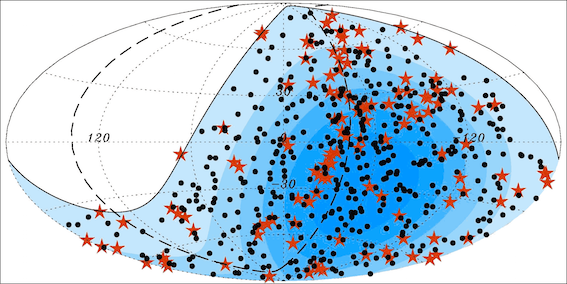}\hfill
\includegraphics[height=\figh\columnwidth]{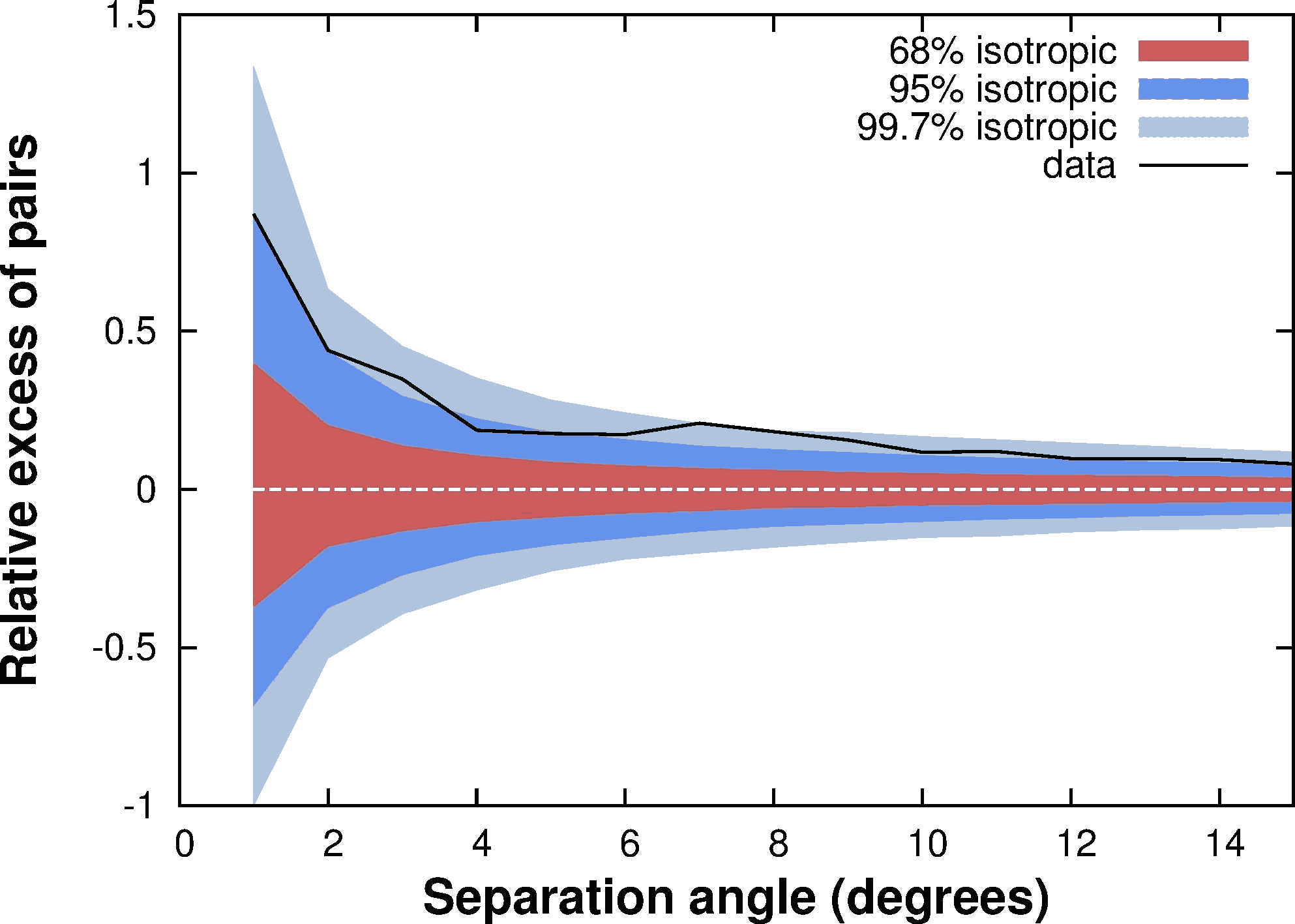}
\\[3mm]
\includegraphics[height=\figh\columnwidth]{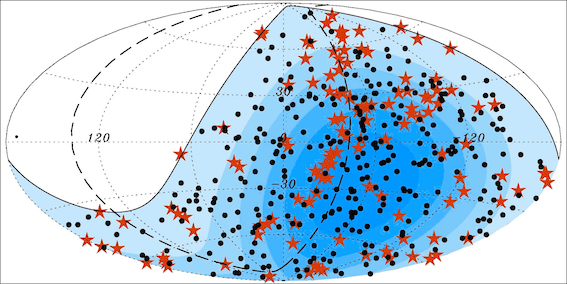}\hfill
\includegraphics[height=\figh\columnwidth]{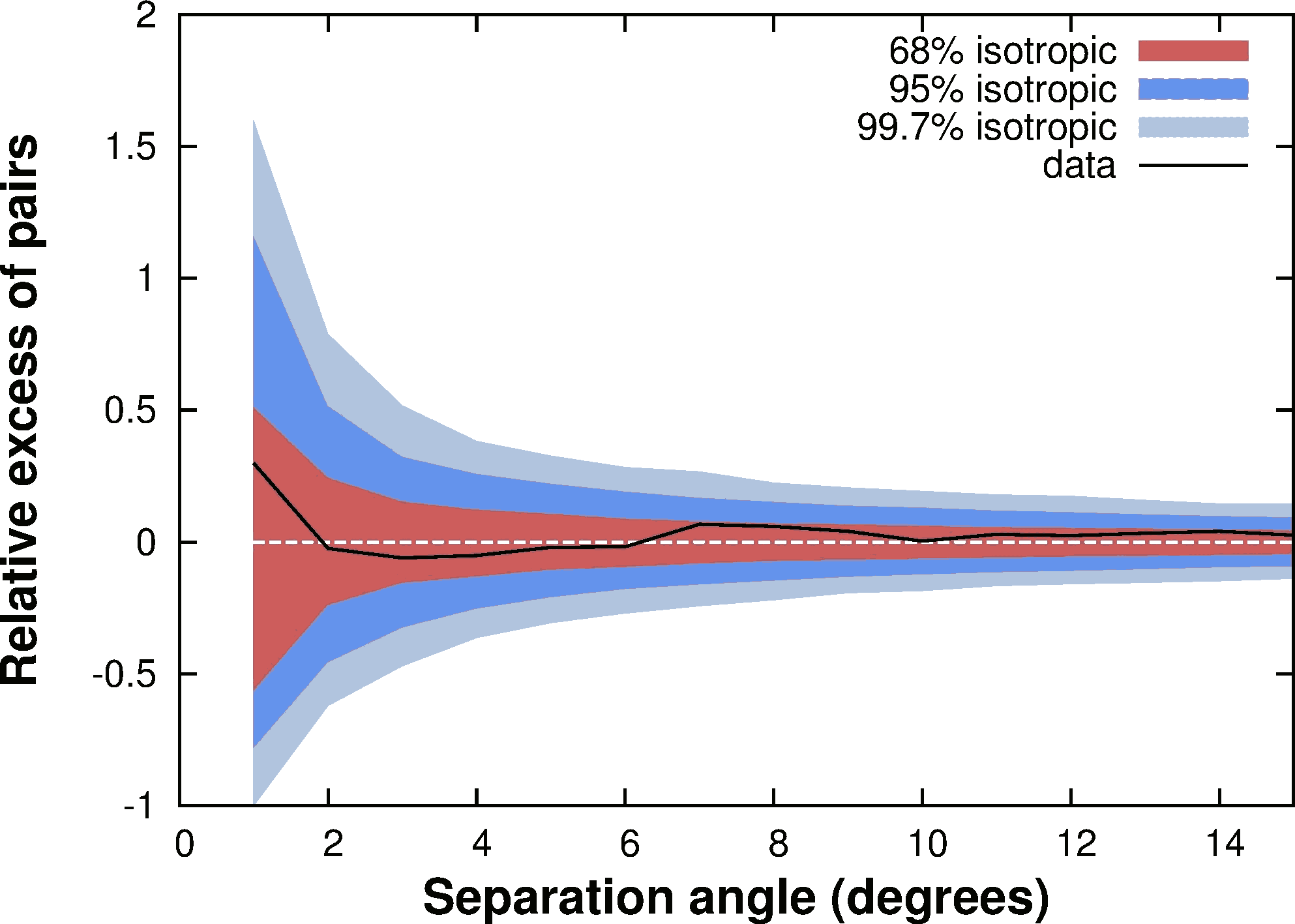}
\\[3mm]
\includegraphics[height=\figh\columnwidth]{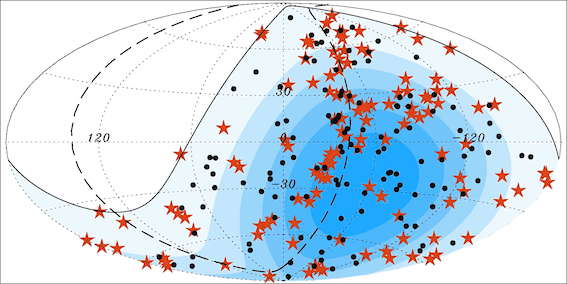}\hfill
\includegraphics[height=\figh\columnwidth]{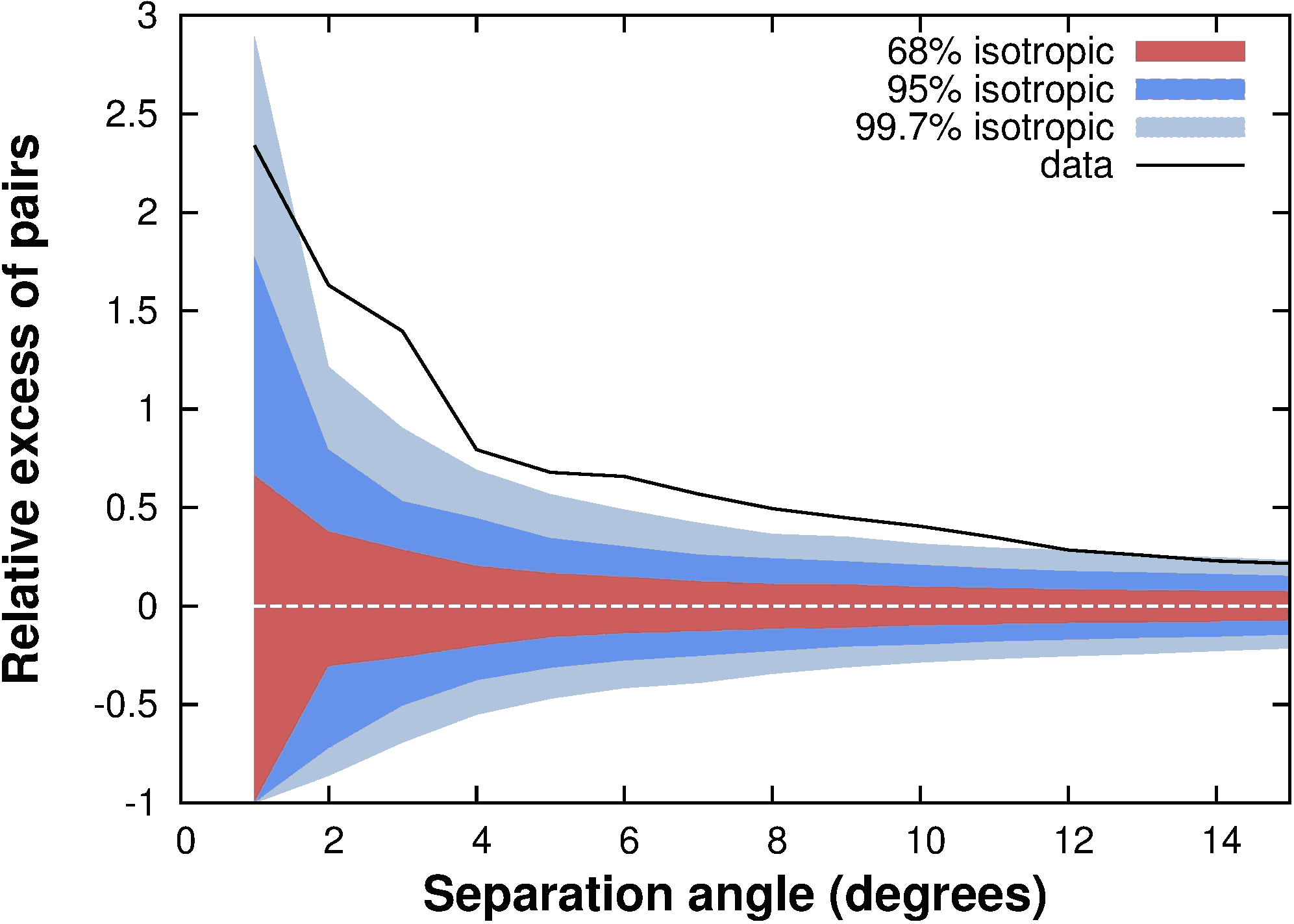}
\caption{ Arrival distribution and angular correlation of cosmic rays
  of the modified Auger data set (black circles) with AGNs of the
  Swift-BAT catalog~\protect\cite{Baumgartner:2012qx} (stars).  Shown
  are events with $\unit[E > 4{\times}10^{19}]{eV}$.  The top row of
  plots show the complete data set (454 events), the middle row the
  selection deprived of light elements (326 events), and the bottom
  row the proton-enriched selection (128 events).  }
\label{fig:Scenario-Swift-BAT-40EeV}
\end{figure}

\begin{figure}[t]
\centering
\def\figh{0.28}
\includegraphics[height=\figh\columnwidth]{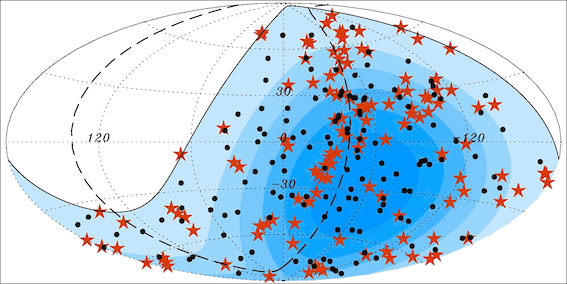}\hfill
\includegraphics[height=\figh\columnwidth]{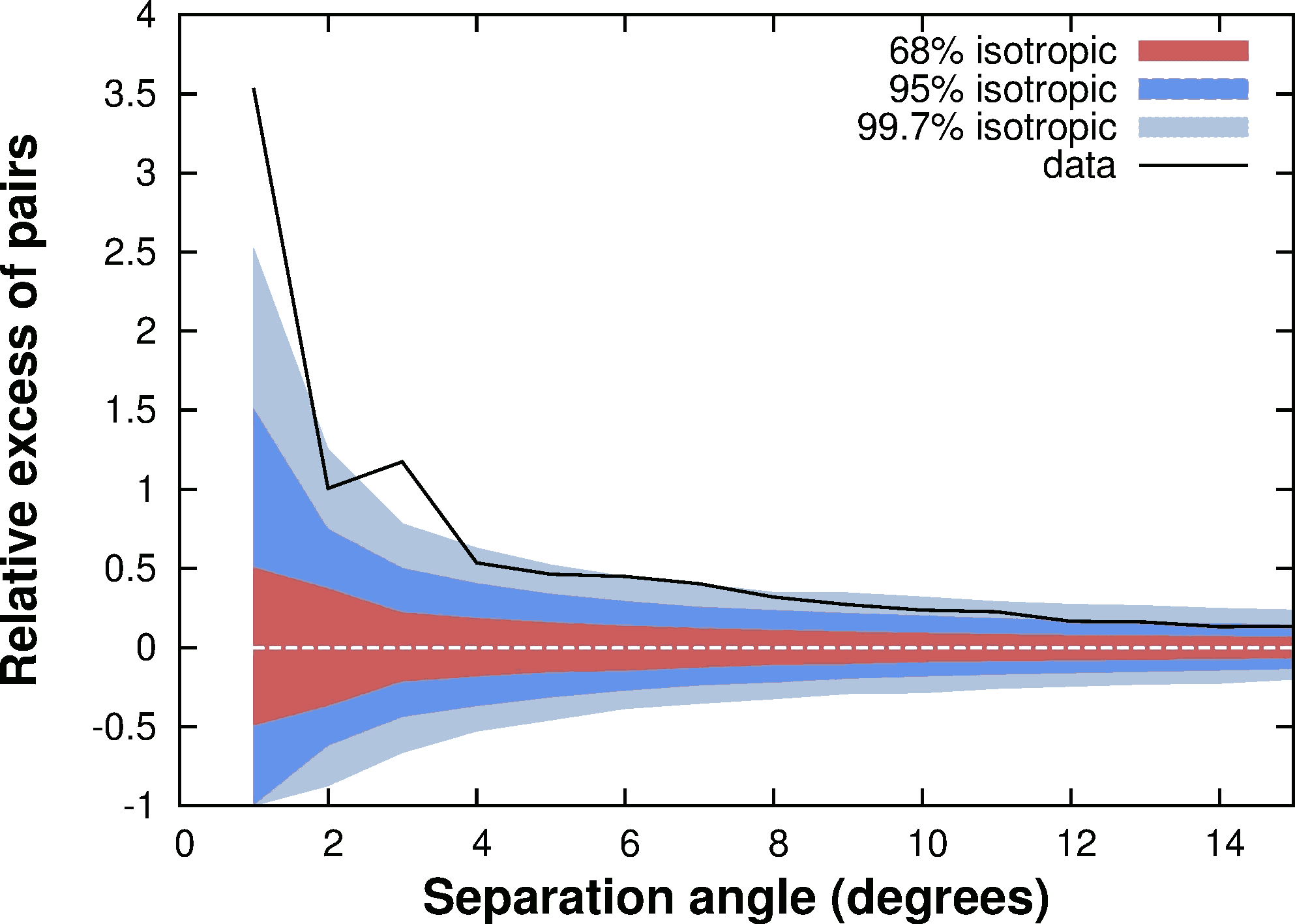}
\\[3mm]
\includegraphics[height=\figh\columnwidth]{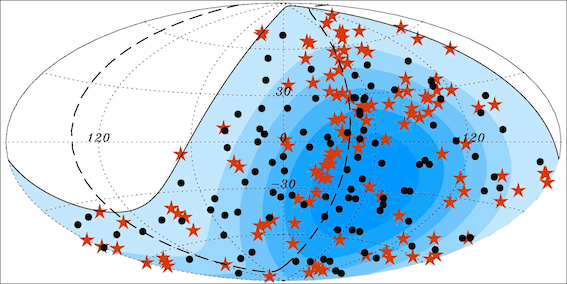}\hfill
\includegraphics[height=\figh\columnwidth]{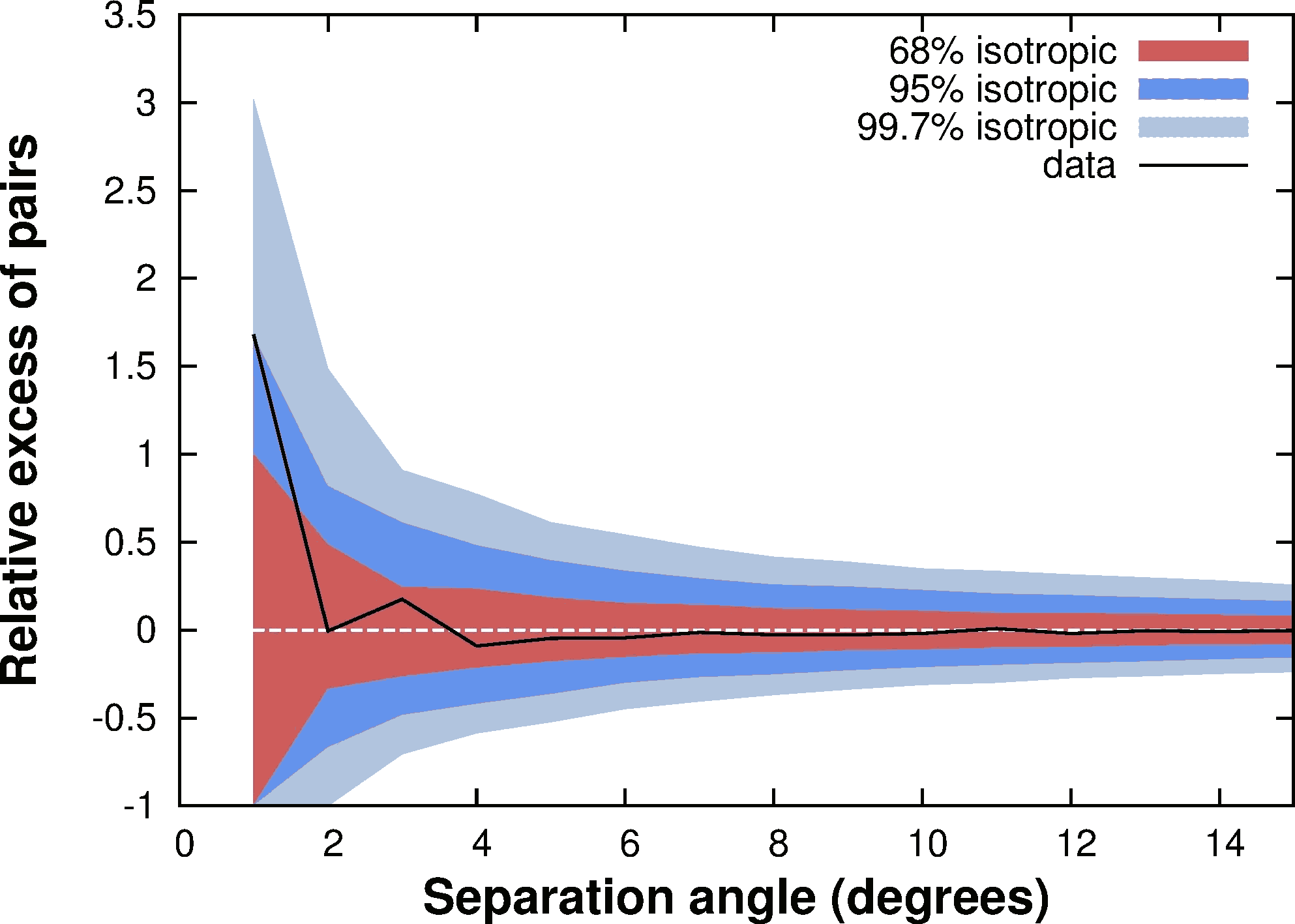}
\\[3mm]
\includegraphics[height=\figh\columnwidth]{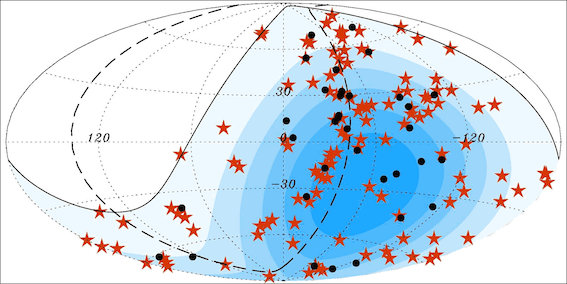}\hfill
\includegraphics[height=\figh\columnwidth]{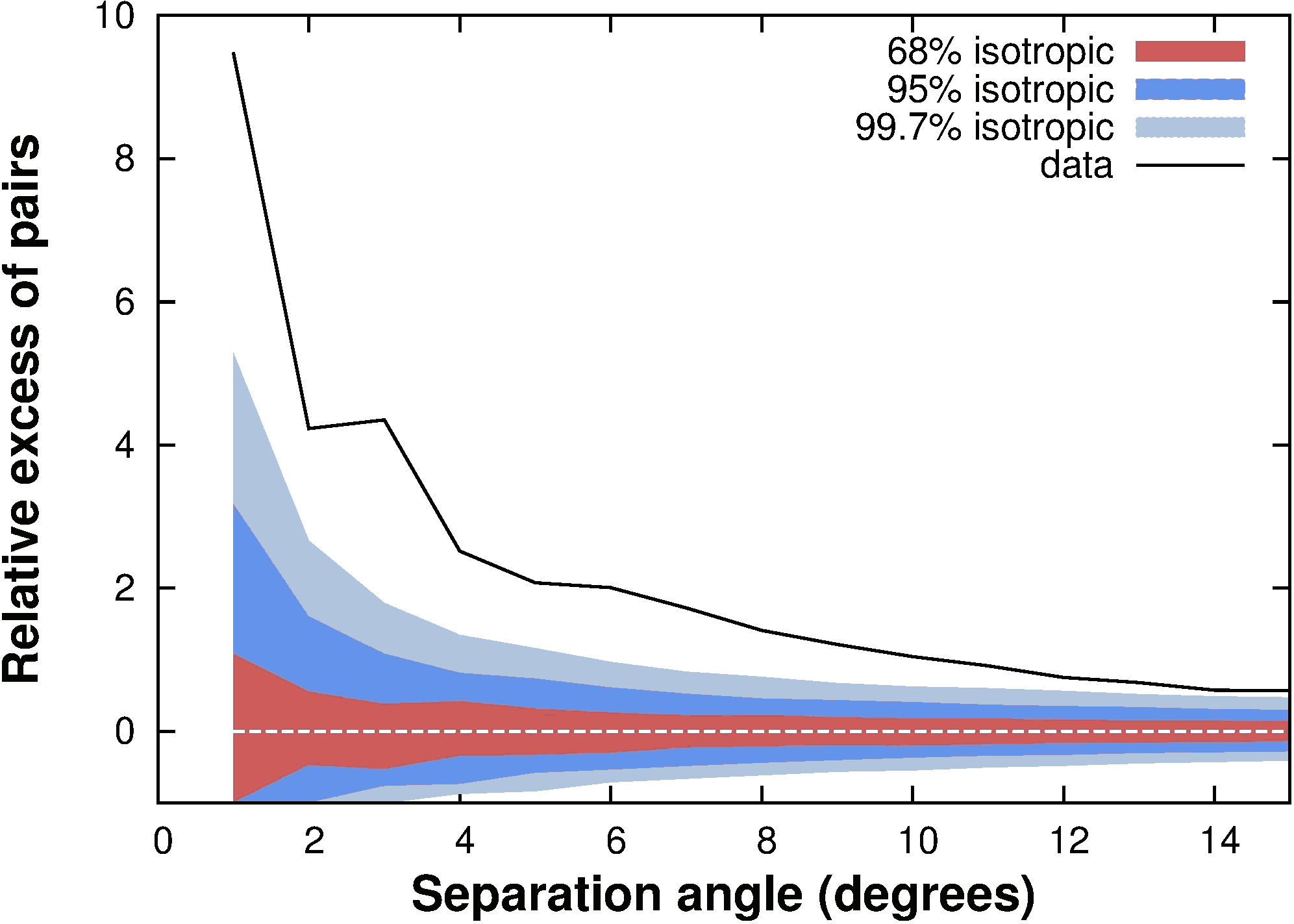}
\\[3mm]
\caption{ Arrival distribution and angular correlation of cosmic rays
  of the modified Auger data set (black circles) with AGNs of the
  Swift-BAT catalog~\protect\cite{Baumgartner:2012qx} (stars).  Shown
  are events with $\unit[E > 5.5{\times}10^{19}]{eV}$.  The top row of
  plots show the complete data set (141 events), the middle row the
  selection deprived of light elements (107 events), and the bottom
  row the proton-enriched selection (34 events).  }
\label{fig:Scenario-Swift-BAT-55EeV}
\end{figure}

Analyzing this data set without using any composition information a
correlation with the AGNs of the Swift-BAT catalog is found at a level
similar to that reported in~\cite{PierreAuger:2014yba}. The
improvement of the sensitivity to find the correlation with AGNs in
this data set is shown in Figs.~\ref{fig:Scenario-Swift-BAT-40EeV} and
\ref{fig:Scenario-Swift-BAT-55EeV}.  The top row of plots shows the
results of the complete data sets with the energy thresholds
$\unit[4{\times}10^{19}]{eV}$ and $\unit[5.5{\times}10^{19}]{eV}$. The
middle row shows what one would obtain in a correlation study if the
proton-like events are removed from the data set. This is done by
selecting events with a reconstructed \xmax of less than
$\unit[770]{g/cm^2}$ at $\unit[10^{19}]{eV}$, adjusted to the event
energies with an elongation rate of $\unit[55]{g/cm^2}$ per decade.
By selecting events with \xmax larger than $\unit[770]{g/cm^2}$ a
proton-enriched data sample is selected.  While the correlation of the
arrival directions with that of AGNs in the Swift-BAT catalog is not
significant for the complete data sets, a correlation well in excess
of $3\sigma$ can be found for the proton-enriched samples. Furthermore
the proton-deprived selections exhibit no correlation with the AGNs at
all, as one would expect for an angular correlation that is just a
statistical fluctuation.

These conservative examples of composition-improved anisotropy studies
underline the large gain of sensitivity achieved by adding
scintillation detectors to the water-Cherenkov detectors of the Pierre
Auger Observatory.

%% file: surface_detector.tex

\chapter{The Surface Detector}

The operating surface detector (SD) of the Pierre Auger Observatory
comprises 1660 stations on a 1500\,m triangular grid covering
3000\,km$^2$.  Each station is a water-Cherenkov detector (WCD) which
samples the particle content of the extensive air showers (EAS) falling
on the array (see Appendix~\ref{currentSD}). 
The SD is overlooked by four fluorescence detector sites (FD), each
with 6 fluorescence telescopes viewing an aggregate $180^\circ$ azimuth by
$30^\circ$ elevation field of view.  Additional detectors and a region
with reduced detector spacing, known as the \emph{infill} array, focus on
lower energies, muon detection, and radio R\&D.

The SD stations will be upgraded with new electronics for faster
sampling of the PMT signals, better timing accuracy and enhanced
triggering and processing capabilities, new light sensors for
increased dynamic range, and improved calibration and monitoring capabilities.

New surface scintillator detectors (SSD) will be placed on top of each
WCD to sample the EAS in another way.  Given the different
sensitivities of plastic scintillators and water-Cherenkov detectors to
the electrons, photons, and muons of EAS, the combination of
measurements will provide EAS muon content information, vital for
cosmic ray mass composition studies and improved energy determination.


\section{The Water Cherenkov Detector}

Each WCD is a rotomolded polyethylene tank filled with purified water that produces
Cher\-enkov light when crossed by energetic charged particles associated with CR showers.
A flexible Tyvek liner inside the tank provides an interface between
the water volume and the light sensors that collect Cherenkov light
(PMTs) and the light sources that produce calibration
pulses (LEDs).  Access to the liner, the PMTs and the LEDs is 
through three hatches located on the top of the tank.  An
electronics box containing front-end charge amplifiers, shapers,
trigger logic, signal buffers, power control, radio transmitter and
receivers is located on the top roof of the tank on one of the hatch-covers
and is protected by a dome.  All the cables connecting the electronics
and the light sensors run inside the tank and connect to the
electronics via feed-throughs in the hatch covers.
Figure~\ref{fig:sd} shows a picture of an operating WCD.

\subsection{Photomultipliers}
\label{sec:pmt}

The WCDs in the field have a redundant set of three identical, large
collecting area (230\,mm diameter) Photonis XP1805 photomultipliers.
We refer to these as the \emph{Large PMTs} (LPMT), for which we are
not envisaging upgrades.

In addition, a fourth, new phototube with a significantly smaller
cathode surface, called the \emph{Small PMT} (SPMT), will be added to
the WCD to extend its dynamic range.  An empty liner window facing the
water volume, located near the center of the tank and originally
planned for a spare LED, can be used for a straightforward
installation of the SPMT, the diameter of which is therefore constrained to
be less than 30\,mm in order to fit in this window.


The new electronics (Section~\ref{sec:sdeu}) will read and digitize
the SPMT anode signals with a dedicated input, analogous to those for
the LPMTs.  As the single muon signal peak (VEM, see
Section~\ref{currentSD}) is not visible in the SPMT signal response,
the scale of the SPMT signal in physical units (VEM) will be
determined by cross-calibrating the SPMT and LPMT signals that occur
in their overlapping region before saturation, and monitored with LED
signals in the same signal region.

By fine tuning the SPMT gain and the signal overlap with the standard
PMTs we expect to be able to extend the WCD dynamic range by an adjustable
factor ranging from ${\sim}10$ to ${\sim}60$.

We summarize in Table~\ref{table-pmt} the basic mechanical and electrical parameters of the SD PMTs.

\begin{savenotes}
\begin{table*}[t]
\caption{Basic properties of the SD photomultiplier tubes}
\label{table-pmt}
\centering
\begin{tabular}{llll}
\toprule
\textbf{Specification} & \textbf{Unit} & \textbf{LargePMT (LPMT)} & \textbf{SmallPMT (SPMT)} \\ 
\midrule
Diameter & mm & 230 & ${<}30$ \\ 
Height (glass) & mm & $ 256 $ & ${<}90$ \\
HV (max) & V & 1950 & 1500 \\
Gain (max)  & $10^6$ & 3 & $40$ \\
QE (peak) & \% & ${>}16\%$ \@420nm & ${>}16\%$ \@420nm \\
Anode dark current & nA & 15 to 50 & 2 to 15 \\
Anode rise time & ns & ${<}5$ & ${<}3$ \\
Anode pulse linearity ($\pm5\%)$ & mA & 50 & 60\footnote{with tapered ratio divider} \\
\bottomrule
\end{tabular}
\end{table*}
\end{savenotes}

We have qualified three models with the required performance for the
SPMT (Hamamatsu R8619, Hamamatsu R6094, ETL9107B).  
The R8619 is our
current baseline SPMT model, as it satisfies all technical requirements
and is industrially produced in large quantities, with benefits on
yield and cost.  We will use the first field tests to verify
the performance of all candidates in realistic conditions.

Prototype mechanics for supporting the SPMT in the LED window were
successfully built with 3D printers and tested in the field.
Molds will be engineered for cost-effective mass fabrication.


\subsection{High Voltage Modules and Control}
\label{sec:SDHV}
The operating LPMTs are equipped with an active base 
which includes the High Voltage (HV) resistive divider, 
a HV DC-DC converter module and a charge amplifier for the dynode readout.  
The base is soldered to the PMT leads and protected with insulating silicone potting.  
Given the high level of moisture and severe temperature cycles inside the WCD, 
the potting proved to be far from ideal, resulting in a low but visible number of failures in the field, 
which we plan to minimize in the upgrade with an improved design for the SPMT, 
with the goal to reducing the necessary maintenance operations to a negligible level.

We therefore designed a simpler, passive resistive base for the new SPMT, 
and moved the HV DC-DC converter to a separate module 
that can be conveniently located outside the WCD and away from the moisture.  
The base will be located inside the WCD and will not be potted, 
but simply coated.
The HV module will be housed in a dedicated box next to the station electronics.

We have qualified a new model for the HV DC-DC converter (CAEN-A7501P), 
with slightly better thermal stability, operational temperature range and power consumption 
with respect to the module used for the LPMTs (SensTech-PS2010/12)~\cite{GAPHVPSTorino}.  
Specifications and test results for the two HV systems considered are illustrated in table~\ref{tab:HV-spmt}.

\begin{table*}[t]
\caption{Basic properties of HV power supplies for the Small PMT and the Scintillator Detector}
\label{tab:HV-spmt}
\centering
\begin{tabular}{llll}
\toprule
\textbf{Specification} & \textbf{Unit} & \textbf{SensTech-PS2010/12} & \textbf{CAEN-A7501P} \\
\midrule
Input Voltage & V & 11.75 to 12.25 & $12\pm5\%$ \\ 
Max High Voltage & V & 2000 & 2100 \\
Output current (at Max HV)& $\upmu$A & 100 & 110 \\ 
Output voltage temperature stability & $10^{-5}/^\circ$C & $4.7$ & $2.5$ \\
Output voltage Ripple & ppm & 15 to 39 & 1.9 to 3.8 \\
Power dissipation (at Max HV) & mW & $400\pm 13$ & $398\pm 4$ \\
Operating Temperature range & $^\circ$C & $-10$ to 60 & $-40$ to 70 \\
\bottomrule
\end{tabular}
\end{table*}

We plan to use the same HV system for the Scintillator Detector.
The LPMTs in the field will not be modified, but only repaired
in case of failures, with replacement bases identical to the operating
ones except for the dynode amplifier, which is no longer
needed.

The Slow Control of the existing Unified Board handles High Voltage (HV)
control and monitoring of the existing standard PMTs.  The Upgraded
Unified Board (UUB) (Section~\ref{sec:UUB}) will implement a new
version of the Slow Control (Section~\ref{sec:SlowControl}) that will
be able to handle up to six HV lines, and will be used to drive the
large PMTs, the new Small PMT and up to two additional PMTs
for the Scintillator Detector (Section~\ref{sec:scintillator}).  
Separate HV boxes for the Small PMT and the Scintillator Detector will
connect to the Slow Control on the one hand and to the PMTs on the
other.

\subsection{Calibration and control system}

\subsubsection{Absolute calibrations with physics events}

With a large number of stations on the field, scattered over a very
large area and often difficult to reach, it is important to be able to
routinely calibrate and monitor each station remotely, with stable and
robust procedures, and ensure a uniform response of the array in terms
of trigger rates and performance.

This is achieved by using trigger rates as proxies of the signal
charge recorded by the tank PMTs.
We vary each LPMT high voltage until we reach a target rate of
$100$\,Hz, with a fixed threshold set at the signal of ${\sim}3$
vertical and central through-going muons.  The Vertical Equivalent
Muon (VEM) signal is the baseline reference unit of the SD
calibrations, and was determined on a test tank with an external
trigger hodoscope to give on average 95 photoelectrons at the cathode
of the LPMTs, corresponding roughly to 150 integrated ADC counts above
pedestal after signal digitization.
We then verified \emph{a posteriori} that the rate proxy is an
accurate approximation at the 5 to 10\% level, and can be further
refined to $3\%$ using off-line analysis of the charge histograms
taken from each PMT with a dedicated high-rate, low-threshold,
short-duration trigger primitive.

When completed, this calibration provides uniform trigger rates, by
definition, and an average LPMT gain of $3.4{\times}10^5$, with a
${\sim}5\%$ spread over the whole array.

This algorithm ensures quick convergence, it does not require complex
algorithms to run on the local station processor, nor transmission of
large data-sets through the communication system and can be easily
repeated when large deviations from the reference VEM signal are
recorded, for example from aging of the PMTs.


The calibration of the large signal range of the LPMTs, given by the
anode readout, is not covered by this procedure due to poor
statistics of large enough signals.  Additionally, in the existing SD,
the dynode signal is amplified on the LPMT base, causing relative
delays with the anode signal which make the signal ratio time
dependent and more complex than a simple scaling factor of 32, which
could be simply calibrated by dividing the anode and dynode peaks of
the signal distribution.

Therefore the anode signal is currently converted to physical units
using the dynode calibration described above and a model of the
relative time development of anode and dynode signals.  Instead, for
the SD Upgrade, a much simpler and direct cross-correlation of the
signals in the overlapping regions will be possible, as the dynode
signal will no longer be used, and the WCD dynamic range will be
spanned by the amplified anode signal (32${\times}$), the anode signal (1${\times}$) and
the Small PMT (we consider a nominal factor of $1/32$ over the anode
signal, but this is adjustable being a signal from an independent
light sensor).

We envisage similar rate-based calibrations for the upgraded WCDs, and
will revisit the target rates with dedicated field measurements.

\subsubsection{Monitor and control with Light Emitting Diodes}

In addition to routine calibrations with physics events, each WCD is
equipped with two Light Emitting Diodes (LEDs).  While these are not
stable sources of calibrated light, they are very useful for
monitoring and control.  We will use the LEDs already installed in the
tanks for several different purposes:

\begin{itemize}
\item setup the initial working point of the SPMT, to be refined later
  with physics events, possibly with rate-based techniques;
\item quickly cross-calibrating the three ranges of the upgraded WCD with arbitrary high rates;
\item testing the linearity of each PMT by recording the difference
  between the signals of the two LEDs, when switched on with the same
  trigger pulse, and the sum of the two LEDs switched on
  independently;
\item creating artificial extensive shower events on the SD array with
  arbitrary topology for verification of the acquisition and
  reconstruction chains.
\end{itemize}

We have integrated a new controller circuit into the Upgraded Unified
Board (Section~\ref{sec:UUB})
that builds on the existing controller for driving separate and
simultaneous triggers to the two LED flashers, but includes better
timing of the trigger signals.  In particular, it offers a hardware
synchronization of the LED trigger pulses with the GPS-PPS signal, to
be able to synchronize LED triggers simultaneously over all the WCDs
in the array.  The new controller drives software triggers with custom
delays between tanks as to emulate any EAS topology over the SD.
Finally, the Slow Control now provides an increased bias voltage on
the LED to give enough light output to sample the full WCD dynamic
range.


\subsection{Solar power system}
Power for the Surface Detector stations is provided by a solar photovoltaic system. The power
system provides  10\,W average power. A 24\,V system has been
selected for efficient power conversion for the electronics. The system consists of  two 55\,Wp solar panels  and two
105\,Ampere-hour (Ah), 12\,V batteries. The batteries  are lead acid batteries
designed for solar power applications. Power is expected to be available over
99\% of the time. Even if after long term operation the capacities of the solar
panels and batteries are degraded to 40\,Wp and 80\,Ah, respectively, power
would be available 97.8\% of the time. 

One of the requirements for the new electronics was to have a sufficiently low power consumption to be able to use the existing power system. Current  estimates of the power consumption  indicate
that the existing power system is sufficient to operate both the WCD and the SSD detectors. An additional power system is needed only in the infill area to accommodate the Underground Muon Detector.

\input{scintillator}
\input{sdeu}
\input{sdeuperformance}


%% file: scintillator.tex
\section{The Scintillator Detector}
\label{sec:scintillator}
\subsection{Introduction and design objectives}
\label{sec:ScinIntro}
The main objective of this additional array of detectors is to add an
extra measurement of the particles in the EAS
independent of the measurements made with the water-Cherenkov
detectors. To achieve the maximum advantage from this additional
measurement, the shower should be sampled in the position of the
WCD with a detector that has a different response to the basic
components of the EAS compared to the present stations. Additionally,
the additional detector has to be reliable, easy to realize and
install, and has to have minimal maintenance.

The design chosen consists of a detector based on a plane of plastic
scintillator positioned on top of the existing surface detectors, and
read in an integral way using only one photo-detector. With this
technique, the signal collected with the scintillator unit can be
compared directly with the signals collected in the WCD. In
particular, the amplitude and time distribution of the collected
signal in the scintillator are different from those coming from the signals collected by
the WCD tanks due to the fact that the signal in the scintillator will be
dominated by signals from electrons while the WCD will be dominated by
photons and muons.

\begin{figure}[t]
\centering
\includegraphics[width=0.6\textwidth]{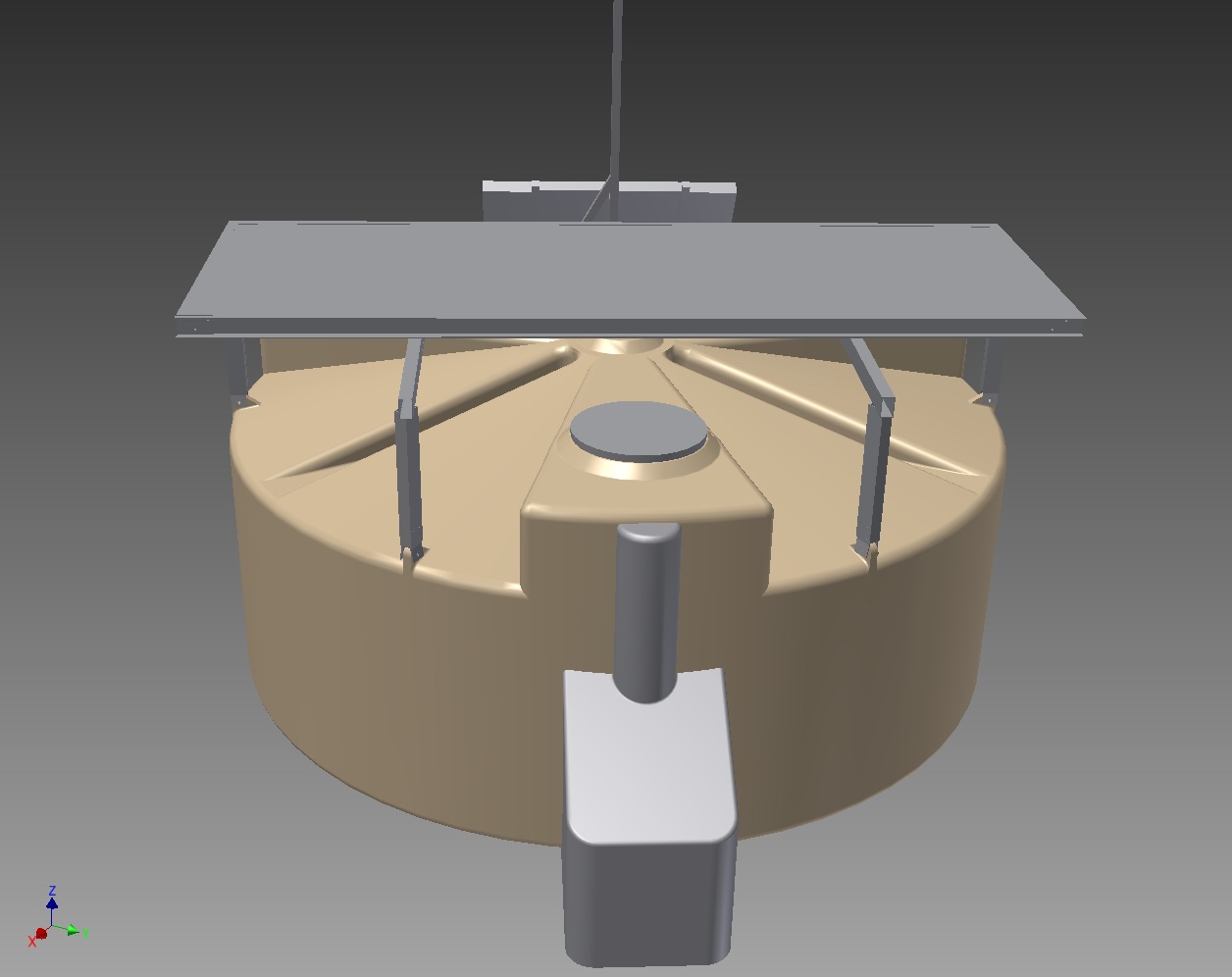}
\caption{3D view of a water-Cherenkov detector with a scintillator unit on top.}
\label{fig:bar}
\end{figure}
The scintillator units have to be precisely calibrated with a technique
similar to the calibration procedure of the WCD (cf.\ section \ref{sec:ScinCali}). The size of the
detector and its intrinsic measurement accuracy should not be the dominant limitations for
the measurement. The dynamic range of the units has to be adequate to
guarantee the physics goals of the proposed upgrade.

The detector will be assembled and tested in parallel in
multiple assembly facilities to reduce the production time and,
therefore, has to be easily transportable. The mechanical robustness
of the scintillator units must be ensured. The units will be
shipped after assembly, and validated at the Malarg\"ue facilities of
the Pierre Auger Observatory before being transported to their final
destination on top of a WCD in the Pampa. They will then have to
operate for 10 years in a hostile environment, with strong winds and
daily temperature variations of up to 30$^\circ$C.

\subsection{Detector design}
\label{sec:ScinDesi}
The baseline design relies only
on existing technology for which performance measurements have been
made.  The Surface Scintillator Detectors (SSD) basic unit consists of two
modules of $\approx\,2$\,m$^2$ extruded plastic scintillator which are read out by
wavelength-shifting (WLS) fibers coupled to a single photo-detector.
Extruded scintillator bars read by wavelength-shifting fibers have already
 been employed in the MINOS detector\cite{minos}.
The active part of each module is a scintillator plane made by
$12$ bars $1.6$\,m long of extruded polystyrene scintillator. Each bar
is 1\,cm thick and 10\,cm wide.
The scintillator chosen for the baseline design is produced by the extrusion line of the
Fermi National Accelerator Laboratory (FNAL)\cite{FNAL_Scint}.

\begin{figure}[!ht]
\centering
\includegraphics[width=0.4\textwidth]{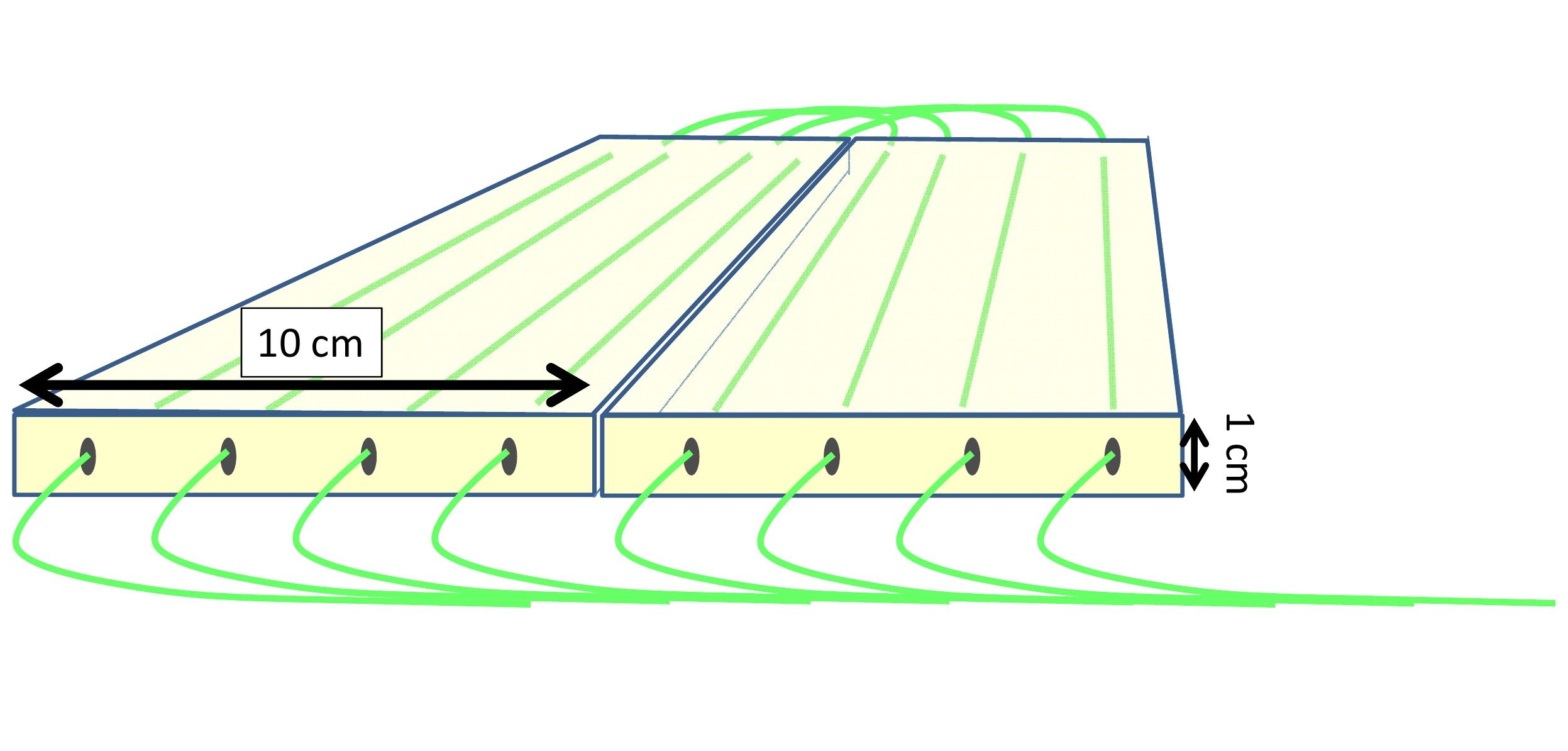}
\caption{Sketch view of bars with the fiber readout.}
\label{fig:bar2}
\end{figure}

The bars are co-extruded with a TiO$_2$
outer layer for reflectivity and have four holes in which the
wavelength-shifting fibers can be inserted.  The fibers are positioned
following the grooves of the routers at both ends, in a ``U''
configuration that maximizes light yield and allows the use of a
single photomultiplier (at the cost of a widening of the time response
of the detector by 5\,ns, which has a totally negligible impact).
The fibers are therefore read out from both ends.  Figure~\ref{fig:bar2}
shows a sketch of two bars with the fiber readout.  The two-ended
readout of the scintillator strips also provides a better longitudinal
uniformity in light response.

Two companies, Kuraray and Saint Gobain, produce suitable WLS
multi-clad optical fibers for our application.  The Kuraray fibers
have a higher light yield and are more readily available. They have
also been used for optical read-out in most large area scintillator
counter experiments. For these reasons they were chosen as the
baseline design option.  However the Saint Gobain WLS fiber may have a
lower cost and the possibility to make use of them is
currently under investigation.

\begin{figure}[!ht]
\centering
\includegraphics[width=0.48\textwidth]{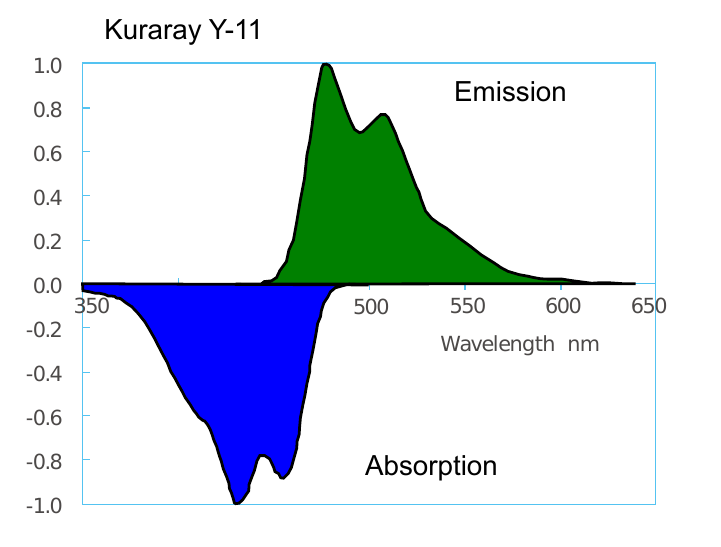}
\includegraphics[width=0.48\textwidth]{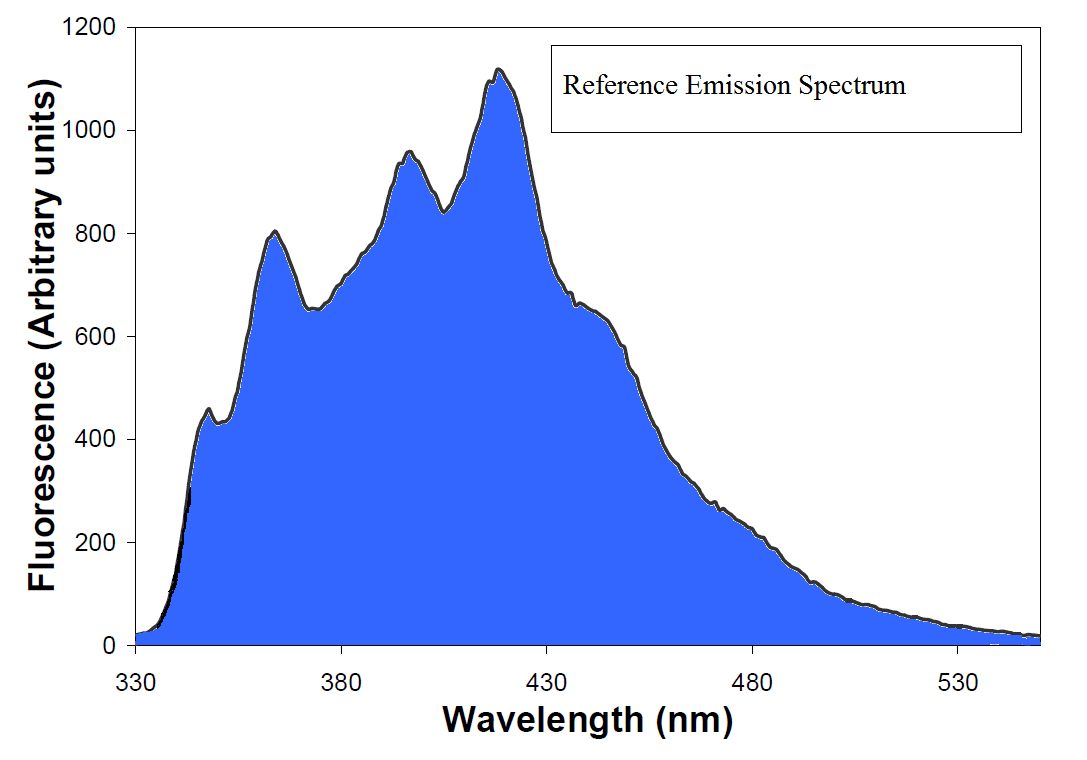}
\caption{Left: emission and absorption spectra for Kuraray Y11 WLS fiber.
Right: Reference emission spectrum of the chosen extruded scintillator. }
\label{fig:scintfiber1}
\end{figure}

For the baseline design, the Kuraray Y11 WLS multi-clad optical
fiber with 1\,mm diameter is chosen, with a concentration of fluorescent dye at either
200 or 300 parts per million. As shown in Figure~\ref{fig:scintfiber1},
the absorption spectrum of the K27 dye (Y11 fiber) matches
perfectly the scintillator emission\cite{Dyshkant:2005my}. On the other hand, the WLS fiber
emission is shifted toward longer wavelengths than the absorption peak
of a standard bialkali photocathode, thus suggesting some caution during
the selection of the read-out photodetector.

\begin{figure}[!ht]
\centering
\includegraphics[width=0.7\textwidth]{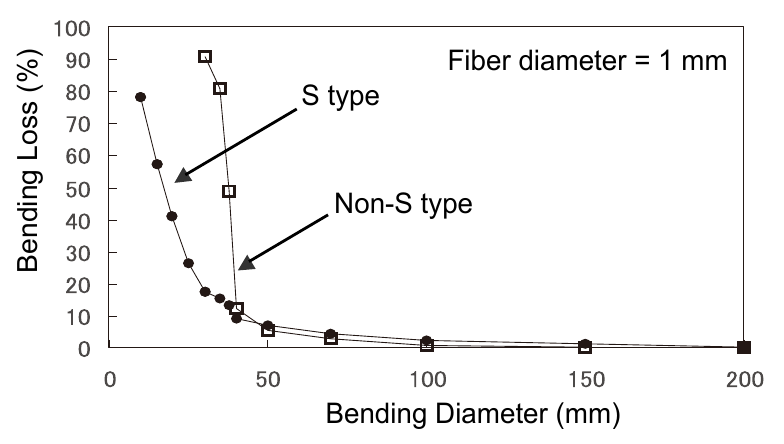}
\caption{Light yield loss due to fiber bending.}
\label{fig:scintfiber2}
\end{figure}

The WLS fibers will be of S-Type to allow shorter bending diameter
(Figure~\ref{fig:scintfiber2}) and minimize the risk of damage
during the detector assembly. In fact, the S-type fiber core has a molecular
orientation along the drawing direction. This fiber is mechanically
stronger against cracking at the cost of transparency; the attenuation
length of this type is nearly $10$\% shorter than the standard
type. Kuraray conservatively recommends a bend diameter 100 times the
fiber diameter. Accordingly, the fiber routers have been designed with
curvature radii of 5\,cm.  The chosen ``U'' shape design means each
fiber goes through two holes of the scintillator bars. To keep the
proper radius of curvature, the same fiber has to pass from one bar to
its neighbor.     

\begin{figure}[!ht]
\centering
\includegraphics[width=0.7\textwidth]{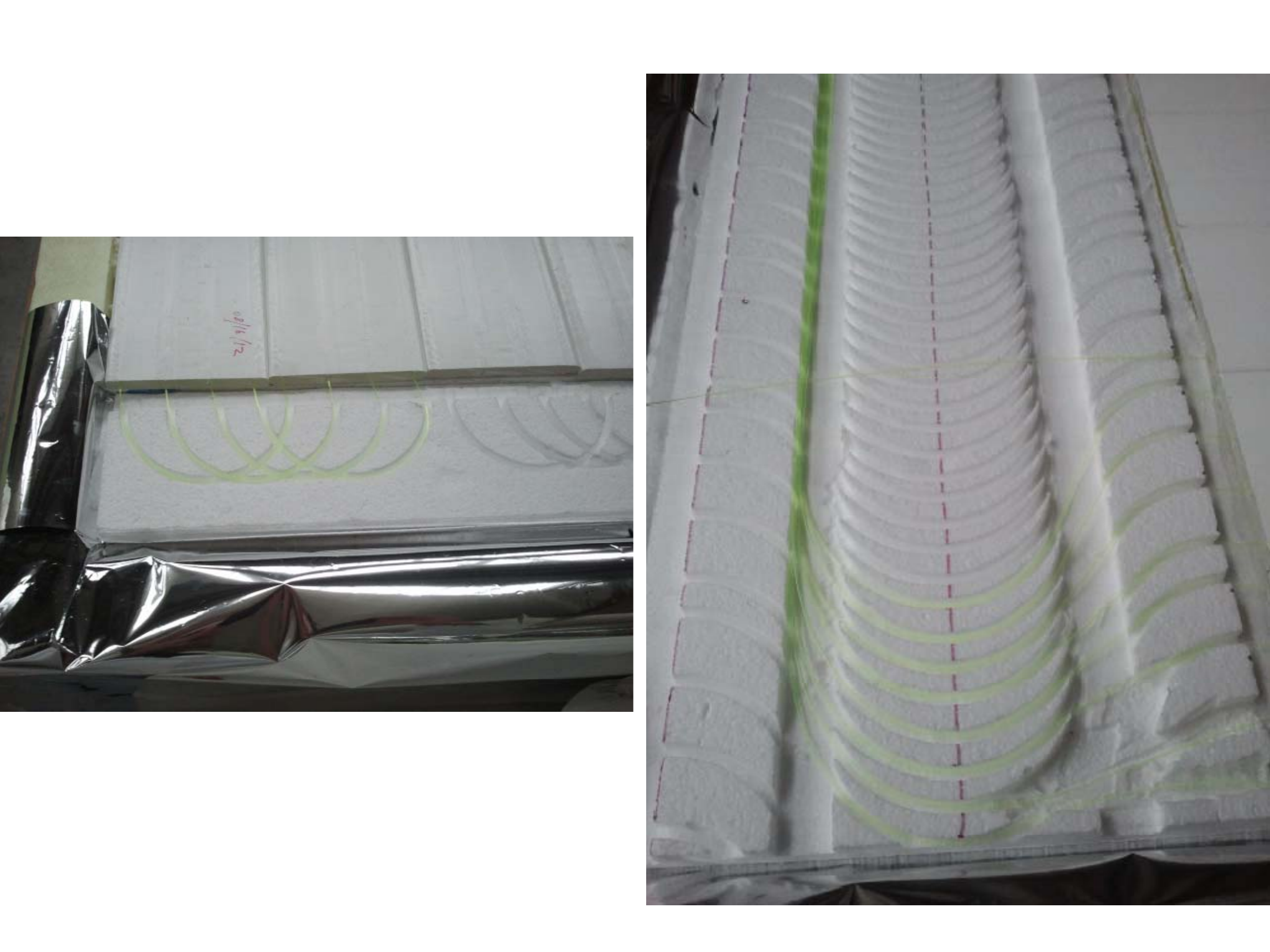}
\caption{Pictures of the extruded polystyrene manifold used in the
  prototype to route the WLS fiber. Left: external side of each module with
  the ``U'' shape. Right: internal side of each module with the ``snake''
  shape.}
\label{fig:manifold}
\end{figure}

A $6$\% non-uniformity\cite{GAPnpeMIPTorino} of the module can 
be obtained by cutting the fibers at the same length and forcing them 
to follow a particular  route (a ``snake'' route) from the bars to the 
photo-detector (see Figure~\ref{fig:manifold}). Given this design, the 
 length of each fiber is $1.6$\,m (bar length) $\times 2 + 1.2$\,m
(module~width)$\times 2 + 0.16$\,m (``U''~shape) $+0.24$\,m $\times 2$
(``snake''~route) $+0.10$\,m (tolerance at the two edges of the
scintillator bars and for merging into the optical connector). 
The total length of each of the $24$ fibers  is therefore $6.3$\,m. 

The performance of the SSD comes mainly from calibration
requirements and, then, from the width of the MIP distribution, which is
dominated by Poisson fluctuations of the number of
photoelectrons (see Sec.~\ref{sec:ScinCali}). 
Therefore, the non-uniformity of the module can be increased up to $10$\%
without deteriorating the performance of the SSD. This permits the possibility
that the fiber lengths in the module need not be all equal, and a consequent
cost saving.

The bars within a module are glued to an extruded polystyrene foam (XPS)
plane, forming a rigid structure for easy handling and mounting. The
ends of the modules have XPS manifolds which route
the WLS fibers to bulk optical connectors.

\begin{figure}[!ht]
\centering
\includegraphics[width=0.9\textwidth]{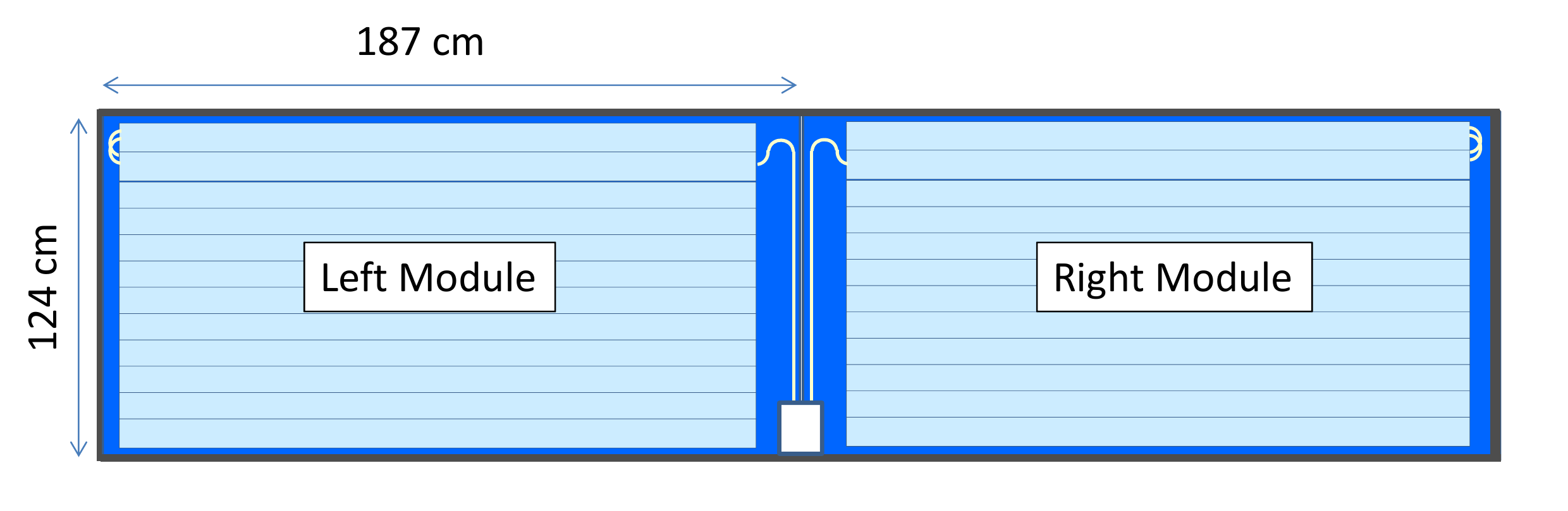}
\caption{Schematic view of the two modules in the extruded polystyrene foam vessel. The dimensions of the vessel are quoted in the figure.}
\label{fig:modules}
\end{figure}

A schematic view of the two basic modules that make up one detector is shown in
Figure~\ref{fig:modules}. The scintillator bars are hosted inside a
vessel that will be provided, ready to use, by specialized companies.
The bottom part will be delivered with the routers for the optical
fibers already embedded to simplify the detector assembly.

The fiber bundle termination of one module is then inserted and glued
inside (one half of) a special optical connector designed to allow the
optical junction of the two modules. After the assembly and testing, the
XPS vessel is hermetically sealed; a short fiber bundle tail is left
free in one corner.
The surface of the optical connector needs to be machined with a
fly-cutter when the epoxy glue used for the fiber bundles is
hardened. Six extra holes for fibers will be machined in the optical
connector to allow replacement of fibers that may be broken during
construction or shipping.

\begin{figure}[!ht]
\centering
\includegraphics[width=0.9\textwidth]{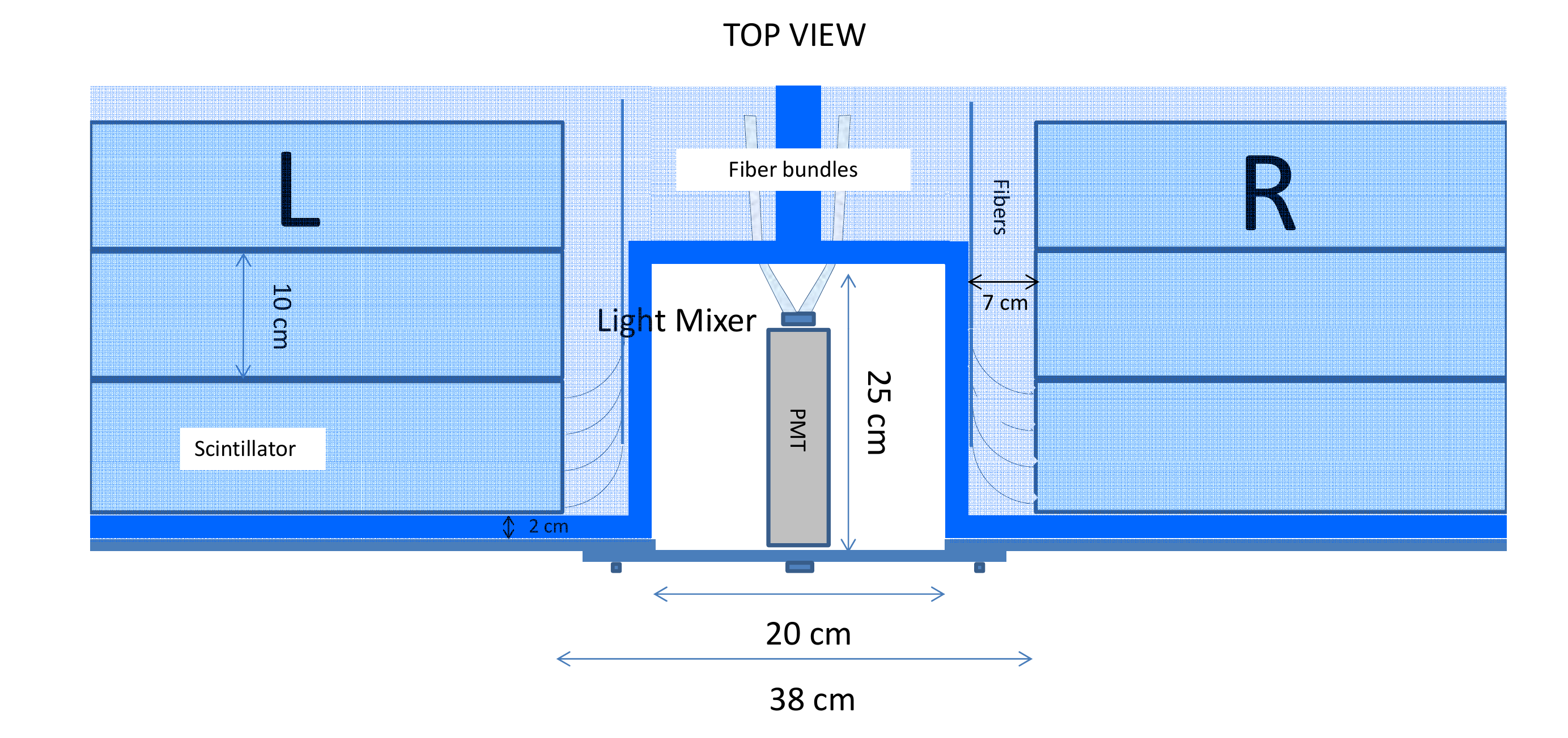}
\caption{The photo-detector that collects the light produced by the two bundles of fibers. }
\label{fig:pmtbox}
\end{figure}

The optical connector with the $48$ fibers of the two modules ($24$
each) is connected by a soft silicon pad to a single photo-detector,
integrating the total charge of all the bars (see
Figure~\ref{fig:pmtbox}).  The signal of the photo-detector is then
split in two: one is attenuated by a factor of $4$, while the other is
amplified by a factor of $32$ to achieve a sufficient dynamic
range. The power needed for operation is close to $1$\,W and can be
taken directly from the current power system without the need of an
extra solar panel.

\begin{figure}[!ht]
\centering
\includegraphics[width=0.5\textwidth]{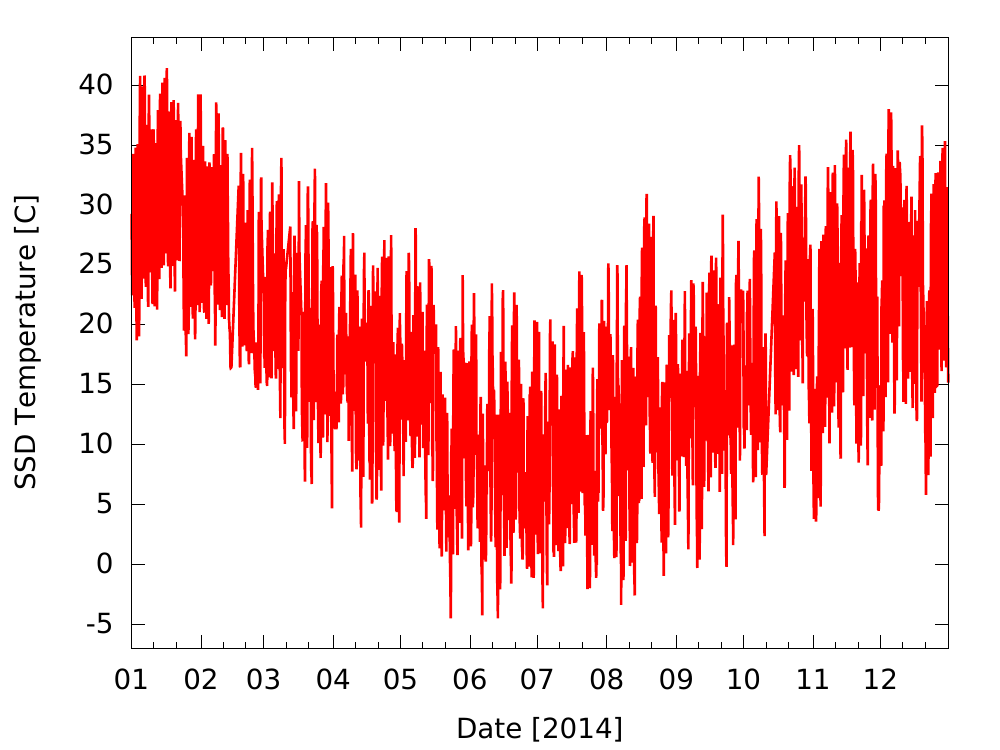}\includegraphics[width=0.5\textwidth]{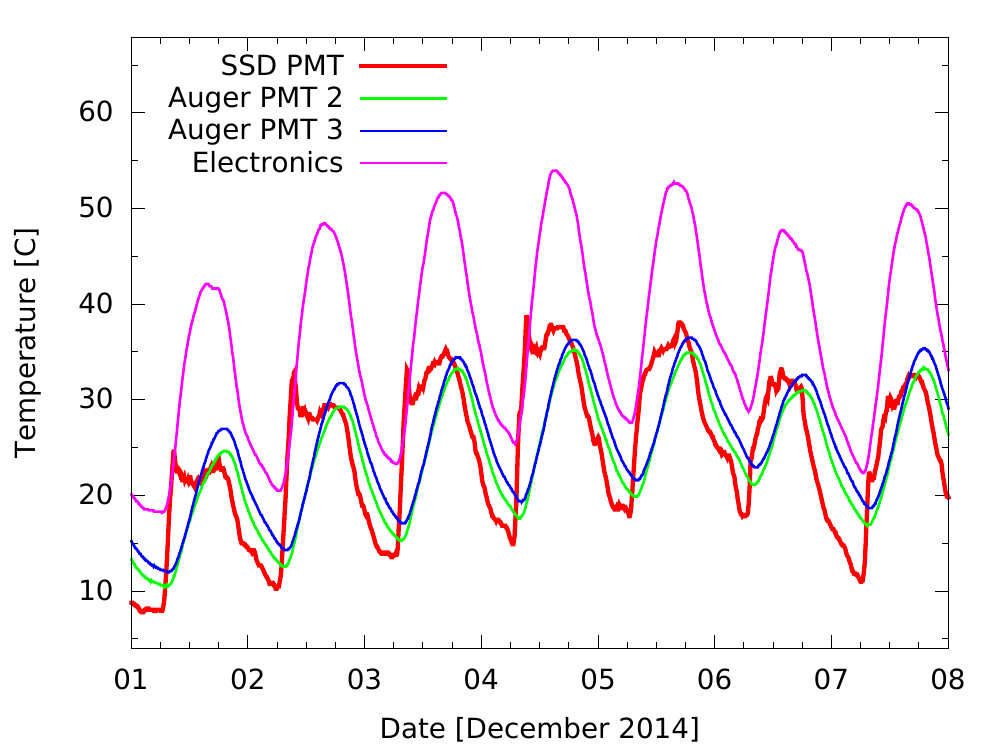}
\caption{Left: one year of temperature measurements inside a 2\,m$^2$
  prototype of SSD in operation in the Pampa. The double roof
  keeps the detector below 40$^\circ$C, even under the harsh direct
  sun of Malarg\"ue. Right: temperature for the first week of December 2014
  compared to the temperature of the WCD PMTs and of the electronics
  enclosure. The small peak in the SSD temperature in some mornings is due
  to direct sunlight reaching the PMT box which was exposed
  in the prototype design.}
\label{fig:1yeartemp}
\end{figure}

As will be discussed in more detail in section \ref{sec:ScinMech}, the external detector
enclosure is made from aluminum to guarantee
light tightness, robustness for 10 years of operation in the field, and
enough rigidity for transportation.  The access to the PMT is obtained
with a mobile door in one side of the detector box.  A double aluminum
roof is installed, separated by 2\,cm, to allow air flow and therefore
reduce the temperature changes.  This design for temperature has been
checked with previous prototypes named ASCII as can be seen in
Figure~\ref{fig:1yeartemp}.  The temperature control is of extreme
importance not only for the correct behavior of the electronics but
also with respect to the aging of the detector. Extensive studies have
been made for the MINOS detector\,\cite{minos}, and the aging was
found to be directly related to the temperature. Using the results
from the MINOS team, which have been validated during 10 years of
operation of their detector, and using the temperatures observed in
2014 from Figure~\ref{fig:1yeartemp}, we obtain an expected light loss
due to aging of 2.8\% per year. For the design we will therefore
assume a 30\% light loss over 10 years of operation. While the MINOS
team reports no effect during tests of temperature cycling, different
groups in the collaboration are repeating these cycling measurements
given the sometimes-strong 30 degree day-night temperature variations observed in
Malarg\"ue.

The total weight of one  SSD unit is about $150$\,kg.

\subsection{PMTs and HV power supply}
\label{sec:ScinPMT}

\begin{figure}[!ht]
\centering
\includegraphics[width=0.4\textwidth]{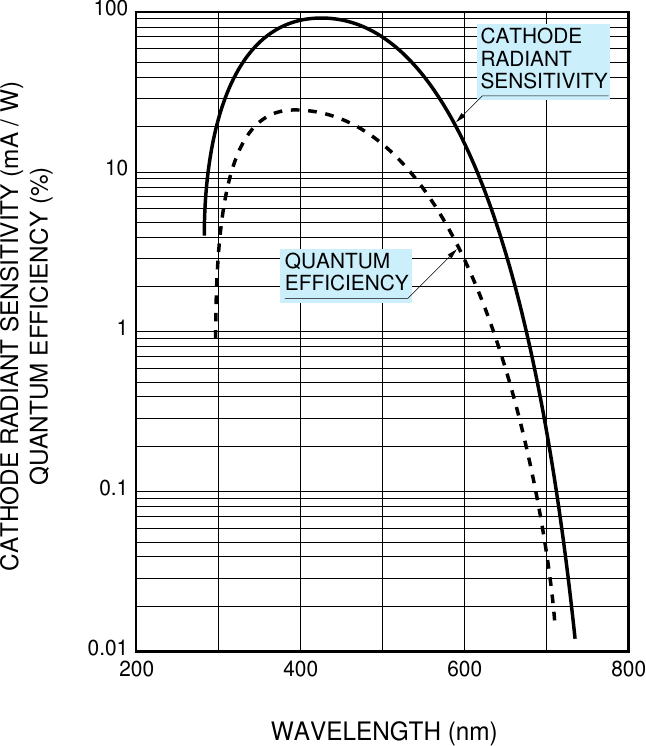}
\includegraphics[width=0.9\textwidth]{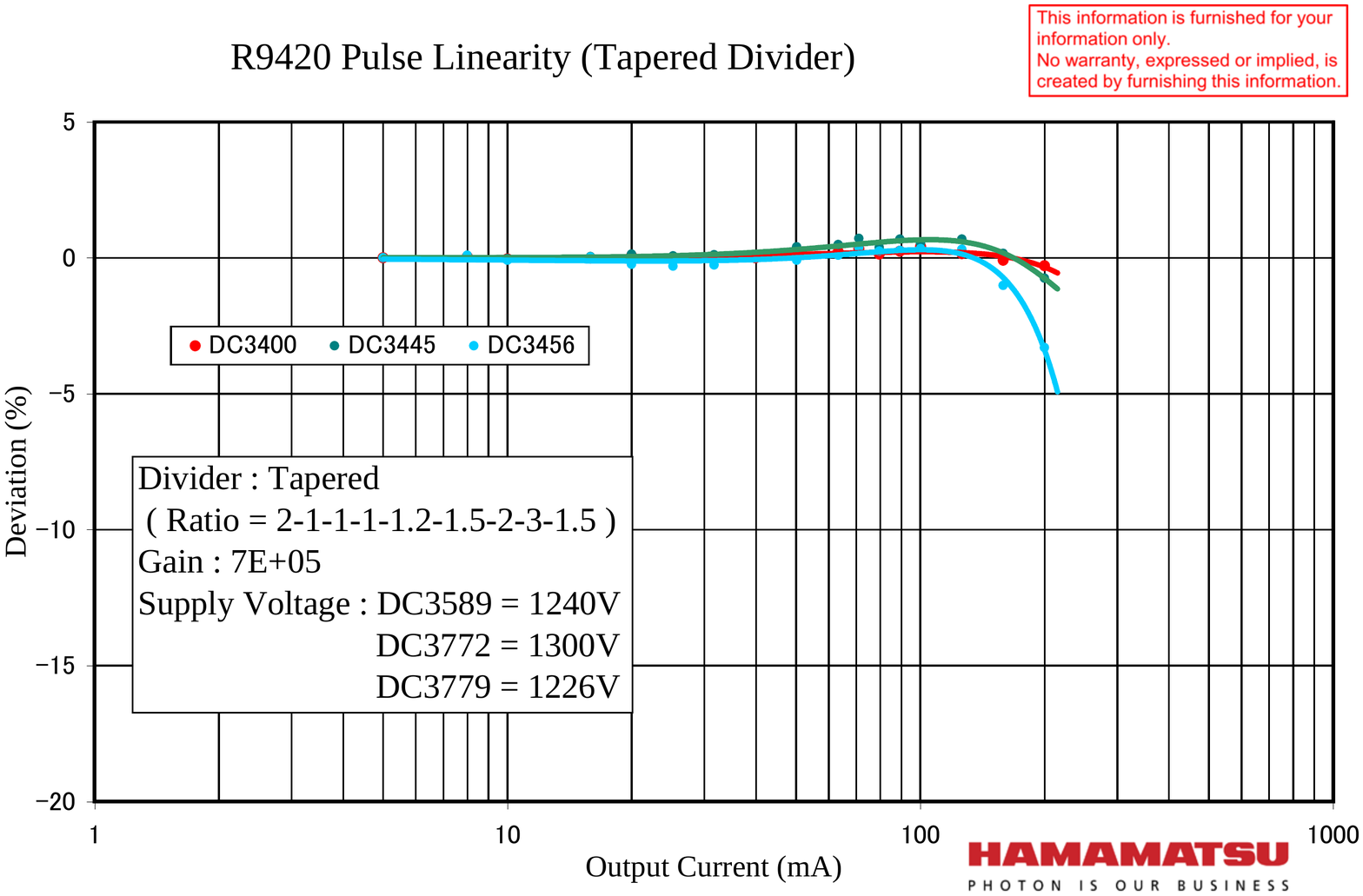}
\caption{Top: Hamamatsu R9420 quantum efficiency.
Bottom: Hamamatsu R9420 linearity.}
\label{fig:scintpmt1}
\end{figure}

The baseline SSD photomultiplier is the Hamamatsu R9420, head-on
type, 8-stage PMT with a $38$\,mm bialkali photocathode. This PMT
shows good quantum efficiency at the wavelength of interest (in the green
region) associated with an excellent linearity range (when
the PMT is supplied through a tapered ratio divider) of up to 200\,mA
of peak anode current for an operating gain of $7{\times}10^5$
(Figure~\ref{fig:scintpmt1}).  As an alternative, the performance of the lower cost Hamamatsu R8619 PMT is being investigated.

The tapered divider has been designed, according to suggestions of Hamamatsu,
with a large value of the total impedance,
$60$\,M$\Omega$, allowed given the very low average signal
current. This design minimizes the consumption of high voltage
power.

The HV module, which is produced by CAEN expressly for the proposal,
has a power dissipation of less that 0.5\,W with a maximum current of 110\,$\upmu$A (see table~\ref{tab:HV-spmt}).
It will located in a small box situated on top of the Upgraded Unified Board  (UUB) enclosure,
and be controlled by the UUB slow control signals.
Cables with length less than 3 meters are required to connect the SSD unit to the companion WCD.
No electronic buffers are required for signal transmissions.
A RG59 type coaxial cable will be used for the PMT high voltage, and RG58 type cables will be used for all the other signals.

%

\subsection{UUB integration}
\label{sec:ScinUUB}
The SSD module will be integrated to the Upgrade Unified Board as an extra
PMT.  The Slow Control of the UUB will be able to handle up to six HV
lines. One of these lines will be used to control the SSD HV with the
same connection as for an existing PMT (including a temperature sensor).

The anode signal of the R9420 photo-detector will be filtered and
split into two in a similar way to the signal from the standard WCD
photo-detectors, Sec.~\ref{sec:FrontEnd}. To achieve the required dynamic
range after the splitting, one of the two signals is attenuated by a
factor of $4$ while the other is amplified by a factor of $32$.
Figure~\ref{fig:scint_dynamics} shows the full dynamic range of the
SSD ranging from 1 to 20,000\,MIP (minimum ionizing particles).

\subsection{Mechanical assembly}
\label{sec:ScinMech}
The availability of important complex components ready to be assembled
is of primary importance to simplify the detector production and
reduce the time needed to complete the project.

One of the detector components that can be manufactured by external
companies is the vessel hosting the scintillators. For its
realization, extruded polystyrene foam (XPS) was preferred because its
mechanical properties well fit our application. Indeed, XPS is
waterproof, sufficiently strong and durable for a long period,
and furthermore, it is very light and easy to model.

Following our design, the first prototype vessels were produced by a
specialized company in Italy. They were built starting from commercial, $10$\,cm
thick XPS slabs, commonly used for thermal insulation.  The vessels
will be delivered to the assembly factories ready to use, saving
work and time in the module construction.

\begin{figure}[!ht]
\centering
\includegraphics[width=0.9\textwidth]{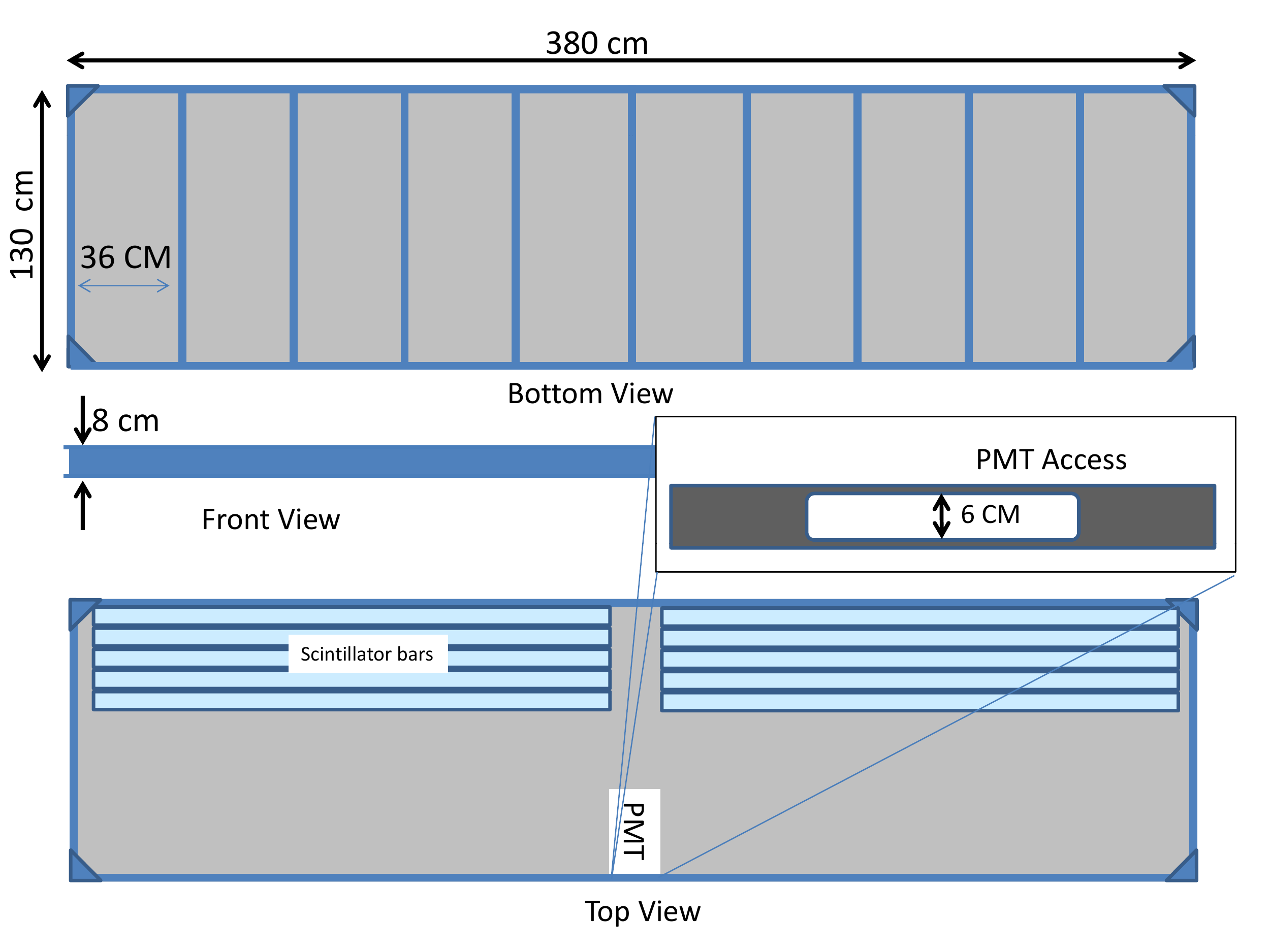}
\caption{Schematic view of the aluminum external box.}
\label{fig:scintfig3}
\end{figure}

The external aluminum structure of the baseline design is realized to
guarantee enough robustness for 10 years of operation in the field and
enough rigidity for transportation. A schematic view of the external
structure is shown in Figure~\ref{fig:scintfig3}. The aluminum box
consists of four ``I'' bars that form the external frame of the box. The
top and bottom surface are made of two sheets of aluminum that are
riveted to the frame. The choice to rivet the aluminum skin to the ``I''
bars gives the advantage of isolating the internal modules from water
and dust and guarantees a better light tightness.  The bottom part of
the box is reinforced with extra aluminum bars (one every $36$\,cm) to
support the scintillators.

The second aluminum roof is installed, separated by 2\,cm, to allow
air flow. The roof consists of a aluminum sheet of the same dimension
of the box that is placed in position on top the scintillator unit with
five square tubes $2$\,cm wide (one every $1$\,m).

\begin{figure}[!ht]
\centering
\includegraphics[width=0.9\textwidth]{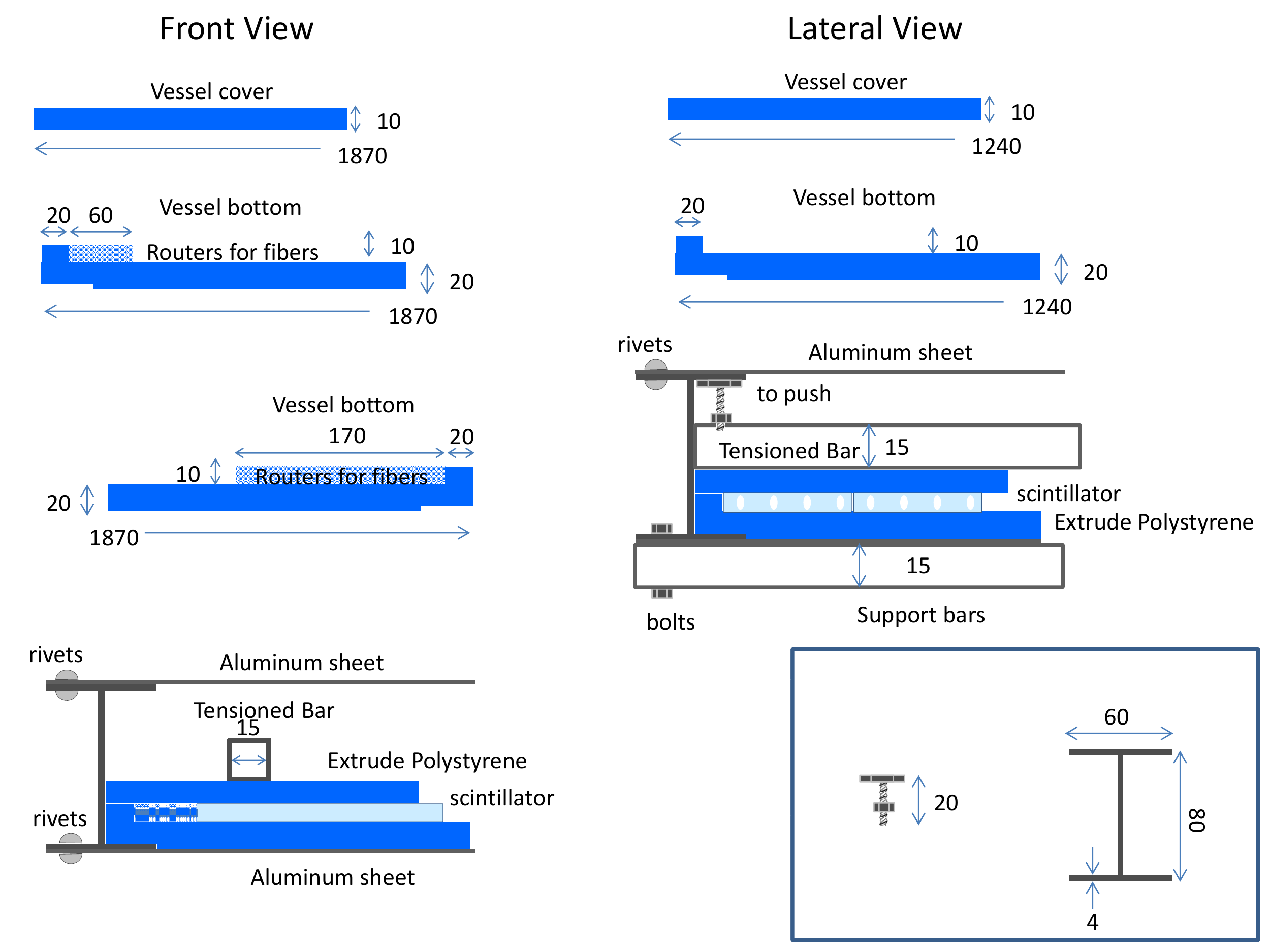}
\caption{Box design of the scintillator unit. Left: detail of the
  front view. The different layers of the unit and the vessel can be
  seen with their relative dimensions. Right: detail of the lateral
  view. The complete unit is shown at the bottom. The module inside
  the box is fixed to the frame with tensioned bars. In the inset
  window the dimension of the ``I'' bars and of the bolt used to push
  the tensioned bars are indicated. All dimensions are in mm.}
\label{fig:scintfig2}
\end{figure}

Inside the box the scintillator bars of each module are firmly fixed
to the external frame with two tensioned bars per module (see
Figure~\ref{fig:scintfig2}).

The two scintillators modules are assembled together in the aluminum
box. The photo-detector will be positioned in the center of the box
between the two modules. A small hatch will guarantee the access to
the photo-detector for maintenance.

\begin{figure}[!ht]
\centering
\includegraphics[width=0.7\textwidth]{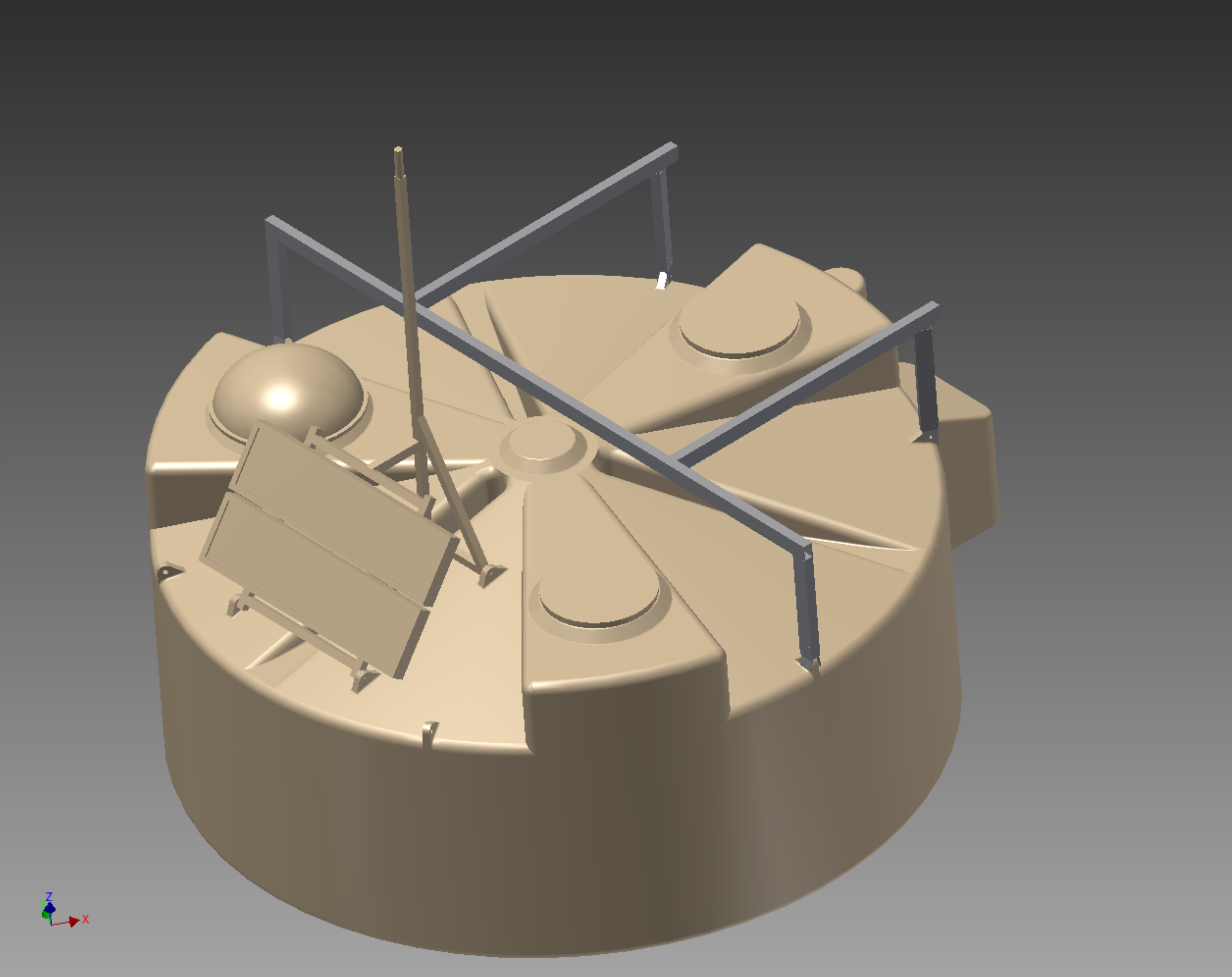}
\caption{3D view of the SSD module with the support bars. The bars are
  connected to the tank using lifting lugs present in the tank
  structure. }
\label{fig:support}
\end{figure}

The rigidity and robustness of the aluminum box reduces the complexity
of the infrastructure needed to fasten the SSD unit onto the water-Cherenkov detector.
Only few aluminum bars are needed because the
box supports itself (see section \ref{sec:deploySSD}). The supporting bars are attached to the WCD
through the lifting lugs present on top of the tank structure (see
Figure~\ref{fig:support}).

\subsection{Detector testing}
\label{sec:ScinTest}

Two steps in testing are foreseen for an SSD unit. The first
upon completion at any assembly site and the second in
Malarg\"ue before deployment. In both cases, as the fibers are not
glued in the detector, the points of failure to explore are simply
broken fibers. This means that no full scanning of the detector is needed,
only a lateral scan of all bars. This can be done with a scanning
table and a radioactive source, but given no longitudinal scan is
needed, it can also be done with a simplified muon telescope, as long
as the track of each muon detected by the SSD unit tested can be
pinpointed to a specific bar.

In order to do such a test, a muon telescope built from 2 RPCs with
cell size smaller than the width of the bars can be used. There is a
great deal of experience in RPCs in the collaboration and similar muon
telescopes are taking data with groups in Portugal and Spain, and are being
built in Brazil. Each SSD unit would be tested by being placed
inside a double muon telescope, each telescope overlooking one half of
the unit. The event rate per bar is expected to be about 5\,Hz.
About 5000 events per bar would be obtained in a 15 minute
data taking run, allowing a proper check of the response of each
individual bar. In case a specific fiber is found to be broken, it can
be identified, removed and replaced. The extra holes in the optical
connector are used in this case.
Where available, a radioactive source scan table could be used instead
of the RPC telescopes. A simple lateral scan of both sides of an SSD
can be done in less than 5 minutes.

All the results of the tests, and of any repairs made, are
kept in a central database. It should be noted that PMT testing (dark
current, afterpulses, linearity) will occur before the final SSD unit
testing, so that after the SSD is tested it can safely be deployed
in the Pampa.

\subsection{Calibration and control system}
\label{sec:ScinCali}

\begin{figure}[!ht]
\centering
\centerline{\includegraphics[width=0.5\textwidth]{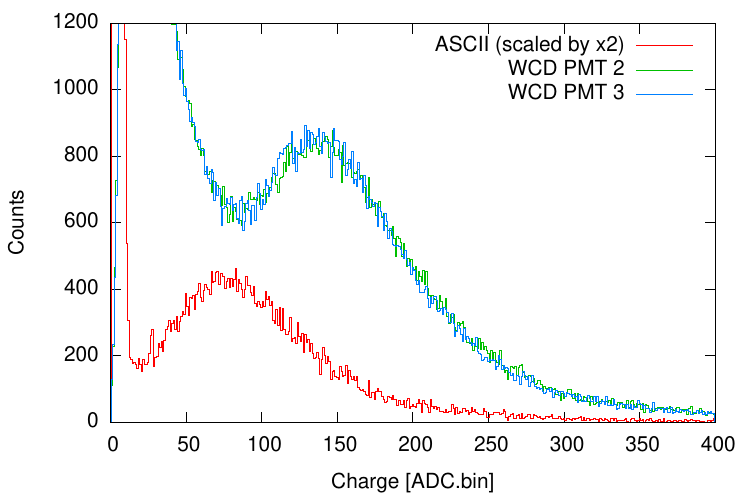}\includegraphics[width=0.5\textwidth]{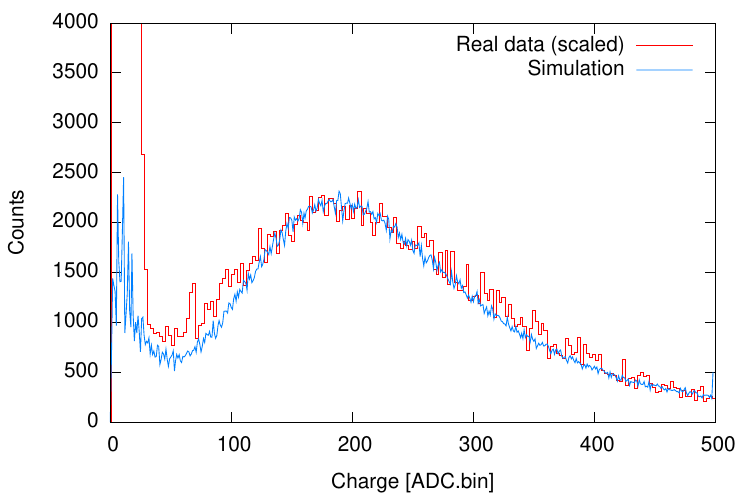}}
\caption{MIP histogram of a 2\,m$^2$ prototype running in the Pampa,
  together with VEM calibration histograms of the WCD over which it is
  installed (left), and comparison to a simplified simulation
  (right). These histograms correspond to one minute of data
  taking. Given the clean separation of the MIP from the low energy
  background, no calibration issue is foreseen. These results were confirmed
  with the detailed codes discussed in section~\ref{sec:offline}, including
  a Geant4~\cite{g4} based simulation of the response of the SSD and WCD to low
  energy showers simulated with CORSIKA~\cite{Heck:1998vt}.}
\label{fig:mip-hist}
\end{figure}

The SSD calibration is based on the signal of a minimum ionizing
particle going through the detector, a MIP.  Since this is a thin
detector, the MIP will not necessarily be well separated from the low energy
background but, being installed on top of the WCD, a cross trigger can
be used to remove all of the background. About
40\% of the calibration triggers of the WCD produce a MIP in the
SSD. The statistics of calibration events recorded in a minute, the
normal WCD calibration period, are therefore enough to obtain a
precise measurement of the MIP. Figure~\ref{fig:mip-hist} shows the
MIP calibration histogram from a 2\,m$^2$ test module, obtained in one
minute of acquisition. The MIP is clearly defined, and will allow an absolute
calibration of the SSD to better than 5\%.

\begin{figure}[!ht]
\centering
\centerline{\includegraphics[width=0.5\textwidth]{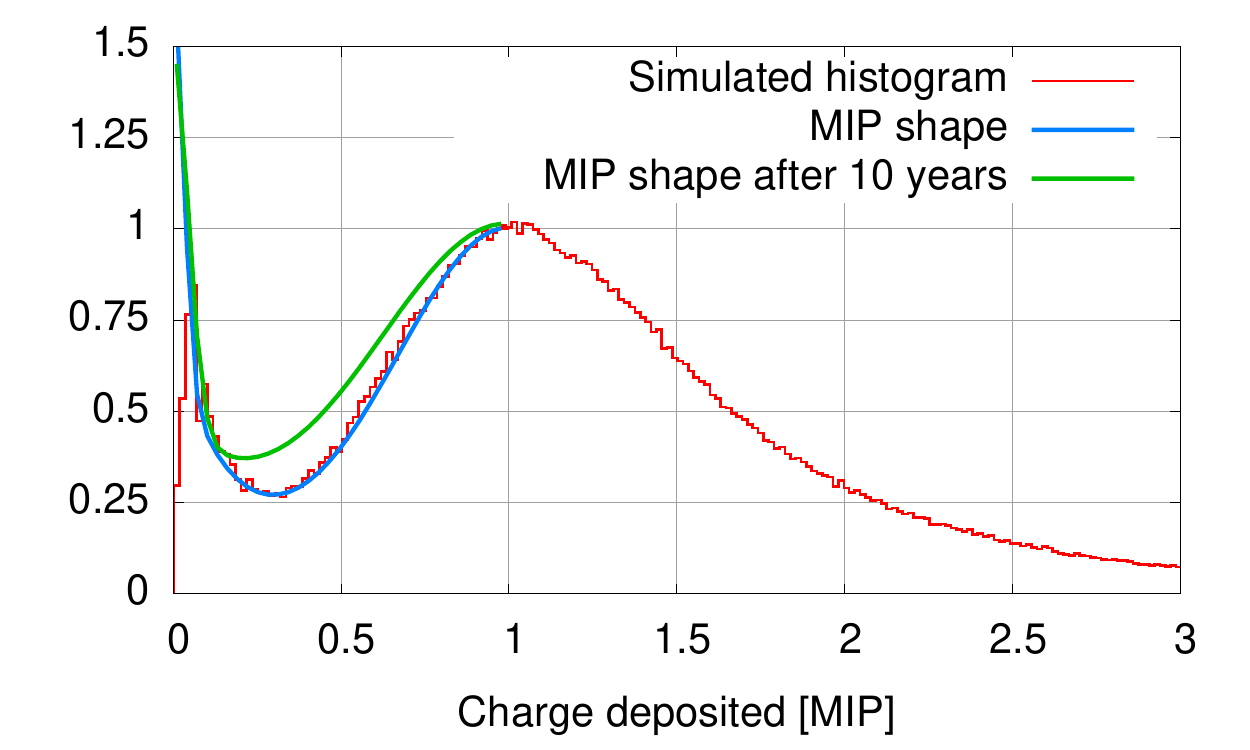}}
\caption{MIP histogram obtained from simulation and extrapolation of
  its shape after 10 years of aging. The MIP can still be easily
  determined with a foreseen 1.5\% statistical accuracy.}
\label{fig:mip-age}
\end{figure}

The performance requirements for the SSD come mainly from calibration
requirements: in shower measurement mode, the dominant measurement
errors are due to Poisson fluctuations of the number of particles
detected, and the overall calibration constant determination.
Detector non-uniformity contributes a small error when compared to the Poisson error,
as long as non-uniformities are below 20\%. While the FWHM of the
WCD calibration histogram will be clearly smaller than that of the
SSD (the calibration unit for the WCD, the VEM, is at about 100\,pe),
the fact that the SSD can be cross-triggered by the WCD means that the MIP is clearly visible against very little background.
The
width of the MIP distribution is mostly determined by Poisson
statistics of the number of photoelectrons per MIP, the
non-uniformity of the detector, and the intrinsic fluctuation of
the response to a single particle, mainly due to different track lengths
in the scintillator. The latter factor was determined from
simulations to be around 18\%. The baseline design chosen for the SSD
produces 12 photoelectrons per MIP\,\cite{GAPnpeMIPTorino}, which
would degrade to 8 photoelectrons after 10 years of operation due to aging. This
amounts to a 35\% contribution to the MIP distribution width.
Non-uniformity is very well controlled with the U-shape for the
fibers and the ``snake'' routing of the fiber up to 6\%\,\cite{GAPnpeMIPTorino}.
The total width expected after 10 years of operation is 40\%,
assuring a MIP determination at better than 1.5\% statistical
accuracy. The expected MIP histogram can be seen in figure~\ref{fig:mip-age}.

In addition to these calibration histograms, which are taken in real
time but used only offline, a rate-based method will be developed to
get an estimate of the value of the MIP at the level of the local
station controller. The advantage of a rate based algorithm,
stabilized with a sigma-delta method, is that it is extremely
robust. It is the same algorithm that has been running in the WCD for
more than 10 years.

Finally, extra calibration information will also be determined for
the SSD as is done for the WCD: high gain/low
gain ratio, baseline noise and average pulse shape. In addition to
these calibration data, monitoring values will also be sent to the central data acquisition system
together with the WCD monitoring block (every 400 seconds). These
monitoring data will include for the SSD unit the DAC settings for the
HV supply of the PMT, the effective voltage at which it is run, the
monitoring current from the PMT, and the temperature inside the SSD
unit. They will all be integrated into the online and offline monitoring
systems.

%% file: sdeu.tex
\section{Surface Detector Electronics}
\label{sec:sdeu}

\subsection{Introduction and design objectives}

The Surface Detector Electronics (SDE) records the tank signals, makes
local triggering decisions, sends timestamps to the central data
acquisition system for the global triggers, and stores event data for
retrieval when a global trigger condition is satisfied (see Sec.~\ref{currentSD}).
Because of the small bandwidth (1200 bits/s) available to each tank,
the station must operate semi-autonomously, performing calibrations
and taking action in response to alarm conditions at the station
level.

The current SDE was designed 15 years ago using the technology
available at that time. Evolution in processors, power consumption of
electronics components, and timing systems make it possible today to
design and implement a higher performance electronics system for the
Surface Detector array. Furthermore, the proposed electronics provides
an interface to allow the scintillator detectors co-located with the
surface detector stations to make use of the data processing and
communications infrastructure of the stations.

The design objectives of the SDE Upgrade (SDEU) globally aim to
increase the data quality (faster sampling for ADC traces, better
timing accuracy, increased dynamic range), to enhance the local
trigger and processing capabilities (more powerful local station
processor and FPGA) and to improve calibration and monitoring
capabilities of the Surface Detector stations.
Backwards-compatibility with the current data-set will be maintained by
retaining the current time span of the PMT traces and providing for
digital filtering and downsampling of the traces to emulate the
current triggers in addition to any new triggers.  The design
objectives also aim for higher reliability and easy maintenance.  An
important feature in the design of the upgraded SD electronics is a
facility for interfacing not only the SSD but also any other
additional detectors.

The proposed upgrade involves the main electronics boards: the Unified
Board (UB) and the front-end board of the current electronics. The
interface board to the power system, the Tank Power Control Board
(TPCB), will not be upgraded, and the interface to the communication
system will also remain unchanged.  Furthermore, new functionalities
will be added to the tank calibration LED system and to the monitoring
system. The dynamic range will be increased by adding a small PMT
(SPMT) to the current 3 large 230\,mm XP1805 PMTs. All the
functionalities will be implemented in a single board, called the Upgraded
Unified Board (UUB). The detailed specifications can be found in
ref. \cite{SDE_specs}.

\subsection{Front-end electronics}
\label{sec:FrontEnd}

\begin{figure}[!ht]
\centering
\includegraphics[width=.9\textwidth,angle=0]{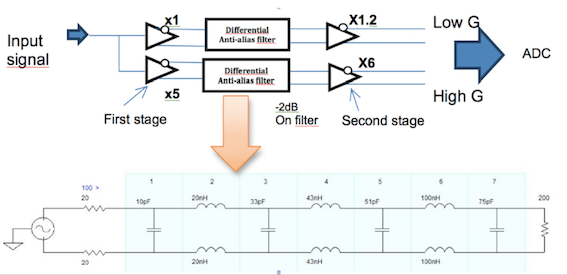}
\caption{Block diagram of the front-end electronics.}
\label{fig:FE_block}
\end{figure}

The signal from the anode of the PMTs is split and the high-gain channel is amplified by 
 using a dual channel ADA4927 Operational Amplifier (OA) yielding an amplification of about a factor of 5. 
Signals are filtered by a 5-pole low-band pass filter using passive components
(inductances and capacitors). Finally the last amplification stage is
implemented by using the same ADA4927 OA to obtain a total amplification of 30\,dB
corresponding to a voltage amplification factor of  about 32 for the high-gain channel.  The
signals are digitized by commercial 12\,bit 120\,MHz AD9628 FADCs,
which achieve this performance with minimal power consumption, an
important consideration due to the 10W station power budget.  The
pulse response of the PMT, when expressed in terms of bandwidth, is
${\sim}70$\,MHz.  This is well matched to a 120\,MHz FADC and
associated 60\,MHz Nyquist filter.  A block diagram of the design is
shown in Fig.~\ref{fig:FE_block}.

\begin{figure}[!ht]
\centering
\includegraphics[width=.7\textwidth,angle=0]{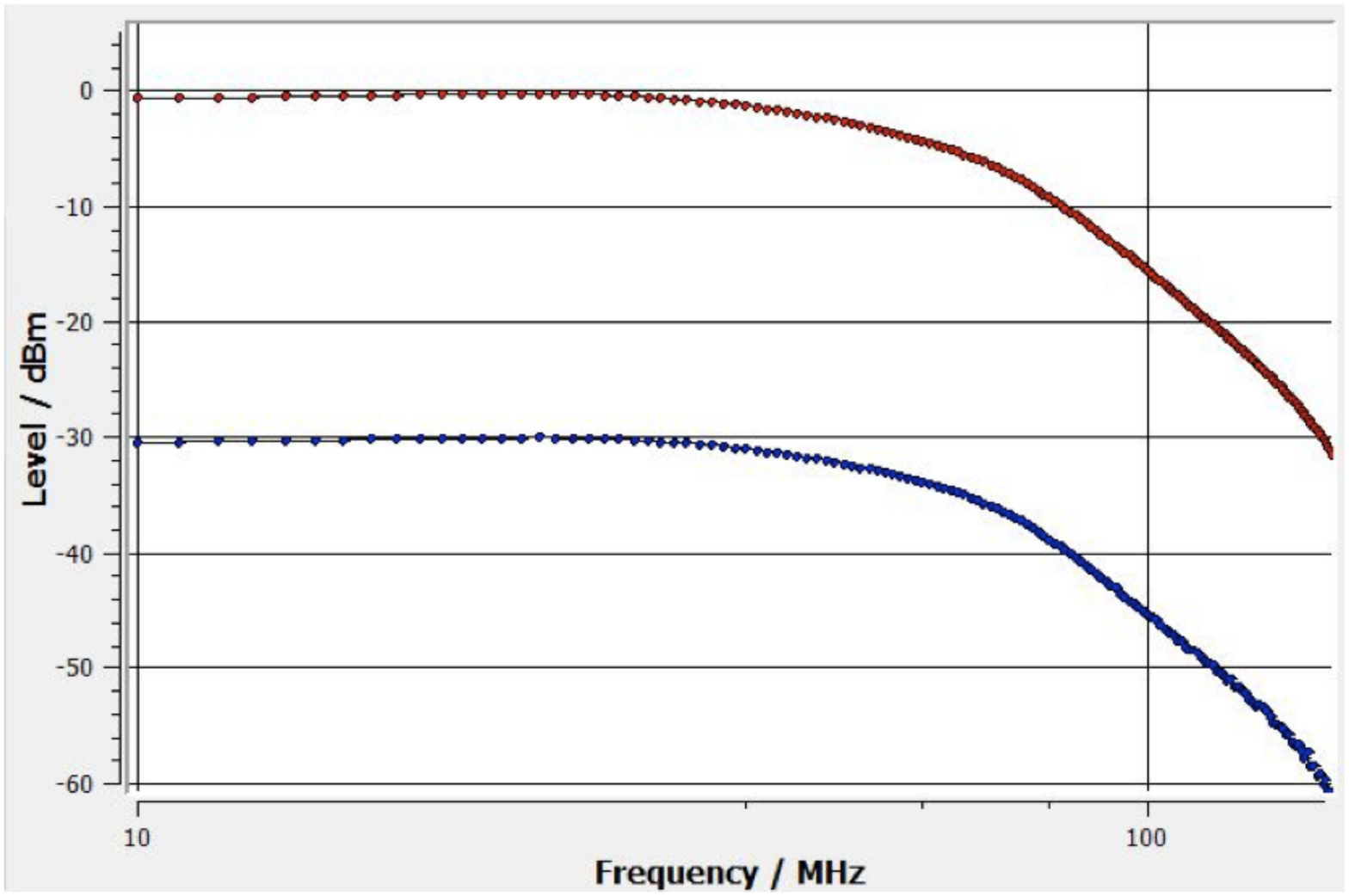}
\caption{Filter response.}
\label{fig:filter}
\end{figure}

The design of the filter was simulated and the response shows the
correct cutoff at 60\,MHz and a noise level of 400\,$\upmu$V RMS. The
measured filter response is shown in Fig.~\ref{fig:filter} for both
the high-gain and low-gain channels. The cut-off frequency is 60\,MHz
and the high gain is 30\,dB.

The large PMTs in the WCD deviate from linearity for peak currents in
excess of ${\sim}50$\,mA.  Conservatively assuming a maximum current
of 40\,mA for any PMT in the field, the corresponding maximum signal
in the WCD before saturation, for an operating gain of $3{\times}10^5$
and a single VEM peak signal of ${\sim}100$ photoelectrons, is
${\sim}600$\,VEM.  This is well matched via standard $50\,\Omega$
termination resistors to the 2V input range of the Front End
digitizers.  In the upgraded WCD, the dynode signal is replaced by the
anode, amplified by a factor of 32.  The full WCD signal range of 600
VEM is then conveniently mapped into a full 17-bits digital range by
the two WCD signals, namely the amplified anode, for single VEM
resolution, and the direct anode signal, each spanning the 12 bits
available in the new electronics and with 7 bits overlap.  In such a
configuration a single bit is worth ${\sim}0.3\,\upmu$A.  The WCD
dynamics is further extended by other 5 bits using the SPMT signal,
the gain of which is tuned to have a signal 32 times smaller with respect
to the anode, corresponding to $600 \times 32 \sim 19$\,kVEM.  As
discussed in Sec.~\ref{sec:pmt}, the SPMT gain and its overlap with the
LPMT signals can be modified, and the overall dynamic range can
comfortably exceed 40\,kVEM.

\begin{figure}[!ht]
\centering
\includegraphics[width=.99\textwidth,angle=0]{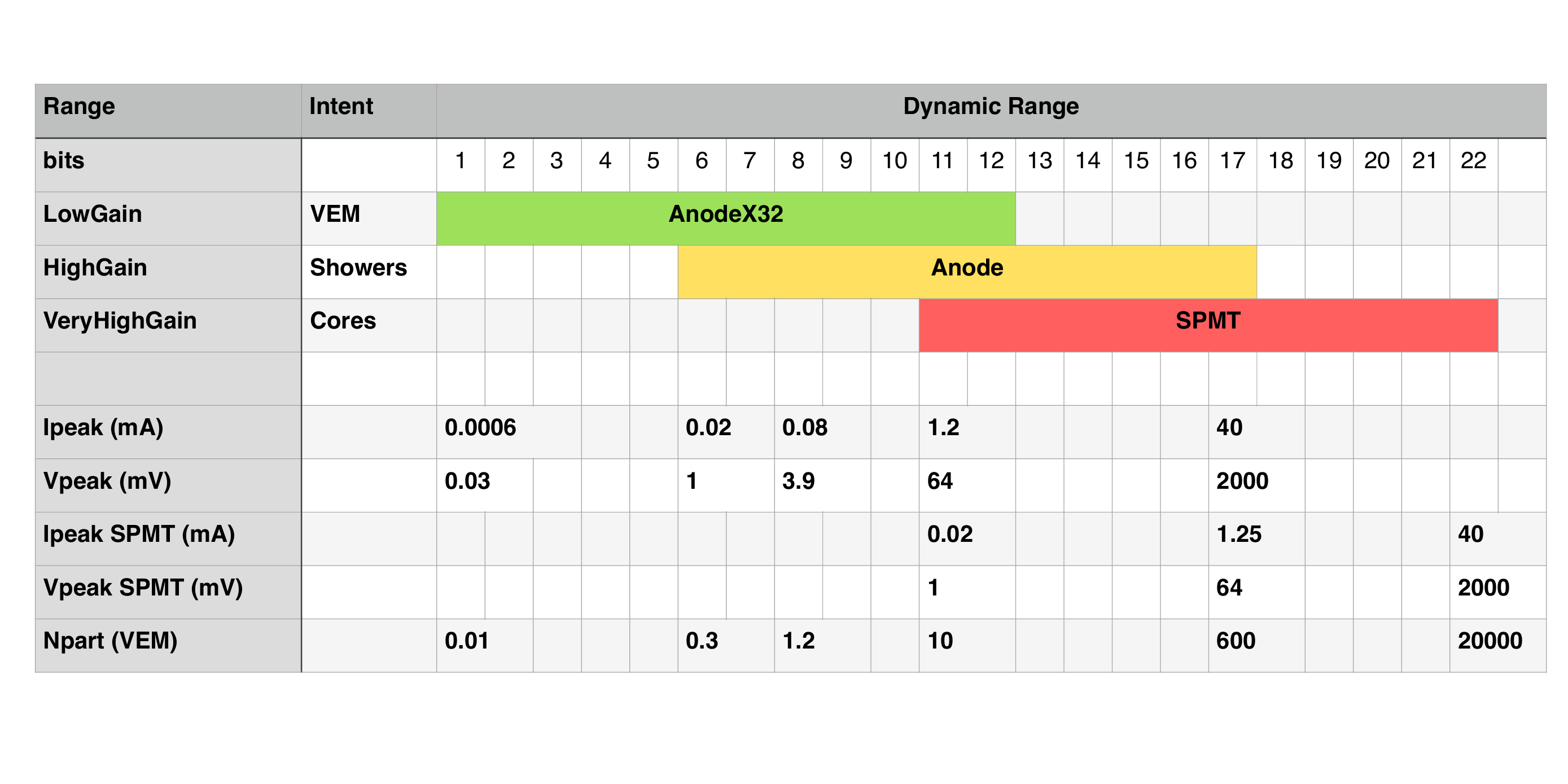}
\caption{WCD dynamic range. The maximum signal before saturation corresponds to 20k particles with the operating settings specified in the text.}
\label{fig:dynamic}
\end{figure}

The global dynamic range for the WCD PMT signals is shown in
Fig.~\ref{fig:dynamic}.
The signal from the large PMT, operated at the current gain of $3 \times 10^5$,
 is split into an amplified {\it LowGain} range for single muon resolution 
and a {\it HighGain} range for measurements of shower signals. 
Events closer to the core have a larger signal that is collected by the  small PMT 
and input into a dedicated {\it VeryHighGain} range.
The dynamic range scheme will allow moving the
trigger threshold two bits higher and increasing the current dynamic 
range by a factor of 32 and up to 20\,kVEM.  The muon peak will be in channel 200.

\begin{figure}[!ht]
\centering
\includegraphics[width=.95\textwidth,angle=0]{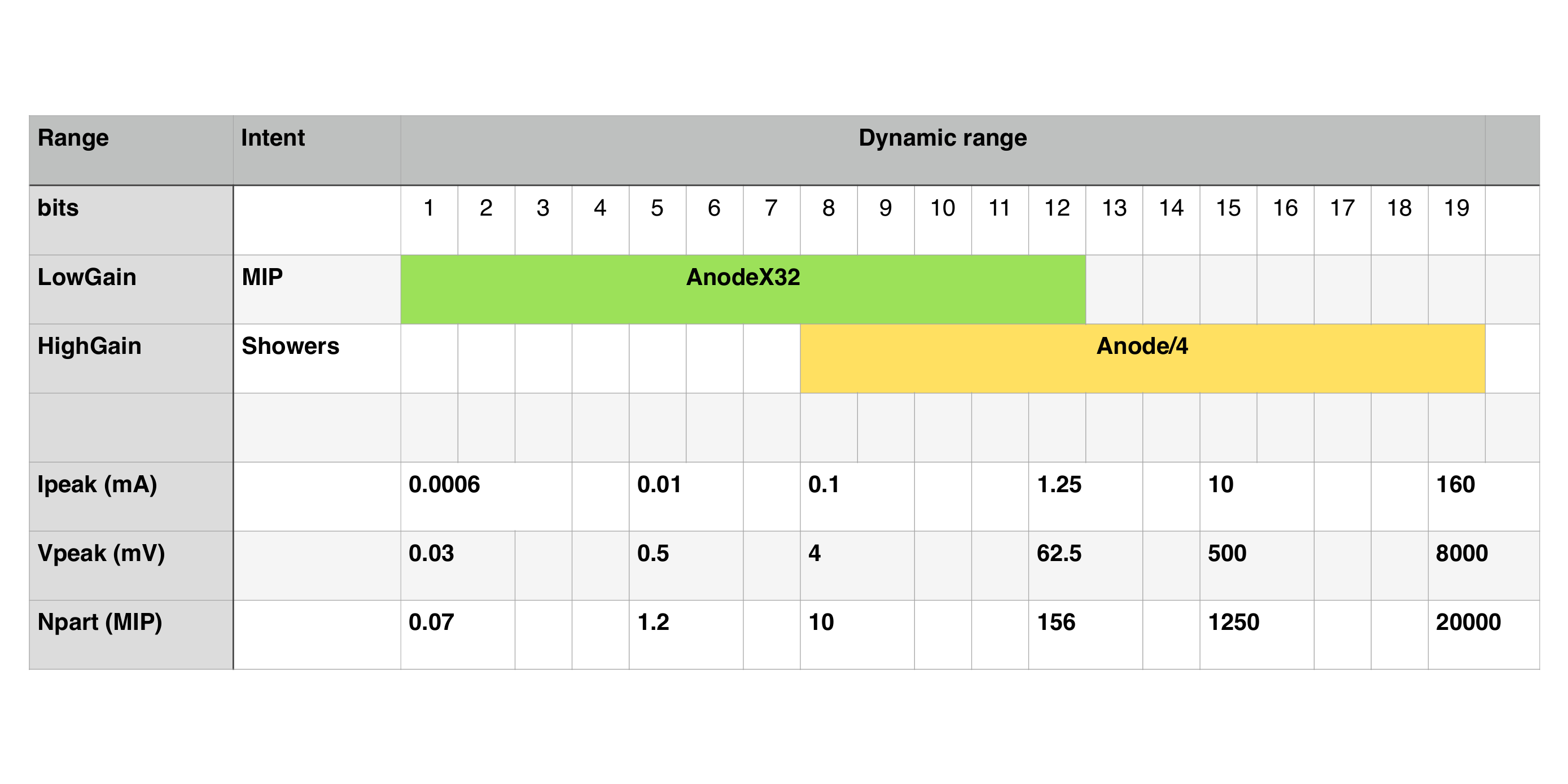} 
\caption{SSD dynamic range. The maximum signal before saturation corresponds to 20k particles with the operating settings specified in the text.}
\label{fig:scint_dynamics}
\end{figure}

The anode signal from the SSD PMT will be split into two ranges ({\it LowGain} and {\it HighGain}), filtered and
sampled in a similar way to the signals from the WCD PMTs.  Like with the
WCD, the SSD dynamic range is determined by the maximum peak current
of the readout PMT, the number of bits available and the amplification factor of
each channel.  Assuming a peak current of 160\,mA, as measured for the proposed R9420 PMT,
we plan to match this to the 2\,V input range of the ADCs by
reducing the signal by a factor of 4.  Reducing the overlap of the low
and high SSD gain ranges to 5 bits, we can count on a full 19 bits
range, with each bit worth $0.6\,\upmu$A.  Using an amplification
factor of 32 for the low gain signal, we place the MIP signal around
30 ADC counts ($\sim 0.5$mV), and stretch the linearity range up to 20\,kMIP for the full
19 bits, matching the WCD range.  
Considering 12 photoelectrons per MIP at the scintillator,
this implies working at a gain of $4{\times}10^4$.  
It should be noted that while the availability of a dedicated PMT for very high range provides further margins to extend the WCD dynamics, 
the comparably large SSD dynamics relies on the properties of the chosen PMT, which exhibits high peak current and low operational gains.
The global dynamic range for the SSD is shown in Fig.~\ref{fig:scint_dynamics}.

\subsection{Timing}
\label{sec:timing}
For the upgraded electronics we have selected the I-Lotus M12M Timing
GPS Receiver manufactured by I-Lotus, LLC
(Singapore)~\cite{ilotus}. The M12M Timing receiver is designed to be
functionally compatible with the Motorola Oncore~UT+ GPS receiver that
is currently used within the Auger SDE Unified Board. Choosing a
compatible unit means that fewer and simpler modifications to the
basic time-tagging system design. Specifically, the M12M provides the
same 1~PPS timing output with serial control and data. The specified
intrinsic device accuracy after the applied ``granularity correction''
(the so-called ``negative saw tooth'') is about 2 nanoseconds.  This
accuracy is very good relative to the UUB specification to achieve
better than 5.0 nanoseconds RMS accuracy.

\begin{figure}[!ht]
\centering
\includegraphics[width=.6\linewidth,angle=0]{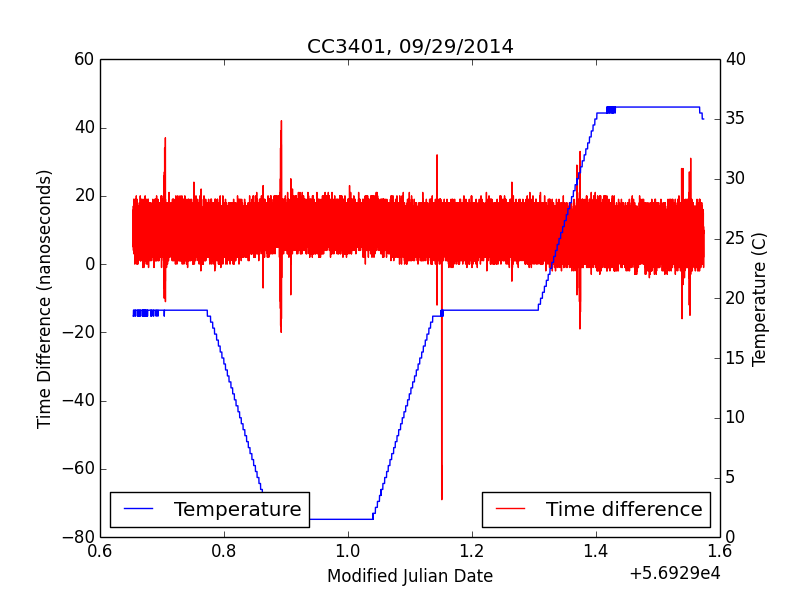}
\includegraphics[width=.39\linewidth,angle=0]{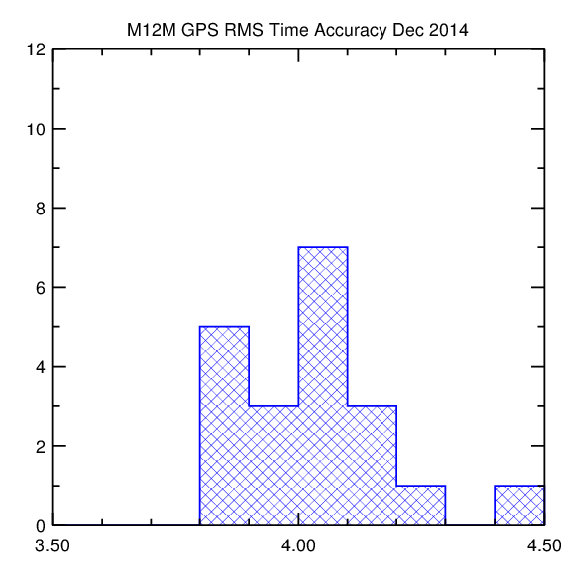}
\caption{{\bf Left:} Timing residuals (red: residuals in nanoseconds)
  for a typical I-Lotus M12M Timing GPS receiver as measured
  second-by-second during a 22 hour thermal test cycle (blue:
  temperature). The offsets are demonstrated here
  to be stable with extreme temperature variations to within
  4.0\,ns. [Note: short red vertical ``spikes'' indicate momentary GPS
    tracking glitches corresponding to less than 0.01~percent
    of all time-stamps which will have no impact on GPS time-tagging accuracy as 
    implemented within
    the UUB.] \hspace*{3mm} {\bf Right:} Histogram showing distribution of measured
  RMS residuals (in nanoseconds) for 20 I-Lotus M12M units tested
  within the thermal chamber. All measurements are well within the
  required specification of 5.0\,ns or better.}
\label{fig:GPS_resolution}
\end{figure}

Fig.~\ref{fig:GPS_resolution} shows a histogram indicating the RMS
timing accuracy as measured for twenty of the newer M12M Timing
receivers during multi-hour thermal test chamber testing meant to
mimic extreme temperate variations recorded on actual SD stations in
the Auger SDE.  No particular dependence of the temperature variation
is observed, and all twenty of the test receivers demonstrate accuracy
better than the 5.0\,ns specification required.

The fundamental architecture of the time-tagging firmware module
parallels the time-tagging design concept used in the original UB 
and is implemented in the UUB board FPGA. The
on-board software for initialization of the time-tagging modules, GPS
hardware control, and timing data is implemented on the original UB as
a framework, forking changes and modifications as needed for the new
UUB.

\subsection{Slow Control}
\label{sec:SlowControl}
\begin{figure}[!ht]
\centering
\includegraphics[width=.8\textwidth,angle=0]{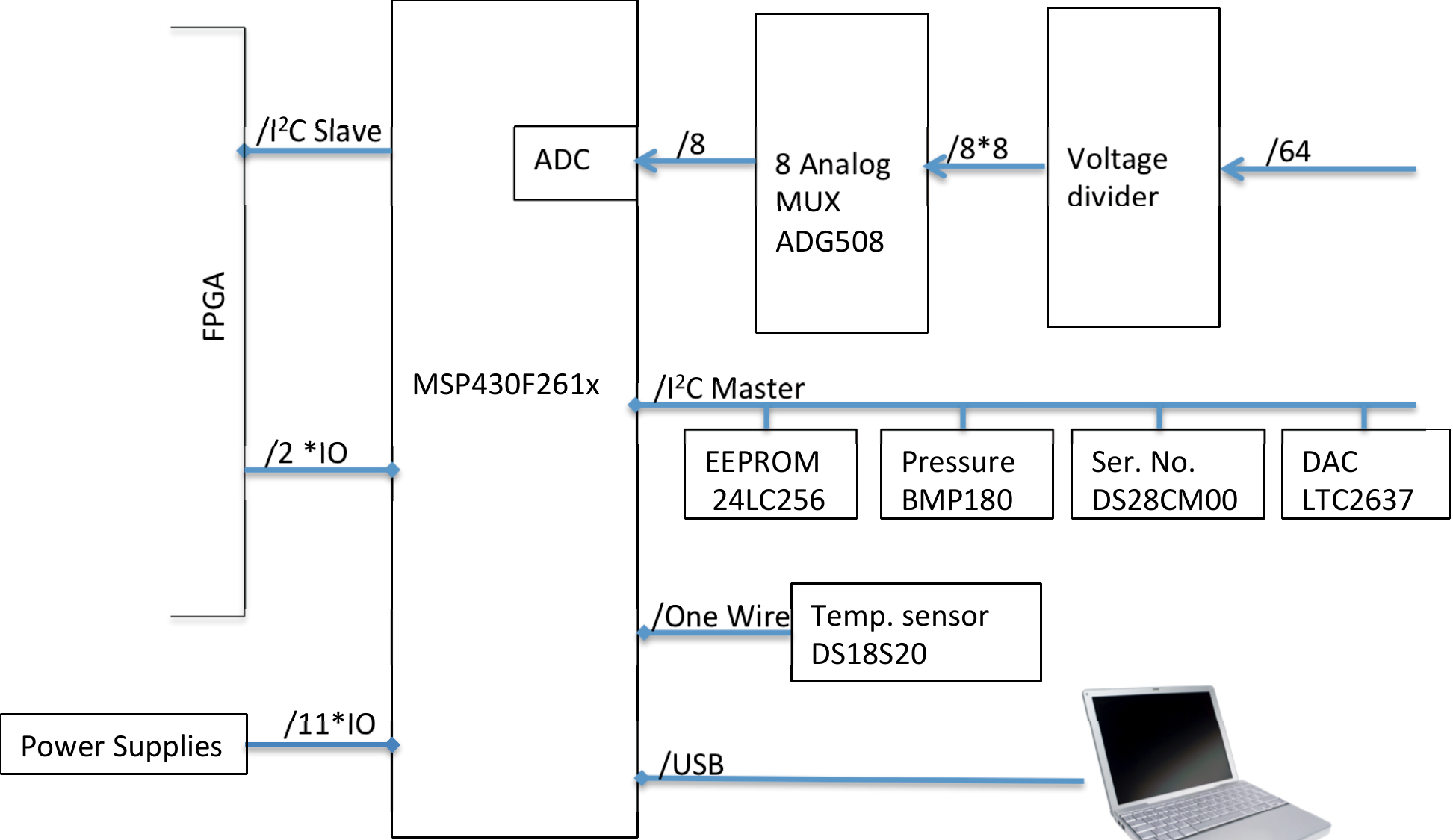}
\caption{Slow control block diagram.}
\label{fig:SC_Block}
\end{figure}

A slow control system similar to that of AERA (The Auger Engineering
Radio Array), incorporating a separate micro-controller (MSP430), will
be used.  There are sixty-four 0 to 5\,V analog inputs, 16 logic IO's and eight
0 to 2.5\,V analog outputs. The module also provides a USB serial
connector.  There are currently several free channels for test
purposes and for additional detectors such as the SSD. Additional
water temperature and pressure sensors will also be implemented.
Fig.~\ref{fig:SC_Block} shows a block diagram of the slow control.

The slow control software provides access to more than 90 monitoring
variables.
These include currents and voltages of the subsystems of the UUB,
environmental sensor values and PMT currents and voltages. Also the
PMT high voltages are controlled by the slow control. For maintenance
there is a human interface implemented via USB serial connection. In
case of trouble, the micro-controller can be instructed by the COMMS to
perform a complete UUB reset. 
Care is taken in monitoring the solar power system,
 and in the case of battery under-voltage, some parts or all components of the UUB
may be switched off.

\subsection{Upgraded Unified Board}
\label{sec:UUB}
\begin{figure}[p]
\centering
\includegraphics[height=.95\linewidth,angle=90]{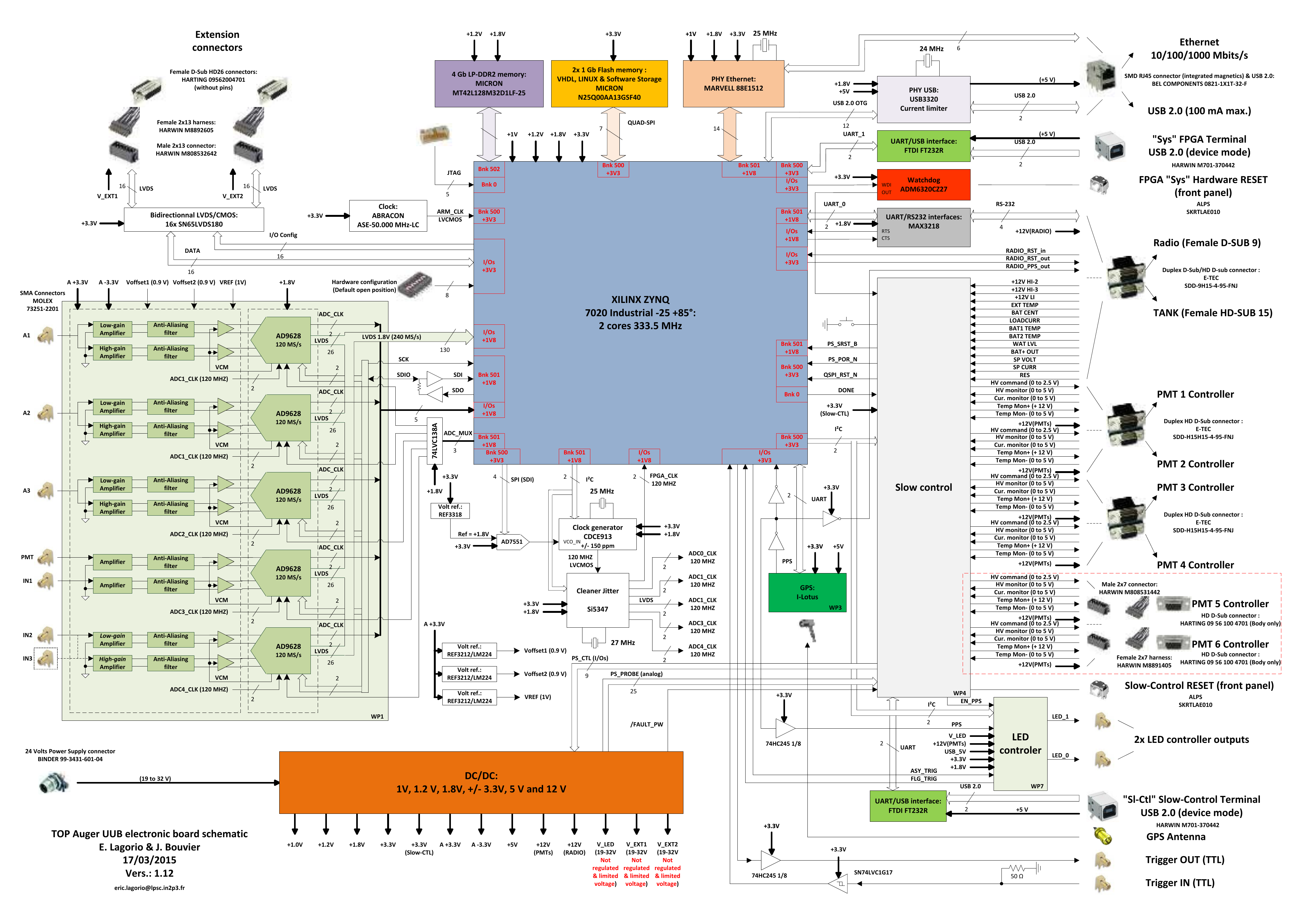}
\caption{General architecture.}
\label{fig:SDE_architecture}
\end{figure}

All the different functionalities described in the previous sections are implemented on a single board,
called the Upgraded Unified Board (UUB). The architecture of the UUB
includes a Xilinx Zynq FPGA with two embedded ARM Cortex A9 333\,MHz
micro-processors, 4\,Gbit LP-DDR2 memory and 2\,Gbit Flash memory
(storage memory).  The general architecture is shown in
Fig.~\ref{fig:SDE_architecture}.

\begin{figure}[!ht]
\centering
\includegraphics[width=.95\textwidth,angle=0]{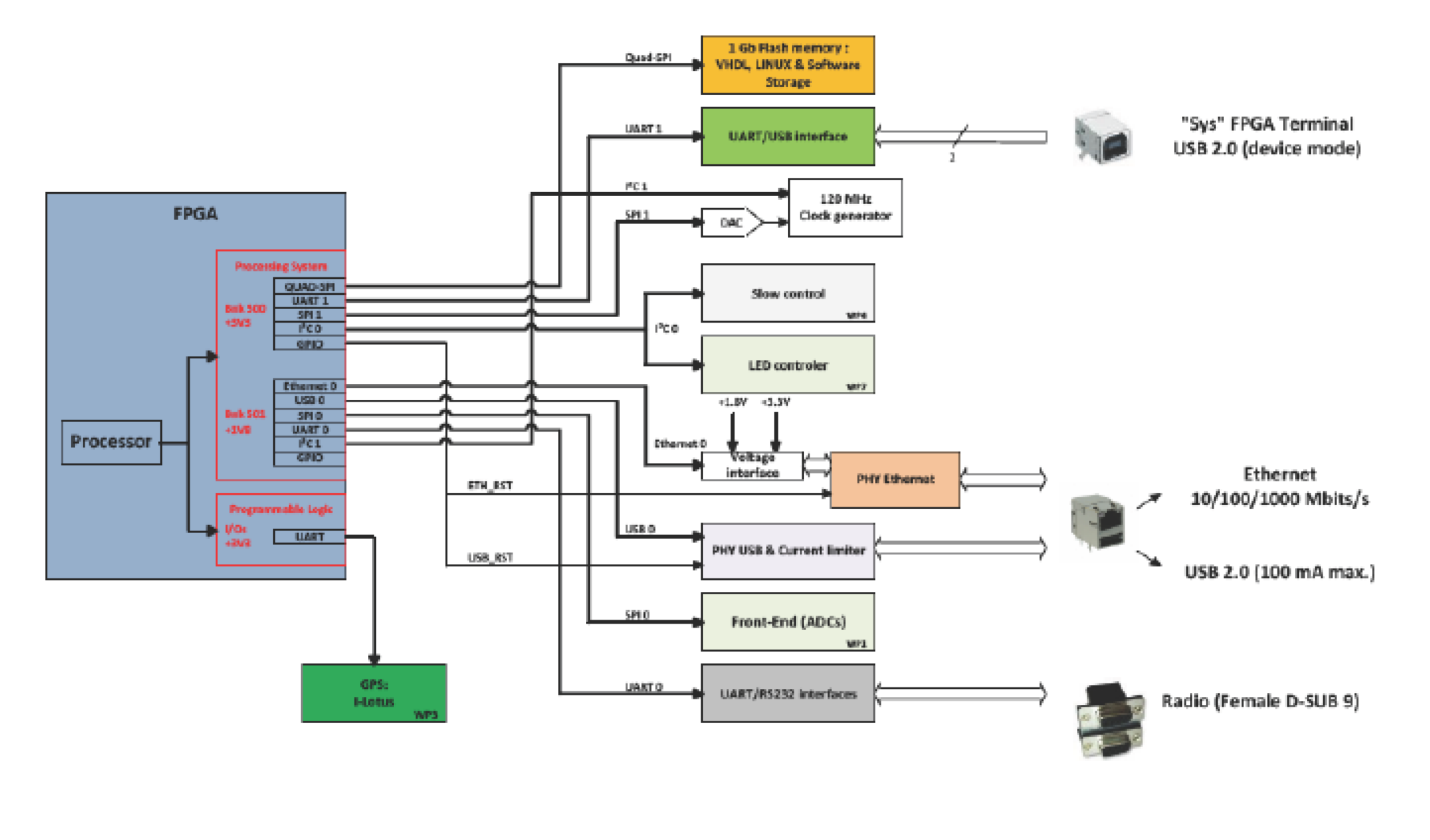}
\caption{Devices managed by the UUB processor.}
\label{fig:UUB_devices}
\end{figure}

The processor manages several devices including front-end electronics,
slow control, LED controller, GPS receiver, clock generator, memory
and various connectors (Fig. \ref{fig:UUB_devices}). The SSD PMT
signal will be connected like the other PMTs by using SMA
connectors. Two digital connectors are provided for future
additional detectors. These connectors provide 8 differential lines,
each of which can be individually defined as input or output in the
FPGA. An example of such allocation could be: Trigger out, Clock out,
PPS out, Busy in, Data in, Sync in, Data out, Sync out, etc.
Moreover, this connector will provide unregulated +24\,V, switched,
limited, with a current monitor.  The addition of accessible trigger
IN/OUT and GPS 1\,PPS signals will simplify time synchronization with
future  detectors. High speed USB interfaces and direct
connection to the trigger FPGA will allow interfacing a variety of
additional detectors.

The currently estimated peak power consumption (including radio and
PMTs) is about 16\,W. This is similar to the peak power of the current
electronics running with an average power below 10\,W. More accurate
estimation of the power consumption will be done with the integrated
prototype.

The design will be implemented on a 10 layer PCB board having the same
size as the current UB board ($340\times240\times1.8$\,mm). Global
specifications for the components are: availability until 2020,
operating temperature range from $-20$ to $+70^\circ$C, and preferably
SMD packaging. A conformal coating will be used to protect the UUB
board.  The UUB will be installed in the current RF-enclosure, only
the front panel will be changed. This will allow a smooth mechanical
integration of the electronics kit between the radio and the tank
power control board under the current weather enclosure.

\subsection{Local trigger and Data Acquisition System}
\label{sec:LocalStationSoftware}

The existing UB software will be ported to LINUX and will be
implemented in the FPGA.  The data acquisition will be simplified by
extending the use of FPGA firmware. The trigger and time tagging
functionalities will also be implemented in the FPGA.  The speed of
the upgraded CPU will be ${>}10$ times faster than the current one,
with a commensurate increase in memory. This will allow much more
sophisticated processing in the local station.

The current local triggers (threshold trigger, time-over-threshold
trigger (ToT), multiplicity of positive steps (MoPS) trigger, etc.) will
be adapted to the 120\,MHz sampling rate. The increased local
processing capabilities will allow new triggers to be implemented such as
asymmetry based triggers, and combined SSD and WCD triggers. 
The current muon memories and scalers will be retained.  
GPS synchronized LED pulses will be implemented which will improve calibration and monitoring of the detectors.
The trigger scheme includes the
ability to downsample and filter the data to the current 40 MHz rate
which will facilitate the detectors to run with new electronics emulating
the current detectors. This will allow deployment of new electronics
during the maintenance of the current system without disturbance to the
data taking.

\begin{figure}[!ht]
\centering
\includegraphics[width=0.8\textwidth,angle=0]{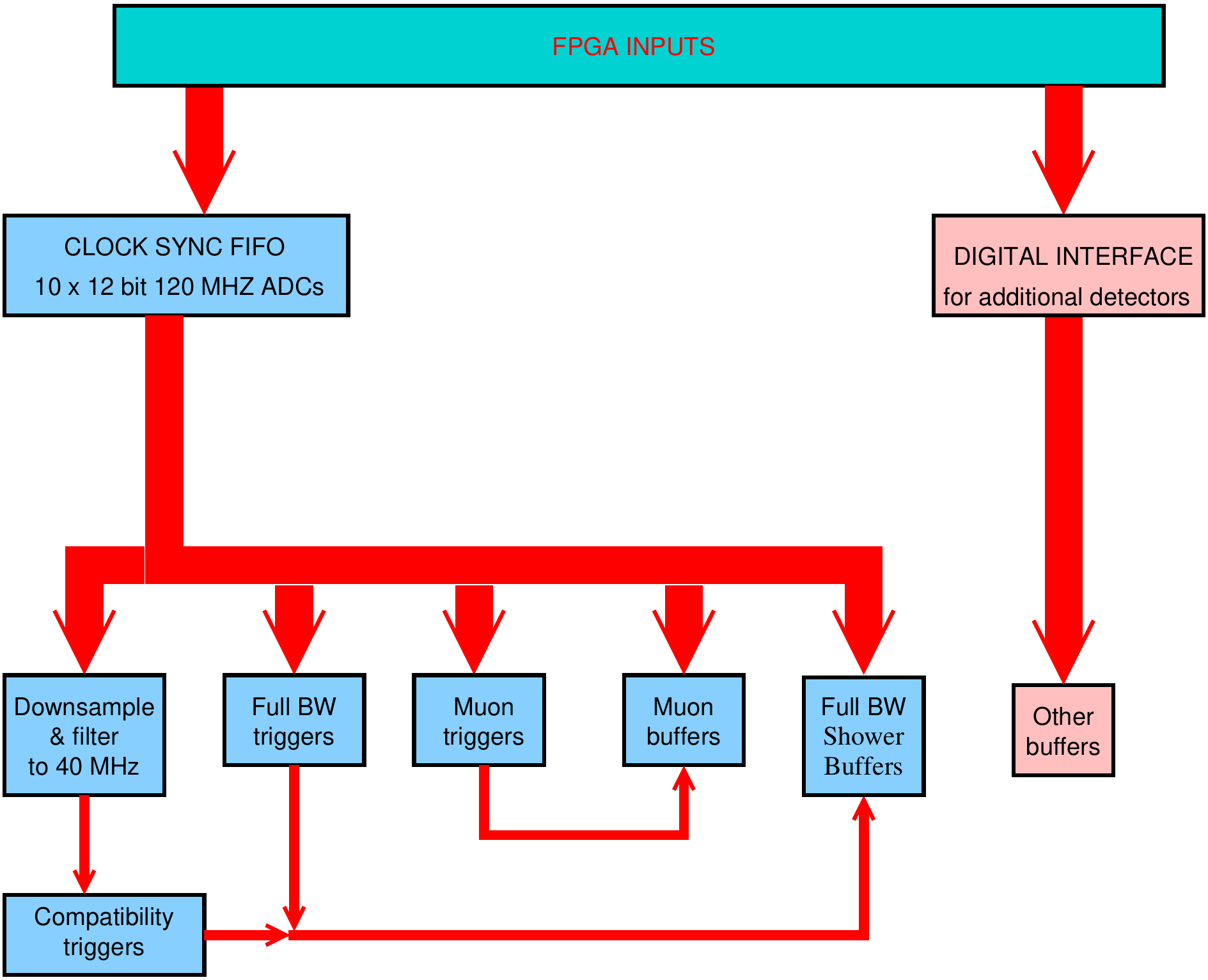}
\caption{Conceptual diagram for local trigger.}
\label{fig:trigger_level}
\end{figure}

\subsection{Manufacturing and tests}

The upgraded electronics will be fabricated and tested following the
Quality Management Plan \cite{SDE_QMP} of the current electronics.
The specific design of the different parts of the SDEU (front-end,
slow control, UUB) have been verified by using evaluation boards. The
first integrated prototypes are currently being fabricated and will be
tested in laboratories. The final validation of the
design will be performed by an Engineering Array of 10 detector
stations on site. This array will allow us to test triggers, validate
various resolutions and test the local data processing and
transmission.

Currently 4 manufacturing sites are foreseen (two in Europe, one in
the USA and one in South America). All fabrication sites need to be
ISO certified. For each production site, a pre-production run of
about 30 to 50 units is planned. This will allow us to
fine-tune the production parameters prior to the final production run.
Component procurement will be centralized and controlled. Only one
manufacturer is foreseen for the PCB procurement.

Each manufacturer will perform electric continuity tests and some
simple functionality tests. The complete functionality tests including
temperature stress testing will be done afterwards in laboratories. A
specific test bench has been developed for this purpose. A
description of the test procedure can be found in
ref. \cite{SDE_test_procedure}. All test results will be stored in a
database. Tested UUB boards will be shipped on site where they will
be assembled into electronics kits (Ekits). The procured GPS receivers
will be tested following similar test procedure and shipped to the site
for assembly. For details of assembly and tests on site see Chapter~\ref{chap:assemblyTests}.

%% file: sdeuperformance.tex

\section{Surface Detector expected performance}

\subsection{Increased dynamic range}

\label{sd-dynamics}

\begin{figure}[!ht]
\centering
\includegraphics[width=.7\textwidth,angle=0]{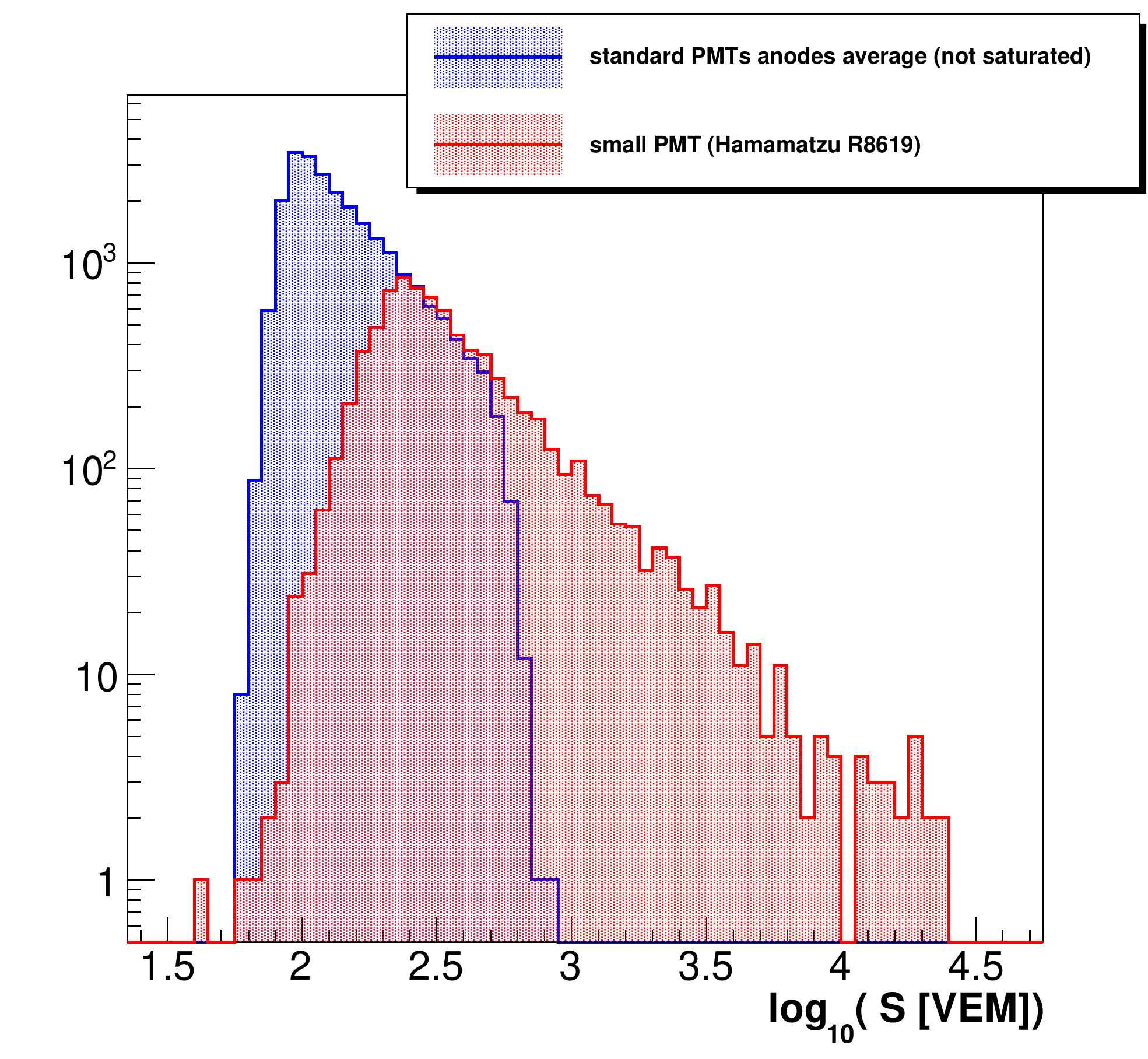} 
\caption{Comparison of the spectra measured by the standard PMT and the small PMT. The small PMT allows to extend the dynamic range above 30\,kVEM.}
\label{fig:WCDspectrum}
\end{figure}

The Auger Upgrade will allow extension of the dynamic range of the SD
enough to measure shower properties as close as about 300m from the core,
both with the WCD and the SSD (see Sec.~\ref{sec:dynamic-resolution-ldf}).
This is achieved in the WCD with the addition of the SPMT, dedicated
to the readout of large signals and providing considerable margin
to extend the dynamics even more. As discussed in Sec.~\ref{sec:pmt},
the SPMT gain and its overlap with the LPMT signals can be modified,
and the overall dynamic range can comfortably exceed 40\,kVEM.
This is shown in Figure~\ref{fig:WCDspectrum}, where the VEM spectrum
of shower particles is shown, measured by a test WCD equipped with a SPMT operated
at a signal ratio of ${\sim}13$ with respect to the anode.
The saturation of the LPMT is clearly visible at the expected value
of ${~\sim}600$ VEMs, while the overall spectrum of the SPMT extends
well beyond 40\,kVEM.

Currently half of events in the energy range $10^{19.5}$ to $10^{19.6}$\,eV
have at least one saturated station. The SDE upgrade extends the linear
non-saturated measurement range to 32 times larger signals than currently
achieved. This reduces the fraction of events in that energy range
that have a saturated station by a factor of 10. Furthermore, the
increased dynamic range will allow measurement of the LDF function
at distances closer to the shower core than is now possible.

In the SSD a single light sensor is used measuring both single particles
for calibration and high density CR shower signals. This is achieved
by taking advantage of the very high-linearity PMT chosen as baseline
for the SSD. This will allow measurement with both WCD and SSD detectors
at distances as close as about 300 m from the shower core.

\subsection{Faster timing and increased processing capability}

The typical time distribution asymmetry between the PMTs of the WCD is of the order of 6\,ns (light transit time between the PMTs).
With the current electronics time bin width of 25\,ns
and $\sigma=\unit[7.2]{ns}$ it is difficult to extract much information
from the time distribution asymmetries between the PMTs. However,
with 8.3\,ns bin width and $\sigma=\unit[2.4]{ns}$
it becomes possible to extract some directional information on a station
by station basis, and to consider improved triggers for horizontal
showers that take advantage of the arrival time differences in the
PMTs.

While the SSD, in combination with the WCD, is required to accurately determine the muon content
of the showers, the existing detector stations can already count muons
in the regions far enough from the shower core where the interval
between muons is more than a few FADC time bins. The various counting
techniques that have been used rely on sensing the fast upward transitions
in the FADC traces. Increasing the FADC sampling rate improves these
techniques, allowing one to more reliably count larger numbers of
muons per station and probe closer to the shower core.

Our current photon limits are no longer background free. Improved
muon discrimination provided by the faster sampling will help reduce
the background of hadronic events in our photon candidate sample.

Substantial additional resources available in the trigger FPGA will
also provide the possibility of more sophisticated triggers tuned
for photon and neutrino searches and may allow us to extend the energy
range to lower energies, e.g.~the expected number of events seen
in the detector given the flux prediction by Waxman and Bahcall is
improved by about 10\% while only readjusting the original trigger
condition to the increased sampling rate.

As the SSD and WCD signals are both provided to the same FPGA, it
also becomes possible to implement dedicated SSD triggers or combined
SSD and WCD triggers. The combination of these triggers may help to
further monitor and understand efficiency and biases of the triggers.

As the PMT signals are AC coupled into the FADCs, the baseline level
can fluctuate on time scales of milliseconds due to the preceding
signals. The current low threshold triggers (ToT, ToTd), are set at
a per bin threshold level of 0.2 VEM. The VEM peak is nominally set
to be 50 counts above baseline, but some in PMTs it may go 
as low as 20-30 counts, before a HV adjustment is made. Thus the
per bin trigger threshold may correspond to only 5 or 6 counts above
baseline, which results in significant trigger rate fluctuations for
those triggers. Baseline tracking by the FPGA has already been tested
in the current electronics, but the coarse trigger threshold granularity
precluded using that information to stabilize the trigger rates. The
addition of 2 more bits of precision in setting the trigger threshold
in the SDEU resolves that issue, and will allow the FPGA to dynamically
track the baseline and make the corresponding trigger threshold adjustment
on the timescale of $\approx100\, \mu \mathrm{s}$, reducing the trigger
rate fluctuations. 

%% file: underground_muon_detector.tex

\chapter{The Underground Muon Detector}
\label{sec:UMD}
\section{Introduction and design objectives}
The Underground Muon Detector (UMD) will provide a direct measurement of
the muon content of a sub-sample of showers observed by the upgraded Auger surface detector.
In the Upgrade plan this serves
as verification and fine-tuning of the methods used to extract shower
muon content using the SSD and WCD stations.  As described
in Sect.~\ref{sec:umdPerformance}, the performance and characteristics of the AMIGA
underground muon detectors match these requirements.  The UMD
will therefore consist of 61 AMIGA muon detectors deployed on a 750\,m
grid in the infill area of the Surface
Detector, instrumenting a total area of 23.5 km$^2$.

The use of the AMIGA muon detectors in this verification role is additional
to the rich physics investigations planned for AMIGA in the ankle-region
of the energy spectrum~\cite{Suarez-ICRC:2013}.

The following sections provide an overview of the design
and the implementation of the AMIGA muon detectors, with more detailed
information given in Appendix~\ref{sec:amiga}.

\section{Detector design}
Each muon detector station will have an area of 30\,m$^2$ and will be buried
at the side of a surface detector station at a depth of approximately 1.3\,m. The
distance to the station will be large enough to avoid shadowing from
the water tank, guaranteeing uniform shielding, but small enough to
represent the same physical point inside the shower front, and allowing shared use of
GPS time signals and telecommunications with its
associated surface detector.

The baseline design for the muon detectors uses the same extruded plastic
scintillators already developed and used by the MINOS experiment. They
will work as counters (i.e. signals above a tunable threshold are
counted) having an appropriate segmentation to prevent pile-up, together
with an integrated signal for large muon densities.
The light
collected in a scintillator will be guided towards a 64 channel
Hamamatsu PMT.  

\begin{figure}[!ht]
\centering
\includegraphics[width=.475\textwidth,angle=0]{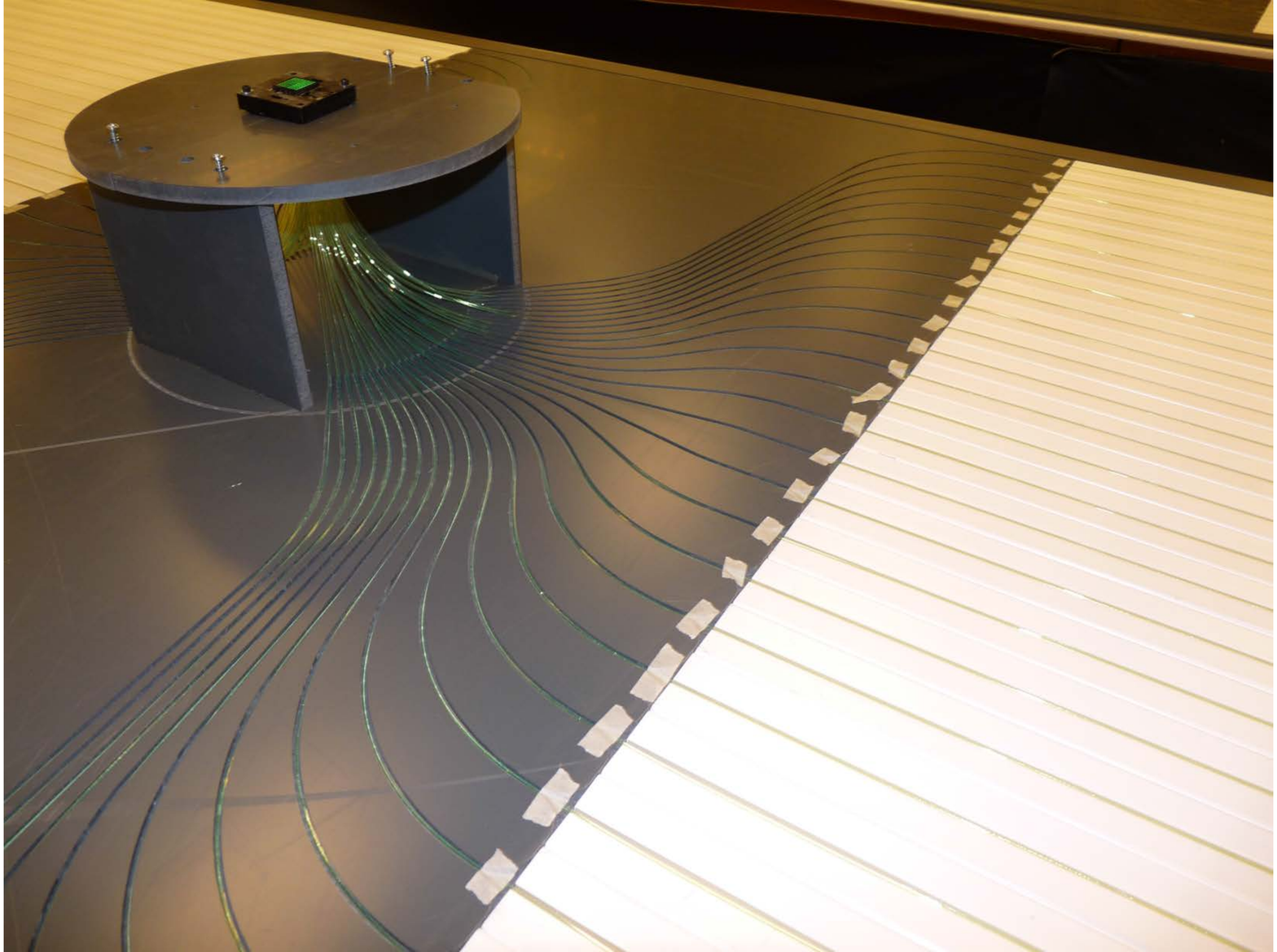}
\includegraphics[width=.45\textwidth,angle=0]{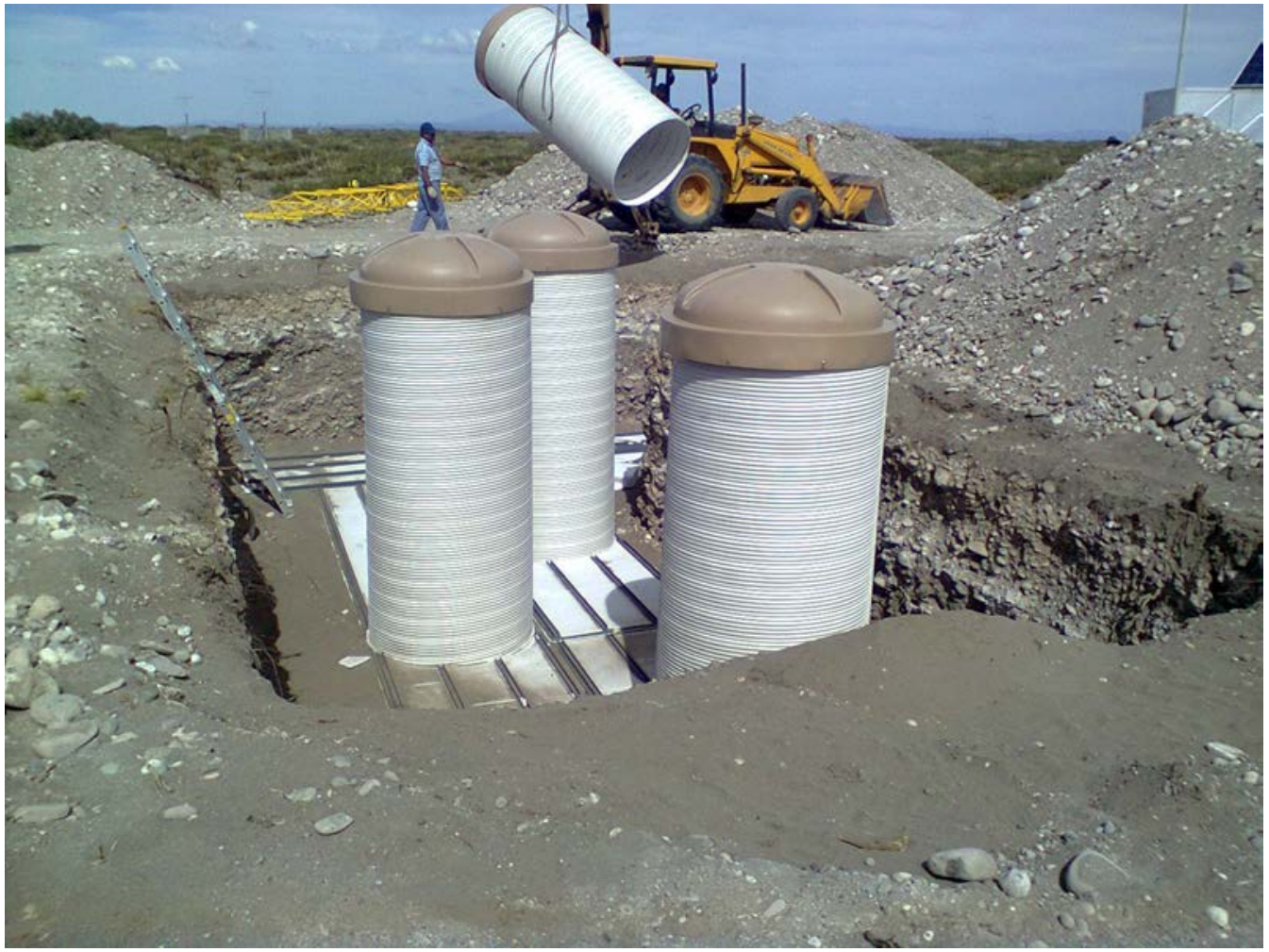}
\caption{Left: An AMIGA 10\,m$^2$ module being manufactured. The black
  optical connector concentrates the 64 optical fibers coming from the
  32 scintillator bars at each side. Right: Deployment of a Unitary
  Cell AMIGA station with two 10\,m$^2$ and two 5\,m$^2$ modules at a depths of 2.3\,m
  underground. The big tubes are to provide access to the electronics for
  development and maintenance purposes, to be replaced by tubes of 31\,cm
  diameter for AMIGA production.}
\label{fig:AMIGA_counter}
\end{figure}

An AMIGA prototype is displayed in
Fig.~\ref{fig:AMIGA_counter}. It consists of 64 4\,m long scintillator strips. The
strips are 4.1\,cm wide and 1.0\,cm thick and therefore the
detector modules have an active area of 10\,m$^2$. 
The strips have a middle
groove which accommodates a wavelength shifter (WLS) fiber of 1.2\,mm
diameter which is glued into the groove and covered with reflective
foil. The scintillator is co-extruded with a TiO$_2$ reflective coating which
prevents light from leaving the material.

The fibers end at an optical connector matched and aligned to a 64
channel multi-anode photomultiplier tube.  Muon counters sample
scintillator signals at a frequency of 320\,MHz, meaning that every
3.125\,ns 64 bits are acquired.  Each bit stores the digitized value
(either a ``{\it 0}'' or ``{\it 1}'' if the signal was below or above
a predefined threshold) associated to one scintillator bar of the
counter. 

Muon counting and the digitization of the integrated signal is implemented
in the AMIGA electronics, including an FPGA with three main functional blocks:
counting, data codification, and external
communications. The data are sent to the CDAS using an independent,
commercial radio communications system.  The power for the counters is provided by a solar panel system similar to that used for the existing SD. The measured power consumption of one
detector station is currently about 37\,W.  For more details about the module design and deployment see Sec.~\ref{sec:receptionUMD} and Appendix~\ref{sec:amiga}.

\section{Implementation}

A unitary cell formed by 7 AMIGA  muon stations and 7 associated surface
detectors has been successfully installed in the field, with elements of
the cell taking data for up to two years.
The performance and  physics results from the unitary cell are very encouraging and the procedure for deploying underground detectors 
in the infill area has been verified. Some R\&D is still in progress,
mainly aimed at reducing costs and increasing the integration of muon detectors with
the SD. In particular, a common communications system is needed for the infill 
area detectors, which will be adapted either 
from the AMIGA or the AERA communication systems.
Furthermore, some R\&D is ongoing on the use of
silicon photomultipliers (SiPM) and electronics improvements. This would allow a reduction in the 
power consumption and the cost of the detectors.

%% file: fluorescence_detector.tex

\chapter{Extending the Duty Cycle of Fluorescence Detector Observations}
\chaptermark{Extending the Duty Cycle of FD Observations}
\label{FDextension}

The fluorescence detector provides exceptional information about extensive air showers
such as a model-independent energy reconstruction and direct measurement of the longitudinal
development profiles. The main limitation of the FD is its duty cycle, currently at the
level of $15\%$. Our goal is to increase the exposure for cosmic ray events above $10^{19}$\,eV
by extending the FD measurement into hours with high night sky background (NSB).
The current setup allows this novel operation and we have performed several tests
that successfully demonstrate that it is feasible.

Safety limits on the long and short term illumination of PMTs by the NSB, and particularly
scattered moonlight, define the data taking period of the current FD operation. The duty cycle
is therefore limited to about $19\%$, which is reduced to $15\%$ by bad weather conditions,
power cuts and malfunctions. A significant increase of the duty cycle is possible by the extension
of the FD operation to times at which a large fraction of the moon in the sky is illuminated. However, during
such operations the PMT gain must be reduced by lowering the supplied high voltage (HV) to avoid
an excessively high anode current leading to an irreversible deterioration of the PMT sensitivity.
The HV power supplies installed in the FD buildings allow switching between two high voltage levels and
the PMTs can be operated at the nominal gain and a lower gain.

\section{Laboratory test measurements}

The FD PMTs (Photonis, model XP~3062) have been extensively tested at the nominal~\cite{Becker:2007zza}
and lowered gain levels in the laboratory. The nominal PMT gain and HV are $5{\times}10^4$ and
850 to 1050\,V, respectively. A special test setup for the planned operation at higher NSB 
has been constructed. A uniform UV light source, which can be simultaneously operated
in the DC and pulsed mode, illuminates a few FD PMTs and the HEAT electronics is used
to read-out measured signals.

\begin{figure}[t]
\def\figh{0.31}
\centering
\includegraphics[height=\figh\textwidth]{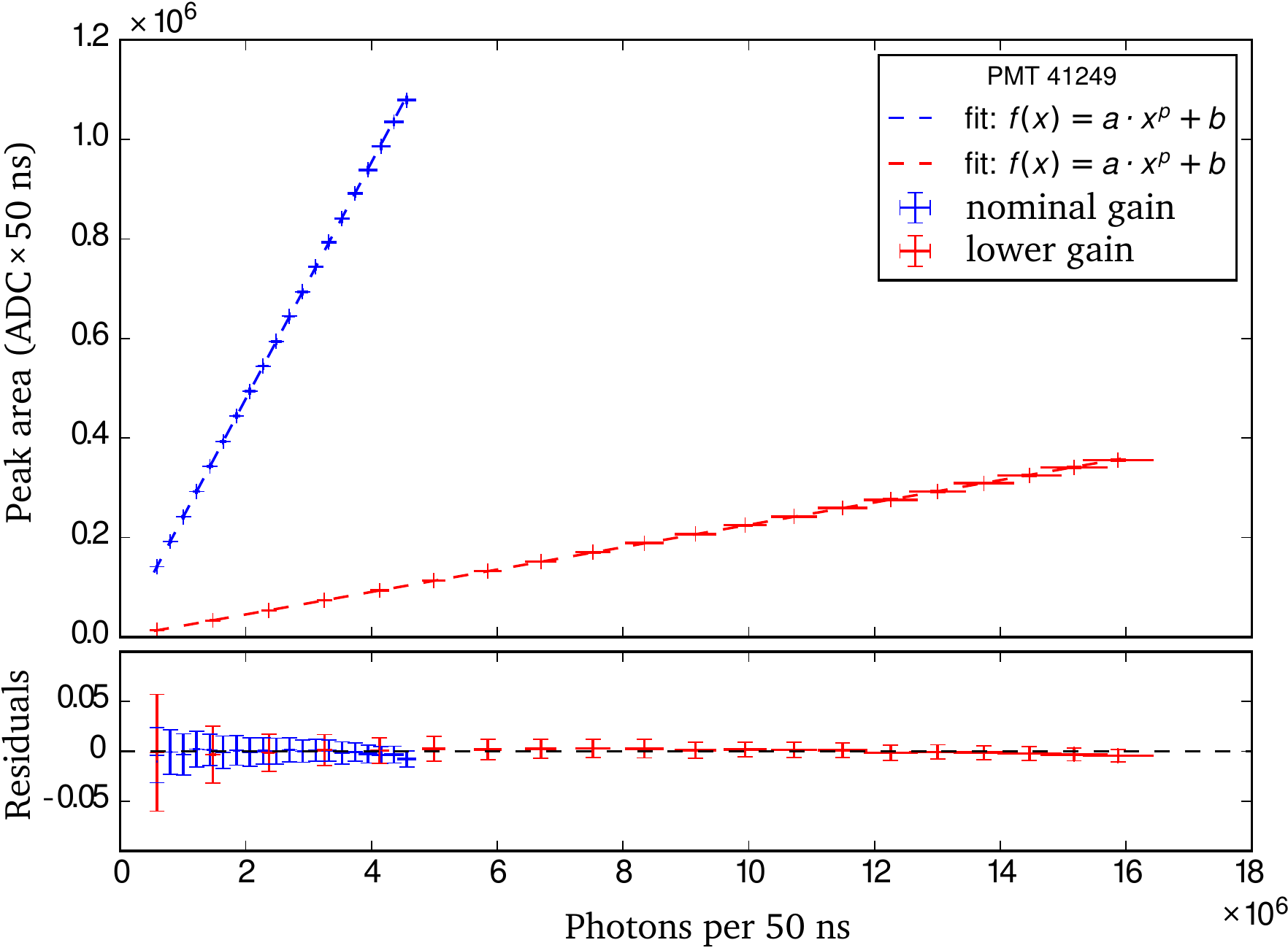}\hfill
\includegraphics[height=\figh\textwidth]{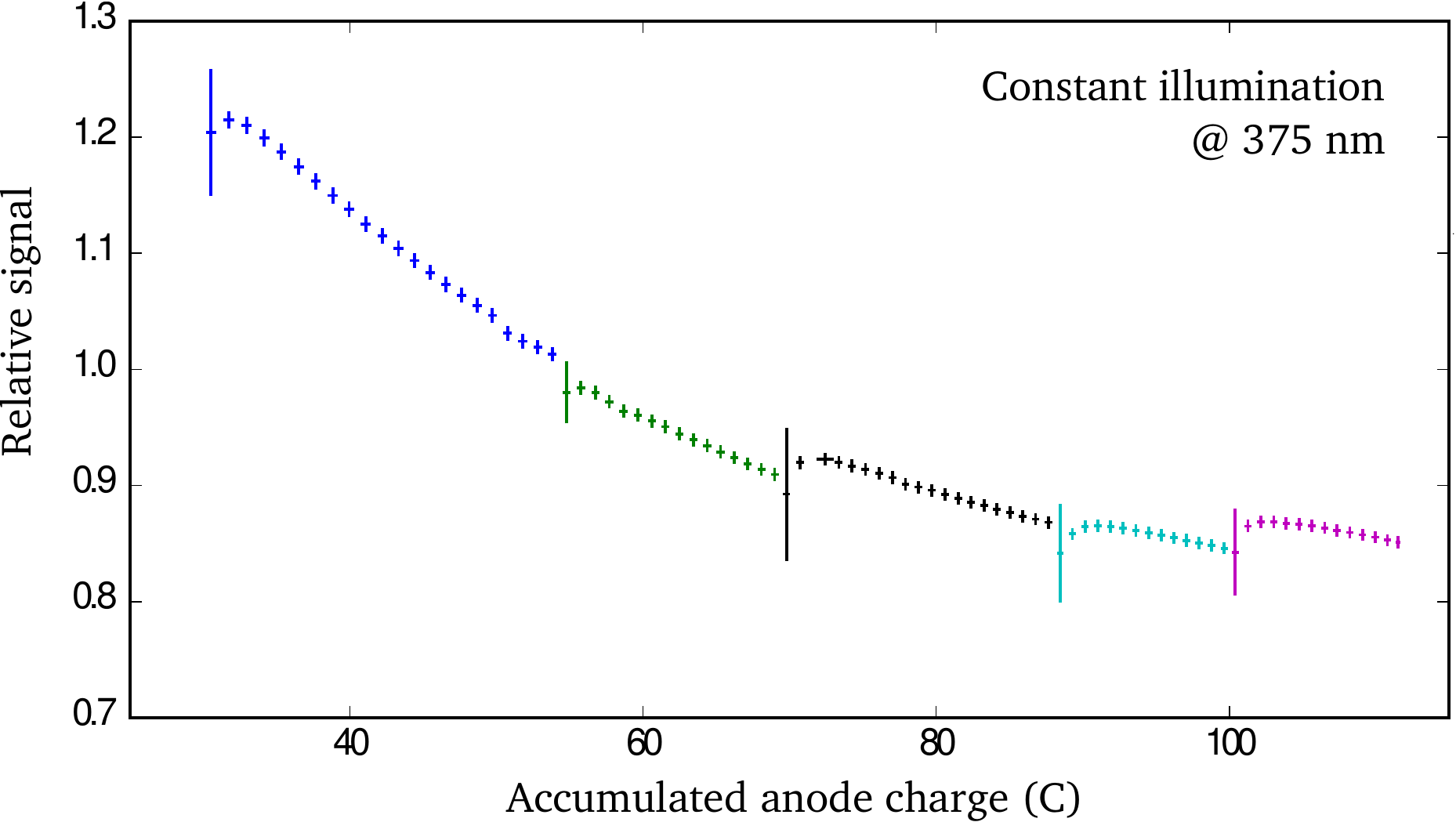}
\caption{
Left: The PMT response to photon flux up to the saturation level measured at the nominal
(blue) and reduced (red) gain. The exponent $p$ is $1.0$ for both gains. Larger error
bars at lower light fluxes are caused by the small amplitude of measured pulses.
Right: The PMT response as a function of the accumulated anode charge measured in a dark
box at ten times lower gain. Jumps correspond to a recovery phase after breaks in our
measurement.
}
\label{fig:FDext_tests}
\end{figure}

The PMTs were tested at HV as low as $400$\,V and their gain changes were measured
as a function
of HV over the whole studied range of HV. The PMT response to different light
fluxes at different gain levels is a linear function of the light flux as is shown in
Fig.~\ref{fig:FDext_tests}. The PMT aging (i.e.\ loss of sensitivity) was also studied
at the lower gain and results for a PMT at a gain of $5{\times}10^3$ (HV of $644$\,V) are shown
in Fig.~\ref{fig:FDext_tests}. After the initial
aging phase the PMT response changes very slowly as has also been measured for PMTs
operated at the nominal gain.

A continuous change of the background light level was used to simulate the expected evolution
of the NSB during nights with a highly illuminated moon moving across the sky. The PMT
gain was automatically changed according to the measured light flux. Because of the AC-coupling of the read-out electronics,
the DC light level can only be obtained from the variances of the measured signal. At a
gain as low as $4000$ the ADC variance is above the electronic noise level and it can provide
information on the DC light level. The DC light level obtained from the PMT has been cross
checked with a photodiode.

Thus, we have verified that the PMTs can be operated at more than ten times lower gain
than the nominal value of $5{\times}10^4$, and that no acceleration of the aging has been detected
at lower gains with high NSB. Moreover, the measurement of the NSB by the FD PMTs
is possible using variances even at lower gain, which is a necessary requirement for automatic
changes of the HV level in the telescopes.

\section{Test measurement with an FD telescope}

The first test measurement outside the standard FD data taking period was performed
with one FD telescope in the austral autumn of 2015~\cite{Zorn2015}.
Telescope 1 at Los Leones was operated
during six nights with a highly illuminated moon above the horizon. We have confirmed
that the HV change can be done remotely and that the HV stabilizes within a few seconds after
its change.

\begin{figure}[t]
\def\figh{0.33}
\centering
\includegraphics[height=\figh\textwidth]{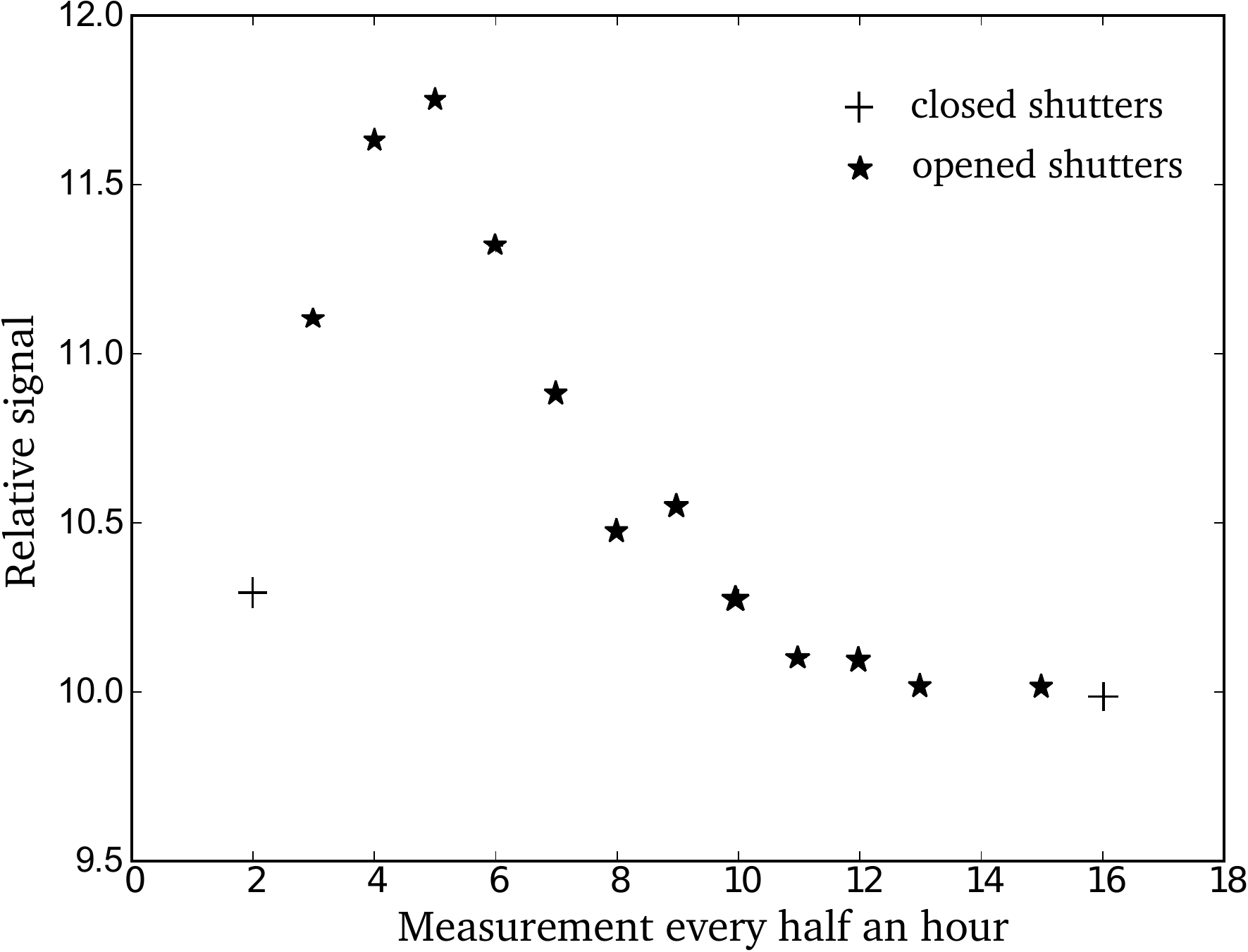}\hfill
\includegraphics[height=\figh\textwidth]{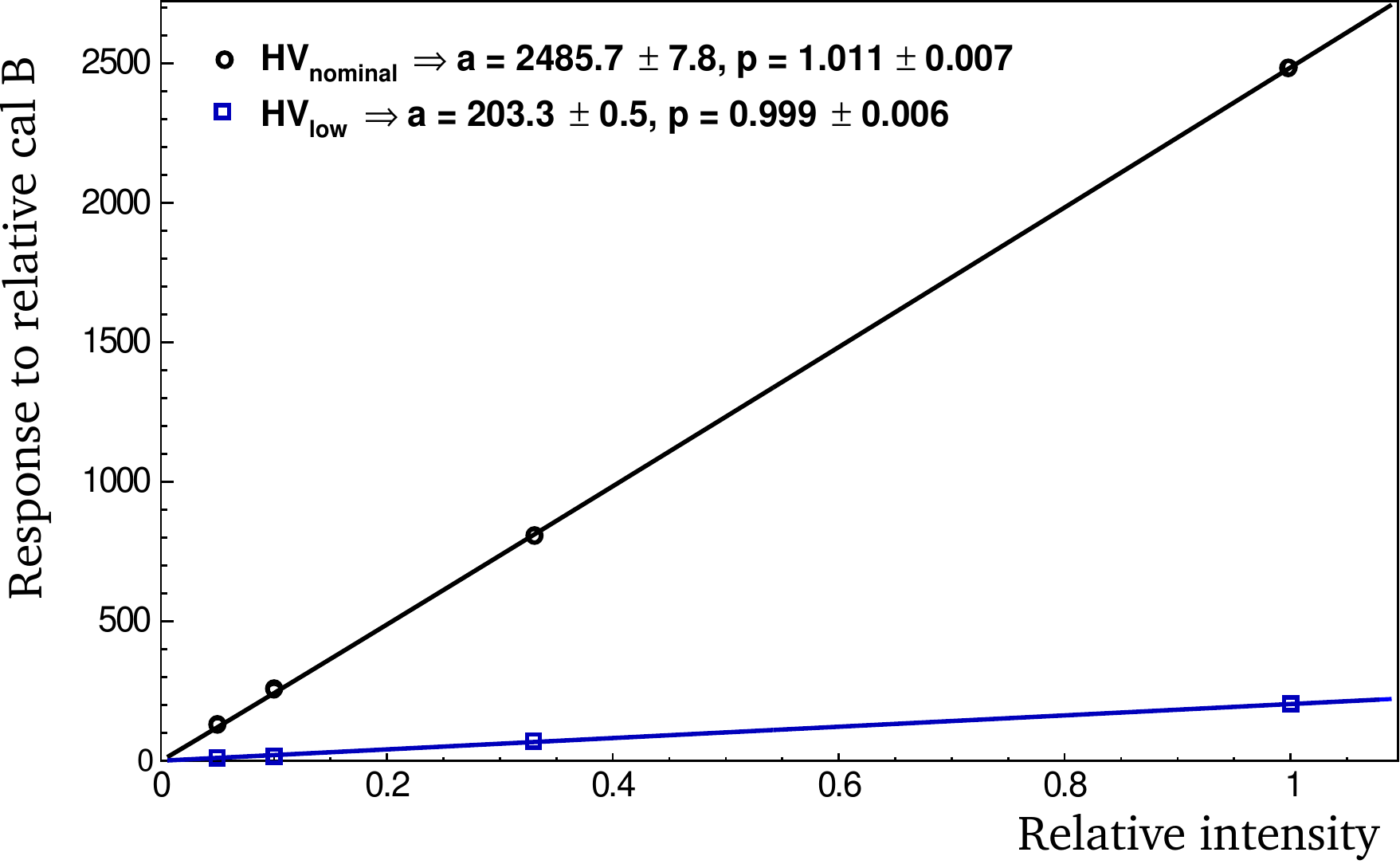}
\caption{
Left:
The response of the PMTs operated at reduced gain during data taking.
Shown is the ratio of the response measured before data taking at the nominal gain,
to the response during data taking.
Crosses and stars indicate closed and open shutters, respectively.
Right:
The camera-averaged response to calibration B as a function of calibration B intensity
for nominal HV (black circles) and reduced HV (blue squares).
Results of both HV settings are fitted with a power law and in both cases the exponent
$p$ is compatible with 1 (i.e.\ a linear fit).
}
\label{fig:FDext_calib}
\end{figure}

The PMT performance can be monitored with the existing calibration setup in exactly the
same way as during the standard data taking, so no change in the system is required.
The PMT performance obtained for one test night is shown in Fig.~\ref{fig:FDext_calib}.
An evolution of the PMT response studied every half an hour is shown in the left panel.
We can see a fast change of the PMT response during the three hours after opening the shutters.
A similar evolution of the PMT response has been observed during standard data taking
at the nominal gain. In the right panel is the
camera-averaged response to four light intensities measured at two gain settings.
A linear fit is compatible with the data within uncertainties.

During these test runs a couple of air showers were measured despite a NSB ten times
higher than is normal during the standard FD operation. The trigger rate was lower,
but this can be explained by a lack of low energy events buried in the higher NSB.

We will continue test measurements during 2015 in order to study the FD
performance in greater detail. We plan to also use a roving laser, which is available
at the Observatory, in front of the telescope.

\section{Air showers measured during high night sky background}

\begin{figure}[t]
\def\figh{0.55}
\centering
\includegraphics[height=\figh\textwidth]{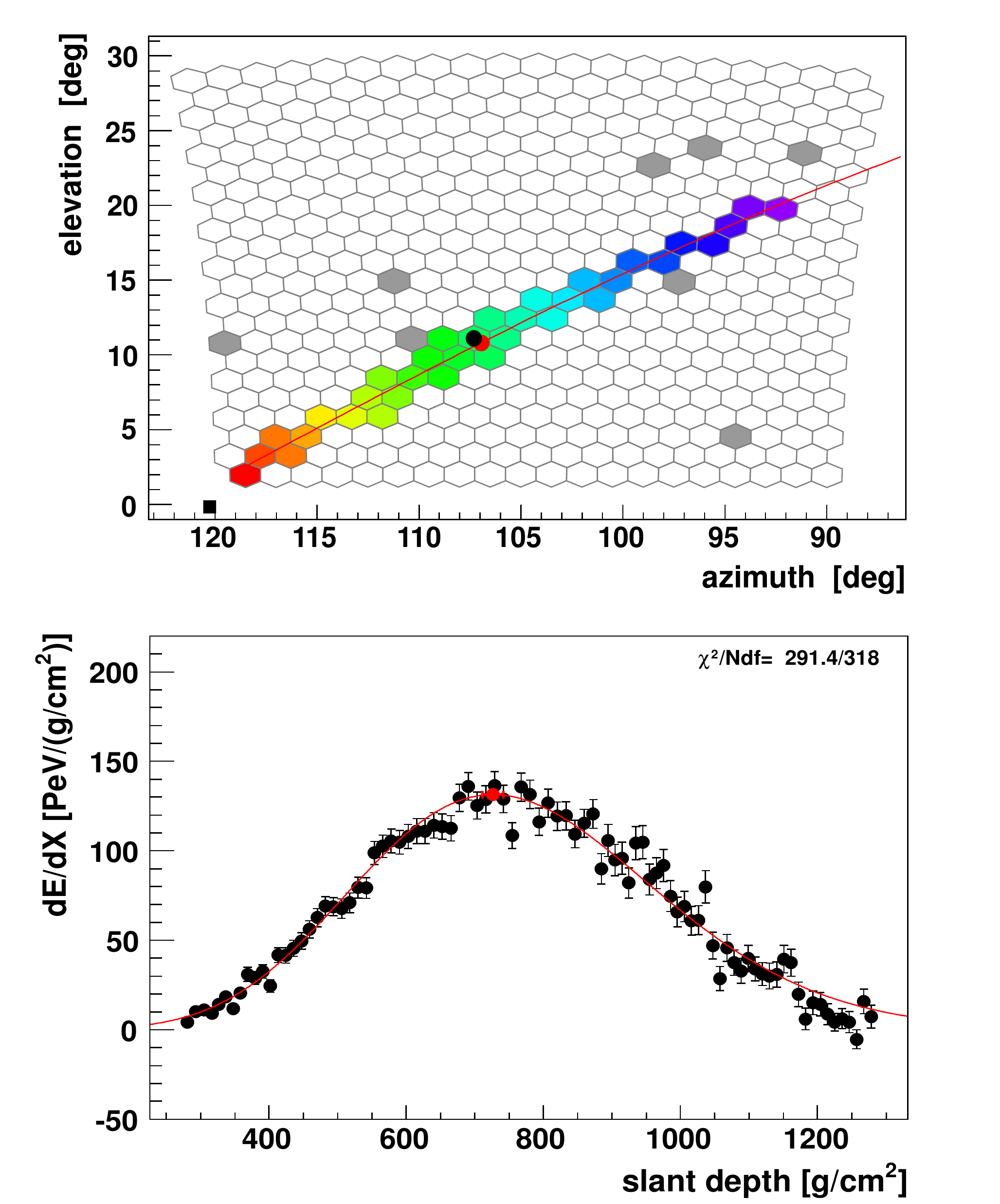}
\includegraphics[height=\figh\textwidth]{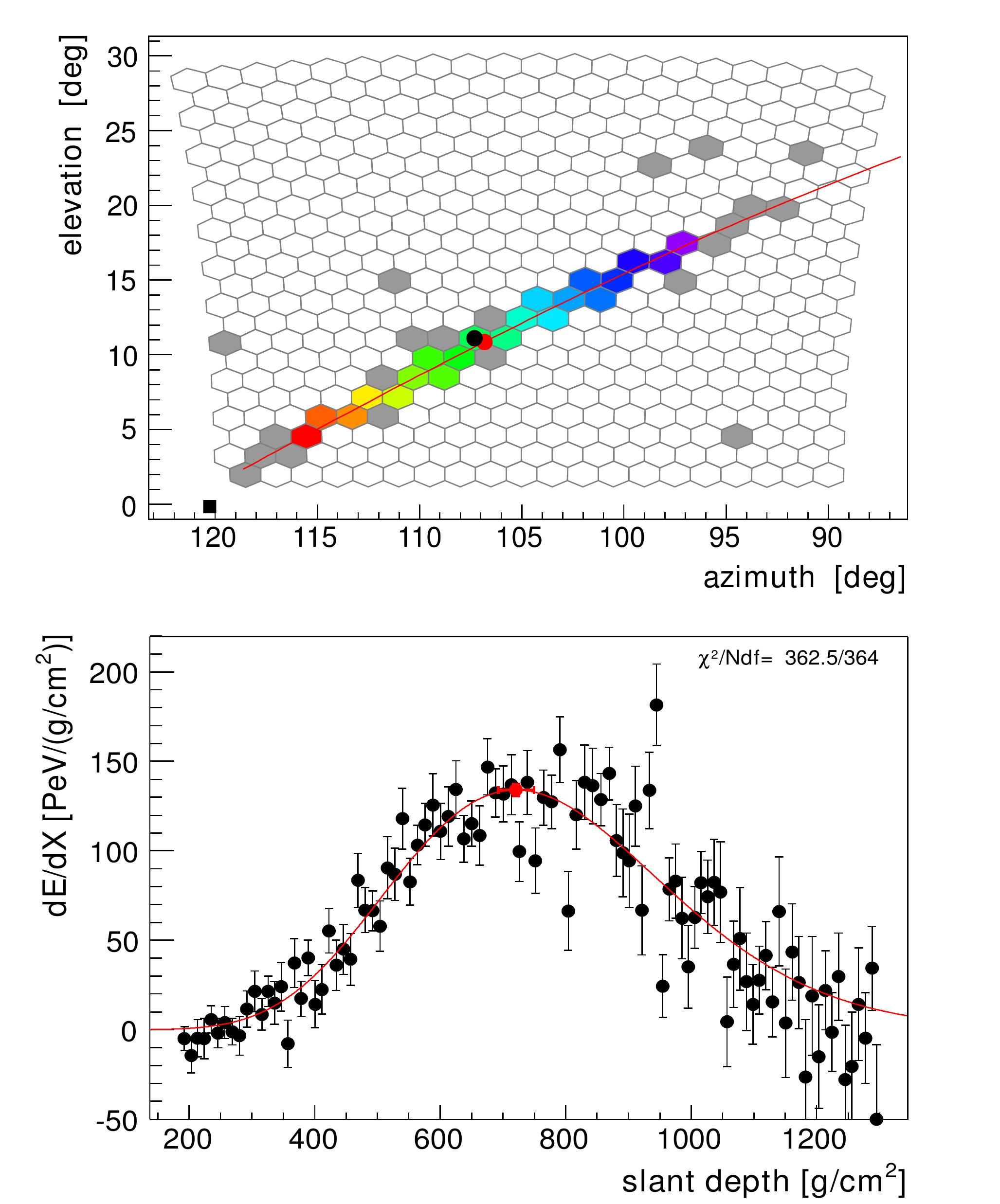}
\caption{
A real FD event with reconstructed energy $7{\times}10^{19}$\,eV. In the left panel are measured data
(clear sky and no scattered moonlight, a baseline variance of $25$\,(ADC counts)$^2$) and in the right panel the same
data after adding random noise corresponding to a $40$ times higher NSB.
}
\label{fig:FDext_RealEvent}
\end{figure}

The effect of a higher NSB on the reconstruction of air showers has been also studied.
Existing measured air showers have been analyzed with the standard reconstruction chain after
adding random noise to the ADC traces. A camera image and a longitudinal profile of
a real FD event before and after the addition of noise is shown in Fig.~\ref{fig:FDext_RealEvent}.

\begin{figure}[t]
\centering
\includegraphics[width=0.6\textwidth]{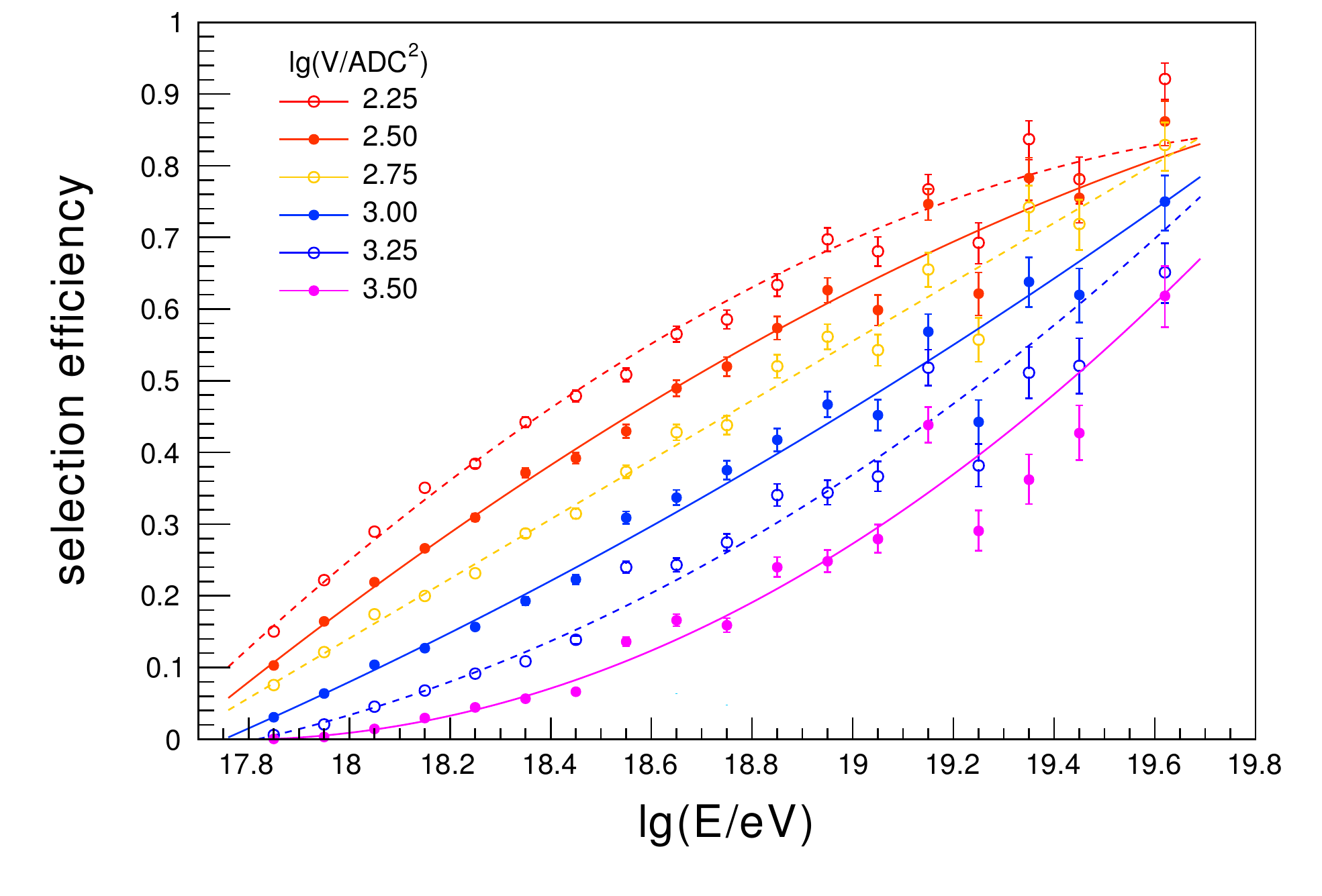}
\caption{
The combined reconstruction and selection efficiency as function of the energy for various levels of the NSB.
The NSB is expressed as the variance of the PMT baseline signal $V$ in (ADC counts)$^2$
and all studied levels are above the current
limit for the standard FD operation which is less than $100$\,(ADC counts)$^2$. The level
of $1000$\,(ADC counts)$^2$ corresponds to the NSB in the presence of a $90\%$ illuminated moon.}
\label{fig:FDext_SelectionEfficiency}
\end{figure}

\begin{figure}[t]
\def\figw{0.48}
\centering
\includegraphics[width=\figw\textwidth]{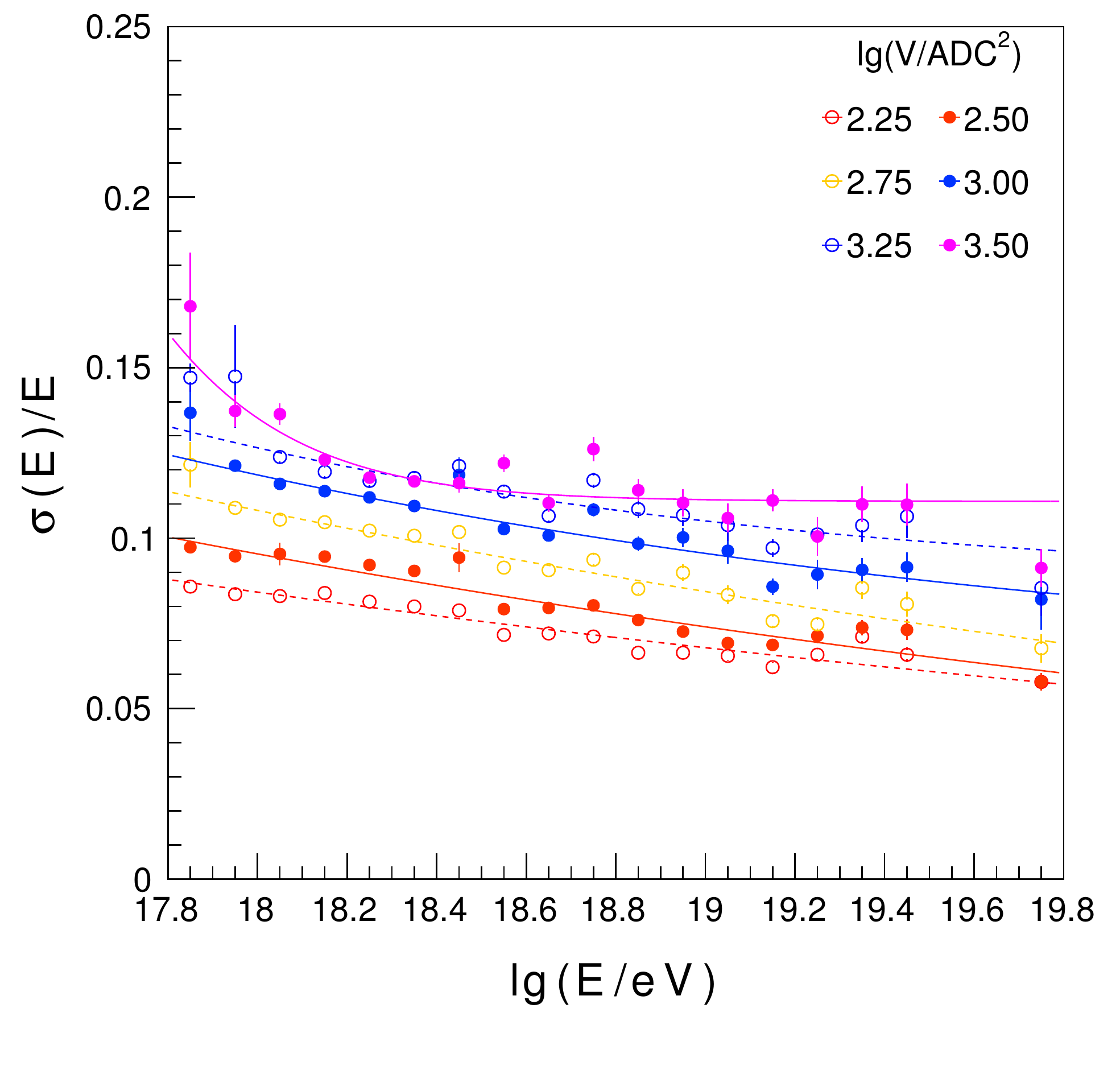}\hfill
\includegraphics[width=\figw\textwidth]{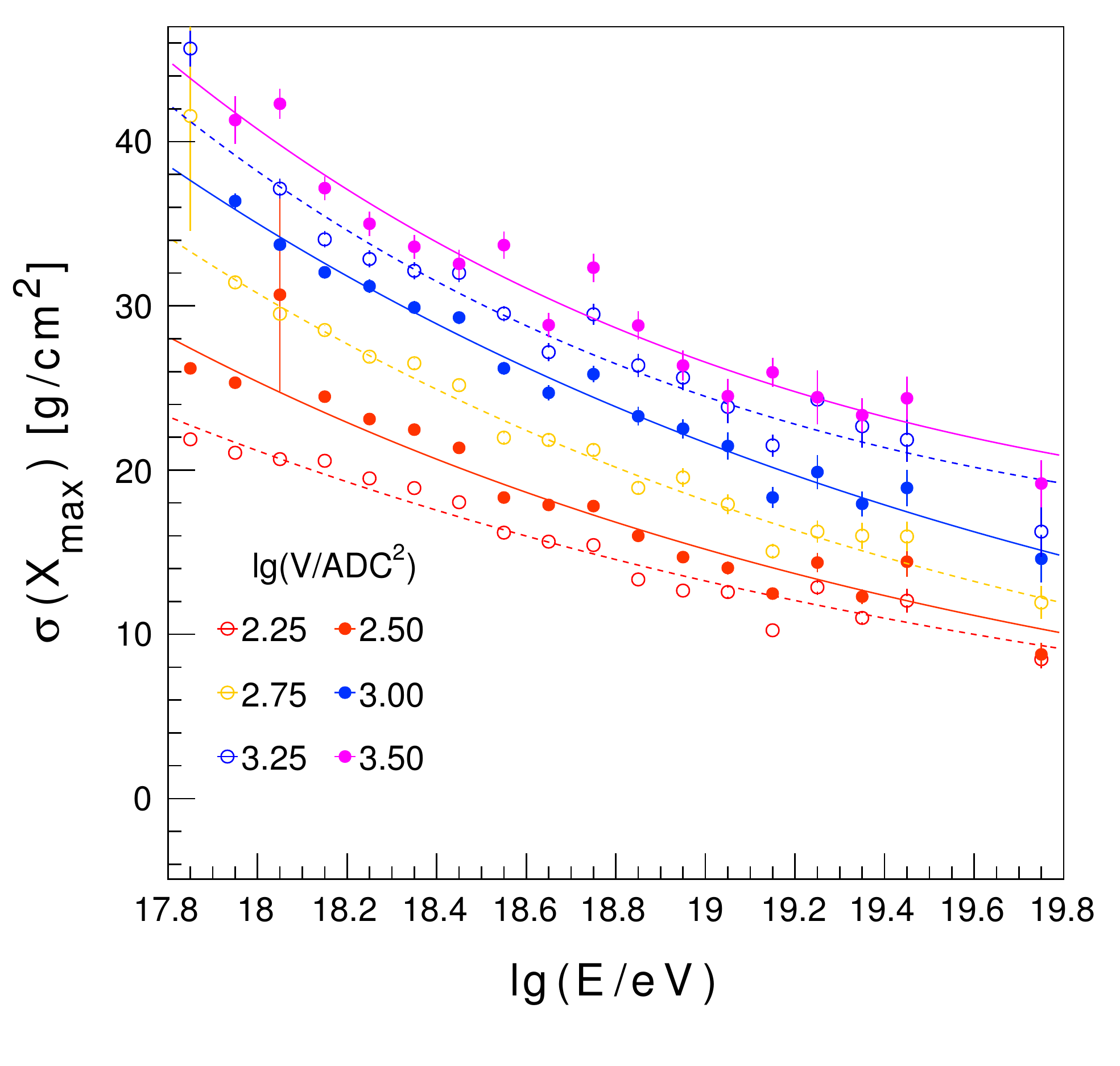}
\caption{
The energy (left) and $X_{\text{max}}$ (right) resolution as function of energy for
various NSB levels. The NSB levels are explained in fig.~\ref{fig:FDext_SelectionEfficiency}.}
\label{fig:FDext_resolutions}
\end{figure}

The reconstruction and selection efficiency, resolutions and biases for various NSB levels
have been evaluated. For this study, the noise in ADC traces of events in our current FD data
set was artificially increased. Modified events were reconstructed in the same way as 
in the standard FD reconstruction and the same selection criteria were applied.
In Fig.~\ref{fig:FDext_SelectionEfficiency} we show that the reconstruction and selection efficiency increases with energy and is $75\%$ at $10^{19.6}$\,eV
even for the most extreme NSB considered for the duty cycle extension (a variance of $1000$\,(ADC counts)$^2$).
In Fig.~\ref{fig:FDext_resolutions} we show the resolution for energy and $X_{\text{max}}$.
Our results show that good quality data can be obtained for air showers above $10^{19}$\,eV
measured in the presence of a NSB which is ten times higher than the maximum level allowed
for the standard FD data taking.

Even though the PMT electronics is AC-coupled, we have studied a possible effect of
a non-uniform NSB on the air shower reconstruction. The non-uniform NSB is caused
by a highly illuminated moon shining in the sky close to the field of view of the
detector.
For an extreme gradient of the NSB the reconstruction biases in energy and $X_{\text{max}}$
were found not to exceed $2\%$ and $5$\,g/cm$^2$, respectively, at energies above
$10^{19}$\,eV. Such biases can be considered acceptable taking into account that the events
observed in the presence of the extreme NSB gradient will not dominate the overall data
sample.

\section{Increase of the duty cycle}

We are currently considering a PMT gain ten times lower than the nominal one for
the FD duty cycle extension. We have verified in all our test measurements that the PMTs
operated at reduced gain satisfy the criteria required for the FD performance (such as
a linearity, stability and lifetime).
At a gain of $5{\times}10^3$ the FD operation can be extended to nights with a moon
fraction of $90\%$, where we expect ADC baseline variances of about $1000$\,(ADC counts)$^2$.
This translates to about $29\%$ FD duty cycle (without accounting for reductions caused by bad
weather conditions and malfunctions), or in other words, a $50\%$ increase of the current
observation period of $19\%$.

Expected number of events for seven years are shown in Table~\ref{tab:FDstat}: for the standard
FD operation we can expect about $(514\pm27)$ and ($52\pm9$) events above $10^{19}$\,eV and $10^{19.5}$\,eV,
respectively. By extending the FD operation to higher NSB we can gain up to $40\%$ more
events above $10^{19.5}$\,eV, see Fig.~\ref{fig:FDext_SelectionEfficiency}. Moreover, by applying
less strict selection cuts for high quality events \cite{Aab:2014kda}, we expect in total about
$102\pm14$ events above $10^{19.5}$\,eV after including the standard and extended FD operation.
This will allow us to improve the cross check of results of the upgraded SD array with
the FD up to the energy of the flux suppression in the cosmic ray flux.

\begin{table}[t]
\centering
\caption{Expected cumulative number of events for a data taking period from 2018 until the end of 2024
  for the FD measurement during the standard operation only, after including the extended operation
  and after applying less strict selection criteria for both operations.
  Compare with the SD Table~\ref{tab:stat}.}
\label{tab:FDstat}
\begin{tabular}{rrrr}
\toprule
$\log_{10}(E/\unit{eV})$ & $\left.N\right|_\textbf{std}$ & $\left.N\right|_\textbf{ext}$ &
$\left.N\right|_\text{\bf cuts}$ \\[8pt]
 & [2018-2024] & [2018-2024] & [2018-2024] \\
\midrule
                    19.0 & 514$\pm$27           & 668$\pm$35           & 1425$\pm$51 \\
\rowcolor[gray]{.9} 19.5 &  52$\pm$9\phantom{7} &  73$\pm$12           &  102$\pm$14 \\
                    19.7 &  11$\pm$4\phantom{7} &  16$\pm$5\phantom{2} &   29$\pm$8\phantom{4} \\
\bottomrule
\end{tabular}
\end{table}


We will maintain precautions to avoid dramatic changes of the PMT sensitivity (long-term
aging) and to have a safe margin for another ten years of operation. The first step is
a continuous monitoring of PMTs during each night. The illumination of the PMTs is also
continuously monitored and the average accumulated anode charge by the PMTs is about $2.7$\,C
per year during standard observations. Our estimate is $5$\,C per year for the extended data
taking during nights with higher NSB. After another ten years the accumulated anode
charge will stay well below the value corresponding to the half-lifetime of the FD PMTs,
which is typically $250$\,C.

%% file: comms_and_daq.tex
\chapter{Communication System and Data Acquisition}

\section{Communications System}
%
%
%

\begin{figure*}[t]
\centering
\includegraphics[width=0.75\textwidth]{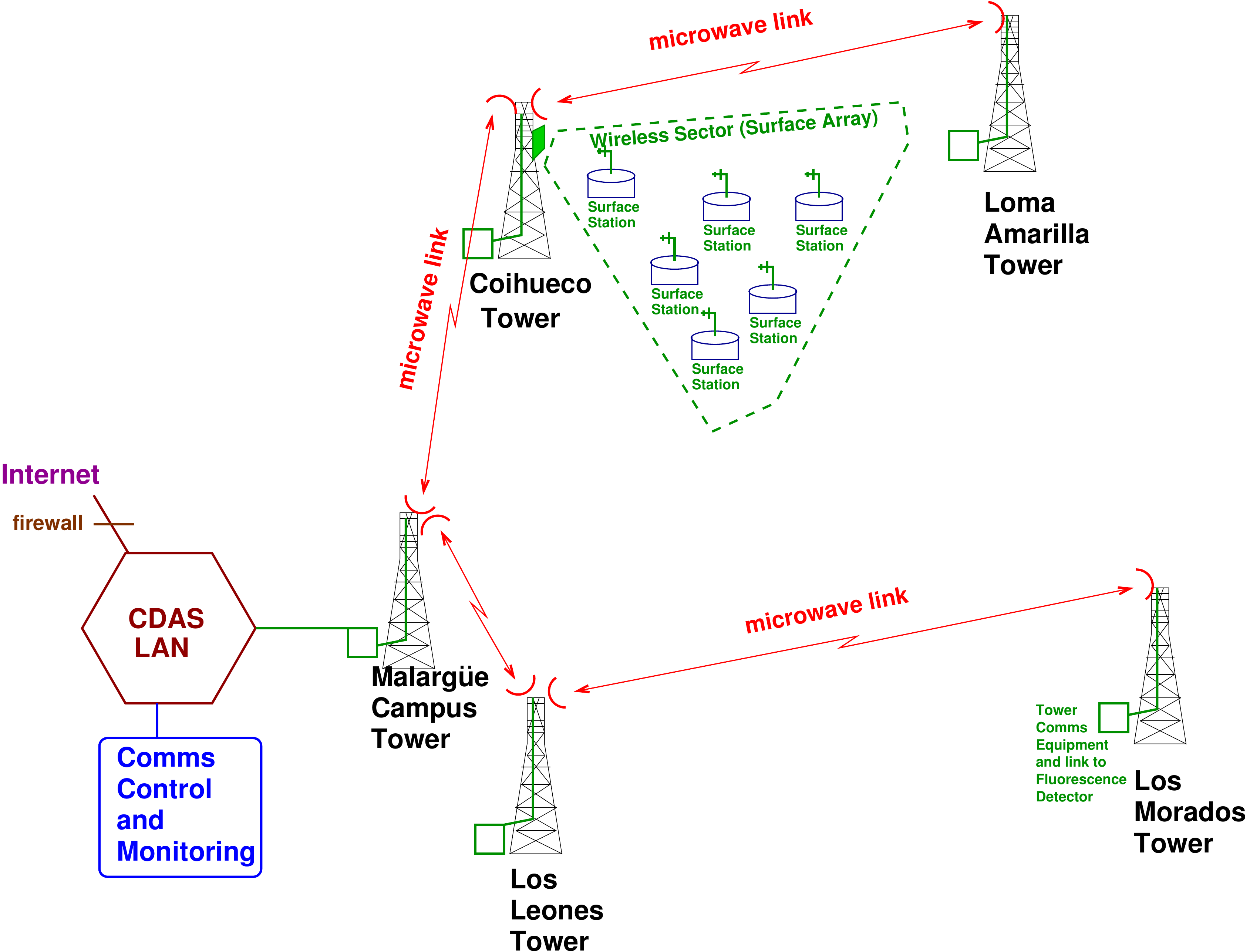}
\caption{Conceptual schematic of the overall radio telecommunications
  system for the Pierre Auger Observatory.}
\label{comms_schematic_inline}
\end{figure*}

The existing two-layer telecommunications system designed by the
University of Leeds that currently provides bi-directional data
transfer and control for both the Fluorescence Detector (FD) and the
Surface Detector (SD) will continue to operate without significant
modification within the Auger Upgrade. See~\ref{s:comms_appendix} for
details on the existing system.  The Leeds communication system will
continue to receive data from the four individual FD sites and from
the approximately 1600 SD stations deployed into the field, exclusive
of those SD stations that are located within the infill region.
\medskip

Individual surface detector stations are connected
by a custom wireless network which is sectorized and supported by four
concentration nodes.  The wireless network is serviced by a high capacity
microwave backbone network which also supports communications between
the four fluorescence detector sites and the main campus data
acquisition and control center.  Figure~\ref{comms_schematic_inline} shows a
conceptual schematic of the overall layout of the data communication
system for the Auger Observatory.  Table~\ref{table:comms_performance}
lists the main performance characteristics.

\begin{table*}[t]
\caption{Performance summary for the radio data communications
system for the Pierre Auger Observatory.}
\centering
\begin{tabular}{ll}
\toprule
\textbf{Microwave backbone network}
\\
\midrule
Links & 4
\\
Frequency & 7\,GHz
\\ 
Data rate & 24\,Mbps
\\
\toprule
\textbf{Wireless LAN}
\\
\midrule
Nodes & 1660
\\
Frequency & 902 to 928\,MHz ISM band
\\
Protocol & TDMA, custom
\\
Subscriber Unit over-air rate & 200\,kbps
\\
Effective payload rate & 1200\,bps uplink
\\
Typical daily data packet loss rate & less than 0.002\%
\\
\bottomrule
\end{tabular}

\label{table:comms_performance}
\end{table*}

\subsection{The microwave backbone network}


The backbone for the Auger data communications system is a 34\,Mbps
network 
operating in the 7\,GHz band.  Receivers and transmitters
are mounted on five communications towers located at the perimeter of
the array.  The microwave backbone provides high speed network
communications to nodes at all four FD sites and the main campus.
\medskip

\begin{figure}[t]
\centering
\includegraphics[width=0.85\textwidth]{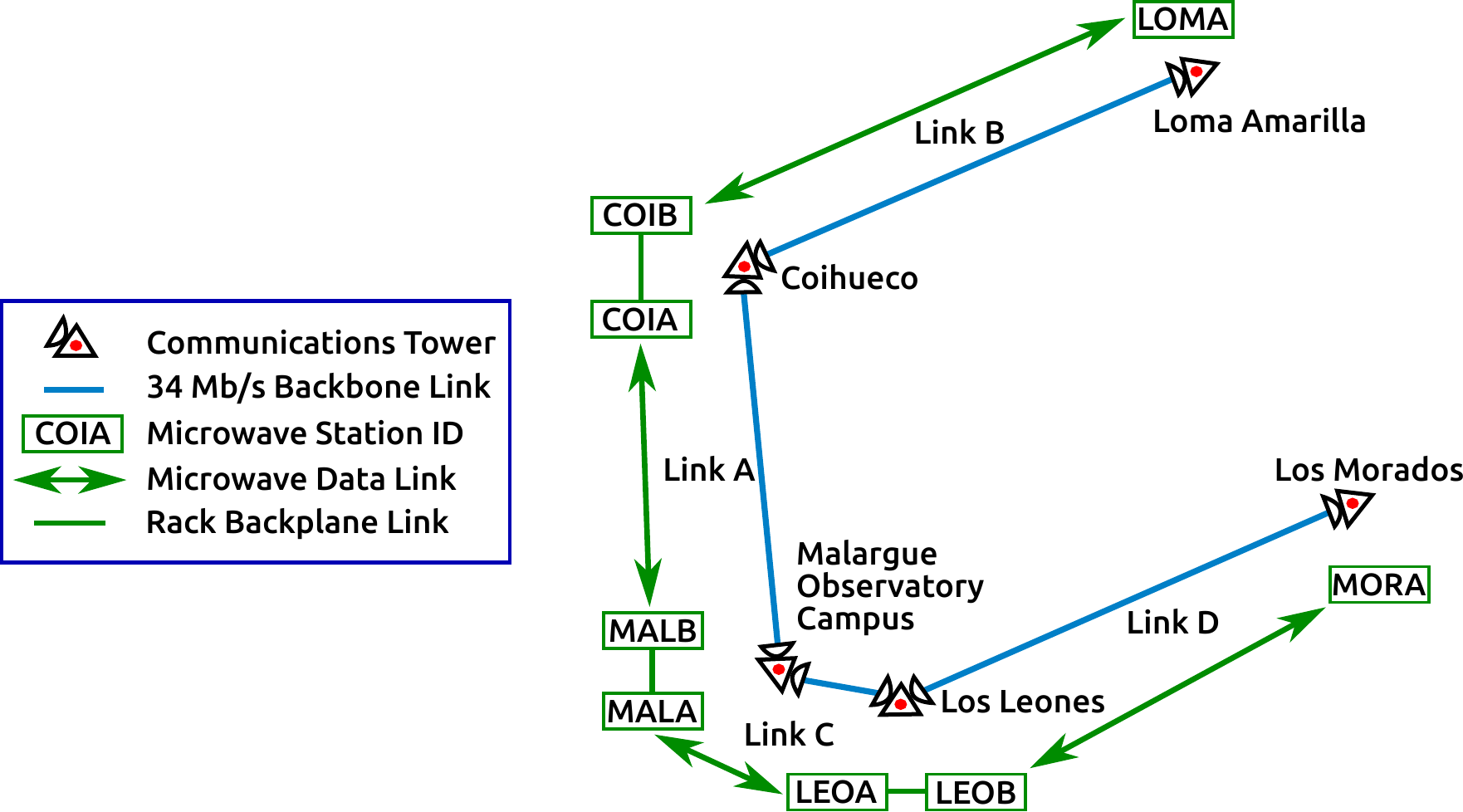}
\caption{Configuration of the high capacity microwave backbone
  network that connects the four FD sites with the main campus control
  and data acquisition center in Malarg\"ue.}
\label{fig:comms_backbone}
\end{figure}

The microwave backbone, depicted schematically in figure~\ref{fig:comms_backbone}, 
consists of a set of paired links providing
sufficient capacity to stream data to and from each of the FD sites
as well as for collecting data from the individual surface stations. \medskip

We note that the current microwave backbone system has worked reliably
for many years, and therefore the baseline design calls for the
existing system to serve for the Auger Upgrade.  However, the specific
transmitter/receiver hardware units deployed in the field are now
obsolete.
Therefore, we are
exploring a range of possible equivalent replacements for
the microwave communication system based on more up-to-date
technology.  Although the exact solution has not been
identified, high-speed tower-to-tower communication links correspond to
a very standard and common communication solution, and there exist
several viable Commercial-Off-The-Shelf (COTS) solutions that can meet
our bandwidth requirements at a cost that is rather lower than the
original implementation.  Our plan is to reduce costs and disruptions
by gradually replacing obsolete units with modern COTS equivalents for
increased reliability and bandwidth.

\subsection{The Leeds Wireless Network}

Tower-to-surface-station communications are accomplished with 
custom designed units operating in the 902 to
928\,MHz ISM band.  A point-to-point bidirectional communications link
is established between each surface detector station and one of four
communication concentration nodes mounted on the four towers located at
each of the fluorescence detector sites.  Communication to the SD
stations is achieved in a manner similar to a cellular telephone
system by dividing the array into 28~sectors, each of which contains
up to 68 stations. \medskip

Communications operations at each surface station are governed by a
custom-built programmable Subscriber Unit (SU) used to mediate the
transmission and reception of digital data between the electronics
board of a surface detector and the concentrator node.  An analogous
custom-built unit, called a Base Station Unit (BSU), controls data
transfer between each concentrator node and the backbone network
connection at each tower.

\subsubsection{Time division multiple access}

\begin{figure}[t]
\centering
\includegraphics[width=0.8\textwidth]{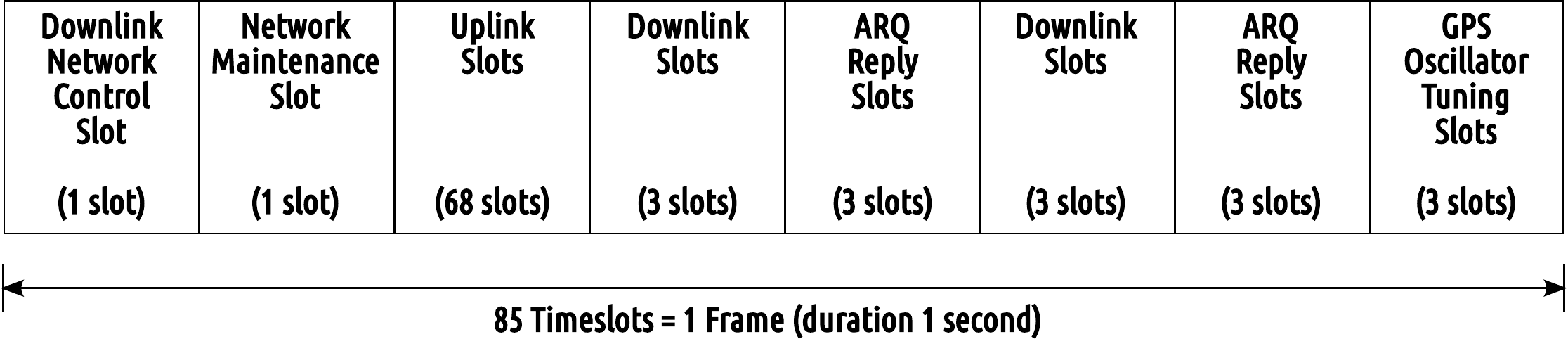}
\caption{A single GPS synchronized one-second TDMA frame is broken in time slots as shown.}
\label{comms_tdma}
\end{figure}

Transmissions to and from the stations are synchronized by GPS timing
so that each station is assigned a particular time slot during which
it is available to send and receive data.  This Time Division Multiple
Access (TDMA) scheme provides a contention free communication
environment within the array. A one-second data frame includes
68~uplink slots for collecting data from the array and 6 downlink
slots for sending trigger requests and other commands to the
stations. An additional 11 slots are reserved for network management,
monitoring, and packet error control. The assignment of individual
time slots within the one-second TDMA frame is shown in
figure~\ref{comms_tdma}.  This provides an effective bandwidth
of at least 1200\,bps uplink for each surface station and a 2400\,bps
for broadcast downlink.

\subsubsection{Error handling}

A central requirement of the Auger WLAN system for collecting data
from the surface detector stations is very high reliability.  In
response to an array trigger, digitized data from PMT traces and other
detector information must be relayed promptly to the central
controller (CDAS) so that the event can be built and recorded. Data
from a single event trigger will be broken into several dozens packets
transmitted by each station on request, a process that can continue
for as long as two minutes.  If even a single data packet is missing
or corrupt, the entire trace from the station is lost.  A custom
packetization protocol that includes Cyclic Redundancy Checking to
detect transmission errors is used at every level.

For the custom wireless network itself, an advanced re-transmit-on-error scheme, commonly
called an Automatic Repeat Request (ARQ), is also employed.  The ARQ
scheme is especially designed to prevent data loss in the case of
variable signal fluctuations, external sources of interference or any
other episodic environmental influences.  If a packet is flagged as
missing or corrupt at the monitoring concentrator of the central
network, a request to re-transmit the packet is automatically initiated
and collected via the subsequent data frame reserved TDMA time slots.
The ARQ request will be sent once per frame and will be repeated so
that at least six attempts are made to retrieve each missing or
corrupt packet.

\subsection{Communications Data Rates for the Auger Upgrade}

The uplink data rate requirements for the Auger Upgrade are based on the 
expected performance of the Upgraded Unified Board (UUB) electronics associated
with each Surface Detector station.  Each station in the field reports data via
uplink in several distinct forms:

\begin{itemize}
\item {\bf T2 time-stamps:} Each individual station in the field has
  set of local second-level trigger conditions (so-called T2
  triggers).  Within the existing array, the T2 trigger rate is
  approximately 25~Hz.  The time-stamps for each of these T2 events
  must be reported via uplink to the central controller (CDAS) which
  decides on array-wide triggers (so-called T3 triggers) corresponding
  to real cosmic ray events. 

  For the Auger Upgrade, we do not anticipate that the rate of T2 triggers will
  increase.   Indeed, given the greater flexibility of the UUB, we anticipate 
  tuning the overall T2 rate down to about 20~Hz. The T2 message has
  a $7~\mbox{byte}$ header, with each individual timestamp requiring
  $3~\mbox{bytes}$. For margin, we
  budget $80~\mbox{bytes}$ for the T2 timestamps.
\item {\bf Array Trigger Event Data:} For array triggers (T3 events)
  participating stations are required to report all station data
  related to the requested trigger event. Due to the increase in the
  number of channels and the increased ADC digitization rate, 
  the upgraded SDs will generate approximately five times as
  much data per T3 array trigger as is currently being generated
  within the existing array.  However, given the very low overall
  event rate for the part of the array that is outside of the infill
  (about 10 events per day per station) the increased data rate can be easily
  handled by the existing Leeds wireless system.  We project each event to
  be approximately $15~\mbox{kB}$ after compression, yielding an average
  event data rate of $2~\mbox{byte/sec}$.
\item{\bf Monitoring/Calibration Data:} Stations also transmit
  calibration data, including the current estimates of the online
  energy calibration, as well as sensor and monitoring information.
  We conservatively budget $1~\mbox{byte/sec}$ for this data, which is more
  than twice the current rate. 
\item{\bf Scaler Rate Data:} In addition to the event trigger data,
  stations also include a very low-threshold scaler trigger which
  provides a raw count of the number of these triggers per second. The
  rate of this data is also estimated to be not more than
  $1~\mbox{byte/sec}$, several times less than the current rate.
\end{itemize} 

In total, the estimated average data rate for the full array is
$84~\mbox{bytes/sec}$, well under the $120~\mbox{bytes/sec}$ uplink
bandwidth of the Leeds system.

We note that during normal operations, the download broadcast link is
used for station control and for event trigger requests from the
central controller for event data (T3 requests).  Generally, the
broadcast rate is much lower than the capacity of 2400 bps for the
Leeds system.  As a rule, the broadcast bandwidth is only filled
during special operations, such as installing new station firmware.
%
%
%
%

\subsection{Data Communication for the Infill Array Region}

The original Leeds wireless communication system for the Surface
Detector system was designed for the original configuration of the
Auger Observatory and will be more than sufficient to handle the
increased bandwidth for the Auger Upgrade within the field for the
1600 surface detectors deployed on the 1.5\,km grid.  However, the
data rates for the region recently occupied by AMIGA, including the
Auger Infill region, result in a substantially higher array trigger
rate (by design).  At present, a combination of re-deployed Leeds
radios, together with a specially configured communications system for
AMIGA are operating within the infill region. Although the Leeds
radios are operating within this region, the substantially increased
infill trigger rates use up virtually all of the spare bandwidth so
that there is no additional margin for additional data.  What this
means is that while the existing Leeds radios will very nicely handle
the increased data rates for the Auger Upgrade associated with the
larger array, they are {\em not} be able to handle the larger rates
within the infill region. \medskip

For the infill/AMIGA region of the array, we will need an alternative
to the exiting Leeds radio communication system.  Specifically, we
will need to identify and implement a communications system that will
independently provide an uplink bandwidth of at least 20~kbps for
anywhere from 100 to 200 stations within the infill. \medskip

Fortunately, the Auger Collaboration already has some experience with
the implementation of alternative communications systems for specific
tests and enhancements.  At present, we can identify at least two
different options for communications systems tailored for the infill
array region: either based on the AMIGA communication system
\cite{bib:amigacomms}, or based on the AERA communication
system~\cite{bib:aeracomms}.

Both options use commercial off-the-shelf 802.11-based wireless links,
and have similar bandwidth ($\sim$\,Mbps) and power consumption
($\sim3$\,W). Technology for both systems should be readily
available, and both are currently deployed and operating well within
the infill region. The exact configuration for each communication system
option has not yet been developed.

\section{Central Data Acquisition System (CDAS)}

The CDAS has been running since March 2001.  The system was designed
to assemble the triggers from the surface array detectors, to allow
control of these detectors and to organize the storage of data.  It is
constructed using a combination of commercial hardware and custom
made, high level, software components. The system is designed to run
continuously, with minimum intervention, with the full 1660 detector
array. Data from the FD are recorded
separately at the FD locations and transferred daily to the computer
center at Malarg\"{u}e, although hybrid coincidences are identified
online within the SD data stream.

The primary role for the CDAS is to combine local trigger information
from the SD stations in order to identify potential physical events
generating an SD higher level trigger (T3). These triggers combined
with the T3 from FD sites (FD T3) are used to generate a request for
the relevant data from SD stations for these events. The CDAS is then
used to combine and store these data to form a shower event. The CDAS
also contains configuration and control mechanisms, the means to
monitor system performance, and the tools to access and download SD
monitoring, calibration, control and configuration data.


The Post Master (Pm) is the interface between the Surface Detector Array 
and the CDAS. It is the end point, at the Observatory Campus, of the 
\emph{communication backbone}, and is aimed at dispatching information 
from the different \emph{data streams} towards the CDAS applications. It 
serves also as a router between the CDAS applications and the Surface 
Detector Array. Pm maintains a routing table by trapping local station 
identifications in the incoming data flow.

The Post Master also implements the \emph{backbone protocol}. Pm is a 
\emph{server} for all the transactions across the data path, 
accepting requests from \emph{clients}. Requests consist of messages. 
The Pm protocol defines how to connect to the server and the content
of the messages to be exchanged for a given \emph{data service.} This 
protocol is also used by the CDAS applications or any external 
client to gain access to the data streams.

The data sent by the local stations belong to several \emph{streams}.
The main streams are:
\begin{itemize}
\item Local triggers (\emph{T2}): higher priority 
stream; time stamps and local trigger type; to be forwarded to the
\emph{Central trigger} (Ct); their occurrence rate is about 20\,Hz 
per station.
\item Shower data and calibration (\emph{Event}): on request from
the CDAS; the events are split into smaller pieces in order to be channeled
into the communication path together with the T2 packets. Once completed, 
the events are forwarded to the \emph{Event Builder (Eb).}
\item Calibration and Monitoring information (\emph{Monitoring}): 
low priority stream; same behavior as Events; to be forwarded 
to the \emph{Monitoring client}.
\end{itemize}
The CDAS transmits to the local stations some commands and configuration 
parameters, event requests (i.e. T3 triggers from Ct), and monitoring requests. 
Software downloads are also possible across the backbone.

Pm is in charge of presenting well formatted, intelligible and complete
data to its clients. It extracts the different data streams from the 
local station messages and processes each of them according to CDAS 
requirements:

\begin{itemize}
\item T2 data from stations present in the run are accumulated. Every 
second, all T2 data which are older than the maximum allowed delay 
(5~s) are sent to the candidate clients (e.g. Ct), then discarded.
\item Event, calibration and monitoring data pieces are reordered 
and concatenated from successive local station messages, then sent to 
clients (e.g. Eb) and discarded. Incomplete data blocks are discarded.
\item Initialization messages from the local stations are formatted, 
forwarded to clients and logged by the Information Kernel (Ik) of the
CDAS.
\item T3 triggers, commands and controls, monitoring requests and 
software downloads are routed to the appropriate Base Station Unit. 
\end{itemize}


As mentioned in Sec.~\ref{sec:LocalStationSoftware}, the local station
acquisition for the Auger Upgrade will be a direct port of the existing code base. Therefore,
the basic messaging protocol between CDAS and the surface detectors
will not change. This has the distinct advantage of allowing easy
operation of a heterogeneous array during the upgrade process, allowing
or both UUB-based surface detectors and UB-based surface detectors
to coexist during data taking.

Data structures themselves, however, will be different, due to the
expanded data format and increased number of detector data channels.
Therefore, to accommodate the new data format, the 'event' portion
of the messaging between the local station and CDAS will be restructured
into a block-based format, allowing the CDAS data acquisition processes
to gracefully handle the presence or absence of data in the event structure.
In addition, this will allow data from the digital expansion connectors
to be eventually added to the event structure.

\section{Monitoring}
The CDAS provides  information to monitor its own operation, the communication status, the trigger rates, the surface detectors 
status, and various environmental parameters. 
As explained above, the CDAS scheme will be easily adapted to the detector upgrade, and this is also the case for the whole monitoring 
procedure. Here we recall the Auger Monitoring scheme,
before pointing out what has to be modified.\\

Besides the information from the data acquisition and the communication systems,
the CDAS handles 3 types of data managed by the water-Cherenkov detector (WCD) acquisition 
software and related to the WCD: event data files, T2 files, monitoring and calibration files.
Moreover, information on the weather conditions at different points in the Observatory are stored in a 4th data stream. \\

Inside the CDAS network,
all these data streams are parsed by programs  to produce XML files, which are copied to the 
monitoring system. The XML files are processed to produce and execute SQL requests in order to fill 
the Auger Monitoring Data Base. The Observatory monitoring web site uses the appropriate scripts (PHP, Ajax, etc.)
developed to display  the information retrieved from the database to control the  online SD status and to raise alarms if needed.\\

The monitoring and calibration files contain information related to the WCD status (solar power system currents and voltages, temperatures)
  PMT status (voltage and current, baseline values, etc.),   calibration procedure (for each PMT : the peak current
corresponding to a VEM; the corresponding charge; the dynode/anode ratio, used to calibrate low gain relative to 
high gain) and the local trigger rates. Each WCD sends this monitoring and calibration information every 6 minutes. 

One "T2" file is produced each day. It provides data on which WCD are participating in data acquisition every second of the day. This is particularly useful for the exposure calculation.

The weather files contain temperature, pressure, and wind speed measurements sent every 5 minutes by different
weather stations located either at the FD sites, or at the Central Laser Facility.

For the Auger Upgrade, the new CDAS version will contain the following
monitoring-related changes:
\begin{itemize}
\item Additional slow control and calibration parameters will be added for the SSD (e.g. PMT voltage, temperature).
\item New sensors available in the upgraded electronics will be recorded (temperature/current measurements).
\item Trigger configuration/versioning information will be included with each event.
\item Compressed calibration histograms will be transmitted periodically ($1/\mbox{hour}$) with calibration data.
\item Data frequency will be reduced for housekeeping parameters (voltage, current) to allow for the additional calibration data.
\end{itemize}

Calibration histograms will be a significant monitoring improvement relative to the existing Auger software.
Currently, these histograms are transmitted only when an event occurs at an individual station, due to the limited bandwidth.
A statistical lossy compression scheme will be used to reduce the histogram size, allowing for a detailed measurement
of the surface detector performance every hour.

These changes do not affect the data treatment between the CDAS frame and the monitoring frame. The different software which produce XML files and send them to the monitoring server, 
and which construct the SQL requests, will be modified to deal with new parameters. Existing DB tables will be
extended to store useful parameters.
The web interface to display monitoring, calibration and triggers rates will be upgraded accordingly. 
During the Engineering Array deployment,  monitoring features developed to consider the status and performance 
of a particular  sub-array (Infill array, test array...) will be used.

%% file: dpa_offline.tex

\chapter{Data Processing and \Offline}
\label{sec:offline}

The \Offline software of the Pierre Auger Observatory
provides both an implementation of simulation and reconstruction algorithms,
discussed later, and an infrastructure to support the development of such
algorithms leading ultimately to a complete simulation, reconstruction
and analysis pipeline. Indeed, when the \Offline code was originally devised,
the only existing systems were the SD and FD. It has since been extended to handle the radio
and AMIGA extensions without requiring dramatic framework
changes. The most recent extensions comprise the
  surface scintillator detector (SSD), the upgraded electronics
  chain for faster sampling, and the small PMT in the water-Cherenkov detector (WCD).
The software has been designed with such flexibility in mind, and is meant to accommodate
contributions from a large number of
physicists developing C++ applications
over the long lifetime of the experiment.  The essential
features include a ``plug-in'' mechanism for physics
algorithms together with machinery which assists users in
retrieving event data and detector conditions from various
data sources, as well as a reasonably straightforward
way of configuring the abundance of different applications and logging all configuration
data. A detailed description of the \Offline software design,
including some example applications, is available in~\cite{Argiro:2007qg}.

\begin{figure}[t]
\centering
\includegraphics[width=0.42\textwidth]{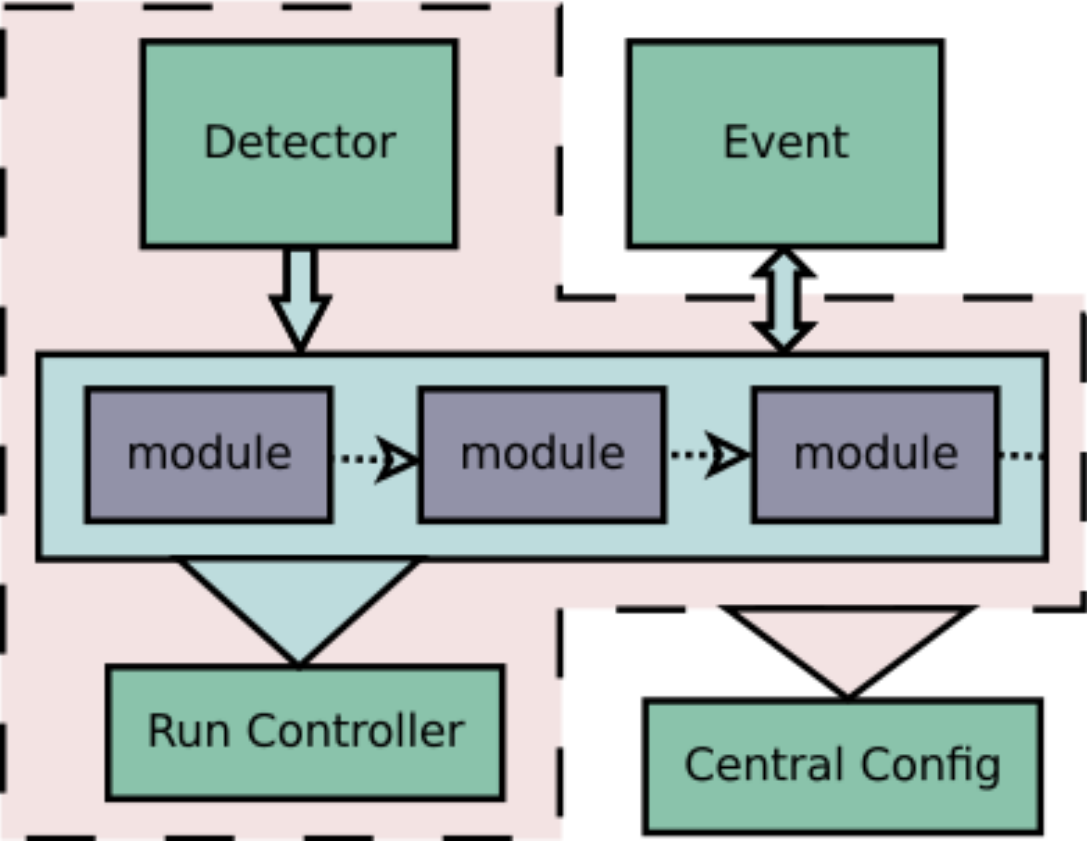}
\caption{General structure of the \Offline framework.
Simulation and reconstruction tasks are encased
in modules.  Each module is able to read information
from the detector description and/or the event,
process the information, and write the results back
into the event under the command of a {\em Run Controller}.
A {\em Central Config} object is responsible for handing
modules and framework components, their configuration
data and for tracking provenance.}
\label{f:general}
\end{figure}

The overall organization of the \Offline framework
is depicted in figure~\ref{f:general}.
A collection of processing {\em modules} is
assembled and sequenced through instructions
contained in an XML file~\cite{xml} or in a Python~\cite{python} script.
An \emph{event} data model allows
modules to relay data to one another,
accumulates all simulation and reconstruction
information, and converts between various formats
used to store data on file.
Finally, a \emph{detector description}
provides a gateway to data concerning detector conditions, including calibration constants and
atmospheric properties as a function of time.

\section{Physics modules}

Simulation and reconstruction tasks can be
factorized into sequences of processing steps which
can simply be pipeli\-ned. Physicists prepare processing algorithms
in modules, which they register with the
\Offline framework via a one line macro.  This modular design allows
collaborators to exchange code, compare algorithms and
build up a variety of applications by combining modules in various
sequences. Run time control over module sequences
is implemented with a {\em Run Controller}, which invokes the various
processing steps within the modules according to a set of
user provided instructions. We devised
an XML-based language as one option for
specifying sequencing instructions; this
approach has proved sufficiently flexible for
the majority of our applications, and it is quite simple
to use.

%
\begin{figure}[hbt]
\centering
\includegraphics[width=0.58\textwidth]{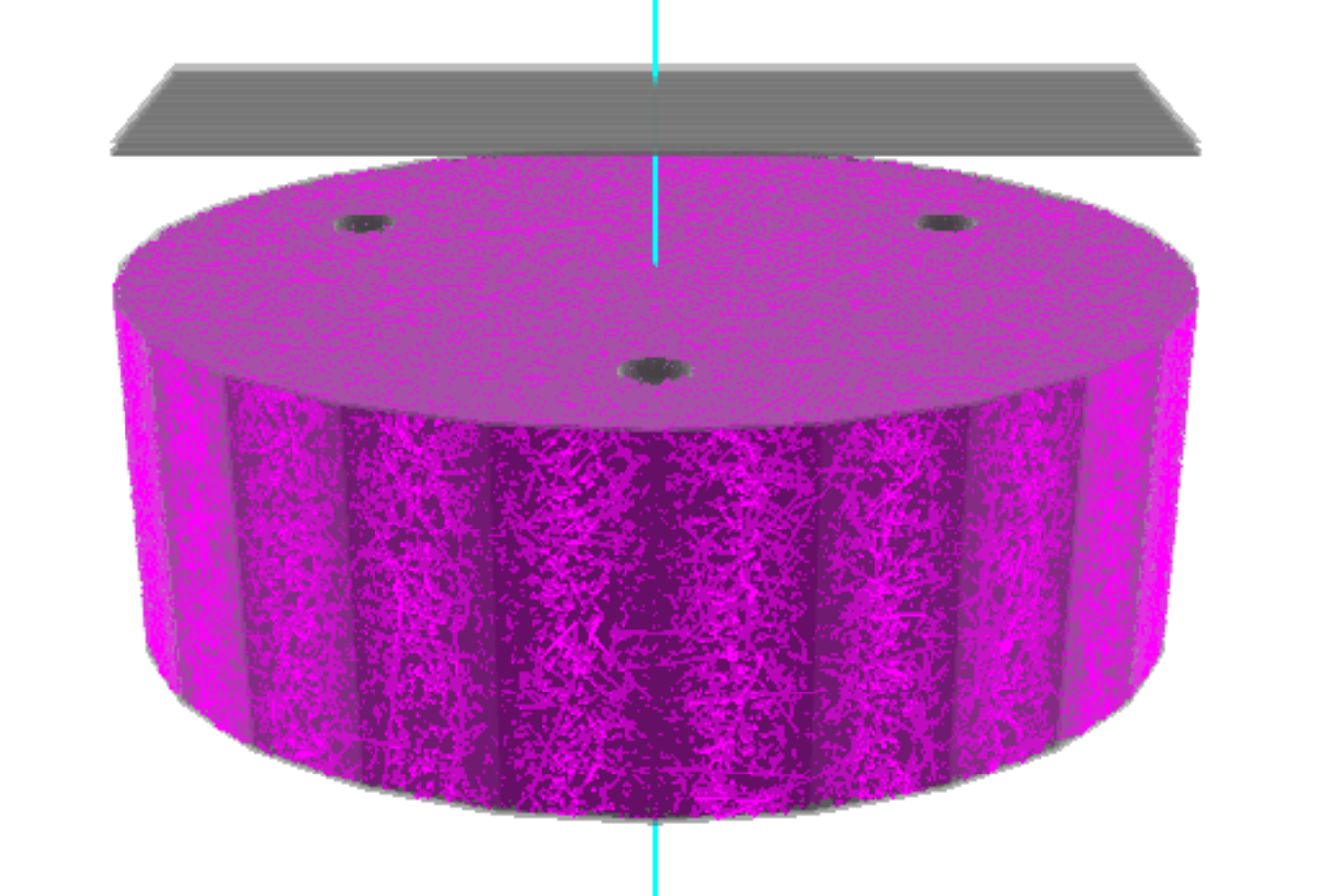}
\caption{Visualization of a Geant4 simulation of the SSD/WCD combination. One can see a muon entering the SSD above
the tank, the three 9-inch PMTs in the station, and the ``gas'' of reflected photons
within the WCD.}
\label{f:station-pic}
\end{figure}

Various simulation chains were prepared in the \Offline framework for the battery of
simulation challenges employed in arriving at the best, most cost-effective
upgrade design. Figure~\ref{f:station-pic} displays an image of a WCD with a
scintillator positioned above it. A single muon is injected vertically for
illustrative purposes. One can see the ``photon gas'' created as Cherenkov light
is emitted in the WCD. The WCD/SSD simulation is based on the Geant4~\cite{g4}
package, which supports detailed simulation of physics processes as well as
relatively straightforward definition of complex detector geometries and
materials.

\section{Data access}

The \Offline framework includes two parallel hierarchies for accessing
data: the detector description for retrieving data about conditions, including
detector geometry, calibration constants, and atmospheric conditions, and
an event data model for reading and writing information
that changes for each event.

\begin{figure}[t]
\centering
\includegraphics[width=0.58\textwidth]{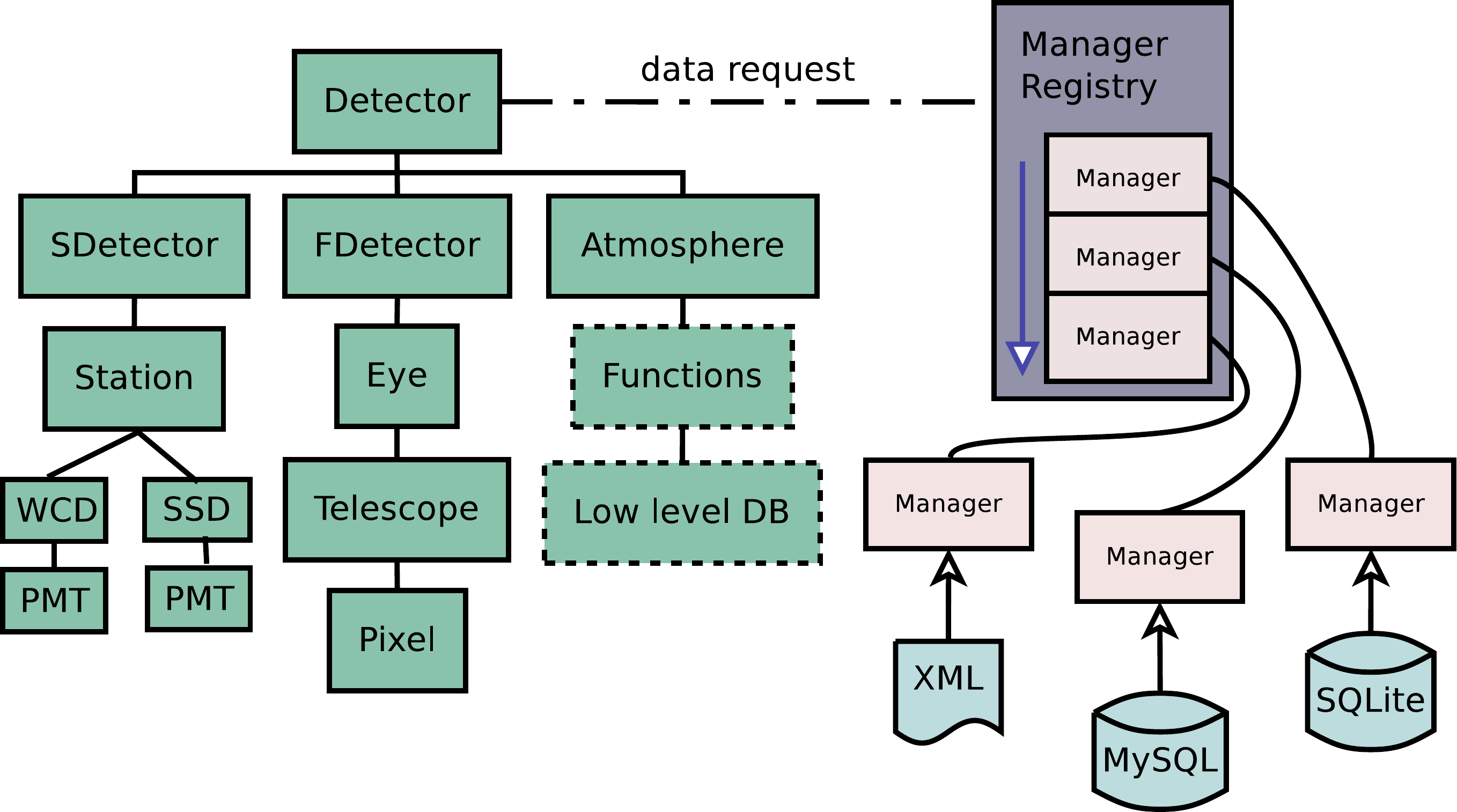}
\caption{
Machinery of the detector description.  The user interface
(left) comprises a hierarchy of objects describing the
various components of the Observatory. These objects relay
requests for data to a registry of managers
(right) which handle multiple data sources and formats.
}
\label{f:detector}
\end{figure}

The {\em detector description} provides a unified interface from which module
authors can retrieve conditions data.  Data requests are passed by this
interface to a back end comprising a registry of so-called {\em managers}, each
of which is capable of extracting a particular sort of information from a
collection of data sources.  The manager mechanism is highly configurable and
relieves authors of the physics code from having to deal with the details of
selecting and reading the correct data source.  The general structure of the
detector description machinery is illustrated in figure~\ref{f:detector}.

The event data model contains raw, calibrated, reconstructed and
Monte Carlo information, and serves as the backbone for communication
between modules.  The event is instrumented with a protocol allowing
modules to discover its constituents at any point in processing, and
thereby determine whether the input data required to carry out the
desired processing are available.

The transient (in memory) and persistent (on disk) event models are
decoupled in order to avoid locking to a single provi\-der solution
for serialization, the process by which C++ objects are converted to a
form that can be written to disk.  When a request is made to write
event contents to file, the data are transferred from the transient
event through a \emph{file interface} to the persistent event, which
is instrumented with serialization machinery.  We currently use the
input/output portion of the ROOT~\cite{root} toolkit to implement
serialization.  Various file formats are interpreted using the file
interface, including numerous raw event and monitoring formats as well
as the different formats employed by the AIRES~\cite{Sciutto:1999jh},
CORSIKA~\cite{Heck:1998vt}, CONEX~\cite{Bergmann:2006yz} and
SENECA~\cite{Drescher:2002cr} air shower simulation packages.

\section{Configuration}
\label{sec:configuration}

The \Offline framework includes a system to organize and track data
used to configure the software for different applications as well as
parameters used in the physics modules.  The \emph{Central Config}
configuration tool (figure~\ref{f:general}) points modules and
framework components to the location of their configuration data, and
connects to Xerces-based~\cite{xerces} XML parsers to assist in reading
information from these locations.  We have wrapped Xerces with our own
interface which provides ease of use at the cost of somewhat reduced
flexibility, and which also adds functionality such as automatic units
conversion and casting to various types, including commonly used
containers.

The {\em Central Config} keeps track of all configuration data
accessed during a run and stores them in an XML log file, which can
subsequently be used to reproduce a run with an identical
configuration. This allows collaborators to easily exchange and use
configuration data for comparing results.  The logging mechanism is
also used to record the versions of modules and external libraries
which are used for each run.

Syntax and content checking of the configuration files is afforded
through W3C XML Schema \cite{xml-schema} standard validation.  Schema
validation is used not only for internal configuration prepared by
framework developers, but also to check the contents of physics module
configuration files.  This approach reduces the amount of code users
and developers must prepare, and supports very robust checking.

\section{Utilities, testing and quality control, and build system}

The \Offline framework is complemented by a collection of utilities, including
an XML parser, an error logger and various mathematics and physics services.  We
have also developed a novel geometry package which allows the manipulation of
abstract geometrical objects independent of coordinate system choice. This is
particularly helpful for our applications since the Observatory comprises many
instruments spread over a large area and oriented in different directions, and
hence possesses no naturally preferred coordinate system. Furthermore, the
Geometry package supports conversions to and from geodetic coordinates.

As in many large scale software development efforts, each low level component of
the framework is verified with a small test program, known as a {\em unit test}.
We have adopted the CppUnit~\cite{cppunit} testing framework as the basis for
implementing these tests.  In addition to unit tests, a set of higher level
acceptance tests has been developed which is used to verify that complete
applications continue to function as expected, within some tolerance, during
ongoing development.  We employ a BuildBot system~\cite{buildbot} to
automatically compile the \Offline software, run the unit and acceptance tests,
and inform developers of any problems each time the code is modified.

The \Offline build system is based on the CMake~\cite{cmake} cross-platform
build tool, which has proven adequate for this project. In order to
ease installation of \Offline and its various external dependencies,
we have prepared a tool known as
APE (Auger Package and Environment)~\cite{ape}. APE is a python-based
dependency resolution engine, which downloads the external packages
required for a particular application, builds them in whatever native
build system applies for each package, and sets up the user's environment
accordingly. APE is freely available, and has been adopted by other
experiments, including HAWC, NA61/SHINE and JEM-EUSO.

\section{Production}

Data production methods are mature, as the Auger Observatory has been
operational for over a decade. We employ GRID resources for simulation
production, from generator-level shower simulation up through simulation of
detector response and reconstruction of the resulting simulated data. The Auger
collaboration has become one of the biggest non-LHC users of the European Grid
Infrastructure (EGI), and at the recommendation of the EGI, a Virtual Research
Community is being established for Auger. Simulations are transferred from GRID
storage elements to the Computation Center - IN2P3 (CC-IN2P3) in Lyon, for easy
availability to collaborators. Raw data from the Observatory site is also
transferred to the CC-IN2P3, where it is available to collaborators for analysis
and further processing.  Reconstruction of real data is performed on a dedicated
cluster located at the Karlsruhe Institute of Technology, and reduced data
summary files are generated for end-user analysis.

The Collaboration is in the process of developing an extended program
to release the full data sample to the general public. The relevant
policies and technical issues are currently being addressed. At the moment
1\% of high-quality data are made public regularly at~\cite{publicEventDisplay}.

%% file: assembly_tests_maintenance.tex


\chapter{Reception Tests, Deployment and Maintenance}
\label{chap:assemblyTests}

This chapter describes the equipment, facilities, and procedures for reception 
at the Central Campus in Malarg\"ue of the detectors and components for the upgrade 
of the Pierre Auger Observatory. It also describes the testing, deployment, and 
maintenance for the next 8-10 years of operation.

\section{Reception and testing of Scintillator Detectors}

The scintillation detector modules will be prepared in distant facilities and 
shipped to Mal\-arg\"ue in crates, with three crates in each 40 foot (42G1) container. 
We presently estimate that we can load 16 modules resting on a side edge per crate 
for a total of 48 per container. 

Upon arrival, the crates, which have a skid base, can be pulled from the container 
and moved by a forklift. The crates, still loaded with modules, 
will be stored in the yard under a protective cover until they are ready to be 
unloaded for testing and deployment. Individual modules can be lifted vertically 
out of the crates and moved in a straightforward manner, by forklift, or by light, 
portable cranes and on wheeled dollies. It will be necessary to buy these light, 
portable cranes and have at least two available to avoid conflicts resulting in 
delays. It may be desirable to rent or purchase an additional forklift 
for the duration of the upgrade activity.

Reception testing of the modules will consist of: 
\begin{itemize}
\item a visual inspection of the exterior of the modules for any shipping damage;
\item measurement of photomultiplier background in light- and dark-conditions to 
identify light leaks;
\item measurement of individual scintillator strip response to cosmic rays.
\end{itemize}
The last measurement will be made using a cosmic ray telescope based on 
Resistive Plate Counters (RPCs) with readout pads sized and positioned to 
match the scintillator strip spacing in the modules. The module is placed 
horizontally on a 3.6m x 1.2m test bench between the RPCs. If four RPCs 
are provided per test bench, then every strip can be tested for efficiency 
and pulse size in a single ``run'' without having to move the module or the 
RPCs. We would speed up the process by having two test benches operating in 
parallel. Based on our long experience in building and running the Auger 
Observatory, we estimate that with one team of two trained technicians we 
can expect a testing rate of 60 modules per week.

Following inspection and testing, the characteristics of the module and 
the serial numbers are recorded in a database.  Cables, connectors, 
and the vent will be protected with plastic covers and the module 
placed in a vertical position, resting again on a side edge. The modules 
can then be stored in ad-hoc racks in the Assembly Building for ready 
deployment or, in the case where testing is well ahead of deployment 
activities, in longer term storage in the original crates in the yard 
under a protective cover. Modules which fail one or more of the incoming 
test procedures will be set aside for repair, to be done as quickly as 
possible in case feedback is needed to improve the manufacturing or 
shipping processes.

Each scintillator module will require a sunroof to be 
installed over it to reduce high temperature fluctuations during 
daytime. The engineering details of these sunroofs are not provided 
as of this writing but they are expected to be simple, corrugated 
sheet metal structures, spaced slightly above the scintillator 
modules to allow air to circulate between the module and the sunroof. 
They will not be installed on the modules when they arrive from the 
factory to reduce shipping costs but will be added in Malarg\"ue or 
in the field during deployment. We may find it most cost effective 
to buy these materials in Argentina.

\section{Deployment of Scintillator Detectors}
\label{sec:deploySSD}

\begin{figure}[!ht]
\centering
\includegraphics[height=0.2\textheight]{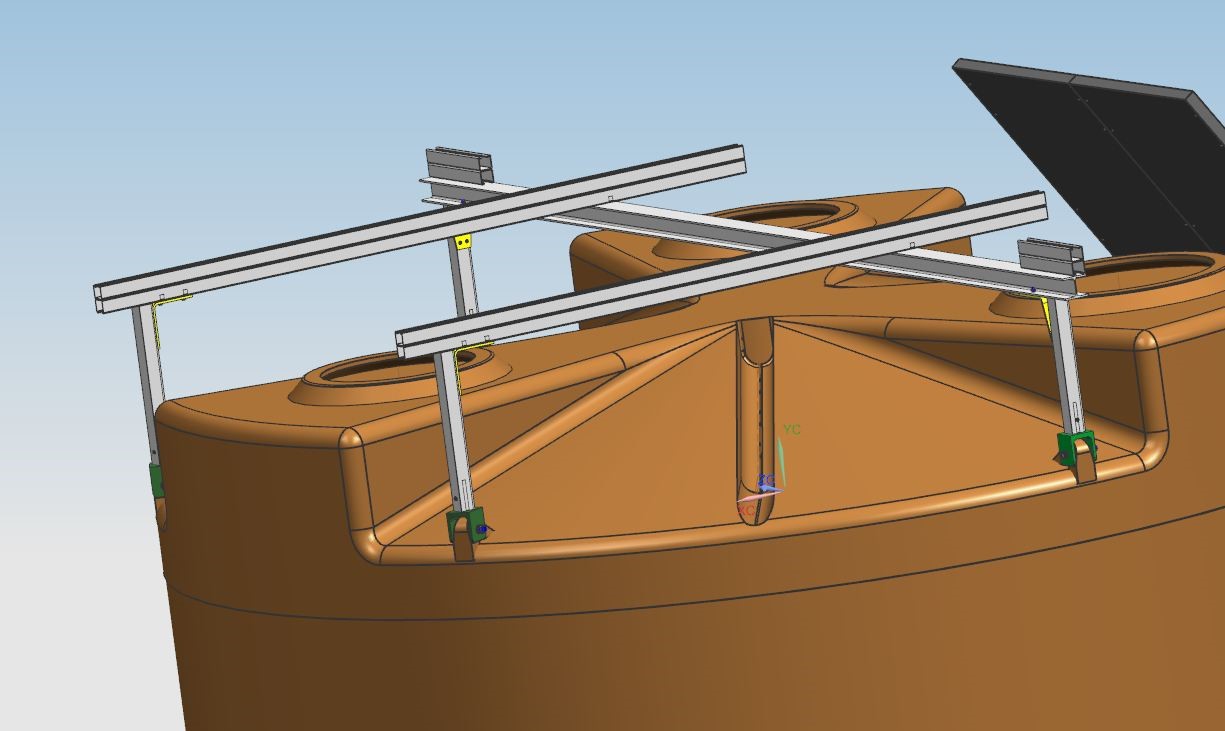}
\includegraphics[height=0.2\textheight]{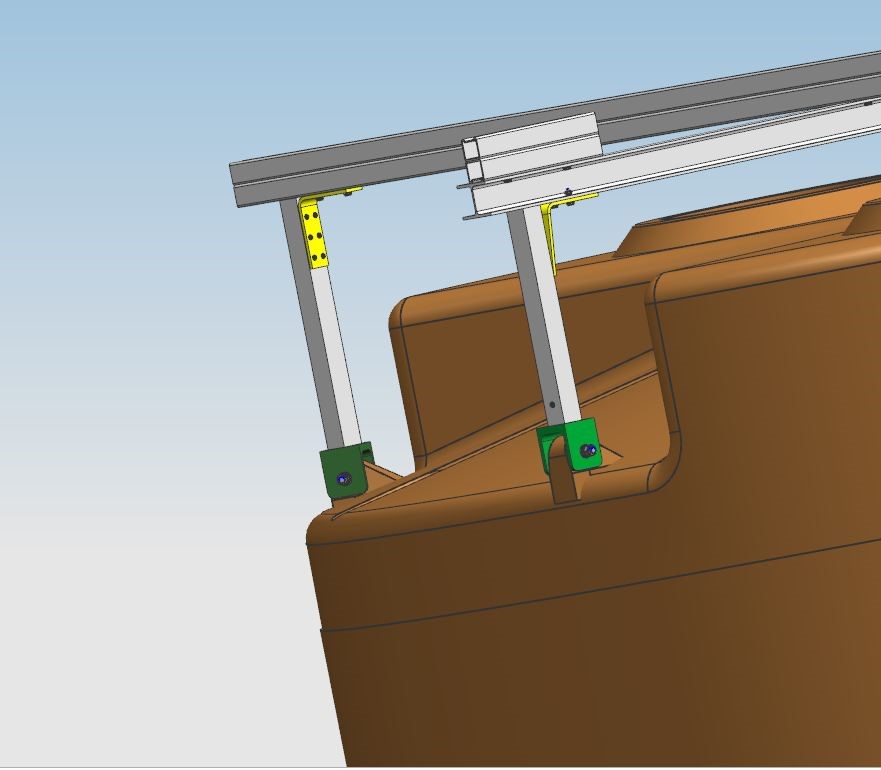}
\caption{The module support structure designed for simplicity, light weight and simple installation. An overview is shown 
in the left panel, while a detail of the structure is shown on the right. An aluminum I-beam and two Unistrut channels are the main structure. Injection molded plastic feet connect the frame to the lifting lugs on the tank.}
\label{fig:frame}
\end{figure}

The modules will be mounted over the existing surface detector water 
tanks on a frame attached to four of the six lifting lugs molded into 
the tank. This frame is shown in the rendering in Figure~\ref{fig:frame}. To reduce corrosion and to minimize the weight loading on the tank and the weight to be lifted during installation, the mounting structure is made from aluminum beams. The structure consists of a main I-beam with vertical support columns at the ends and of two Unistrut\textsuperscript{\textregistered} cross beams with one vertical column each. Although the nominal dimensional tolerances on the tanks is $\pm$1.5\%, in practice the tanks are more uniform, with a few centimeters variation in the lug position being more typical. The mounting frame will have enough adjustment to allow for this variation by using the easy positioning ability of 
the Unistrut\textsuperscript{\textregistered} 
cross beams and with slotted mounting holes in the attachment brackets at the ends of the main beam. Connection of the support frame to the tank is done with injection molded plastic feet at the four mounting lugs. These parts are inserted into vertical columns and their depth of insertion into the columns is controlled by thick plastic spacer washers, allowing the plane of the support frame to be adjusted to allow for warping of the tank top and subsequent non-planarity of the lugs. The scintillator module is fixed to the top of the cross beams and is supported at 25\% and 75\% of the length of the module. 
Additional support of the module at the ends of the main beam will be added to increase stability and resistance to wind-driven resonances in the structure. 

The desired deployment rate requires that 1660 detectors be installed in 
200 working days per year. A rate of six per day will allow us to complete 
installation in 1.5 years with margin for lost time due to weather, 
technician illness, and maintenance of the deployment equipment. Two 
teams of two people each will be able to deploy three detectors each 
day based on two hours of driving to/from the local site and 1.5 hours 
of work at each site. We considered the possibility of delivering a 
large number of modules to the field in a container, then deploying 
them from that ``base''. It was concluded that the crew could easily 
take as many modules as required into the field each day, and that
there was no advantage to providing a large local storage of modules 
in the field. The scintillator deployment, the electronics deployment 
with the small PMT, and the underground muon detector deployment will all be carried 
out independently.  The electronics deployment requires a different 
skill set, a different deployment rate, and different equipment, so 
maximizing efficiency suggests that this deployment be done separately. 
The underground muon detectors are deployed in a very localized area 
of the detector array and require a very different technique and time 
scale from either the electronics or scintillator deployments. 

Each scintillator deployment team will have a truck with a trailer. The trailer 
will be fitted with a rack that can hold five modules installed 
vertically, i.e., resting on the long edge. Each trailer will also 
be able to carry the mounting structure (in three pieces) and the 
sunroof, as well as any small parts and tools required for 
installation. Although the design deployment rate is three modules 
per day per team, it may be possible on some days to install more, 
so the trailer capacity was set at five modules to make this possible. 
Total loaded weight of the trailer will not exceed 1000kg, which can 
be pulled by a large pickup truck. If the sunroofs are mounted on the 
detectors in the Assembly Building, the trailer design would then be 
such that the modules are supported in a horizontal position on racks. 
As each module is removed the supporting cross-beams for the rack for 
that module are also removed. 

The pickup truck will have a jib crane installed on the back with a 
capacity of 200 kg and a reach of at least 2.5 m. It will consist of 
a horizontal I-beam with a trolley from which is suspended a rolling 
electric hoist, powered by the truck electrical system. The horizontal 
beam is attached to a vertical column which pivots about a vertical 
axis, allowing the beam to swing over the trailer, pick up the module 
with the hoist, raise it and swing it over the tank and lower the module 
in position. There are other possibilities for a crane, such as the common 
``hydrogrua'' hydraulic crane that was used for the deployment of the tanks, 
or other commercially available cranes designed for mounting in pickup trucks. 
Outriggers will be added to the truck to maintain stability during the lift. 
The mounting structure, especially considering that it will be in three large 
pieces (the beams) that can be lifted into position by two technicians by hand, 
will not require the crane for deployment. 

Upon arrival at the site, the support structure is installed first. 
The three large components, the main beam and the two cross beams, 
will be lifted by hand onto the tank and fitted to the lifting lugs. 
Adjustments will be made as needed for irregularities in tank dimensions. 
One vertical column needs to have adjustability (by selection of bolt holes, 
for example, or by selection from a set of vertical columns of varying lengths) 
to maintain the module in a plane. This plane may not be quite horizontal and 
we will have to determine based on physics analysis requirements what an 
acceptable deviation from horizontal might be. 

The module is then lifted from the trailer using the crane on the truck. 
The module perimeter consists of a wide-flange I-beam so this is a strong 
point which can allow lifting fixtures to be attached for lifting in either 
a horizontal or a vertical orientation, and which can be used to transition 
between one orientation and the other. If the module is horizontal with the 
sunroof attached, it is simply lifted into position on the mount and connected 
with the attachment brackets to the beams. If the module is vertical without 
the sunroof attached, it needs to be rotated to a horizontal position, either 
by setting the module down on the ground on a temporary bench, or by setting 
it on the already-installed mounting structure and lowering to a horizontal 
position. After fixing the module to the mount, the sunroof can be installed.

The installation can then be inspected visually and manually to verify 
the proper installation of the module, mount, and connections for proper 
installation and stability. The cables can be connected to the module and 
sealed (if a separate weather/light seal is provided) and the CDAS operator 
contacted. The CDAS operator will verify proper operation of the new module. 
It is expected that the number of defective installations will be very low, 
and modules that fail will be retrieved from the field and repaired in the 
Central Campus later. Calibration of the detectors occurs remotely by CDAS. 
If the CDAS operator is not able to give a quick evaluation of performance, 
the installation crew will move on to the next station for the next deployment. 
In those rare occasions when the CDAS operator later reports a problem, it 
is a simple matter to return to the site and retrieve the module. We hope 
to learn a lot about deployment rates early in the Engineering Array 
deployment experience and adjust the plan and equipment requirements 
accordingly.

%

\section{Reception and testing of Surface Detector Electronics kits}


The integrated electronics board will be fabricated in 3-4 fabrication sites by following the  the Quality Management Plan of the current electronics~\cite{SDE_QMP}. The boards will be conformally coated and tested (including thermal stress testing) following the test plan described in~\cite{SDE_test_procedure} after which the boards will be shipped to the Auger Observatory. The GPS receivers will be procured and tested and also shipped to the Observatory. The front panel of the electronics enclosure and the cables and other loose parts will be fabricated and shipped to the Observatory. 

In the Auger Observatory the various electronics parts are stored in a specific building called SDEco. They will first be visually inspected and then assembled into the electronics enclosure (Ekit). The final end-to-end tests are performed following the SDE test plan \cite{SDE_test_procedure} after which the Ekits are ready for deployment.  All test results will be stored in a database, as with the previous electronics test procedure.

\section{Deployment of new SDE kits and small PMTs}


As mentioned earlier, the electronics and small PMT deployment will be
separate from scintillator deployment because the time scale and
talent requirements are different.  The electronics deployment is very
similar to what is currently done for the electronics
maintenance. Therefore, Ekits with new electronics can be deployed
during the regular maintenance of the Surface Detector, when the old
electronics Ekit will be simply replaced by a new Ekit. The new
electronics can be downgraded to the function of the current
electronics. In particular, the data can be compressed to 40\,MHz
traces which would allow application of the same trigger and data analysis
routines as with the current electronics. This would allow a continued
maintenance of the Surface Detector by using new electronics with no
noticeable effect on data. The small PMT can be very easily deployed
whenever a new Ekit is installed.  On average, about 18 detector
stations are maintained per week or about 800 per year. This would
allow an upgrade of the array in two years.

In addition to the maintenance deployment, a dedicated deployment of
electronics will be done. The strategy is to gradually increase the
area of surface detectors with new electronics. In particular, this
deployment will include pre-production phases corresponding to
different electronics production sites (a total of about 100 detectors).
The experience from the previous deployment of the Surface Detector
electronics suggests that new electronics and small PMT installation
will require 35 minutes. Therefore, typically 8 Ekits and 8 small PMTs
can be deployed during one trip with a crew of two technicians. This
would allow a deployment of about 100 Ekits and small PMTs per month and
shorten the total deployment time to about one year. The addition of 
these 100 Ekits and small PMTs per month in a contiguous pattern will 
allow the subsequent installation of scintillator modules to also occur 
in a contiguous pattern, therefore maximizing the number of showers 
that can be detected with all triggered stations belonging to the 
upgraded array.

It is estimated that the additional deployment of the new electronics and small PMTs  would require the hiring of one additional technician for the period of two years (see the WBS). Additional help will be provided by students as was done in the previous electronics deployment.

\section{Reception and testing of Underground Muon Detectors}
\label{sec:receptionUMD}
\begin{figure}[!th]
\centering
\includegraphics[width=0.48\textwidth]{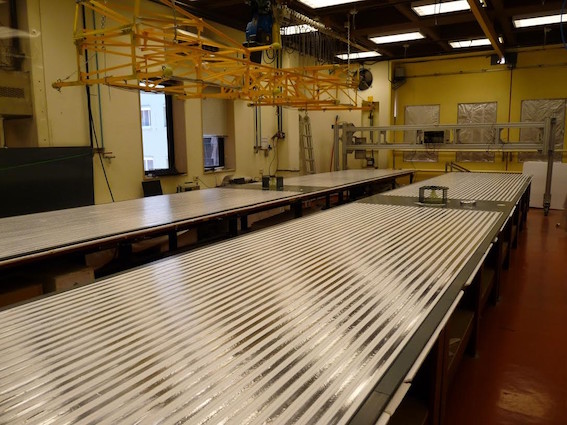}
\includegraphics[width=0.2\textwidth]{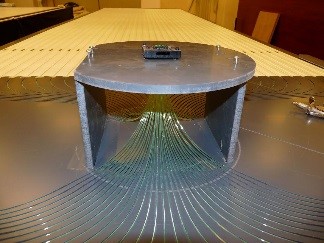}
\caption{Two 10 m$^2$ AMIGA modules being manufactured. The 64 scintillator bars are grouped into two groups
of 32 bars each, with the optical fibers curved up in the middle (see insert) onto an optical coupling
device for the PMT.}
\label{fig:scint_modules}
\end{figure}

The underground detectors have a modular design in order to speed up the assembly process and to allow easy handling 
and transportation~\cite{AMIGA_prototype}. The modules are divided into three main parts: 
i) WLS fibers and optical connector, ii) scintillators and fiber routing components, and iii) PVC casing 
(see Fig.~\ref{fig:scint_modules}).

The fiber package has 64 WLS fibers threaded into the optical connector, which is identical in all detector modules. 
Fibers have different lengths (varying from approximately 4.3 m to 4.9 m) according to their routing in the 
PVC manifold towards the optical connector (see Fig.~\ref{fig:scint_modules}). Also, each fiber far-end 
(i.e., away from the optical connector and PMT, see Fig.~\ref{fig:scint_modules} upper corner) is painted black 
to avoid reflections that would impoverish both timing and pile-up accuracies. The nearside is leveled and polished 
with a fly-cutter which is instrumented with two diamond tool bits. This fiber package is the most sensitive part 
of the detector to mechanical stress, so it is to be built in laboratories with testing facilities. It might be packed 
into a special transportation tube if the module is finally to be assembled at the Observatory campus. 
The manufacturing laboratories must be clean and with a quality-assurance procedure plan. The testing of this 
critical package will include mechanical verification of the fibers (diameter, lengths, and fly-cutting quality) 
and of the optical connector dimensions and alignment. Light-transmission tests could be performed with a LED source 
fired directly onto the fibers and/or with a radioactive source on an already assembled module. A final quick test 
with background radiation is to be performed after the modules are deployed in the field for both quality assurance 
and calibration.

The fiber routing parts, the scintillating bars, the fiber package, and the PVC casing sheets may be transported 
to the module assembly facility. This final assembling of the underground detector modules may be performed 
at different laboratories and at the Auger Observatory facilities. All parts are finally glued together to form 
a solid and robust plastic detector. The optical cement on the scintillator groove which glues fibers to scintillators 
is to be well controlled and inspected before closing the module in order to ensure no detector sectors have
bad optical coupling.

\section{Deployment of Underground Muon Detectors} 

The AMIGA engineering array has shown that an underground detector (three scintillating modules of 10 m$^2$ each) is 
easily deployed in two days, including the hole filling, cabling, electronics deployment, and final testing. 
In production, two machines might be required to perform in parallel the mentioned tasks in order to streamline the job.
Deployment of a 10 m$^2$ module requires a 15 m$^2$ hole (5 m$^2$ are for maneuvering), and 10 m$^2$ for the excavated 
soil to be deposited alongside the hole. This hole would amount to 21 m$^3$ of removed soil.

\begin{figure}[!ht]
\centering
\includegraphics[width=0.35\textwidth]{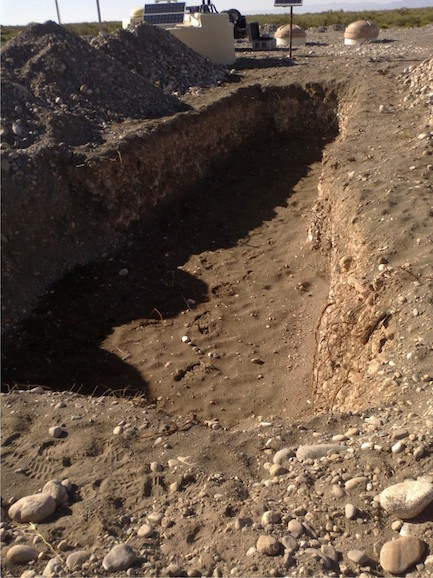}
\includegraphics[width=0.55\textwidth]{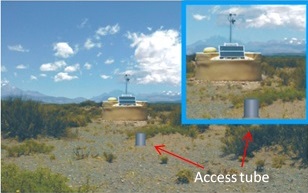}
\caption{Left: 1.3 m deep hole for a 10 m$^2$ module; Right: photograph of final deployment.}
\label{fig:scint_deployment}
\end{figure}  

The production deployment strategy consists of placing the three modules into a transportation container, 
mounting the container onto a cart provided with its own crane, and driving to the deployment destination with a pick-up 
truck. The three modules are lifted and deployed by a team of three technicians while the retro-digger machine 
opens another pit. After the modules are buried, the cabling is performed, electronics deployed, and a first diagnostic 
and calibration program is run. The experience with the engineering array has also shown that this diagnostic program 
is not needed to be run before the detector is buried.   However, it could be performed for the first production modules, 
and then in a sampling mode, to check if good deployment conditions are maintained in the long term.

\section{Maintenance of Surface Detector System}

Maintenance of the present surface detector, 
surface detector electronics, and communications systems is 
done by a local crew of technicians at the Observatory. 
Two to three field trips per week are required, mainly to 
replace batteries, repair PMTs and electronics, to occasionally 
replace solar power regulators or solar panels, and to clear 
the detectors of (potentially flammable) vegetation and birds' 
nests.

The expected failure rate of the scintillation detectors 
is very low. A more precise estimate of the maintenance 
requirements will be available once some experience has been 
gained and the common failure modes have been identified. 

It is expected that scintillator modules with a failure will 
be brought back to the Central Campus for repair, using the 
same equipment that was used for deployment. Perhaps the 
addition of one technician to the presently available crew 
will be an adequate allowance for the additional components 
(additional scintillator module, additional PMT in the tank) 
as well as for the aging of the existing equipment.

\section{Maintenance of Underground Muon Detectors}

%

\begin{figure}[!ht]
\centering
\includegraphics[width=0.33\textwidth]{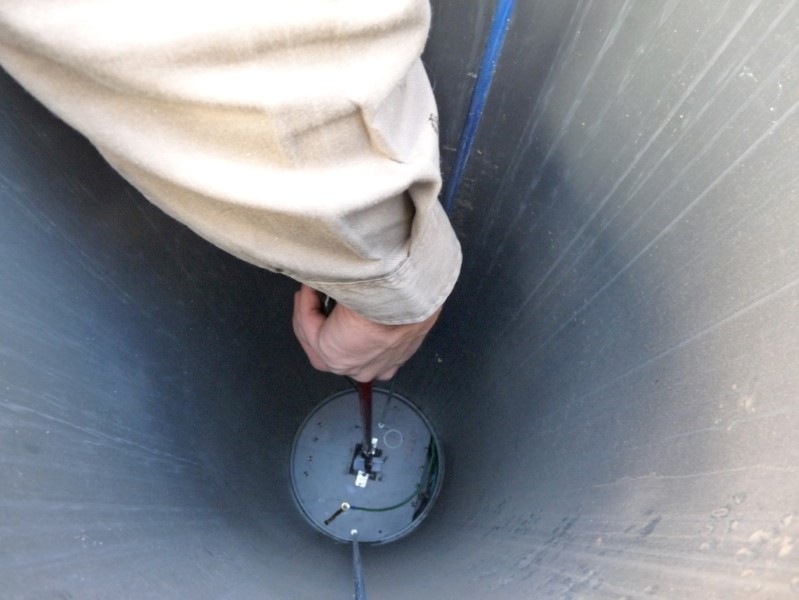}
\includegraphics[width=0.3\textwidth]{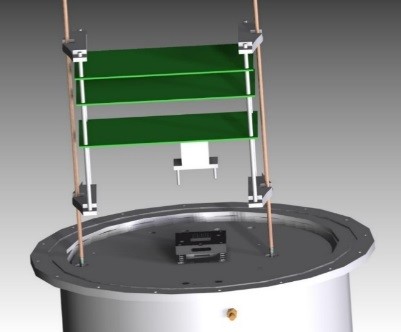}
\includegraphics[width=0.332\textwidth]{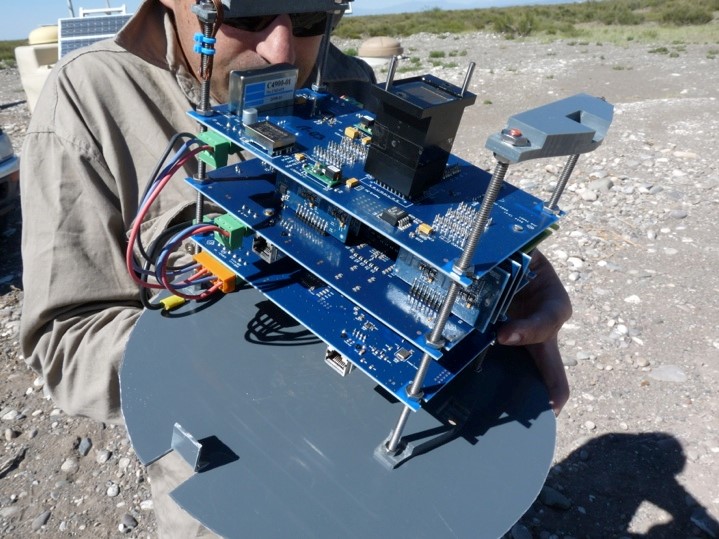}
\caption{Electronics deployment and maintenance. Left: hook; center: diaphragm; right: electronics with PMT and pins.}
\label{fig:electronics_deployment}
\end{figure}

Maintenance would mainly be performed on the electronics since the underground conditions at the Auger Observatory 
site seem to be convenient for the modules. The PVC casing does a very good job in providing water seal while the 
ground above the modules is a very good thermal insulator and a perfect light seal. Electronics is not to be repaired 
in the field but rather in trained laboratories. Removal of the electronics kit and PMT is performed from the surface 
with a probe which hooks on to the system (Fig.~\ref{fig:electronics_deployment}, left). The electronics kit slides out on two 
thin metal rods (see Fig.~\ref{fig:electronics_deployment}, center). The PMT has two pins 
(Fig.~\ref{fig:electronics_deployment}, center and right) to match two conical holes bored into the optical connector. 
This procedure has already been practiced in the engineering array.

The refilling of this small access tube can be done with small soil bags introduced into the tube. They would be light 
enough to be lifted without a crane. 

Very few problems were found in both electronics kits and cabling since the first module was deployed in November 2009. 
No problems were found concerning PMTs, probably due to the favorable operating conditions of temperature and humidity.

\section{Maintenance of the Upgraded FD}

Four members of a local crew of technicians at the Observatory provide
support for the operation and maintenance of the fluorescence detector
including atmospheric monitoring devices. Currently, the regular
maintenance work is done outside data taking periods and repairs
are performed after each night in the case of a malfunction. An expert is
needed during the data taking period to monitor the FD performance and
provide support to shifters. We don't foresee any increase of manpower
needed for the extended FD operation.

Travel to FD buildings by two crew members will be required in the case
of a serious malfunction and a regular schedule of visits to the buildings
will be maintained. The failure rate of various electronic components (typically
moving mechanical parts and aging because of drying out capacitors), having
been installed a decade ago, has been increasing in recent years and reparations
are required by consuming still available spare parts. We expect
higher failure rates after extending the FD operation, until most of the failing
components have been renewed.

Very few problems have been experienced with PMTs and head
electronics, but this system will be monitored closely
during the longer data taking period. In addition, the slow control system
will also require more attention as it almost doubles its operation time.

The costs of implementation of the FD extended operations do not appear
in the WBS costing of the upgrade because no extra equipment is needed
to extend operations into times of increased night sky background. Maintenance
of the existing FD systems is covered by continuing memoranda of understanding
with collaborating institutions.

%% file: organization_and_management.tex

\chapter{Upgrade Organization and Management}


\section{Introduction}
An International Agreement, {\em The Agreement for the Organization, Management and Funding of the Pierre Auger 
Observatory}~\cite{Agreement}, approved in 1999, provides the basis for the organizational structure of the Auger Project.  An extension of the International Agreement to 2025 has been approved by the Auger Finance Board and awaits final signatures.

The Auger Project Management Plan~\cite{PMP}  has provided the basis for the detailed organization and management both 
for construction and operation of the Auger Observatory since its adoption by the Collaboration in 2001.  
The Auger Project employs the tools of management that have been learned over recent years from industry and 
large science projects like those at Fermilab and CERN.  These include cost and schedule tracking, elements 
of systems engineering, quality assurance, risk management and ES\&H programs.  These tools were adapted to 
the particular needs of Auger, an international project based on an equal partnership of participating countries.
Having successfully 
guided the Auger Project since its inception, the essential elements of the management structure will remain in place 
for the Observatory Upgrade.  An updated version of the Auger Project Management Plan describes organizational and 
management features particular to the Upgrade.

\section{Organization}
\begin{figure}[t]
\centering
\includegraphics[width=0.95\textwidth]{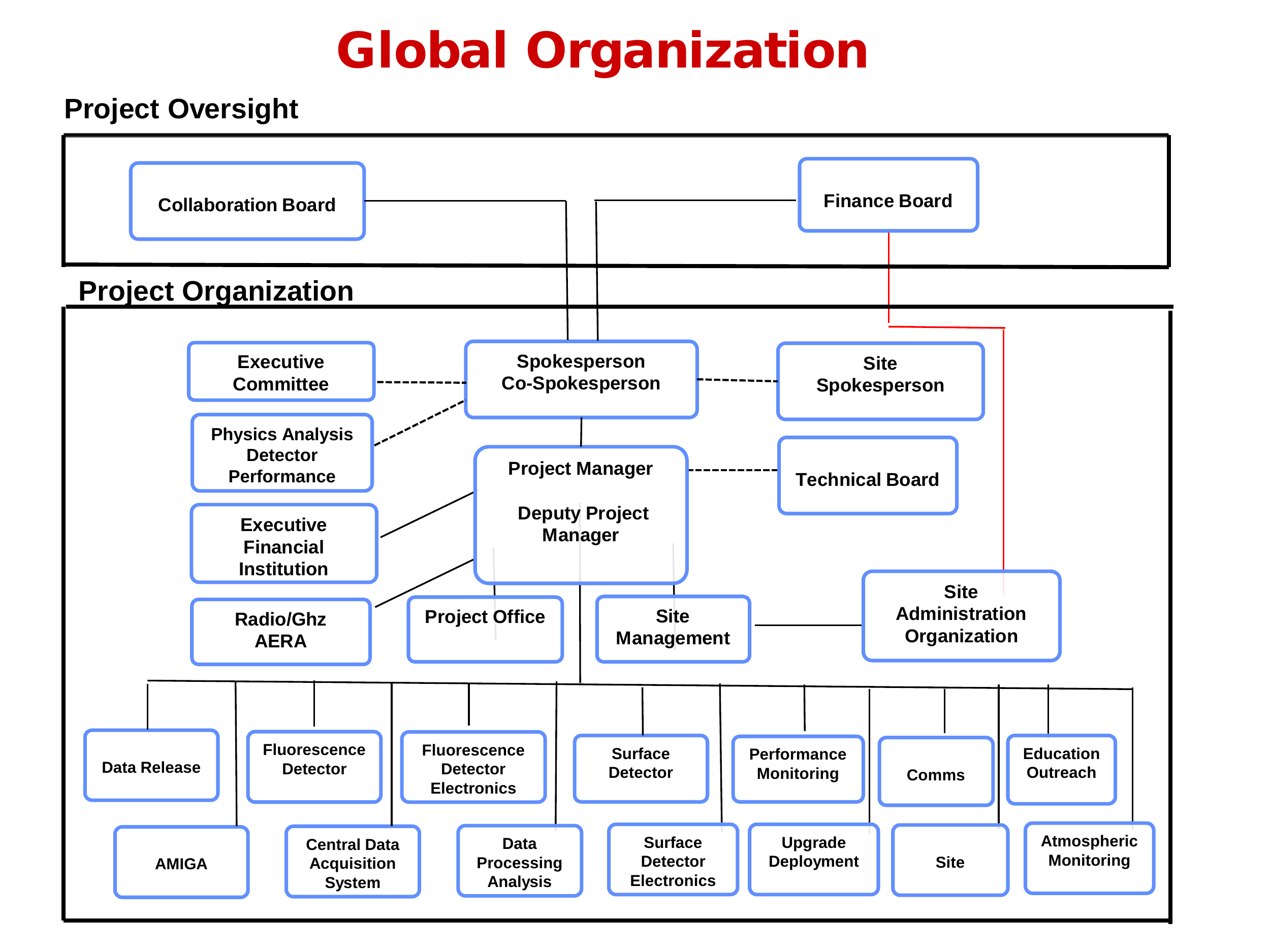}
\caption{Auger Project organization and oversight}
\label{fig:organization}
\end{figure}

The Pierre Auger Project has administrative oversight, both by the scientific collaboration through the Collaboration Board (CB),
and by the funding agencies of 
the participating countries through the Finance Board (FB).  
The global 
project organization has been established by the International Agreement, a non-legally binding agreement among the 
funding agencies of the countries committing support to the construction, operation and upgrade of the Auger Observatory.  
The Agreement contains a statement of continuing support of the Auger Project, rules for the movement of experimental 
equipment through customs, a statement regarding ownership of Observatory property, tax status and related issues.  
The Pierre Auger global organization is shown in Figure~\ref{fig:organization}.

\section{Management Structure}

\subsection{Project Oversight}

The Finance Board is the oversight body for the financial aspects of the Project and is comprised of representatives 
of the funding agencies or their designees and works by consensus.  The Finance Board assures the respective governments 
that the project is meeting its funding goals and furthermore provides a mechanism whereby funding problems associated 
with differences in accounting procedures, currency fluctuations, short term availability of funds within each 
country, etc. can be addressed.
The Finance Board approves a yearly budget and financial reports presented by the Project Manager.  It also receives 
status reports from the Project Manager and Spokesperson to ensure that schedule and budgetary goals are met.  
The Finance Board meets in-person on an annual basis and by phone conference as needed.

The Collaboration Board is the principal oversight body concerned with the scientific and technical aspects of the Project.  
It deals with issues including governance, scientific policy, admission policy of new members and institutions, 
publication policy and monitoring of the construction and operation of the Pierre Auger Observatory to ensure 
that the scientific objectives of the project are met.
The members of the Collaboration Board are appointed by their home institutions among the scientists participating 
in the Pierre Auger Collaboration.  Each institution with three or more active collaborators has one representative 
on the Collaboration Board.  Institutions having fewer than three collaborators may join with other institutions and 
have a common representative.  Details on organization of the Collaboration Board are described in its bylaws that can be 
found in the Project Management Plan.

\subsection{Project Organization}

The scientific and technical direction of the project is invested in the Spokesperson 
and the Co-Spokesperson by the Collaboration Board.  
The Project Manager is responsible for the operation of the Observatory and reports to the Spokesperson.  
The Project Manager is supported by the Project Office staff and by the Technical Board.  A Site Manager directs 
operations on-site and reports directly to the Project Manager.  Task Leaders are responsible for the major components 
of the detectors.

The Auger Project Spokesperson and Co-Spokesperson are elected by the Collaboration via the CB
and represent the Collaboration 
in scientific, technical, and management concerns.  The Spokesperson speaks and negotiates on behalf of the Collaboration.  
The Spokespersons are responsible for establishing the scientific goals and the means for the Collaboration to pursue these 
goals successfully.  They are also expected to pursue the identification of resources needed by the Auger Project and 
to seek the commitment of such resources toward the operation of and upgrades to the Observatory.  These resources 
come from the scientific groups and institutions that collaborate in the Auger Project, as well as their various 
sources of funding for that purpose.  The Spokespersons serve renewable three-year terms.

The Executive Financial Institution maintains an account for the operating funds and the operations reserve.  
The current Executive Financial Institution is the Karlsruhe Institute of Technology (KIT) in Karlsruhe, Germany.  
The operating and 
reserve funds are invoiced by KIT and disbursed at the request of the Project Manager.

The site spokesperson is the liaison between the Project Spokesperson and the host country.

The Project Manager is responsible for the operation of the Pierre Auger Observatory.  The Project Manager's duties 
and responsibilities for operations include oversight of the day-to-day activities of the Observatory, preparation 
of budgets and tracking of expenditures, monitoring the environment, safety, and health program, 
serving as chair the Technical Board, 
carrying out technical reviews and preparation of MOUs for Operations with collaborating institutions.  
The Project Manager serves an indefinite term.  The Project Manager reports to the Project Spokesperson.
The Deputy Project Manager works with the Project Manager in carrying out Project Management responsibilities.
They are assisted by personnel that make up the Project Office.  
These include a cost and schedule officer, an ES\&H officer and clerical support.  The Project Office is currently 
located at KIT. 

The Task Leaders (level 2 managers) played a fundamental role in the construction and operation of 
the Observatory, and will also do so for the Upgrade.  
All tasks are broken down into a number of subtasks each with its own leader.  These Subtask Leaders 
support the Task Leader in carrying out his/her duties.  Task Leaders report to the Project Manager.  

The Technical Board consists of scientists and engineers involved in leadership roles in the various technical areas 
of the Auger Project.  The members of the Technical Board are the Spokesperson, Task Leaders and others appointed 
by the Spokesperson and the Project Manager.  The Technical Board is chaired by the Project Manager.  
The Technical Board advises the Project Manager on technical issues pertaining to operation, maintenance and upgrades 
to the Observatory systems. The Technical Board also serves as the Change Control Board.

The Site Management staff is headed by the Site Manager.  Operating funds for the 
Site Management are provided by the participating countries. 

\begin{figure}[t]
\centering
\includegraphics[width=0.85\textwidth]{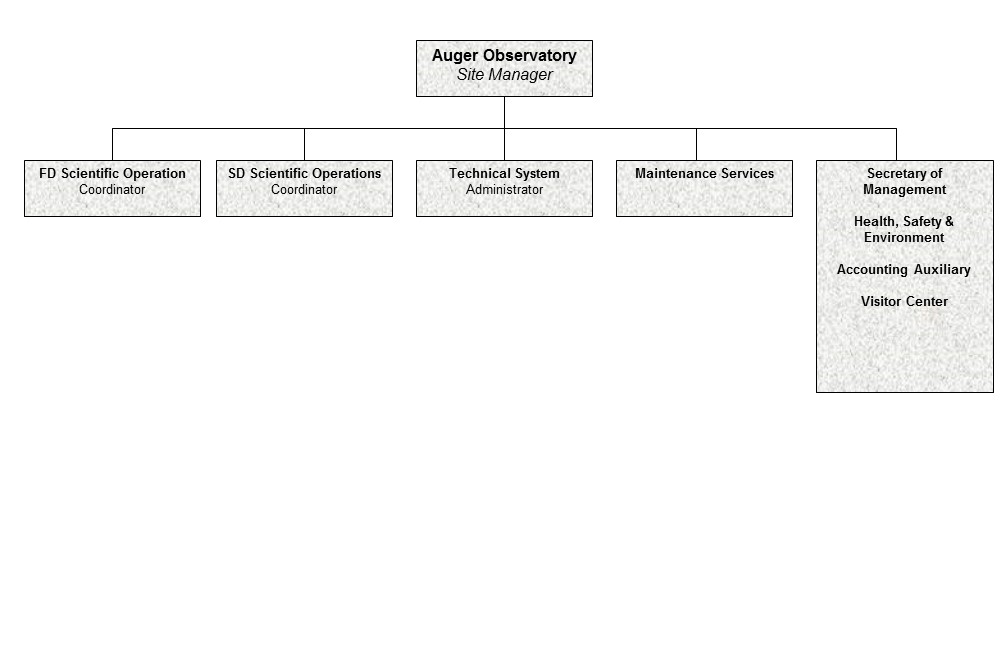}
\caption{The Site Management}
\label{fig:site_org}
\end{figure}

The responsibilities of the Site Manager include supervision of the site staff, writing contracts for local services 
needed for operations, maintenance of the buildings and infrastructure, assisting the Task Groups in receiving and 
storage of equipment, maintenance of the Visitor Center, serving as point of contact with land owners, local 
and state officials and ensuring that the environment is protected and the safety and health of staff and visitors is assured.  
The Auger Site Manager reports to the Pierre Auger Project Manager.  
The site organization is shown in Figure~\ref{fig:site_org}.

The Observatory Site Administrative Organization is responsible for contracting local personnel and for the 
local disbursement of funds in conformity with the laws of Argentina. The original Site Administrative Organization, 
Fundaci\'{o}n Observatorio Pierre Auger Argentina (FOPAA), a non-profit foundation, was established in 2003, with 
bylaws as agreed to by the Pierre Auger Finance Board.  The Site Administrative Organization is the only 
legally recognized entity directly related to the Auger Observatory, other than participating institutions and 
funding agencies in their respective countries. The Site Administration Organization will transition during 2015 to another 
not-for-profit foundation, {\em Ahuekna}, formed by a group of institutions in Argentina that support the Observatory.

In its role as part of the Auger Collaboration, the Site Administrative Organization reports annually to the 
Finance Board on its activities and funding. It handles operating funds on the basis of a budget approved by the 
Finance Board and the instructions of Auger Management. It also handles money and legal issues in consultation with 
the Project Management.  Its responsibilities on behalf of the Auger Collaboration include employment of personnel, 
signing of contracts for services and maintenance, agreements for land usage and the handling of operating funds.

The Site Administrative Organization maintains legally required accounting and record of meeting books, hires lawyers, 
as needed, for labor and other legal issues, and utilizes computerized accounting and salary processing. It carries 
out the legal requirements from the tax and social services authorities and holds bank accounts and insurance policies.

\section{Upgrade Management}
The Auger Upgrade will be managed within the basic organizational structure under which the Observatory was 
successfully constructed.  Minor changes will be made to the organization to reflect the specific responsibilities 
associated with the new elements of the Upgrade. 

The revised Project Management Plan describes the management of the Upgrade.  
Associated management documents including Performance Requirements, Integrated Project Schedule, Work Breakdown 
Structure (WBS), WBS Dictionary, Quality Assurance Plan, Risk Management Plan, ES\&H plan will be updated to 
reflect the particular needs of the Upgrade.

The Project Office will be strengthened to accommodate increased responsibilities of the Upgrade.  In particular, 
project engineering, cost and schedule tracking, quality assurance, risk management and safety will require additional effort.

The management of the construction of the new scintillator detectors will take place within the existing Surface Detector Task.  
New subtasks will be added as necessary.  Upgrades to the Surface Detector electronics will be carried out within 
the existing Surface Detector Task.  Subtasks will be added as appropriate.  

The MOUs with each institution participating in the Upgrade will include the commitment of institutional 
collaborators to the upgrade effort, their deliverables, and delivery schedule.

\section{Quality Management}
The Pierre Auger Project has a Quality Assurance Plan~\cite{QAP} to ensure the performance and reliability of the Observatory systems.  Quality Assurance  
was an integral part of the Observatory design, procurement, assembly and test processes and the responsibility resides 
at all levels of the organization.  The Auger Quality Assurance Plan was developed and will be updated by the 
Project Manager and the Project Quality Assurance Manager to accommodate the specific quality assurance/quality 
control requirements of the Upgrade.

Using the Quality Assurance requirements, each Task Leader has developed a quality system or plan that documents 
their approach and methods for achieving quality components and services.  The Quality Assurance Manager assists 
the Project Manager and Task Leaders in developing their quality systems and audits their implementation.

As part of the Upgrade QA Plan a standard suite of internal technical reviews will be initiated early to ensure that 
the design and fabrication processes will achieve the physics goals of the Upgrade.  The review panels will be made up 
of Auger collaborators, the project office (Project Manager, Project Engineer, QA manager, ES\&H officer and 
Cost/Schedule officer) and outside experts as needed.  The details of the each of the reviews below are contained 
in the Project Management Plan.

\begin{itemize}
  \item Design Requirements Review (DRR) 
  \item Preliminary Design Review (PDR)
  \item Critical Design Review (CDR)
  \item Production Readiness Review (PRR)
  \item Operations Readiness Review (ORR) 
\end{itemize}

\section{Risk Management}
Risk management is a process of the Auger Upgrade activities project. It includes processes for risk management 
planning, identification, analysis, monitoring and control. It is the objective of risk management to 
decrease the probability and impact of events adverse to the project.

The Risk Management Plan (RMP) documents the processes, tools and procedures that will be used to manage and control 
those events that could have a negative impact on the Auger Upgrade activities. It is the controlling document for 
managing and controlling all project risks. This plan will follow the risk policy and will address:

\begin{itemize}
  \item Risk Identification
  \item Risk Assessment
  \item Risk Response
  \item Risk Tracking and Reporting
\end{itemize}

The risk policy, described in the Risk Management Plan, defines the level-values scale for the risk impact on 
schedule, resources and performances, the scoring scheme for the likelihood of occurrence, and a risk index scheme to 
denote the magnitudes of the risks of the various risk scenarios.

For the Auger Upgrade project, the risks are defined in four classes:

\begin{itemize}
  \item External 
  \item Technical and Science 
  \item Sub-Contractor and Industry 
  \item Human and Organization 
\end{itemize}

Risks can be identified from a number of different sources. Some may be quite obvious and are identified prior 
to project kick-off. Others are identified during the project life cycle, and can be identified by 
anyone associated with the project. 

For each risk scenario, the risk assessment process should determine:
\begin{itemize}
  \item The impact on the schedule, resources and performances
  \item The likelihood
  \item The risk index
  \item The risk magnitude (number of risk) per index
\end{itemize}

For each identified risk, a response is identified. The possible response options are:
\begin{description}
  \item[Avoidance:]  Change the project to avoid the risk. 
  \item[Transference:] Shift the impact of a risk to a third party 
  \item[Mitigation:] Take steps to reduce the probability and/or impact of a risk. 
	\item[Acceptance:] Accept the impacts of the risk. 
\end{description}

For each risk that will be mitigated, ways to prevent the risk from occurring or reduce its impact or probability 
of occurring are identified and defined in a mitigation plan. This may include prototyping, adding tasks to 
the project schedule, adding resources, etc.
Note that even if a risk is acceptable, a reducing solution must be studied.

The level of risk on the Auger Upgrade project is tracked, monitored and reported throughout the project life cycle.  
A list of the major risks is maintained and is reported as a component of the project status reporting process.  
All major project change requests are analyzed for their possible impact on the project risks.

The results of the risk analysis for different upgrade components can be found in the 
risks analysis report~\cite{Risk_analysis}. No major risks have been identified. Component procurement and manufacturing will be done typically in at least two sites reducing the risk related to manufacturing. The design requirements for Surface 
Scintillator Detector (SSD), Surface Detector Electronics Upgrade (SDEU), and Underground Muon Detector (UMD)
include easy deployment and maintenance reducing the risk related to the field access. The most important risk is related to funding.

%% file: cost_schedule_funding.tex

\chapter{Cost, Schedule, and Funding}

\section{Introduction}
The cost estimate and schedule for the Upgrade will be based on the Upgrade work breakdown structure (WBS).  
Task Leaders and, in turn, subtask leaders provide the details of the WBS for their subsystems from 
which the cost estimates and schedule are derived.

\section{Cost estimation}
The steps in cost estimate development are:

\begin{enumerate}
  \item The WBS, a list of all components and tasks organized by subsystem, is developed and constitutes the work to complete the project.  Each component and task is a WBS element.
	\item A description for each WBS element is entered in the WBS dictionary.
	\item The estimated cost of the components and activities which comprise the lowest level of the WBS is prepared and a basis of estimate document (BOE) is completed for each component and activity.
\end{enumerate}

The total cost, including infrastructure costs, is included in the project cost.  The cost of project scientists 
and their support (travel etc.) is borne by their home institution and {\em not} included in the project cost.  Data analysis and associated hardware are {\em not} included in the project cost.

Labor rates are established for the institutions at which significant labor will be performed.  For others, 
generic labor rates may be used.  The labor rates are fully burdened.  Typically a burdened labor rate 
includes direct labor, fringe, overhead, vacation, sick leave, and general and administrative costs.

Often labor estimates do not include all labor associated with manufacturing a product (e.g., manufacturing support, 
facility maintenance, etc.) because some of this effort is included in the overhead rate for that institution.  
A description of what is included in overhead at a given location should be provided in the BOEs.

Material costs include the purchase of raw materials for fabrication and the procurement of components, sub-assemblies, 
and tooling from outside sources, or items estimated in such a way that only a total dollar amount can be identified.  
This includes detector hardware, equipment, fixtures, tooling, utilities, test and assembly equipment, computer hardware 
and software, raw materials, and procurement processing.   The BOEs should indicate 
the basis for arriving at the materials cost estimate.

The Auger Cost and Schedule Officer is responsible for assisting the Project Manager in tracking costs and progress 
in the Upgrade to the Auger Observatory.  The principal tools for tracking cost are the Project Schedule and 
the Work Breakdown Structure (WBS). Each task will be broken up 
into a suitable number of intervals or subtasks and milestones to allow effective tracking. 



A preliminary WBS  for the completion of the AMIGA muon counters that will be used for the Underground Muon Detector of the Auger upgrade is contained in Appendix~\ref{app:WBS}.
 

\section{Upgrade Construction Funding}
Each institution makes commitments for deliverables to the Upgrade project by way of a Memorandum of Understanding.  
Each country and the institutions within that country make commitments for deliverables with resources available 
within that country.  Each institution then reports on a monthly basis to the Task Leaders.  The Task Leaders will 
in turn report progress to the Cost and Schedule Officer.  The Cost and Schedule Officer prepares reports for the 
Project Manager, Spokesperson, Collaboration and Finance Board.  The Cost and Schedule Officer flags deviation from 
the project plan for corrective action by the Project Manager.

Country contributions will be primarily in-kind.   Each country will also contribute to a common fund for 
such large procurements as scintillator and PMTs.

\section{Schedule}
The Upgrade project Work Breakdown Structure is the basis for the Upgrade schedule.  Each task will be broken up 
into a suitable number of intervals or subtasks and milestones to allow effective tracking.  
Each Task Leader will be responsible for tracking scheduled activities within his/her own task using the same 
scheduling tool.  Progress reports from the Task Leaders at the WBS level 4 and higher will be transmitted to the 
Project Cost and Schedule Officer on a monthly basis. The Cost and Schedule Officer will, in turn, prepare progress 
reports for use by the Spokesperson, Project Manager, Collaboration Board and Finance Board.

A preliminary  schedule is contained in Appendix~\ref{app:WBS}. 
The components of the baseline design have been tested for their
suitability for the upgrade.  Some specific R\&D is still in progress
to optimize the performance/cost ratio, in particular for the scintillator
detectors and the underground muon detectors. The final validation of the SSD and SDEU
designs will be undertaken in an Engineering Array of 10 detector stations
at the end of 2015. The production and deployment of the SSD, the SDEU
and the UMD will be done in parallel and will extend over the period 2016-18. The
production schedule is mainly driven by the funding profiles in
the countries of the collaboration, not by the production or deployment rates.

\section{Operating Funds}
Contributions by participating countries to Observatory operating costs are based on the Operations Cost List (OCL) 
as is current practice.  The OCL consists of authors of scientific papers excluding students.  The Upgrade scintillation detectors will require little additional maintenance effort as they are simple in design.  Based on previous experience we expect an upper limit of 10 PMT failures per year which can be repaired as part of our regular surface detector maintenance.  Beyond gradual (acceptable) degradation of the scintillator and fiber performance, no other failures of the Upgrade scintillator detectors are expected.  Additional maintenance for the revised operation of the FD is expected to be negligible.  The overall increase in operating costs as a result of the Upgrade is expected to be less than five percent.

%% file: outreach_and_education.tex

\chapter{Outreach and Education}

The scale and scope of the physics explored at the Pierre Auger
Observatory offer significant opportunities for outreach both to the
local community and to the collaborating countries. Education,
outreach and public relations have been an integral part of the Auger
Observatory organization from the beginning when these activities were
included as a level two management task group. The goals of the
Outreach and Education Task are to encourage and support a wide range
of efforts that link Auger scientists and the science of particle
astrophysics, particle physics and related technologies to the public
and especially to schools.  Outreach focused on the communities
surrounding the Observatory has fostered a remarkable amount of
goodwill, which has contributed significantly to the success of the
project. The Auger Collaboration initiated outreach first locally as a
way to become better integrated into the community during the
construction phase of the Observatory. Later outreach activities
spread to the participating institutions but on a larger scale and to
the Internet.

\begin{figure}[!th]
\begin{center}
\hspace*{-2ex}\includegraphics[width=0.517\textwidth]{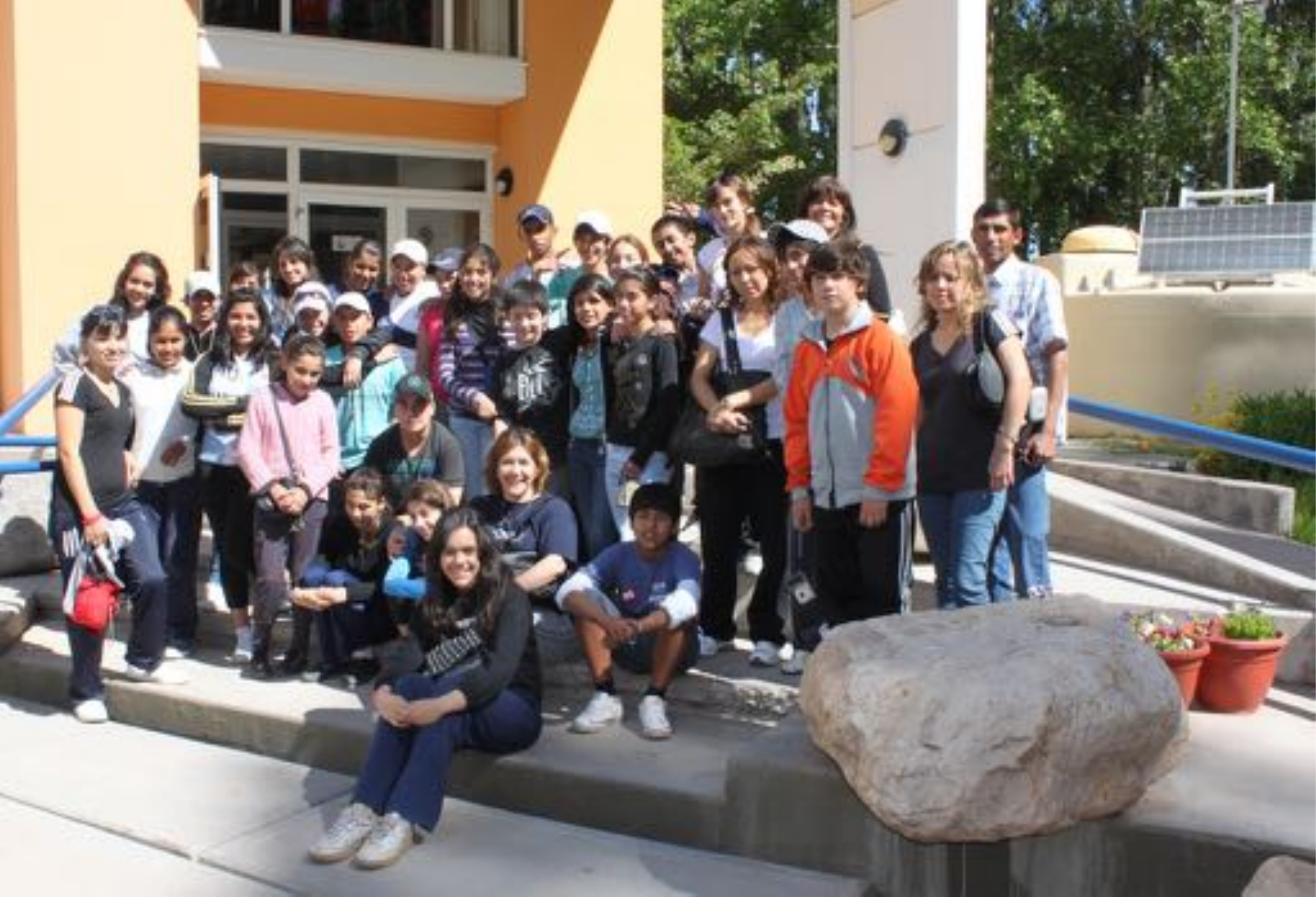}
\includegraphics[width=0.47\textwidth]{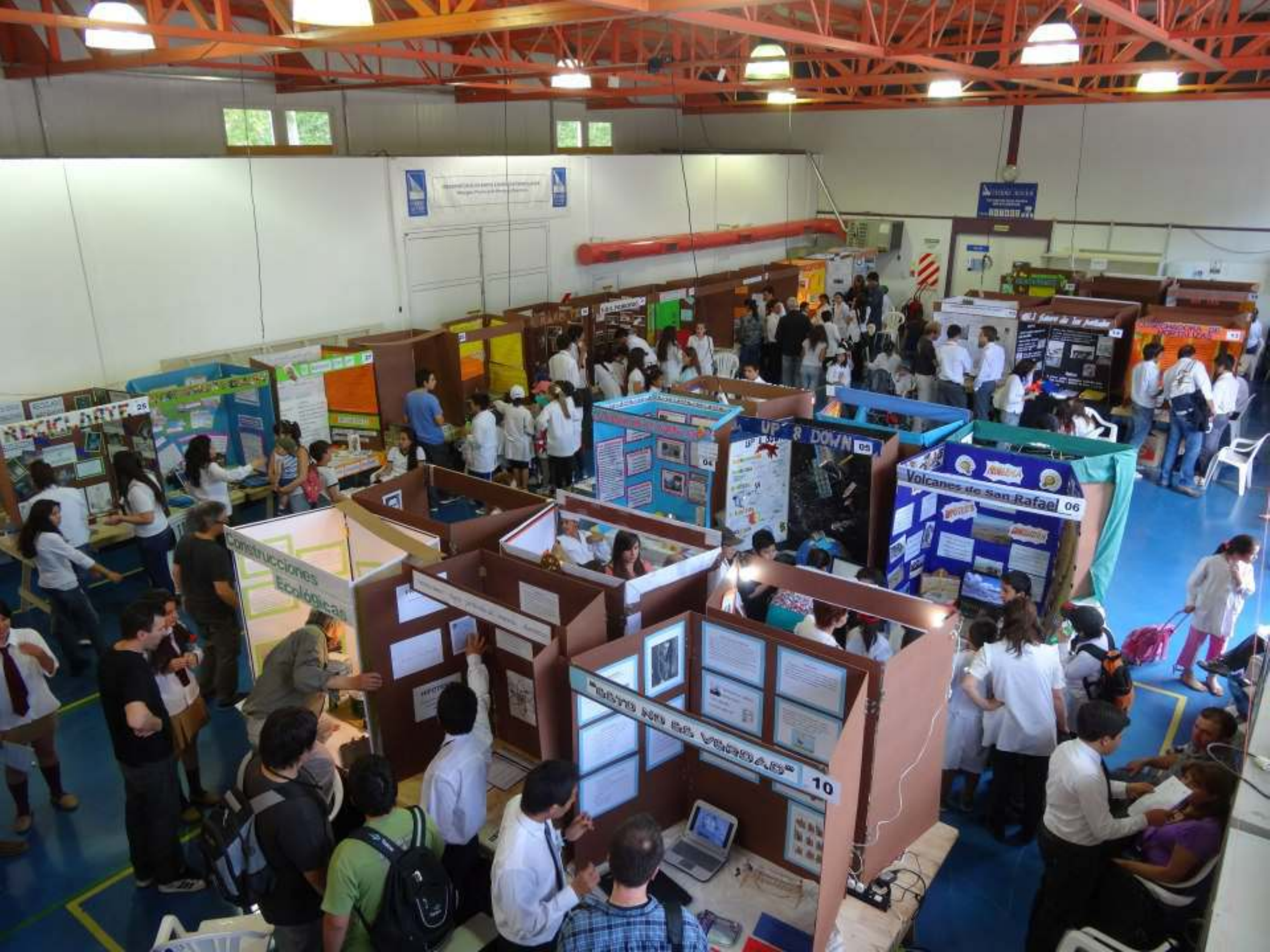}
\caption{Left panel: A group of middle school students and teachers
  after their Visitor Center tour. A surface detector station is
  visible in the upper right which visitors can inspect closely. Right
  panel: A recent Science Fair in the Assembly Building on the Auger
  campus.}
\label{fig:outreach_photos}
\end{center}
\end{figure}

The heart of local outreach activities is the Auger Visitor Center
(VC), located in the central office and data acquisition building on
the Observatory campus in Malarg\"{u}e.  A staff member dedicated to
outreach gives presentations and tours to visitors that are mostly
from the area but often from all over Argentina and even from other
countries worldwide. Many of the visitors are in the area because of
the proximity of the Las Le\~{n}as ski area and other area tourist
attractions. Almost 100,000 people have attended the lectures in the
Visitor Center since it opened in 2001. The impact of these visits can
be seen from the continuing interest and the comments in the guest
book. The VC, which seats up to 50 people, is outfitted with
multimedia equipment and contains a number of displays illustrating
features of the Observatory.  These displays include a full size SD
station, a quarter sized model of an FD mirror, a spark chamber, a
Geiger counter, a number of posters that explain the science and
detectors of the Observatory and a library of books in several
languages. Fig.~\ref{fig:outreach_photos}(left) shows a visiting group
of middle school students and teachers outside the Auger office
building after their Visitor Center tour.

The Auger upgrade provides an excellent opportunity to modernize the
VC, where the emphasis can be moved from a fully organized tour to a
hands-on multimedia experience. Such a new setting will allow a
continuously changing exhibition in which the physics of cosmic rays
and the motivation for the upgrade of the Auger detector can be
explained to the public at large. A more modern VC provides motivation
to the local community to re-acquaint itself with the Observatory.

Every two years a Science Fair is organized by the Observatory which overlaps with a collaboration meeting in Malarg\"{u}e. The
fair targets both elementary schools and high schools, and is still
growing. The latest fair hosted 33 entries with schools from all over
Mendoza Province participating. The exhibits and presentations of the
participants were judged by international members of the Auger
Collaboration. The interaction of the participants with the Auger
scientists reinforces the connection between Auger and surrounding
communities. Fig.~\ref{fig:outreach_photos}(right) shows a recent
Science Fair in the Assembly Building.

The collaboration realizes the importance of making data available to
the general public as well as other scientists. For this purpose a
special Data Release task has been established. For outreach purposes,
the Collaboration makes 1\% of recorded surface detector events
available on its public web page (www.auger.org), and this fraction is
foreseen to increase to 10\% in the near future.  The current public
data set contains over 38,000 events and has been analyzed by students
worldwide for science fairs, classroom activities, and
research projects. An explanation of how to handle this wealth of
data, aimed at the high school level and beyond, is provided at the
same web location. The online event display, coupled to the public
data set, is a useful tool to provide insight to students on what is
measured and how it is interpreted. Furthermore, an online analysis
interface called VISPA~\cite{vispa} has been developed, which allows
students to work with and analyze these data. The explanation, event
display, and VISPA form a complete set of tools that allow students to
work as scientists on the largest cosmic ray data set to become
available to the general public.

Outreach has been an important part of the activities of the Auger
Observatory that will continue through its lifetime. Our close relationship with the people of Malarg\"{u}e
and the other local communities as a result of our outreach activities
has not only made our work comfortable and rewarding but has, indeed,
contributed to the success of the Observatory. Among the collaborating
institutions many innovative outreach ideas have sprung from our
research, leading to, for example, institute open houses, public
exhibitions to large audiences, and teacher workshops featuring the
Observatory, its science and accomplishments. Because we can easily
show their continuous presence around us, cosmic rays provide an
effective vehicle to excite young people about the wonders and science
of the cosmos.

%% file: acknowledgements.tex
\chapter{Acknowledgements}

The successful installation, commissioning, and operation of the
Pierre Auger Observatory would not have been possible without the
strong commitment and effort from the technical and administrative
staff in Malarg\"{u}e.  We are very grateful to the following agencies
and organizations for financial support:

\begin{flushleft}
 Comisi\'{o}n Nacional de Energ\'{\i}a At\'{o}mica, 
 Fundaci\'{o}n Antorchas, Gobierno de la Provincia 
 de Mendoza, Municipalidad de Malarg\"{u}e, 
 NDM Holdings and Valle Las Le\~{n}as, in gratitude 
 for their continuing cooperation over land access, 
 Argentina; the Australian Research Council; Conselho 
 Nacional de Desenvolvimento Cient\'{\i}fico e 
 Tecnol\'{o}gico (CNPq), Financiadora de Estudos e 
 Projetos (FINEP), Funda\c{c}\~{a}o de Amparo \`{a} 
 Pesquisa do Estado de Rio de Janeiro (FAPERJ), 
 S\~{a}o Paulo Research Foundation (FAPESP) 
 Grants No. 2010/07359-6 and No. 1999/05404-3, 
 Minist\'{e}rio de Ci\^{e}ncia e Tecnologia (MCT), 
 Brazil; Grant No. MSMT-CR LG13007, No. 7AMB14AR005, 
 and the Czech Science Foundation Grant No. 14-17501S, 
 Czech Republic;  
 Centre de Calcul IN2P3/CNRS, Centre National de la 
 Recherche Scientifique (CNRS), Conseil R\'{e}gional 
 Ile-de-France, D\'{e}partement Physique Nucl\'{e}aire 
 et Corpusculaire (PNC-IN2P3/CNRS), D\'{e}partement 
 Sciences de l'Univers (SDU-INSU/CNRS), Institut 
 Lagrange de Paris (ILP) Grant No. LABEX ANR-10-LABX-63, 
 within the Investissements d'Avenir Programme  
 Grant No. ANR-11-IDEX-0004-02, France; 
 Bundesministerium f\"{u}r Bildung und Forschung (BMBF), 
 Deutsche Forschungsgemeinschaft (DFG), 
 Finanzministerium Baden-W\"{u}rttemberg, 
 Helmholtz Alliance for Astroparticle Physics (HAP), 
 Helmholtz Gemeinschaft Deutscher Forschungszentren (HGF), 
 Ministerium f\"{u}r Wissenschaft und Forschung, Nordrhein Westfalen, 
 Ministerium f\"{u}r Wissenschaft, Forschung und Kunst, Baden-W\"{u}rttemberg, Germany; 
 Istituto Nazionale di Fisica Nucleare (INFN), Istituto Nazionale di Astrofisica (INAF), Ministero dell'Istruzione, dell'Universit\'{a} 
 e della Ricerca (MIUR), Gran Sasso Center for Astroparticle Physics (CFA), CETEMPS Center 
 of Excellence, Ministero degli Affari Esteri (MAE), Italy; 
 Consejo Nacional de Ciencia y Tecnolog\'{\i}a (CONACYT), Mexico; 
 Ministerie van Onderwijs, Cultuur en Wetenschap, 
 Nederlandse Organisatie voor Wetenschappelijk Onderzoek (NWO), 
 Stichting voor Fundamenteel Onderzoek der Materie (FOM), Netherlands; 
 National Centre for Research and Development, 
 Grants No. ERA-NET-ASPERA/01/11 and 
 No. ERA-NET-ASPERA/02/11, National Science Centre,
 Grants No. 2013/08/M/ST9/00322, No. 2013/08/M/ST9/00728 
 and No. HARMONIA 5 - 2013/10/M/ST9/00062, Poland; 
 Portuguese national funds and FEDER funds within 
 Programa Operacional Factores de Competitividade 
 through Funda\c{c}\~{a}o para a Ci\^{e}ncia e a  Tecnologia (COMPETE), Portugal; 
 Romanian Authority for Scientific Research ANCS, 
 CNDI-UEFISCDI partnership projects Grants No. 20/2012 
 and No. 194/2012, Grants No. 1/ASPERA2/2012 ERA-NET, 
 No. PN-II-RU-PD-2011-3-0145-17 and No. PN-II-RU-PD-2011-3-0062, 
 the Minister of National  Education, Programme  
 Space Technology and Advanced Research (STAR), 
 Grant No. 83/2013, Romania; 
 Slovenian Research Agency, Slovenia; 
 Comunidad de Madrid, FEDER funds, Ministerio de Educaci\'{o}n y Ciencia, 
 Xunta de Galicia, European Community 7th Framework Program, 
 Grant No. FP7-PEOPLE-2012-IEF-328826, Spain; 
 Science and Technology Facilities Council, United Kingdom; 
 Department of Energy, 
 Contracts No. DE-AC02-07CH11359, No. DE-FR02-04ER41300, 
 No. DE-FG02-99ER41107 and No. DE-SC0011689, 
 National Science Foundation, Grant No. 0450696, 
 The Grainger Foundation, USA; 
 NAFOSTED, Vietnam; 
 Marie Curie-IRSES/EPLANET, European Particle Physics 
 Latin American Network, European Union 7th Framework 
 Program, Grant No. PIRSES-2009-GA-246806; and UNESCO.
\end{flushleft}

%% file: appendix-wbs_schedule.tex
\chapter{Work Breakdown Structure (WBS) and Schedule}
\label{app:WBS}

\section{WBS}

\begin{figure}[!ht]
\centering
\includegraphics[width=1.\textwidth]{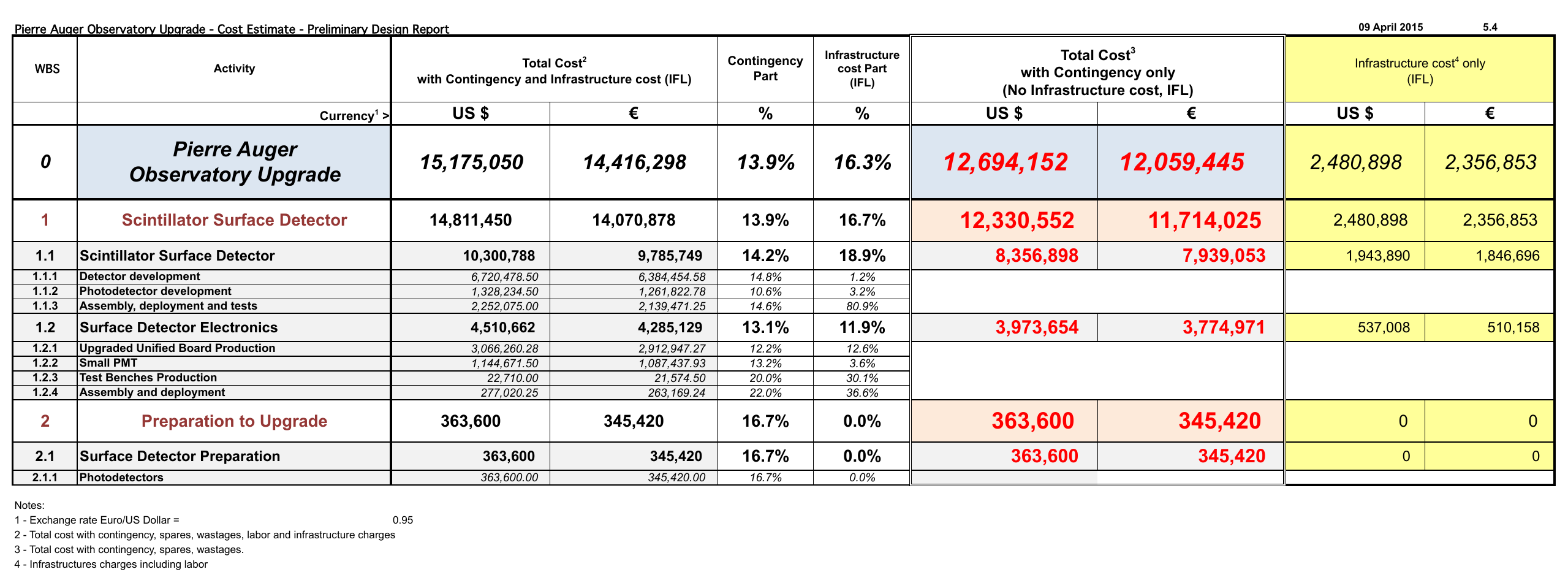}
\caption{Summary of the SDEU and SSD WBS.}
\label{fig:WBS2}
\end{figure}

A summary of a preliminary  WBS for the Auger upgrade including SDEU and SSD is shown in Fig.~\ref{fig:WBS2}. 

The total cost with contingency, including infrastructure costs, is
estimated to be US \$15.2\,M, and the total cost with contingency, but without
infrastructure costs (the so-called European accounting), is estimated to be
US\$12.7\,M. The overall increase in operating costs as a result of
the upgrade is expected to be less than five percent.  A more
detailed WBS is available in ref.~\cite{WBS}.

The costs of implementation of the extended FD duty-cycle
(Chap.~\ref{FDextension}) do not appear in the WBS because no extra
equipment is needed to extend operations into times of increased night
sky background. Maintenance of the existing FD systems is covered by
continuing memoranda of understanding with collaborating institutions.
However, provision is made for purchasing 300 water-Cherenkov detector
(WCD) photomultipliers as spares for future operation (Item 2.1.1 in
Figure~\ref{fig:WBS2}).  The WCD PMTs were originally purchased with
the collaboration's Common Fund, and hence no single institution has
explicit responsibility for their maintenance.

\begin{figure}[!ht]
\centering
\includegraphics[width=.9\textwidth]{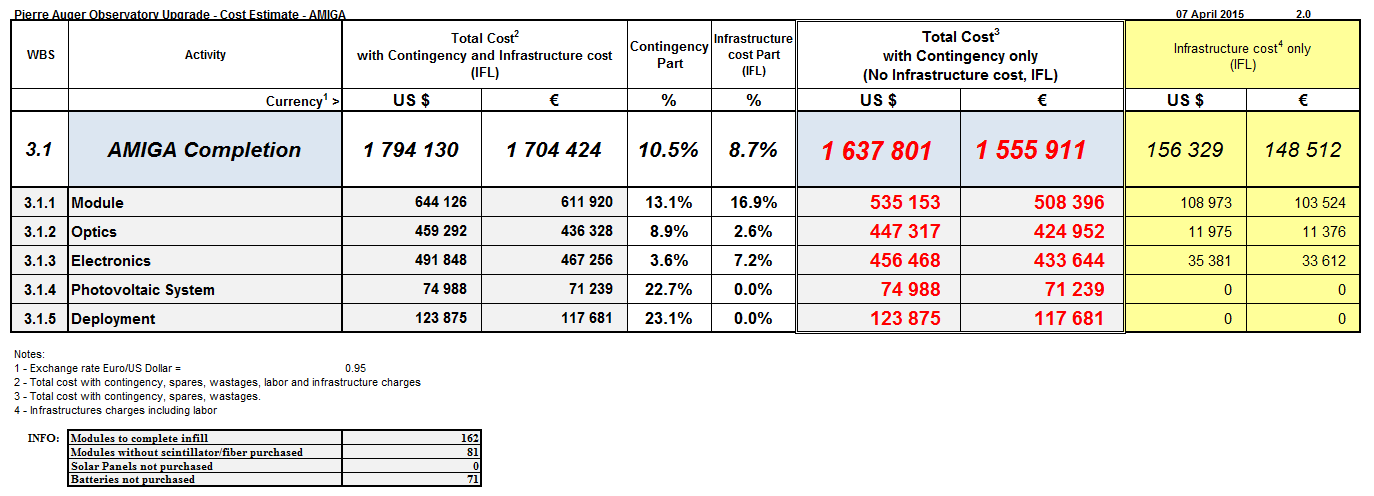}
\caption{Summary of the AMIGA WBS.}
\label{fig:WBS_AMIGA}
\end{figure}

A summary of a preliminary  WBS for the completion of the AMIGA muon
counters that will be used for the Underground Muon Detector of the
Auger upgrade is shown in Fig.~\ref{fig:WBS_AMIGA}.

The current cost estimate for the completion of the AMIGA counters, without the infrastructure cost, is
US\$1.6\,M.  A more detailed WBS is available in ref. \cite{WBS}.

\section{Schedule}

\begin{figure}[!ht]
\centering
\includegraphics[width=.8\textwidth,angle=0]{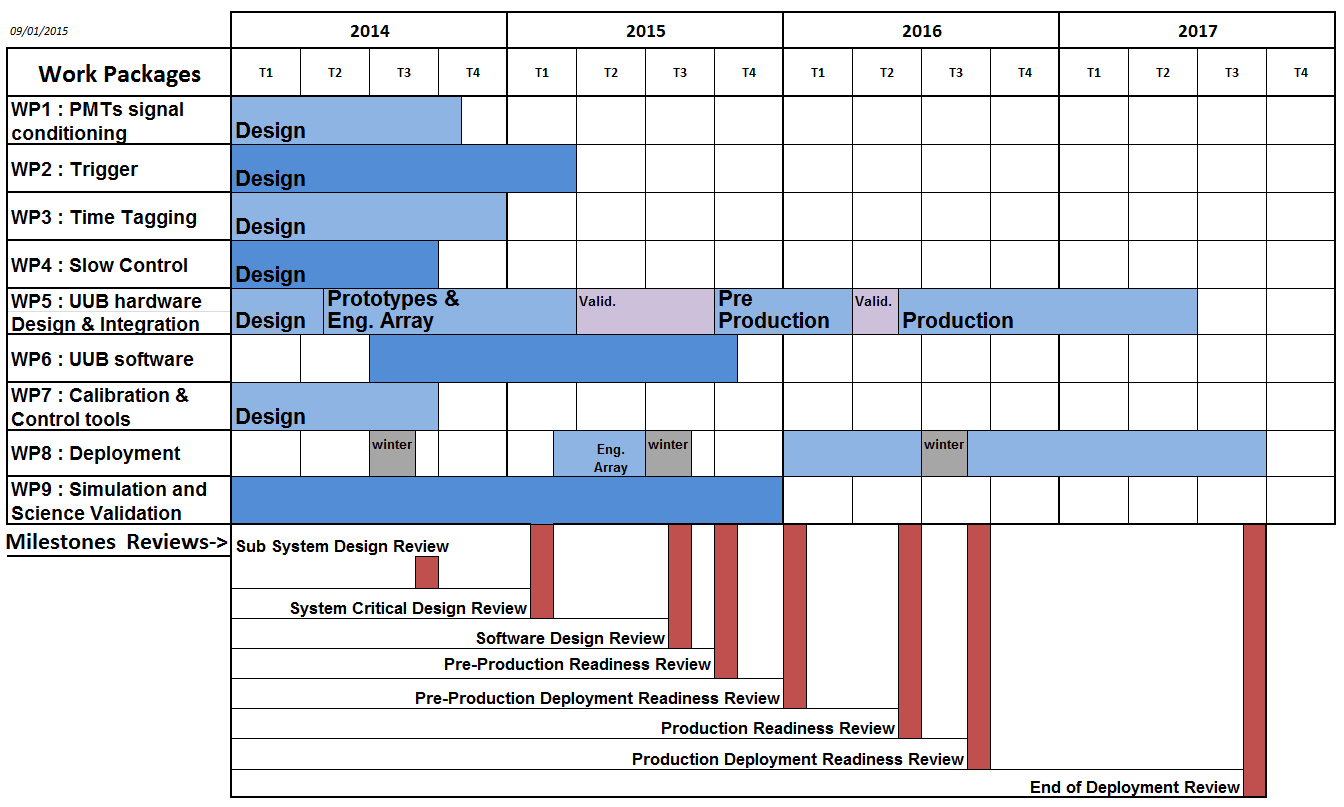}
\caption{SDEU schedule.}
\label{fig:Schedule_SDE}
\end{figure}

The schedule for the Surface Detector Electronics upgrade is shown in
Fig.~\ref{fig:Schedule_SDE}.
The final validation of the SDEU design will be done in an Engineering
Array of 10 detector stations at the end of 2015. The production and
deployment of the SDEU can done in 2016-17. However, the production
schedule is mainly driven by the funding profiles in different
countries which typical extend over 3 years, and will therefore be
extended up to 2018.

\begin{figure}[!ht]
\centering
\includegraphics[width=.8\textwidth,angle=0]{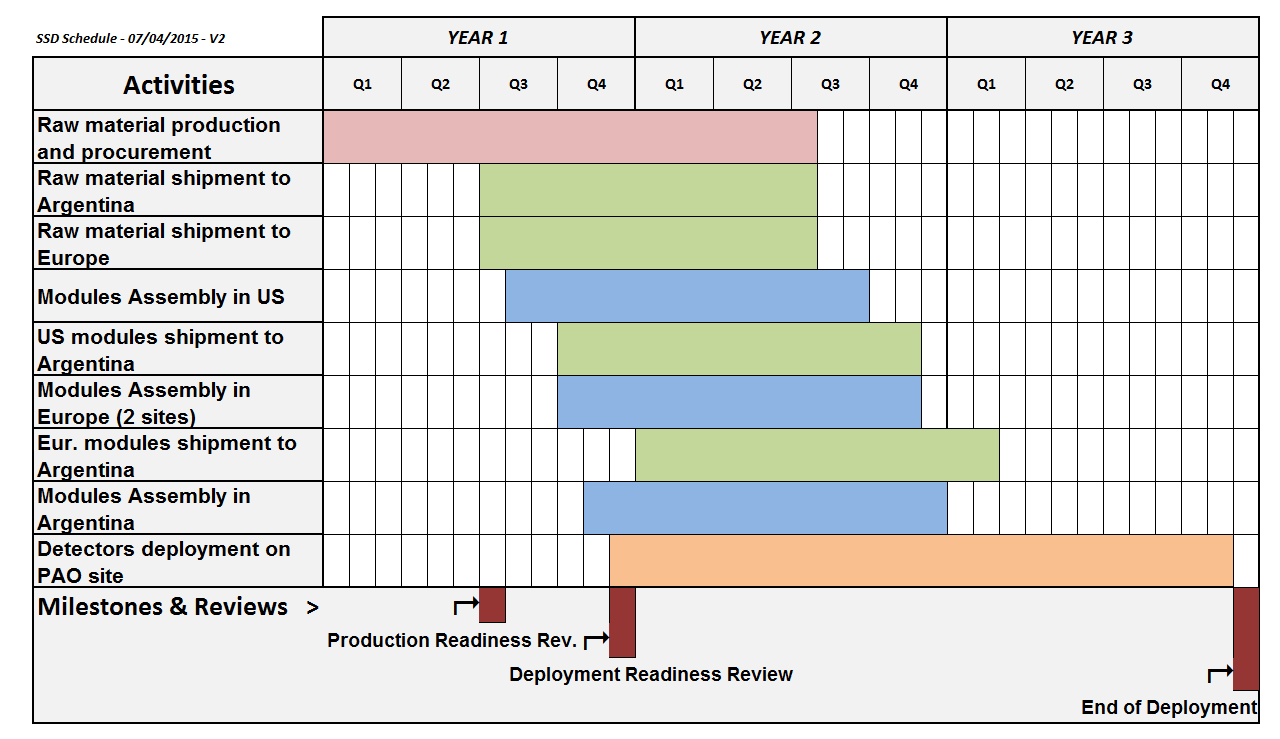}
\caption{SSD schedule.}
\label{fig:Schedule_SSD} 
\end{figure}

The schedule for the Scintillator Surface Detector is shown in Fig.~\ref{fig:Schedule_SSD}.
Similarly to the SDEU, the final validation of the SSD design will be
done in an Engineering Array of 10 detector stations at the end of
2015. The production and deployment of the SSD will done in
2016-18. Like the SDEU, the production and deployment schedule
is mainly driven by the funding profiles.

%% file: appendix-current_state.tex

\chapter{The Current Pierre Auger Observatory}
\label{chap:appendix_currentObservatory}

\section{Introduction}

The Pierre Auger Project was proposed in 1998 ``to discover and
understand the source or sources of cosmic rays with energies
exceeding $10^{19}$\,eV.'' A unique partnership of 17 countries came
together to pursue this science.  To achieve its goals, the
collaboration designed an experiment to achieve high-quality data in
a high-statistics study of the most extreme cosmic rays.  The Pierre
Auger Observatory in the Province of Mendoza, Argentina, has been
taking data since 2004, adding detectors as they became active until
completion in 2008.  Measured properties of air showers are used
to determine the cosmic ray energy, direction and composition.

A powerful feature of the Auger design is the capability of observing
air showers simultaneously by two different but complementary
techniques.  On dark moonless nights, air fluorescence telescopes
record the development of what is essentially the electromagnetic
shower that results from the interaction of the primary particle with
the upper atmosphere.  The surface array measures particle
signals as the shower strikes the earth just beyond its maximum
development.  By recording the light produced by the developing air
shower, fluorescence telescopes can make a nearly calorimetric
measurement of the energy.  This energy calibration can then be
transferred to the surface array with its 100\% duty factor and large
event-gathering power.  The energy conversion and subsequent
determination of the spectrum can be done with minimal reliance on
numerical simulations or on assumptions about the composition or
interaction models.

\begin{figure}[t]
\centering
\includegraphics[width=0.6\textwidth]{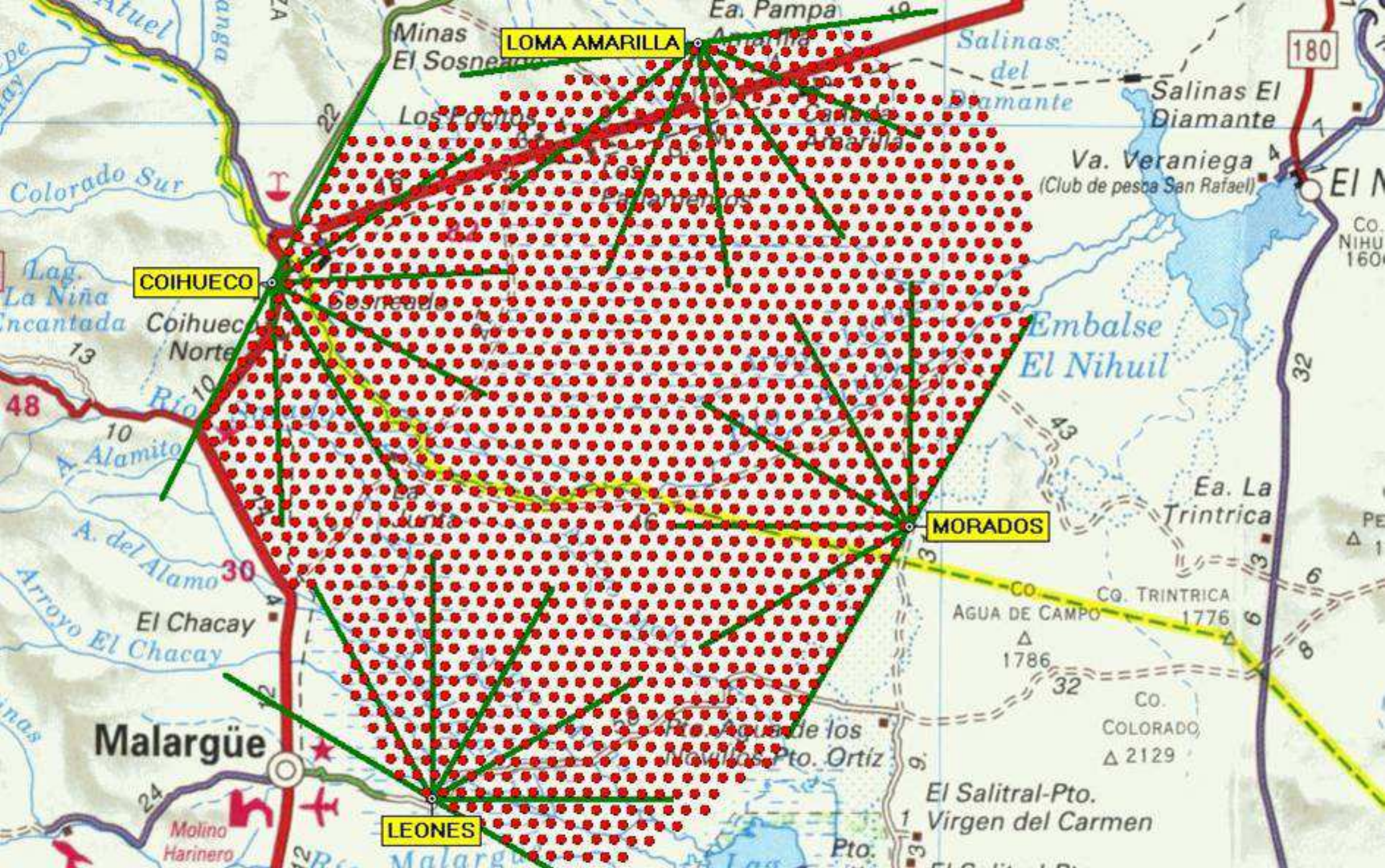}
\caption{The Auger Observatory layout.  Each dot corresponds to one of the
1660 surface detector stations.  The four fluorescence detector
enclosures are shown, each with the field of view of its six
telescopes.}
\label{southern_site}
\end{figure}

The Observatory design features an array of 1660 water-Cherenkov
surface detectors spread over 3000\,km$^2$ and arranged on a triangular
grid, with the sides of the triangles being 1.5\,km (see Figure
\ref{southern_site}).  Four fluorescence detector stations, each
containing six fixed telescopes designed to detect air-fluorescence
light, overlook the surface array.  (An additional three telescopes
view higher elevations for lower energy air showers.)  The surface
detector stations measure the density distribution of the air shower
cascade as it strikes the surface while the fluorescence telescopes
measure the light produced by atmospheric nitrogen excited by the
cascading shower. This dual approach is called the \textit{hybrid
  technique}.

A water-Cherenkov particle detector was chosen for use in the surface
array because of robustness, low cost, and sensitivity to showers at
high zenith angles.  A surface detector station (SD) consists of a
12,000\,liter polyethylene water tank containing a sealed laminated
polyethylene liner with a reflective inner surface.  Cherenkov light
from the passage of charged particles is collected by three 230\,mm
photomultiplier tubes (PMTs) that look through windows of clear polyethylene
into highly purified water.  The surface detector station is self
contained.  A solar power system provides power for the PMTs and
electronics package.  The electronics package, consisting of a
processor, GPS receiver, radio transceiver and power controller, is
mounted on the tank.

The Auger fluorescence detector (FD) operates in conjunction with the
surface detector (SD).  Its primary purpose is to measure the
longitudinal profile of showers recorded by the SD whenever it is dark
and clear enough to make reliable measurements of atmospheric
fluorescence from air showers.  The integral of the longitudinal
profile is used to determine the shower energy, and the speed of
the shower development is indicative of the primary particle's mass.  The
hybrid detector has better angular resolution than the surface array
alone.

The Auger Project was conceived during the International Cosmic Ray
Conference in Dublin in 1991 by J.W.~Cronin of the University of
Chicago and A.A.~Watson of the University of Leeds.  It had become
clear to them that only the construction of a very large air shower
array in both the northern and southern hemispheres would yield the
statistical power and complete sky coverage necessary to address the
question of the origin of the highest energy cosmic rays.

A six month long design workshop was held in early 1995 that produced
a design report with a discussion of the science, a conceptual design
and cost estimate.  The design report became the basis for funding
proposals by the Auger collaborators to their funding
agencies.  

Subsequent to the workshop, a team of scientists evaluated numerous
prospective sites in both hemispheres.  Site considerations included
altitude, latitude, topography, and sky clarity.  In 1995 and 1996
preferred sites were selected by the collaboration in the Southern and
Northern hemispheres respectively.  At the direction of the funding
agencies, the project began by building the Auger Observatory
in the southern hemisphere.

\begin{figure}[t]
\centering
\includegraphics[height=0.33\textwidth]{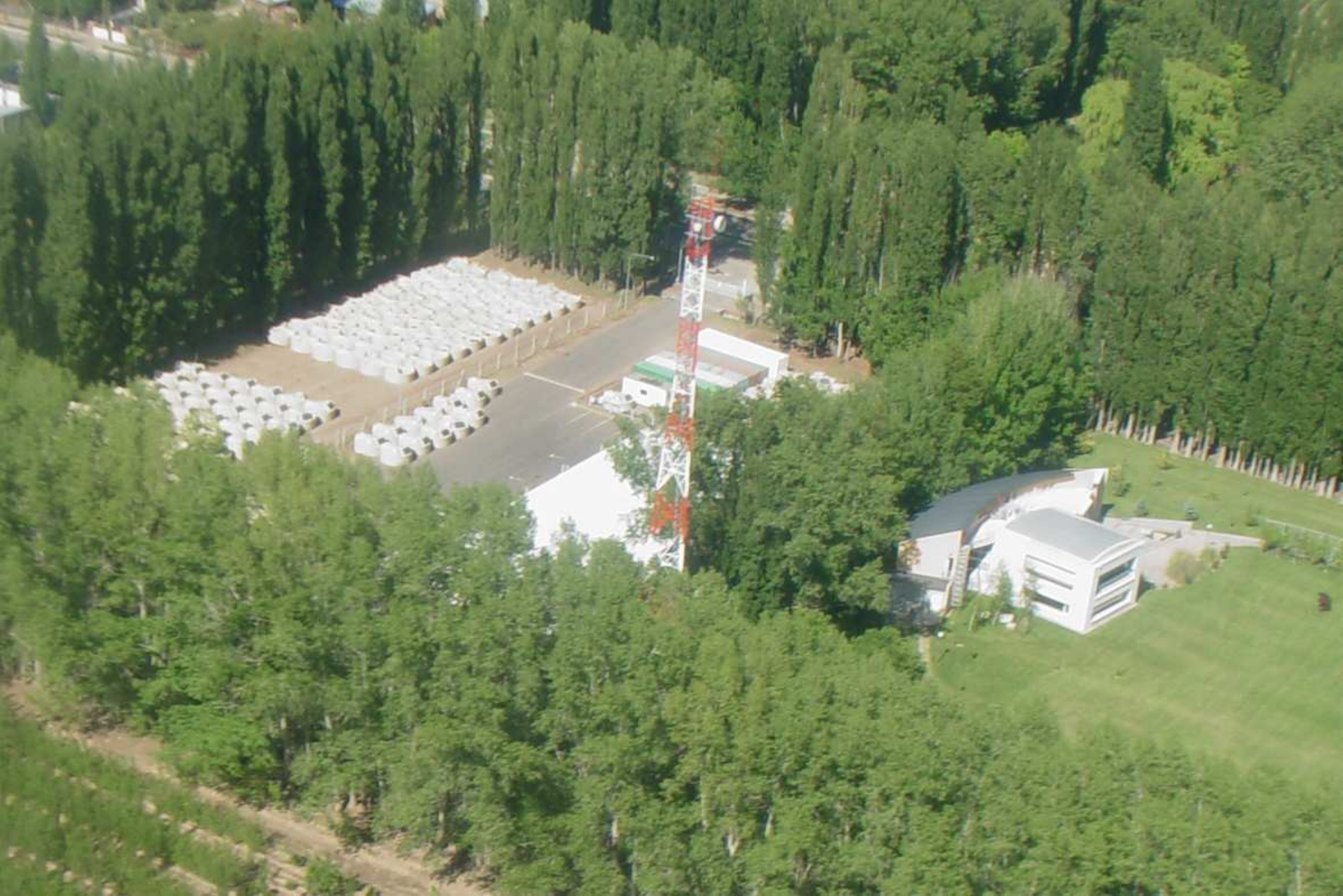}
\hfill
\includegraphics[height=0.33\textwidth]{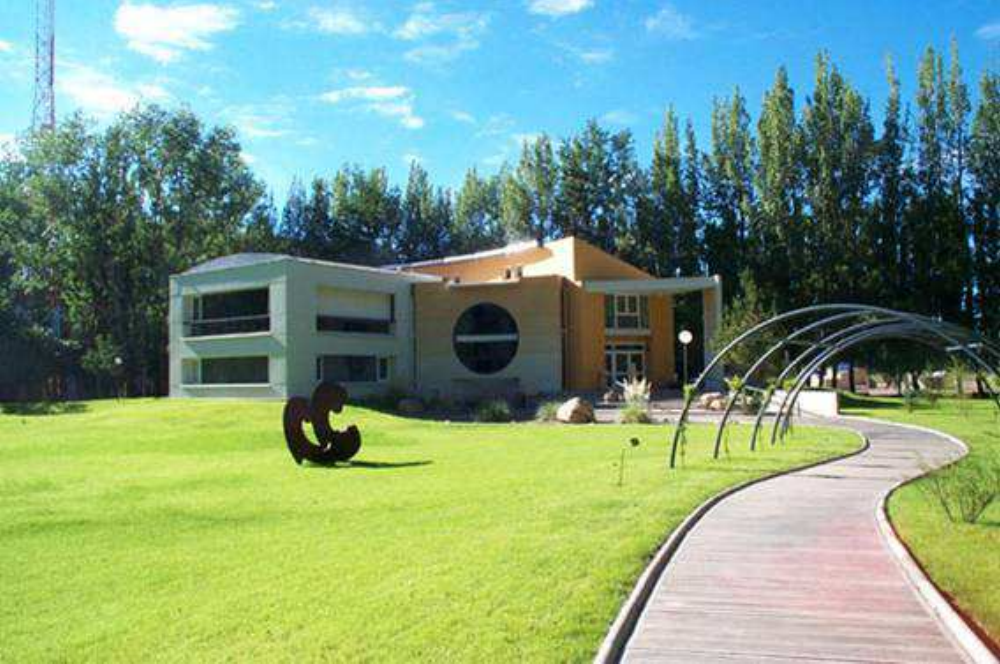}
\caption{\textit{Left}: an aerial view of the Auger campus in Malarg\"{u}e. 
\textit{Right}: the office building and visitor center.}
\label{fig:south_campus}
\end{figure}

The site is in the Province of Mendoza near the city of
Malarg\"{u}e (pop.~18,000) and 180\,km south-west of the city of San
Rafael (pop.~100,000).  The site is located at about latitude
35$^\circ$ south with a mean altitude of 1400\,m a.s.l.  The site is a
relatively flat alluvial plain, sufficiently large to easily encompass
the required 3000\,km$^2$ footprint of the array.  There are convenient
elevated positions on the edge of the array that allow placement of
the four fluorescence telescope enclosures slightly above ground
level.  A campus area in Malarg\"{u}e includes an office building with
a visitor center, a detector assembly area, and a staging area for
detectors (See Figure~\ref{fig:south_campus}).

\begin{floatingfigure}[r]{0.5\textwidth}
\includegraphics[width=0.45\textwidth]{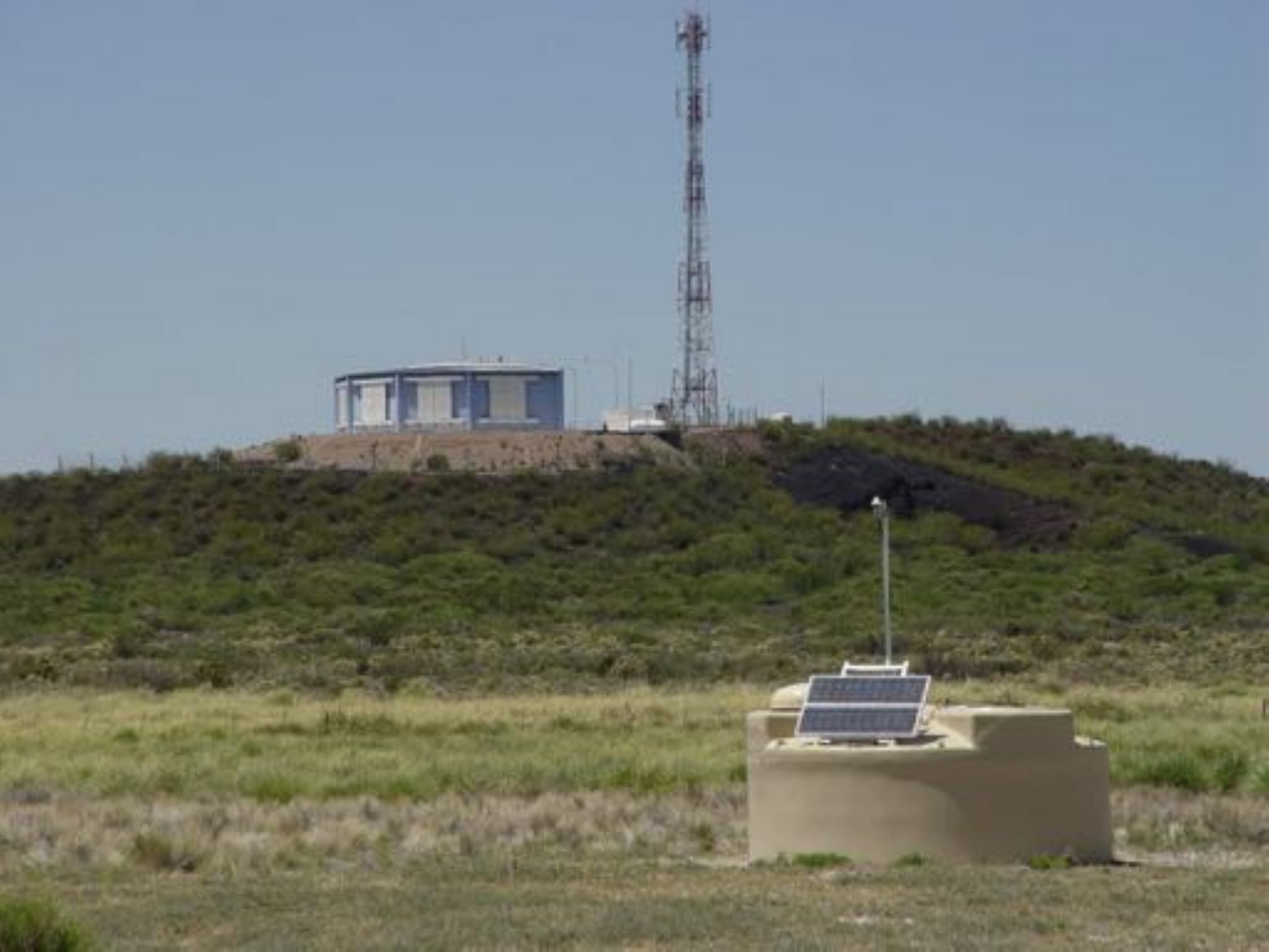}
\caption{A fluorescence telescope enclosure, seen on the hill top, and
a surface detector station, below.}
\label{fig:hybrid_detectors}
\end{floatingfigure}

After a period of research and development, the Engineering Array,
consisting of 32 prototype surface array detectors and two prototype
fluorescence telescopes, was built to validate the design
\cite{Abraham:2004dt}.  At the end of 2001, before the end of the
scheduled two years, the Engineering Array was able to record and
reconstruct air shower events simultaneously with both the surface
array and the fluorescence detectors.  The Engineering Array was able
to demonstrate the validity of the design and the performance of all
of the detectors, communications equipment and data systems as well as
the deployment methodology.  Indeed, we found that the detectors
performed even better than expected, substantially increasing our
physics reach.

Installation of production detectors was started in 2002.  While the
Engineering Array was assembled and deployed almost completely by
Auger collaborators, production deployment was transferred to trained
Observatory staff.  The scientists oversaw the quality of the work and
carried out the commissioning of the completed detectors.  The
Observatory started collecting data in January 2004.  The first
physics results were presented during the 2005 conference
season.  

Many important results have now been published by the Auger
Collaboration that have had a major impact on the field of cosmic ray
physics. As of this writing, 55 full author list papers have been
published or accepted, with another 2 submitted and about 15 more in
preparation. The Auger Collaboration is also training a cadre of
future scientists, with 218 students granted PhDs based on their work
on Auger. Another 157 PhD students are in the pipeline.

\section{The Surface Detector}
\label{currentSD}
\subsection{Detector hardware}

Each surface detector station of the Observatory consists of a 3.6\,m
diameter water tank containing a sealed liner with a reflective
inner surface. The liner contains 12,000\,l of ultra-high purity water.
Three 230\,mm diameter photomultiplier tubes are
symmetrically distributed at a distance of 1.20\,m from the center of
the tank and look downwards through windows of clear polyethylene
into the water to collect the Cherenkov light produced by the
passage of relativistic charged particles through the water. The
water height of 1.2\,m makes it also sensitive to high energy
photons, which convert to electron-positron pairs in the water
volume.

\begin{figure}[t]
\centering
\includegraphics[width=0.6\textwidth]{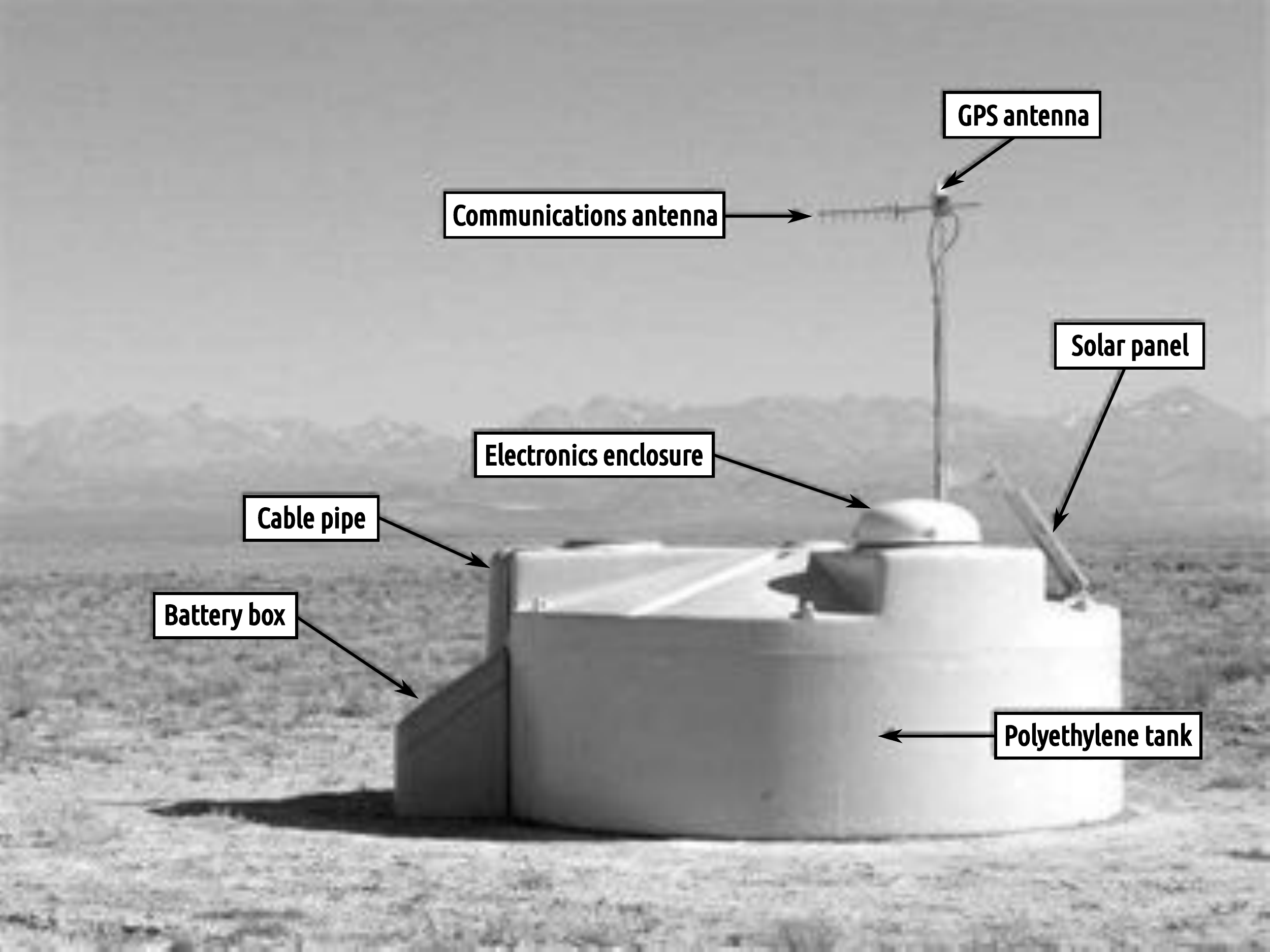}
\caption{A pictorial view of a surface detector
station in the field, showing its main components.}
\label{fig:sd}
\end{figure}

The surface detector station is self-contained. A solar power system
provides an average of 10\,W for the PMTs and the electronics package
consisting of a processor, GPS receiver, radio transceiver and power
controller. The components of the surface detector station are shown
in Fig.~\ref{fig:sd}. The hardware of the surface detector is
described extensively in \cite{Abraham:2004dt,AugerSouthNIM}.

The tanks are made of high-density polyethylene by the rotomolding
process.  The exterior is colored beige to minimize the visual impact.
The resins are compounded with additives to enhance ultraviolet
protection.  The interior has added carbon-black to guarantee
light-tightness.  The tanks have a nominal wall thickness of 1.3\,cm and a
weight of 530\,kg.

Three hatches, located above the PMTs,  provide access to the
interior  of the tank for assembly, water filling and eventual
servicing of the interior parts. The hatches are covered with light-
and water-tight polyethylene hatch covers. Hatch cover 1 is larger and
accommodates the electronics on its top. The electronics is
protected by an aluminum dome. The tanks also possess lugs for
lifting and four additional lugs to support the solar panel and
antenna mast assembly.

Electrical power for the electronics is provided by two 55\,Wp solar
panels which feed two 12\,V, 105\,Ah lead-acid low maintenance
batteries.  Batteries are charged through a commercial charge
controller.  The electronics assembly possesses a Tank Power Control
Board (TPCB) which also monitors the charging and discharging of
batteries and sets the system to hibernation mode if the charge of the
batteries falls below a critical level.  The batteries are
accommodated in a thermally insulated battery box which is installed
next to the tank at the shaded southern side.  

The solar panels are mounted on aluminum brackets, which also
support a mast of a height of 2.15\,m. The communications and GPS
antennas are mounted at the top of this mast.

The tank liners are right circular cylinders made of a flexible
plastic material conforming approximately to the inside surface of
the tanks. They enclose the water volume, provide a light-tight
environment and diffusively reflect the Cherenkov light produced in
the water volume. The liners are produced from a laminate composed
of an opaque three-layer co-extruded low-density polyethylene (LDPE)
film bonded to a layer of DuPont
Tyvek\textsuperscript{\textregistered} 1025-BL
by a layer of Titanium-dioxide pigmented LDPE. The three-layer
co-extruded film consists of a carbon black loaded LDPE formulated
to be opaque to single photons, sandwiched between layers of clear
LDPE to prevent any carbon black from migrating into the water
volume.

The liner has 3 windows through which the PMTs look into the water
volume from above. These windows are made of UV-transparent linear
low-density polyethylene. The PMTs are optically coupled to the
windows with optical silicone and protected with a light-tight
plastic cover.

Once deployed in their correct position in the field, the tanks are
filled with high purity water produced at a water plant owned by the
Auger Project. Water quality exceeds 10\,M$\Omega$\,cm at the output of the
water plant and is transported in clean ad-hoc transport tanks. The
water is expected to maintain its quality without degradation for
the lifetime of the Observatory.

\subsection{The existing surface detector electronics} \label{subsec:AS-sde}

{\em The Auger Upgrade includes a major overhaul of the electronics of the
surface detectors, as described in the body of this Report.  Here we
outline the existing SD electronics.}

To collect the Cherenkov light produced in the water volume of the
detectors by the air showers, three PMTs look at the water volume from
the top. The PMTs (Photonis XP1805/D1) have a 230\,mm diameter
photocathode and eight dynodes, with the chemical composition of the
dynode surfaces optimized by the manufacturer to maximize
linearity. Due to their proximity to water they are operated with a
positive anode voltage, the photocathode being grounded. The high
voltage is provided locally from a module integrated in the PMT base,
and is proportional to a DC control voltage provided by the slow
control system. The PMTs are operated at a nominal gain of $2{\times}
10^{5}$, and are specified for operation at gains up to $10^6$.  The
PMTs are required to be linear within 5\% up to 50\,mA anode
current.  The base, including the high
voltage supply, is attached to the tube by soldering to flying leads
and is potted in GE silicone RTV-6136 to protect it from the high
humidity present in the tank.

Each PMT has two outputs. An AC coupled anode signal is provided.  In
addition, the signal at the last dynode is amplified and inverted by
the PMT base electronics to provide a signal with 32 times the charge
gain of the anode.  No shaping of the signal is performed on the PMT
base.

Six identical channels of electronics are provided to digitize
the anode and amplified dynode signals from each of the PMTs.  Each
channel consists of a 5-pole Bessel filter with a $-3$\,dB cutoff at
20\,MHz and a voltage gain of $-0.5$.  This filter is implemented using a
pair of Analog Devices AD8012 current feedback op-amps.  The filtered
analog signals are fed to Analog Devices AD9203 10\,bit 40\,MHz
semi-flash ADCs.  The ADC negative inputs are biased to $-50$\,mV to
bring the input pedestal on scale and allow for amplifier section
offsets.  The choice of filter cutoff results in 5\% aliasing noise
while preserving the time structure of the signals.  The use of two
10\,bit ADCs with a gain difference of 32 extends the dynamic range of the
system to 15\,bits with a 3\% precision at the end of the overlap region.

An LED flasher is mounted in a test port of the water tank liner.  The
LED flasher incorporates two LEDs which can be pulsed independently or
simultaneously and with variable amplitude.  This allows testing of
the linearity of the photomultipliers to be conducted remotely.

Each SD station contains a GPS receiver with its corresponding antenna
mounted at the top of the communications mast for event timing and
communications synchronization. The receiver is a Motorola (OEM)
Oncore UT+. This receiver outputs a timed one-pulse-per-second (1\,PPS).
The GPS 1\,PPS signal is offset from the true GPS second by up to
50\,ns, and a correction for this offset is provided periodically by
the receiver.  Event timing is determined using a custom ASIC which
references the timing of shower triggers to the GPS 1\,PPS clock.  The
ASIC implements a 27\,bit clock operating at 100\,MHz.  This clock is
latched on the GPS 1\,PPS signal at the time of each shower
trigger.  A counter operating at the 40\,MHz ADC clock is also latched
on the GPS 1\,PPS clock.  These data, together with the timing
corrections provided by the GPS receiver, are used to calibrate the
frequencies of the 40\,MHz and 100\,MHz clocks and to synchronize the
ADC data to GPS time within 10\,ns RMS.

The digital data from the ADCs are clocked into a programmable logic
device (PLD).  In the first half of the deployment, we employed two
ALTERA ACEX PLDs (model EP1\-K100QI208-2) with 16k $\times$ 36 bits
additional external static RAM.  In later stations, an Altera Cyclone
FPGA replaced the two ACEX devices and external memory.  The PLD
implements firmware that monitors the ADC outputs for interesting
trigger patterns, stores the data in a buffer memory, and informs the
station micro-controller when a trigger occurs.  There are two local
trigger levels (T1 and T2) and a global third level trigger, T3.
Details of the local triggers are described in section~\ref{subsec:triggers}.

The front end is interfaced to a unified board which implements the
station controller, event timing, and slow control functions,
together with a serial interface to the communications system.
The slow control system consists of DACs and ADCs used to measure
temperatures, voltages, and currents relevant to assessment of the
operation of the station.

The station controller consists of an IBM PowerPC 403 GCX-80\,MHz, with
a 32\,MB DRAM bank to store data and executable code, and a 2\,MB Flash
EPROM for the bootstrap and storing of the OS9 operating system.  The
data acquisition system implemented on the station controller transmits
the time stamps of the ${\sim}20$ T2 events collected each second to CDAS
(Central Data Acquisition System; see section~\ref{subsec:cdas}).
CDAS returns T3 requests to the station within ${\sim}8$ seconds of the event
(including communications delays due to re-transmission).  The station
controller then selects the T1 and T2 data corresponding to the T3
requests and builds it into an event for transmission to
CDAS. Calibration data are included in each transmitted event.

\subsection{Surface Detector Calibration} \label{subsec:calibration}

The detector calibration is inferred from background muons. The
typical rise time for a muon signal is about 15\,ns with a decay time
of the order of 60 to 70\,ns. The average number of photoelectrons per
muon collected by one PMT is 95. By adjusting the trigger rates, the
gains of the three PMTs are matched within 6\%. The measurement of
the muon charge spectrum allows us to deduce the charge value for
the signal produced by a single, central, vertical muon, $Q_\text{VEM}$,
from which the calibration is inferred for the whole dynamic range.
The cross calibration between the anode and dynode output channels
is performed by using small shower signals in the overlap region
\cite{Bertou:2005ze}.

\begin{figure}
\centering
\includegraphics[width=0.52\textwidth]{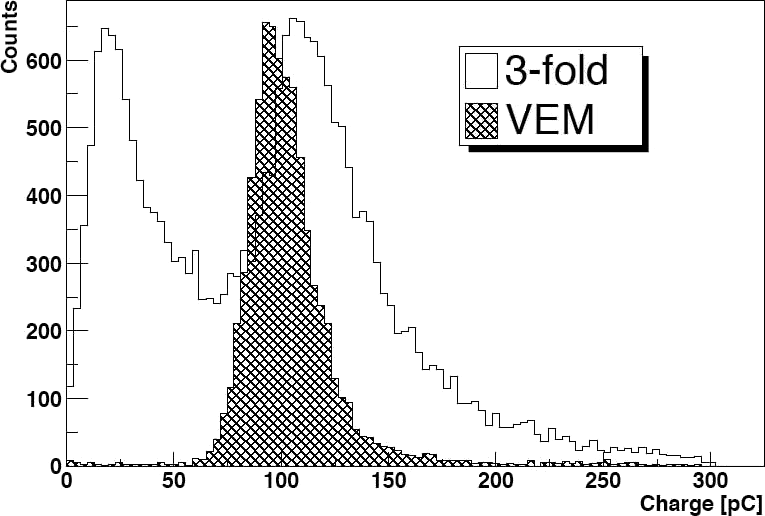}
\caption{Muon peak.}
\label{fig:Muonpeak}
\end{figure}

The decay constant of the muon signal is related to the absorption
length of the light produced. This depends of various parameters
such as the Tyvek\textsuperscript{\textregistered} reflectivity and the purity of
the water. The signal decay constant correlates with the so called
area-to-peak (A/P) ratio of the signal:
\begin{equation}
{\rm A/P} = \frac{Q_\text{VEM}}{I_\text{VEM}}
\end{equation}
where $I_\text{VEM}$ is the maximum current of the muon signal. This
area-to-peak ratio is a routine monitoring quantity that is directly
available from the local station software.

\subsection{The SD local triggers} \label{subsec:triggers}

The front-end electronics implements three types of trigger functions.  Shower triggers
result in the recording of 768 samples (19.2\,$\upmu$s) of the six ADCs.
Muon triggers result in the recording of 24 samples of the three
high gain dynode channels for use in calibration.  Double buffered
scalers are also implemented for use in monitoring rates and for
auxiliary physics purposes.

A shower trigger is generated when one of several conditions is
satisfied.  A single bin threshold trigger is generated when at least
a specified number of the high gain signals (amplified dynode) are each above a
threshold level.  The standard trigger condition is $2\,I_\text{VEM}$
on at least two of the high gain signals.  The rate of this trigger is
about 100\,Hz, and is sensitive to showers near the core but subject
to contamination due to small showers.  A software selection of this
trigger with a higher threshold at $3.2\,I_\text{VEM}$ is also
performed.  In addition, a time-over-threshold (ToT) trigger is
implemented. This trigger requires that single bin threshold trigger
be satisfied for at least a minimum number of samples within a sliding
time window.  A ToT trigger is generated when at least 12 samples
within a 3\,$\upmu$s window (120 samples) exceed a threshold of
$0.2\,I_\text{VEM}$ on at least two out of the three tubes.  The ToT
trigger is efficient for shower signals far from the core. The rate of
the ToT trigger is a few Hz, and depends on the shape of the muon
pulse in the tank.  The shower triggers implemented in the PLD are
collectively referred to as T1. The station controller (see below)
transmits timestamps for the ToT and $3.2\,I_\text{VEM}$ threshold
triggers (collectively referred to as T2) to the CDAS 
for global (T3) trigger determination.  

Two additional sophisticated triggers were introduced in 2013 to
somewhat lower the energy threshold of the array, and improve
sensitivity to photon and neutrino initiated showers.  See
\cite{AugerSouthNIM} for details.

\section{The Fluorescence Detector} 

\begin{figure}[t]
\centering
\includegraphics[width=0.48\textwidth]{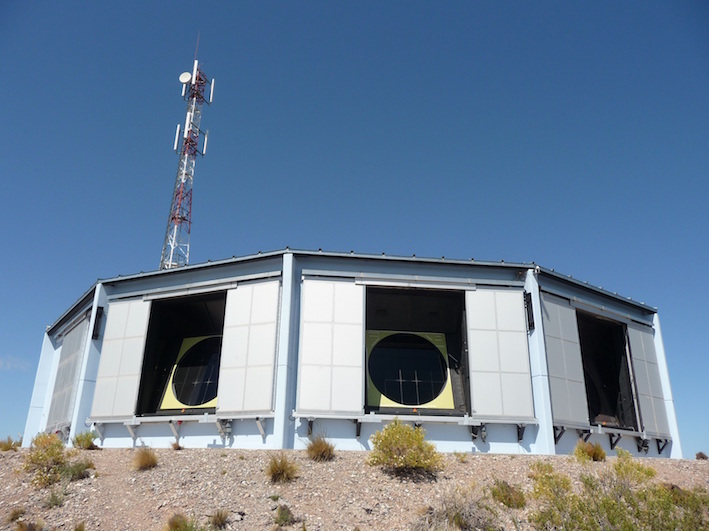}  
\caption{FD building at Los Leones during the day. Behind the building
is a communication tower. This photo was taken during daytime when
shutters were opened because of maintenance.}
\label{fig:FD-losleones-building}
\end{figure}

The $24$ telescopes of the FD overlook the SD array from four sites --
Los Leones, Los Morados, Loma Amarilla and Coihueco
\cite{Abraham:2009pm}. Six independent telescopes are located at each
FD site in a clean climate-controlled building~\cite{Abraham:2004dt},
an example of which is seen in figure~\ref{fig:FD-losleones-building}.
A single telescope has a field of view of $30^\circ\times30^\circ$ in
azimuth and elevation, with a minimum elevation of $1.5^\circ$ above
the horizon. The telescopes face towards the interior of the array so
that the combination of the six telescopes provides $180^\circ$
coverage in azimuth.

\subsection{FD telescopes}

\begin{figure}[t]
\centering
\includegraphics[width=0.4\textwidth]{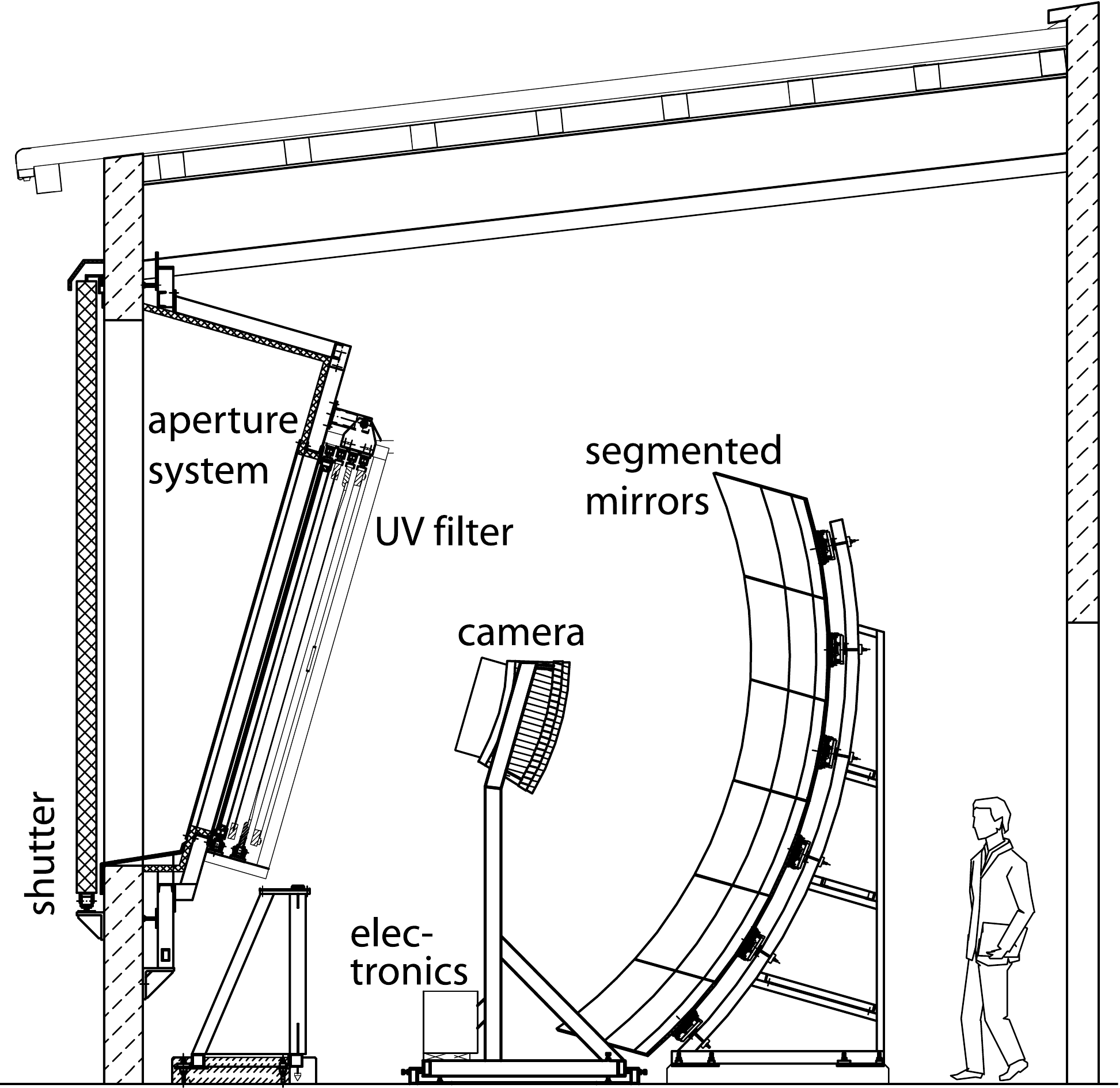}\hfill
\includegraphics[width=0.48\textwidth]{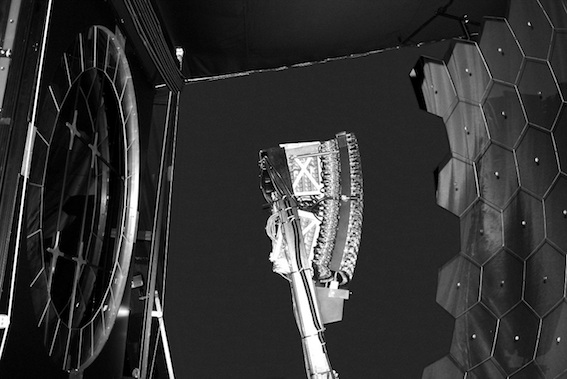}  
\caption{\textit{Left}: Schematic view of a fluorescence telescope with a description of its main components. \textit{Right}: Photograph of a fluorescence telescope at Coihueco.}
\label{fig:Telescope}
\end{figure}

The details of the fluorescence detector telescope are shown in
figure~\ref{fig:Telescope}. The telescope design is based on Schmidt
optics because it reduces the coma aberration of large optical
systems.  Nitrogen fluorescence light, emitted isotropically by an air
shower, enters through a circular diaphragm of $1.1$\,m radius covered
with a Schott MUG-6 filter glass window. The filter transmission is
above $50$\% between $310$ and $390$\,nm in the UV range.  The filter
reduces the background light flux and thus improves the
signal-to-noise ratio of the measured air shower signal. It also
serves as a window over the aperture and thus keeps the space
containing the telescopes and electronics clean and climate
controlled.  The shutters seen in figure~\ref{fig:Telescope} are
closed during daylight and also close automatically at night when the
wind becomes too high or rain is detected.  In addition, a fail safe
curtain is mounted behind the diaphragm to prevent daylight from
illuminating a camera in case of a malfunction of the shutter or a
failure of the Slow Control System.

A simplified annular lens, which corrects spherical aberration
and eliminates coma aberration,
is mounted in the outer part of the
aperture. The segmented corrector ring has inner and outer radii
of $850$ and $1100$\,mm, respectively. Six corrector rings were
made from Schott BK7 glass and Borofloat was used for the rest.
More details about the corrector ring can be found
in~\cite{Abraham:2004dt,deOliveira:2004dh}.

The light is focused by a spherical mirror of ${\sim}3400$\,mm
radius of curvature onto a spherical focal surface with radius of curvature
${\sim}1700$\,mm.
Due to its large area (${\sim}13$\,m$^2$), the primary mirror 
is segmented to reduce the cost and weight of the
optical system. Two alternative segmentation configurations
are used: one is a tessellation of $36$ rectangular anodized
aluminum mirrors of three different sizes; the other is a
structure of $60$ hexagonal glass mirrors (of four shapes and
sizes) with vacuum deposited reflective coatings~\cite{Abraham:2004dt}.
The average reflectivity of cleaned mirror segments at a wavelength $\lambda=370$\,nm
is more than $90$\%. Measurements have shown that  dust layer deposits
could reduce the mirror reflectivity by about $5$\% in the bottom
part of the spherical mirror, where the segments are turned slightly
upward (see, e.g., figure~\ref{fig:Telescope}). Therefore,
careful mirror cleaning is performed as needed.

The camera body is machined from a single aluminum block of
$60$\,mm thickness, with an outer radius of curvature of
$1701$\,mm and an inner curvature radius of $1641$\,mm. The
hexagonal photomultiplier tubes, model XP3062 manufactured by
Photonis, are positioned inside $40$\,mm diameter holes drilled
through the camera block at the locations of the pixel centers. The
pixels are arranged in a matrix of $22$ rows by $20$ columns.

The PMT boundaries are approximate hexagons with a side to side
distance of $45.6$\,mm. The PMTs are separated by simplified
Winston cones secured to the camera body which collect the light
to the active cathode of the photomultiplier tube. The light
collectors serve to prevent photons from landing in the dead
spaces between the PMT cathodes. The upper edge of the light
collectors lie on the focal surface of $1743$\,mm radius. The
pixel field of view defined by the upper edges corresponds
to an angular size of $1.5^\circ$.

All support structures and cables are distributed so as to
minimize any obscuration in the light path.
The contribution of reflection and scattering inside the optical
system of the telescope has been measured in situ and with
an airborne remotely controlled platform carrying an isotropic
and stabilized UV light source~\cite{Baeuml-ICRC:2013}.
The measured point spread function of the light distribution
in pixels has been implemented in the software used in the air
shower reconstruction.

Cleaning and maintenance work has been required during
years of detector operation. The cleaning of the UV
filter from outside has been performed several times
because of deposited dust layers. Less frequently, the inner
side of the filter and the corrector ring were washed.
Dry and wet methods of mirror cleaning have been adopted
over the years and they both improve the reflectivity of mirrors
by $\leq1$\% (in the case of mirror segments in the upper rows)
up to about $5$\% for mirror segments in the bottom rows.

Alignment of individual mirror segments was cross-checked
with a laser on site. Moreover, additional methods using
data measured by telescopes were used, such as star tracking,
Central Laser Facility (CLF) and eXtreme Laser Facility (XLF)
shots (section~\ref{subsec:atmos}), or a comparison of FD and SD geometry
reconstruction. Only in two cases were a realignment of a telescope
and a readjustment of camera position needed.

\subsection{FD Electronics}
 
The FD electronics must provide a large dynamic range and
strong background rejection, while accepting any physically
plausible air shower. Moreover, the electronics is responsible
for anti-alias filtering, digitizing, and storing signals
from the PMTs.

The XP3062 photomultiplier tube is an 8-stage unit with a bi-alkaline
photocathode with quantum efficiency of about $25$\% in the
wavelength range $350$ to $400$\,nm.
The PMT high voltage is provided by a HV divider
chain which forms a single physical unit together with the signal
driver circuitry. This head electronics unit is soldered to the 
flying leads of the PMT \cite{Becker:2007zza}.

The nominal gain for standard operation of the FD is set to
$5{\times}10^4$. Stabilization of the HV potential for large
pulses, and in the presence of the low but not negligible light
intensity of the dark sky background, is realized by employing an active
network that uses bipolar transistors in the last three stages
of the PMT. The active divider ensures that the gain shift due
to the divider chain is less than $1$\% for anode currents up to
about $10$\,mA. The normal dark sky background on moonless
nights induces an anode current of about $0.8\,\upmu$A on each PMT.

The head electronics for each PMT is connected to a distribution
board located just behind the camera body. Each board serves 44
PMTs, providing high and low voltage and receiving the output
signals. The signal is then shaped and digitized in the front-end
electronics (FE) unit, where  threshold and geometry
triggers are also generated. Analog boards in the FE unit are designed
to handle the large dynamic range required for air fluorescence
measurements; this means a range of $15$\,bits and $100$\,ns timing.

As the PMT data are processed, they are passed through a flexible
three-stage trigger system implemented in firmware and software.
The trigger rate of each pixel in a camera (first level trigger)
is kept around $100$\,Hz by adjusting the pixel threshold level.
The algorithm of the second level trigger searches for track
segments at least five pixels in length within a camera. The typical
trigger rate per camera fluctuates between $0.1$ and $10$\,Hz. The third level
trigger is a software algorithm designed to clean the air shower
data stream of noise events that survive the low-level hardware
triggers. It is optimized for the fast rejection of triggers caused
by lightning, triggers caused by cosmic ray muon impacts on the camera and
randomly triggered pixels.

The events surviving all trigger levels are sent to the computer,
which builds an event from the coincident data in all telescopes
and generates a hybrid trigger (T3) for the surface array. The event
rate is about $0.012$\,Hz per building for the $24$ baseline telescopes.

\subsection{FD Calibration}
The reconstruction of air shower profiles and the ability to determine
the total energy of a reconstructed shower depend on the conversion of
ADC counts to light flux at the telescope aperture for each channel
that receives a portion of the signal from a shower.  To obtain this
important relation, it is necessary to evaluate the response of each
pixel to a given flux of incident photons from the solid angle covered
by that pixel, including the effects of aperture projection, optical
filter transmittance, reflection at optical surfaces, mirror
reflectivity, pixel light collection efficiency and area, cathode
quantum efficiency, PMT gain, pre-amp and amplifier gains, and digital
conversion.  This response is measured in a single end-to-end
calibration.

\begin{figure}[!t]
    \subfigure[Schematic view.]
    {\label{asfd_fig8}
    \includegraphics[width=0.46\textwidth]{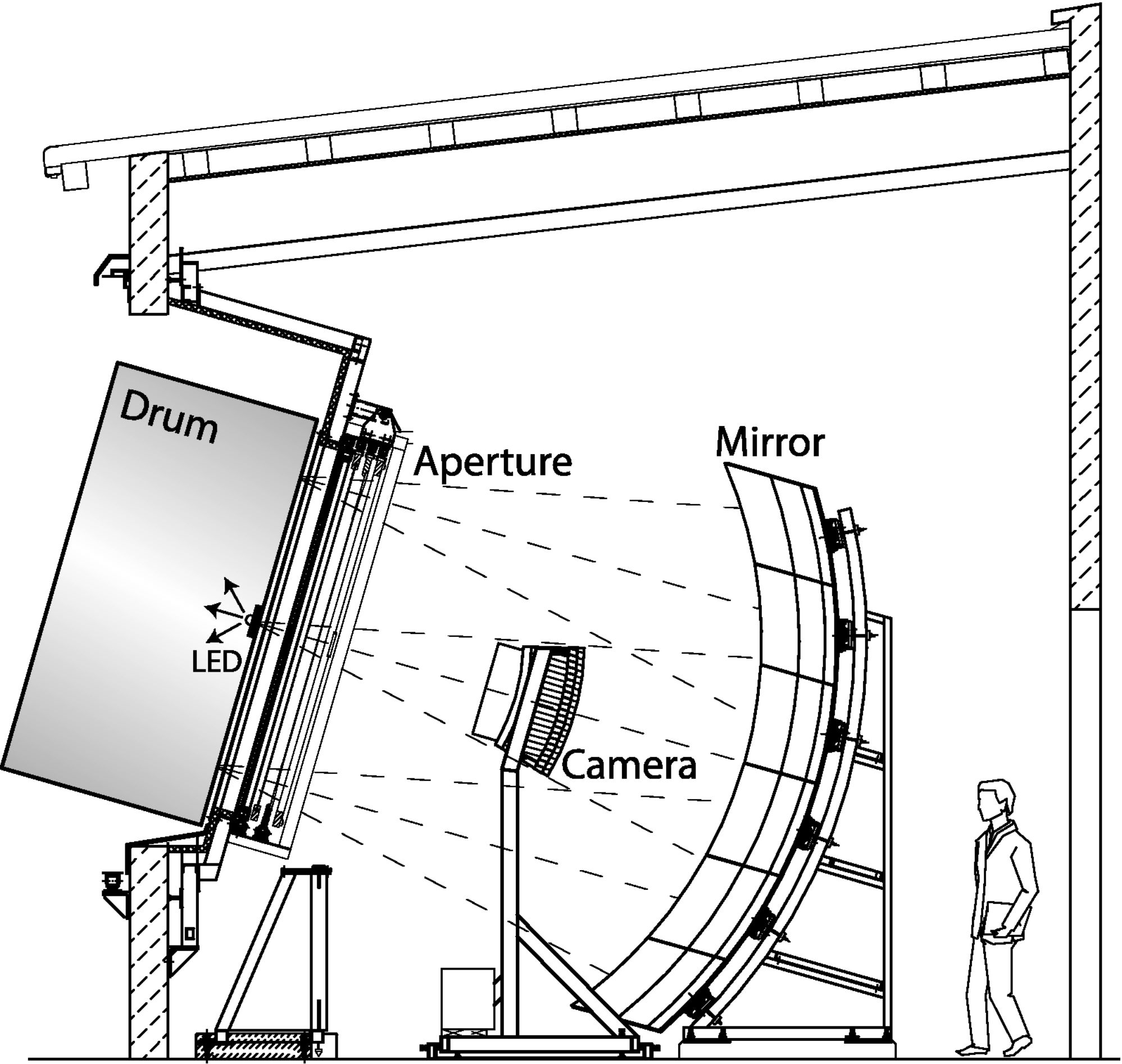}}\hfill
    \subfigure[multi-wavelength measurements, normalized at 375\,nm.]
    {\label{asfd_fig9}
    \includegraphics[clip, bb=0 -40 537  357,width=0.49\textwidth]{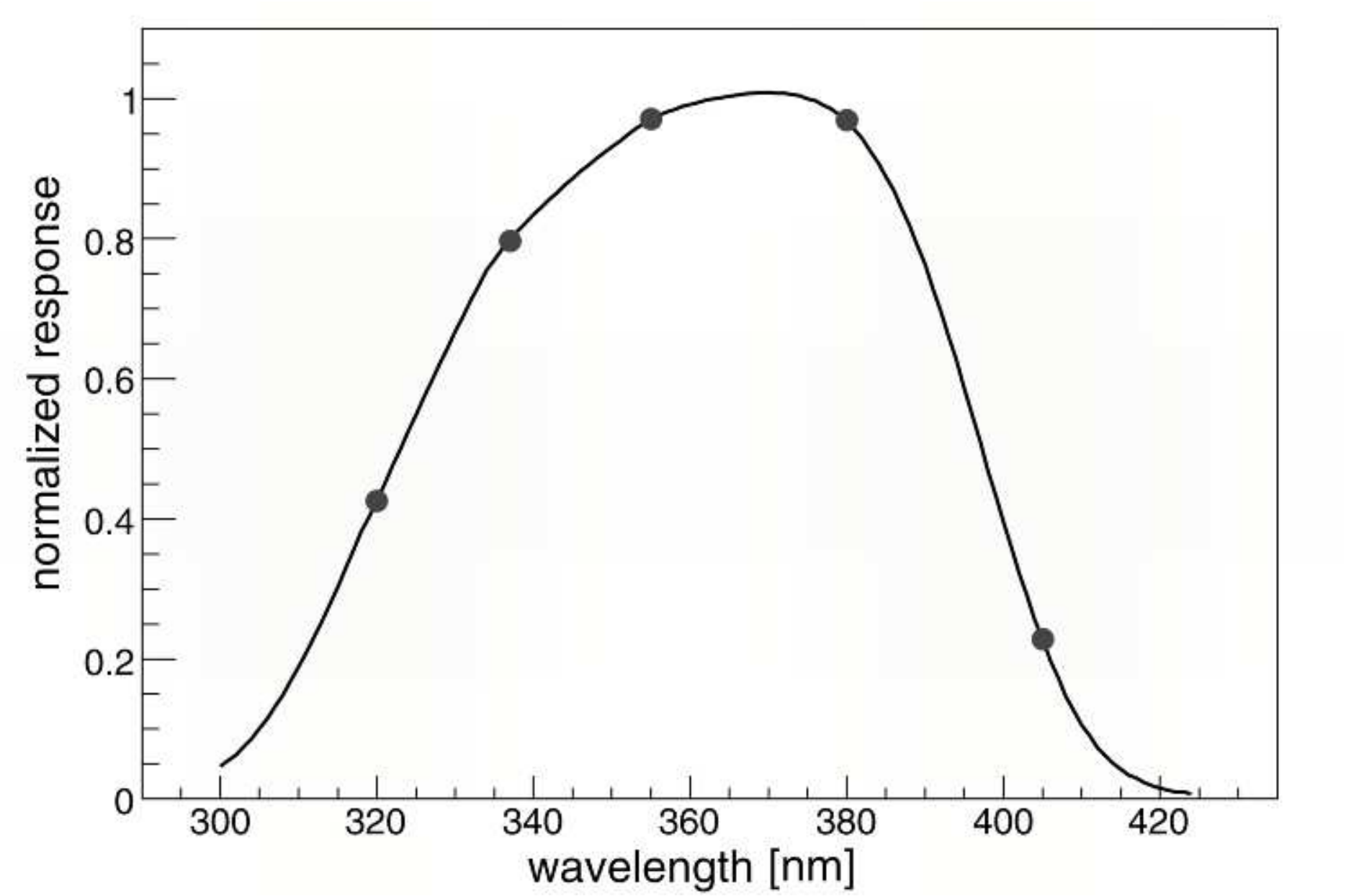}}
    \caption{Detector calibration with the 'drum'.}
    \label{fig:drum}
\end{figure}

The absolute calibration of the fluorescence detectors uses a portable
drum shaped calibrated light source at the telescope aperture,
providing uniform illumination to each pixel.  The technique
\cite{Brack:2013bta} is based on a 2.5\,m diameter, 1.4\,m deep,
drum-shaped light source which mounts on the exterior of the FD
apertures (see Figure \ref{asfd_fig8}).  The source provides a pulsed
photon flux of known intensity and uniformity across the aperture, and
simultaneously triggers all the pixels in the camera.  In the lab,
light source uniformity is studied using CCD images and the intensity
is measured relative to NIST calibrated photodiodes.  Use of the drum
for gain adjustment and calibration provides a known, uniform response
for each pixel in a detector.

For calibration at wavelengths spanning the FD acceptance, a xenon
flasher is mounted at the back of the drum, with a filter wheel
containing 5 notch filters for selection of wavelengths.  The xenon
flasher \cite{Rovero2009305} provides 0.4\,mJ optical output per pulse
covering a broad UV spectrum, in a time period of a few hundred
nanoseconds.  Relative drum intensity measurements at wavelengths of
320, 337, 355, 380 and 405\,nm have been made with the same reference
PMT used in the absolute measurements.  The signals detected at the
various wavelengths combine with the lab work to form a curve of
relative camera response shown in Figure~\ref{asfd_fig9}.  A new
detailed measurement procedure was developed that utilized a
monochromator and UV light source to measure the FD efficiency in
5\,nm steps and found efficiencies consistent with the curve in
Figure~\ref{asfd_fig9}~\cite{Gookin2014}.

Three additional calibration tools are used at Auger.
First, before and after each night of data taking a relative
calibration of the PMTs is performed \cite{Abraham:2009pm}.  This
relative calibration is used to track both short and long term changes
in detector response.  Secondly, the relative FD response has been
measured at wavelengths of 320, 337, 355, 380 and 405\,nm, defining a
spectral response curve that has been normalized to the absolute
calibration.  Thirdly, an independent check of the calibration in some
phototubes is performed using vertical shots from a portable laser in
the field.

\subsection{Atmospheric monitoring}
\label{subsec:atmos}

The exploitation of the calorimetric measurement of the fluorescence
signal in the atmosphere depends essentially on the efficiency of
fluorescence light production and subsequent transmission to an FD
telescope.  In particular, the aerosol content of the atmosphere, in
the form of clouds, dust, smoke and other pollutants, needs to be well
characterized.  The aerosol content of the atmosphere can be variable
on short time-scales necessitating the routine monitoring of light
transmission conditions in the atmospheric volume above the Pierre
Auger Observatory.  To account for possible horizontal
non-uniformities in the aerosols the area enclosed by the observatory
is divided into 5 sub-regions within which only the vertical
characteristics of the aerosols are described.  Within each region the
aerosols are characterized in vertical slices of 200\,m thickness, up
to a height of 10 km.  The aerosol parameters that are important for
EAS reconstruction are the $\text{VAOD}(h)$, the vertical aerosol optical
depth as a function of height, $\alpha(h)$, the aerosol scattering
coefficient as a function of height and $\mathrm{d}\sigma/\mathrm{d}\Omega$, the
aerosol differential cross section.  The wavelength dependence of
these parameters in the 300 to 400\,nm sensitivity range of the FDs is
also measured.  Aerosol parameters are updated hourly during the
periods of FD operation.

\begin{figure}[htb]
\centering
\includegraphics[width=0.48\textwidth]{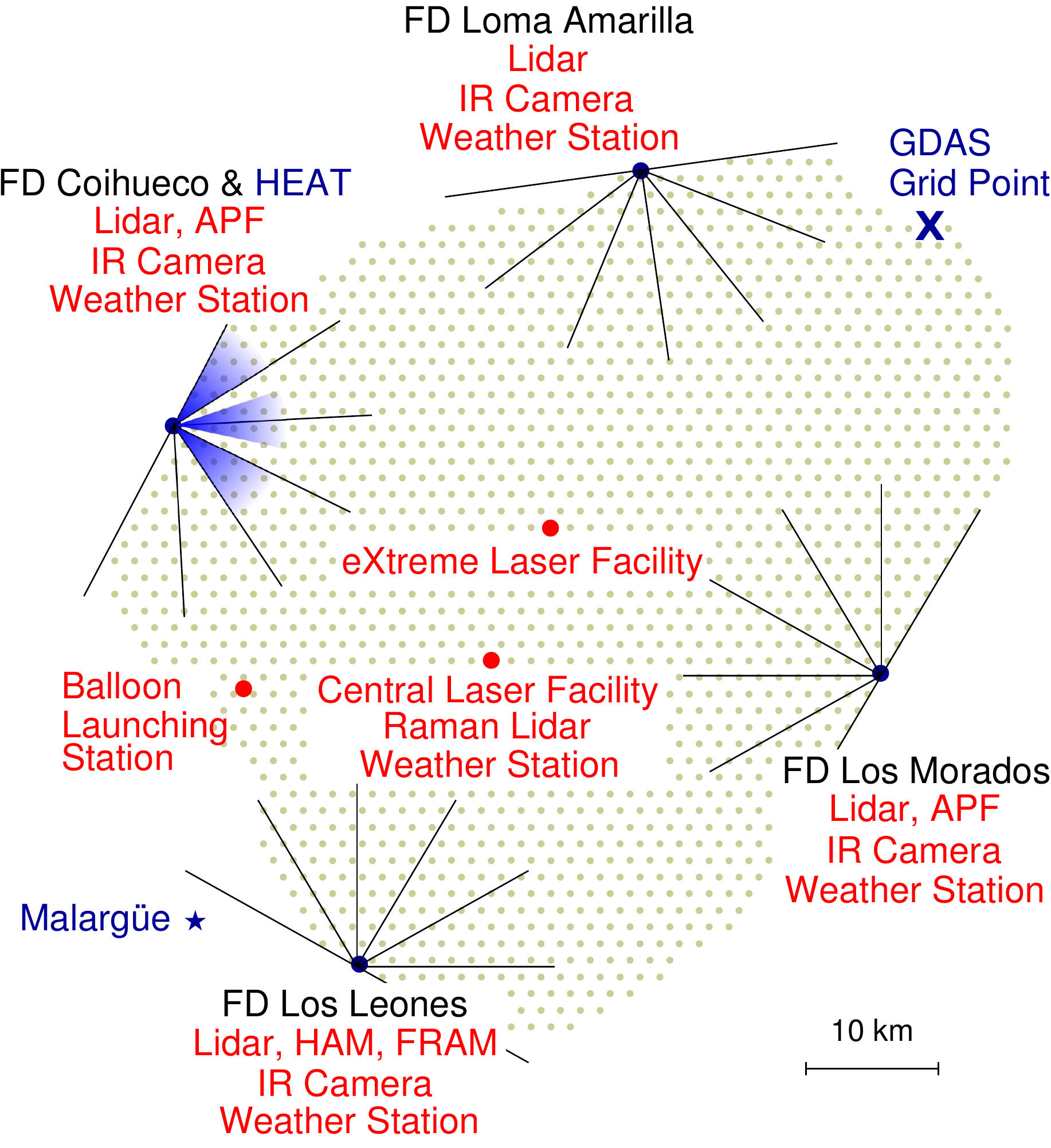}

\caption{Schematic overview of the atmospheric monitoring devices
  installed at the Pierre Auger Observatory. At each FD site, there is
  a lidar station, a ground-based weather station, and an infra-red
  camera for cloud cover detection. In addition, there are devices for
  measuring the Aerosol Phase Function (APF) at FD Coihueco and Los
  Morados, a Horizontal Attenuation Monitor at FD Los Leones, and a
  ph(F)otometric Robotic Atmospheric Monitor also at Los Leones.  A
  steerable backscatter elastic lidar system is installed at each of
  the 4 FD sites to measure aerosols and the positions of clouds near
  each site.  At central positions within the surface detector array,
  two laser facilities are installed (CLF and XLF) to measure
  $\tau_{\rm aer}(h)$ in the line of sight of each FD telescope 4
  times per hour.  In 2013 the CLF was upgraded with a Raman lidar.
  At the western boundary of the array, the Balloon Launching Site has
  been assembled together with a weather station.  From this station,
  the weather balloons were launched so that they were typically
  carried across the entire array by westerly winds.}
\label{fig:atmosOverview}
\end{figure}

These measurements are accomplished using a complex set of instruments
including backscatter \textsc{lidar}s, two laser facilities (the
Central Laser Facility, CLF and XLF) near the middle of the array,
horizontal attenuation monitors, Aerosol Phase Function monitors, star
monitors and cloud cameras.  The location of these components is shown
in Figure \ref{fig:atmosOverview} and are described in more detail in
\cite{AugerSouthNIM}.

\section{Communications System and CDAS}

\subsection{Communications system} \label{sec:as-comms}
\label{s:comms_appendix}

Due to the large coverage area and widely dispersed nature of the 1660
Cherenkov detectors that makes up the surface detector array, a
communications system based on radio technology was deemed to be the
only economically viable solution for the Pierre Auger Observatory.
The system consists of two integrated radio networks organized as a
2-layer hierarchy: the individual detectors are serviced by the
surface detector wireless LAN (WLAN), which is a sectorized network
supported by 4 data-concentration nodes.  These WLAN nodes are
serviced by a high capacity microwave backbone network.  The backbone
also supports communications from the Fluorescence Detector sites.

The data-rate requirements of the surface detector array are
determined primarily by the T2 triggering rate of the individual
surface detectors.  The presence of an analysis computer at each
detector greatly reduces the required bandwidth as local events within
the detector must pass through several stages of discrimination before
they need to be communicated to the Central Data Acquisition System
(CDAS) which is described below.

The design specification for the uplink from each detector to CDAS is
a continuously available capacity of 1200 bps.  A reverse downlink
data path is also required so that the CDAS system may request full T3
trigger readouts from those detectors that have collected relevant
data.  A downlink broadcast capacity of 2400 bps to all detectors is
sufficient to meet this requirement.

A bi-directional 2.048\,Mbps link is available to each fluorescence
detector building via direct connection to the microwave backbone
network.

\begin{figure}[t]
\centering
\includegraphics[width=0.60\textwidth]{comms_overall_schematic_revised_fixed}\hfill
\includegraphics[width=0.3\textwidth]{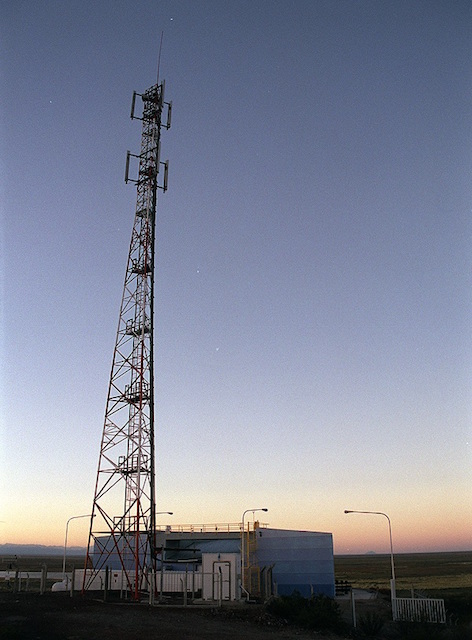}
\caption{\textit{Left:} Conceptual schematic of the overall radio
  telecommunications system for the Pierre Auger
  Observatory. \textit{Right:} One of the five communications towers:
  the one shown is deployed at the Los Leones site. }
\label{comms_schematic}
\end{figure}

\paragraph{Microwave Backbone Network }

The backbone network uses a standard 34\,Mbps telecommunications
architecture based on commercially available microwave point-to-point
equipment operating in the 7\,GHz band.  The equipment consists of
dish-mounted microwave transceivers installed on communications
towers, together with secondary units located in shelters at the base
of each tower.  As shown in Fig.~\ref{comms_schematic} (left), the backbone
network consists of two arms, both of which terminate at the
Observatory Campus in Malarg\"{u}e, at which point the data are routed to
the central data acquisition system.  The backbone network has
sufficient capacity for the transfer of both the surface detector WLAN
data and all FD data to and from the fluorescence detector sites.

\paragraph{Surface Detector WLAN }

The surface detector WLAN has been specially designed for the Auger
project using custom radio hardware running proprietary network access
protocols.  This network operates in the 902-928~MHz industrial,
scientific and medical (ISM) radio band and provides data
communications to and from each of the 1660 surface detectors over a
3000\,km$^{2}$ area.  This is achieved in a manner similar to a
cellular telephone system, whereby the area containing the detectors
is divided into a number of sectors, and communications within each
sector are coordinated by a base-station.  Factorization is required
in order to meet legislation pertaining to maximum transmitter powers
and frequency re-use within the ISM band.  It also greatly distributes
the data processing load of the array, and reduces the possibility
that a failure at a data collection node will cause a total
Observatory outage.  A single sector typically contains 57 surface detectors. %

\paragraph{Data Path from the Surface Detectors to Campus }

WLAN air-interface and related functions at each surface detector are
performed by a subscriber unit.  This unit communicates with the main
surface detector electronics module via a serial link, and
incorporates a proprietary digital radio transceiver running
appropriate firmware on its control processor.  The subscriber unit is
connected to a 12\,dBi Yagi antenna via a short low-loss feeder, with
the antenna mounted on a short communications mast integrated into the
detector's solar panel support.  This can be seen in
Fig.~\ref{fig:sd}, which shows a typical detector tank installation.
The antenna is mounted at a height of 3\,m above the ground.

Data are transmitted over a path of up to 30\,km to a local data
concentration tower, where the signal is received via a high-gain
cellular-style panel antenna.  The antenna is connected via very
low-loss feeders to a base-station unit located in a shelter at the base
of the tower.  A single base-station can serve up to 68 detector
tanks.  A base-station incorporates the same digital radio transceiver
platform employed in the subscriber units, with additional processing
capability.  

At each tower, data from several base-station units is concentrated
onto an E1-ring and then processed by a custom interface before being
passed on to the backbone microwave network for transmission to the
Observatory campus in Malarg\"{u}e.  There the data passes from the E1
microwave network into the central data acquisition system via TCP/IP
running on a conventional Ethernet network.

\paragraph{Digital Radio Transceiver Development }

The need to provide many highly robust data links over distances of
30\,km and beyond using a minimal amount of power has presented some
unique equipment performance challenges that could not be met with
existing communications equipment.  To meet the requirements, a
low-power digital radio transceiver platform has been designed, the
functionality of which is determined by re-configurable firmware
stored in Flash memory and loaded into a digital signal processor
(DSP) when the unit powers up.  This reconfigurability not only allows
the functionality of the transceiver to evolve and be refined over
time, it also allows a common radio transceiver platform to be used
for both subscriber units and base-station units, thereby reducing the
hardware development time and simplifying long-term hardware support.

The transceiver uses very low power devices and a highly flexible
architecture to provide reliable long range digital communications
within the strict power budget of the solar-powered surface detectors.
Power consumption is less than 1.1\,W of DC power.

\subsection{Central Data Acquisition System, CDAS}\label{subsec:cdas}

The CDAS has been running since March 2001.  The system was designed
to assemble the triggers from the surface array detectors, to allow
control of these detectors and to organize the storage of data.  It is
constructed using a combination of commercial hardware and custom
made, high level, software components. The system is designed to run
continuously, with minimum intervention, with the full 1660 detector
array, and can manage many more. Data from the FD are recorded
separately at the FD locations and transferred daily to the computer
center at Malarg\"{u}e, although hybrid coincidences are identified
on-line within the SD data stream.

The primary role for the CDAS is to combine local trigger information
from the SD stations in order to identify potential physical events
generating an SD higher level trigger (T3). These triggers combined
with the T3 from FD sites (FD T3) are used to generate a request for
the relevant data from SD stations for these events. The CDAS is then
used to combine and store these data to form a shower event. The CDAS
also contains configuration and control mechanisms, the means to
monitor system performance, and the tools to access and download SD
monitoring, calibration, control and configuration data.

Except for triggering information (see
section~\ref{event-triggering}), the CD\-AS and the FD data
acquisition systems are completely independent. The merging of FD and
SD data is made off-line during the daytime following an FD run.  Data
are synchronized on the central storage hardware after each night of
observation.  The newly acquired data within the central storage are
mirrored at the primary data mirror located at the Lyon HEP Computer
Center (France) every 3 hours; later these data can be transferred to
secondary mirror sites such as Fermilab.  The data may then be
transferred from a convenient mirror site to over 50 participating
institutions.

The communication between applications within the CDAS is controlled
using a central message routine manager called the `Information
Kernel'. This manager allows formatted messages to be broadcast by
producer applications (applications that need to advertise their
status), and for consumer applications (applications that need to know
about the status of others) to receive them on demand.  All data, with
one exception, are exchanged between the CDAS applications in human
readable formatted ASCII and go through the `Information Kernel'
manager.  The exception is the large binary block of raw data coming
from the SD stations.  Data exchanged in raw format are calibration
blocks and FADC traces (these comprise the event data), data from
local triggers as well as the monitoring data.

\paragraph{Data Collection}

The data flow over the radio network, from individual SD stations to
the central campus, is controlled by a dedicated application called
the 'Post Master'. The Post Master is the end point of the
communication backbone at the Observatory Campus, and is designed to
dispatch information extracted from the different data streams of a
local station to the other applications of the CDAS. As its name
suggests, the Post Master application is used to read the data type
contained in a radio frame and to forward it to the proper application
within the CDAS so that specific data can be handled. When the data
received from individual SD stations are split into several radio
frames, they are reassembled and forwarded to clients by the Post
Master after all the frames have been received.

The Post Master is used also to route data between the applications of
the CDAS and the SD. Commands and configuration parameters can be
transmitted, along with event requests, such as the level 3 trigger
identified by the 'Central Trigger' processor. Software downloads over
the communications link are also possible, thus enabling upgrades of
the local DAQ software at the stations without the need to travel many
kilometers to each one.

Data received from each SD station belong to different data streams:
Local triggers, forwarded to the Central Trigger application; Shower
and calibration data, forwarded to the Event Builder application;
Control data, forwarded to the Information Kernel application; and
Calibration and monitoring data, forwarded to the Monitoring Recorder
application.

\paragraph{Event Triggering System}\label{event-triggering}

The triggering system of the Observatory fulfills two important
conditions.  First it must be able to detect showers with high
efficiency across the SD, namely 99\% efficiency for showers with
energy above $3{\times}10^{18}$\,eV and zenith angle less than
60$^\circ$.  Secondly, it allows and identifies cross-triggers (hybrid
events) between the FD and SD systems.

The local DAQ system of each SD station is designed to generate low
level triggers (T2) as described in section~\ref{subsec:triggers}. The
time stamps of these triggers are sent every second to the CDAS. The
T2 requirements are such that the average rate per station is always
around 20 to 25\,Hz so that at least 50\% of the bandwidth is free for
data transmission.

The CDAS Central Trigger processor is used to identify groups of
stations that are clustered in time and space as SD events.  These are
T3s, and are created if they fulfill one of the following conditions:

\begin{enumerate}
\item {The main trigger condition is based on both the temporal and
the spatial clustering of the local triggers (T2) received from each
station.  Basically, temporal clusters are sought by centering a time
window on each T2.  Clusters, with multiplicity of three or more, are
then examined for spatial coincidences.  For 3-fold coincidences, the
triggered stations must lie within the first two crowns (hexagons)
centered on the station whose T2 served as the center of the time
window.  For a 4-fold coincidence, one station with a T2 may be as far
away as in the fourth crown.  This condition is, however, stronger
than the 3-fold coincidence requirement.  Thus, all 4-fold triggers
are also 3-fold triggers.  Once the spatial coincidence is verified,
final timing criteria is imposed: each T2 must be within
$(3+8n)\,\upmu$s of the central one where $n$ represents the crown
number.}
\item {A random trigger is generated every $N$ minutes (with
$3<N<30$) by selecting one of the T2s in an arbitrary manner and
promoting it to a T3.  The purpose of this trigger is to monitor
randomly the FADC traces that satisfy the local trigger conditions and
thus to verify the efficiency of the global trigger processor.}
\end{enumerate}
Once a trigger has been identified, a message requesting all FADC
trace information recorded within a certain time of the central T2 sent to all
stations in the array.

The DAQ system of the fluorescence detector is completely independent
of the CDAS. Local triggers are generated at each camera and those
identified as T3 FD event triggers are logged by a local processor if
a shower track can be found.  After each night of operation, details
of events recorded at the FD telescopes are transferred to the CDAS.

To build the hybrid event set, the T3 information from the FD local
DAQ system is transmitted in real time to the CDAS. This trigger
information includes an estimate of the geometry of the shower
candidate, including the time of arrival of the light front of the
shower at the camera.  The time of the shower impact at a ground
position in the region of the SD stations is computed from this
information, and a corresponding SD event T3 is constructed.  All FADC
traces recorded within a time window of the computed time are
assembled as a normal ``SD-only'' event, with the addition of the
identification of the corresponding FD T3 trigger.  Data from these
triggers forms the hybrid data set and is merged with the data
collected by the FD DAQ and analyzed offline.  

\paragraph{Monitoring}\label{subsec:cdasmonitoring}

The CDAS provides monitoring information for its own operation, as
well as the slow control information of the SD stations and various
environmental parameters.  The operation of the CDAS is monitored
using a low level application that routinely checks that all software
components are running correctly.  This ``watch dog'' system is used
to re-initialize and re-launch any application that may have failed.

SD station monitoring information is sent by the stations at regular
intervals.  This information consists mainly of ``slow control'' data,
PMT and CPU board voltages and environmental parameters.  Calibration
data are also collected by the CDAS.

\section{Data processing and \Offline Software}
\label{appendix:offline}

The Pierre Auger Observatory \Offline software provides an
infrastructure to support development of hybrid event simulation and
reconstruction.  The software has been designed to accommodate
contributions from a large number of physicists developing C++
applications over a long experimental run.  The essential features
include a ``plug-in'' mechanism for physics algorithms together with
machinery which assists users in retrieving event and detector
conditions data from various data sources.  A detailed description of
the \Offline software design, including some example applications, is
available in~\cite{Argiro:2007qg}; additional information is also
given in~\cite{AugerSouthNIM}.

\begin{figure}[t]
\centering
\includegraphics[width=0.4\textwidth]{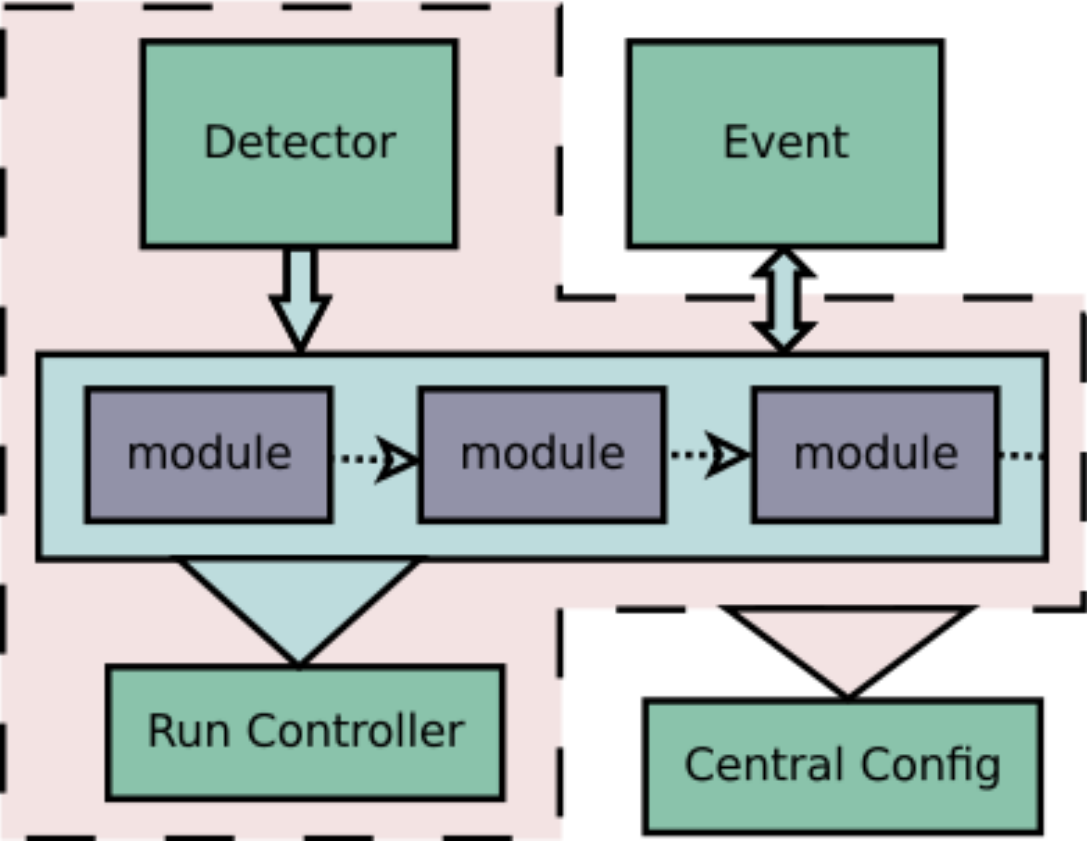}
\caption{General structure of the \Offline framework.
Simulation and reconstruction tasks are encased
in modules.  Each module is able to read information
from the detector description and/or the event,
process the information, and write the results back
into the event under command of a {\em Run Controller}.
A {\em Central Config} object is responsible for handing
modules and framework components their configuration
data and for tracking provenance.}
\label{offline_overview}
\end{figure}

The overall organization of the \Offline framework is depicted in
Fig.~\ref{offline_overview}.  A collection of processing {\em modules}
are be assembled and sequenced through instructions contained in an
XML file~\cite{xml}.  An {\em event} data model allows modules to
relay data to one another, accumulates all simulation and
reconstruction information, and converts between various formats used
to store data on file.  Finally, a {\em detector description} provides
a gateway to detector conditions data, including calibration constants
and atmospheric properties as a function of time.

Simulation and reconstruction tasks are factorized into sequences of
processing steps which can simply be pipelined.  Physicists prepare
processing algorithms in modules, which they register with the \Offline
framework.  This modular design allows collaborators to exchange code,
compare algorithms and build up a variety of applications by combining
modules in various sequences.  Run-time control over module sequences
is obtained through a {\em run controller}, which invokes the various
processing steps within the modules according to a set of
user-provided instructions written in XML.

The \Offline framework includes two parallel hierarchies for accessing
data: the detector description for retrieving conditions data,
including detector geometry, calibration constants, and atmospheric
conditions; and an event data model for reading and writing
information that changes for each event.

The {\em detector description} provides a unified interface from which
module authors can retrieve conditions data.  Data requests are passed
by this interface to a back end comprising a registry of so-called
managers, each of which is capable of extracting a particular sort of
information from a given data source.

The transient (in memory) and persistent (on disk) event models are
decoupled.  When a request is made to write event contents to file,
the data are transferred from the transient event through a so-called
{\em file interface} to the persistent event, which is instrumented
with serialization machinery, currently using ROOT~\cite{root}.
Various file formats are interpreted using the file interface,
including raw event and monitoring formats as well as the different
formats employed by the AIRES~\cite{Sciutto:1999jh}, CORSIKA~\cite{Heck:1998vt},
CONEX~\cite{Bergmann:2006yz} and SENECA~\cite{Drescher:2002cr} air shower simulation
packages.

The \Offline framework includes a system to organize and track data
used to configure of the software for different applications as well
as parameters used in the physics modules.  A {\em central
configurator} points modules and framework components to the location
of their configuration data, and creates Xerces-based~\cite{xerces}
XML parsers to assist in reading information from these locations.

The central configurator keeps track of all configuration data
accessed during a run and stores them in an XML log file, which can
subsequently be used by the central configurator to reproduce a run
with an identical configuration.  The logging mechanism is also used
to record the versions of modules and external libraries which are
used for each run.  Syntax and content checking of the configuration
files is afforded through W3C XML Schema~\cite{xml-schema} standard
validation.  The configuration machinery can also verify configuration
file contents against a set of default files by employing MD5
digests~\cite{md5}.

The \Offline framework is complemented by a collection of utilities,
including an XML parser, an error logger and various mathematics and
physics services.  We have also developed a novel geometry package
which allows the manipulation of abstract geometrical objects
independent of coordinate system choice.

Low-level components of the framework are verified with a small test
program, known as a unit test, while full applications are vetted with
more detailed acceptance tests.  We employ a BuildBot
system~\cite{buildbot} to automatically compile the \Offline software,
run the unit and acceptance tests, and inform developers of any
problems each time the code is modified.

\section{Event Reconstruction and Aperture}

Extensive air showers in the Auger energy range are such dramatic and
large scale events that there is essentially no background to both SD
and FD measurements.  Triggers are easily set up to exclude virtually
any possibility of chance coincidences of triggers of the individual
SD stations and/or FD pixels that would mimic a real cosmic-ray
shower.  Therefore, the performance of the detector and its ability to
produce high-quality data depend solely on the accuracy of the
cosmic-ray shower reconstruction and of the computation of the
acceptance of the detector.

\subsection{Surface Detector}\label{sec:SDreco}
Several experiments have proved successful in measuring extensive air
shower parameters by use of a surface array. The quantities that can
be measured directly are the geometry of the shower axis and the
lateral distribution function (LDF), or the particle signal as a
function of distance from the core. The primary energy can be inferred
from the LDF, or, more specifically, from $S(1000)$, the detector
signal at 1000\,m from the core. At smaller distances, close to the
core, fluctuations due to the nature of the first interactions of the
primary with the atmosphere are dominating, while at larger distances
statistical fluctuations become important. The relation between
$S(1000)$ and the primary energy established by using shower
simulations is therefore model dependent. The calibration of the of a
zenith angle independent measure of $S(1000)$, $S_{38}$, with the FD
energy is developed and used instead. The SD-only reconstruction takes
place in three steps: event and station selection; determination of
the shower geometry; and measurement of the shower lateral distribution
function (LDF).

\begin{figure}[!t]
  \center
  \subfigure[Footprint of the shower.]
  {\label{fig:Footprint}
    \includegraphics[width=0.37\textwidth]{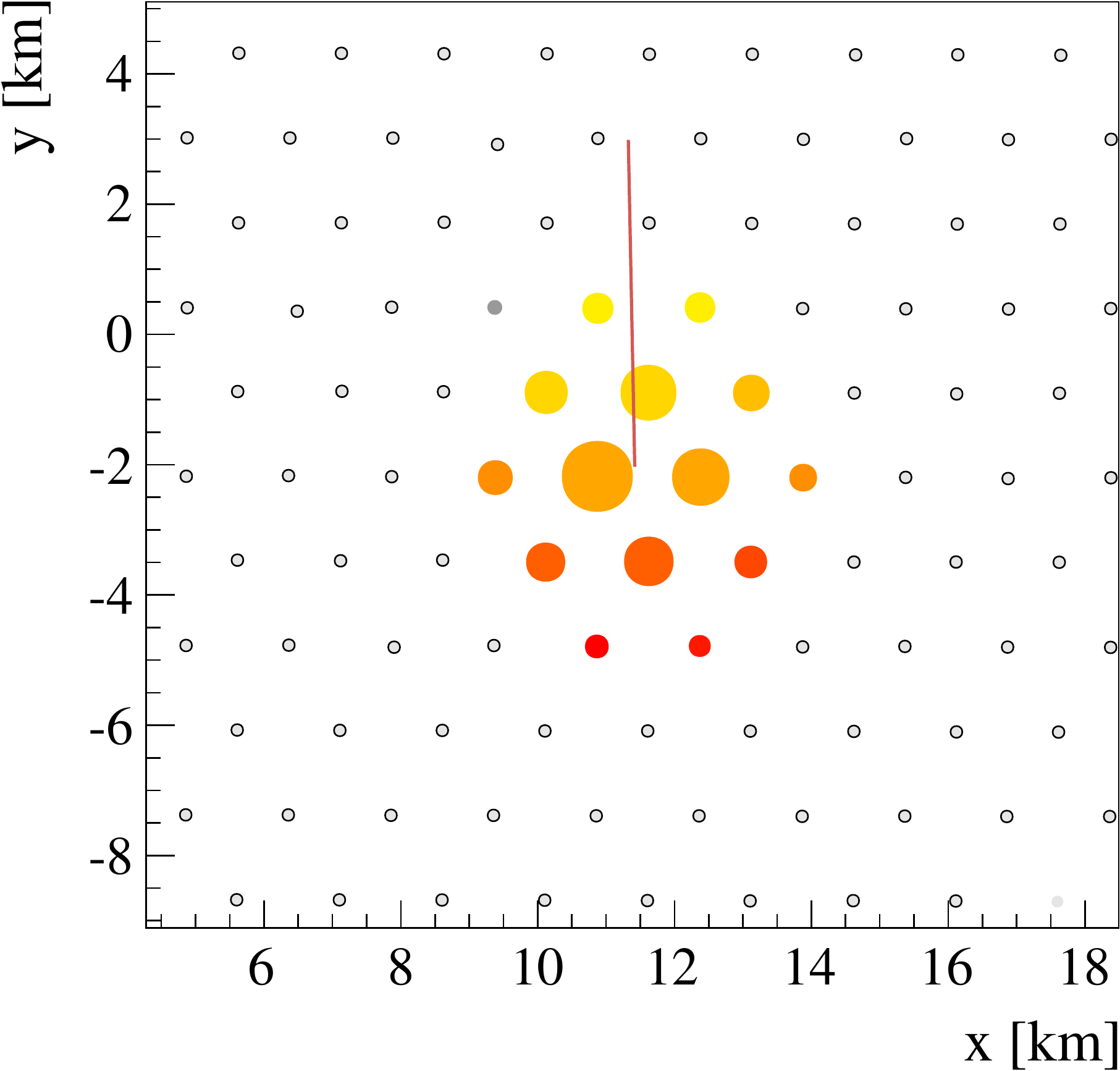}}\hfill
  \subfigure[Lateral distribution. Dependence of the signal size on distance from the shower core.]
  {\label{fig:LDF}
    \includegraphics[width=0.50\textwidth]{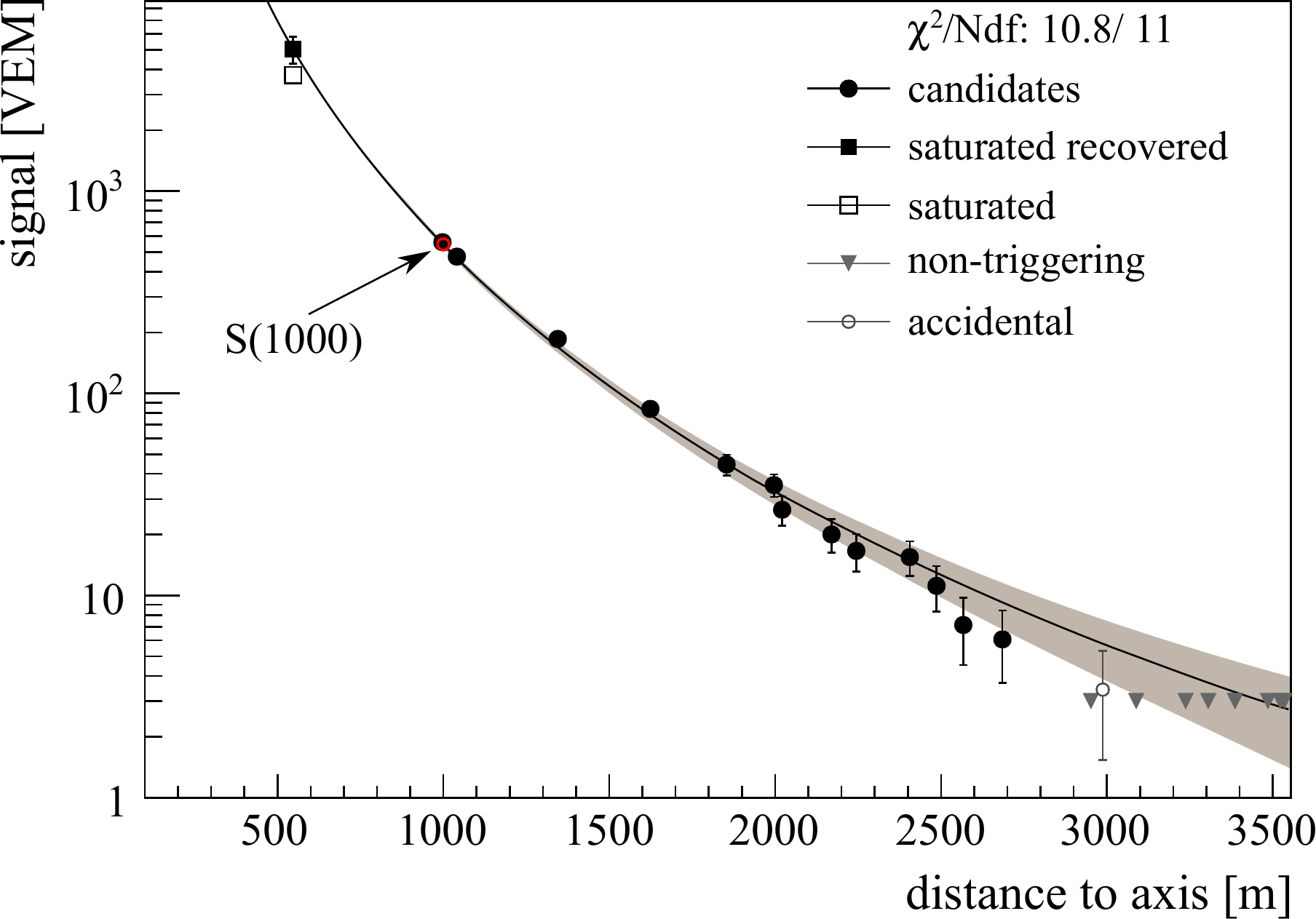}}
  \caption{Footprint and lateral distribution of a reconstructed SD
    event ($E=104$\,EeV, $\theta=25.1^\circ$).  In the
    footprint plot, colors represent the arrival time of the shower
    front from early (yellow) to late (red) and the size of the
    markers is proportional to the logarithm of the signal. The line
    represents the shower arrival direction.}
  \label{fig:SDEventAS}
\end{figure}

\paragraph{Event selection} 
To ensure good data quality for physics analysis there are two
additional off-line triggers.  The physics trigger, T4, is needed to
select real showers from the set of stored T3 data (see
section~\ref{event-triggering}) that also contain background signals
from low energy air showers.  This trigger is mainly based on a
coincidence between adjacent detector stations within the propagation
time of the shower front.  In selected events, random stations are
identified by their time incompatibility with the estimated shower
front. Time cuts were determined such that 99\% of the stations
containing a physical signal from the shower are kept. An algorithm
for the signal search in the time traces is used to reject signals
produced by random muons by searching for time-compatible peaks.

To guarantee the selection of well-contained events, a fiducial cut
(called the 6T5 trigger) is applied so that only events in which the
station with the highest signal is surrounded by all 6 operating
neighbors (i.e., a working hexagon) are accepted. This condition
assures an accurate reconstruction of the impact point on the ground,
and at the same time allowing for a simple geometrical calculation of
the aperture/exposure \cite{Abraham:2010zz}, important for, e.g., the
spectrum analysis \cite{Abraham:2010mj}. For arrival-direction studies
a less strict cut can be used (5T5 or even 4T5).

\paragraph{Geometry and Energy measurement}
An approximate shower geometry solution can be obtained from the
simplified linear model assuming that all stations lie within some
plane, i.e.\ here the tangential plane on the reference ellipsoid that
contains the signal-weighted barycenter is chosen. In such a case one
can expect $z_i \ll x_i,\,y_i$ for the station position
$(x_i,y_i,z_i)$. The $z$-component is neglected and the linear
$\chi^2$ is obtained,
\begin{equation}
  \chi^2=\sum_i\frac{[ct_i-ct_0+ux_i+vy_i]^2}{\sigma^2_i},
  \label{chi2-simple}
\end{equation}
where $t_i$ is the signal start time in tank $i$ and $t_0$ is the
time when the shower passes the barycenter.
Equation~\ref{chi2-simple} can be expressed as a set of linear
equations and is analytically solved. The approximate solution serves
as starting point to more elaborate 3D-fitting attempts taking the
varying altitude of the stations, and a more accurate core location
from the LDF fit (below), into account.

An example of the footprint on the array of an event produced by a
cosmic ray with an energy of (104$\pm 11$)\,EeV and a zenith angle of
$(25.1\pm 0.1)^\circ$ is shown in figure~\ref{fig:SDEventAS}. The
lateral distribution of the signals is depicted in
figure~\ref{fig:LDF}. The function employed to describe the lateral
distribution of the signals on the ground is a modified
Nishimura-Kamata-Greisen function
\cite{Kamata:1958,Greisen:1956},
\begin{equation}
S(r) = S(r_\text{opt})
  \left(\frac{r}{r_\text{opt}}\right)^\beta
  \left(\frac{r+r_1}{r_\text{opt}+r_1}\right)^{\beta+\gamma}
\end{equation}
where $r_\text{opt}$ is the optimum distance, $r_1=\unit[700]{m}$ and
$S(r_\text{opt})$ is an estimator of the shower size used in an energy
assignment. For the SD array with station spacing of 1.5\,km the
optimum distance \cite{Newton:2007} is $r_\text{opt}=1000$\,m and the
shower size is thus $S(1000)$.  The parameter $\beta$ depends on the
zenith angle and shower size. Events up to zenith angle $60^\circ$ are observed at an
earlier shower age than more inclined ones, thus having a steeper LDF
due to the different contributions from the muonic and the
electromagnetic components at the ground.  For events with only 3
stations, the reconstruction of the air showers can be obtained only by
fixing the two parameters, $\beta$ and $\gamma$ to a parameterization
obtained using events with a number of stations larger than 4.

\begin{figure}[t]
\centering
\includegraphics[width=0.4\textwidth]{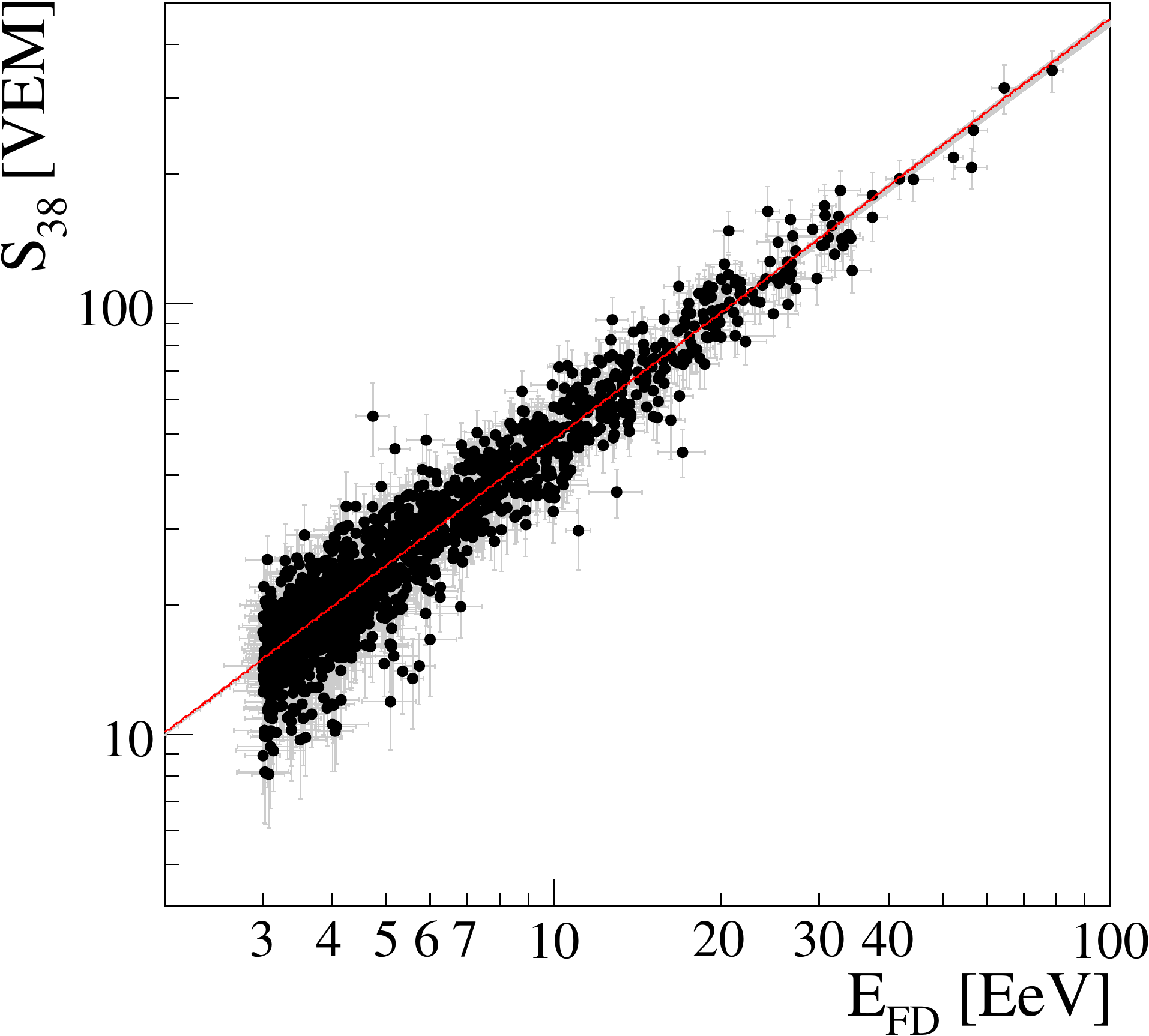}
\caption{Correlation between $S_{38}$ and $E_\text{FD}$
\cite{Schulz-ICRC:2013,Pesce-ICRC:2011}.}
\label{fig:energy_calibration}
\end{figure}

The primary particle energy is determined from $S(1000)$ and the
shower zenith angle $\theta$.  For a given energy, the value of
$S(1000)$ decreases with $\theta$ due to the attenuation of the shower
particles and geometrical effects. Assuming an isotropic flux of
primary cosmic rays at the top of the atmosphere, we extract the shape
of the attenuation behavior from the data using the Constant
Intensity Cut (CIC) method \cite{Hersil:1961zz}. An attenuation curve
$f_\text{CIC}(\theta)$ has been fitted with a third degree polynomial
in $x=\cos^2\theta-\cos^2\bar\theta$,
i.e.,\ $f_\text{CIC}(\theta)=1+a\,x+b\,x^2+c\,x^3$, where
$a=0.980\pm0.004$,
$b=-1.68\pm0.01$, and $c=-1.30\pm0.45$~\cite{Schulz-ICRC:2013}.

The median angle, $\bar\theta=38^\circ$, is taken as a reference point to
convert $S(1000)$ to \\
$S_{38}\equiv S(1000)/f_\text{CIC}(\theta)$.  $S_{38}$ may
be regarded as the signal a particular shower with size $S(1000)$ would have
produced had it arrived at $\theta=38^\circ$.

High quality hybrid events, events seen by both the SD and FD, are
used to calibrate $S_{38}$ with the near-calorimetric measurement of
the primary energy by the FD, $E_\text{FD}$.  The 1475 high quality
hybrid events recorded between Jan 2004 and Dec 2012 which have an
energy above the SD full efficiency trigger threshold
\cite{Abraham:2010zz} are used in the calibration.  The correlation
between the two variables is obtained from a maximum likelihood method
\cite{Pesce-ICRC:2011,Dembinski-ICRC:2011} which takes into account
the evolution of uncertainties with energy, as well as event
migrations due to the finite energy resolution of the SD. The relation
between $S_{38}$ and $E_\text{FD}$ is well described by a single
power-law function,
\begin{equation}
E_\text{FD} = A\,(S_{38}/\text{VEM})^B
\end{equation}
where the resulting parameters from the data fit are
$A=(1.90\pm0.05){\times}10^{17}$\,eV and
$B=1.025\pm0.007$~\cite{Schulz-ICRC:2013,Verzi-ICRC:2013}.  As can be seen in
figure~\ref{fig:energy_calibration}, the most energetic event used in this
analysis has an energy of 79\,EeV.

The resolution of the final SD energy estimator,
\begin{equation}
E_\text{SD} =
  A(S(1000)/f_\text{CIC}(\theta)/\text{VEM})^B,
\end{equation}
can be inferred from the distribution of the ratio
$E_\text{SD}/E_\text{FD}$.  Using the FD energy resolution of 7.6\%,
the resulting SD energy resolution with its statistical uncertainty is
$\sigma_{E_\text{SD}}/E_\text{SD}=(16\pm1)$\% at the lower energy
edge in figure~\ref{fig:energy_calibration} and $(12\pm1)$\% at the
highest energies. Due to the large number of events accumulated until
December 2012, the systematic uncertainty on the SD energy due to the
calibration is better than 2\% over the whole energy range. The systematic
uncertainties are dominated by the FD energy scale uncertainty of
14\%~\cite{Verzi-ICRC:2013}.
The main contributions to this uncertainty are related to
the knowledge of the fluorescence yield (3.6\%), the atmospheric
conditions (3.4 to 6.2\%), the absolute calibration of the telescopes
(9.9\%), the shower profile reconstruction(6.5 to 5.6\%) and the
invisible energy (3 to 1.5\%).

\subsection{Hybrid Reconstruction}

\begin{figure}[!t]
\subfigure[Hybrid geometry variables.]
{\label{fig:sketch}\includegraphics[width=0.32\textwidth]{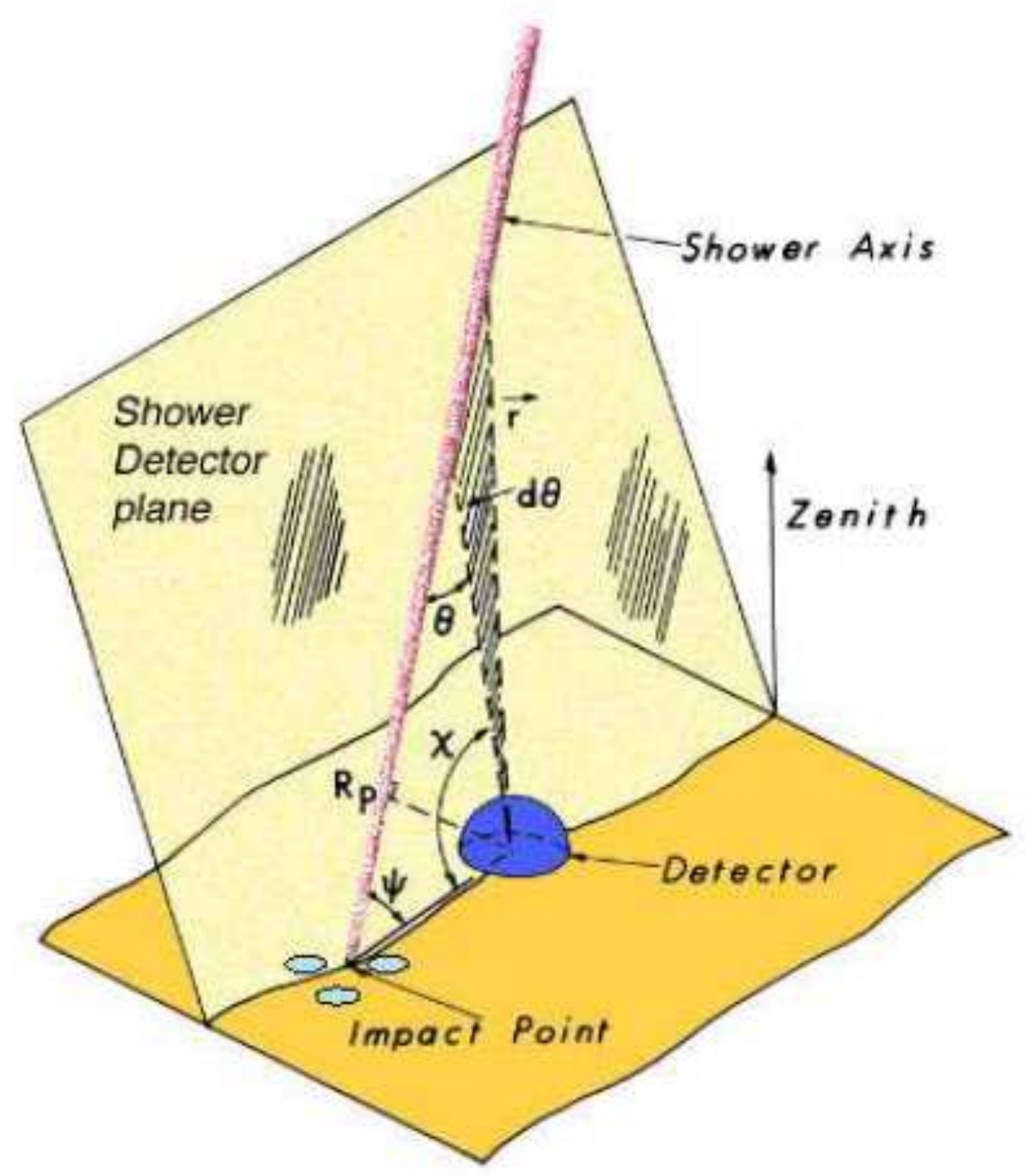}}
\subfigure[Core resolution.]
{\label{RESOLVRP}
\includegraphics[width=0.32\textwidth]{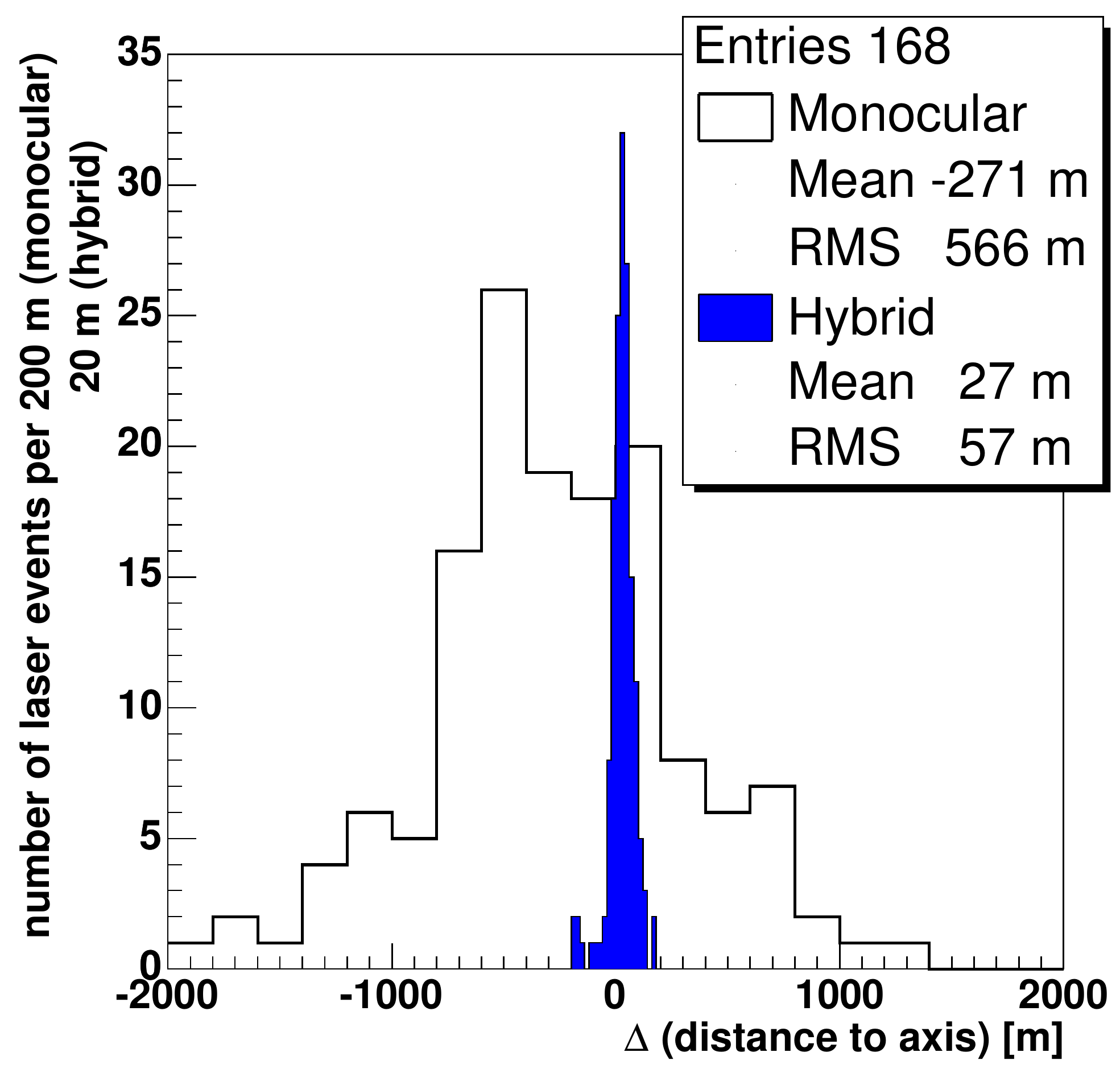}}
\subfigure[Angular resolution.]
{\label{RESOLVCHI0}
\includegraphics[width=0.32\textwidth]{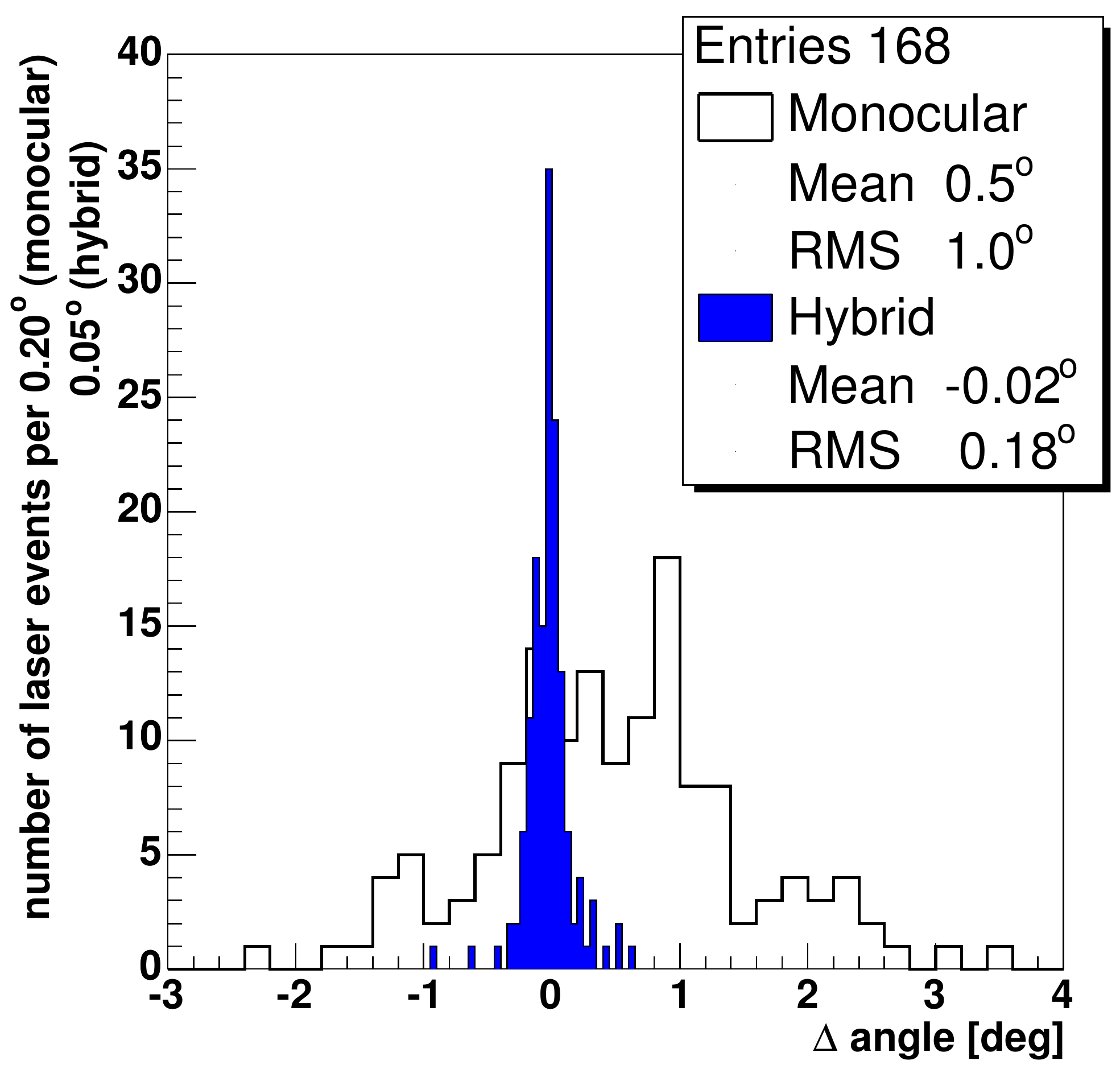}}
\caption{Illustration of the geometrical shower reconstruction from
  the observables of the fluorescence detector (left) and Comparison
  of mono- and hybrid geometry reconstruction of vertical laser beams
  (right)}
        \label{fig:geomReco}
\end{figure}

\paragraph{Geometry Reconstruction}
A hybrid detector achieves the best geometrical accuracy by using
timing information from all the detector components, both FD pixels
and SD stations.  Each element records a pulse of light from which one
can determine the central time of the pulse and its uncertainty.  Each
trial geometry for the shower axis yields a prediction for the times
at each detector component.  Differences between actual and predicted
times are weighted using their corresponding uncertainties, squared,
and summed to construct a $\chi^2$ value.  The hypothesis with the
minimum value of $\chi^2$ is the reconstructed shower axis.  In the
FD, cosmic ray showers are detected as a sequence of triggered pixels
in the camera.  The first step in the analysis is the determination of
the shower-detector plane (SDP) that is the plane that includes
the location of the eye and the line of the shower axis 
(cf.\ Fig.~\ref{fig:sketch}). Experimentally, it is determined
by minimizing the signal weighted sum of scalar product of its normal and
the pixel pointing directions. 

Next, the timing information of the pixels is used for reconstructing
the shower axis within the SDP.  As illustrated in
Fig.~\ref{fig:sketch}, the shower axis can be characterized by two
parameters: the perpendicular distance $R_\text{p}$ from the eye to the
track and the angle $\psi$ that the track makes with the horizontal
line in the SDP.  Each pixel which observes the track has a pointing
direction which makes an angle $\chi_i$ with the horizontal line.
If $t_0$ is the time when the shower front on the axis passes the
point of closest approach $R_\text{p}$ to the eye, then the light arrives at
the $i$th pixel at the time
\begin{equation}
	t_i = t_0 + \frac{R_\text{p}}{c} \cot{[(\psi+\chi_i)/2]}.
	\label{eq:timefit}
\end{equation}
The shower parameters are then determined by fitting the data points
to this functional form.  The accuracy of the monocular (FD-only)
reconstruction is limited when the measured angular speed $\mathrm{d}\chi/\mathrm{d}t$
does not change much over the observed track length.  For such
showers, degeneracy in the fitting parameters can be broken by
combining the timing information from the SD stations with that of the
FD telescopes.  This is called the \textit{hybrid} reconstruction.
Example results are shown in Fig.~\ref{RESOLVRP} and
Fig.~\ref{RESOLVCHI0} for reconstruction of a vertical laser beam at
the Central Laser Facility (CLF) where some laser light is also
injected into a neighboring SD station.  There we compare the mono
and hybrid reconstructions of the distance to the laser and the zenith
angle.  With the monocular reconstruction, the location of the CLF can
be determined with a resolution of ${\sim}500$\,m.  After including the
timing information of the single SD station, the resolution improves
by one order of magnitude with no systematic shift.

\paragraph{Profile Reconstruction and Energy Determination}
\label{sec::profileReconstruction}
Once the geometry of the shower is known, the light collected at the
aperture as a function of time can be converted to energy deposit,
$\mathrm{d}E/\mathrm{d}X$, at the shower as a function of slant depth. For this purpose,
the light attenuation from the shower to the detector needs to be
corrected for and all contributing light sources need to be
disentangled: fluorescence light~\cite{Arqueros:2008cx,Ave:2008zza,Ave:2012ifa}, direct and scattered Cherenkov
light~\cite{Giller:2004,Nerling:2006yt} as well as multiply scattered
light~\cite{Roberts:2005xv, Pekala:2009fe, Giller:2012tt}.  Since the
Cherenkov and fluorescence light produced by an air shower are
connected to the energy deposit by a linear set of equations, the
shower profile is obtained by an analytic linear least square
minimization~\cite{Unger:2008uq}.  Due to the lateral extent of air
showers, a small fraction of shower light is not contained within the
optimal light collection area. This is corrected for by taking into
account the universal lateral fluorescence~\cite{Gora:2005sq}
and Cherenkov light~\cite{Dawson-Giller:2007} distributions.  The
calorimetric energy, $E_\mathrm{cal}$, of a shower is given by the
integral over the longitudinal energy deposit profile,
\begin{equation} 
  E_\mathrm{cal} =
  \int_0^\infty \mathrm{d}E/\mathrm{d}X(X)\, \mathrm{d}X .  
\end{equation}

\begin{figure}[!t]
\subfigure[Light at aperture.]
 {\label{fig:lightAtAperture}
\includegraphics[width=0.49\textwidth]{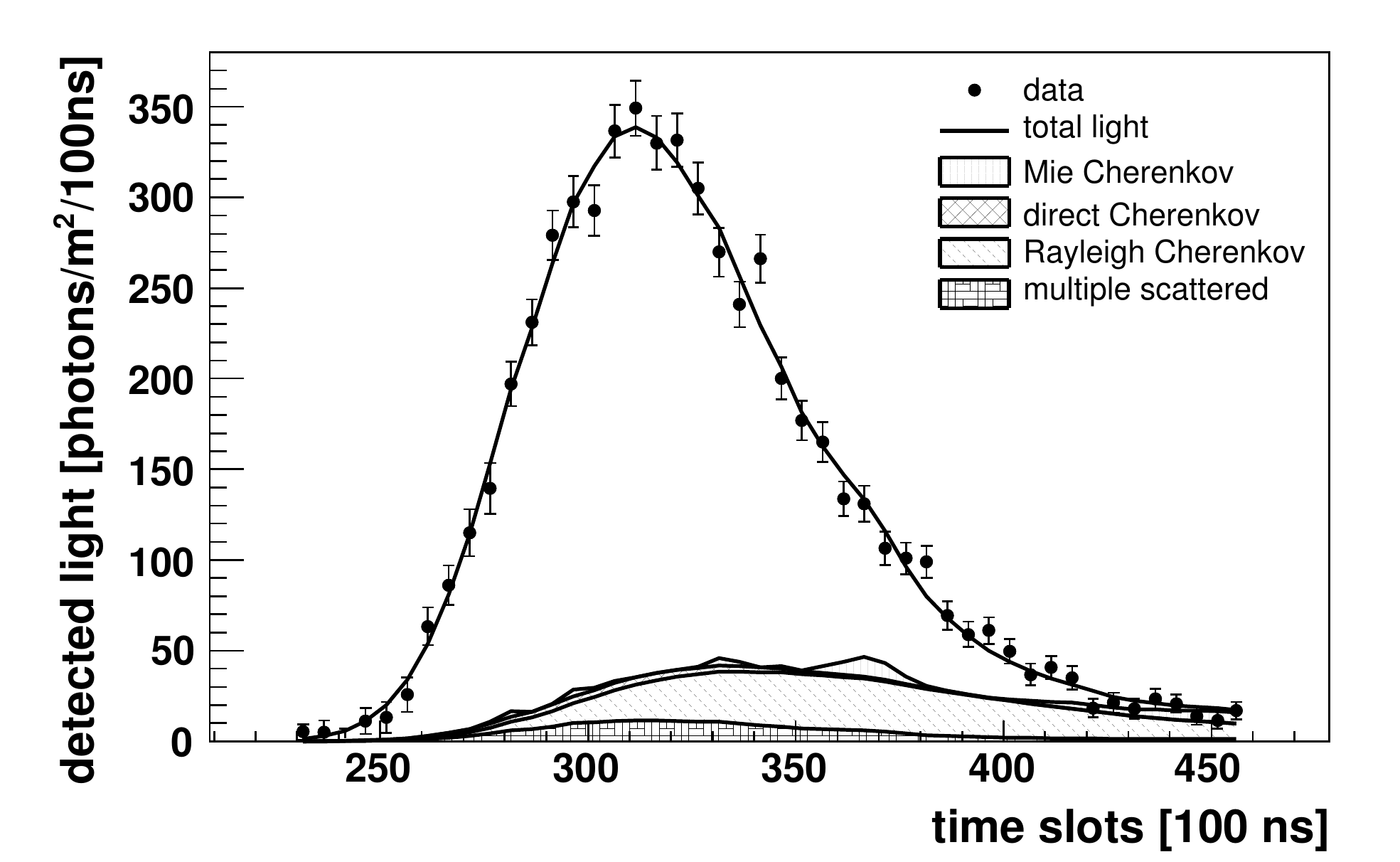}}
\subfigure[Energy deposit.]
 {\label{fig:dEdXProfile}
 \includegraphics[width=0.49\textwidth]{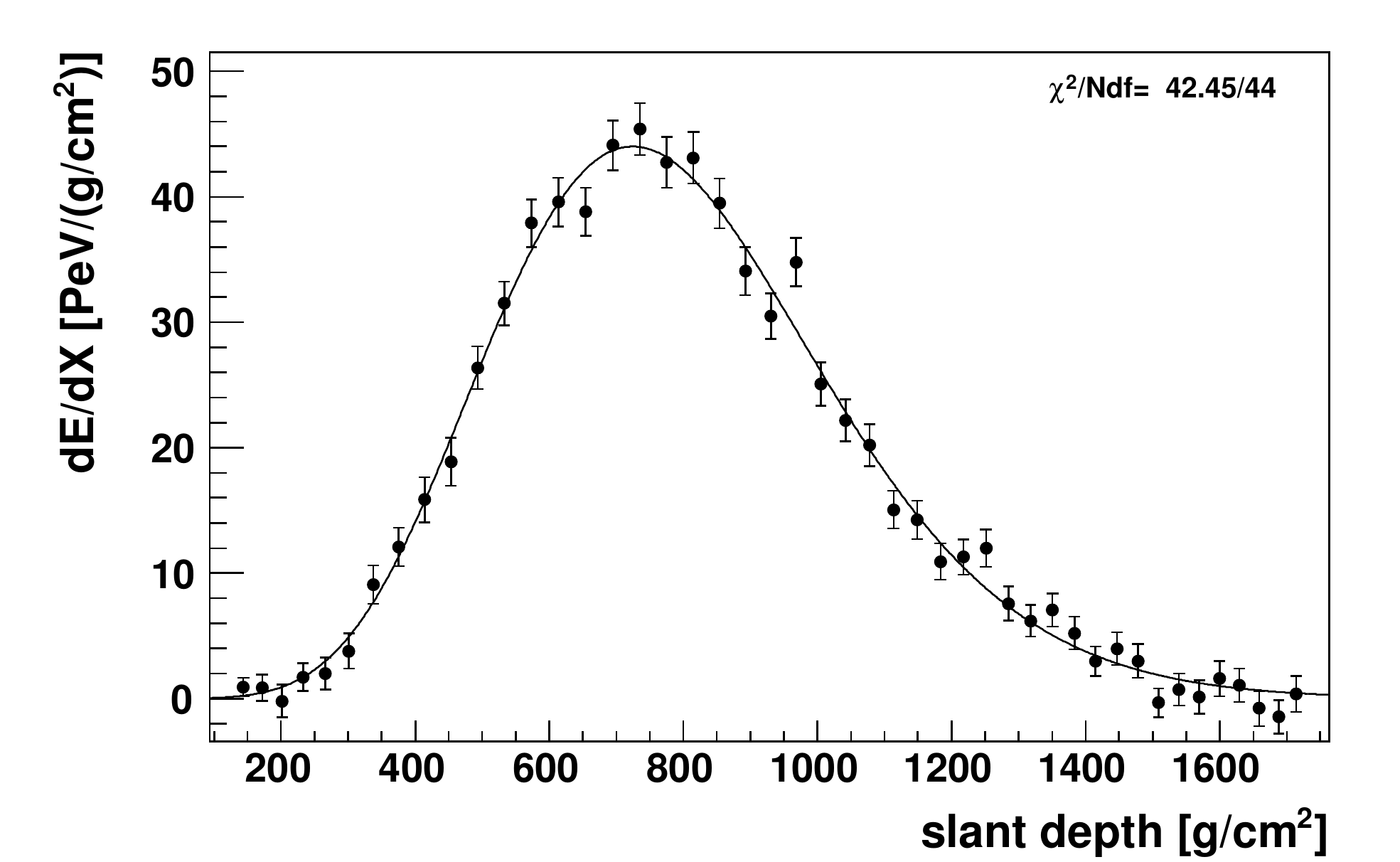}}
  \caption{Example of a reconstructed shower profile.}
        \label{fig:lateral}
\end{figure}

Since usually the full profile cannot be observed within the field of
view of the FD, this integral is evaluated from a Gaisser-Hillas
function~\cite{Gaisser-Hillas:1977} that is fitted to the
reconstructed energy deposit.  In addition, this fit yields an
estimate of $X_{\rm max}$, the mass sensitive position of the shower
maximum. An example of the measured light at aperture and the
reconstructed light contributions and energy deposit profile is shown
in Figs.~\ref{fig:lightAtAperture} and~\ref{fig:dEdXProfile}.  The
total energy of the shower is obtained from $E_\mathrm{cal}$ by
correcting for the 'invisible energy' carried away by neutrinos and
high energy muons~\cite{Tueros-ICRC:2013}.

The resolution of the profile measurement can be determined by
reconstructing simulated showers, that have passed a full detector
simulation~\cite{PradoJr2005632}. Moreover, at the high energies it can
be determined from the data itself by comparing independent
measurements of the same shower by different eyes (stereo
events). Both studies show that the energy of a shower can be
determined with a precision of 8\% above 10\,EeV.  For the shower
maximum, the resolution is 20\,g/cm$^2$~\cite{Dawson:2007di}.

\subsection{Aperture}

\begin{figure}[!t]
\centering
\includegraphics[width=0.5\textwidth]{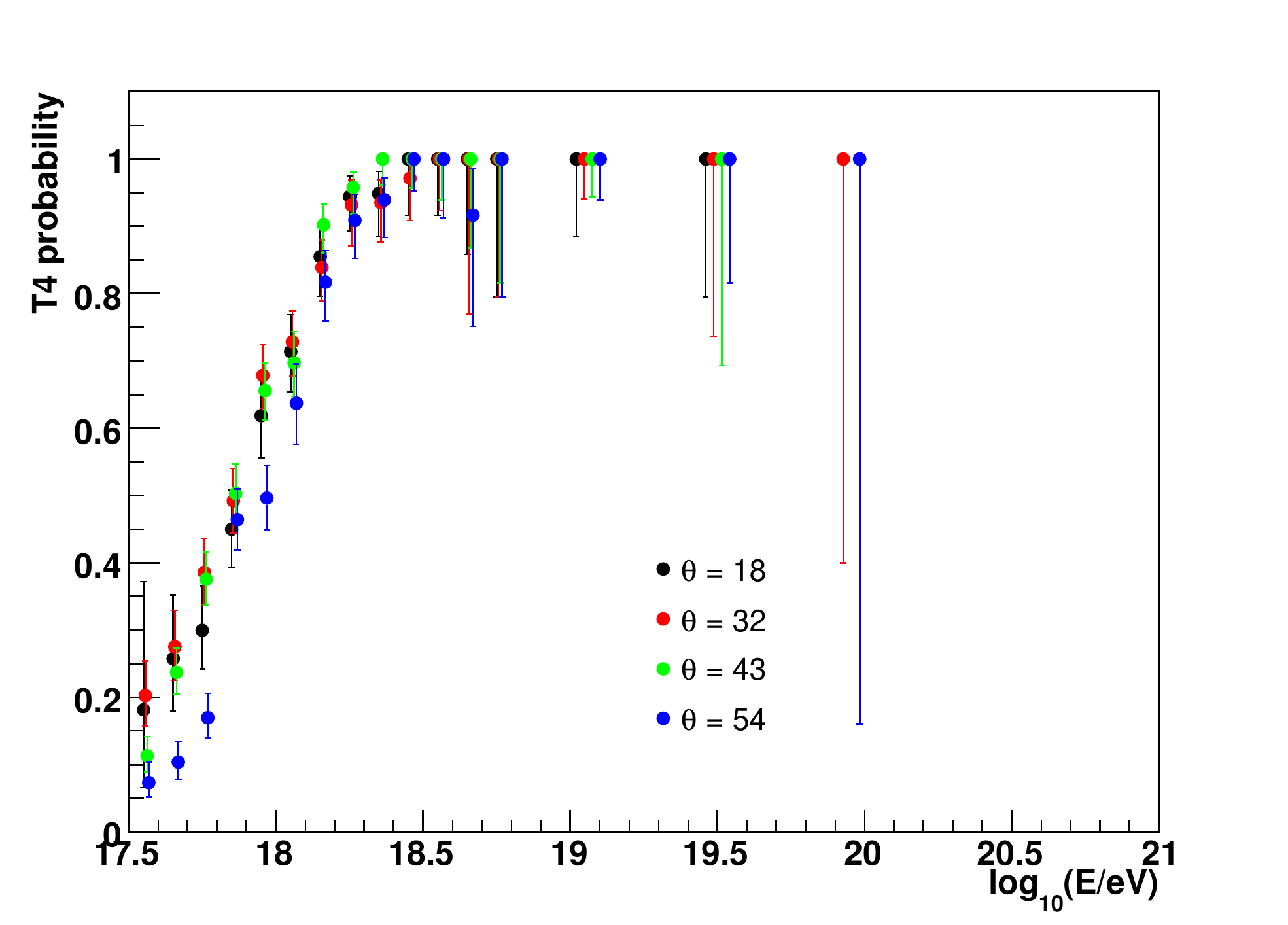}
\caption{Detection efficiency of the SD array, as a function of
energy, as measured from the data using the hybrid data set, for
different zenith angles (averaged over azimuths).}
\label{fig:SDDetectionEfficiency}
\end{figure}

An important feature of the SD is that it allows for a straightforward
control of the shower detection volume.  With the requirement that the
shower core of the events be reconstructed within the limits of the
region covered by the SD stations, the SD aperture can be obtained
from a simple geometric calculation of the actual size of the active array on
the ground, at any given time.  This sets the effective detection
surface on the ground, to be weighted by the energy-dependent
detection efficiency of cosmic ray showers.  The latter can be
measured directly from the hybrid data: restricting oneself to
conditions where the showers are known to be detectable with 100\%
efficiency in the 1-tank hybrid mode, one derives the SD detection
efficiency at a given energy as the fraction of the corresponding
showers that do trigger the SD, at the ``physics trigger'' level (T4)
described above.  This is shown in
Fig.~\ref{fig:SDDetectionEfficiency}, where the detection efficiency
is seen to reach 100\% at the \emph{saturation energy}
$E_{\mathrm{sat}} \simeq 3{\times}10^{18}$\,eV, in very good agreement with
simulations and measurements based on the SD trigger probability and
signal fluctuations as a function of distance to the shower axis.

\begin{figure}[!t]
\centering
\includegraphics[width=0.7\textwidth]{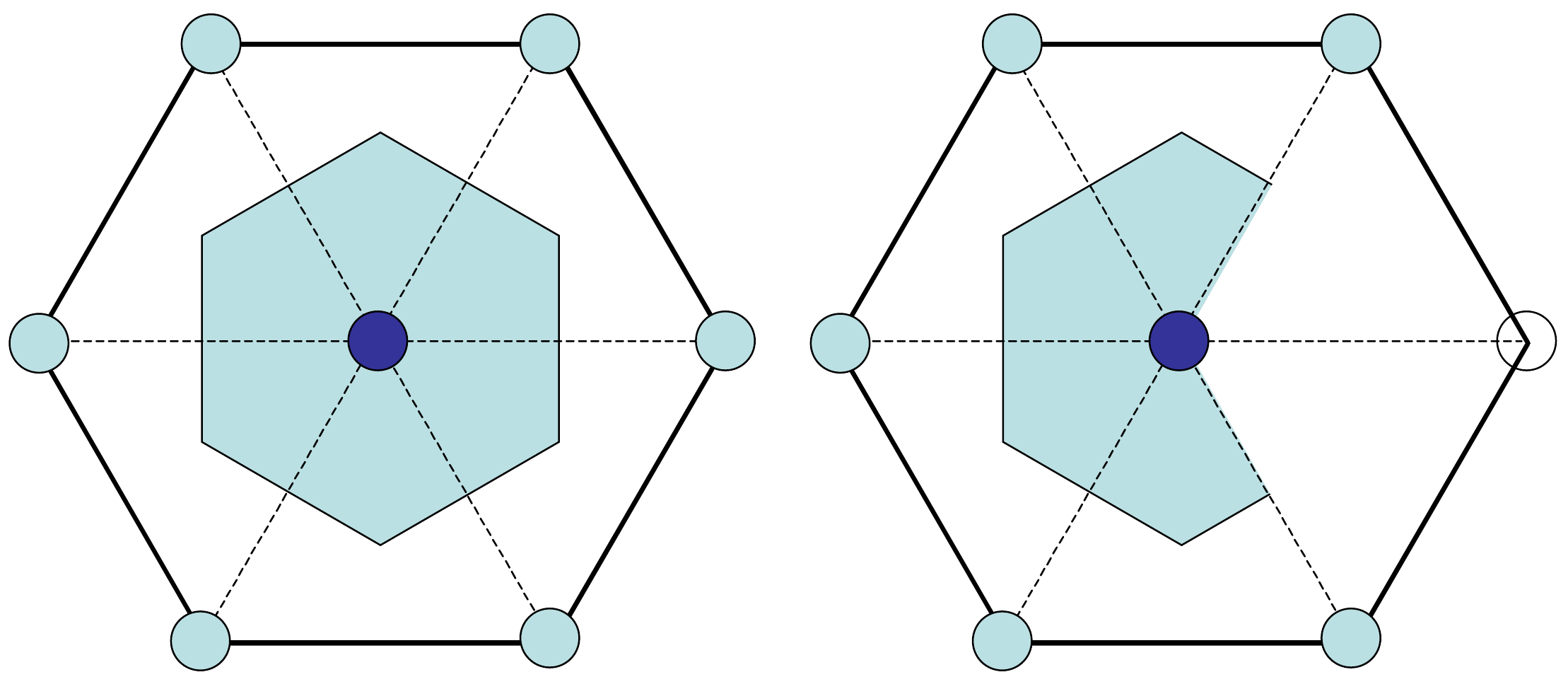}
\caption{Schematic view of the area (shaded region) where the core of
a vertical shower must be located inside an elementary hexagonal cell
of the SD array to pass the quality trigger (T5).  Left: for a
complete hexagon with 7 active tanks.  Right: for a hexagon with one
missing tank.}
\label{fig:effectiveDetectionAreaPerTank}
\end{figure}

Above $E_{\mathrm{sat}}$, the instantaneous aperture of the SD is
derived from the total surface covered by the array.  In order to
avoid border effects and a potentially degraded energy reconstruction
for showers hitting the ground close to an edge of the array or in a
region where some tanks are momentarily inactive, an additional cut is
applied to the data (referred to as the \emph{quality trigger}, or T5)
to ensure nominal reconstruction accuracy.  This T5 trigger requires
that the station recording the highest signal in a given CR event be
surrounded by at least 5 active stations (out of the 6 nearest
neighbors).  In the case when only 5 neighboring stations are active,
the shower core must also be reconstructed inside an elementary
triangle of stations that were active at that time.
Figure~\ref{fig:effectiveDetectionAreaPerTank} illustrates this
requirement, showing the core positions allowed for vertical showers
arriving well inside the array (left) or near a missing station
(right).  The total detection area associated with the central station
is then seen to be $D^2\sqrt{3}/2\simeq 1.95\,\mathrm{km}^2$ in
the former case, and 2/3 of this in the latter.  A final integration
over solid angle for showers with zenith angles between 0 and 60$^\circ$
gives the nominal aperture per active station: $A_0\simeq4.59$\,km$^2$\,sr.

The computation of the total SD aperture at any given time is then
obtained by multiplying the elementary aperture by the number of
active stations (with the required number of active neighbors),
obtained from a simple census using the SD monitoring data, which give
the list of active stations on a second-by-second basis.  Finally, the
integrated SD exposure in \emph{linsleys} ($1\,\mathrm{L} =
1$\,km$^2$\,sr\,yr) is obtained by the time
integration of the instantaneous aperture of the SD array, taking into
account any changes in the array configuration, using the same
monitoring data.  In this way, the growth of the array during the SD
deployment period could be automatically included in the exposure
calculation, whatever the shape and duration of the intermediate
configurations.

Overall, the above-mentioned technique provides a very accurate
determination of the SD acceptance, with an uncertainty of ${\sim}3$\%,
which can be considered as negligible with respect to the uncertainty
on the energy reconstruction.

\section{Enhancements}

Instrumental enhancements have been installed close to the Coihueco FD
station.  These include underground muon detectors, additional water
Cherenkov detectors, and high-elevation fluorescence telescopes for a
larger field-of-view.  Also, research programs are underway to assess
the utility of radio and microwave emission from air showers.

\subsection{High Elevation Auger Telescopes (HEAT)}
\label{sec:HEAT}

\begin{figure}[t]
  \subfigure[Horizontal (downward) mode for service and cross-calibration.]{
    \includegraphics[width=0.48\textwidth]{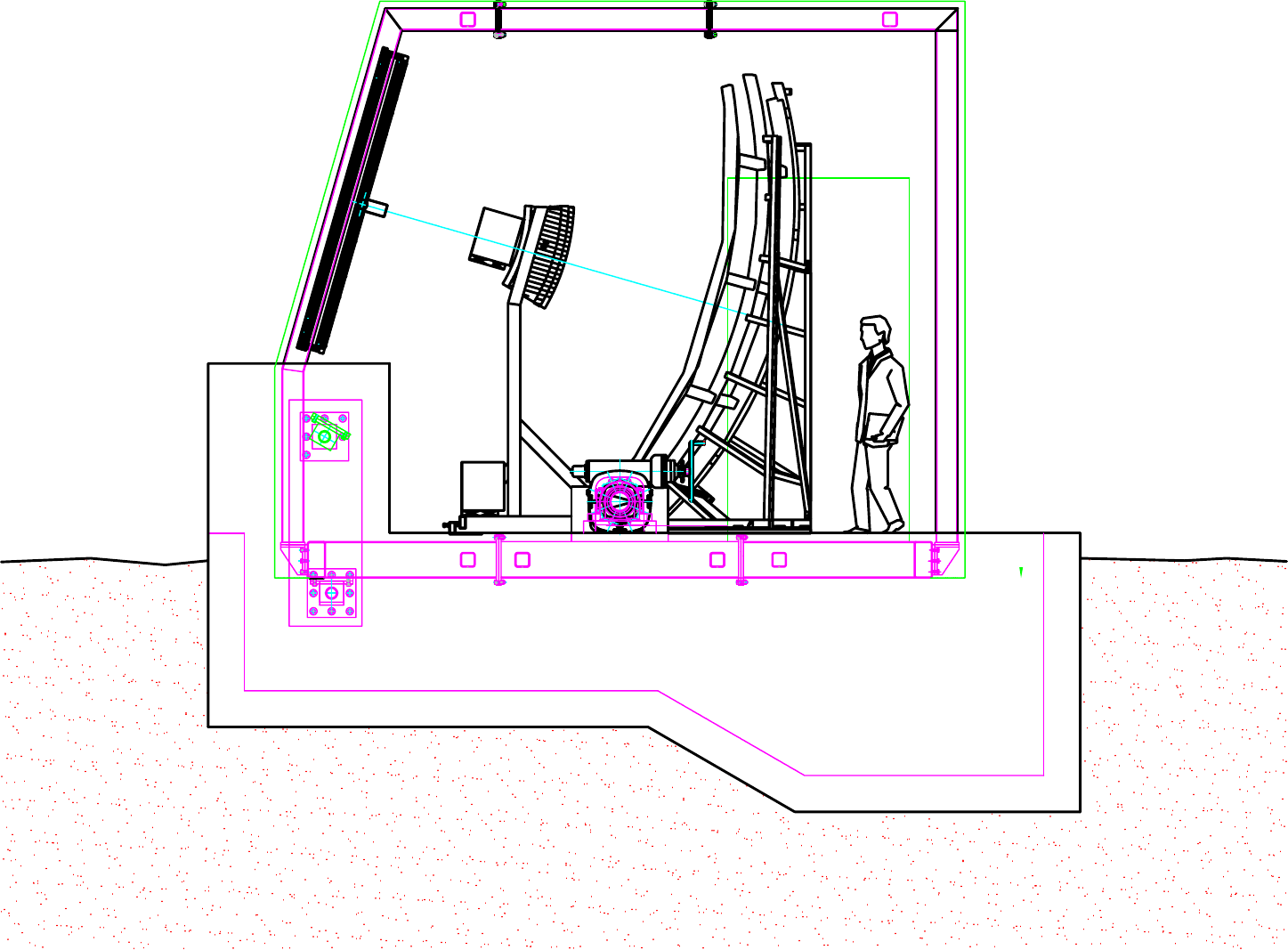}
  }
  \subfigure[Data taking (upward) mode in tilted orientation.]{
    \includegraphics[width=0.48\textwidth]{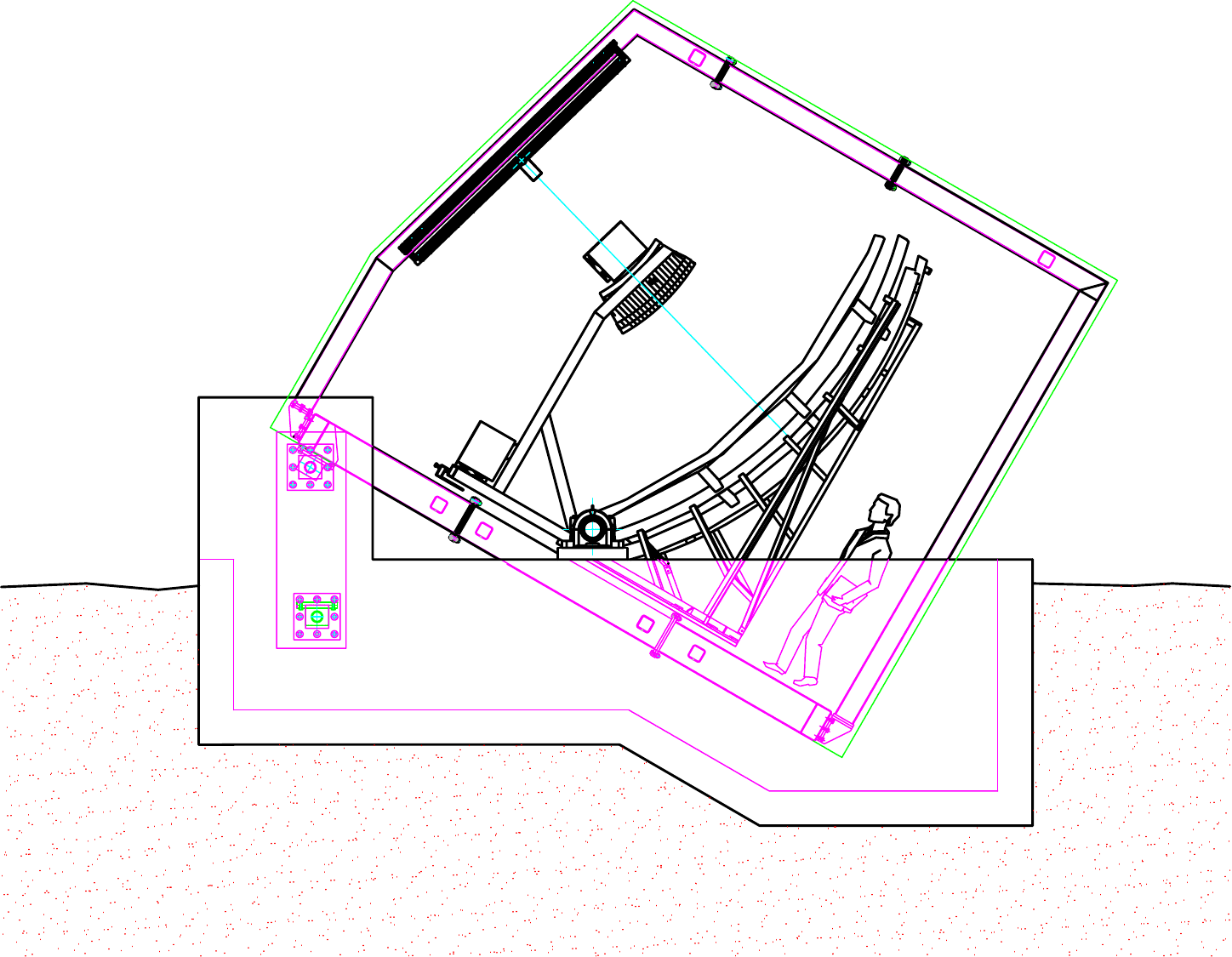}
  }
\caption{Schematic view of the cross-section of one of
the HEAT telescopes.}
\label{fig:Heat-modes}
\end{figure}

Three additional fluorescence telescopes with an elevated field of
view were built at the FD site at Coihueco~\cite{Mathes-ICRC:2011}.
These telescopes are very similar to the original fluorescence
telescopes but can be tilted by $29^\circ$ upward with an electrically
driven hydraulic system. These three telescopes work independently of
other FD sites and form the ``fifth site" of the Observatory.  The
HEAT telescopes were designed to cover the elevation range from
$30^\circ$ to $58^\circ$, which lies above the field of view of the
other FD telescopes.  The HEAT telescopes allow a determination of the
cosmic ray spectrum and $X_\text{max}$ distributions in the energy
range from below the second knee up to the ankle. The design of the HEAT telescopes
is depicted in Figure~\ref{fig:Heat-modes}.

The main objective of this extension is to lower the energy threshold
of hybrid data to enable an unbiased detection of nearby low-energy
showers. In combination with the SD information from an infill array
of water-Cherenkov detectors on a 750\,m grid (see
section~\ref{sec:amiga}) close to the HEAT site, the energy range of
high quality hybrid air shower measurements has been extended down to
$10^{17}$\,eV.  

The HEAT telescopes can be tilted using the hydraulic mechanism.
The telescopes are parked in the horizontal position between
the FD data taking periods to be accessible for maintenance.
The same position is used for the absolute calibration of the HEAT
telescopes and also for the cross-calibration with telescopes at Coihueco.

The Schmidt optics of the HEAT telescopes, camera body, PMTs, light
collectors, etc., are the same as in the other sites. All three spherical
mirrors are built up from hexagonal glass mirrors with vacuum-deposited
reflective coatings.

A feature that sets HEAT apart from the classic Auger telescopes is its
new electronics kit that can sample up to 40\,MHz instead of 
10\,MHz. 
In practice, a sampling rate of 20\,MHz (corresponding
to a 50\,ns FADC bin size) was chosen. The higher rate improves the measurement
for close showers that have a correspondingly larger angular velocity --
precisely the showers we are interested in observing with HEAT. From this it 
follows that the first level trigger  interval was reduced to 50\,ns,
whereas the second level trigger  continues to operate every 100\,ns.
The length (in time) of the FADC traces remains the same, so the number
of bins doubles.

The trigger rate of the HEAT telescopes is high, particularly because of the
Cherenkov light from low energy showers. Therefore the T4 trigger 
has been implemented to reduce the readout of the SD array for these
low energy showers.

\subsection{Auger Muon and Infill Ground Array (AMIGA)}
\label{sec:amiga}

\begin{figure}[t]
\centering
\includegraphics[width=0.48\textwidth]{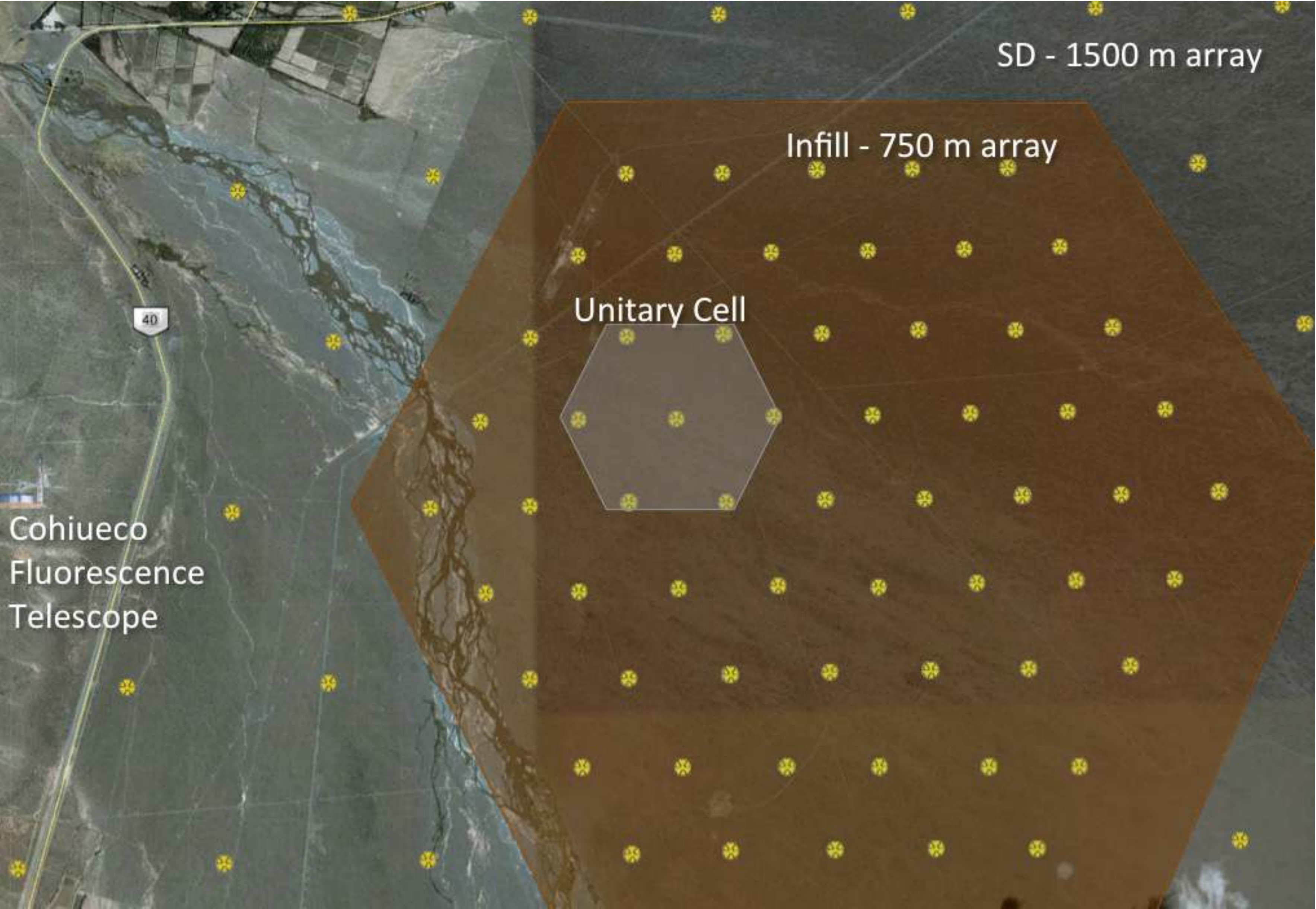}
\caption{%
          AMIGA layout: an infill of surface stations with an inter-detector spacing of 750\,m.
          Plastic scintillators of 30\,m$^2$ are buried under $\approx 280\,\mathrm{g/cm^2}$ of vertical mass to measure the muon component of the showers.
          The small shaded area indicates the prototype hexagon ({\it Unitary Cell}) of the muon detector.
        }%
\label{fig:AMIGAScheme}
\end{figure}

A dedicated detector to directly measure the muon content of air
showers is being built.  The AMIGA enhancement
\cite{Suarez-ICRC:2013,Sanchez-ICRC:2011,Platino:2011zz} is a joint
system of water-Cherenkov and buried scintillator detectors that spans
an area of 23.5\,km$^2$ in a denser array with 750\,m spacing nested
within the 1500\,m array (see figure~\ref{fig:AMIGAScheme}).  The area
is centered 6\,km away from the Coihueco fluorescence site.  The
infill array is fully efficient from $3{\times}10^{17}$\,eV onwards for
air showers with zenith angle $\leq 55^\circ$ \cite{Maris-ICRC:2011}.

The SD infill array was completed in September 2011 while the first
prototype hexagon of buried scintillators, the {\it Unitary Cell}, was
fully operational at the end of 2014.  This engineering array consists
of seven water-Cherenkov detectors paired with 30\,m$^2$
scintillators segmented in two modules of 10\,m$^2$ plus two
of 5\,m$^2$ in each position.  In addition, two positions of
the hexagon were equipped with {\it twin} detectors (extra
30\,m$^2$ scintillators) to allow the accuracy of the muon
counting technique to be experimentally assessed
\cite{Maldera-ICRC:2013} and one position has 20\,m$^2$ of
extra scintillators buried at a shallower depth to analyze the
shielding features.  The proven tools and methods used for the
analysis of the 1500\,m SD array data have been extended to
reconstruct the lower energy events.  The angular resolution for
$E\geq4{\times}10^{17}$\,eV is better than $1^\circ$ and the energy
reconstruction is based on the lateral density of shower particles at
the optimal distance of 450\,m from the core
\cite{Ravignani-ICRC:2013}.

\begin{figure}[t]
\centering
\includegraphics[width=0.48\textwidth]{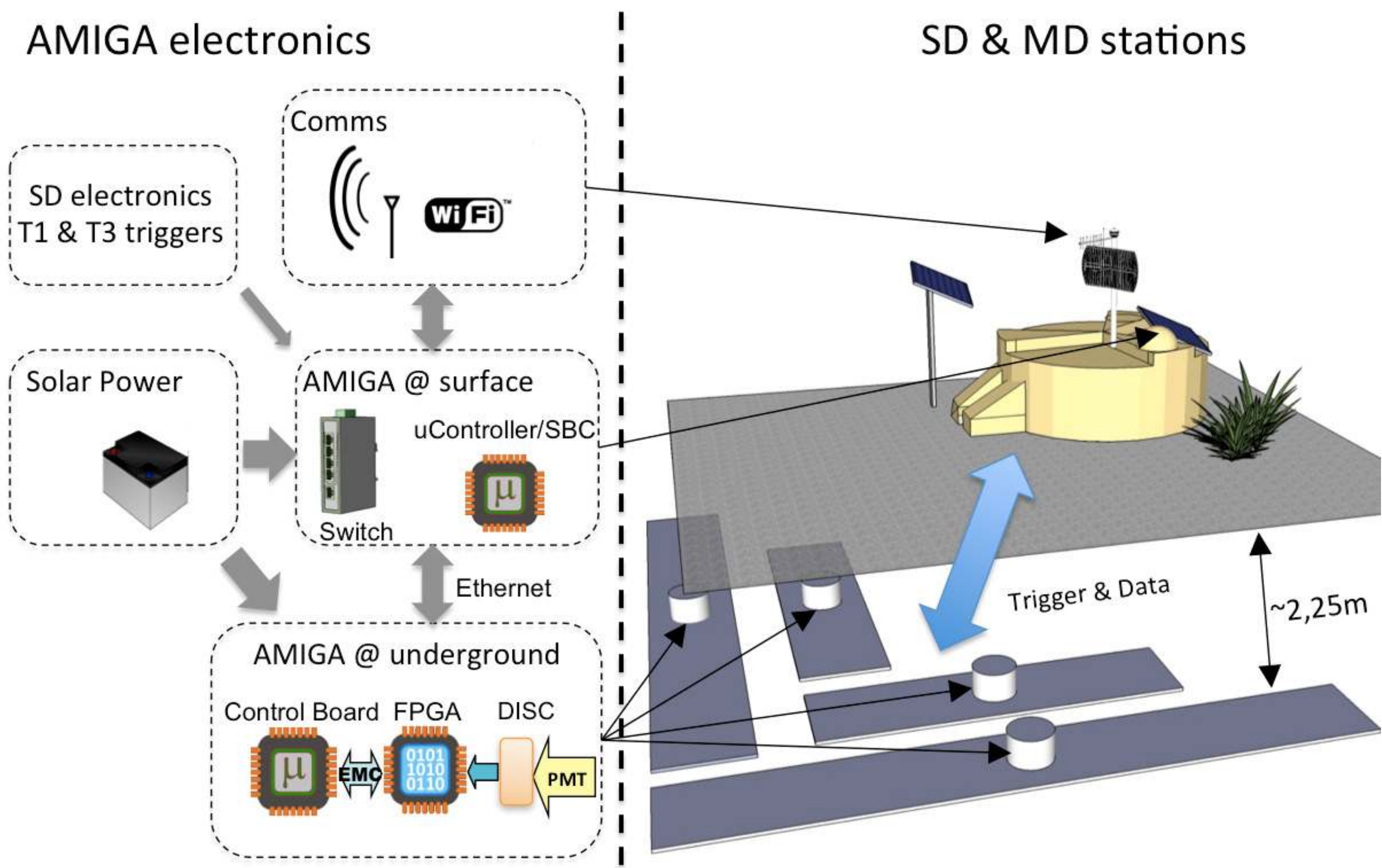}
\caption{ AMIGA station: SD+MD paired detectors.  During the MD
  prototype phase the 30\,m$^2$ buried scintillators are
  segmented into 4 modules, $2\times 10\,\mathrm{m}^2 + 2\times
  5\,\mathrm{m}^2$. For the production phase, only 3 modules of
  10\,m$^2$ will be used.  To avoid shadowing effects by the
  water-Cherenkov detector, there are 5\,m of sideways clearance.  The
  buried front end electronics is serviceable by means of an access
  pipe which is filled with local soil bags.  Data are sent by a
  dedicated WiFi antenna.  }
\label{fig:AMIGA_Layout}
\end{figure}

The buried scintillators are the core of the detection system for the
muonic component of air showers (the muon detector, MD).  To
effectively shield the electromagnetic component, the MD is placed
under ${\approx}280\,\mathrm{g/cm^2}$ of vertical mass corresponding to
a depth of 1.3\,m in the local soil.  This shielding imposes a cutoff for vertical muons of around
1\,GeV. The layout of SD+MD paired stations is shown in
figure~\ref{fig:AMIGA_Layout}.  The scintillator surface of each MD
station is highly segmented. It consists of modules made of 64 strips
each.  Strips are 4.1\,cm wide $\times$ 1.0\,cm thick and 400\,cm 
long for the 10\,m$^2$ modules.
They consist of extruded Dow Styron 663W polystyrene
doped by weight with 1\% PPO (2,5-diphenyloxazole) and 0.03\% POPOP
(1,4-bis(5-phenyloxazole-2-yl)benzene).  They are completely wrapped
with a thin white reflective layer of titanium dioxide (TiO$_2$)
except for a central groove into which a wavelength shifting (WLS)
optical fiber is installed.  The light output uniformity is
$\pm$5\%. Because the scintillators have an attenuation length of
${\sim}(55\pm5)$\,mm, light is transported to a photomultiplier tube
using the WLS fiber.  The manifold of fibers of each module ends in an
optical connector matched to a 64 multi-anode PMT from the Hamamatsu
H8804 series. 

The bandwidth of the front end electronics is set to 180\,MHz to
determine the pulse width.  Signal sampling is performed by a Field
Programmable Gate Array (FPGA) from the ALTERA Cyclone III series at
320\,MHz.  MD scintillator modules receive the trigger signal from
their associated SD station.  The lowest level trigger (T1) of the
surface detectors is used.  Once a T1 condition is fulfilled on the
surface, its MD companion freezes a 6.4\,$\upmu$s data sample into a
local buffer -- 1.6\,$\upmu$s before and 4.8\,$\upmu$s after the T1.
Data are then moved to an external RAM capable of storing 1024
triggers \cite{Wainberg:2014}.

Incoming analog signals from each pixel of the PMT are digitized with
a discriminator that provides the input to the FPGA.  Samples can be
either a logical ``1'' or ``0'' depending on whether the incoming
signal was above or below a given (programmable) discrimination
threshold.  This method of {\it one-bit} resolution is very robust for
counting muons in a highly segmented detector. This avoids missing muons due to
simultaneous particle arrivals \cite{Supanitsky:2008dx}.  It relies
neither on deconvolving the number of muons from an integrated signal,
nor on the PMT gain or its fluctuations, nor on the point of impact of
the muon and the corresponding light attenuation along the fiber.  It
also does not require a thick scintillator to control Poissonian
fluctuations in the number of single photoelectron pulses per
impinging muon \cite{Wundheiler-ICRC:2011}.  The MD station power is
supplied by an additional solar panel and battery box (see
figure~\ref{fig:AMIGA_Layout}).

\subsection{Radio and Microwave Research Programs}

\paragraph{Radio Research Program}
\label{sec:developments_radio}

The observation of air showers with radio detection techniques can be
done at all times (day and night).  Moreover, radio signals are
sensitive to the development of the electromagnetic component of
particle showers in the atmosphere of the Earth and, in particular, to
the depth of the shower maximum or mass of the incoming cosmic
ray~\cite{LOFAR:2013a}.  Radio detection of air showers started in the
1960s, and the achievements in those days have been presented in
reviews by Allan \cite{Allan:1971} and Fegan \cite{Fegan:2011fb}.
More recent developments are based on initial studies performed by the
LOPES \cite{Falcke:2005tc} and the CODALEMA \cite{Ardouin:2006nb}
collaborations and the LOFAR radio telescope~\cite{Schellart:2013bba}.
In the last 10 years the radio detection technique in the MHz region
has been revived and the present radio detector arrays for cosmic ray
research are equipped with low noise and high rate analog-to-digital
converters.  

\begin{figure}[t]
\centering
\includegraphics[width=0.48\textwidth]{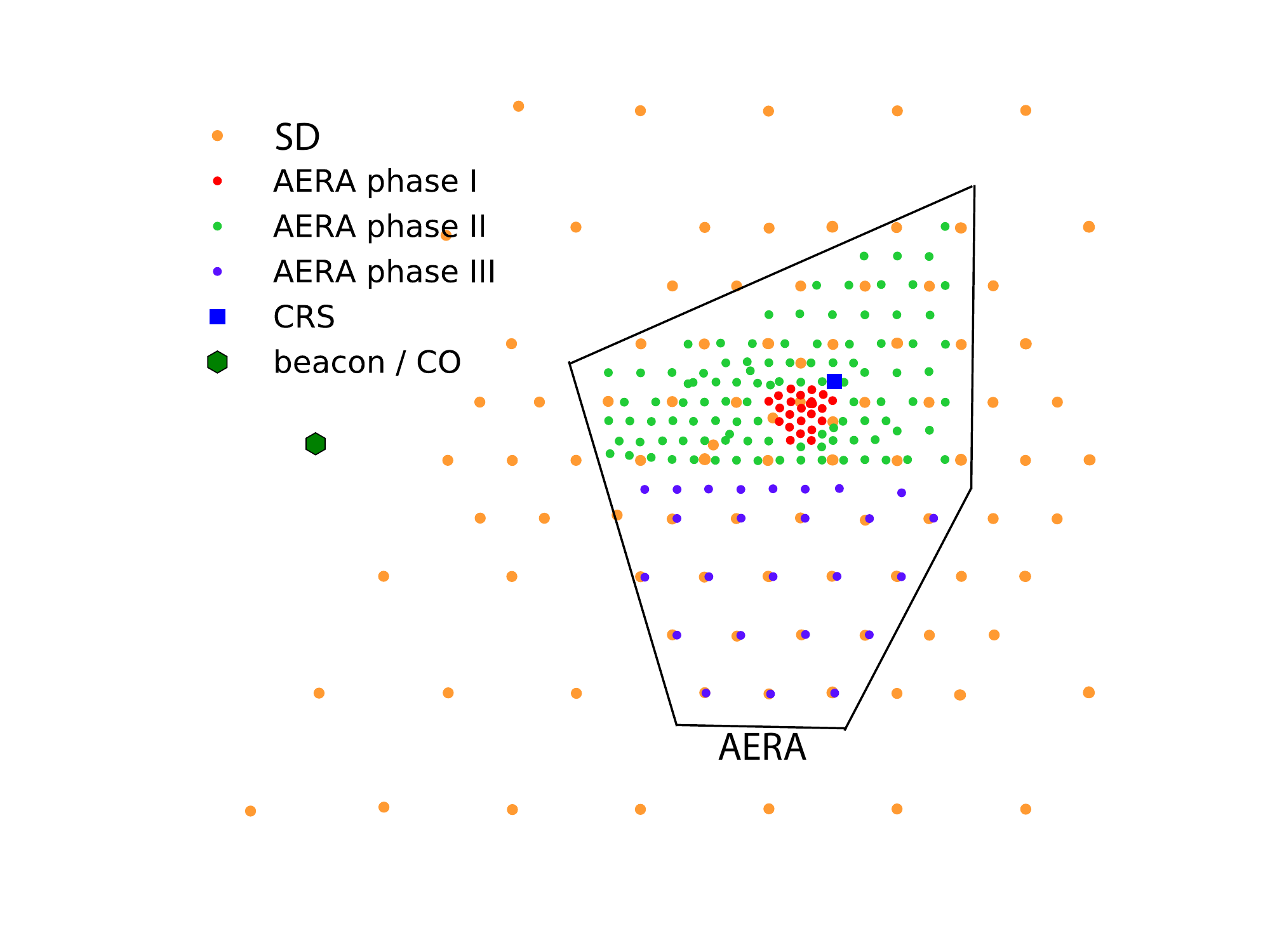}
\includegraphics[width=0.48\textwidth]{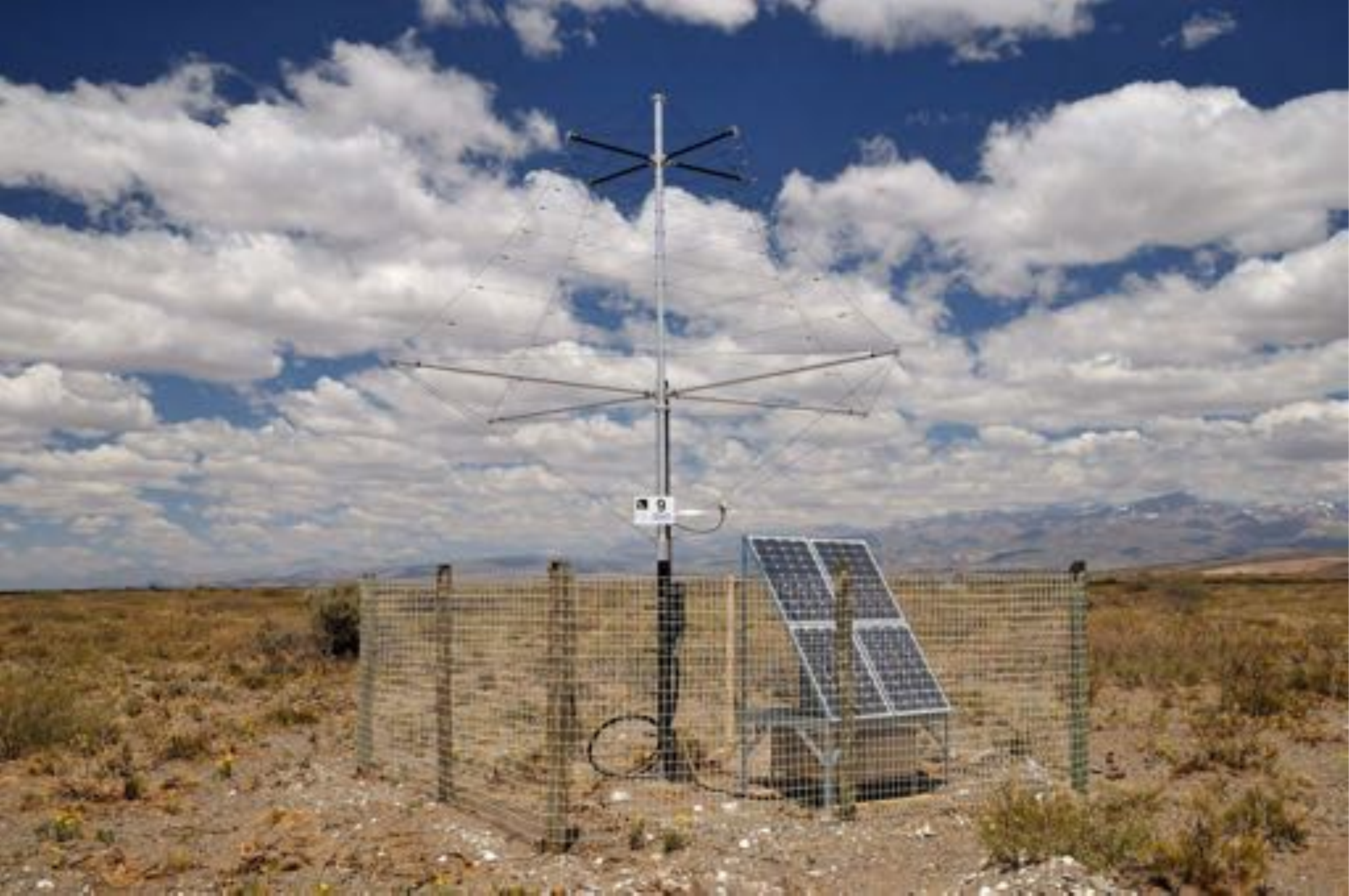}
\caption{Map of the Auger Engineering Radio Array (left) and a photograph 
of one station. AERA consists of 153 antenna stations at the Auger Infill 
array covering an area of $\approx 17\,$km$^2$.}   
\label{fig:aera}
\end{figure}

The Pierre Auger Collaboration has started a research program to
examine the feasibility and quality of radio observations of air
showers.  Since 2009 the activities are coordinated within the Auger
Engineering Radio Array (AERA), which is based on work within the
Collaboration using various prototypes at the site of the Pierre Auger
Observatory~\cite{Revenu-ICRC:2011, Kelley-ICRC:2011, Acounis:2012dg}.
As a first step the emission mechanisms need to be understood.
Recently, AERA has measured the relative contribution of the two main
emission processes in air showers~\cite{Aab:2014esa}.  As a second
step the data obtained with radio detection stations deployed at the
Observatory will be used to check their sensitivity with respect to
the determination of the air shower parameters.

The scientific goals of the AERA project are as follows: 1)
calibration of the radio emission from the air showers, including
sub-dominant emission mechanisms; 2) demonstration at a significant
scale of the physics capabilities of the radio technique, e.g.,
energy, angular, and mass resolutions; and 3) measurement of the
cosmic ray composition from 0.3 to 5\,EeV, with the goal of
elucidating the transition from galactic to extragalactic cosmic rays.

At present, AERA consists of 153 radio detection stations. Each
station is comprised of a dual polarization antenna, sensing the
electric field in the north/south and east/west directions, associated
analog and digital readout electronics, an autonomous power system and
a communication link to a central data acquisition system. 
Nine of the stations are additionally equipped with a third, vertically 
aligned antenna to measure the full electric field. The
antennas are sensitive between 30 and 80\,MHz, chosen as the relatively
radio quiet region between the shortwave and FM bands. 
AERA deployment began in 2010 with 24 stations. Stable physics data
taking started in March 2011, and the first hybrid detection of cosmic
ray events by radio, fluorescence, and surface particle detectors was
recorded in April 2011. In May 2013 additional 100, and in March 2015 
further 25 stations were installed (see~figure~\ref{fig:aera}), 
where the  stations are optimized compared to the first phase, 
in particular related to the antenna type used~\cite{Abreu:2012pi}.
AERA covers an area of $\approx 17\,$km$^2$ and operates in self-trigger 
and at the same time in external (SD, FD, and minimum bias) trigger mode. 
The AERA data is merged with those of the other detector components and 
analyzed by \Offline which enables detailed comparisons and hybrid 
reconstruction on a single event basis. Several thousands hybrid events 
are presently analyzed including tens of so-called super-hybrid 
(SD - FD - AERA - AMIGA-$\mu$) events.

\paragraph{Microwave research program}

Recent results of a test beam experiment at SLAC \cite{Gorham:2007af}
showed that it could be possible to use microwave radiation to detect
extensive air showers. This radiation, expected to be isotropic and
broad in frequency, is interpreted \cite{Gorham:2007af} as molecular
bremsstrahlung (MBR) produced by the scattering of low energy
electrons in the weakly ionized plasma produced by the shower with the
neutral molecules of the atmosphere. The Auger collaboration is
pursuing an active R\&D program to determine if a detector sensitive
to MBR would be a suitable alternative for the study of ultra-high
energy cosmic rays.

This R\&D program \cite{Allison-ICRC:2011, FacalSanLuis:2013qza}
consists of three different setups installed at the Observatory. The
AMBER and MIDAS experiments use radio-telescope style detectors
intended for the observation of the shower longitudinal development in
the same manner as an FD. In the EASIER setup on the other hand, SD
tanks are instrumented with smaller radio receivers that take
advantage of the enhancement of the signal when the shower is observed
close to its axis.

Installation of the microwave detectors was finalized in Sep\-tember
2012. A previous result by the MIDAS detector
\cite{AlvarezMuniz:2012ew}, obtained in Chicago, places tight
constraints on the amount of microwave signal emitted and its scaling
with the energy of the shower \cite{AlvarezMuniz:2012dx}. The ongoing
work to identify showers detected at the same time in the SD and in
one of the microwave detectors already yielded the fist unambiguous
detection of a cosmic ray shower in the EASIER setup in June 2011
\cite{FacalSanLuis:2013qza}.

\section{Performance of the Observatory}
\label{sec:performance}

\subsection{Key performance parameters}

In Table~\ref{tab:key_perf} are summarized some of the important
parameters that characterize the performance of the Observatory. These
parameters include the event rate of the detectors and the resolutions
of the different reconstructed observables.

\begin{table*}
\caption{Key performance parameters for the Auger Observatory}
\label{tab:key_perf}
\begin{center}
\begin{tabular}{ll}
\toprule
\multicolumn{2}{c}{\bf SD}
\\
\midrule
SD Annual Exposure               & ${\sim}5500$\,km$^2$\,sr\,yr
\\
\midrule
T3 rate                   & 0.1\,Hz
\\
T5 events/yr, $E>3$\,EeV & ${\sim}14,500$
\\
T5 events/yr, $E>10$\,EeV & ${\sim}1500$
\\
Reconstruction accuracy ($S(1000)$) & 22\% (low $E$) to 12\% (high $E$)
\\
Angular resolution        & $1.6^{\circ}$ (3 stations)
\\
                          & $0.9^{\circ}$ (${>}5$ stations)
\\
Energy resolution         & 16\% (low $E$) to 12\% (high $E$)
\\
\midrule
\multicolumn{2}{c}{\bf FD}
\\
\midrule
Duty cycle                    & ${\sim}15$\%
\\
Rate per building               & 0.012\,Hz
\\
Rate per HEAT  & 0.026\,Hz
\\
\midrule
\multicolumn{2}{c}{\bf Hybrid}
\\
\midrule
Core resolution           & 50\,m
\\
Angular resolution        & $0.6^\circ$
\\
Energy resolution (FD)    & 8\%
\\
$X_\text{max}$ resolution      & ${<}20$\,g/cm$^2$
\\
\bottomrule
\end{tabular}
\end{center}
\end{table*}

\subsection{Surface detector performance}

Stable data taking with the surface detector array started in January
2004 and the Observatory has been running in its full configuration
since 2008. As described in section~\ref{subsec:cdasmonitoring}, various
parameters are continuously monitored to optimize the performance of
the detectors and ensure reliable data.

\begin{figure}[t]
\centering
\includegraphics[width=0.8\columnwidth]{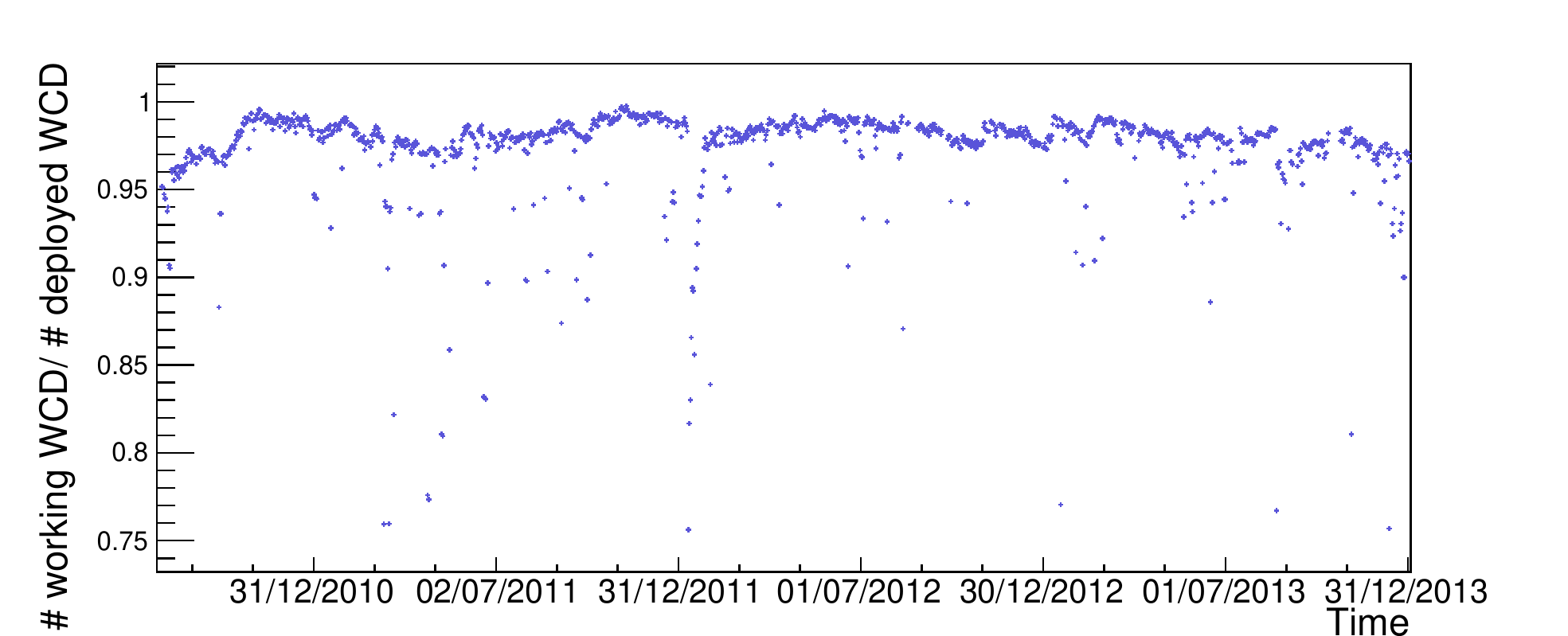}
\caption{Number of active SD stations normalized to the nominal number of SD stations in the array, as a function of time. }
\label{fig:ratiotank}
\end{figure}

The monitoring tool includes so-called performance metrics to monitor
the overall performance of the surface detector array.  Relevant data
useful for long term studies and for quality checks are stored in the
Auger Monitoring database on a one-day basis.  For example, mean
values over one day of the number of active SD detectors and the
number of active hexagons as well as the nominal value (expected value
if all the detectors deployed were active) are available.  As an
example, figure~\ref{fig:ratiotank} shows the number of active SD
stations normalized to the nominal number of stations in the array for
the last 4 years. This plot is a convolution of the status of the
active stations and of the efficiency of the CDAS, which since the
beginning is better than 99.5\%.

\begin{figure}[t]
\centering
\includegraphics[width=0.8\columnwidth]{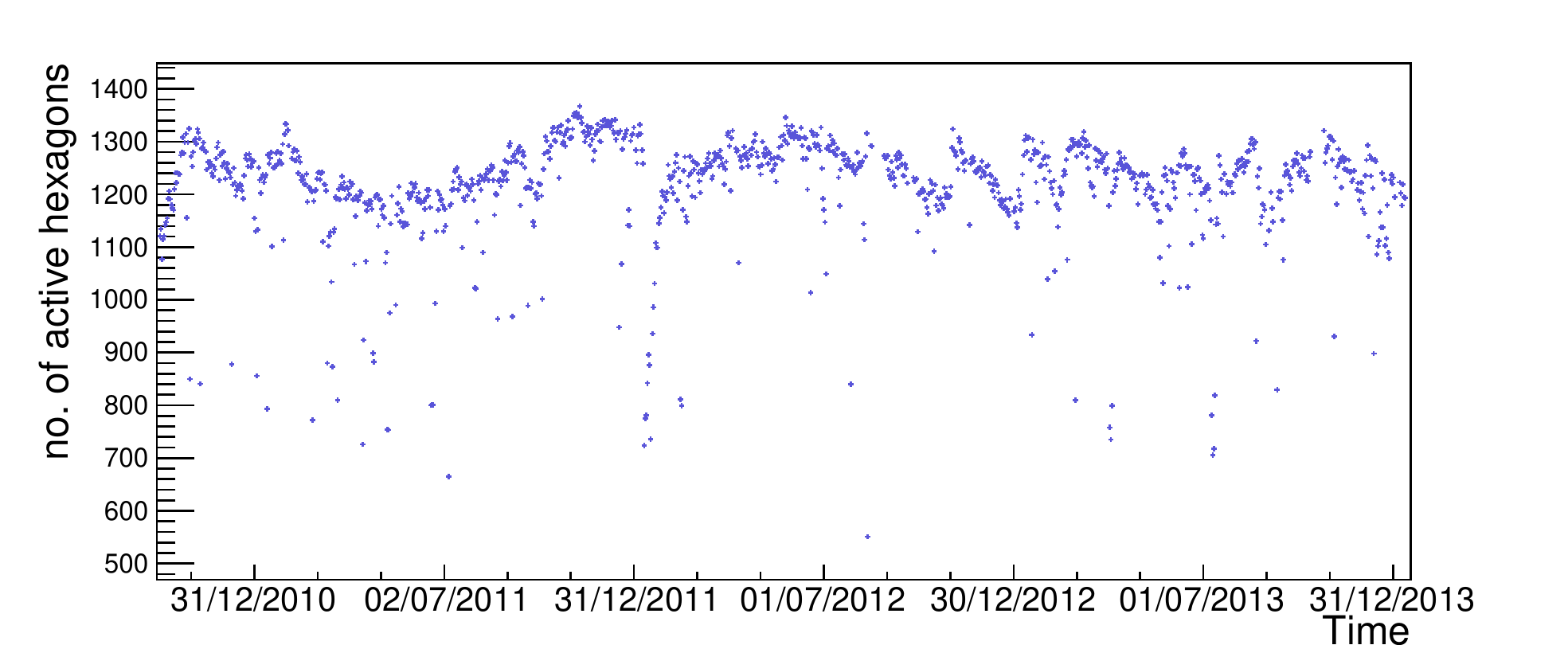}
\caption{Number of active hexagons as a function of time}
\label{fig:hexa}
\end{figure}

Figure~\ref{fig:hexa} shows the number of active hexagons for the same
period. This variable is a key parameter since it is the basis of the
exposure evaluation. Indeed, the off-line T5 fiducial trigger,
described in section \ref{sec:SDreco} selects only events for which
the hottest station is surrounded by an active hexagon. Thus, above
$3{\times}10^{18}$\,eV, when the full efficiency of detection of the
array is reached (at least three triggered tanks ), the exposure is
simply proportional to the integrated number of active hexagons during
the period.

\begin{figure}[t]
\centering
\includegraphics[width=0.8\columnwidth]{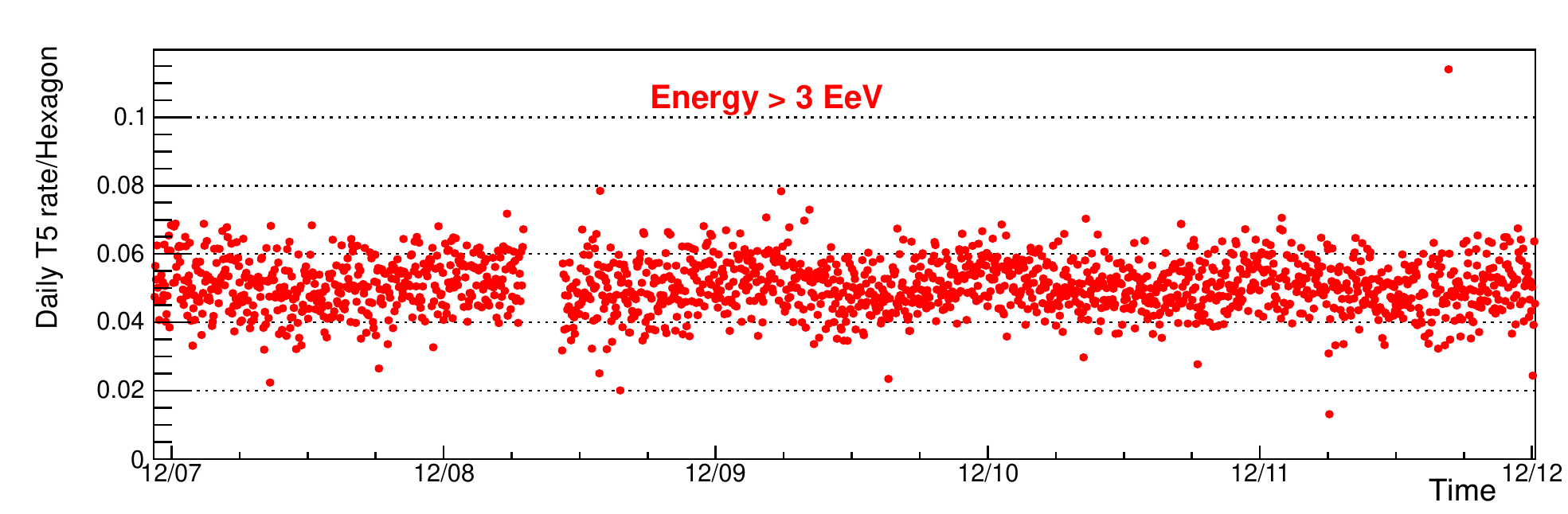}
\caption{Evolution of the daily T5 rate normalized to the number of hexagons for the period 2008 to 2012.}
\label{fig:t5rate3}
\end{figure}

The rate of events (T5 events) normalized to the average number of
active hexagons is expected to be stable in time above the energy
threshold of $3{\times}10^{18}$\,eV, which can be seen in figure
\ref{fig:t5rate3}.

\begin{figure}[t]
\centering
\includegraphics[width=0.55\columnwidth]{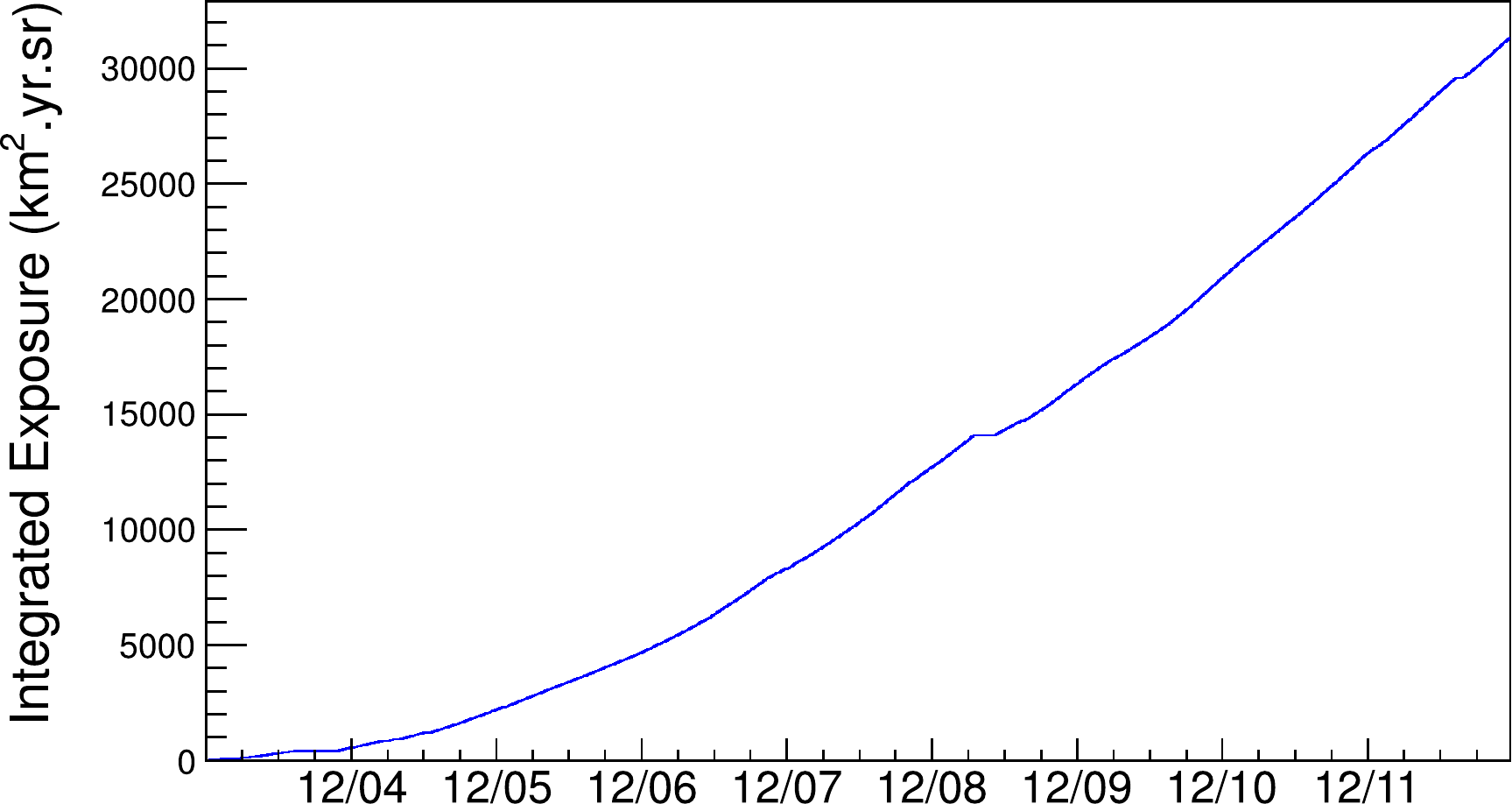}
\caption{Evolution of the exposure between 1 January 2004 and 31 December 2012.}
\label{fig:Exposure}
\end{figure}

Finally the integrated exposure between 1 January 2004 and 31 December
2012 is shown in Fig.~\ref{fig:Exposure}. Since completion of the array
in 2008, the increase of the exposure has been about 5500\,km$^2$\,sr per
year.

\subsection{Fluorescence detector performance}

The data taking of the FD can only take place under specific
environmental conditions and is organized in night shifts.  The
telescopes are not operated when the weather conditions are
unfavorable (high wind speed, rain, snow, etc.) and when the observed
sky brightness (caused mainly by scattered moonlight) is too high. As
a consequence, the shifters have to continuously monitor the
atmospheric and environmental conditions and judge the operation mode
on the basis of the available information.

The performance of the fluorescence and hybrid data taking is then
influenced by many effects. These can be external, e.g., lightning or
storms, or internal to the data taking itself, e.g., DAQ
failures. For the determination of the \emph{on-time} of the Pierre
Auger Observatory in the hybrid detection mode it is therefore crucial
to take into account all of these occurrences and derive a solid
description of the data taking time sequence.

Data losses and inefficiencies can occur on different levels, from the
smallest unit of the FD, i.e., one single photomultiplier (pixel)
readout channel, up to the highest level, i.e., the combined SD/FD
data taking of the Observatory.

\begin{figure}[t]
\centering
\includegraphics[width=0.45\textwidth]{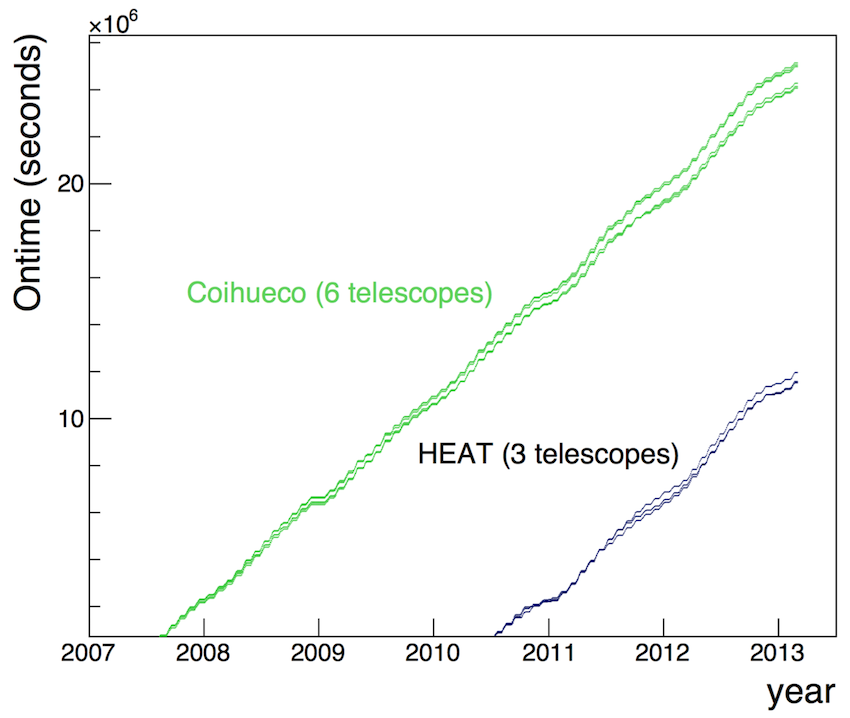}\hfill
\includegraphics[width=0.45\textwidth]{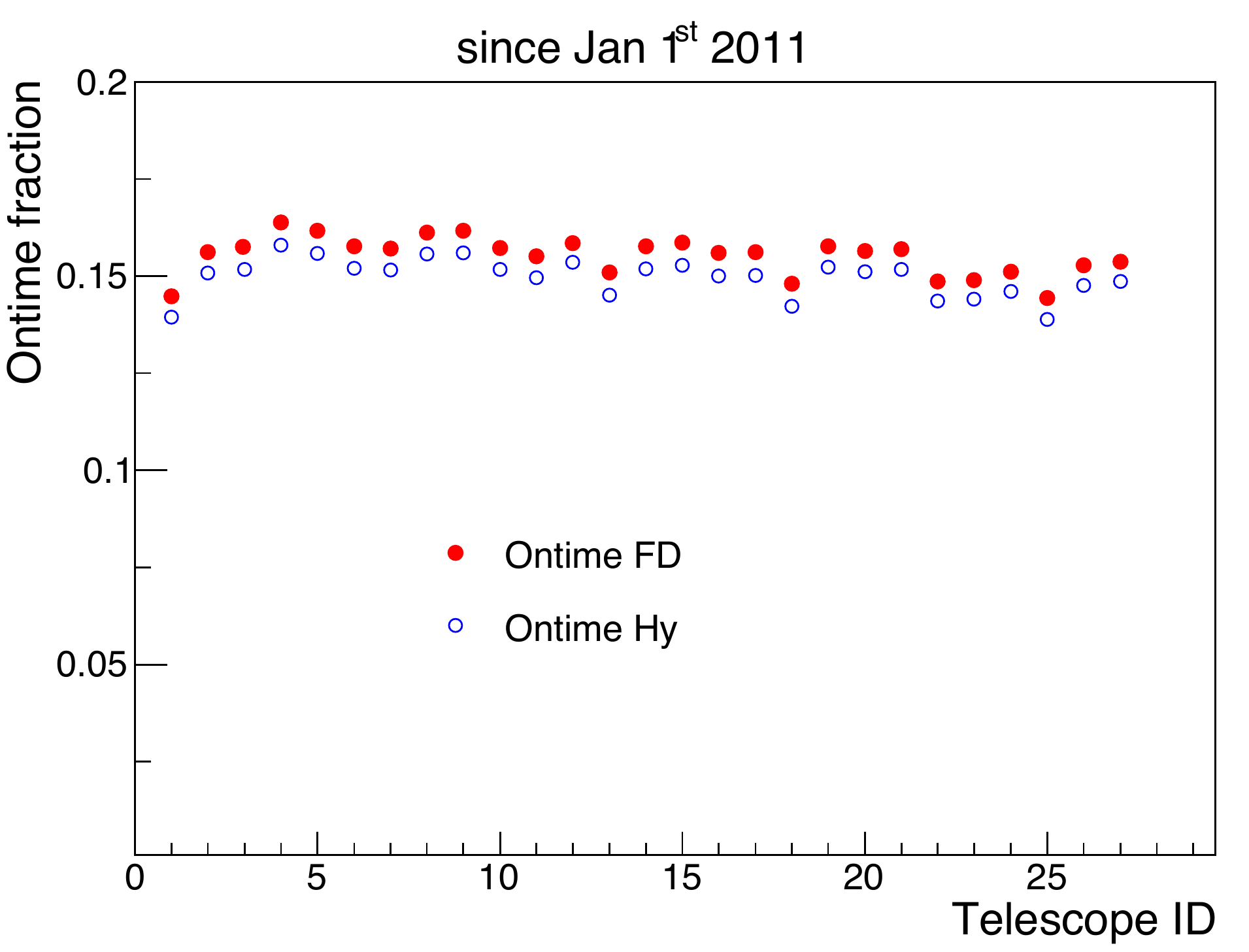}
\caption{\textit{Left:} accumulated on-time
since 1 Jul 2007 for the Coihueco and HEAT telescopes. 
\textit{Right:} On-time of individual telescopes since 1 Jan 2011.
(1--6), (7--12), (13--18), (19--24), (25--27) for the sites of Los Leones, Los
Morados, Loma Amarilla, Coihueco and HEAT, respectively.}
\label{fig:on-time}
\end{figure}

\begin{figure}[t]
\centering
\includegraphics[width=0.49\textwidth]{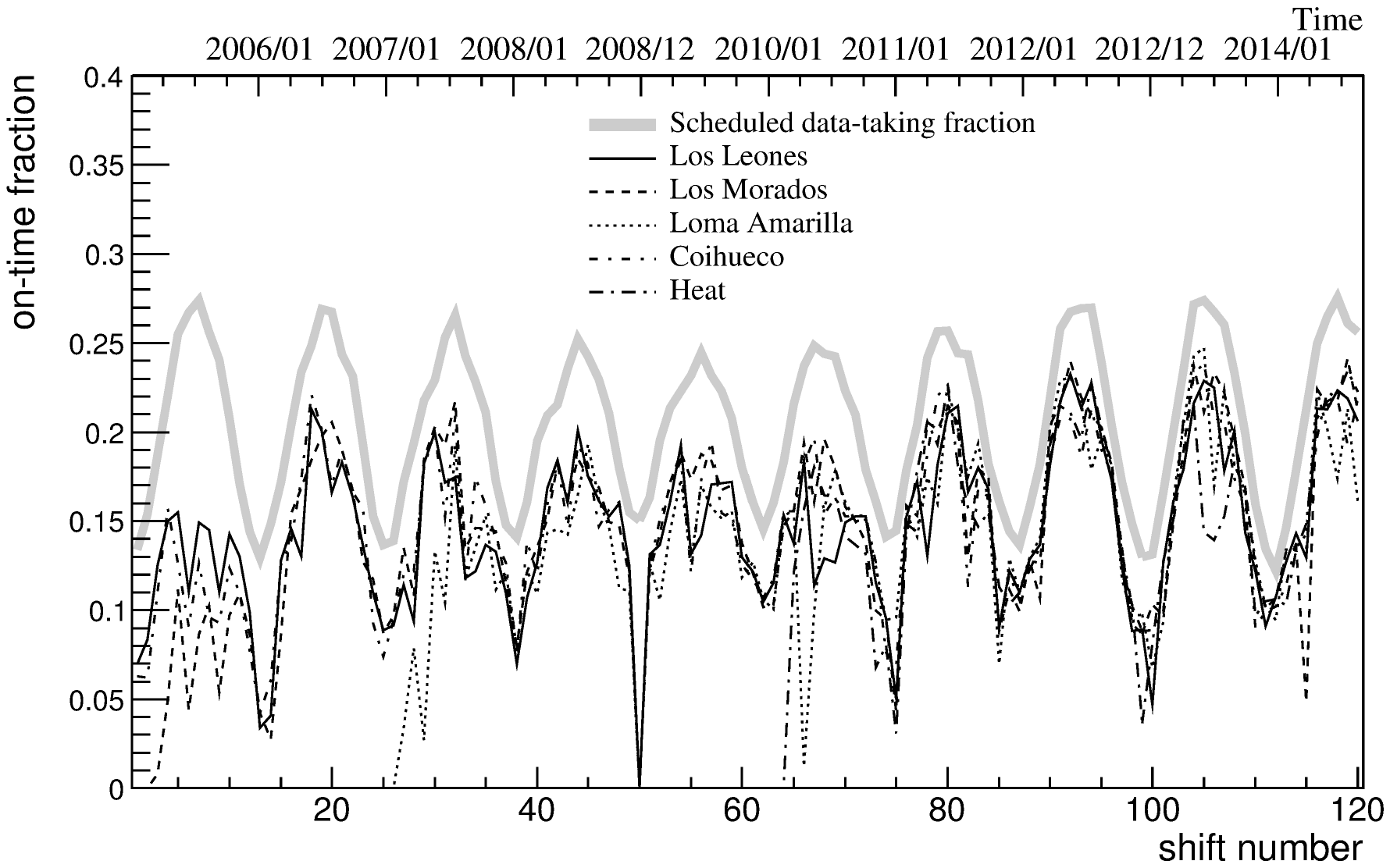}\hfill
\includegraphics[width=0.49\textwidth]{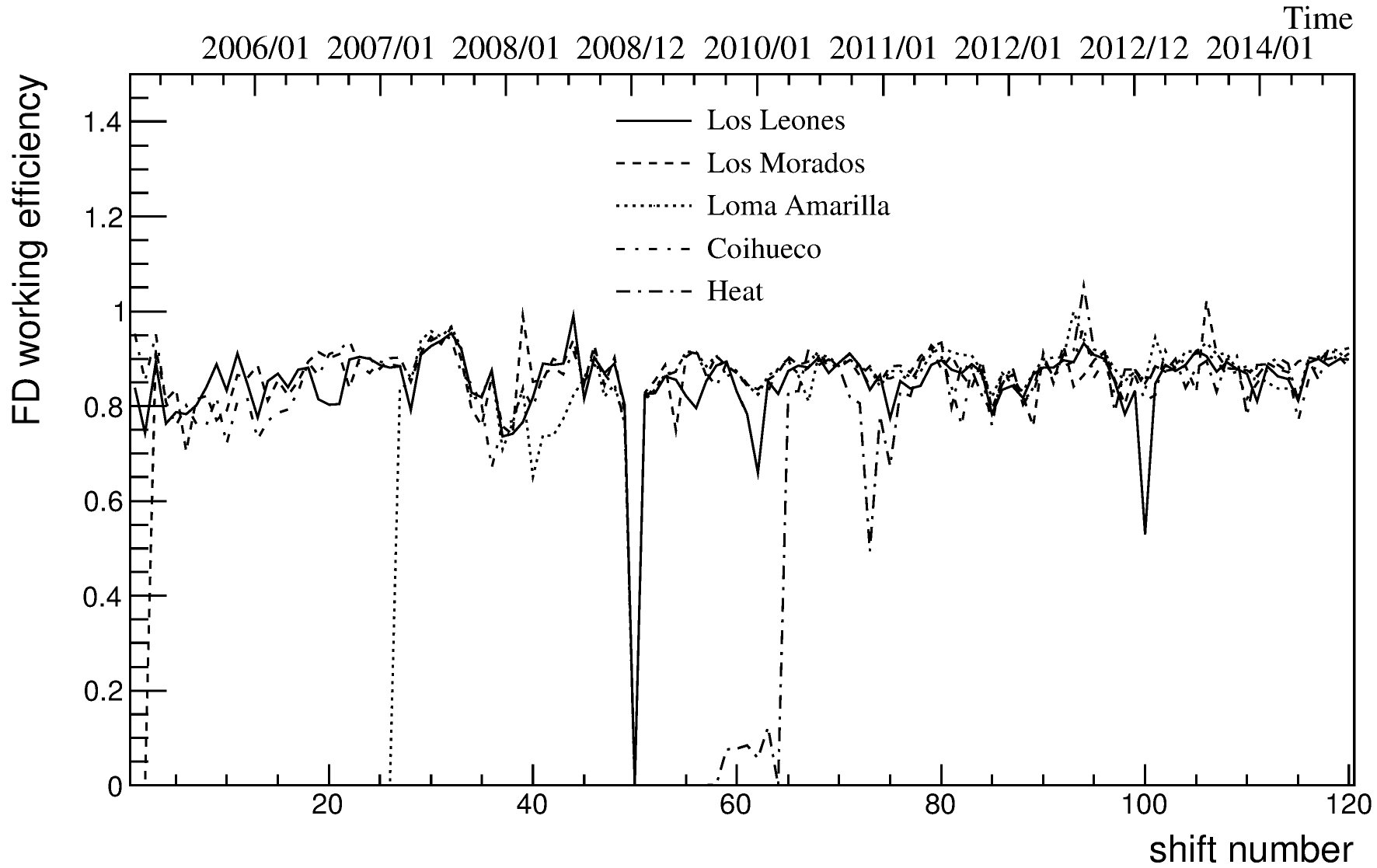}
\caption{\textit{Left:} time evolution of the average hybrid on-time fraction over 9
years of operation of the Pierre Auger Observatory.
The thick gray line defines the scheduled
data taking time fraction defined as the time periods with
moon fraction lower than 70\% and with the moon being below the horizon for
more than 3 hours.
\textit{Right}: readiness of the FD
detector (see text for details).}
\label{fig:metrics}
\end{figure}

The active time of the FD data acquisition is calculated using a
minimum bias data stream with a less restrictive trigger condition.
Since July 2007, the relevant information concerning the status of the
FD detector has been read out from the Observatory monitoring system.
An on-time dedicated database has been set up by storing the average
variances and the on-time fraction of individual telescopes in time
bins of 10 minutes. The information on the veto due to the operation
of the lidar or to an anomalous trigger rate on FD together with the
status of the CDAS needed to form a hybrid event are also recorded.
The method to calculate the on-time of the hybrid detector is
described in detail in~\cite{Abreu:2010aa}.

The accumulated on-time
is shown in figure~\ref{fig:on-time} (left), for the six telescopes at Coihueco
and for the three HEAT telescopes.
The average FD on-time (full circles) of individual telescopes since 1 January
2011 is shown in figure~\ref{fig:on-time} (right).  Requiring that the CDAS 
is active defines the hybrid on-time (empty circles).

The time evolution of the full hybrid duty cycle over 9 years of
operation is shown in figure~\ref{fig:metrics} (left), for all FD
sites.  Time bins are taken as the time intervals elapsed between two
subsequent FD data taking shifts.  The performance of the hybrid
detector is compared to the nominal DAQ time in the top panel of
figure~\ref{fig:metrics}.  In the right-hand panel, the FD on-time is
normalized to the time with high voltage ON, leading to an average FD
detector readiness of about 85\% for all telescopes.  The remaining
inefficiency can be ascribed to different factors such as bad weather
conditions (high wind load and/or rain) or high variances due to
bright stars/planets crossing the field of view of the FD.

It should be noted that the FD site of Los Morados became operational in
May 2005, Loma Amarilla starting from March 2007 and HEAT since September 2009.
After the initial phase due to the start up of the running operations, the mean
on-time is about 15\% for all of the FD sites.  Additionally, a seasonal
modulation is visible, since higher on-time fractions are observed in the
austral winter during which the nights are longer.

\subsection{Time stability of the hybrid detector response}

\begin{figure}[t]
\centering
\includegraphics[width=0.49\textwidth]{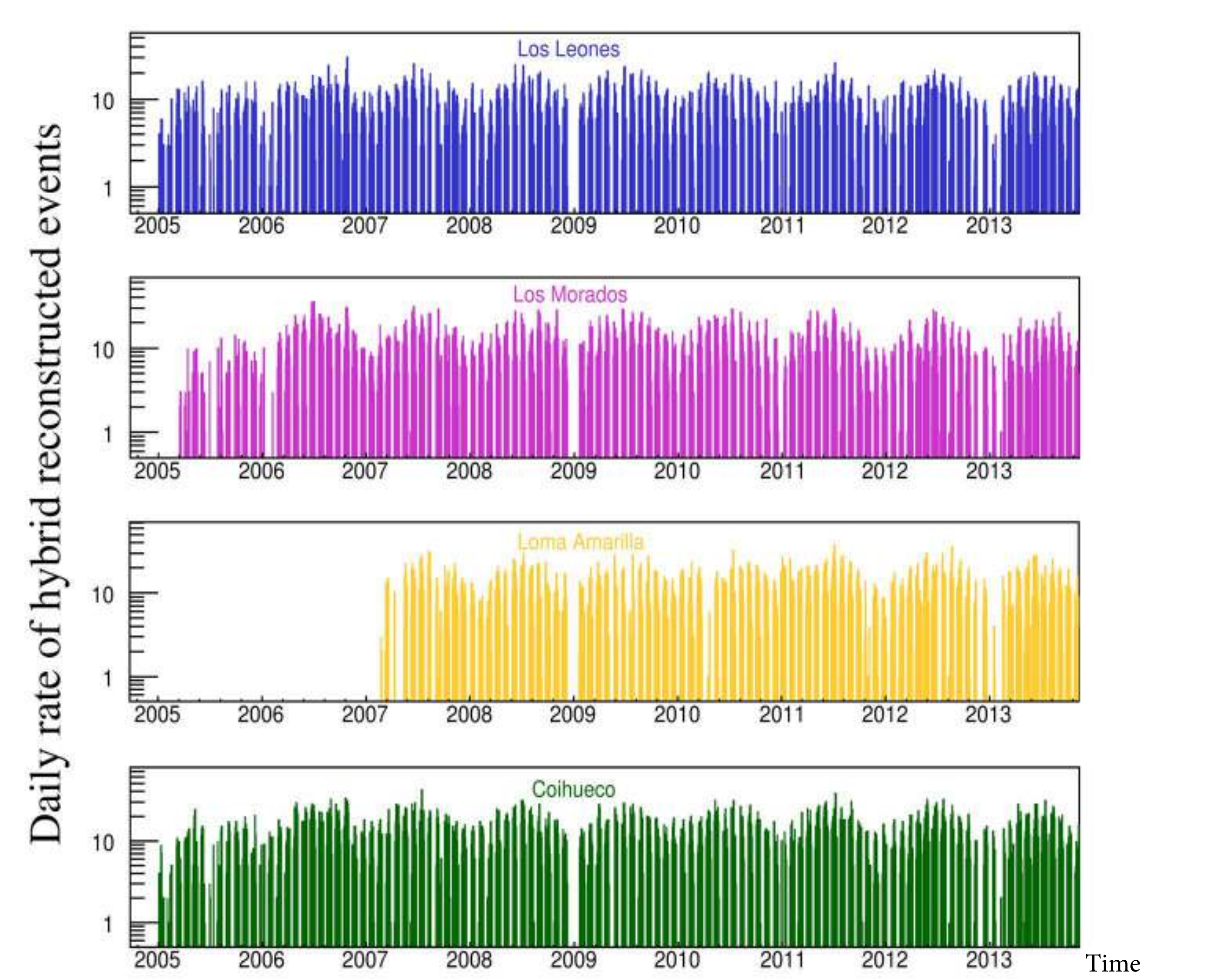}
\caption{Daily rate of hybrid reconstructed events as a function of year, starting in 2005, for (from top to bottom) Los Leones, Los Morados, Loma Amarilla and Coihueco, respectively.}
\label{fig:rate}
\end{figure}

The performance of the hybrid detector is demonstrated as a function
of time using a sample of events fulfilling basic reconstruction
requirements, such as a reliable geometrical reconstruction and
accurate longitudinal profile and energy measurement.  The daily rate
of well-reconstructed hybrid events observed by individual FD sites is
shown in figure~\ref{fig:rate} as a function of time, starting in
2005.

\begin{figure}[t]
\centering
\includegraphics[width=0.46\textwidth]{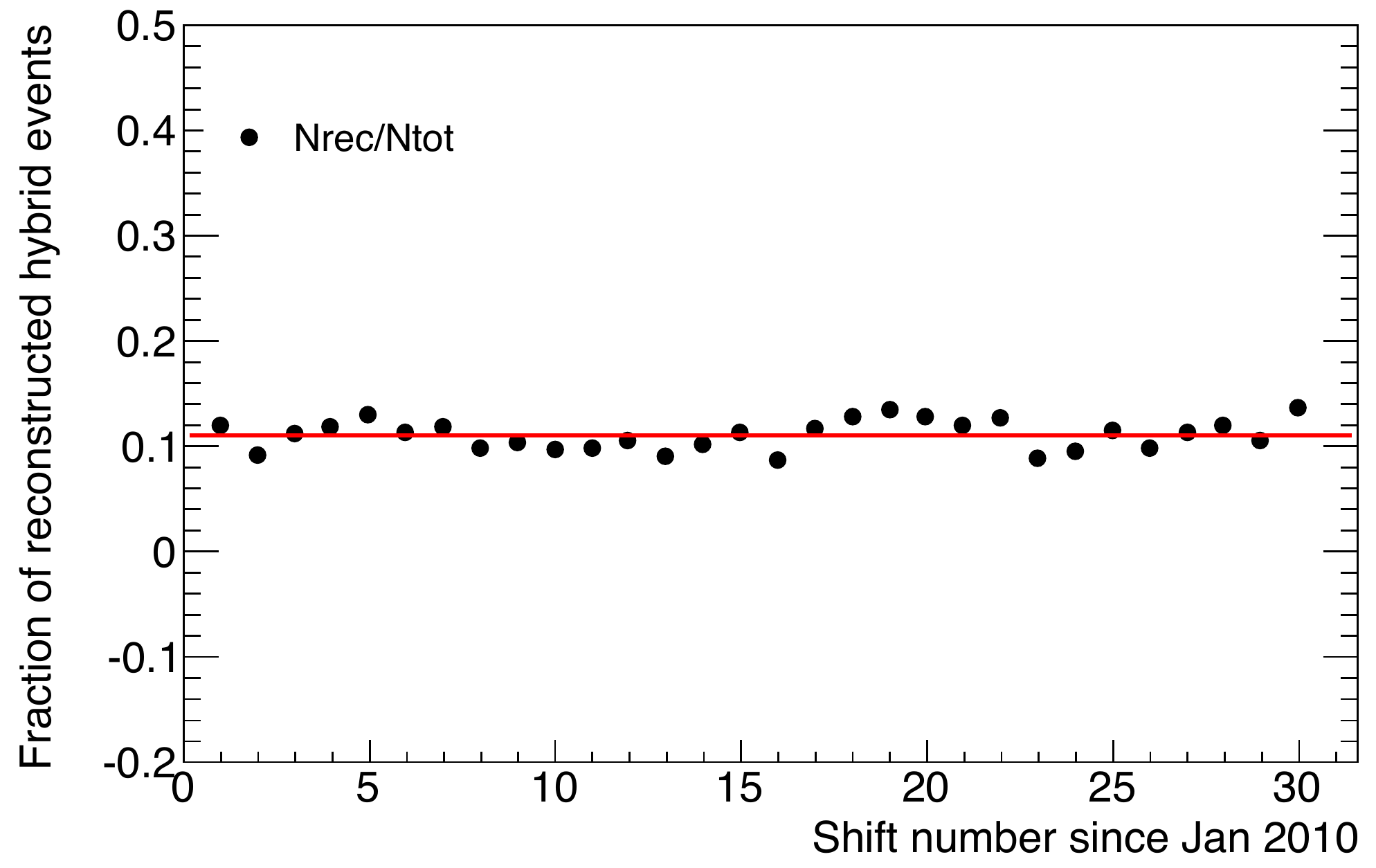}\hfill
\includegraphics[width=0.46\textwidth]{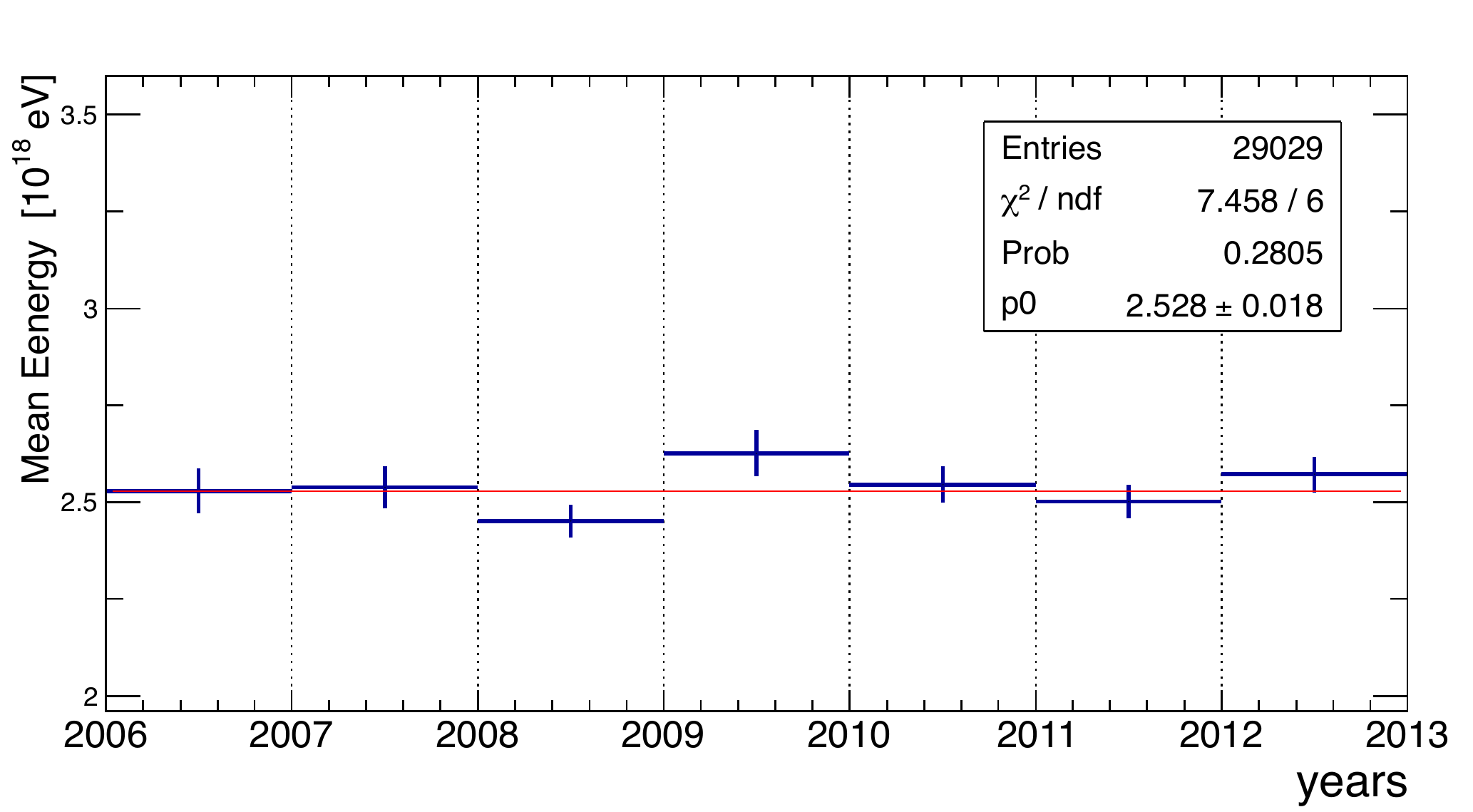}
\caption{\textit{Left:} fraction of all events that are well reconstructed hybrids since 2010.
\textit{Right:} mean energy for reconstructed hybrid events.}
\label{fig:effontime}
\end{figure}

An important benchmark for the time stability of the hybrid detector
response is the study of the effective on-time, defined as the
fraction of all events that are well reconstructed hybrids.  Its time
evolution, shown in figure~\ref{fig:effontime} (top), exhibits quite a
stable behavior over time.  Moreover the mean energy of the hybrid
events above $10^{18}$\,eV, with distance to the shower maximum
between 7 and 25\,km (corresponding to the 90\% of the entire hybrid
data sample), is shown as a function of time in
figure~\ref{fig:effontime} (bottom).  All these features demonstrate
the quality of the collected hybrid data and directly assess their
long term stability.

\section{Maintenance}

Currently more than 1660 surface detector stations are operational. Concerning
the water-Cherenkov detectors themselves, very few failures have been detected.  Only a few liners
were observed to leak shortly after installation.  In this case, which
constitutes the worst failure mode, the tank is emptied and brought back to the
Assembly Building for replacement of the interior components. Similarly, only a
few  solar panels have been damaged.  Solar power system
parameters are recorded and analyzed using the central data acquisition system.
The average battery lifetime is 4 years, and batteries are changed during
regular maintenance trips.

The PMTs and electronic boards are the most critical elements of the Surface
Detector stations. They are subject to very severe environmental conditions:
temperature variations, humidity, salinity and dust.  The failure rates of the
PMTs are about 20 per year (about 0.5\%). Some HV module and base problems
have been detected as well as some problems due to bad connections. All other
failures except those concerning the  PMTs (such as  broken photocathode)  can
be repaired on site.  It is currently estimated that the number of spare PMTs
is sufficient for about 10 to 15 more years of operation.  The failure rate of
electronic boards is about 1\% per year. Some of the problems are repaired
simply by reflashing the software.  Most of the electronic problems can also
be repaired on site. All the spare parts are stored on site.

The operation of the array is monitored on-line and alarms are set on various
parameters. The maintenance goal is to have no more that 20 detector stations
out of operation at any time. Currently the achieved number is less that 10
detector stations out of operation. It is currently estimated that the
long-term maintenance (including the battery change) requires about 3 field
trips per week. This maintenance rate is within the original expectations.  The
maintenance is organized by the Science Operation Coordinator and performed by
local technicians. The Surface Detector Array does not require a permanent presence
of physicists from other laboratories on site. However, remote shifts for the
data quality monitoring will be implemented.